\documentclass[11pt,a4paper,titlepage]{book}
\usepackage{amsfonts}
\usepackage{amsmath}
\usepackage{amssymb}
\usepackage{caption}
\usepackage{colortbl}
\usepackage[countmax]{subfloat}
\usepackage[english]{babel}
\usepackage{enumerate}
\usepackage{epsfig}
\usepackage{exscale}
\usepackage{fancyhdr}
\usepackage{graphicx}
\usepackage{indentfirst}
\usepackage{lscape}
\usepackage{microtype}
\usepackage{multirow}
\usepackage{natbib}
\usepackage{relsize}
\usepackage{rotating}
\usepackage{setspace}
\usepackage{subfig}
\usepackage{tikz}
\usepackage{txfonts}
\usepackage{ulem}
\usepackage{url}
\usepackage{wrapfig}

\usepackage[hyperindex]{hyperref}

\definecolor{viola2}{RGB}{187,51,255}
\definecolor{violac}{RGB}{158,50,249}
\definecolor{viola3}{RGB}{178,49,243}
\definecolor{lilla}{RGB}{230,204,255}
\definecolor{darkblue}{RGB}{0,0,153}
\definecolor{lightblue}{RGB}{204,228,255}
\definecolor{lightblue_1}{RGB}{51,149,255}
\definecolor{blue_2}{RGB}{102,153,255}
\definecolor{green2}{RGB}{53,173,120}
\definecolor{green3}{RGB}{141,235,175}
\definecolor{green4}{RGB}{43,214,171}
\definecolor{green5}{RGB}{0,153,153}
\definecolor{yellow2}{RGB}{253,187,63}
\definecolor{yellow3}{RGB}{255,241,153}
\definecolor{orange2}{RGB}{255,199,51}
\definecolor{lightgrey}{RGB}{195,195,195}
\definecolor{lightgrey_1}{RGB}{211,211,211}
\newcommand\blfootnote[1]{%
  \begingroup
  \renewcommand\thefootnote{}\footnote{#1}%
  \addtocounter{footnote}{-1}%
  \endgroup
}

\begin{document}
    \begin{titlepage}
        \pagenumbering{alph}  \thispagestyle{empty}
        \begin{center}
            \Huge Universit\`a degli Studi di Napoli \\
            Federico II \\
            \begin{figure}[!h] \begin{center}
                \includegraphics[width=5cm]{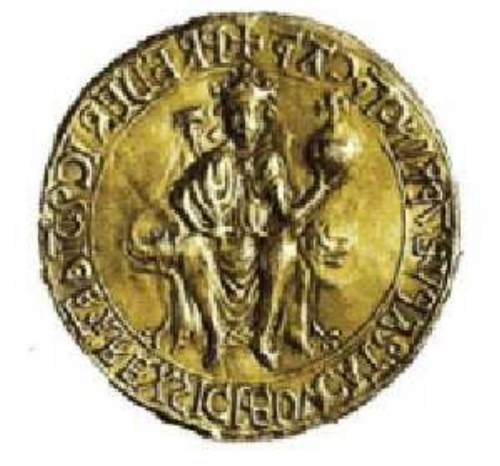}\end{center}
            \end{figure}
            \vskip 1cm
            \Huge Data-rich astronomy: mining synoptic sky surveys
            \\
            \vskip 1cm \normalsize Stefano Cavuoti
            \vskip 1cm \normalsize PhD Thesis XXV cycle \vskip 1cm
        \end{center}

\vskip 1cm

Supervisors:
\vskip 0.5cm
Ch.mo Prof. Giuseppe Longo

Dott. Massimo Brescia

        \vfill
        \begin{center}
            Academic Year 2012$/$2013
        \end{center}
    \end{titlepage}

    \newpage
    \thispagestyle{empty}
    $\phantom{fantasminopawa}$ \\
    \clearpage

\thispagestyle{empty}    
$\phantom{fantasminopawa}$ \\
\vskip 10cm

\noindent \Large \textit{Ph. D. Committee:}
\vskip 1.5cm
\noindent \normalsize Prof. George S. Djorgovski\\
\indent \textit{California Insitute of Technology}\\
\vskip 0.5cm
\noindent Prof. Fabio Pasian\\
\indent \textit{Istituto Nazionale di Astrofisica, Osservatorio Astronomico di Trieste}\\
\vskip 0.5cm
\noindent Prof. Dario Trevese\\
\indent \textit{Universit\`a di Roma ``La Sapienza"} \\
\vfill        
        \begin{center}
            May 29, 2013
        \end{center}

    \newpage
    \setcounter{page}{0} \thispagestyle{empty}
    $\phantom{fantasminopawa}$ \\
    \clearpage

    \pagenumbering {Roman}

    \tableofcontents
        \newpage   $\phantom{fantasminopawa}$ \\

    \chapter*{Preamble}\addcontentsline{toc}{chapter}{Preamble}  \markboth{\MakeUppercase{Preamble}}{} \pagenumbering {arabic}
        \hfill\begin{tabular}{@{}p{.6\linewidth}@{}}
\textit{``One of the greatest challenges for 21st-century science is how we respond to this new era of data-intensive science. This is recognized as a new paradigm beyond experimental and theoretical research and computer simulations of natural phenomena - one that requires new tools, techniques, and ways of working."}\\ \href{http://research.microsoft.com/en-us/collaboration/fourthparadigm/}{Douglas Kell, University of Manchester}.\\ \phantom{aaa}
\end{tabular}

In the last decade a new generation of telescopes and sensors has allowed the production of a very large amount of data and astronomy has become, a data-rich science; this transition is often labeled as: ``data revolution" and ``data tsunami". The first locution puts emphasis on the expectations of the astronomers while the second stresses, instead, the dramatic problem arising from this large amount of data: which is no longer computable with traditional approaches to data storage, data reduction and data analysis.
In a new, age new instruments are necessary, as it happened in the Bronze age when mankind left the old instruments made out of stone to adopt the new, better ones made with bronze. Everything changed, even the social structure. In a similar way, this new age of Astronomy calls for a new generation of tools and, for a new methodological approach to many problems, and for the acquisition of new skills. The attems to find a solution to this problems falls under the umbrella of a new discipline which originated by the intersection of astronomy, statistics and computer science: \textit{Astroinformatics}, \citep{borne2009,djorgovski2006}.

The various topics which I addressed during Ph. D.  fall exactly in this intersection and explore some new possibilities offered by this new discipline.

The present work is structured as follow: in Chapter \ref{chap:introduction}, I present the scientific and methodological background of my work, in Chapter \ref{chap:dm}, I give a short overview about data mining and I present three different methods involved in this work; in Chapter \ref{chap:scientificgateway} I present the two scientific gateways that I contributed to implement; in section  \ref{chap:dame}, I give an overview about the DAME infrastructure; while in section \ref{sec:STRADIWA}, I describe the STRADIWA project. In chapters: \ref{chap:GAME}, \ref{chap:QNA} and \ref{chap:SVM}, I present three different methods involved in this work.
Chapter {part:science}  presents three application to different classification problems: the Globular Cluster classification (sec. \ref{chap:GC}) while  section \ref{chap:comparison} contains a comparison to different extraction software and in section \ref{chap:agn}, I present an attempt to automatically disentangle different types of AGN;  and in chapter \ref{chap:photoz}, I show two applications to the estimation of photometric redshift.
In chapter \ref{chap:transients} I show the very preliminary results of our work on transients detection.

Conclusions will be presented in chapter \ref{chap:conclusionstech}.

The structure of the thesis reflects the fact that it has been largely extracted from the following papers which I completed during my PhD.
\begin{enumerate}
    \item  \textbf{Cavuoti, S.}; Brescia, M.; D'Abrusco, R.; Longo, G.; Photometric AGN Classification in the SDSS with Machine Learning Methods \textbf{to be Submitted to MNRAS}
    \item Brescia, M.; \textbf{Cavuoti, S.}; Garofalo, M.; Guglielmo, M.; Longo, G.; Nocella, A.; Riccardi, S.; Vellucci, C.;  Djorgovski, G.S.; Donalek, C.; Mahabal, A. Data Mining in Astronomy with DAME. \textbf{to be Submitted to PASP}

    \item Brescia, M.; \textbf{Cavuoti, S.}; D'Abrusco, R.; Longo, G.; Mercurio, A.; 2013, Photo-z prediction on WISE - GALEX - UKIDSS - SDSS Quasar Catalogue, based on the MLPQNA model, \textbf{Submitted to ApJ}

    \item \textbf{Cavuoti, S.}; Garofalo, M.; Brescia , M.; Paolillo,  M.;  Pescape', A.; Longo, G.; Ventre, G.; GPUs for astrophysical data mining. A test on the search for candidate globular clusters in external galaxies, \textbf{Submitted to New Astronomy, accepted}
    \item Annunziatella, M.; Mercurio, A.; Brescia, M.; \textbf{Cavuoti, S.}; Longo, G.; 2013, Inside catalogs: a comparison of source extraction software, \textbf{PASP, 125, 68}
    \item \textbf{Cavuoti, S.}; Brescia, M.; Longo, G.; Mercurio, A.; 2012, Photometric Redshifts with Quasi Newton Algorithm (MLPQNA). Results in the PHAT1 Contest, \textbf{A\&A, Vol. 546, A13, pp. 1-8}
    \item Brescia, M.; \textbf{Cavuoti, S.}; Paolillo, M.; Longo, G.; Puzia, T.; 2012, The detection of Globular Clusters in galaxies as a data mining problem, \textbf{MNRAS, Volume 421, Issue 2, pp. 1155-1165}, available at \href{http://arxiv.org/abs/1110.2144v1}{arXiv:1110.2144v1}.
    \item \textbf{Cavuoti, S.}; Brescia, M.; Longo, G., 2012, Data mining and Knowledge Discovery Resources for Astronomy in the Web 2.0 Age, Proceedings of SPIE Astronomical Telescopes and Instrumentation 2012, Software and Cyberinfrastructure for Astronomy II, Ed.(s): N. M. Radziwill and G. Chiozzi, Volume 8451, RAI Amsterdam, Netherlands, July 1-4 \textbf{refereed proceeding}
    \item \textbf{Cavuoti, S.}; Garofalo, M.; Brescia, M.; Pescape', A.; Longo, G.; Ventre, G., 2012, Genetic Algorithm Modeling with GPU Parallel Computing Technology, 22nd WIRN, Italian Workshop on Neural Networks, Vietri sul Mare, Salerno, Italy, May 17-19 \textbf{refereed proceeding}
    \item \textbf{Cavuoti, S.}; Brescia, M.; Longo, G.; Garofalo, M.; Nocella, A.; 2012, DAME: A Web Oriented Infrastructure for Scientific Data Mining and Exploration, Science - Image in Action. Edited by Bertrand Zavidovique (Universite' Paris-Sud XI, France) and Giosue' Lo Bosco (University of Palermo, Italy) . Published by World Scientific Publishing Co. Pte. Ltd., 2012. ISBN 9789814383295, pp. 241-247
    \item Djorgovski, S. G.; Longo, G., Brescia, M., Donalek, C., \textbf{Cavuoti, S.}, Paolillo, M., D'Abrusco, R., Laurino, O., Mahabal, A., Graham, M., ``DAta Mining and Exploration (DAME): New Tools for Knowledge Discovery in Astronomy". American Astronomical Society, AAS Meeting \#219, \#145.12, Tucson, USA, January 08-12
    \item Brescia M., \textbf{Cavuoti, S.}, Djorgovski, G.S., ,Donalek, C., Longo, G., Paolillo, M., 2011, Extracting knowledge from massive astronomical data sets, \href{http://arxiv.org/abs/1109.2840}{arXiv:1109.2840}, Springer Series in Astrostatistics, Volume 2, Springer Media New York, ISBN 978-1-4614-3322-4  15 pages \textbf{[invited review]}.
    \item Brescia, M.; \textbf{Cavuoti, S.}; D'Abrusco, R.; Laurino, O.; Longo, G.; 2010, DAME: A Distributed Data Mining \& Exploration Framework within the Virtual Observatory, INGRID 2010 Workshop on Instrumenting the GRID, Poznan, Poland, in Remote Instrumentation for eScience and Related Aspects, F. Davoli et al. (eds.), Springer Science+Business Media, LLC 2011, DOI 10.1007/978-1-4614-0508-5\_17
    \item Djorgovski, S. G.; Longo, G., Brescia, M., Donalek, C., \textbf{Cavuoti, S.}, Paolillo, M., D'Abrusco, R., Laurino, O., Mahabal, A., Graham, M., 2012, DAta Mining and Exploration (DAME): New Tools for Knowledge Discovery in Astronomy. American Astronomical Society, AAS Meeting \#219, \#145.12, Tucson, USA, January 08-12
    \item Brescia, M.; Longo, G.; Castellani, M.; \textbf{Cavuoti, S.}; D'Abrusco, R.; Laurino, O., 2012, DAME: A Distributed Web Based Framework for Knowledge Discovery in Databases, 54th SAIT Conference, Astronomical Observatory of Capodimonte, Napoli, Italy, May 6, Mem. S.A.It. Suppl. Vol. 19, 324
\end{enumerate}
Whereas possible I tried to avoid repetitions but, being thesis largely assembled from the above papers some might have escaped my attention.
During my PhD work I produced also the following technical documents:

\begin{itemize}
\item Brescia, M.; Annunziatella, M.; 	\textbf{Cavuoti, S.};  Longo, G.;  Mercurio, A.; STraDiWA Project Sky Transient Discovery Web Application SOFTWARE Documentation DAME-DOC-NA-0003-Rel1.0
\item 	\textbf{Cavuoti, S.}; Riccardi, S.;  Guglielmo M.; DAMEWARE Installation and Deployment Developer Manual DAME-MAN-NA-0019-Rel1.0
\item Fiore, M.;  	\textbf{Cavuoti, S.}; Data Mining Plugin User/Administration Manual VONEURAL-MAN-NA-0005-Rel1.6
\item Fiore, M.;  	\textbf{Cavuoti, S.}; Data Mining Plugin Wizard User Manual VONEURAL-MAN-NA-0004-Rel1.3
\item 	\textbf{Cavuoti, S.}; Mercurio, A.; Annunziatella, M.; Brescia, M.; Variable Sky Objects Simulation and Detection Workflow Simulation Package  Procedure  DAME-PRO-NA-0010Rel2.0
\item Brescia, M.; 	\textbf{Cavuoti, S.}; Garofalo, M.; Nocella, A.;  Riccardi S.; DAME Web Application REsource Design Summary DAMEWARE-SDD-NA-0018-Rel1.0
\item 	\textbf{Cavuoti, S.}; Di Guido, A.; Data Mining Suite 2.0 Software Design Description IEEE 1016 Component Data Mining Model VONEURAL-SDD-NA-0008-Rel2.0
\item Brescia, M.; Annunziatella, M.; 	\textbf{Cavuoti, S.};  Longo, G.;  Mercurio, A.; STraDiWA Sky Transient Discovery Web Application  Description of the Workflow SOFTWARE Specifications DAME-SPE-NA-0011-Rel1.0
\item Di Guido, A.;  Fiore, M.;  	\textbf{Cavuoti, S.};  Brescia M.; DMPlugin Description Report  Beta release of Web Application Data Mining Model Technical Report DAME-TRE-NA-0016-Rel1.0
\item Brescia, M.; 	\textbf{Cavuoti, S.}; DAMEWARE Web Application REsource Internal Test Report  DAME-TRE-NA-0019Rel1.0
\item Brescia, M.; 	\textbf{Cavuoti, S.};  Photo-z prediction on PHAT1 Catalogue, based on MLPQNA regression model DAMEWARE-VER-NA-0008-Rel1.0
\end{itemize}

    \chapter{Introduction}\label{chap:introduction}
        \hfill\begin{tabular}{@{}p{.6\linewidth}@{}}
\textit{``... while data doubles every year, useful information seems to be decreasing, creating a growing gap between the generation of data and our understanding of it..."}\\ \cite{dunham2002}.\\ \phantom{aaa}
\end{tabular}

\blfootnote{this chapter is largely extracted from: \tiny
\begin{itemize}
\item  \textbf{Cavuoti, S.}; Brescia, M.; Longo, G., 2012, Data mining and Knowledge Discovery Resources for Astronomy in the Web 2.0 Age, Proceedings of SPIE Astronomical Telescopes and Instrumentation 2012, Software and Cyberinfrastructure for Astronomy II, Ed.(s): N. M. Radziwill and G. Chiozzi, Volume 8451, RAI Amsterdam, Netherlands, July 1-4 \textbf{refereed proceeding}
\item Brescia, M.; \textbf{Cavuoti, S.}; Garofalo, M.; Guglielmo, M.; Longo, G.; Nocella, A.; Riccardi, S.; Vellucci, C.; Djorgovski, G.S.; Donalek, C.; Mahabal, A. Data Mining in Astronomy with DAME. \textbf{to be Submitted to PASP}
\item  \textbf{Cavuoti, S.}; Brescia, M.; Longo, G.; Garofalo, M.; Nocella, A.; 2012, DAME: A Web Oriented Infrastructure for Scientific Data Mining and Exploration, Science - Image in Action. Edited by Bertrand Zavidovique (Universite' Paris-Sud XI, France) and Giosue' Lo Bosco (University of Palermo, Italy) . Published by World Scientific Publishing Co. Pte. Ltd., 2012. ISBN 9789814383295, pp. 241-247
\item  Djorgovski, S. G.; Longo, G., Brescia, M., Donalek, C., \textbf{Cavuoti, S.}, Paolillo, M., D'Abrusco, R., Laurino, O., Mahabal, A., Graham, M., ``DAta Mining and Exploration (DAME): New Tools for Knowledge Discovery in Astronomy". American Astronomical Society, AAS Meeting \#219, \#145.12, Tucson, USA, January 08-12
\item Brescia M., \textbf{Cavuoti, S.}, Djorgovski, G.S., ,Donalek, C., Longo, G., Paolillo, M., 2011, Extracting knowledge from massive astronomical data sets, \href{http://arxiv.org/abs/1109.2840}{arXiv:1109.2840}, Springer Series in Astrostatistics, Volume 2, Springer Media New York, ISBN 978-1-4614-3322-4 15 pages \textbf{[invited review]}.
\item Brescia, M.; \textbf{Cavuoti, S.}; D'Abrusco, R.; Laurino, O.; Longo, G.; 2010, DAME: A Distributed Data Mining \& Exploration Framework within the Virtual Observatory, INGRID 2010 Workshop on Instrumenting the GRID, Poznan, Poland, in Remote Instrumentation for eScience and Related Aspects, F. Davoli et al. (eds.), Springer Science+Business Media, LLC 2011, DOI 10.1007/978-1-4614-0508-5\_17
\item Djorgovski, S. G.; Longo, G., Brescia, M., Donalek, C., \textbf{Cavuoti, S.}, Paolillo, M., D'Abrusco, R., Laurino, O., Mahabal, A., Graham, M., 2012, DAta Mining and Exploration (DAME): New Tools for Knowledge Discovery in Astronomy. American Astronomical Society, AAS Meeting \#219, \#145.12, Tucson, USA, January 08-12
\item Brescia, M.; Longo, G.; Castellani, M.; \textbf{Cavuoti, S.}; D'Abrusco, R.; Laurino, O., 2012, DAME: A Distributed Web Based Framework for Knowledge Discovery in Databases, 54th SAIT Conference, Astronomical Observatory of Capodimonte, Napoli, Italy, May 6, Mem. S.A.It. Suppl. Vol. 19, 324
\end{itemize}}
As it was already mentioned in the preamble, my thesis spans a quite variegate spectrum of topics: from algorithms to information and communication technologies (ICT), to observational astronomy and cosmology;
the main drivers being the interest in cosmology and the need to cope with the methodological revolution that is currently taking place in astronomy.

\begin{figure*}
\centering
    \includegraphics[width=12.cm]{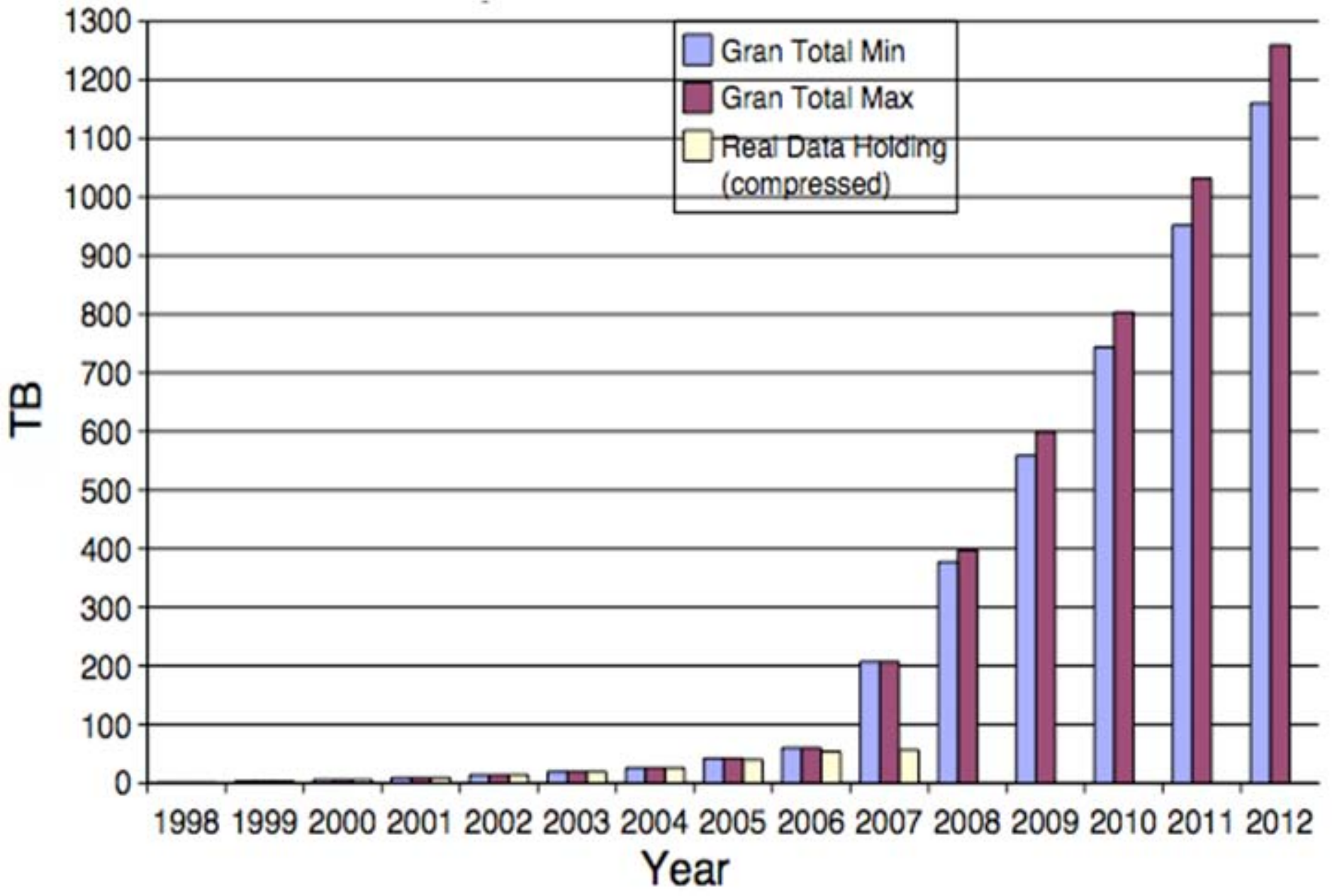}
\caption[The Data Growth.]{The Data Gap: Data growth in the ESO case, credit of ESO.}\label{INTRO:figESO}
\end{figure*}

\begin{figure*}
\centering
    \includegraphics[width=12.cm]{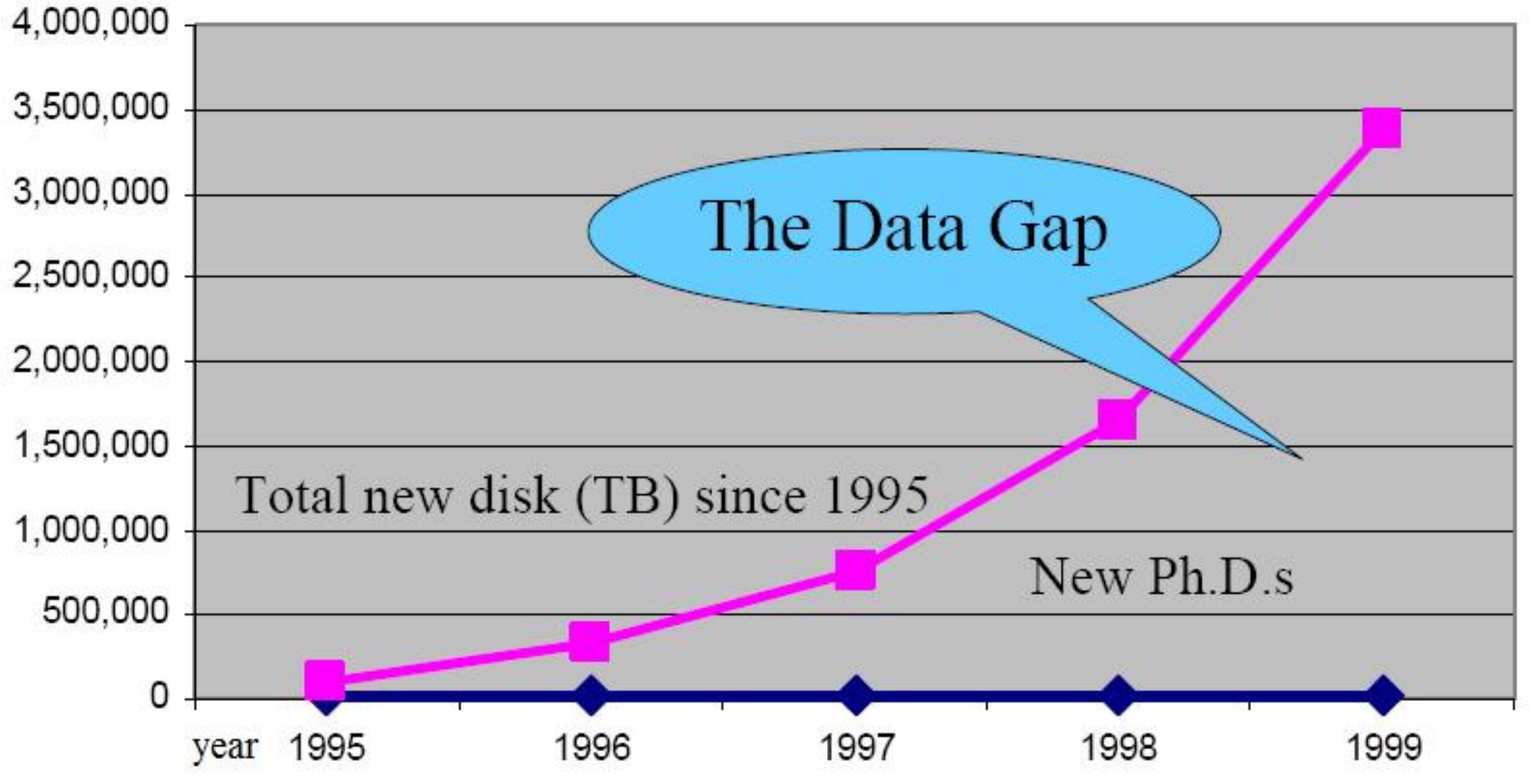}
\caption[The Data Gap: Data growth against the number of analyst]{The Data Gap: Data growth against the number of analyst, from \cite{grossman2001}. In spite of the large increase in data, the number of FTE (Full Time Equivalent)  involved in data analysis remains in practice constant.}\label{INTRO:fig0}
\end{figure*}

Astronomical data originate from sensors and telescopes operating in some wavelength regime, either from the ground or from the space. These data come in one or more of the following forms: images,
spectra, time series, or data cubes \citep{brunner2001b, djorgovski2012d}.
Data typically represent signal intensity as a function of the position on the sky, wavelength or
energy, and time.  The bulk of the data are obtained in the form of images (in radio astronomy, as interferometer fringes, but those are also converted
into images).
The sensor output is then processed by the appropriate custom pipelines, that remove instrumental signatures and perform calibrations.
In most cases, the initial data processing and analysis phase segments the images into catalogs of detected discrete sources (e.g., stars, galaxies, etc.), and their
measurable attributes, such as their position on the sky, flux intensities in different apertures, morphological descriptors of the light distribution, ratios of
fluxes at different wavelengths (colors), and so on.
These first order data products are then stored in local (instrument or mission based) or national archives hosting raw and processed sensor data, and the initial derived data products such as source catalogs with their measured attributes, which are provided through dedicated archives and are accessible online.\\
Since almost thirty years scientific analysis proceeds from such first-order data products and, in this respect, not much would be changed if it were not for the data size,  data quality and data complexity. The trend in figure \ref{INTRO:figESO} shows how much a typical astronomical archive has increases in size over the last thirty years. Such exponential growth is not matched by an equivalent increase in the number of data analyst (figure \ref{INTRO:fig0}) and already now data analysis requirements  have largely exceeded the power of dedicated human brains, thus pushing  astronomy into the rather exclusive club of \textit{data intensive sciences}.\\
Even more complex appears to be the near future where other challenges are waiting. Think for instance of the Large Synoptic Survey Telescope (LSST) which will likely become operational in 2016 and will produce a data flow of ca. 16 TB per observing night, or many PB/year \citep{ivezic2011}, and the EUCLID space mission, foreseen to be operational in 2019, where a complete data release is estimated to have a size of  more than 13 PB of data, mixed between catalogues and images \citep{brescia2011c}... but we shall come back to this later on.

Nowadays existing national and mission based archives have been federated into the Virtual Observatory and more and more large datasets keep being added to it every day. By incorporating the subtleties of data modeling and understanding provided by the domain experts into complex data models, these modern archives allow everyone to pursue scientific projects of unprecedent complexity (multiwavelenght, multiepoch, ...) and size (millions of objects rather than the few dozens was in the past) creating a potential cornucopia for discoveries.\\

Computer assisted decision making, statistical pattern recognition, data mining, machine learning, web 2.0 technologies... are just a few of the many new words and locutions with which the present and future generations of astronomers need to become acquainted.

This change must not be underestimated since it affects deeply not only the every day praxis of scientific research, but
also the underlaying methodology and the type of science which is enabled.
In 2009 Tony Hey analysed the problem of data rich sciences in a seminal book named The Fourth Paradigm \citep{hey2009} and he explained
why data analysis needs to be considered the fourth independent methodological pillar of modern science after experiment, theory and simulations.\\
When the amount of data exceeds the human capability to see, evaluate and understand each data point, scientists need
to rely more and more upon automated machine driven procedures capable to isolate significant from redundant features, to identify correlations and patterns of high dimensionality, to identify rare or peculiar behaviours.\\ 
This, however, is only a part of the story, since also astrophysical and cosmological understanding come, in fact, more and more from complex numerical simulations producing results in the form of multiTera or Peta byte data sets.
The falsification of these theories calls for the comparison of these huge simulated datasets  with the even larger data sets from observations, and for the compression of the resulting datasets to a size and a level of complexity understandable for a human brain.

In other words, astronomers are, as it has never happened before tied to ICT, and the new generations will need to know more and more about data fusion and federation, about virtual working environments, about web 2.0 technologies, machine learning and data mining, advanced visualization, etc...
a large set of disciplines which since a few years falls under the umbrella of the emerging field of Astroinformatics \citep{borne2009}: a new discipline placed at the crossroad between traditional astronomy, applied mathematics, computer science and ICT technologies.

 As it always happens, innovations do not come without a price and this thesis is about this price, about the need to change the traditional and comfortable environment of old fashioned astronomy, to enter a new, un-familiar era based on a more extreme exploitation the possibilities enabled by ICT. \\

\section{From Data-Archives to Users}

Already in the late nineties the astronomical community realized the complexity of the problems they were facing, and began to think about a ``Virtual Observatory", (VO). The concept was strongly endorsed by the influential NSF ``decadal" report \citep{mckee2001} and further explored in a seminal meeting held the same year in Pasadena \citep{brunner2001a}. VO was imagined as a complete, distributed (Web-based) research environment for astronomy, with large and complex data sets to be implemented by federating
geographically distributed data and computing infrastructures, and the necessary tools and expertise for their use ~\citep{brunner2001a,djorgovski2002a}.
VO was also supposed to facilitate the transition from the old data poverty regime, to the overwhelming data abundance, and to be a
mechanism by which the progress in ICT could  easily be used to solve scientific challenges.  The concept immediatly lead to a number of national and international VO organizations, now federated through the International Virtual Observatory Alliance (IVOA; \url{http://ivoa.net}). In Italy, the VO is currently embodied as Italian Virtual Observatory (\url{http://vobs.astro.it/}) while in Europe and US the VO initatives are under the umbrella of Euro-VO (\url{http://euro-vo.org}) and Virtual Astronomical Observatory (VAO; \url{http://usvao.org}) respectively.

\begin{figure*}
\centering
    \includegraphics[width=12.cm]{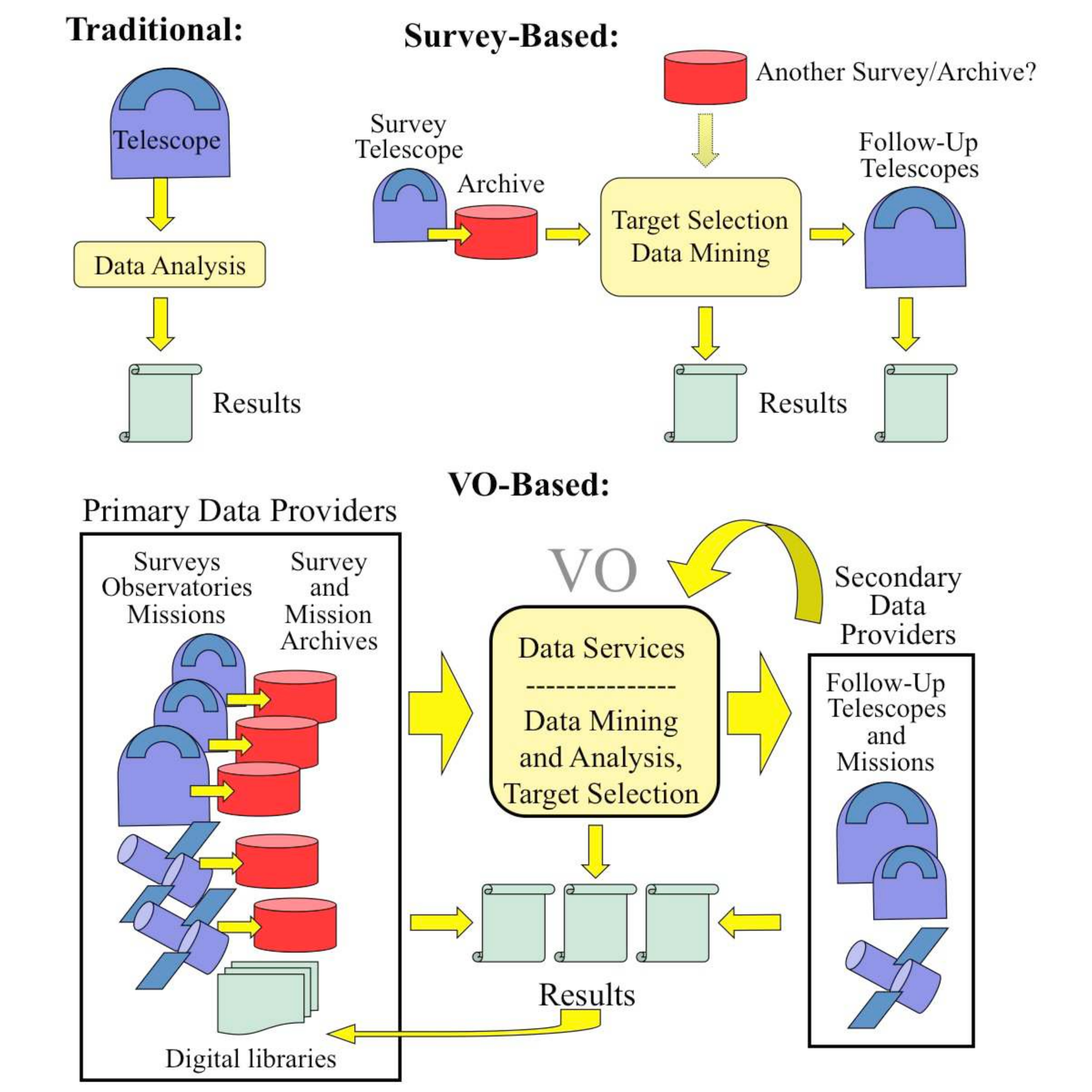}
\caption[The evolving modes of observational astronomy.]{The evolving modes of observational astronomy.
{\em Top left:}  In the traditional approach, targeted observations from a single telescope (sensor), sometimes combined with other data, are used to
derive science.  This mode is typical of Megabyte to Gigabyte-scale data sets.
{\em Top right:}  In the survey mode, data from a given survey are stored in an archive, and may be used to produce science on its own.
Sometimes, they may be matched to another survey.  Selection of interesting targets using data mining can then lead to new targeted observations,
and new results.  This mode is characterized by Terabyte scale data sets.
{\em Bottom:}  In the VO mode, a large variety of surveys, space missions, and ground-based observatory archives are federated in the VO framework.
Data fusion can lead to new science, or can be used to select targets for follow-up observations, that themselves contribute to the evolving data grid.
This mode is characteristic of Terabyte to Petabyte-scale data sets.
A new generation of synoptic sky surveys imposes a requirement that the data-to-research cycle happens in a real time.  In practice, all three modes
continue to coexist (courtesy of G. S. Djorgovski).}\label{INTRO:fig1}
\end{figure*}

In other words, one can regard the VO as a meta-infrastructure gathering of heterogeneous data streams from a global network of telescopes and space missions, enabling data access and
federation, and making such value-added data sets available for a further analysis, as it is schematically illustrated in Fig.\ref{INTRO:fig1}. \\
The implementation of the VO framework over the past decade was focused on the production of the necessary data infrastructure, interoperability,
standards, protocols, middleware, data discovery services, and even a few useful data federation and analysis services, that we describe below;
see \citep{djorgovski2005,hanisch2007,graham2007}, for quick summaries and examples of practical tools and services implemented under the VO umbrella.

While much still remains to be done, data discovery and access in astronomy have
never been easier, and the established infrastructure can at least in principle expand and scale up to the next generation of sky surveys, space missions, etc.


%

Even before the VO astronomers had already done very successful attempts toward standardization see, for instance the fact that they adopted early universal standards for data exchange, such as the Flexible Image Transport System (FITS; \citealt{wells1981}). \\

Within the VO, a common set of data access protocols ensures that the same interface is employed across all data archives, no matter where they are located, to perform the
same type of data query (see Table \ref{INTRO:TAB:1}  for a summary of those defined).

\begin{table*}
\centering
\small
\setlength{\tabcolsep}{2pt}
\begin{tabular}{|c|c|}
\hline
\hline
{\bf        Name}                     &  {\bf    Description} \\
\hline
Simple Cone Search (SCS)     & Retrieve all objects within a circular region on the sky\\
Simple Image Access (SIA)    & Retrieve all images of objects within a region on the sky\\
Simple Spectral Access (SSA)& Retrieve all spectra of objects with a region on the sky\\
Simple Line Access (SLA)      & Retrieve spectral line data\\
Simulations (SIMDAL)           & Retrieve simulation data\\
Table Access (TAP)               & Retrieve tabular data\\
\hline
\end{tabular}
\caption{Different types of data access protocol defined by the IVOA.}\label{INTRO:TAB:1}
\end{table*}

Although common data formats may be employed in transferring data, e.g., VOTable for tabular data, individual data providers usually represent
and store their data and metadata in their own way.
Common data models define the shared elements across data and metadata collections and provide a framework for describing relationships between
them so that different representations can interoperate in a transparent manner.
Most of the data access protocols have an associated data model, e.g., the Spectral data model defines a generalized model for spectrophotometric
sequences and provides a basis for a set of specific case models, such as Spectrum, SED and TimeSeries.
There are also more general data models for spatial and temporal metadata, physical units, observations and their provenance, and characterizing
how a data set occupies multidimensional physical space.

When individual measurements of arbitrarily named quantities are reported, either as a group of parameters or in a table, their broader context within a standard data model can be established through the IVOA Utypes mechanism.

These strings act as reference pointers to individual elements within a data model thus identifying the concept that the reported value represents, e.g., the UType ``Data.FluxAxis.Accuracy. StatErrHigh" identifies a quantity as the upper error bound on a flux value defined in the Spectral data model.
Namespaces allow quantities/concepts defined in one data model to be reused in another one.

Data models can only go so far in tackling the heterogeneity of data sources; they provide a way to identify and refer to common elements but not to
describe how these are defined or related to each other. Concept schemes, from controlled vocabularies to thesauri to ontologies, specify in increasing
levels of detail the domain knowledge that is ultimately behind the data models. It then becomes possible, for example, to automatically construct a set
of extragalactic sources with consistent distances, even if each initially has it specified in a different way; the Tully-Fisher relationship can be used
with those with HI line widths whereas surface brightness and velocity dispersion can be used for elliptical galaxies.

Working with large amounts of data also requires certain infrastructure components:

The VO provides a  lightweight common interface ``VOSpace" to the host of data storage solutions that are available, ranging in scale from a local
filesystem on a laptop to a data farm in the cloud. It does not define how data is stored or transferred, only the control messages to gain access to
data and manage data flows, such as online analysis of large distributed data sets. VOSpace can also be used to support data caches for temporary
interprocess results, such as checkpoints, and as staging areas for both initial data and final results, where permanent storage is not required.

The Universal Worker Service (UWS) defines a design pattern for asynchronous services and a security framework supports authentication and credential
delegation to allow a chain of secure services all working together, e.g., in a workflow with proprietary data. Finally, the IVOA provides a Registry
tool where descriptions of available data archives and services can be found, e.g., catalogs of white dwarfs or photometric redshift services.

%
The key to further progress  is the availability of data exploration and analysis tools capable to operate on the Terascale data sets and beyond.
Progress in this arena is being made mainly by individual research groups in universities, or associated with particular observatories and surveys... and this is where Astroinformatics comes into the game.\\

The idea behind Astroinformatics is, in fact, that of providing an informal, open environment for the exchange of ideas, software, etc., and to act as a ``connecting tissue"
between the researchers working in this general arena.  The motivation is to engage a broader community of researchers, both as contributors and as
consumers of the new methodology for data-intensive astronomy, thus building on the data-grid foundations established by the VO framework.
The field is still young, and a good
introduction to it are the talks and discussions at the series of the international Astroinformatics conferences, starting with \url{http://astroinformatics2010.org} and continuing through the 2012 edition.

\section[Beyond the VO]{Beyond the VO into the exascale regime}
%
Over the past several years, improvements in detector technology enabled a new generation of synoptic sky surveys that cover large swaths
of the sky repeatedly; they represent a panoramic cosmic cinematography.  Some recent and current examples include:
\begin{itemize}
\item Palomar-Quest (PQ; \citealt{mahabal2005,djorgovski2008},
\item Catalina Real-Time Transient Survey (CRTS; \citealt{drake2009,mahabal2011,djorgovski2012a};
\url{http://crts.caltech.edu}),
\item Palomar Transient Factory (PTF, \citealt{rau2009,law2009}, \url{http://www.astro.caltech.edu/ptf/}),
\item PanSTARRS (\citealt{kaiser2004}; \url{http://pan-starrs.ifa.hawaii.edu/}),
\item VST VOICE \\(\url{http://people.na.infn.it/~covone/vst_project/VOICE_letter_of_intent.pdf}),
\item VST KIDS (\url{http://www.astro-wise.org/projects/KIDS/}),
\item VISTA VIKING (\url{http://www.astro-wise.org/projects/VIKING/}).
\end{itemize}
The Large Synoptic Sky Survey (LSST; \citealt{tyson2002,ivezic2009}; \url{http://lsst.org}) will be an 8.4-m wide-field telescope that will be located
at Cerro Paranal in Chile. It will take more than 800 panoramic images each night, with 2 exposures per field, covering the accessible sky twice
each week.  The data (images, catalogs, alerts) will be continuously generated and updated every observing night.  In addition, calibration and
co-added images, and the resulting catalogs, will be generated on a slower cadence, and used for data quality assessments. The final object catalog
is expected to have more than 40 billion rows, comprising 30 TB of data per night, for a total of 60 PB over the envisioned duration of the survey.
Its scientific goals and strategies are described in detail in the LSST Science Book \citep{ivezic2009}.
Processing and analysis of this huge data stream poses a number of challenges in the arena of real-time data processing, distribution, archiving, and analysis.

The Square Kilometer Array (SKA; \url{http://skatelescope.org}) will be the world's largest radio telescope, hoped to be operational in the mid-2020's,
divided into two facilities, one in Australia, and one in South Africa.  It will consist of thousands of individual radio dishes, with a total
collecting area of $\sim 1 \ km^2$,  with a continuous frequency coverage from 70 MHz to 30 GHz.  The data processing for the SKA poses significant challenges,
even if we extrapolate Moore's law to its projected operations.  The data will stream from the detectors into the correlators at a rate of $\sim 4.2 \  PB/s$,
and then from the correlators to the visibility processors at rates between 1 and 500 TB/s, depending on the observing mode, which will require processing
capabilities of $\sim 200 \  Pflops$ to $\sim 2.5 \ Eflops$. Subsequent image formation needs $\sim 10 \ Pflops$ to create data products ($\sim 0.5 - 10 \ PB/day$),
which would be available for science analysis and archiving, the total computational costs of which could easily exceed those of the pipeline.  Of course,
this is not just a matter of hardware provision, even if it is specially purpose built, but also high computational complexity algorithms for wide field
imaging techniques, deconvolution, Bayesian source finding, and other tasks. Each operation will also place different constraints on the computational
infrastructure, with some being memory bound and some CPU bound that will need to be optimally balanced for maximum throughput.  Finally the power required
for all this processing will also need to be addressed – assuming the current trends, the SKA data processing will consume energy at a rate of $\sim 1 \ GW$.
These are highly non-trivial hardware and infrastructure challenges.



With the new scientific opportunities come new challenges.  Processing and analysis of these massive data streams inherits all of the same
challenges pertaining to the single-pass surveys, described above, but with larger data volumes, and with new ones brought by the time axis,
that describe the variations in brightness as pointed out by \citealt{djorgovski2001a,djorgovski2001b,djorgovski2002b,djorgovski2006}. In addition to the traditional notion of sources on the sky, i.e., a flux distribution in a spatial
sense, we now also have events, that are spatio-temporal in nature.  Moreover, most of the phenomena of interest in the time domain, e.g.,
supernova explosions, are highly perishable, and must be followed up with other observations as soon as possible.  Thus, there is a need
for real-time processing and analysis of massive data streams from the sky, and discovery and characterization of detected events; this
urgency sharpens many of the challenges.  Some of the TDA studies are focused on such transient events, but others are not time-critical,
e.g., studies of variability of sources of some astrophysical type: a supernova can explode only once, but a variable star can be pulsing,
or a black hole accreting for many millions of years; yet there is unique information in their temporal variability.



%

We are therefore entering the Petascale regime in terms of the data volumes, but the exponential growth continues.  As already mentioned one important recent development is
the advent of synoptic sky surveys, which cover large areas of the sky repeatedly, thus escalating the challenges of data handling and analysis from
massive data sets to massive data streams, with all of the added complexities.  This trend is likely to continue, pushing astronomy towards the
Exascale regime.  Two major upcoming facilities deserve a special mention:

\section{Outstanding Challenges}
It is not just the data abundance that is fueling this ongoing revolution, but also Internet-enabled data access, and data re-use.
The informational content of the modern data sets is so high as to make archival research and machine learning not merely profitable, but practically obligatory:
in most cases, researchers who obtain the data can only extract a small fraction of the science that is enabled by it.
Furthermore, numerical simulations are no longer just a crutch of an analytical theory, but are increasingly becoming the dominant or even the only way in
which various complex phenomena (e.g., star formation or galaxy formation) can be modeled and understood, often hand in hand with traditional analytics.
These numerical simulations also produce copious amounts of data as their output; in other words, theoretical statements are expressed not as formulae, but
as data sets.
Since physical understanding comes from the confrontation of experiment and theory, and both are now expressed as ever larger and more complex data sets,
science is truly becoming data-driven in the ways that are both quantitatively and qualitatively different from the past.

%
%
%
%

Many good statistical and data mining tools and methods exist, and are gradually permeating the practicing science communities, astronomy included,
although their uptake has been slower than what may be hoped for.  Social issues aside, one tangible technical problem is the scalability of DM tools:
most of the readily available ones do not scale well to massive data sets.  The key problem is not so much the data volume, but the dimensionality (expressible, e.g., as a number of feature vectors in some data set): most algorithms may work very well in 2 or 3 or 6 dimensions,
but are simply impractical when the intrinsic dimensionality of the data sets is measured in tens, hundred, or thousands. Effective, scalable software and
a methodology needed for knowledge discovery in modern, large and complex data sets typically do not exist yet.

A closely related, but even more difficult problem is the effective visualization of hyper-dimensional data sets.  Visual data examination and exploration
is valuable in itself, and it is also necessary to guide the data mining process.  Finally, visualization is usually the way we reach an intuitive understanding
of some phenomenon.  Here we run into the intrinsic limitations of the human perception: we are biologically optimized for 3D, and we can encode up to a dozen
dimensions in a graphical display.  Our modern, complex data sets may, and probably do contain meaningful structures in $\gg 3$ dimensions, representing new knowledge
to be discovered, that cannot be projected to some humanly comprehensible display without a loss of information.  This, for instance, could also explain why all empirical law known so far depend at most on three parameters only. Improving our ability to visualize highly dimensional
data structures is a key challenge for ``big data" science.

As the data streams – such as those from synoptic sky surveys – replace stationary data sets, new, additional data analysis challenges arise,
especially if there are perishable, short-lived phenomena that need to be addressed or followed up with further measurements in a time-critical manner.
In astronomy, those could be various types of stellar explosions, flares, etc., but one can easily see that equivalent situations may arise in other fields,
e.g., environmental monitoring, security, etc.  While the process of detection of transient events (e.g., by comparing the new data with some sliding average
baseline) is relatively straightforward, their characterization or classification is a much harder problem.  In contrast to most ``textbook" ML applications
for automated classification, here the data are generally very sparse, incomplete, and heterogeneous, and have to be supplemented by generally heterogeneous
archival data, and contextual information that is hard to capture in a quantitative manner.

    \section{Euclid}\label{sec:euclid}
        Before closing this introductory chapter I feel the need to spend a few words about the EUCLID MISSION which I joined during the $3^{rd}$ year of my PhD and will likeley represent the arena where in the near future I shall apply my ``know how".
In the Euclid mission I joined two groups: the Data Quality Common Tools, and the Science Working group for the Legacy Science requirements definitions dedicated to transient objects detection and classification.

Euclid, is a mission devised to provide insight into the nature of dark energy and dark matter by accurate measurement of the accelerated expansion of the Universe, emerged from two mission concepts that were proposed in response to the ESA Cosmic Vision 2015-2025 Call for Proposals, issued in March 2007: the DUNE - Dark Universe Explorer - mission proposed to measure the effects of weak gravitational lensing; the SPACE - Spectroscopic All Sky Cosmic Explorer - mission, aimed at measuring the baryonic acoustic oscillations and redshift-space distortion patterns in the Universe.

In October 2007 the ESA advisory structure selected both proposals to enter the assessment study phase, considering them as equally relevant to the investigation of dark energy. ESA then appointed a Concept Advisory Team with the task of identifying the best possible concept for the dark energy mission to be studied during this phase. This team recommended a combined mission which could simultaneously measure weak lensing and baryonic acoustic oscillations (BAOs). The new mission concept was called Euclid, honouring the Greek mathematician Euclid of Alexandria (~300 BC) who is considered as the father of geometry.

The ESA internal pre-assessment phase for Euclid ran from November 2007 until May 2008. The outcome of this study was a preliminary design for the Euclid mission and its payload which formed the basis for the Invitation to Tender that was issued to Industry in May 2008. A parallel competitive contract was awarded to EADS Astrium Friedrichshafen (Germany) and Thales Alenia Space (Italy); these industrial activities were concluded in September 2009.

Two instrument consortia responded to ESA's call for Declaration of Interest for payload studies issued in May 2008. These studies ran from October 2008 until August 2009.

The report of the assessment study, which includes the Euclid science case together with a synthesis of the industrial and instrument consortium studies, was presented to the scientific community in December 2009. In addition, an independent technical review of the assessment study was conducted by ESA. The recommendations of the review board were presented to the scientific community also in December 2009.

In early 2010 ESA's Science Programme Committee recommended that Euclid, along with two other M-class candidate missions (PLATO and Solar Orbiter) proceed to the next phase: a more detailed definition phase during which period the cost and implementation schedule for the mission must be established. This detailed definition phase was completed in mid 2011.

In October 2011, Euclid was selected by the SPC as one of the first two medium-class missions of the Cosmic Vision 2015-2025 plan; Solar Orbiter was the other mission selected at the time.

Euclid received final approval to move into the full construction phase at the SPC meeting in June 2012. At this meeting, the SPC also formalised an agreement between ESA and fundings agencies in a number of its Member States to develop Euclid's two scientific instruments (a visible-wavelength camera and a near-infrared camera/spectrometer) and the large distributed processing system needed to analyse the data they produce. Nearly 1000 scientists from more than 100 institutes form the Euclid Consortium building the instruments and participating in the scientific harvest of the mission. The consortium comprises scientists from 13 European countries: Austria, Denmark, France, Finland, Germany, Italy, the Netherlands, Norway, Portugal, Romania, Spain, Switzerland and the UK. It also includes a number of US scientists, including 40 nominated by NASA. The Consortium is led by Yannick Mellier, Institut d'Astrophysique de Paris, France.

In December 2012, Astrium SAS (Toulouse) has been contracted to design and build the payload module, which includes the telescope and the accommodation for the instruments, which are to be delivered by the Euclid Consortium. The Prime Contractor, the overall responsible for the building of Euclid satellite will be selected in June 2013.

The Euclid mission has been adopted with launch planned for 2020

\chapter{Astronomical Data Mining}\label{chap:dm}
        \hfill\begin{tabular}{@{}p{.6\linewidth}@{}}
\textit{``With four parameters I can fit an elephant, and with five I can make him wiggle his trunk.
"} \\John von Neumann.\\ \phantom{aaa}
\end{tabular}

From the previous discussion it is  apparent that nowadays, and even more in the future, the most interesting problems will call for the use of complex, multi-wavelength, multi-epoch data collected with heterogeneous instruments. But data -- no matter how great -- are just incidental to the real task of scientists: knowledge discovery.  Unfortunately, the extraction of useful and relevant knowledge from such datasets is still a highly non trivial task which requires a new generation of software tools: automatic, scalable and highly reliable.
Traditional methods of data analysis typically do not scale to the data sets in the Terascale regime, and/or with a high dimensionality.
Thus, the adoption of modern data mining (DM) and Knowledge Discovery in Databases (KDD) techniques becomes a necessity.
Large data volumes tend to preclude direct human examination of all data, and thus an automatization of these processes is needed,
requiring use of Machine Learning (ML) techniques. This fact has been recently recognized by the implementation of a specific Interest Group on Knowledge Discovery in Databases within the IVOA \citep{pasian2011}, focusing on recent developments in the field of astronomical Data Mining  (hereafter DM) or ``Knowledge Discovery in Databases" (KDD) as it is also often called. Some reviews of these topics can be found in \cite{djorgovski2012d}, \cite{djorgovski2006}, \cite{dunham2002} and \cite{brescia2012d}.

In its widest meaning, Knowledge Discovery in Databases or Data Mining regards the discovery of ``models" for data. There are, however, many different methods which can be used to discover these underlying models: statistical pattern recognition, machine learning, summarization, etc. and an extensive review of all these models would take us far beyond the purposes of this work. In what follows we shall therefore summarize only the main methodological aspects.   \\ Machine learning, which is sometimes considered to be a branch of artificial intelligence, is a scientific discipline concerned with the design and development of algorithms that allow computers to evolve behaviors based on empirical data. A ``learner" can take advantage of examples (data) to capture characteristics of interest of their unknown underlying probability distribution (cf. \citealt{bishop2006}).
These data form the so called Knowledge Base (KB): a fairly large set of examples to be used for training and to test the performances. The difficulty lies in the fact that often, if not always, the set of all possible behaviors given all possible inputs is too large to be covered by the KB.  Hence the learner must possess some generalization capabilities in order to be able to produce useful outputs when presented new instances.
The use of a DM application requires a good understanding of the mathematics underlying the methods, of the computing infrastructures, and of the complex workflows which need to be implemented.

Strictly speaking, the KDD discipline is about algorithms for inferring knowledge from data and ways of validating the obtained results, as well as about running them on infrastructures capable to match the computational demands. In practice, whenever there is too much data or, more generally, a representation in more than 5 dimensions \citep{odenwald1987}, there are basically three ways to make learning feasible. The first one is trivial: applying the training scheme to a decimated dataset. The second method relies on parallelization techniques, the idea being to split the problem into smaller parts, then solve each using a separate CPU and finally combine the results together \citep{paliouras1993}. Sometimes this is feasible due to the intrinsic natural essence of the learning rule (such as genetic algorithms, \citealt{goldberg1988}). However, even after parallelization, the algorithm's asymptotic time complexity cannot be improved. The third and more challenging way to enable a learning paradigm to deal with Massive Data Sets (MDS) is to develop new algorithms of lower computational complexity but in many cases this is simply not feasible.\\

Astronomical applications of ML are still relatively recent and restricted to a handful of problems.
This is surprising, given the data richness and a variety of possible applications in the data-driven astronomy.
DM can enable multiple uses of the same data by many different groups for different applications.
The comparison of two pertinent reviews (\citealt{tagliaferri2003a}, \citealt{ball2010}) shows a relatively slow growth in both the number
and the variety of ML and DM applications in astronomy.  Some of this slow growth can be accounted for by a reluctance to adopt
new methods that are still not adequately taught in the advanced astronomy curriculum, despite their great potential and even necessity.
But sociological challenges aside, there are some technical ones that need to be addressed.

First, a large family of ML methods (the so called supervised ones) require the availability of relatively large and well characterized knowledge bases (KB),
e.g., reliable (``ground truth") training data sets of examples from which the ML methods can learn the underlying patterns and trends.
Such KBs are relatively rare and are available only for a few specific problems.

Second, most ML algorithms used so far by the astronomers cannot deal well with missing data (i.e., no measurement was obtained for a given attribute)
or with upper limits (a measurement was obtained, but there is no detection at some level of significance).  While in many other fields
(e.g., market analysis and many bioinformatics applications) this is only a minor problem since the data are often redundant and/or can
be cleaned of all records having incomplete or missing information, in astronomy this is usually not so, and all data records, including
those with an incomplete information, are potentially scientifically interesting and cannot be ignored.

Examples of early uses of modern ML tools for analysis of massive astronomical data sets include automated classification of sources detected in sky surveys as
stars (i.e., unresolved) vs. galaxies (resolved morphology), using Artificial Neural Nets (ANN) or Decision Trees (DT), e.g., by Weir (1985) or
Odewahn et al. (1992, 2004).
Further improvements include \citep{donalek2006}, who introduced external a priori constraints in the classification, and \cite{russo2008} who used the so-called Bregman co-clustering \citep{bregman1967} to reduce the weight of missing or incomplete information.  \cite{brescia2012a} have recently used several ML method for a different type of
resolved/unresolved objects separation, namely the identification of globular clusters in external galaxies, other details can be found in section \ref{chap:GC}.

Another set of ML applications is in classification or selection of objects of a given type in some parameter space, e.g., colors (ratios of fluxes measured at different
wavelengths, expressed logarithmically).  This is particularly well suited for the identification of quasars and other active galactic nuclei, which are morphologically
indistinguishable from normal stars, but represent vastly different physical phenomena.
A novel approach to this problem includes hybrids of supervised and unsupervised classifiers (D'Abrusco et al. 2009, 2012).
In a complementary approach, \cite{richards2009} used Bayesian techniques for a selection of quasars in an 8-dimensional color parameter space.

Yet another type of interesting scientific applications of that methods is the are estimate of the so-called photometric redshifts (measures of distances in cosmology), that are derived from
colors (available for most detected sources in the large imaging surveys) rather than from spectroscopy (much more costly in terms of the observing time, and thus
available for a much smaller subset of sources).
ANN have performed very well in this task \citep{tagliaferri2002,firth2003,hildebrandt2010,cavuoti2012b}.
\\ \cite{laurino2011} implemented a hybrid procedure based on a
combination of unsupervised clustering and several independent classifiers that has improved the accuracy, for both normal galaxies and quasars.

Below we discuss a new set of ML challenges in the context of automated classification of transient events found in synoptic sky surveys.
A very different approach to astronomical object classification through crowdsourcing or ``citizen science" is exemplified by the ``Galaxy Zoo" project
\citep{lintott2008}, that harvests human pattern recognition for a visual morphological classification.  Such projects represent an excellent public outreach,
and can produce useful scientific results.  While their utility depends on the exact problem to be addressed and the specific implementation,
we note that this approach does not scale, due to the limited resources
of human time and attention.

For good recent reviews of ML applications in astronomy, see, e.g., \cite{ball2010}, or the volume edited by Way et al. (2012).

In a way, a lot of DM can be seen as algorithmic applications of statistics, and numerous statistical tools have been used in astronomy since its beginnings.
A useful Astro-Statistics portal is maintained by E. Feigelson and G.J. Babu at \url{http://astrostatistics.psu.edu}; see also Feigelson \& Babu (2012ab).

In the VO framework, a web service, VOStat \citep{graham2005} offers a set of accessible statistical tools for data analysis.
There are currently two different versions of it, \url{http://astrostatistics.psu.edu:8080/vostat/} and \url{http://vo.iucaa.ernet.in/$\sim$voi/VOStat.html},
developed initially as collaborative efforts between the groups at PSU, Caltech, and IUCAA.
While these services address a growing need for the use of advanced statistical methods in astronomical data analysis,
their uptake by the community has been relatively slow.
Another useful guide to available tools is at \url{http://wiki.ivoa.net/twiki/bin/view/IVOA/IvoaKDDguide}.


\section{Data Mining Functionalities}\label{sec:functionality}
Hereinafter we shall adopt a definition of Data Mining based on the fact that in most real life applications several different methods (functional domains), need to be combined in more complex and often hierarchical workflows to produce a reliable and robust result. The taxonomy of data mining functionalities which will be adopted through this work is:
 \begin{itemize}
\item	Dimensional reduction;
\item	Classification;
\item	Regression;
\item	Clustering;
\item	Segmentation;
\item	Filtering;
\end{itemize}
Single or groups of such functionalities can be associated with a variety of models and algorithms (e.g. Neural Networks, Support Vector Machines, Bayesian networks, Genetic Algorithms etc.) and specific use cases are therefore defined by a proper association ``functionality-model".

More in detail:
\begin{itemize}
\item \textit{classification} is a procedure in which individual items are placed into groups based on quantitative information (referred to as features, synonym of parameters in the problem domain) using the knowledge contained in a training set of previously labeled items (known also as Knowledge Base or KB).  A classifier is therefore a system that performs a mapping from a feature space $X$ to a set of labels $Y$. Classification may be either ``crispy" or ``probabilistic". In the first case, given an input pattern $x$ the classifier returns its computed label $y$. In the second, given an input pattern $x$ the classifier returns a vector $y$ which contains the probability of $y_i$ to be the ``right" label for $x$. Both types of classification can be applied to both ``two-class" and ``multi-class" cases. Typical astrophysical problems which have been addressed with this functionality are the so called ``star/galaxy" separation (which would be better called resolved-unresolved objects separation), morphological classification of galaxies, classification of stellar spectra, etc.\\
\item \textit{Regression} is instead generally intended as the supervised search for a mapping from a domain in $R^n$ to a domain in $R^m$, where $m < n$. Also in this case, one can distinguish between two different types of regression:
\begin{itemize}
\item	\textit{Data-table statistical correlation}: in which the user tries to find a mapping without any prior assumption on the functional form of the data distribution;
\item	\textit{function fitting}: in which the user tries to validate the hypothesis, suggested by some theoretical framework, that the data distribution follows a well-defined, and known, function.
\end{itemize}
The most common astrophysical example of a regression problem is the evaluation of photometric redshifts of galaxies from a limited but statistically sufficient KB based on spectroscopic redshift samples.
\textit{Dimensional reduction} is the process of reducing the number of random variables under consideration, and can be divided into feature selection and feature extraction.
Feature selection approaches try to find a subset of the original variables (also called features or attributes), \cite{guyon2003}. Two strategies are filter (e.g. information gain) and wrapper (e.g. search guided by the accuracy) approaches.
Feature extraction transforms the data in the high-dimensional space to a space of fewer dimensions. The data transformation may be linear, as in Principal Component Analysis (PCA), but many non-linear techniques also exist \citep{guyon2006}.\\
\item \textit{Clustering techniques} apply when there is no class to be predicted but rather when the instances need to be divided into natural groups. From the self-adaptive computing point of view, clustering models are also referred to as ``unsupervised methods", since they do not require the use of an extensive KB \citep{jain1999}.
In general, there are different ways in which the results of clustering can be expressed: for instance the identified groups can be exclusive or overlapping. But they may be also probabilistic, whereby an instance belongs to each group with a certain probability. Other clustering algorithms produce a hierarchical structure of clusters, so that at the top level the instance space divides into just a few clusters, each of which divides into its own sub-clusters at the next level, and so on. Clustering is often followed by a stage where a decision tree or ``set of rules" is inferred in order to allocate each instance to the cluster to which it belongs. The choice between these different models is dictated by the nature of the specific problem to be tackled.
In spite of the enormous potentialities (think for instance to the identification of unknown types of objects in the parameter space), the application of clustering methods to astrophysical MDS is still in a very early stage even though, in many cases they are embedded into complex DM workflows.
\textit{Segmentation}, synonym of ``image processing", in the DM with machine learning context is strictly correlated with image clustering functional domain. More in general, in computer vision, segmentation refers to the process of partitioning a digital image into multiple segments (sets of pixels, also known as superpixels). The goal of segmentation is to simplify and/or change the representation of an image into something that is more meaningful and easier to analyze \citep{bishop2006}. Image segmentation is typically used to locate objects and boundaries (lines, curves, etc.) in images. More precisely, image segmentation is the process of assigning a label to every pixel in an image such that pixels with the same label share certain visual characteristics. The result of image segmentation is a set of segments that collectively cover the entire image, or a set of contours extracted from the image \citep{lindeberg2001}. Each of the pixels in a region are similar with respect to some characteristic or computed property, such as color, intensity, or texture. Adjacent regions are significantly different with respect to the same characteristics.
\item Finally, data-based model \textit{filtering} helps to create complex architectures based on different and multiple mining models that use subsets of data in a filtered mining structure. A useful way to have a right vision of data-driven model filtering is: \textit{Model filtering operates without altering the underlying model data. This allows one set of data to be shared among multiple components, each of which may interpret the data in a different manner. Filters can be layered, enabling model data to be interpreted through several different filter objects} \citep{goldstein2001}.
Filtering gives you flexibility when you design your mining structures and data sources, because you can create a single mining structure, based on a comprehensive data source view. You can then create filters to use only a part of that data for training and testing a variety of models, instead of building a different structure and related model for each subset of data.
For example, it is possible to develop specialized DM models on a selected part of the data sets. Next, you define a multiple mining structure that includes all the features coming out the previous filtered ensemble of models. Finally, you create a model that is filtered on a particular customer attribute, such as a specific wavelength band. You can then easily make a copy of that model, and change just the filter condition to generate a new model based on a different spectrum region.
Some real-life scenarios where you might benefit from this feature include the following:
\begin{itemize}
\item	Creating separate models for discrete values such as wavelength, regions, and so forth;
\item	Experimenting with models by creating and then testing multiple groupings of the same data;
\item	Specifying complex filters on nested data contents.
\end{itemize}
Data-based model filtering greatly simplifies the task of managing mining structures and mining models, because you can easily create multiple models that are based on the same structure. You can also quickly make copies of existing mining models and then change only the filter condition. Good examples of such filtered mining models are the Gated Experts (GE; \citealt{weigend1995}).
\end{itemize}

\section[Technological requirements]{Technological requirements for effective Data Mining}

As mentioned before computing has started to change how science is done, enabling new scientific advances through enabling new kinds of experiments. These experiments are also generating new kinds of data of increasingly exponential complexity and volume. Achieving the goal of being able to use, exploit and share these data most effectively is a huge challenge.
It is necessary to merge the capabilities of a file system to store and transmit bulk data from experiments, with logical organization of files into indexed data collections, allowing efficient query and analytical operations. It is also necessary to incorporate extensive metadata describing each experiment and the produced data. Rather than flat files traditionally used in scientific data processing, the full power of relational databases is needed to allow effective interactions with the data, and an interface which can be exploited by the extensive scientific toolkits available, for purposes such as visualization and plotting.
Different disciplines require support for much more diverse types of tasks than we find in the large, very coherent and stable virtual organizations. Astronomy, for example, has far more emphasis on the collation of federated data sets held at disparate sites. There is less massive computation, and large-scale modeling is generally done on departmental High Performance Computing (HPC) facilities, where some communities are formed of very small teams insisting relatively undeveloped computational infrastructure. In other cases, such as life sciences, the problems are far more related to heterogeneous, dispersed data rather than computation. The harder problem for the future is heterogeneity, of platforms, data and applications, rather than simply the scale of the deployed resources. The goal should be to allow scientists to explore the data easily, with sufficient processing power for any desired algorithm to process it. \\
Our convincement is  that most aspects of computing will see exponential growth in bandwidth, but sub-linear or no improvements at all in latency. Moore's Law will continue to deliver exponential increases in memory size but the speed with which data can be transferred between memory and CPUs will remain more or less constant and marginal improvements can only be made through advances in caching technology. Certainly Moore's law will allow the creation of parallel computing capabilities on single chips by packing multiple CPU cores onto it, but the clock speed that determines the speed of computation is constrained to remain limited by a thermal wall \citep{sutter2005}. We will continue to see exponential growth in disk capacity, but the factors which determine latency of data transfer will grow sub-linearly at best, or more likely remain constant. Thus computing machines will not get much faster. But they will have the parallel computing power and storage capacity that we used to only get from specialist hardware. As a result, smaller numbers of supercomputers will be built but at even higher cost. From an application development point of view, this will require a fundamental paradigm shift from the currently sequential or parallel programming approach in scientific applications to a mix of parallel and distributed programming that builds programs that exploit low latency in multi core CPUs. But they are explicitly designed to cope with high latency whenever the task at hand requires more computational resources than can be provided by a single machine. Computing machines can be networked into clouds or grids of clusters and perform tasks that were traditionally restricted to supercomputers at a fraction of the cost. A consequence of building grids over wide-area networks and across organizational boundaries is that the currently prevailing synchronous approach to distributed programming will have to be replaced with a fundamentally more reliable asynchronous programming approach. A first step in that direction is Service-Oriented Architectures (SOA) that have emerged and support reuse of both functionality and data in cross-organizational distributed computing settings. The paradigm of SOA and the web-service infrastructures facilitate this roadmap \citep{shadbolt2006}.
Traditionally, scientists have been good at sharing and reusing each other's application and infrastructure code. In order to take advantage of distributed computing resources in a grid, scientists will increasingly also have to reuse code, interface definitions, data schemas and the distributed computing middleware required to interact in a cluster or grid. The fundamental primitive that SOA infrastructures provide is the ability to locate and invoke a service across machine and organizational boundaries, both in a synchronous and an asynchronous manner. The implementation of a service can be achieved by wrapping legacy scientific application code and resource schedulers, which allows for a viable migration path \citep{taylor2007}. Computational scientists will be able to flexibly orchestrate these services into computational workflows. The standards available for service design and their implementation support the rapid definition and execution of scientific workflows. With the advent of abstract machines, it is now possible to mix compilation and interpretation as well as integrate code written in different languages seamlessly into an application or service. These platforms provide a solid basis for experimenting with and implementing domain-specific programming languages and we expect specialist languages for computational science to emerge that offer asynchronous and parallel programming models while retaining the ability to interface with legacy Fortran, C, C++ and Java code.
Finally, scalability of algorithms can be an issue.
Most existing ML methods scale badly with both increasing number of records and/or of dimensionality (i.e., input variables or features):
the very richness of our data sets makes them difficult to analyze.
This can be circumvented by extracting subsets of data, performing the training and validation of the methods on these manageable data subsets,
and then extrapolating the results to the whole data set.  This approach obviously does not use the full informational content of the data sets,
and may introduce biases which are often difficult to control.
Typically, a lengthy fine tuning procedure is needed for such sub-sampling experiments, which may require tens or sometimes hundreds of experiments to be
performed in order to identify the optimal DM method for the problem in hand, or, a given method, the optimal architecture or combination of parameters.
DAMEWARE (see \citealt{brescia2010b} and section \ref{chap:dame} for more details) resource was designed by taking all these issues into account.

\subsection{Data Mining Packages}
There are also various free DM/KDD packages commonly used in the academic community that would be suitable for adoption by the astronomical community,
although their uptake has also been relatively slow.  Several of them have been evaluated in this context by \cite{donalek2011}, including Orange,
Rapid Miner, Weka, VoStat and DAME (see below).

Orange\footnote{\url{http://orange.biolab.si}} is an intuitive data mining desktop application;
most standard data mining techniques have been implemented such as decision trees, kNN, SVM, random forests, K-means, etc.
The ``Orange Canvas" UI is quite intuitive.
All tasks are performed as schemas constructed using widgets that can be individually configured.
This interface is quite convenient for people who run at the thought of programming since it allows a more natural click-and-drag connection flow between widgets.
Widgets can be thought of as black boxes that take in an input connection from the socket on their left and output their results to the socket on their right.
Workflows can thus be easily constructed between data files, learning algorithms and evaluation routines. However, although it is quite straightforward
to setup experiments in the UI, their successful execution is not always guaranteed.

Weka\footnote{\url{http://www.cs.waikato.ac.nz/$\sim$ml/weka}} is a cross-platform DM package, written in Java. Most standard methods have been implemented.
There is also a wide range of more classification algorithms available as plug-ins to Weka including learning vector
quantization, self-organizing maps, and feed-forward ANNs.

Rapid Miner\footnote{\url{http://rapid-i.com/content/view/181/196}} also has most standard DM/KDD methods implemented. There are plug-ins available to interface with Weka,
R and other major DM packages, so all operations from these can be integrated as well.
DAME is described in more detail below. We note that given the increasing volumes and complexity of data sets in astronomy, it is inevitable that these
modern DM/KDD tools will be increasingly more used by the community.

Data Mining and Exploration\footnote{\url{http://dame. dsf.unina.it/} or \url{http://dame.caltech.edu/}} (DAME) web application \citep{brescia2010b,brescia2012c},
a joint effort between the Astroinformatics groups at University Federico II, the Italian National Institute of Astrophysics,
and the California Institute of Technology.
DAME aims at solving in a practical way some of the DM problems, by offering a completely transparent architecture,
a user-friendly interface, and the possibility to seamlessly access a distributed computing infrastructure.
It adopts VO standards in order to facilitate interoperability of data; however, at the moment, it is not yet fully VO compliant.
This is partly due to the fact that new standards need to be defined for data analysis, DM methods and algorithm development.
In practice, this implies a definition of standards in terms of an ontology and a well-defined taxonomy of functionalities
to be applied to the astrophysical use cases.

DAME offers asynchronous access to the infrastructure tools, thus allowing the running of jobs and processes
outside the scope of any particular web application, and independent of the user connection status.
The user, via a simple web browser, can access application resources and can keep track of his jobs
by recovering related information (partial/complete results) without having to keep open a communication socket.
Furthermore, DAME has been designed to run both on a server and on a distributed computing infrastructure (e.g., Grid or Cloud).

A detailed technical description of the other components can be found in \cite{brescia2010b} and in section \ref{chap:dame}.

In the following sections I shall outline three data mining models which I either implemented or helped to integrate, in DAMEWARE during my PhD work.

    \section[Model 1: SVM]{Model 1: SVM - Support Vector Machines}\label{chap:SVM}
        \blfootnote{this section is largely extracted from: \tiny
\begin{itemize}
\item Brescia, M.; \textbf{Cavuoti, S.}; Paolillo, M.; Longo, G.; Puzia, T.; 2012, The detection of Globular Clusters in galaxies as a data mining problem, \textbf{MNRAS, Volume 421, Issue 2, pp. 1155-1165}, available at \href{http://arxiv.org/abs/1110.2144v1}{arXiv:1110.2144v1}.
\item \textbf{Cavuoti, S.}; Brescia, M.; D'Abrusco, R.; Longo, G.; Photometric AGN Classification in the SDSS with Machine Learning Methods \textbf{to be Submitted to MNRAS}	
\end{itemize}}
Support vector machines (SVMs, \citealt{boser1992}, also support vector networks) are supervised learning models with associated learning algorithms that analyze data and recognize patterns, used for classification and regression analysis; I applied SVM in two classification cases described in sections: \ref{chap:agn} and \ref{chap:GC}; SVM were also ported into DAMEWARE.

\begin{figure*}
   \centering
   \includegraphics[width=12cm]{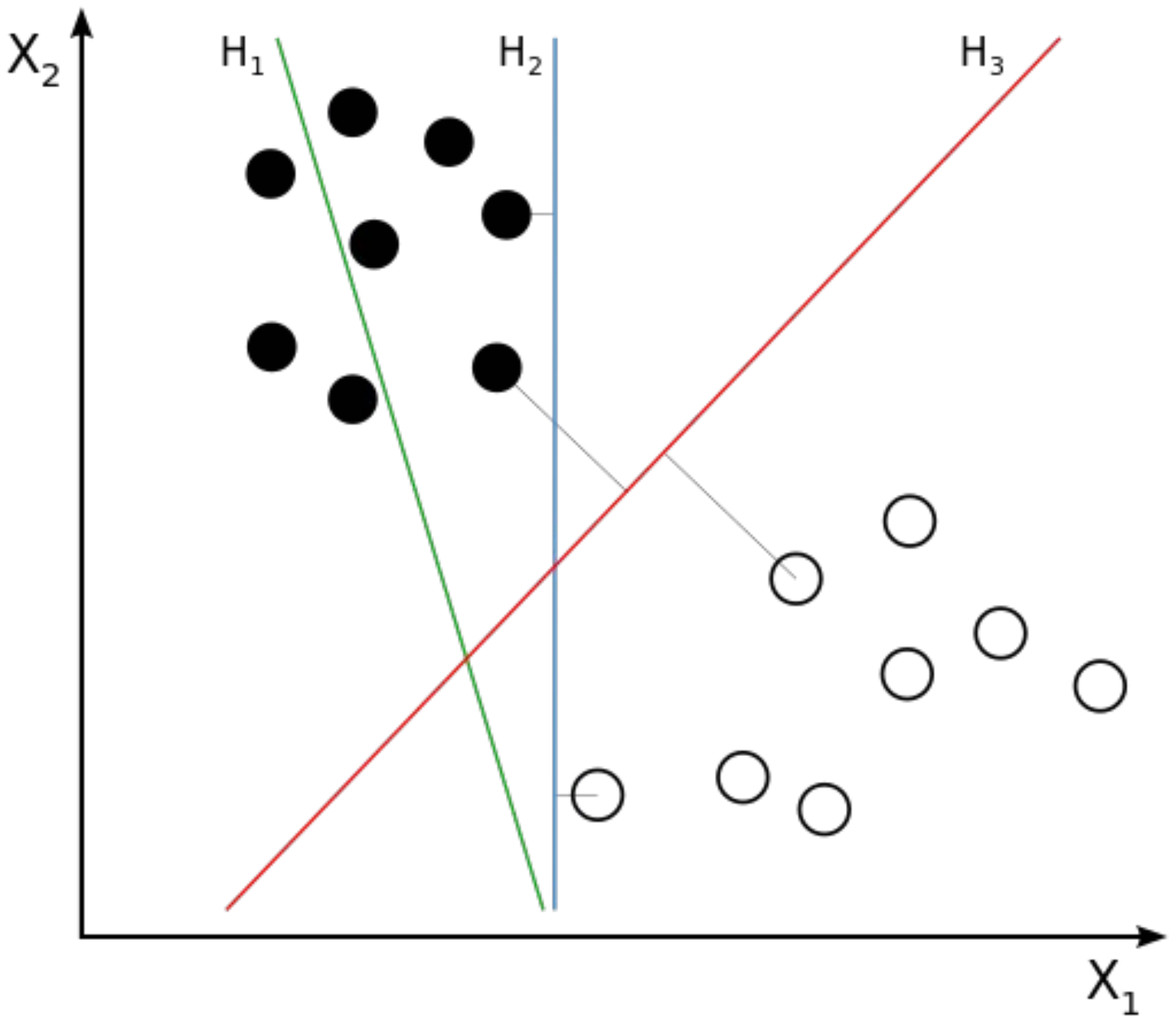}
   \caption[Three possible hyperplanes.]{Three possible hyperplanes, H1 does not separate the classes. H2 does, but only with a small margin. H3 separates them with the maximum margin.}\label{svm:hyperplanes}
   \end{figure*}

The basic SVM takes a set of input data and predicts, for each given input, which of two possible classes forms the output, making it a non-probabilistic binary linear classifier. Given a set of training examples, each marked as belonging to one of two categories, a SVM training algorithm builds a model that assigns new examples into one category or the other. A SVM model is a representation of the examples as points in space, mapped so that the examples of the separate categories are divided by a clear gap that is as wide as possible. New examples are then mapped into that same space and predicted to belong to a category based on which side of the gap they fall on.
SVM models were originally defined for the classification of two classes of objects linearly separable by identifying the hyperplane with the \textit{best} margin, in figure \ref{svm:hyperplanes} we can see what we mean with \textit{best}.

 \begin{figure*}
   \centering
   \includegraphics[width=12cm]{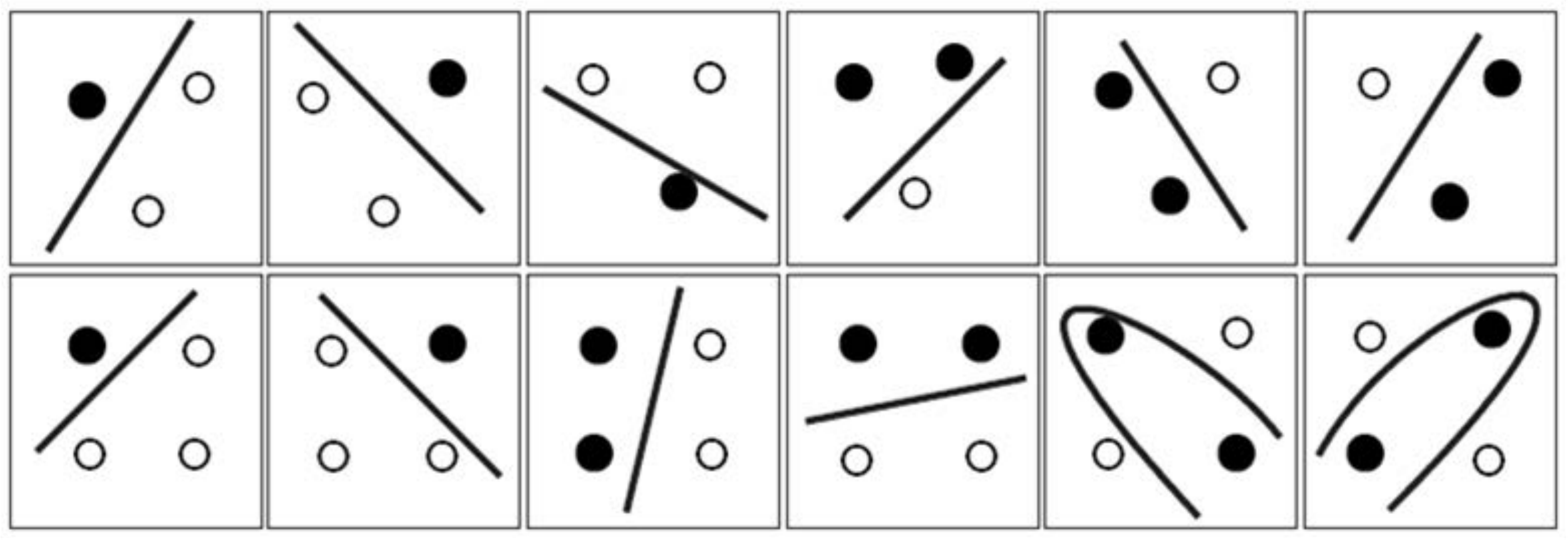}
   \caption{In a plane three points are always linearly separable, four are not ever separable, credit of \cite{ivanciuc2007}.}\label{svm:twodimension}
   \end{figure*}
Obviously SVM can be used also to separate classes that are not separable by a linear classifier, otherwise their application in real cases will be not feasible.

\begin{figure*}
   \centering
   \includegraphics[width=12cm]{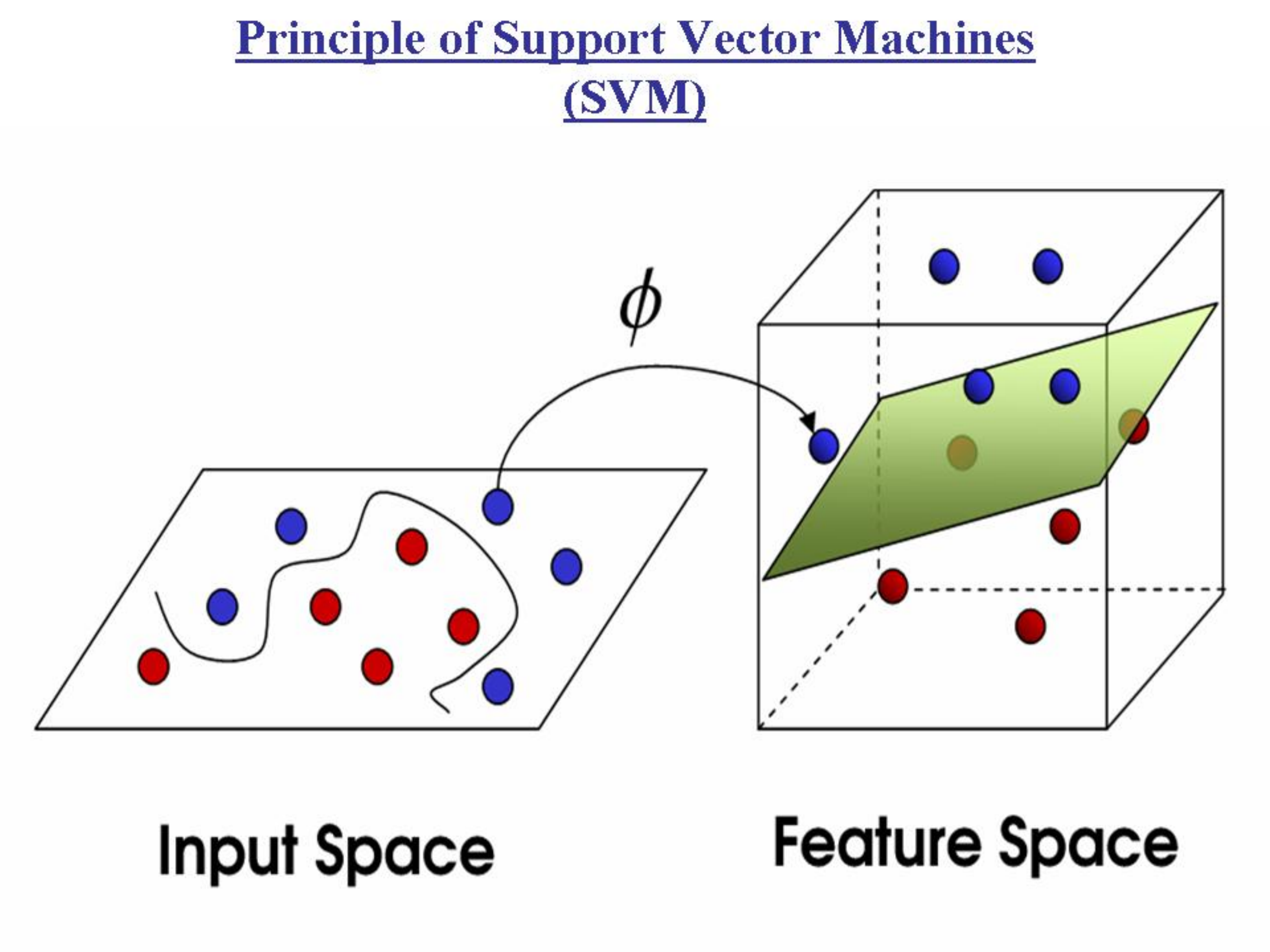}
   \caption{Original points are mapped in the \textit{feature space} via a \textit{feature function} where the solution is found.}\label{svm:mapping}
   \end{figure*}

In addition to performing linear classification, SVMs can efficiently perform non-linear classification using what is called the kernel trick, implicitly mapping their inputs into high-dimensional feature spaces, see figure \ref{svm:mapping}.

More formally, a support vector machine constructs a hyperplane or set of hyperplanes in a high- or infinite-dimensional space, which can be used for classification, regression, or other tasks. Intuitively, a good separation is achieved by the hyperplane that has the largest distance to the nearest training data point of any class (so-called functional margin), since in general the larger the margin the lower the generalization error of the classifier.

Given a training set of instance-label pairs $(x_i,y_i),i=1...l$ where $x_i\in R^n$ and $y_i\in \{1,-1\}^l$, , the support vector
machines require the solution of the following optimization problem:

\begin{equation}\label{AGN:eqSVM1}
\mathop {\min }\limits_{w,b,\xi } \frac{1}{2}w^T w + C\sum\limits_{i= 1}^l {\xi _i }
\end{equation}

subjet to the following conditon:

\begin{equation}\label{AGN:eqSVM2}
y_i (w^T \phi (x_i ) + b) \ge 1 - \xi _i
\end{equation}
\begin{equation}\label{AGN:eqsvmdimenticata}
\xi \geq 0
\end{equation}

Here training vectors xi are mapped into a higher (maybe infinite) dimensional space by the function $\phi$. Then SVM finds a linear
separating hyperplane, represented by its support vectors, with the maximal margin in this higher dimensional space. $C > 0$ is the penalty parameter of the error
term. Furthermore, $K(x_i; x_j)\equiv \phi(x_i)^T(x_j)$ is called the kernel function. The usual four basic kernels are:

\begin{itemize}
\item linear: $K(x_i,x_j) = x_i ^T x_j$
\item polynomial: $K(x_i,x_j) = (\gamma x_i ^T x_j + r)^d,\gamma >0$
\item radial basis function (RBF): $K(x_i,x_j) = \exp (-\gamma\parallel x_i - x_j\parallel^2),\gamma>0$
\item sigmoid: $K(x_i,x_j) = \tanh (\gamma x_i ^T x_j + r)$
\end{itemize}

Here, $\gamma$, $r$, and $d$ are kernel parameters. The RBF kernel nonlinearly maps samples into a higher dimensional
space, so it, unlike the linear kernel, can handle the case when the relation between class labels and attributes is nonlinear.
Furthermore, the linear kernel is a special case of RBF as Keerthi and Lin \citep{keerthi2003} shows that the linear kernel with a penalty
parameter $C$ has the same performance as the RBF kernel with some parameters ($C,\gamma$). In addition, the sigmoid
kernel behaves like RBF for certain parameters.

\subsection{SVM parameter determination strategy}\label{svmstrategy}

There are two parameters while using RBF kernels with the C-SVC model: $C$ (from the model) and $\gamma$ (from the kernel). It is
not known in principle which C and $\gamma$ are the best for one problem; consequently some kind of model selection (parameter
search) must be done. The goal is to identify good values of the parameters ($C$; $\gamma$) so that the classifier can accurately
predict unknown data (i.e., testing data). Note that it may not be useful to achieve high training accuracy (i.e., classifiers
accurately predict training data whose class labels are indeed known).
Therefore a common way is to split the training samples into two groups, respectively, the training set, used to train the model, and the test set, used a posteriori to validate the trained
model performances.
Then the prediction accuracy on this set can more precisely respect
the performance on classifying unknown data. An improved version of this procedure is a technique known as cross validation. In v-fold cross-validation, we
first divide the training set into v subsets of equal size. Sequentially one subset is tested using the classifier trained on
the remaining v-1 folder. Thus, each instance of the whole training set is predicted once so the cross-validation accuracy is the
percentage of data which are correctly classified. The cross-validation procedure can prevent the overfitting problem. The
experiments that we are going to present are made with five folders. $C, \gamma \in \Re^+$ need to bee evalutated by finding the maximum
on a grid of values wich is usually defined by letting $C$ and $\gamma$ vary as: $C=2^{-5}, 2^{-3}, ... 2^{15}, \gamma  = 2{-15},2^{-13}...2^{3}$.
A practical application will be discussed in section \ref{chap:agn}.

    \section[Model 2: MLPQNA]{Model 21: MLPQNA - Multi Layer Perceptron trained with Quasi Newton Algorithm}\label{chap:QNA}
        \blfootnote{this section is largely extracted from: \tiny
\begin{itemize}
			\item Brescia, M.; \textbf{Cavuoti, S.}; D'Abrusco, R.; Longo, G.; Mercurio, A.; 2013, Photo-z prediction on WISE - GALEX - UKIDSS - SDSS Quasar Catalogue, based on the MLPQNA model, \textbf{Submitted to ApJ}
			\item \textbf{Cavuoti, S.}; Brescia, M.; Longo, G.; Mercurio, A.; 2012, Photometric Redshifts with Quasi Newton Algorithm (MLPQNA). Results in the PHAT1 Contest, \textbf{A\&A, Vol. 546, A13, pp. 1-8}
			\item Brescia, M.; \textbf{Cavuoti, S.}; Paolillo, M.; Longo, G.; Puzia, T.; 2012, The detection of Globular Clusters in galaxies as a data mining problem, \textbf{MNRAS, Volume 421, Issue 2, pp. 1155-1165}, available at \href{http://arxiv.org/abs/1110.2144v1}{arXiv:1110.2144v1}.
\end{itemize}}
From a technical point of view, the MLPQNA method, is a Multi Layer Perceptron (MLP; \citealt{bishop2006}) implemented with a learning rule based on the Quasi Newton Algorithm (QNA); in other words and as it is synthesized in the acronym, MLPQNA differs from more traditional MLPs implementations in the way the optimal solution of the regression problem is
found.
\\The algorithm was involved in several experiments on astronomical datasets, both in regression (photometric redshift on galaxies, sec \ref{sec:phat}, and quasar, sec \ref{sec:qso}) and classification (active galactic nuclei, section \ref{chap:agn}, globular clusters, section \ref{chap:GC}, and transients, section \ref{chap:transients}) with remarkable results.\\
According to \cite{bishop2006}, feed forward neural networks (in their various implementations) provide a general framework for representing non linear functional mappings between a set of input variables (also called features) and a set of output variables (the targets).\\
The MLP architecture is one of the most typical feed-forward neural network model. The term feed-forward
is used to identify basic behavior of such neural models, in which the impulse is propagated always in the
same direction, e.g. from neuron input layer towards output layer, through one or more hidden layers (the
network brain), by combining weighted sum of weights associated to all neurons (except the input layer).
As easy to understand, the neurons are organized in layers, with proper own role. The input signal, simply
propagated throughout the neurons of the input layer, is used to stimulate next hidden and output neuron
layers. The output of each neuron is obtained by means of an activation function, applied to the weighted
sum of its inputs. Different shape of this activation function can be applied, from the simplest linear one up
to sigmoid. The number of hidden layers represents the degree of the complexity achieved for the energy
solution space in which the network output moves looking for the best solution. As an example, in a typical
classification problem, the number of hidden layers indicates the number of hyper-planes  used to split the
parameter space (i.e. number of possible classes) in order to classify each input pattern. What is different in
such a neural network architecture is typically the learning algorithm used to train the network. It exists a
dichotomy between supervised and unsupervised learning methods.

In the first case, the network must be firstly trained (training phase), in which the input patterns are
submitted to the network as couples (input, desired known output). The feed-forward algorithm is then
achieved and at the end of the input submission, the network output is compared with the corresponding
desired output in order to quantify the learning quote. It is possible to perform the comparison in a batch way
(after an entire input pattern set submission) or incremental (the comparison is done after each input pattern
submission) and also the metric used for the distance measure between desired and obtained outputs, can be
chosen accordingly problem specific requirements (in the MLP-BP the MSE, Mean Square Error, is used).
After each comparison and until a desired error distance is unreached (typically the error tolerance is a precalculated value or a constant imposed by the user), the weights of hidden layers must be changed
accordingly to a particular law or learning technique.
After the training phase is finished (or arbitrarily stopped), the network should be able not only to recognize
correct output for each input already used as training set, but also to achieve a certain degree of
generalization, i.e. to give correct output for those inputs never used before to train it. The degree of
generalization varies, as obvious, depending on how “good” has been the learning phase. This important
feature is realized because the network does not associates a single input to the output, but it discovers the
relationship present behind their association. After training, such a neural network can be seen as a black box
able to perform a particular function (input-output correlation) whose analytical shape is a priori not known.
In order to gain the best training, it must be as much homogeneous as possible and able to describe a great
variety of samples. Bigger the training set, higher will be the network generalization capability.
Despite of these considerations, it should been always taken into account that neural networks application
field should be usually referred to problems where it is needed high flexibility (quantitative result) more than
high precision (qualitative results).
Concerning the hidden layer choice, there is the possibility to define zero hidden layers (SLP, Single Layer
Perceptron, able to solve only linear separation of the parameter space), 1 or 2 hidden layers, depending on
the complexity the user wants to introduce in the not linear problem solving experiment.

Second learning type (unsupervised) is basically referred to neural models able to classify/cluster patterns
onto several categories, based on their common features, by submitting training inputs without related
desired outputs.  This is not the learning case approached with the MLP architecture, so it is not important to
add more information in this document.

The training of a neural network can be in fact seen as the search for the function which minimizes the errors of the predicted values with respect to the true values available for a small but significant subsample of objects in the same data set. This subset is also called \textit{training set} or \textit{knowledge base}.  The final performances of a specific Neural Network (NN) depend on many factors, such as topology, the way the minimum of the error function is searched and found, the way errors are computed, as well as the intrinsic quality of training data.\\
The formal description of a feed-forward neural network with two computational layers is given in the Eq. \ref{qso:eq0}:

\begin{equation} \label{qso:eq0}
y_k = \sum_{j=0}^{M} w^{(2)}_{kj} g \left( \sum_{i=0}^{d} w^{(1)}_{ji} x_i \right)
\end{equation}

\begin{figure}
\centering
\includegraphics[width=6.cm]{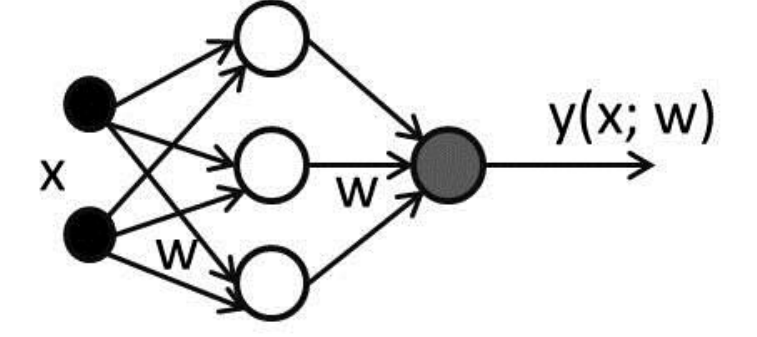}
\caption{Scheme of a Multi Layer Perceptron general architecture for two input variables and one output value.}\label{qso:mlp}
\end{figure}

Equation \ref{qso:eq0} can be better understood by using a graph like the one which is shown in Figure \ref{qso:mlp}.
The input layer $\left( x_i \right)$ is made of a number of neurons (also known as perceptrons) equal to the number of input variables $\left( d \right)$; the output layer, on the other hand, will have as many neurons as the output variables $\left( k \right)$.
In the general case, the network may have an arbitrary number of hidden layers (in the depicted case there is just one hidden layer as in most real implementations) each of one can be formed by an arbitrary number of neurons $\left( M \right)$.
In a fully connected feed-forward network each node of a layer is connected to all the nodes in the adjacent layers.
Each connection is represented by an adaptive weight $\left( w^{l}_{kj} \right)$  which can be regarded as the the strength of the synaptic connection between neurons $k$ and $j$, while the response of each perceptron to the inputs is represented by a non-linear function $g$, referred to as the activation function.

Eq. \ref{qso:eq0} assumes a linear activation function for the neurons in the output layer.

All the above characteristics, the topology of the network and the weight matrix of its connections, define a specific implementation and are usually called \textit{model}.
The model, however, is only part of the story. In fact, in order to find the model that best fits the data in a specific problem, one has to provide the network with a set of examples, id est of objects for which the final output is known by independent means. These data form the so called \textit{training set} or \textit{Knowledge Base} (KB) and through a learning rule are used by the network to find the optimal model.\\
In our implementation we choose as learning rule the Quasi Newton Algorithm (QNA), which differs from the Newton Algorithm in how the Hessian of the error function is computed.
Newtonian models are variable metric methods used to find local maxima and minima of functions \citep{davidon1968} and, in the case of MLPs, they can be used to find the stationary (i.e. the zero gradient) point of the learning function. The complete mathematical description of the MLP with QNA model is reported in the section \ref{qso:appendixA}.\\
The model is publicly made available through the DAMEWARE (DAta Mining \& Exploration Web Application REsource\footnote{\textit{http://dame.dsf.unina.it/beta\_info.html}}; \citealt{cavuoti2012c}).\\

\subsection{The implementation of MLPQNA}\label{qso:sec:mlpqna}
In this work we use our implementation of the QNA based on the limited-memory BFGS (L-BFGS; \citealt{byrd1994}), where BFGS is the acronym composed of the names of the four inventors \citep{broyden1970, fletcher1970, goldfarb1970, shanno1970}.\\

Summarising, the algorithm for MLP with QNA is the following:

Let us consider a generic MLP with $w^{(t)}$ being the weight vector at time $(t)$.

\begin{enumerate}
\item Initialize all weights $w^{(0)}$ with small random values (typically normalized in $[-1, 1]$), set constant $\varepsilon$ and $t = 0$;
\item present to the network all training set and calculate $E(w^{(t)})$ as the error function for the current weight configuration;
\item if $t = 0$ then $d^{(t)} = -\nabla E^{(t)}$
\item else $d^{(t)}=-\nabla E^{(t-1)}+Ap+B\nu$, where $p = w^{(t+1)} - w^{(t)}$ and $\nu = g^{(t+1)} - g^{(t)}$;
\item calculate $w^{(t+1)} = w^{(t)} - \alpha d^{(t)}$, where $\alpha$ is obtained by line search equation (see Eq. \ref{qso:eq6} in the Appendix);
\item calculate $A$ and $B$ for the next iteration, as reported in Eq. \ref{qso:eq19};
\item if $E(w^{(t+1)}) > \varepsilon$ then $t = t+1$ and goto (ii), else STOP.
\end{enumerate}

As it is known, all \textit{line search} methods, being based on techniques which search for the minimum error by exploring the error function surface, are likely to get stuck in a local minimum and many solutions have been proposed \citep{floudas2005}.
In order to accelerate the convergence of Gradient Descent Analysis (GDA, see Appendix), Newton's method uses the information on the second-order derivatives. QNA is able to better optimize the convergence time by approximating second-order information with first-order terms \citep{shanno1990}.

By having information on the second derivatives, QNA is able to avoid local minima of the error function and to be more precise in the error function trend follow-up, thus revealing a \textit{natural} capability to find  the absolute minimum error of the optimization problem.

In the L-BFGS version of the algorithm, in the case of high dimensionality (i.e. input data with many parameters), the amount of memory required to store the Hessian is too big, along with the machine time required to process it. Therefore, instead of using a complete number of gradient values to generate the Hessian, we can use a smaller number of values.

On the one hand, the convergence slows down, performances could even increase. A statement which only a first sight might seem paradoxical but, while the convergence is measured by the number of iterations, the performance depends on the number of processor's time units spent to calculate the result.

Related to the computational cost there is also the strategy adopted in terms of stopping criteria for the method. As it is known, the process of adjusting the weights based on the gradients is repeated until a minimum is reached. In practice, one has to decide the stopping condition of the algorithm. Among the possible criteria, the algorithm could be terminated when: (i) the gradient becomes sufficiently small (by definition the gradient will be zero at a minimum); (ii) based on the error to be minimized, in terms of a fixed threshold; (iii) based on the cross validation.\\
The cross validation can be used to monitor generalization performance during training and to terminate the algorithm when there is no more improvement. Statistically significant results come out by trying multiple independent data partitions and then averaging the performances.
There are several variants of cross validation methods \citep{sylvain2010}. We, in particular, have chosen the k-fold cross validation, particularly suited in presence of a scarcity of known data samples \citep{geisser1975}. The mechanism, also known as \textit{leave-one-out},  is quite simple, since it consists in dividing the training set of $N$ samples into $k$ subsets ($k > 1$). The model is then trained on $N-1$ subsets and validated by testing it on the left out subset. This procedure is then iterated leaving out each time a different subset for validation and its squared error is averaged on all cycles.

\subsection{The Quasi Newton learning rule}
\label{qso:appendixA}

Most Newton methods use the Hessian of the function to find the stationary point of a quadratic form. It needs to be stressed, however, that the Hessian of a function is not always available and in many cases it is far too complex to be computed in an analytical way.
More often it is easier to compute the function gradient which can be used to approximate the Hessian via $N$ consequent gradient calculations.
In order to better understand why QNA are so powerful, it is convenient to start from the classical and quite common  Gradient Descent Algorithm (GDA) used for Back Propagation \citep{bishop2006}. In GDA, the direction of each updating step for the MLP weights is derived from the error descent gradient, while the length of the step is determined from the learning rate. This method is inaccurate and ineffective and therefore may get stuck in local minima.
A more effective approach is to move towards the negative direction of the gradient (\textit{line search direction}) not by a fixed step, but by moving towards the minimum of the function along that direction.  This can be achieved by first deriving the descent gradient and then by analyzing it with the variation of the learning rate \citep{brescia2012d}.
Let us suppose that at step $t$, the current weight vector is $w^{(t)}$, and let us consider a search direction $d^{(t)} =  - \nabla E^{(t)}$.
If we select the parameter $\lambda$ in order to minimize $E(\lambda) = E({w^{(t)} + \lambda d^{(t)} })$, the new weight vector can be expressed as:

\begin{equation}
w^{({t + 1})}  = w^{(t)}  + \lambda d^{(t)}
\end{equation}\label{qso:eq1}

\noindent and the problem of \textit{line search} becomes a  1-dimensional minimization problem which can be solved in many different ways.
Simple variants are: i) to move $E(\lambda)$ by varying $\lambda$ by small intervals, then evaluate the error function at each new position,
and stop when the error begins to increase, or ii) to use the parabolic search for a minimum and compute the parabolic curve crossing pre-defined learning rate points. The minimum $d$ of the parabolic curve is a good approximation of the minimum of $E(\lambda)$ and it can be derived by means of the parabolic curve which crosses the fixed points with the lowest error values.

Another approach makes instead use of \textit{trust region} based strategies which minimize the error function, by iteratively growing or contracting the region of the function by adjusting a quadratic model function which best approximates the error function.
In this sense this technique can be considered as a dual to line search, since it tries to find the best size of the region by fixing the step size (while the line search strategy always chooses the step direction before selecting the step size), \citep{celis1985}.
All these approaches, however, rely on the assumption that the optimal search direction is given at each step by the negative gradient: an assumption which not only is not always true, but can also lead to serious wrong convergence.
In fact, if the minimization is done along the negative gradient direction, the subsequent search direction (the new gradient) will be orthogonal to the previous one: in fact, note that when the line search founds the minimum, it is:
\begin{equation}
\frac{{\partial E}}{{\partial \lambda }}({w^{(t)} + \lambda d^{(t)} }) = 0
\label{qso:eq2}
\end{equation}
and hence,
\begin{equation}
g^{({t + 1})T} d^{(t)} = 0
\label{qso:eq3}
\end{equation}
where $g \equiv \nabla E$.
The iteration of the process therefore leads to oscillations of the error function which slow down the convergence process.

The method implemented here relies on selecting other directions so that the gradient component, parallel to the previous search direction, would remain unchanged at each step.
Suppose that you have already minimized with respect to the direction $d^{(t)}$ starting from the point $w^{(t)}$ and reaching the point $w^{(t+1)}$, where Eq. \ref{qso:eq3} becomes:
\begin{equation}\label{qso:eq4}
g({w^{({t + 1})}})^T d^{(t)} = 0
\end{equation}
\noindent by choosing $d^{(t+1)}$ so to preserve the gradient component parallel to $d^{(t)}$ equal to zero, it is possible to build a sequence of directions $d$ in such a way that each direction is conjugated to the previous one on the dimension $|w|$ of the search space (this is known as conjugate gradients method; \citealt{golub1999}).

In presence of a squared error function, the update weights algorithm is:
\begin{equation}
w^{({t + 1})} = w^{(t)} + \alpha ^{(t)} d^{(t)}
\label{qso:eq5}
\end{equation}
with:
\begin{equation}
\alpha ^{(t)} = - \frac{{d^{(t)T} g^{(t)} }}{{d^{(t)T} Hd^{(t)} }}
\label{qso:eq6}
\end{equation}

Furthermore, $d$ can be obtained for the first time via the negative gradient and in the subsequent iterations,  as a linear combination of the current gradient and of the previous search directions:
\begin{equation}
d^{({t + 1})} = - g^{({t + 1})} + \beta ^{(t)} d^{(t)}
\label{qso:eq7}
\end{equation}
with:
\begin{equation}
\beta ^{(t)} = \frac{{g^{({t + 1})T} Hd^{(t)} }}{{d^{(t)T} Hd^{(t)} }}
\label{qso:eq8}
\end{equation}

This algorithm finds the minimum of a square error function at most in $|w|$ steps but at the price of a high computational cost, since in order to determine the values of $\alpha$ and $\beta$, it makes use of that \textit{hessian matrix H} which, as we already mentioned, is very demanding in terms of computing time. A fact which puts serious constraints on the application of this family of methods to medium/large data sets.  Excellent approximations for the coefficients $\alpha$ and $\beta$ can, however, be obtained from analytical expressions that do not use the Hessian matrix explicitly.
For instance, $\beta$ can be calculated through any one of the following expressions (respectively \citealt{hestenes1952, fletcher1964, polak1969}):

\begin{equation}
Hestenes-Sitefel: \beta ^{(t)} = \frac{{g^{({t + 1})T}( {g^{( {t + 1})} - g^{(t)}})}}{{d^{(t)T}({g^{({t + 1})} - g^{(t)}})}}
\label{qso:eq9}
\end{equation}
\begin{equation}
Fletcher-Reeves: \beta ^{(t)} = \frac{{g^{({t + 1})T} g^{({t + 1})} }}{{g^{(t)T} g^{(t)} }}
\label{qso:eq10}
\end{equation}
\begin{equation}
Polak-Ribiere: \beta ^{(t)} = \frac{{g^{({t + 1})T} ( {g^{({t + 1})} - g^{(t)} })}}{{g^{(t)T}g^{(t)}}}
\label{qso:eq11}
\end{equation}

\noindent These expressions are all equivalent if the error function is square-typed, otherwise they assume different values. Typically the Polak-Ribiere equation obtains better results because, if the algorithm is slow and subsequent gradients are quite alike between them, its equation produces values of $\beta$ such that the search direction tends to assume the negative gradient direction \citep{vetterling1992}.

Concerning the parameter $\alpha$, its value can be obtained by using the line search method directly. The method of conjugate gradients reduces the number of steps to minimize the error up to a maximum of $|w|$ because there could be almost $|w|$ conjugate directions in a $|w|$-dimensional space. In practice however, the algorithm is slower because, during the learning process, the property \textit{conjugate} of the search directions tends to deteriorate.
It is useful, to avoid the deterioration, to restart the algorithm after $|w|$ steps, by resetting the search direction with the negative gradient direction.

By using a local square approximation of the error function, we can obtain an expression for the minimum position. The gradient in every point $w$ is in fact given by:
\begin{equation}
\nabla E = H \times ( {w - w^* })
\label{qso:eq12}
\end{equation}
where $w^*$ corresponds to the minimum of the error function, which satisfies the condition:
\begin{equation}
w^* = w - H^{ - 1} \times \nabla E
\label{qso:eq13}
\end{equation}

The vector $- H^{ - 1}  \times \nabla E$ is known as Newton direction and it is the base for a variety of optimization strategies, such as for instance the QNA, which instead of calculating the $H$ matrix and then its inverse, uses a series of intermediate steps of lower computational cost to generate a sequence of matrices which are more and more accurate approximations of $H^{ - 1}$.

From the Newton formula (Eq. \ref{qso:eq13}) we note that the weight vectors on steps $t$ and $t+1$ are correlated to the correspondent gradients by the formula:
\begin{equation}
w^{({t + 1})} - w^{(t)} = - H^{({ - 1})}( {g^{({t + 1})} - g^{(t)}})
\label{qso:eq14}
\end{equation}
which is known as \textit{Quasi Newton Condition}. The approximation $G$ is therefore built in order to satisfy this condition. The formula for $G$ is:
\begin{equation}
G^{({t + 1})} = G^{(t)} + \frac{{pp^T }}{{p^T \nu }} - \frac{{({G^{(t)} \nu })\nu ^T G^{(t)}}}{{\nu ^T G^{(t)} \nu }} + ( {\nu ^T G^{(t)} \nu })uu^T
\label{qso:eq15}
\end{equation}
where the vectors are:
\begin{equation}
p = w^{({t + 1})} - w^{(t)}; \nu = g^{( {t + 1})} - g^{(t)}; u = \frac{p}{{p^T \nu }} - \frac{{G^{(t)} \nu }}{{\nu ^T G^{(t)} \nu }}
\label{qso:eq16}
\end{equation}

Using the identity matrix to initialize the procedure is equivalent to consider, step by step, the direction of the negative gradient while, at each next step, the direction $-Gg$ is for sure a descent direction. The above expression could carry the search out of the interval of validity for the squared approximation. The solution is hence to use the \textit{line search} to found the minimum of function along the search direction.
By using such system, the weight updating expression (Eq. \ref{qso:eq5}) can be formulated as follows:
\begin{equation}
w^{({t + 1})} = w^{(t)} + \alpha ^{(t)} G^{(T)} g^{(t)}
\label{qso:eq17}
\end{equation}
where $\alpha$ is obtained by the \textit{line search}.

One of the main advantage of QNA, compared with conjugate gradients, is that the \textit{line search} does not require the calculation of $\alpha$ with a high precision, because it is not a critical parameter. Unfortunately, however, again, it requires a large amount of memory to calculate the matrix $G$ ($|w| \times |w|$), for large $|w|$.
One way to reduce the required memory is to replace at each step the matrix $G$ with a unitary matrix. With such replacement and after multiplying by $g$ (the current gradient), we obtain:
\begin{equation}
d^{({t + 1})} = - g^{(t)} + Ap + B\nu
\label{qso:eq18}
\end{equation}

Note that if the line search returns exact values, then the above equation produces mutually conjugate directions. $A$ and $B$ are scalar values defined as:
\begin{equation}
\begin{array}{l}
 A =  - ({1 + \frac{{\nu ^T \nu }}{{p^T \nu }}})\frac{{p^T g^{({t + 1})} }}{{p^T \nu }} + \frac{{\nu ^T g^{({t + 1})} }}{{p^T \nu }} \\
 \\
 B = \frac{{p^T g^{({t + 1})} }}{{p^T \nu }} \\
 \end{array}
\label{qso:eq19}
\end{equation}

    \section[Model 3: GAME]{Model 3: GAME - Genetic Algorithm Model Experiment}\label{chap:GAME}
        \blfootnote{this section is largely extracted from: \tiny
\begin{itemize}
\item \textbf{Cavuoti, S.}; Garofalo, M.; Brescia , M.; Paolillo, M.; Pescape', A.; Longo, G.; Ventre, G.; GPUs for astrophysical data mining. A test on the search for candidate globular clusters in external galaxies, \textbf{Submitted to New Astronomy, accepted}
			\item Brescia, M.; \textbf{Cavuoti, S.}; Paolillo, M.; Longo, G.; Puzia, T.; 2012, The detection of Globular Clusters in galaxies as a data mining problem, \textbf{MNRAS, Volume 421, Issue 2, pp. 1155-1165}, available at \href{http://arxiv.org/abs/1110.2144v1}{arXiv:1110.2144v1}.
			\item \textbf{Cavuoti, S.}; Garofalo, M.; Brescia, M.; Pescape', A.; Longo, G.; Ventre, G., 2012, Genetic Algorithm Modeling with GPU Parallel Computing Technology, 22nd WIRN, Italian Workshop on Neural Networks, Vietri sul Mare, Salerno, Italy, May 17-19 \textbf{refereed proceeding}
\end{itemize}}
%
For what scalability is concerned,  there are good reasons to believe that in the near future most aspects of computing will see exponential growth in bandwidth, but sub-linear or no improvements at all in latency. Moore's Law \citep{moore1965} will continue to deliver exponential increases in memory size but the speed with which data can be transferred between memory and Central Processing Units (CPUs) will remain more or less constant and marginal improvements can only be made through advances in
caching technology \citep{akhter2006}.
Certainly Moore's law still leaves room for the creation of parallel computing capabilities on single chips by packing multiple CPU cores onto it, but the clock speed that determines the speed of computation is constrained to remain limited by a thermal wall \citep{sutter2005}.
Thus it has to be expected that individual CPUs will not become much faster in the near future, even though they will have the parallel computing power and storage capacity that today can be provided only via specialized hardware. From the application development point of view, to build programs capable to exploit low latency
in multi core CPUs will require a fundamental shift from the currently sequential or parallel programming, to a mixture of parallel and distributed programming.
In recent years a few groups have began to test the possibility to solve some of the above problems by means of  General Purpose Graphical Processing Unit (known as GPGPU; \citealt{nvidia2012}) parallel technologies. Among the most known parallel programming models, we quote the OpenCL\footnote{\url{http://www.khronos.org/registry/cl/sdk/1.0/docs/man/xhtml/}} (Open Computing Language), which is a foundation layer of platform-independent tools; CUDA\footnote{\url{https://developer.nvidia.com/what-cuda}} (Compute Unified Device Architecture): a general purpose architecture that leverages the parallel compute engine within GPUs to solve many complex computational problems in a more efficient way than on a single CPU and, finally, OpenACC\footnote{\url{http://www.openacc-standard.org/}}, a set of compiling directives allowing programmers to create high-level CPU+GPU applications without the need to explicitly manage the parallel architecture configuration and its communication with the CPU host.
In astronomy, pioneering attempts to use the Graphical Processing Unit (GPU) technology have been made in an handful of applications. GPUs were attempted mainly to speed up numerical simulations (e.g. \citealt{nakasato2011, valdez2012}) of various types from traditional N-body to fluid dynamics; but also to perform complex repetitive numerical tasks such as computation of correlation functions \citep{ponce2012}, real time processing of data streams \citep{barsdell2012}, and complex visualization \citep{hassan2012}, to quote just a few.\\
In this section I present the implementation on GPUs of the multi-purpose Genetic Algorithm Model Experiment (GAME), capable to deal with both regression and classification problems. The serial version of this algorithm was implemented by our groups to be deployed on the DAME hybrid distributed infrastructure and made available to the community through the DAMEWARE (DAME Web Application REsource; \citealt{brescia2010b,djorgovski2012d}) web application.
The present paragraph is structured as follows: Sec. \ref{game:game} presents the design and serial implementation of the GAME model; describes the knowledge base (astrophysical dataset), used for the experiments; Sec. \ref{game:GPU} is dedicated to the overview of the GPU technology in the astrophysical context, Sec \ref{sec:parallelization} present the parallelization strategy while the experiments and results are described in Sec. \ref{game:experiment}, just before our conclusions.

\subsection{GAME}\label{game:game}
Genetic algorithms \citep{mitchell1998} are derived from Darwin's evolution law \citep{darwin1859} and are intrinsically parallel in their learning evolution rule and processing data patterns. This implies that the parallel computing paradigm can lead to a strong optimization in terms of processing performances.
Genetic and evolutionary algorithms  are an important family of supervised Machine Learning models which tries to emulate the evolutionary laws that  allow living organisms to adapt their life style to the outdoor environment and survive.\\
In the optimization problems which can be effectively tackled with genetic / evolutionary models, a key concept is the evaluation function (or \textit{fitness} in evolutionary jargon), used to determine which possible solutions get passed on to multiply and mutate into the next generation of solutions. This is usually done by analyzing the \textit{genes}, which refer to
some useful information about a particular solution to the problem under investigation.
The fitness function looks at the genes, performs some qualitative assessment and then returns the fitness value for that solution. The remaining parts of the genetic algorithm discard all solutions having a poor fitness value, and accept only those with a good fitness value. Frequently, the fitness function has not a closed mathematical expression and, for instance, in some cases it may be given under the form of the simulation of a real physical system.\\
GAME is a pure genetic algorithm specially designed to solve supervised optimization problems related with regression and classification functionalities.  It is scalable to efficiently manage massive datasets and is based on the usual genetic evolution methods which will be shortly outlined in what follows.\\
As a supervised machine learning technique, GAME requires to select and create the data parameter space, i.e. to create the data input patterns to be evaluated. It is important in this phase to build homogeneous patterns, i.e. with each pattern having the same type and number of parameters (or features), as well as to prepare the datasets which are needed for the different experiment steps: training, validation and test sets, by splitting the parameter space into variable subsets to be submitted at each phase. The dataset must include also target values for each input pattern, i.e. the desired output values, coming from any available knowledge source.\\
From an analytic point of view, in the data mining domain, a pattern is a single sample of a real problem, composed by a problem-dependent number of features (sometimes called attributes), which is characterized by a certain S/N (Signal-to-Noise) ratio. In other words, such pattern contains an unspecified amount of information, hidden among its correlated features, that an unknown analytical relation is able to transform into the correct output. Usually, in real cases (such as the astronomical ones), the S/N ratio is low, which means that feature correlation of a pattern is \textit{masked} by a considerable amount of noise (intrinsic to the phenomenon and/or due to the acquisition system); but the unknown correlation function can ever be approximated with a polynomial expansion.

The generic function of a polynomial sequence is based on these simple considerations:\\
Given a generic dataset with N features and a target $t$, let $pat$ be a generic input pattern of the dataset, $pat = (f_1, ..., f_N, t)$ and $g(x)$ a generic real function.
The representation of a generic feature $f_i$ of a generic pattern, with a polynomial sequence of degree $d$ is:

\begin{equation}
G({f_i}) \cong a_0 + a_1g({f_i}) + ... + a_dg^d(f_i )
\label{game:eq1}
\end{equation}

Hence, the k-th pattern $(pat_k)$ with N features may be written as:

\begin{equation}
Out({pat_k}) \cong \sum\limits_{i = 1}^N {G({f_i}) \cong a_0 + \sum\limits_{i = 1}^N {\sum\limits_{j = 1}^d {a_j g^j({f_i})} } }
\label{game:eq2}
\end{equation}

and the target $t_k$, related to pattern $pat_k$, can be used to evaluate the approximation error (the fitness) of the input pattern to the expected value:

\begin{equation}
E_k  = ({t_k - Out({pat_k})})^2
\label{game:eq3}
\end{equation}

If we generalize eq.~\ref{game:eq2} to an entire dataset, with NP number of patterns $(k = 1,..., NP)$, the \textit{forward} phase of the GA (Genetic Algorithm) consists of the calculation of NP expressions of the eq.~\ref{game:eq2}, which represent the polynomial approximation of the dataset.\\
In order to evaluate the fitness of the NP patterns as extension of eq.~\ref{game:eq3}, the Mean Square Error (MSE) or Root Mean Square Error (RMSE) may be used:

\begin{equation}
MSE  = \frac{\sum\limits_{k = 1}^{NP} {(t_k - Out(pat_k))^2 }} { NP }
\label{game:eq4}
\end{equation}

\begin{equation}
RMSE  = \sqrt{\frac{\sum\limits_{k = 1}^{NP} {(t_k - Out(pat_k))^2 }} { NP }}
\label{game:eq5}
\end{equation}\\

On the basis of the above, we obtain a GA with the following characteristics:

\begin{itemize}\addtolength{\itemsep}{-0.5\baselineskip}
\item The expression in eq.~\ref{game:eq2} is the \textit{fitness} function;
\item The array $(a_0,..., a_M)$ defines M genes of the generic chromosome (initially they are generated random and normalized between -1 and +1);
\item All the chromosomes have the same size (constrain from a classic GA);
\item The expression in eq.~\ref{game:eq3} gives the standard error to evaluate the fitness level of the chromosomes;
\item The population (genome) is composed by a number of chromosomes imposed from the choice of the function $g(x)$ of the polynomial sequence.
\end{itemize}

The number of chromosomes is determined by the following expression:

\begin{equation}
NUM_{chromosomes}  = (B \cdot N) + 1
\label{game:eq6}
\end{equation}

where $N$ is the number of features of the patterns and $B$ is a multiplicative factor that depends on the $g(x)$ function, which in the simplest case is just 1, but can be even 3 or 4 in more complex cases. The parameter $B$ also influences the dimension of each chromosome (number of genes):

\begin{equation}
NUM_{genes}  = (B \cdot d) + 1
\label{game:eq7}
\end{equation}

where $d$ is the degree of the polynomial. In more general terms, the polynomial expansion within the fitness function, defined in eq.~\ref{game:eq2}, could be arbitrarily chosen and we conceived the implementation code accordingly. In the specific context of this work we decided to adopt the trigonometric polynomial expansion, given by the following expression (hereinafter \textit{polytrigo}):

\begin{equation}
g(x) = a_0  + \sum\limits_{m = 1}^d {a_m \cos (mx) + } \sum\limits_{m = 1}^d {b_m \sin (mx)}
\label{game:eq8}
\end{equation}

In order to have $NP$ patterns composed by $N$ features, the expression using eq.~\ref{game:eq2} with degree $d$, is:

\begin{eqnarray}\label{game:eq9}\nonumber
Out(pat_{k = 1...NP} ) \cong \sum_{i = 1}^{N} G(f_i ) \cong a_0  + \\
+ \sum_{i = 1}^{N} \sum_{j = 1}^{d} a_j \cos (jf_i ) +
\sum_{i = 1}^{N}
\sum_{j= 1}^d b_j \sin (jf_i )
\end{eqnarray}

In the last expression we have two groups of coefficients (sine and cosine), so $B$ will assume the value $2$.
Hence, the generic genome (i.e. the population at a generic evolution stage), will be composed by $2N+1$ chromosomes, given by eq.~\ref{game:eq6}, each one with $2d+1$ genes $[a_0, a_1,...,a_d, b_1, b_2,...,b_d]$, given by eq.~\ref{game:eq7}, with each single gene (coefficient of the polynomial) in the range $[-1, +1]$ and initially randomly generated.\\

By evaluating the goodness of a solution through the MSE or RMSE metrics, it may happen that a better solution in terms of MSE corresponds to a worse solution for the model, for example when we have a simple crispy
(jargon for classification with sharp boundaries between adjacent classes) classification problem with two patterns (class types 0 and 1).\\
As an example, if the solutions are, respectively, 0.49 for the class 0 and 0.51 for the class 1, the efficiency (i.e. the percentage of objects correctly classified) is 100\%, with a $MSE = 0.24$. But a solution of 0 for the class 0 and 0.49 for the class 1 (efficiency of 50\%), gives back a $MSE = 0.13$ and  consequently the model will prefer the second solution, although with a lower efficiency.

In order to circumvent this problem, we decided to implement in GAME what we could call a \textit{convergence tube} or \textit{Threshold MSE} (TMSE):\\

\begin{equation}
\left\{ \begin{array}{l}
 if ( {t_k  - Out( {pat_k } )} )^2  > R \Rightarrow E_k  = ( {t_k  - Out( {pat_k } )} )^2  \\
 if ( {t_k  - Out( {pat_k } )} )^2  \le R \Rightarrow E_k  = 0 \\
 \end{array} \right.
\label{game:eq10}
\end{equation}\\

where $R$ is a user defined numerical threshold in the $]0,1[$ range.\\
In the previous example, by using $R = 0.5$ in eq.~\ref{game:eq10}, we obtain a $TMSE = 0$ and $TMSE = 0.13$, respectively in the first and second cases, thus indicating that the first solution is better than the second one, as expected.\\
Through several cycles, the population of chromosomes, originally started as a random generation (typically following a normal distribution), is replaced at each step by a new one which, by having higher fitness value, represents a step forward towards the best population (i.e. the optimal solution).
Typical representations used are the binary code or the real values in [-1, 1] for each element (gene) of a chromosome. \\

As it is typical for the supervised machine learning models, the classification experiment with GAME is organized in a sequence of functional use cases:

\begin{itemize}\addtolength{\itemsep}{-0.5\baselineskip}
\item \textit{train}: the first mandatory case, consisting into submitting training datasets in order to build and store the best GA population, where best is in terms of its problem solving capability;
\item \textit{test}: case to be used in order to verify and validate the learning performance of the trained GA;
\item \textit{run}: the normal execution mode after training and validation;
\item \textit{full}: a workflow case, including \textit{train} and \textit{test} cases in sequence.
\end{itemize}

{The training phase is the most important use case, and is based on a user defined loop of iterations (schematically shown in Fig.~\ref{game:figure1}), whose main goal is to find the population of chromosomes (i.e. the solutions to the classification problem), having the maximum fitness.\\
\textbf{Three types of random generation criteria can be adopted:}
\begin{itemize}\addtolength{\itemsep}{-0.5\baselineskip}
\item pseudo-random values from [-1, +1];
\item random values from the normal distribution in [-1, +1];
\item pseudo-random values from [0, 1].
\end{itemize}

\begin{figure}
\centering
\includegraphics[height=12cm]{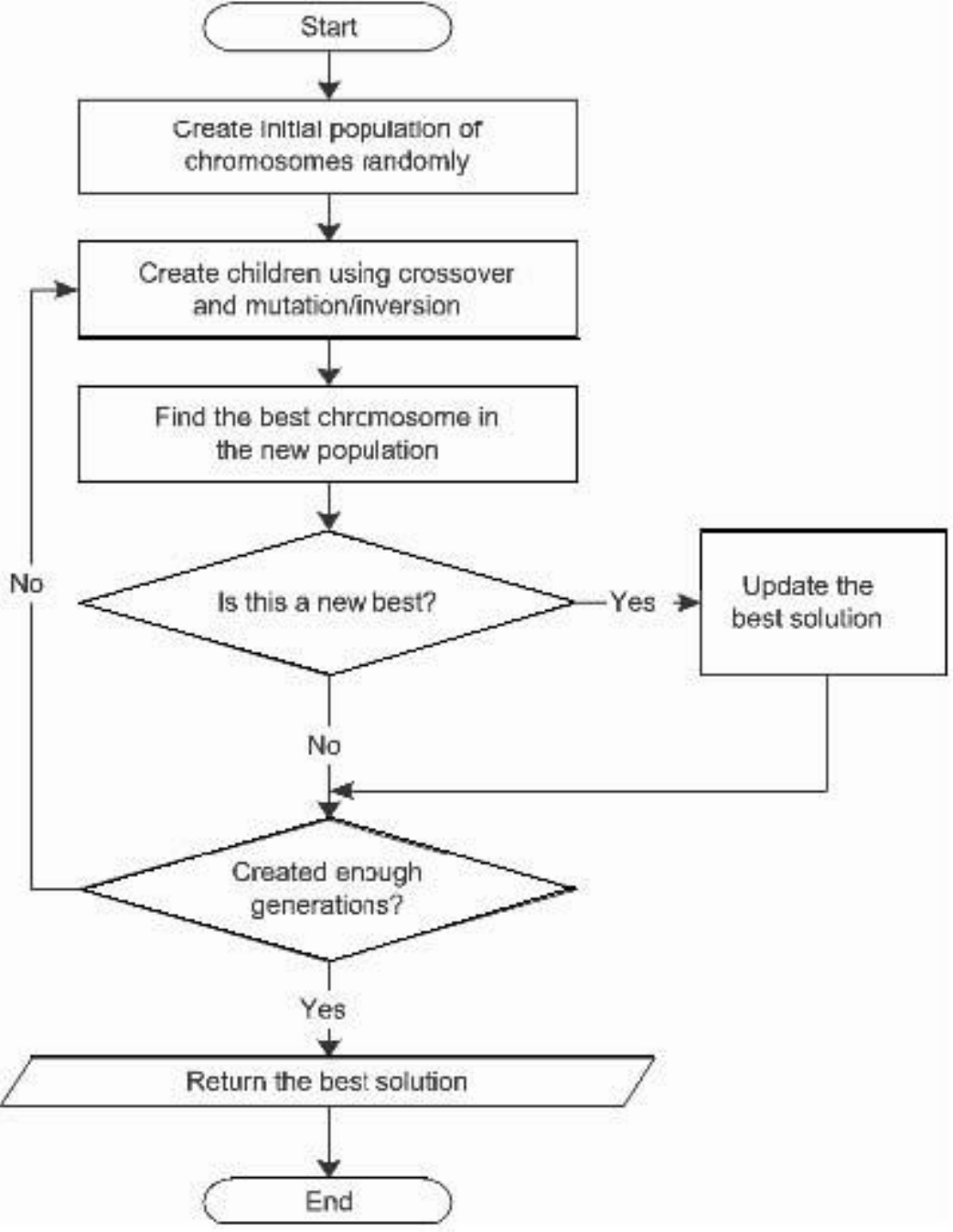}
\caption{The flow-chart of the CPU-based GAME model implementation.}
\label{game:figure1}
\end{figure}

At each training step there are the following options, which can be chosen by the user during the configuration phase:

\begin{itemize}\addtolength{\itemsep}{-0.5\baselineskip}
\item \textit{error function} type: MSE, TMSE or RMSE;
\item \textit{selecting criterion}: the genetic selection function to evolve the population of chromosomes at each training iteration. It is possible to choose between the classical \textit{ranking} and \textit{roulette} types \citep{mitchell1998};
\item \textbf{\textit{Elitism}, which is a further mechanism to evolve the population.} The user has to choose the number of copies of the winner chromosome, to be maintained unchanged in the next generation. Main role of elitism is indeed to preserve the best fitness along evolution \citep{mitchell1998}.
\end{itemize}

In the present work, the idea was to build a GA able to solve supervised crispy classification and regression problems, typically related to a high-complexity parameter space where the background analytic function is not known, except for a limited number of couples of input-target values, representing a sample of known solutions for a physical category of phenomena (also indicated as the Knowledge Base, or KB). The goal is indeed to find the best chromosomes so that the related polynomial expansion is able to approximate the correct classification of the input patterns. Therefore in our case the fitness function is evaluated in terms of the training error, obtained as the absolute difference between the target value and the polynomial expansion output for all patterns.

The training phase is the most important use case, and is based on a user defined loop of iterations (schematically shown in Fig.~\ref{game:figure1}), whose main goal is to find the population of chromosomes (i.e. the solutions to the classification problem), having the maximum fitness. At each iteration, the polynomial expansion (whose coefficients are the genes of each chromosome in the current population), is applied to all input patterns (\textit{forward} phase). Then the error between the output of the polynomial expansions and the pattern targets is evaluated by comparing it with the user defined threshold.
Until the derived error remains higher than the threshold, a further evolution of the genetic population is achieved  and each chromosome is updated by means of the application of classical genetic operators, i.e. crossover and mutation (\textit{backward} phase) \citep{mitchell1998}.\\
At the end of each iteration, the new (evolved) genetic population is ready to be used in the next loop.

\subsubsection{GPU}\label{game:GPU}
As mentioned before whenever there is a large quantity of data, there are two approaches to making learning feasible. The first one is trivial, consisting of applying the training scheme to a decimated data set. Obviously, in this case, the information may be easily lost and there is no guarantee that this loss is negligible in terms of correlation discovery. This approach, however, may turn very useful in the lengthy optimization procedure that is required by many machine learning methods (such as neural networks or genetic algorithms). The second method relies in splitting the problem in smaller parts (parallelization) sending them to different CPUs (Central Processing Units) and finally combine the results together. However, implementation of parallelized versions of learning algorithms is not always easy \citep{rajaraman2010}, and this approach should be followed only when the learning rule, such as in the case of genetic algorithms \citep{meng2009}, is intrinsically parallel.
GPGPU is an acronym standing for General Purpose Computing on Graphics Processing Units. It was invented by Mark Harris in 2002 \citep{harris2003} by recognizing the trend to employ GPU technology for not graphic applications.
In general the graphic chips, due to their intrinsic nature of multi-core processors (many-core) and being based on hundreds of floating-point specialized processing units, make many algorithms able to obtain higher (one or two orders of magnitude) performances than usual CPUs. They are also cheaper, due to the relatively low price of graphic chip components.
Particularly useful for super-computing applications, often requiring several execution days on large computing clusters, the GPGPU paradigm may drastically decrease execution times, by promoting research in a large variety of scientific and social fields (such as, for instance, astrophysics, biology, chemistry, physics, finance, video encoding and so on).
The critical points for traditional multi-core CPU architecture come out in case of serial programs. In this case, in the absence of the parallel approach, the processes are scheduled in such a way that the full load on the CPU is balanced, by distributing them over the less busy cores each time. However many software products are not designed to fully exploit the multi-core features, so far the micro-processors are designed to optimize the execution speed on sequential programs.
The choice of graphic device manufacturers, like NVIDIA Corp., was the many-core technology (usually many-core is intended for multi-core systems over 32 cores). The many-core paradigm is based on the growth of execution speed for parallel applications. Began with tens of cores smaller than CPU ones, such kind of architectures reached hundreds of core per chip in a few years. Since 2009 the throughput peak ratio between GPU (many-core) and CPU (multi-core) was about 10:1. It must be issued that such values are referred mainly to the theoretical speed supported by such chips, i.e. 1 TeraFLOPS against 100 GFLOPS. Such a large difference has pushed many developers to shift more computing-expensive parts of their programs on the GPUs.
Therefore, astronomers should perform a cost-benefit analysis and some initial development to investigate whether their code could benefit from running on a GPU. Used in the right way and on the right applications, GPUs will be a powerful tool for astronomers processing huge volumes of data.
In a recent paper, Barsdell, Barnes and Fluke \citep{barsdell2010} have analyzed astronomy algorithms to understand which algorithms can be best engineered to run on GPUs. The authors used algorithm analysis to understand  how algorithms can be optimized to run in parallel in a GPU environment (as opposed to implementation optimization).

Broadly speaking, the following are features for high efficiency parallelized algorithms:
\begin{itemize}
\item	Repetitive operations can be parallelized into many fine-grained elements;
\item	Standardized mechanisms to access locations in memory;
\item	Minimize processing threads which execute different instructions;
\item	Parallelize high intensity arithmetic operations;
\item	Minimize memory transfers between GPU and host CPU.
\end{itemize}

%

However, before 2006, GPUs were rather difficult to use, mainly because in order to access the graphic devices, programmers were conditioned to use specific APIs (Application Programming Interfaces), based on methods available through libraries like OpenGL\footnote{\url{http://www.opengl.org}} and Direct3D\footnote{\url{http://msdn.microsoft.com}}.
The limits and poor documentation of these APIs were strongly undermining the design and development of applications.\\
In order to better address the GPU-based design, starting from the CPU-based serial implementation of the GAME model, we chose the \textit{ad hoc} software development methodology \textit{APOD} (Assess, Parallelize, Optimize, and Deploy), see \citet{nvidia2012}.
The APOD design mechanism consists first in quickly identifying the portions of code that
can easily exploit and benefit from GPU acceleration, then in exploiting the optimization
task as fast as possible, see Fig.~\ref{game:figure2}. The
APOD is a cyclical process: initial speedups can be achieved, tested, and deployed quickly,
so that the cycle can start over to identify further optimization opportunities.
At each APOD cycle the identification of the parts of the code responsible for most of the execution time
is done by investigating the parallelized GPU accelerations.
An upper limit to performance improvement can be derived considering requirements and constraints,
and by applying the Amdahl \citep{amdahl1967} and Gustafson \citep{gustafson1988} laws.

\begin{figure}
\centering
\includegraphics[width=7cm]{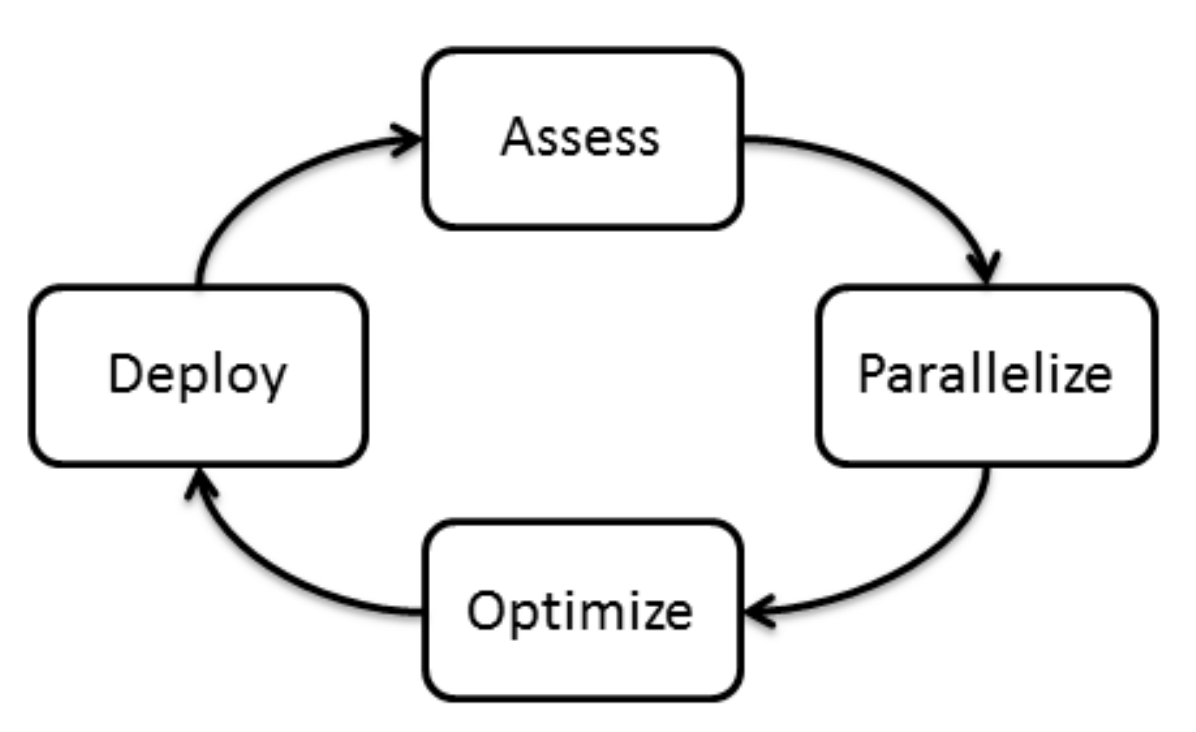}
\caption{The APOD parallel programming cyclic design mechanism.}
\label{game:figure2}
\end{figure}

By exposing the parallelism to improve performance and simply maintaining the code of sequential applications, we are able to ensure also the maximum parallel throughput on GPU platform. This could be as simple as adding a few preprocessor directives, such as OpenMP and OpenACC, or it can be done by calling an existing GPU-optimized library such as cuBLAS, cuFFT, or Thrust. Specifically, Thrust \citep{bell2011} is a parallel C++ template library resembling the C++ STL (Standard Template Library), described in \citet{stepanov1995}, which provides a rich collection of data parallel primitives such as scan, sort, and reduce, that can be composed together to implement complex algorithms with concise, readable source code.
After the parallelization step is complete, we can optimize the code to improve the performance. This phase is based on (i) maximizing parallel execution, identifying the most computationally expensive parts of the code, (ii) optimizing memory usage and (iii) minimizing low bandwidth host-to-device data transfers.\\

In this context we designed and developed a multi-purpose genetic algorithm (see section \ref{chap:GAME}) implemented with GPGPU/CUDA parallel computing technology. The model derives from the paradigm of supervised machine learning, addressing both the problems of classification and regression applied on massive data sets \citep{cavuoti2012a}. Since GAs are embarrassing parallel, the GPU computing paradigm has provided an exploit of the internal training features of the model, permitting a strong optimization in terms of processing performances. The use of CUDA translated into a 75x average speedup, by successfully eliminating the largest bottleneck in the multi-core CPU code. Although a speedup of up to 200X over a modern CPU is impressive, it ignores the larger picture of use a Genetic Algorithm as a whole. In any real-world the dataset can be very large (those we have previously called Massive Data Sets) and this requires greater attention to GPU memory management, in terms of scheduling and data transfers host-to-device and vice versa. Moreover, the identical results for classification and regression functional cases demonstrated the consistency of the implementation, enhancing the scalability of the proposed GAME model when approaching massive data sets problems.

\subsection{The parallelization of GAME}\label{sec:parallelization}

In all execution modes (use cases), GAME exploits the \textit{polytrigo} function defined by eq.~\ref{game:eq8}, consisting in a polynomial expansion in terms of sum of sines and cosines.

In particular, this calculation involves the following operations:
\begin{enumerate}[1.]
\item randomly generate the initial population of chromosomes;
\item calculate the fitness functions to find and sort the best chromosomes in the population;
\item evaluate the stop criteria (error threshold or maximum number of iterations);
\item stop or use the genetic operators to evolve the population and go to 2.
\end{enumerate}

Specifically in the training use case, the \textit{polytrigo} is used at each iteration to evaluate the fitness for all chromosomes of the population. It is indeed one of the critical aspects to be investigated in the parallelization process.\\
\begin{sidewaysfigure}
\centering
\includegraphics[height=11cm]{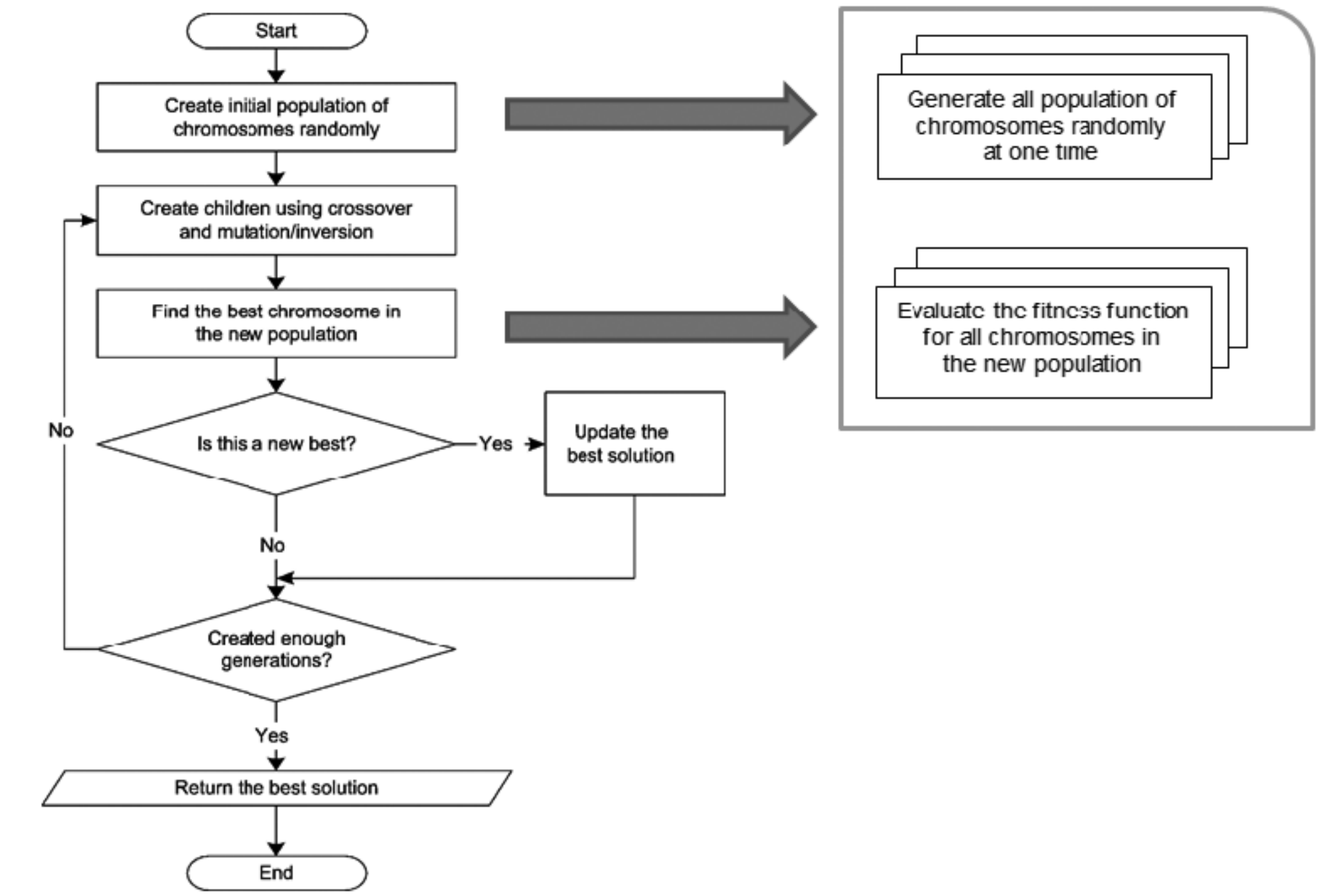}
\caption{The flow-chart of the GPU-based GAME model implementation.}
\label{game:figure3}
\end{sidewaysfigure}

The main concern in designing the software architecture for the GPUs is to analyse the partition of work: i.e. which part of the work should be done on the CPU rather than on the GPU.
As shown in Fig.~\ref{game:figure3}, coherently with the APOD process we have analyzed and identified the time consuming critical parts to be parallelized by executing them on the GPU. They are the generation of random chromosomes and the calculation of the fitness function of chromosomes.
The key principle is that we need to perform the same instruction simultaneously on as much data as possible. By adding the number of chromosomes to be randomly generated in the initial population as well as during each generation, the total number of involved elements is never extremely large but it may occur with a high frequency. This is because also during the population evolution loop a variable number of chromosomes are randomly generated to replace older individuals. To overcome this problem we may generate a large number of chromosomes randomly \textit{una tantum}, by using them whenever required. On the contrary, the evaluation of fitness functions involves all the input data, which is assumed to be massive datasets, so it already has an intrinsic data-parallelism. The function \textit{polytrigo} takes about $\sim75\%$ of the total execution time of the application, while the total including child functions amounts to about $\sim87.5\%$ of total time execution. This indeed has been our first candidate for parallelization.\\

In order to give a practical example, for the interested reader, we report the source code portions related to the different implementation of the \textit{polytrigo} function, of the serial and parallelized cases.

\medskip

\noindent
{\it C++ serial code for \textit{polytrigo} function (eq.~\ref{game:eq8}):}
\begingroup
\fontsize{7pt}{7pt}
\begin{verbatim}
for (int i = 0; i < num_features; i++) {
   for (int j = 1; j <= poly_degree; j++) {
      ret +=  v[j] * cos(j * input[i]) + v[j + poly_degree] *
              * sin(j * input[i]); } }
\end{verbatim}
\endgroup

\medskip

\noindent
{\it CUDA C (Thrust) parallelized code for \textit{polytrigo} function (eq.~\ref{game:eq8}):}
\begingroup
\fontsize{7pt}{7pt}
\begin{verbatim}
struct sinFunctor {  __host__ __device__
 double operator()(tuple <double, double> t) {
    return sin(get < 0 > (t) * get < 1 > (t)); }};
struct cosFunctor {  __host__ __device__
 double operator()(tuple <double, double> t) {
    return cos(get < 0 > (t) * get < 1 > (t)); }};
thrust::transform(thrust::make_zip_iterator(
    thrust::make_tuple(j.begin(), input.begin())),
    thrust::make_zip_iterator(
       thrust::make_tuple(j.end(), input.end())),
       ret.begin(), sinFunctor(), cosFunctor());
\end{verbatim}
\endgroup

In the serial code the vector $v[]$ is continuously updated in the inner loop by means of the polynomial degree, while the vector $input[]$ (i.e. the elements of the input dataset) is used in each calculation but never altered. Therefore, in the parallelized version we rewrite the function by calculating in advance the sums of sines and cosines, storing the results in two vectors that are used in the function  \textit{polytrigo} at each iteration. This brings huge benefits because we calculate the time consuming trigonometric functions only once rather than at every iteration, so exploiting the parallelism on large amount of data.\\
From the time complexity point of view, by assuming to have as many GPU cores as population chromosomes, the above CUDA C code portion would take a constant time, rather than  the polynomial time required by the corresponding C++ serial code.\\
The part of the code which remains serialized runs on the host unless
an excessive host-to-device transfer slows down the process, in which case
also the serial part of the code runs on the device.
Having analyzed the application profile, we apply either Amdahl or Gustafson law to estimate an upper limit of the achievable speedup.
Once we have located a hotspot in our application's profile assessment, we used Thrust library to expose the parallelism in that portion of our code as a call to an external function. We then executed this external function onto the GPU and retrieved the results without requiring major changes to the rest of the application.\\
We adopted the Thrust code optimization in order to gain, at the cost of lower speedup, a rapid development and a better code readability. There are three high-level optimization techniques that we employed to yield significant performance speedups when using Thrust:

\begin{enumerate}[1.]
\item Fusion: In computations with low arithmetic intensity, the ratio of calculations per memory access are constrained by the available memory bandwidth and do not fully exploits the GPU. One technique for increasing the computational intensity of an algorithm is to fuse multiple pipeline stages together into a single one. In Thrust a better approach is to fuse the functions into a single operation $g(f(x))$ and halve the number of memory transactions. Unless f and g are computationally expensive operations, the fused implementation will run approximately twice as fast as the first approach. Fusing a transformation with other algorithms is a worthwhile optimization. Thrust provides a transform iterator which allows transformations to be fused with any algorithm;
\item Structure of Arrays (SoA): An alternative way to improve memory efficiency is to ensure that all memory accesses benefit from coalescing, since coalesced memory access patterns are considerably faster than non-coalesced transactions. The most common violation of the memory coalescing rules arises when using an Array of Structures (AoS) data layout. An alternative to the AoS layout is the SoA approach, where the components of each structure are stored in separate arrays. The advantage of the SoA method is that regular access to its components of a given vector is fusible;
\item Implicit Sequences: the use of implicit ranges, i.e. those ranges whose values are defined programmatically and not stored anywhere in memory. Thrust provides a counting iterator, which acts like an explicit range of values, but does not carry any overhead. Specifically, when counting iterator is de-referenced it generates the appropriate value \textit{on the fly} and yields that value to the caller.
\end{enumerate}

\subsection{The Experiment}
\label{game:experiment}

The dataset used will be described in section \ref{chap:GC} and consists of 2100 rows (input patterns) and 12 columns (features), 11 as input and the latter class target (class labels, respectively, 0 for not GC and 1 for GC objects). The performance was evaluated on several hardware platforms. We compared our production GPU code with a CPU implementation of the same algorithm. The benchmarks were run on a 2.0 GHz Intel Core i7 2630QM quad core CPU running 64-bit Windows 7 Home Premium SP1. The CPU code was compiled using the Microsoft C/C++ Optimizing Compiler version 16.00 and GPU benchmarks were performed using the NVIDIA CUDA programming toolkit version 4.1 running on a NVIDIA GPUs GeForce GT540M device.\\

As execution parameters were chosen combinations of:

\begin{itemize}\addtolength{\itemsep}{-0.5\baselineskip}
\item Max number of iterations: 1000, 2000, 4000, 10000, 20000 and 40000;
\item Order (max degree) of polynomial expansion: 1, 2, 4 and 8;
\end{itemize}

The other parameters remain unchanged for all tests:

\begin{itemize}\addtolength{\itemsep}{-0.5\baselineskip}
\item Random mode for initial population: normal distribution in [-1, +1];
\item Type of error function (fitness): Threshold Mean Square Error (TMSE);
\item TMSE threshold: 0.49;
\item Selection criterion: both \textit{ranking} and \textit{roulette};
\item Training error threshold: 0.001, used as a stopping criteria;
\item Crossover application probability rate: 0.9;
\item Mutation application probability rate: 0.2;
\item Number of tournament chromosomes at each selection stage: 4;
\item Elitism chromosomes at each evolution: 2.
\end{itemize}

The experiments were done by using three implementations of the GAME model:

\begin{itemize}\addtolength{\itemsep}{-0.5\baselineskip}
\item serial: the CPU-based original implementation (serial code);
\item Opt: an intermediate version, obtained during the APOD process application (optimized serial code);
\item GPU: the parallelized version, as obtained at the end of an entire APOD process application.
\end{itemize}

For the scope of the present experiment, we have preliminarily verified the perfect correspondence between CPU- and GPU-based implementations in terms of classification performance.
 In fact, the scientific results for the serial version have been already evaluated and documented in a recent paper \citep{brescia2012a}, where the serial version of GAME was originally included in a set of machine learning models, provided within our team and compared with the traditional analytical methods.\\
Referring to the best results for the three versions of GAME implementation, we obtained the percentages shown in Tab. \ref{game:comp}.

\begin{table*}
\centering
\setlength{\tabcolsep}{3pt}
\begin{tabular}{|l|lrccc|}
\hline
type of experiment &	 missing features & figure of merit & serial & Opt & GPU \\
\hline
complete patterns  & --	
                  & $\textit{class.accuracy}$		        &    $82.1$ & $82.2$  & $81.9$  \\
 &                  & $\textit{completeness}$		        &    $73.3$ & $73.0$  & $72.9 $ \\
  &                 & $\textit{contamination}$		    &    $18.7$ & $18.5$  & $18.8$  \\
\hline
without feat. 11         & 11
                 & $\textit{class.accuracy}$		        &    $81.9$ & $82.1$  & $81.7$  \\
  &                 & $\textit{completeness}	$	        &    $79.3$ & $79.1$  & $78.9$  \\
   &                & $\textit{contamination}$		    &    $19.6$ & $19.5$  & $19.8$  \\
\hline
only optical &   8, 9, 10, 11
 	    & $\textit{class.accuracy}$		                &    $86.4$ & $86.3$  & $86.0$  \\
  &                 & $\textit{completeness}$		        &    $78.9$ & $78.6$  & $78.4 $ \\
   &                & $\textit{contamination}$		    &    $13.9$ & $13.7$  & $14.1$  \\
\hline
mixed      &  5, 8, 9, 10, 11
 	    & $\textit{class.accuracy}$		                &    $86.7$ & $86.9$  & $86.5 $ \\
  &                 & $\textit{completeness}$		        &   $ 81.5$ & $81.4 $ & $ 81.2 $ \\
   &                & $\textit{contamination}$		    &    $16.6$ & $16.2$  & $16.7$  \\
\hline
\end{tabular}

\caption[Summary of the performances (in percentage) of the three versions of the GAME classifier.]{Summary of the performances (in percentage) of the three versions of the GAME classifier. There are reported results for the four main dataset feature pruning experiments, respectively with all 11 features, without the last structural parameter (tidal radius), with optical features only and the last one without 5 features (mixed between optical and structural types).}
\label{game:comp}
\end{table*}

The results were evaluated by means of three statistical figures of merit, for instance \textit{completeness, contamination and accuracy}. However, these terms are differently defined by astronomers and \textit{data miners}.
In this case, for \textit{classification accuracy} we intend the fraction of patterns (objects) which are correctly classified (either GCs or non-GCs) with respect to the total number of objects in the sample; the \textit{completeness} is the fraction of GCs which are correctly classified as such and finally, the \textit{contamination} is the fraction of non-GC objects which are erroneously classified as GCs. In terms of accuracy and completeness, by using all available features but the number 11 (the tidal radius), we obtain marginally better results, as can be expected given the high noise present in this last parameter, which is affected by the large background due to the host galaxy light. In terms of contamination, better results are obtained by removing structural parameters, demonstrating the relevance of information carried by optical features in the observed objects (in particular, the isophotal and aperture magnitudes and the FWHM of the image were recognized as the most relevant by all pruning tests). Moreover, these experiments have also the advantage to reduce the number of features within patterns, without affecting  the classification performance. The less numerous are the patterns, the shorter the execution time of the training phase, thus providing a benefit to the overall computational cost of the experiments. \\
Finally, it is worth pointing out that the performance results quoted in Tab. \ref{game:comp} are all referred to the test samples (without considering the training results), and do not include possible biases affecting the KB itself. Hence they rely on the assumption that the KB is a fair and complete representation the \textit{real} population that we want to identify.\\
From the computational point of view, as expected, for all the three versions of GAME, we obtained consistent results, slightly varying in terms of less significant digits, trivially motivated by the intrinsic randomness of the genetic algorithms and also by the precision imposed by different processing units. We recall in fact, that the GPU devices used for experiments are optimized for single precision and they may reveal a little lack of performance in the case of double precision calculations. GPU devices, certified for double precision would be made available for testing in the first quarter of 2013, when optimized \textit{Kepler} GPU class type devices will be commercially distributed.\\
After having verified the computing consistency among the three implementations, we investigated the analysis of performance in terms of execution speed.\\
We performed a series of tests on the parallelized version of the model in order to evaluate the scalability performance with respect to the data volume. These tests make use of the dataset used for the scientific experiments (see Sec. \ref{chap:GC}), extended by simply replicating its rows several times, thus obtaining a uniform group of datasets with incremental size.\\
By considering a GPU device with $96$ cores, $2GB$ of dedicated global memory and a theoretical throughput of about $14GB/sec$ on the PCI-E bus, connecting the CPU (hereinafter Host) and GPU (hereinafter Device), we compared the execution of a complete training phase, by varying the degree of the polynomial expansion (function \textit{polytrigo}) and the size of the input dataset.\\
Tab. \ref{game:benchgpu} reports the results, where the training cycles have been fixed to $4000$ iterations for simplicity. The derived quantities are the following:

\begin{itemize}
\item HtoD (Host to Device): time elapsed during the data transfer from Host to Device;
\item DtoH (Device to Host): the opposite of HtoD;
\item DtoD (Device to Device): time elapsed during the data transfers internally to the device global memory;
\item DP (Device Processing): processing time on the GPU side;
\item HP (Host Processing): processing time on the CPU side;
\item P (Processing): total duration of the training processing phase (excluding data transfer time);
\item T (Transfer): total duration of various data transfers between Host and Device;
\item TOT (Total): Total time duration of a training execution;
\end{itemize}

The relationships among these terms are the following:

\begin{equation}
P = DP + HP
\label{game:eq11}
\end{equation}

\begin{equation}
T = HtoD + DtoH + DtoD
\label{game:eq12}
\end{equation}

\begin{equation}
TOT = P + T
\label{game:eq13}
\end{equation}

The total execution time of a training phase, given by Eq. \ref{game:eq13}, can be obtained by adding the processing time (on both Host and Device), given by Eq. \ref{game:eq11}, to the data transfer time (Eq. \ref{game:eq12}). However, in principle, in Eq. \ref{game:eq12} it would be necessary to calculate also the time elapsed during data transfers through the \textit{shared} memory of the Device, i.e. the data flow among all threads of the GPU blocks. In our case this mechanism is automatically optimized by Thrust itself, which takes care of the flow, thus hiding the thread communication setup to the programmer.\\
By comparing the time spent to transfer data between Host and Device (sum of columns HtoD and DtoH) with the processing time (column P), it appears an expected not negligible contribution of transfer time, well known as one of the most significant bottlenecks for the current type of GPUs (i.e. before the Kepler technology). In this case, a particular effort is required in the code design to minimize as much as possible this data flow. In our case, as the data size increases, such contribution remains almost constant (approximately $9\%$), thus confirming the correctness of our approach in keeping this mechanism under control.\\
In terms of work distribution between Host and Device, by comparing the processing time between Host and Device (respectively, columns HP and DP), for all data size the percentage of GPU calculation time remains almost the same (approximately $70\%$ of the whole processing time, given in column P).
Furthermore, by evaluating the average time needed to process one MB of data at different polynomial degrees, when the size grows from $0.15$ to $512$ MB, we obtain on average $\sim0.058$ MB/sec with degree = $1$, $\sim0.055$ MB/sec with degree = $2$, $\sim0.046$ MB/sec with degree = $4$, and $\sim0.036$ MB/sec with degree = $8$. Of course, as it was expected, the increase in the polynomial degree affects the average time to process a single MB of data, but of an amount which is small in respect of the growing size of data.\\
In conclusions, by considering both scientific and computing performances of the GPU based model, we can conclude that a polynomial expansion of degree $8$ is a good compromise between the approximation of the desired training error and the processing time needed to reach it, still maintaining good performances also in the case of large volumes of training data.

\begin{table*}
\centering
\small
\setlength{\tabcolsep}{2pt}
\begin{tabular}{|c|r|rrrrrrrr|}
\hline
 degree                 & data size  & $DP$       & $HP$       & $HtoD$    & $DtoH$    & $DtoD$   & $P$        & $T$       & $TOT$      \\
\hline\hline
$1$               &	            &            &            &           &           &          &            &           &          \\
                  & $0.15MB$    & $9.59$     & $5.16$     & $0.11$    & $0.11$    & $0.03$   & $14.75$    & $0.25$    & $15.00$    \\
                  & $16MB$	    & $152.12$   & $65.20$    & $10.00$   & $10.00$   & $2.68$   & $217.32$   & $22.68$   & $240.00$   \\
                  & $128MB$	    & $1191.26$  & $510.54$   & $82.00$   & $81.20$   & $5.50$   & $1701.80$  & $168.70$  & $1870.50$  \\
                  & $512MB$	    & $4919.60$  & $2108.40$  & $322.00$  & $322.00$  & $8.00$   & $7028.00$  & $652.00$  & $7680.00$  \\
\hline
$2$               &	            &            &            &           &           &          &            &           &          \\
                  & $0.15MB$    & $10.38$    & $5.35$     & $0.12$    & $0.12$    & $0.04$   & $15.73$    & $0.27$    & $16.00$    \\
                  & $16MB$	    & $161.74$   & $69.32$    & $11.00$   & $11.00$   & $2.95$   & $231.05$   & $24.95$   & $256.00$   \\
                  & $128MB$	    & $1265.59$  & $542.39$   & $90.20$   & $89.32$   & $6.05$   & $1807.98$  & $185.57$  & $1993.55$  \\
                  & $512MB$	    & $5232.36$  & $2242.44$  & $354.20$  & $354.20$  & $8.80$   & $7474.80$  & $717.20$  & $8192.00$  \\
\hline
$4$               &	            &            &            &           &           &          &            &           &          \\
                  & $0.15MB$    & $12.72$    & $5.98$     & $0.13$    & $0.13$    & $0.04$   & $18.70$    & $0.30$    & $19.00$    \\
                  & $16MB$	    & $193.59$   & $82.97$    & $12.10$   & $12.10$   & $3.24$   & $276.56$   & $27.44$   & $304.00$   \\
                  & $128MB$	    & $1517.58$  & $650.39$   & $99.22$   & $98.25$   & $6.66$   & $2167.98$  & $204.13$  & $2372.11$  \\
                  & $512MB$	    & $6257.36$  & $2681.72$  & $389.62$  & $389.62$  & $9.68$   & $8939.08$  & $788.92$  & $9728.00$  \\
\hline
$8$               &	            &            &            &           &           &          &            &           &          \\
                  & $0.15MB$    & $16.33$    & $7.34$     & $0.14$    & $0.14$    & $0.04$   & $23.67$    & $0.33$    & $24.00$    \\
                  & $16MB$	    & $247.67$   & $106.14$   & $13.31$   & $13.31$   & $3.57$   & $353.81$   & $30.19$   & $384.00$   \\
                  & $128MB$	    & $1947.10$  & $834.47$   & $109.14$  & $108.08$  & $7.32$   & $2781.58$  & $224.54$  & $3006.12$  \\
                  & $512MB$	    & $7994.13$  & $3426.06$  & $428.58$  & $428.58$  & $10.65$  & $11420.19$ & $867.81$  & $12288.00$ \\
\hline
\end{tabular}
\caption[Summary of the time performances of the GPU version of the GAME classifier.]{Summary of the time performances of the GPU version of the GAME classifier. The figures were obtained by varying the degree of the polynomial function \textit{polytrigo} (first column) and the size of input dataset (second column). The training cycles have been fixed to $4000$ iterations.
For each degree value, the first row (with datasize = $0.15MB$) is referred to the original size of the dataset used for scientific experiments. The other rows are referred to the same dataset but artificially extended by replicating its rows a number of times. All other columns 3-10 are time values expressed in seconds. Their meaning is defined as follows: $DP$ (Device Processing) is the processing time on the GPU side; $HP$ (Host Processing) is the processing time on the CPU side; $HtoD$ (Host to Device) is the time elapsed during the data transfer from Host to Device; $DtoH$ (Device to Host) is the opposite of $HtoD$; $DtoD$ (Device to Device) is the time elapsed during the data transfers internally to the device global memory; $P$ is the sum of columns $DP$ and $HP$; $T$ is the sum of columns $HtoD$, $DtoH$ and $DtoD$ and finally the last column $TOT$ is the sum of columns $P$ and $T$.}
\label{game:benchgpu}
\end{table*}

The three versions of GAME, respectively the serial one, the serial optimized and the parallel, have been tested under the same setup conditions. As expected, while the two CPU-based versions, serial and Opt, appear comparable, there is an evident difference with the GPU version. The diagrams shown in Fig.~\ref{game:figure4} report the direct comparisons among the three GAME versions, by setting an incremented degree of the polynomial expansion which represents the evaluation function for chromosomes.

\begin{figure}
\centering
\includegraphics[width=7.5cm]{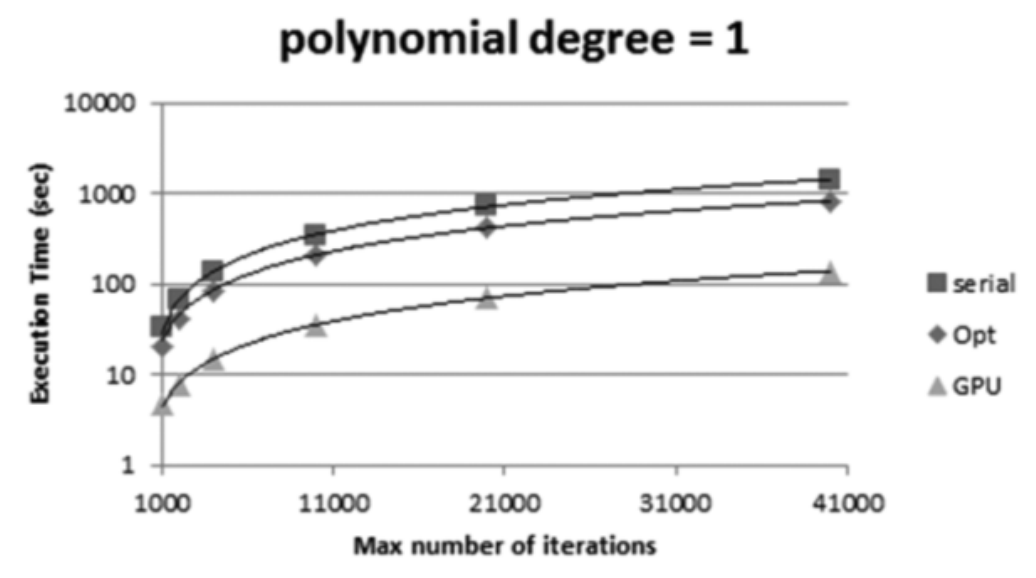}(a)
\includegraphics[width=7.5cm]{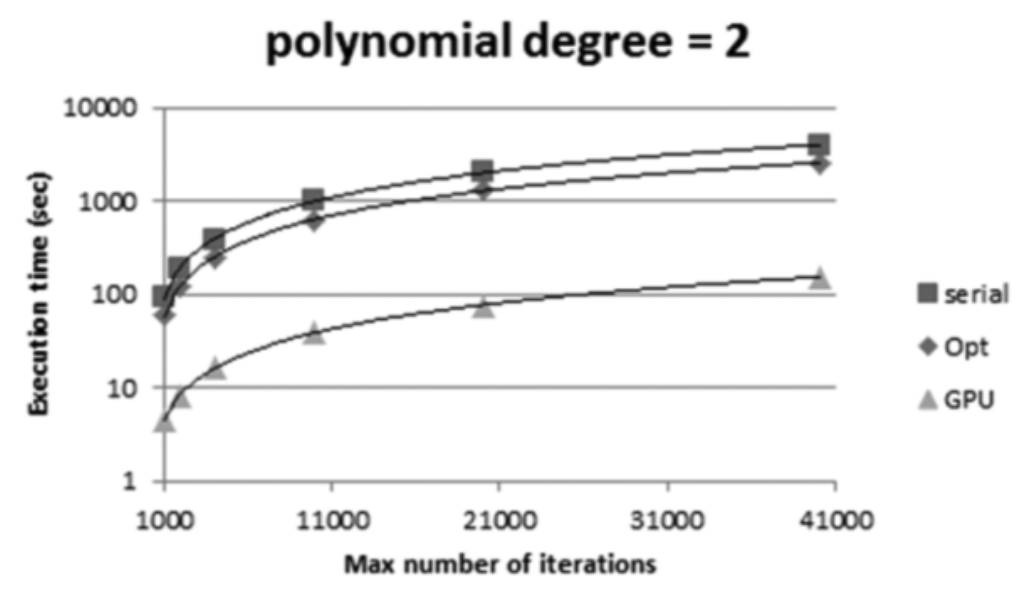}(b)
\includegraphics[width=7.5cm]{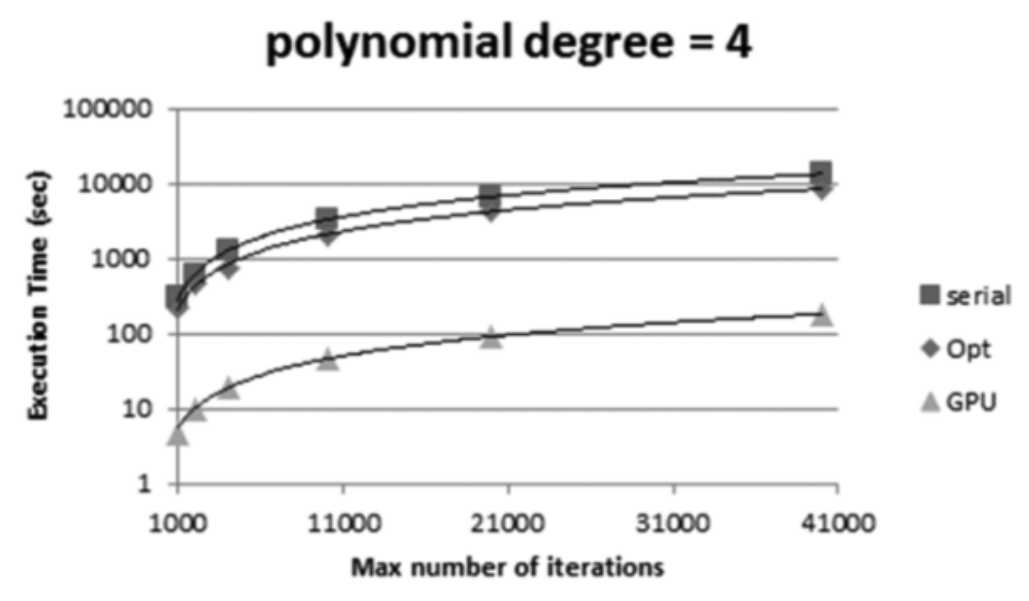}(c)
\includegraphics[width=7.5cm]{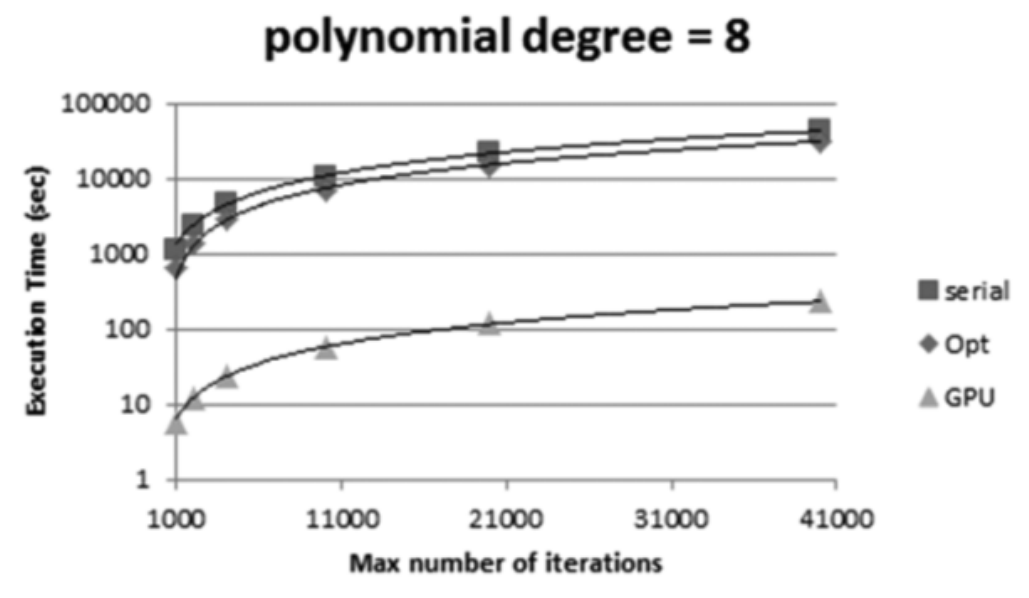}(d)
\caption[Comparison among the three GAME implementations with the polynomial degree 1, 2, 4 and 8]{Comparison among the three GAME implementations with the polynomial degree 1 (a), 2 (b), 4 (c) and 8 (d). The squares represent the serial version; rhombi represent the Opt version, while triangles are used for the GPU version.}
\label{game:figure4}
\end{figure}

The trends show that the execution time increases always in a linear way with the number of iterations, once fixed the polynomial degree. This is what we expected because the algorithm repeats the same operations at each iteration. The GPU version speed is always at least one order of magnitude less than the other two implementations. We remark also that the classification performance of the GAME model increases by growing the polynomial degree, starting to reach good results from a value equal to 4. Exactly when the difference between CPU and GPU versions starts to be 2 orders of magnitude.

\begin{figure}
\centering
\includegraphics[width=7.5cm]{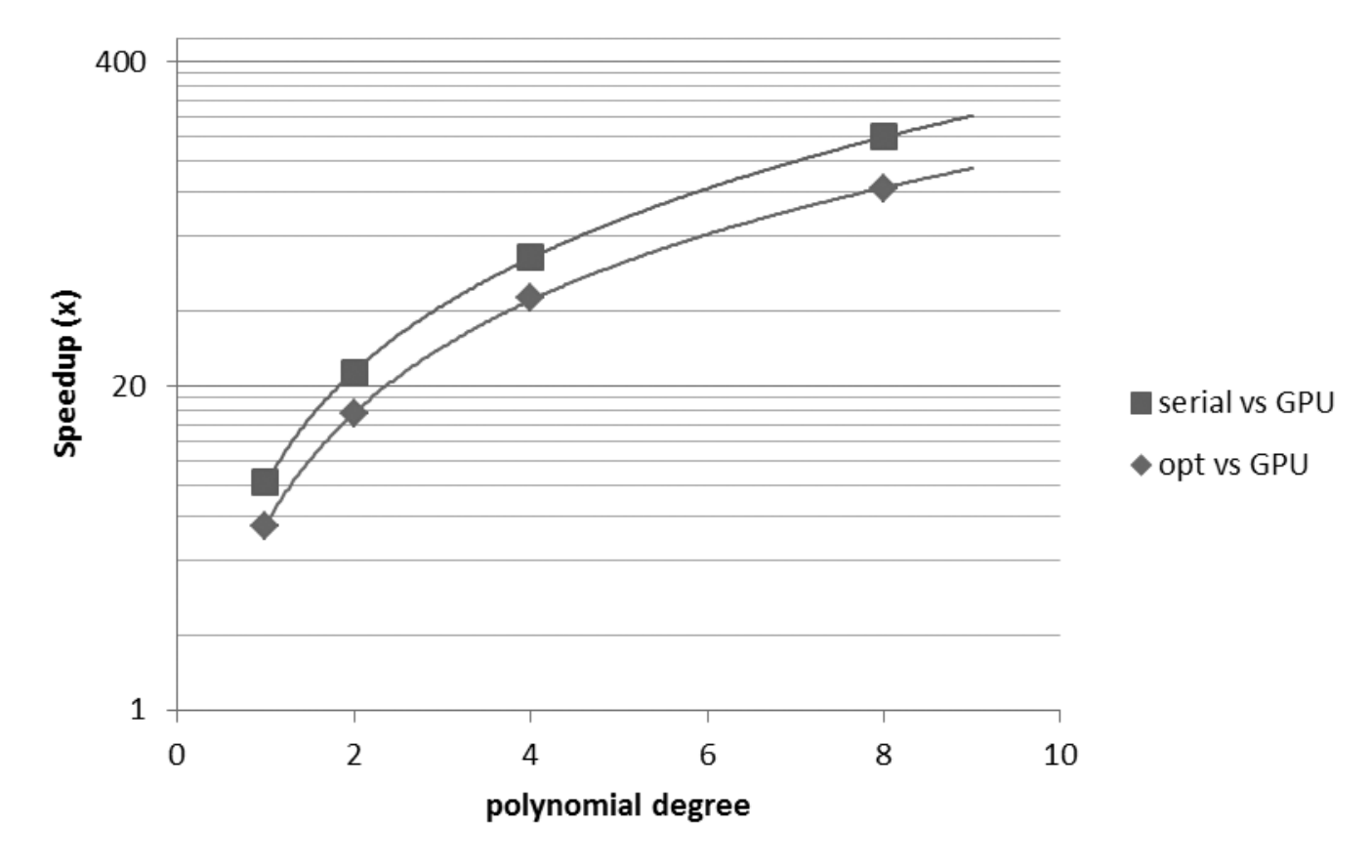}
\caption{Speedup comparison among GAME CPU implementations (serial and Opt) against the GPU version.}
\label{game:figure5}
\end{figure}

In the diagram of Fig.~\ref{game:figure5}, the GPU version is compared against the CPU implementations. As shown, the speedup increases proportionally with the increasing of the polynomial degree. The diagram shows that for the average speed in a range of iterations from 1000 to 40000, the GPU algorithm exploits the data parallelism as much data are simultaneously processed. As previously mentioned, an increase of maximum degree in the polynomial expansion leads to an increase in the number of genes and consequently to a larger population matrix. The GPU algorithm outperforms the CPU performance by a factor ranging from $8\times$ to $200\times$ against the serial version and in a range from $6\times$ to $125\times$ against the Opt version, enabling an intensive and highly scalable use of the algorithm that were previously impossible to be achieved with a CPU.

\subsection{Discussion}

A multi-purpose genetic algorithm implemented with GPGPU/CUDA parallel computing technology has been designed and developed. The model comes from the paradigm of supervised machine learning, addressing both the problems of classification and regression applied on medium/massive datasets.\\
The GPU version of the model is foreseen to be deployed on the DAMEWARE web application, presented in \citet{brescia2010b} and \citet{brescia2011a}. The model has been successfully tested and validated on astrophysical problems \citep{brescia2012a}.\\
Since genetic algorithms are inherently parallel, the parallel computing paradigm provided an exploit of the internal training features of the model, permitting a strong optimization in terms of processing performance. In practice, the usage of CUDA translated into a $75\times$ average speedup, by successfully eliminating the largest bottleneck in the multi-core CPU code. Although a speedup of up to $200\times$ over a modern CPU is impressive, it ignores the larger picture of using a Genetic Algorithm as a whole.\\

Real datasets can be very large (those we have previously called Massive datasets) and this requires greater attention to GPU memory management, in terms of scheduling and data transfers host-to-device and \textit{vice versa}.

We wish also to remark that the fact that the three different implementations (serial, serial optimized and GPU)  of the GAME method lead to identical scientific results, implies that the three versions are fully consistent. Furthermore, by considering the good results of a deep analysis of the processing time profile, as expected, the GPU version largely improves the scalability of the method in respect of massive data sets.\\
Finally, the very encouraging results suggest to investigate further optimizations, like: (i) moving the formation of the population matrix and its evolution in place on the GPU. This approach has the potential to significantly reduce the number of operations in the core computation, but at the cost of higher memory usage; (ii) exploring more improvements by mixing Thrust and CUDA C code, that should allow a modest speedup justifying development efforts at a lower level; (iii) use of new features now available on NVIDIA Kepler architecture, such as OpenACC directives, able to achieve faster atomics and more robust thread synchronization and multi GPUs capability.

\chapter{Scientific Gateways and web 2.0}\label{chap:scientificgateway}

\hfill\begin{tabular}{@{}p{.6\linewidth}@{}}
\textit{``Technology is an inherent democratizer. Because of the evolution of hardware and software, you're able to scale up almost anything. It means that in our lifetime everyone may have tools of equal power."}\\ Sergey Brin.\\ \phantom{aaa}
\end{tabular}

A Science Gateway is, in general, a community-developed set of tools, applications or data that are integrated via a web portal or a suite of applications, usually (but not always) under a graphical user interface that is further customized to meet the needs of a specific community. The aim of Gateways is to enable entire communities to use common resources through an interface that is configured for optimal use, so that researchers can focus on their scientific goals and less on assembling the cyber-infrastructure they require. The cyber-infrastructure and all the technological requirements become for the researcher just a powerful and simple black box.
In this chapter I present three different gateways which I contributed to develop during my PhD: DAME, STraDiWA and the Euclid Data Quality.

    \section[DAME]{DAME - DAta Mining \& Exploration}\label{chap:dame}
        \blfootnote{this section is largely extracted from: \tiny
\begin{itemize}
\item Brescia, M.; \textbf{Cavuoti, S.}; Garofalo, M.; Guglielmo, M.; Longo, G.; Nocella, A.; Riccardi, S.; Vellucci, C.; Djorgovski, G.S.; Donalek, C.; Mahabal, A. Data Mining in Astronomy with DAME. \textbf{to be Submitted to PASP}
			\item \textbf{Cavuoti, S.}; Brescia, M.; Longo, G., 2012, Data mining and Knowledge Discovery Resources for Astronomy in the Web 2.0 Age, Proceedings of SPIE Astronomical Telescopes and Instrumentation 2012, Software and Cyberinfrastructure for Astronomy II, Ed.(s): N. M. Radziwill and G. Chiozzi, Volume 8451, RAI Amsterdam, Netherlands, July 1-4 \textbf{refereed proceeding}
			\item \textbf{Cavuoti, S.}; Brescia, M.; Longo, G.; Garofalo, M.; Nocella, A.; 2012, DAME: A Web Oriented Infrastructure for Scientific Data Mining and Exploration, Science - Image in Action. Edited by Bertrand Zavidovique (Universite' Paris-Sud XI, France) and Giosue' Lo Bosco (University of Palermo, Italy) . Published by World Scientific Publishing Co. Pte. Ltd., 2012. ISBN 9789814383295, pp. 241-247
			\item Djorgovski, S. G.; Longo, G., Brescia, M., Donalek, C., \textbf{Cavuoti, S.}, Paolillo, M., D'Abrusco, R., Laurino, O., Mahabal, A., Graham, M., ``DAta Mining and Exploration (DAME): New Tools for Knowledge Discovery in Astronomy". American Astronomical Society, AAS Meeting \#219, \#145.12, Tucson, USA, January 08-12
			\item Brescia M., \textbf{Cavuoti, S.}, Djorgovski, G.S., ,Donalek, C., Longo, G., Paolillo, M., 2011, Extracting knowledge from massive astronomical data sets, \href{http://arxiv.org/abs/1109.2840}{arXiv:1109.2840}, Springer Series in Astrostatistics, Volume 2, Springer Media New York, ISBN 978-1-4614-3322-4 15 pages \textbf{[invited review]}.
			\item Brescia, M.; \textbf{Cavuoti, S.}; D'Abrusco, R.; Laurino, O.; Longo, G.; 2010, DAME: A Distributed Data Mining \& Exploration Framework within the Virtual Observatory, INGRID 2010 Workshop on Instrumenting the GRID, Poznan, Poland, in Remote Instrumentation for eScience and Related Aspects, F. Davoli et al. (eds.), Springer Science+Business Media, LLC 2011, DOI 10.1007/978-1-4614-0508-5\_17
			\item Djorgovski, S. G.; Longo, G., Brescia, M., Donalek, C., \textbf{Cavuoti, S.}, Paolillo, M., D'Abrusco, R., Laurino, O., Mahabal, A., Graham, M., 2012, DAta Mining and Exploration (DAME): New Tools for Knowledge Discovery in Astronomy. American Astronomical Society, AAS Meeting \#219, \#145.12, Tucson, USA, January 08-12
			\item Brescia, M.; Longo, G.; Castellani, M.; \textbf{Cavuoti, S.}; D'Abrusco, R.; Laurino, O., 2012, DAME: A Distributed Web Based Framework for Knowledge Discovery in Databases, 54th SAIT Conference, Astronomical Observatory of Capodimonte, Napoli, Italy, May 6, Mem. S.A.It. Suppl. Vol. 19, 324
\end{itemize}}
As already mentioned in the previous chapter, many scientific areas share the same need to deal with massive and distributed datasets and to perform on them complex knowledge extraction tasks. DAME (DAta Mining \& Exploration) is an innovative, general purpose, Web-based, VObs compliant, and distributed data mining infrastructure specialized in Massive Data Sets exploration with machine learning methods. \\ Initially fine-tuned to deal with astronomical data only, DAME has evolved in a general purpose platform able to find applications also in other domains of human endeavor. The present section is focused on the e-science context and on the DAME products description. I feel obliged to add that my specific contribution to the development of DAME were the following: I developed the first prototype of DAME, I was and still am the person in charge for the DMM (Data Mining Models) package design and implementation. Moreover I supervised, checked and tested each released plugin, finally, I was involved in the ideation, implementation and test of several models, such as GAME (described in section \ref{chap:GAME}) MLPQNA (see section \ref{chap:QNA}) and SVM (see section \ref{chap:SVM}) and in the new plugin procedure (described in section \ref{sec:dame:design}) and in the ``Hydra Lernaen" idea (see section \ref{sec:hydra}).

\subsection{Generalities}

The main feature of DAME is its usability and scalability which address the well-known fact that Knowledge Discovery in Databases is a complex process. In most cases, in fact, the optimal results can be found only on a trial and error base by comparing the outputs of different methods and different experiments with the same method. This implies that, for a specific problem, data mining requires a lengthy fine tuning phase which is often not easily justifiable to the eyes of a non-experienced user.
Such complexity is one of the main explanations for the gap still existing between the new outcoming methodologies and the huge community of potential users which fail to adopt them. In order to be effective, in fact, KDD requires a good understanding of the mathematics underlying the methods, of the computing infrastructures and of the complex workflows which need to be implemented, and most users (even in the scientific community) are usually not willing to make the effort to understand the process and prefer to recur to traditional approaches which are far less powerful but much more user friendly \citep{hey2009}.
DAME (Data Mining \& Exploration) copes with the afore mentioned data-deluge by implementing the KDD models under an hybrid distributed computing infrastructure, also exploiting the the S.Co.P.E. grid facility \citep{merola2008,brescia2009,deniskina2009} and, by taking into account that background knowledge, it can  reduce the amount of data needed to be processed by adopting a learning rule based on the fact that most of the attributes (e.g. domain parameters) often turn out to be irrelevant when background knowledge is considered \citep{paliouras1993}.

DAME, by making use of the web application and service paradigm, of extensive and user friendly documentation, and of a sample of documented real use cases, represents a first attempt to bring the KDD models to the user hiding most of their complexity behind a well-designed infrastructure.
The DAME program is based on a platform which allows the scientific community to perform data mining and exploratory experiments on massive data sets, by using a simple web browser. It offers several tools, by means of the state of the art Web 2.0 technologies (for instance web applications and services), which can be seen as working environments where to choose data analysis functionalities such as clustering, classification, regression, feature extraction etc., together with models and algorithms. The latter being derived by machine learning paradigms. \\
The user is indeed able to setup, configure and execute experiments on his own data on top of a virtualized computing infrastructure, without the need to install any software on his local machine. Furthermore, the DAME infrastructure offers the possibility to extend the original library of available tools, by allowing end users to plug-in and execute his own code in a simple way, by uploading his program, without any restriction about the native programming language, into our infrastructure and automatically installing it through a simple guided interactive procedure.
Moreover, DAME platform offers a variety of computing facilities, organized as a cloud of versatile architectures, from the single multi-core processor to a grid (sub-network of S.Co.P.E. project) farm, automatically assigned at runtime to the user task, depending on the specific problem computing and storage requirements
In particular we refer to the beta release of the DAMEWARE (DAME Web Application Resource) web application, nowadays publicly available\footnote{\url{http://dame.dsf.unina.it}} , which addresses many of the above issues and aims at providing the scientific community with a user friendly and powerful data mining platform.

\subsection{DAMEWARE design and architecture}\label{sec:dame:design}
The early design stage of the DAMEWARE resource had as main goal to provide a web application for data mining on MDS. For the computational infrastructure two main requirements emerged: a-synchronous access and scalability.
Most available web based data mining services run synchronously. This means that they execute jobs during a single HTTP transaction. This might be considered useful and simple, but it does not scale well when it is applied to long-run tasks. With synchronous operations, all the entities in the chain of command (client, workflow engine, broker, processing services)  must remain up for the duration of the activity. If any component stops, the context of the activity is lost.
DAME was conceived to provide the scientific community with an extensible, integrated environment for data mining and exploration. In order to do so, DAME had to:
\begin{itemize}
\item	Support the VObs standards and formats, especially for data interoperability;
\item	To abstract the application deployment and execution, so to provide the VObs with a general purpose computing platform exploiting modern technologies.
\end{itemize}
With web applications, a remote user does not require to install program clients on his desktop, and has the possibility to collect, retrieve, visualize and organize data, configure and execute mining applications through the web browser and in an asynchronous way. An added value of such approach being the fact that the user does not need to directly access large computing and storage  facilities. He can transparently make his experiments by exploiting computing networks and archives located worldwide, requiring only a local laptop (or a smartphone) with a network connection.
For what the DAMEWARE functionalities are concerned, we need to distinguish between supervised and unsupervised algorithms. In the first case, we have a set of data points or observations, for which we know the desired output, class, target variable, or outcome. The outcome may take one of many values called ``classes" or ``labels". The target variable provides some level of supervision in that it is used by the learning algorithm to adjust parameters or make decisions that allow it to predict labels for new data. Finally, when the algorithm is predicting labels to observations, we call it a classifier. Some classifiers are also capable of providing a probability for a data point to belong to a given class and are often referred to as probabilistic models or regression, not to be confused with statistical regression models (cf. \citealt{duda2001}).\\

In the unsupervised case, instead of trying to predict a set of known classes, we are trying to identify the patterns inherent to the data, without outcomes or labels. In other words, unsupervised methods try to create clusters of data that are inherently similar.
In the DAME data mining infrastructure, the choice of any machine learning model is always accompanied by the functionality domain. To be more precise: several machine learning models can be used in the same functionality domain, since it represents the experiment domain context in which  the exploration of data is performed.
In what follows we shall therefore adopt the following terminology:
\begin{itemize}
\item	Data mining model: any of the data mining models integrated in the DAME suite. It can be either a supervised machine learning algorithm or an unsupervised one, depending on the available Knowledge Base (KB), i.e. the set of training or template cases available, and the scientific target of the experiment;
\item	Functionality: one of the already discussed functional domains, in which the user wants to explore the available data (like for example, regression, classification or clustering). The choice of the functionality target can limit the choice of the data mining model;
\item	Experiment: it is the scientific workflow (including optional pre-processing or preparation of data) and includes the choice of a combination of, a data mining model and a functionality;
\item	Use Case: for each data mining model, different running cases are exposed to the user . These can be executed singularly or in a prefixed sequence. Being the models derived from the machine learning paradigm, each may have training, test, validation and run use cases, in order to, respectively, perform learning, verification, validation and execution phases. In most models there is also the ``full" use case that automatically executes all listed cases as a sequence.
\end{itemize}
\begin{table}

\centering
\begin{tabular}{|c|c|c|}
\hline
\textbf{Model}	& \textbf{Category}	&\textbf{Functionality}\\   \hline
MLPBP 	& Supervised &	Classification, regression\\
MLPQNA & Supervised &	Classification, regression\\
MLP 	& Supervised &	Classification, regression\\
SVM	& Supervised &	Classification, regression\\
SOFM	& Unsupervised &	Clustering\\
K-Means	& Unsupervised &	Clustering\\
PPS	& unsupervised &	Dimensional Reduction\\
GAME	& supervised &	Classification, regression\\
\hline
\end{tabular}
\caption[The available data mining models and functionalities in the DAMEWARE web application.]{The available data mining models and functionalities in the DAMEWARE web application, MLPBP stands for: Multi Layer Perceptron (MLP) trained by Back Propagation learning rule; MLPQNA stands MLP for trained by Quasi Newton learning rule; MLPGA stands for MLP trained by Genetic Algorithm; SVM stands for Support Vector Machines; SOFM stands for Self Organizing Features Maps; PPS stands for Probabilistic Principal Surfaces; GAME stands for  Genetic Algorithm Model Experiment. }\label{DAME:tab1}
\end{table}

The functionalities and models currently available in DAMEWARE, beta release, are listed in Table \ref{DAME:tab1}.
Depending on the specific experiment and on the execution environment, the use of any of the above models can take place with a more or less advanced degree of parallelization. All the models require some parameters that cannot be defined a priori, thus causing the necessity of iterated experiment interactive sessions in order to find the best tuning. For this reason not all the models can be developed under the MPI (Message Passing Interface) paradigm \citep{chu2007}. But the possibility to execute more jobs at once (specific grid case) intrinsically exploits the multi-processor architecture. As for all data mining models integrated in the DAME framework, there is instanced a specialized plugin (called DMPlugin) for each couple model-functionality, that includes the specific parameter configuration. In this way, it is also possible to easily replace and/or upgrade models and functionalities without any change or modification to the DAME software components \citep{brescia2010b}. \\The plugin extension facility represents also a practical baseline prototype for a more reliable and performing approach to handling massive datasets, by truly overcoming the data transfer bandwidth limitations.\\
In order to  deal with MDS, DAMEWARE offers asynchronous access to the infrastructure tools, thus allowing the running of activity jobs and processes outside the scope of any particular web-service operation and without depending on the user connection status. The user, via a simple web browser, can access the application resources and has the possibility to keep track of his jobs by recovering related information (partial/complete results) without having the need to maintain open the communication socket. Furthermore DAMEWARE has been designed to run both on a server and on a hybrid distributed computing infrastructure.
From the technological point of view, DAME consists of five main components: Front End (FE), Framework (FW), Registry and Data Base (REDB), Driver (DR) and Data Mining Models (DMM).

\begin{figure*}
\centering
    \includegraphics[width=12.cm]{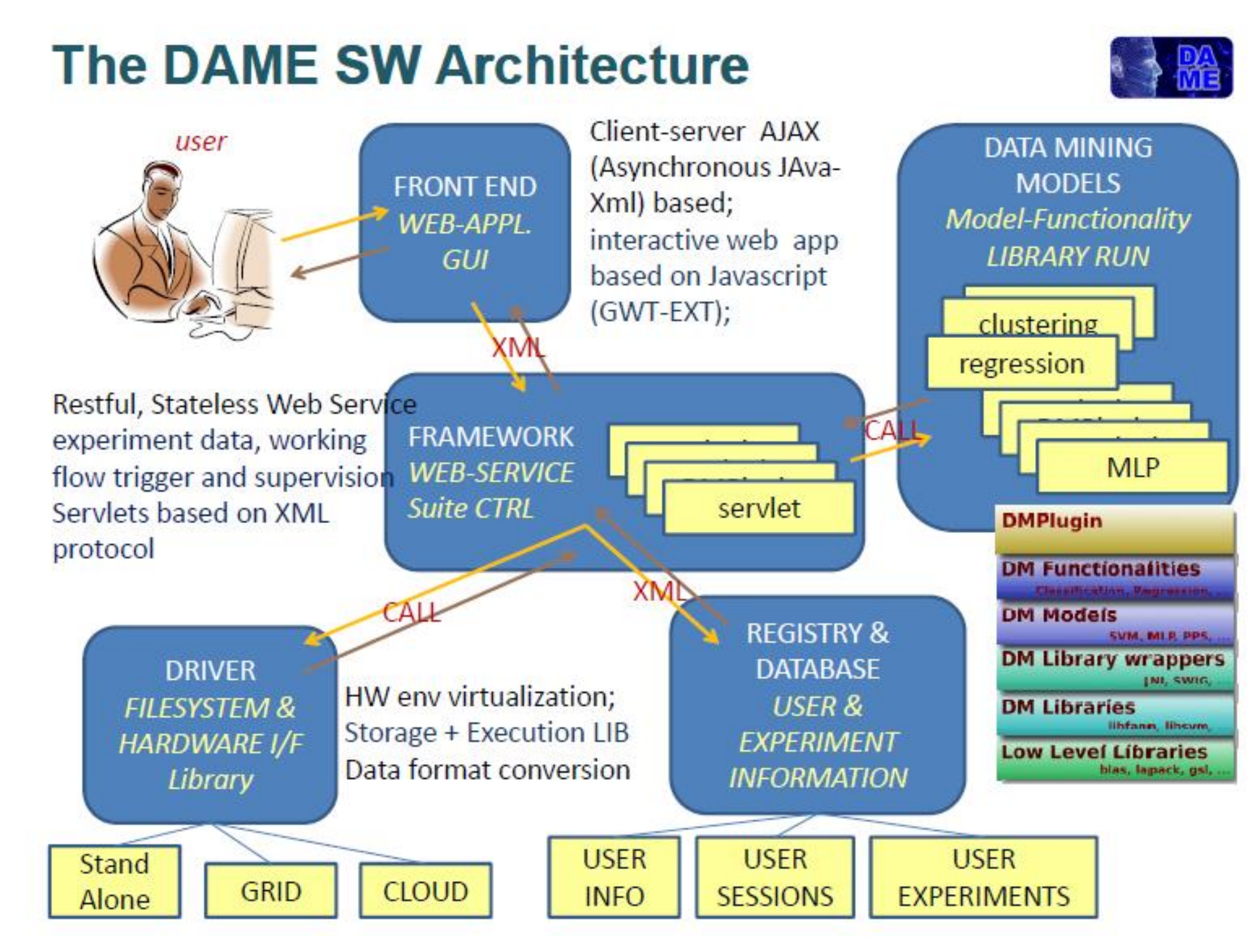}
\caption{The general Software Architecture of DAMEWARE.}\label{DAME:fig1}
\end{figure*}

The scheme in Fig. \ref{DAME:fig1} shows the component diagram of the entire suite with their main interface / information exchange layout. While details on the DAME infrastructure can be found in \citep{brescia2010b} and the documentation available on the DAME web site, here we shall just outline a few relevant features.
The DAME design architecture is implemented following the standard LAR (Layered Application Architecture) strategy, which leads to a software system based on a layered logical structure, where different layers communicate with each other via simple and well-defined rules:
\begin{itemize}
\item	Data Access Layer (DAL): the persistent data management layer, responsible of the data archiving system, including consistency and reliability maintenance.
\item	Business Logic Layer (BLL): the core of the system, responsible of the management of all services and applications implemented in the infrastructure, including information flow control and supervision.
\item	User Interface (UI): responsible of the interaction mechanisms between the BLL and the users, including data and command I/O and views rendering.
\end{itemize}
A direct implication of the LAR strategy adopted in DAME is the Rich Internet Application (RIA; \citealt{bozzon2010}), consisting in applications having traditional interaction and interface features of computer programs but usable via simple web browsers, i.e. not needing any installation on user local desktop. RIAs are particularly efficient in terms of interaction and execution speed.
By keeping this in mind, the main concepts behind the distributed data mining applications implemented in the DAME Suite are based on three issues:
 \begin{itemize}
\item	Virtual organization of data: this is the extension of the already remarked basic feature of the VObs;
\item	Hardware resource-oriented: this is obtained by using computing infrastructures, like grid, which enable parallel processing of tasks, using idle capacity. The paradigm in this case is to obtain large numbers of instances running for short periods of time;
\item	Software service-oriented: this is the base of usual cloud computing paradigm \citep{shende2010}. The data mining applications implemented runs on top of virtual machines seen at the user level as services (specifically web services), standardized in terms of data management and working flow.
\end{itemize}
The complete Hardware infrastructure of the DAME Program is shown in Fig. \ref{DAME:fig2}, where the grid sub-architecture provided by the S.Co.P.E. supercomputing facility \citep{merola2008,brescia2010b} is incorporated into the more general cloud scheme, including a network of  multi-processor and multi-HDD PCs and workstations, each of them dedicated to a specific function.

\begin{figure*}
\centering
    \includegraphics[width=12.cm]{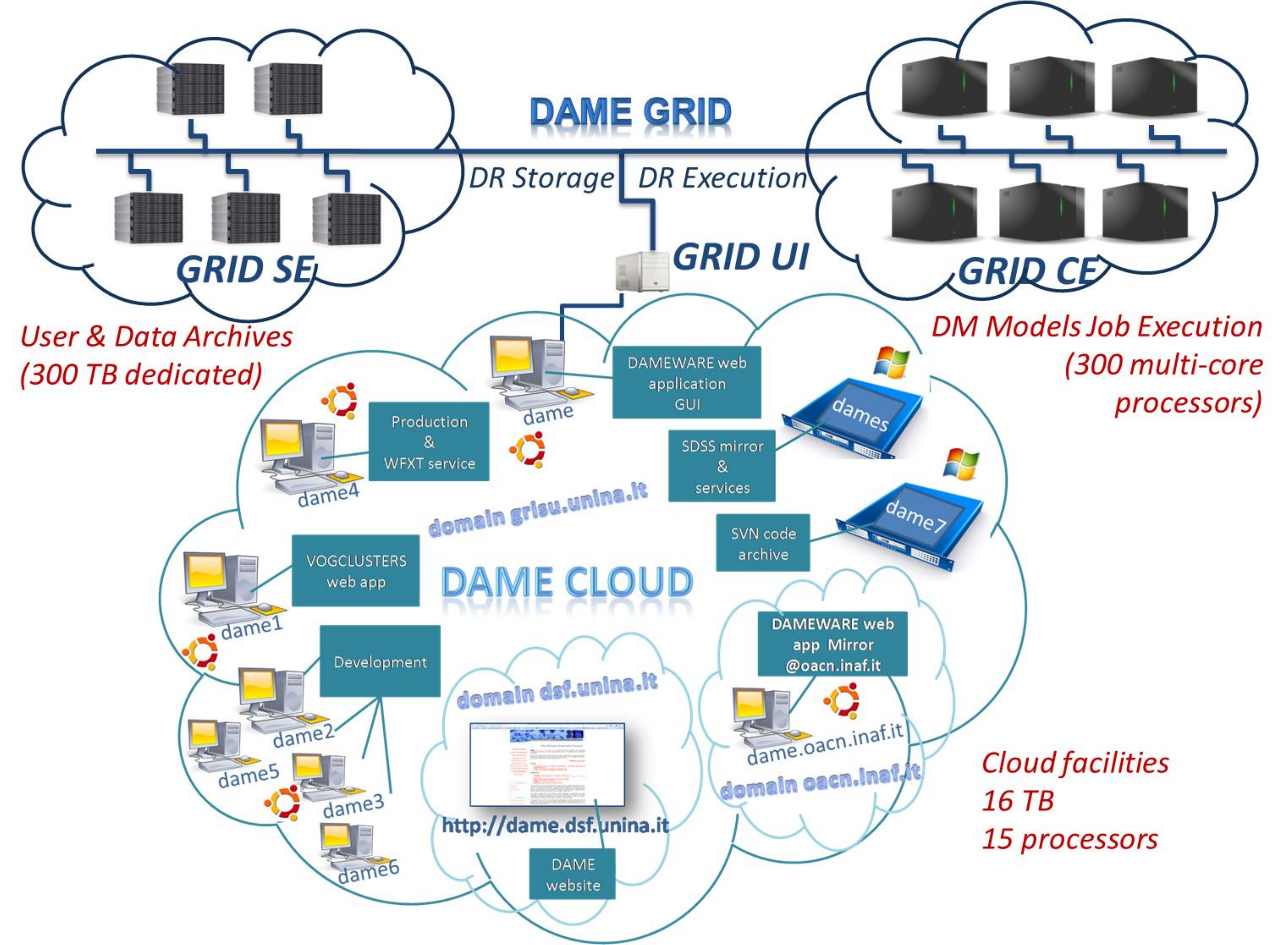}
\caption{The general Hardware Architecture of DAME applications}\label{DAME:fig2}
\end{figure*}

There are two sub-networks addressable from an unique access point, the website, which provides an embedded access to the user to all DAME web applications and services. The integrity of the system, including the grid public access, is guaranteed by a registration procedure, which gives the possibility to access all facilities from just one account. In particular, a robot certificate is automatically handled by the DAME system to provide transparent access to the S.Co.P.E. grid resources \citep{deniskina2009}.
Depending on the computing and storage power requested by the job and by the processing load currently running on the network, an internal mechanism redirects the jobs to a job-queue in a pre-emptive scheduling scheme. The interaction with the infrastructure is completely asynchronous and a specialized software component (DR, DRiver) has the responsibility to store off-line  job results in the user storage workspaces, that can be retrieved and downloaded in subsequent accesses.\\
This hybrid architecture, renders it possible to execute simultaneous experiments that gathered all together, bring the best results.\\ Even if the individual job is not parallelized, we obtain a running time improvement by reaching the limit value N of the Amdahl's law \citep{amdahl1967}, shown in equation \ref{DAME:eq1}:
    \begin{equation}\label{DAME:eq1}
\frac{1}{{(1 + P) + \frac{P}{N}}}
    \end{equation}

\noindent where $P$ is the fraction of a program that can be made parallel (i.e. which can benefit from parallelization), and $(1 - P)$ is the fraction that cannot be parallelized (remains serial), then the resulting maximum speed-up that can be achieved by using $N$ processors is obtained by the law expressed above.
For instance, in the case of the AGN (Active Galactic Nucleus) classification experiment \citep{cavuoti2008,brescia2009}, which used SVM (Support Vector Machines) each of the 110 jobs runs for about a week on a single processor. By exploiting the GRID S.Co.P.E.\footnote{S.Co.P.E. is a general purpose GRID infrastructure of the University Federico II in Naples
funded through the Italian National Plan (PON) by the Italian Government to support
both fundamental research and small/medium size companies. The infrastructure has been
conceived as a metropolitan GRID, embedding different (and in some cases pre-existing) and
heterogeneous computing centers each with its specific vocation: high energy physics, astrophysics, bioinformatics, chemistry and material sciences, electric engineering, social sciences. Its intrinsically multi-disciplinary nature renders the S.Co.P.E. an ideal test bed for
innovative middleware solutions and for interoperable tools and applications finely tuned on
the needs of a distributed computing environment \citep{brescia2009}.}, the experiment running time was reduced to about one week instead of more than 2 years.
From the software point of view, the baselines behind the engineering design of DAME Suite were:
\begin{itemize}
\item	Modularity: software components with standard interfacing, easy to be replaced;
\item	Standardization: basically, in terms of information I/O between user and infrastructure as well as between software components (in this case based on XML-schema);
\item	Hardware virtualization: i. e. independent from the hardware deployment platform (single or multi processor, grid etc.);
\item	Interoperability: by following VO requirements;
\item	Expandability: many parts of the infrastructure require to be increased and updated along its lifetime. This is particularly true concerning computing architecture, framework capabilities, GUI (Graphical User Interface) features, data mining functionalities and models (this also includes the integration within the system of end user proprietary algorithms);
\item	Asynchronous interaction: the end user and the client-server mechanisms do not require a synchronous interaction. In other words, the user is not constrained to remain connected after launching an experiment in order to wait for the end of execution. Moreover, by using the Ajax (Asynchronous Javascript and XML described in \citealt{garrett2005}) mechanism, the web applications can retrieve data from the server asynchronously in  background without interfering with the display and behavior of the existing page.
\item	Language-independent Programming: this basically concerns the API (Application Programming Interface) forming the data mining model libraries and packages. Although most of the available models and algorithms were internally implemented, this is not considered as mandatory (it is possible to re-use existing tools and libraries, integration of end user tools etc.). So far, the Suite provided a Java based standard wrapping system to achieve the standard interface with multi-language APIs;
\item	Distributed computing: the components can be deployed on the same machine as well as on different networked computers;
\item	Pluggable: with the new plugin procedure users can extend the data mining model library integrated into the web app, by simply download and run a Java applications, which through a driven procedure generate source code to be integrated into the web app software infrastructure. This procedures are the first prototype of our proposal (IVOA Interoperabilty Workshop 2011, Naples) of a new vision of the KDD App approach: shared apps that can move on demand to the various datacenter plugged in each other, that could be extended to every VO App, in order to obtain a new generation of instruments, based on the minimization of data transfer and maximization of interoperability in the VO community.
\end{itemize}
\subsection{Dame Scalability}
For what scalability is concerned, whenever there is too much data, there are three approaches at making learning feasible. The first one is trivial and consists in applying the training scheme to a decimated dataset. Obviously, in this case, the information may be easily lost and this loss needs to be negligible in terms of correlation discovery. This approach, however, may turn very useful in the lengthy optimization procedure which is required by many machine learning methods (such as neural networks or genetic algorithms).
The second method relies in splitting the problem in smaller parts (parallelization), by sending them to different CPUs and finally combining the results together. However, to implement parallelized versions of learning algorithms is not always easy, and this approach should be followed only when the learning rule (such as in the case of genetic algorithms, \citep{mitchell1998} or Support Vector Machines  \citep{chang2011} is intrinsically parallel. The parallelization could be intended not only in terms of the use of parallel programming paradigm for a single algorithm, but also by considering a resource oriented platform (such as a Grid computing farm), on which to process in parallel more instances of a serial algorithm applied on ad hoc selected subsets of input data. However, even after parallelization, the asymptotic time complexity of the algorithms cannot be improved.
A third and more challenging way to enable a learning paradigm to deal with MDS is to develop new algorithms of lower computational complexity but in many cases this is simply not feasible.\\
Also background knowledge can make possible to reduce the amount of data that needs to be processed by adopting a learning rule since in many cases most of the attributes turn out to be irrelevant when background knowledge is taken into account.
In many exploration cases, however, such background knowledge simply does not exists, or it may introduce biased knowledge in the discovery process
The concept of distributed archives is familiar to most scientists. In the astronomical domain, the leap forward was the possibility to organize through the Virtual Observatory (VObs)  the data repositories to allow efficient, transparent and uniform access \citep{genova2002}. In other words, the VObs have been intended to be a paradigm to use multiple archives of astronomical data in an interoperating, integrated and logically centralized way, so to be able to ``observe" and analyze a virtual sky by position, wavelength and time. In spite of some underscoping, the VObs can still be considered as an extension of the classical computational grid, in the sense that it fits the data grid concept, being based on storage and processing systems, together with metadata and communications management services. The link between data mining applications and the VObs data repositories is currently still under discussion since it requires (among the other things) the harmonization of many recent achievements in the fields of VObs, grid, cloud, HPC (High Performance Computing), and Knowledge Discovery in Databases. Our first attempt to tackle the problem of scalability of massive datasets was done with the GRID via the e-token, but with the technological evolution we are approaching the GPU solution with massive parallelization.

\subsection{Evolution and further developments}

One of the key aspects of our strategy is to achieve the exploration of massive data sets (MDS) coming out from ground segments of large telescopes or space mission facilities, in such a way that it is not required to physically move huge data over the network from their original repositories. First of all because there should be any restrictive policy to prevent such action. Second, because it can be simply impossible, due to the  huge amount of data to be moved respect of the available network channel bandwidth.
In these cases, the problem is indeed to be able to move data mining toolsets directly to the locations hosting data centers. Current solution strategies, under investigation in some communities, such as the VO in the astronomical context where it is under design a web based protocol for application interoperability (named Web Samp Connector), result only partial, because they always require to exchange data over the web between application sites. In other words, any of such interoperability scenarios, based on data communication between all possible combinations of desktop and web applications, results not able to process MDS in a realistic and affordable way.
\\Our approach is completely different. We do not move data between sites. The strategy is based on the concept of creating a standardized web application repository cloud, named as ``Hydra Lernaen" as described in sec. \ref{sec:hydra} from the name of the ancient nameless serpent-like a water beast, with reptilian traits that possessed many independent but equal heads.

Other foreseen improving features are:
\begin{itemize}
\item	Parallelization: Introduction of MPI (Message Passing interface) technology in the  Framework and  DBMS components of DAME, by investigating its deployment on a multi-core platform, based on GPU+CUDA computing technique \citep{maier2009}. This could improve computing efficiency in data mining models, such as Genetic Algorithms, naturally implementable in a parallel way.
\item	Scripting and workflows: A workflow is a sequence of ``nuclear" tasks, that could include ``loops", ``case of" and  so on. At the moment this can be done executing all the task manually using the GUI, as soon as possible it will be feasible to write a script using our API.
\item	Statistical and Visualization tools: our models already produce some graphics, but at the moment  they cannot be produced by the user independently from the execution of a job: We are working to implement statistical and visualization tools that can be performed as single task.
\item	More editing tools: we want to enlarge the set of dataset editing tools for pre and post 	processing.
\end{itemize}

\subsubsection{Lernaean Hydra}\label{sec:hydra}

There are at least two reasons for not moving data over the network from their original repositories to the user's computing infrastructures. First of all the fact that the transfer could be impossible due to the available network bandwidth and, second, because there could be restrictive policies to data access.

 \begin{figure*}
   \centering
   \includegraphics[width=9cm]{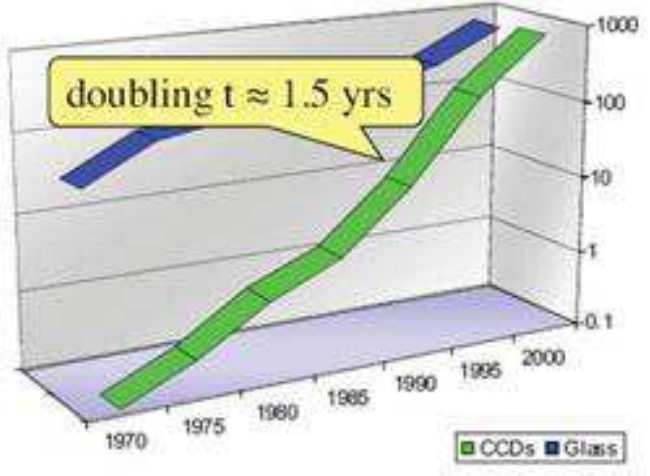}
   \caption[Data in astrophysics are growing exponentially, with a doubling time of 1.5 years.]{According to the Nielsen's Law Network bandwidth double every 21 month; instead data in astrophysics are growing exponentially, with a doubling time of 1.5 years (courtesy G. S. Djorgovski).}\label{hydra:fig1}
   \end{figure*}

In these cases, the problem is to move the data mining toolsets to the data centers. Current strategies, under investigation in some communities such as the VObs, are based on implementing web based protocols for application interoperability (\citealt{derrierre2010} and \citealt{goodman2012a}). Possible interoperability scenarios could be:

\begin{enumerate}
\item	DA1 $\Leftrightarrow$ DA2 (data + application bi-directional flow)
\begin{enumerate}
\item	Full interoperability between DA (Desktop Applications);
\item	Local user desktop fully involved (requires computing power);
\end{enumerate}
\item	DA $\Leftrightarrow$ WA (data + application bi-directional flow)

\begin{enumerate}
\item   Full WA $\Rightarrow$ DA interoperability;
\item	Partial DA $\Rightarrow$ WA (Web Applications) interoperability (such as remote file storing);
\item	MDS must be moved between local and remote apps;
\item	user desktop partially involved (requires minor computing and storage power);
\end{enumerate}

\item	WA $\Leftrightarrow$ WA (data + application bi-directional flow)
\begin{enumerate}
\item   Except from URI exchange, no interoperability and different accounting policy;
\item	MDS must be moved between remote apps (but larger bandwidth);
\item	No local computing power required;
\end{enumerate}

\noindent All of these mechanisms are only a partial solution since they still require to exchange data over the web between application sites. Since the International Virtual Observatory Alliance (IVOA) Interop Meeting, held in Naples in 2011, we proposed a different approach\footnote{\url{http://www.ivoa.net/cgi-bin/twiki/bin/view/IVOA/InterOpMay2011KDD}} :

\item	WA $\Leftrightarrow$ WA (plugin bi-directional exchange)
\begin{enumerate}
\item	All DAs must  become Was;
\item	Unique accounting policy (google/Microsoft like);
\item	To overcome MDS flow, applications must be plug \& play (e.g. any WAx feature should be pluggable in WAy on demand);
\end{enumerate}

\end{enumerate}

The plugin exchange mechanism foresees a standardized web application repository cloud named ``Lernaean Hydra" (from the name of the ancient snake-like monster with many independent but equal heads).
A HEAD (Hosting Environment Application Dock) cloud is in practice a bunch of software containers of data mining model and tool packages, to be installed and deployed in a pre-existing data warehouse.
In such cloud, the HEADs can be in principle different in terms of number and type of available models, originally provided at the installation time. This is essentially because any hosting data center could require specific kinds of data mining and analysis tools, strictly related with specific data and knowledge search types.
But all such HEAD would be practically the same in terms of internal structure and I/O interfaces, being based on a pre-designed set of standards, which completely describe their interaction with external environment and application plugin and execution  procedures.
If two generic data warehouses host a HEAD on their site, they are able to engage the mining application interoperability mechanism by exchanging algorithms and tool packages on demand.
On a specific request the mechanism could indeed start a very simple automatic procedure which moves applications, organized under the form of small packages (some MB in the worst case), through the Web from a HEAD source to a HEAD destination, install them and makes the receiving HEAD able to execute the imported model on local data. Further refinements of the above mechanism could be introduced at the design phase, such as for instance to expose, by each HEAD, a public list of available models, in order to inform other sites about services which could be imported.
Of course, such strategy requires a well-defined design approach, in order to provide a suitable set of standards and common rules to build and codify the internal structure of HEADs and data mining applications (for example any kind of rules like PMML, Predictive Model Markup Language). These standards can be in principle designed to maintain and preserve the compliance with data representation rules and protocols already defined and currently operative in a particular scientific community (such as VO in Astronomy).
Considering also that HPC (High Performance Computing) resources started to be no more prohibitive and more easily approachable by data warehouses, our scenario seems to be really feasible. Any data Center can provide a suitable computing infrastructure hosting a HEAD and become a sort of mirror site of data mining application repository.
The first step towards the realization of a prototype of such strategy is the currently available resource, named DMPlugin. It consists of an SW package useful to import any third-party routine/program into the DAMEWARE platform, according to the standard definition of plugin in the Web terminology. By following graphical wizard steps, the user is indeed able to configure and define all the I/O requirements to wrap his own package, ready to be automatically embedded into the Web App REsource.

It consists in Web Application Repositories (WAR) of data mining model and tool packages, to be installed and deployed in a generic data warehouse. Different WARs may differ in terms of available models since any hosting data center might require specific kinds of data mining and analysis tools. If the WARs are structured around a pre-designed set of standards which completely describe their interaction with the external environment and application plugin and execution procedures, two generic data warehouses can exchange algorithms and tool packages on demand. On a specific request the mechanism starts a very simple automatic procedure which moves applications, organized under the form of small packages (some MB in the worst case), through the Web from a WAR source to a WAR destination, install them and makes the receiving WAR able to execute the imported model on local data. More refinements of the above mechanism can be introduced at the design phase, such as for instance to expose, by each WAR, a public list of available models, in order to inform other sites about services which could be imported. Such strategy requires a standardized design approach, in order to provide a suitable set of standards and common rules to build and codify the internal structure of WARs and of the data mining applications themselves, such as, for example, any kind of rules like Predictive Model Markup Language (PMML \citealt{guazzelli2009}). These standards should be designed to maintain and preserve the compliance with data representation rules and protocols already defined and currently operative in a particular scientific community (such as the VObs in Astronomy).
In case of scheme 4, no local computing power is required. Also smartphones can run DM applications. Then it descends the following series of requirements:

\begin{itemize}
\item	Standard accounting system;
\item	 No more MDS moving on the web, but just moving applications, structured as plugin repositories and execution environments;
\item	 standard modeling of WA and components to obtain the maximum level of granularity;
\item	 Evolution of existing architectures to extend web interoperability (in particular for the migration of the plugins);
\end{itemize}

 \begin{figure*}
   \centering
   \includegraphics[width=12cm]{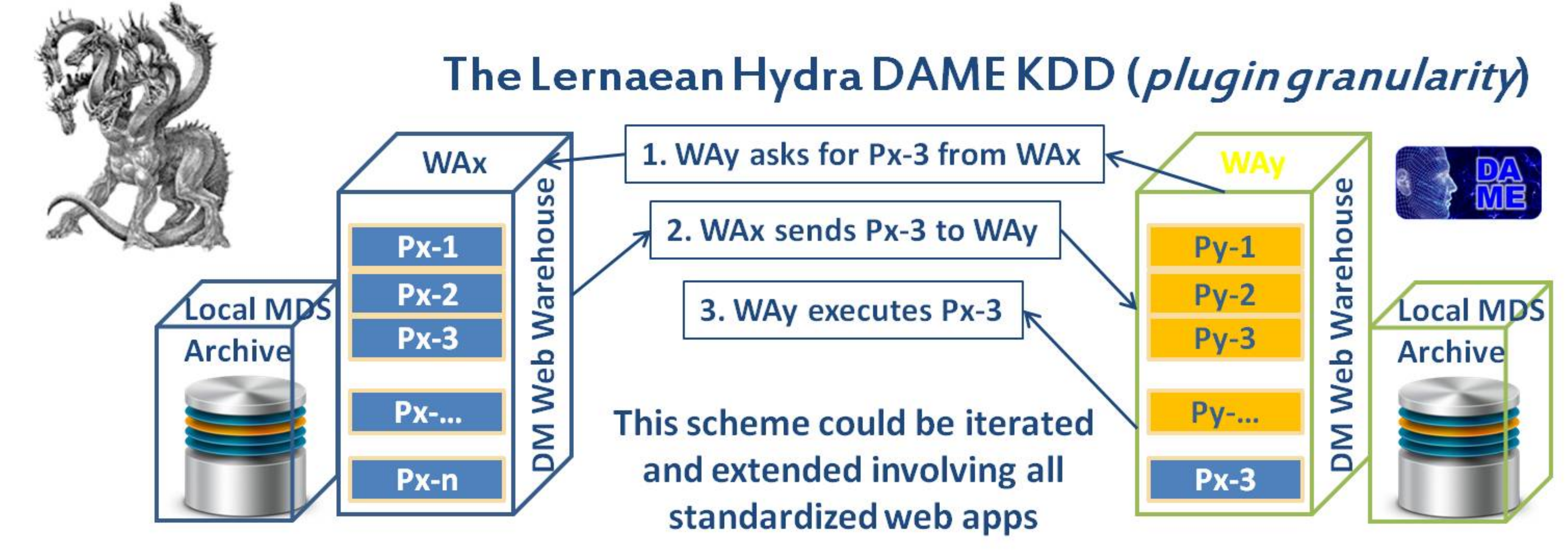}
   \caption{The main steps of the application plugins exchange mechanism among data mining web warehouses.}\label{hydra:fig2}
   \end{figure*}
 \begin{figure*}
   \centering
   \includegraphics[width=10cm]{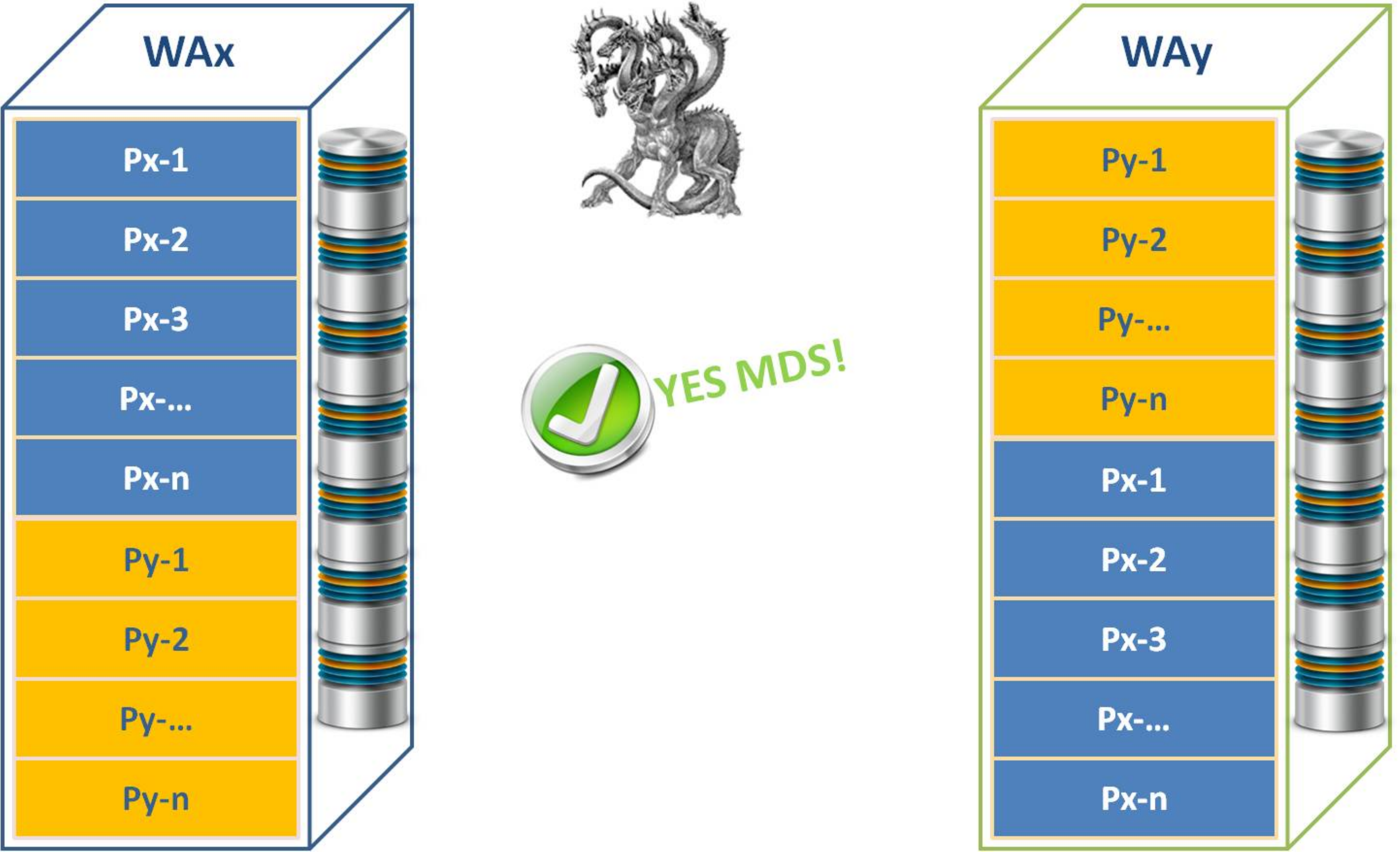}
   \caption[The scenario after several cycles of the application plugins exchange mechanism.]{The scenario after several cycles of the application plugins exchange mechanism. At the end, all web repositories are equivalent mirrors, able to perform data mining experiments on local massive data archives. }\label{hydra:fig3}
   \end{figure*}

   After a certain number of such iterations the scenario will become:
\begin{itemize}
\item	No different WAs, but simply one WA with several sites (eventually with different GUIs and computing environments);
\item	All WA sites can become a mirror site of all the others;
\item	The synchronization of plugin releases between WAs is performed at request time;
\item	Minimization of data exchange flow (just few plugins in case of synchronization between mirrors).
\end{itemize}

Any data center could implement a suitable computing infrastructure hosting the WAR and thus become a sort of mirror site of a world-wide cross-sharing network of data mining application repository in which it could be engaged a virtuous mechanism of a distributed multi-disciplinary data mining infrastructure, able to deal with heterogeneous or specialized exploration of MDS. Such approach seems the only effective way to preserve data ownership and privacy policy, to enhance the e-science community interoperability and to overcome the problems posed by the present and future tsunami of data. By following this approach, the DAMEWARE web application, described in the previous chapters, represents a first prototype towards a WAR mechanism.

\subsection{Present Status}

So far DAME has been an evolving platform and new modules and specific workflows as well as additional features have been continuously added. The modular architecture of DAME can also be exploited to build applications, finely tuned to specific needs. Examples available so far and accessible through the DAME website, being VOGCLUSTERS (Virtual Observatory Globular Clusters)\footnote{\url{http://dame.dsf.unina.it/vogclusters.html}} , a VObs web application aimed at collecting and make available all existing data on galactic globular clusters for data and text mining purposes, and NExt-II (Neural Extractor) for the segmentation of wide field astronomical images. All  models reported in Tab. 1 have already been tested and validated on scientific cases. They have been made available with the official $\beta$ release, of May 2011\footnote{\url{http://dame.dsf.unina.it/beta_info.html}}. In may 2013, due to the continuous technological evolution the, by now, six years old DAME project will be considered officially closed and the work to prepare a new DM platform will begin. This new platform will exploit both the new possibilities offered by ICT and the experience gained through the DAME framework.

For what concerns the ``Hydra Lernaen" approach, its harmonization process into the data centers can indeed be part of a world-wide cross-sharing network, in which it could be engaged a virtuous mechanism of a distributed multi-disciplinary data mining infrastructure, able to deal with heterogeneous or specialized exploration of MDS. It can result at the same time able to preserve data ownership and privacy policy, to enhance the e-science community interoperability and to overcome (not circumventing) the problem to explore present and future tsunami of outcoming data.

%
%

    \section[STraDiWA]{STraDiWA: a simulation environment for transient discovery}\label{sec:STRADIWA}
        \label{marianna:2}

As mentioned in the introduction in the last five years new problems have arisen from the analysis of data-streams, i.e. from the time domain astronomy; while the scientific background will be descrived in chapter \ref{chap:transients}, here we describe the technical aspects of the DAME web application used as a simulation environment for variable objects and transients: STraDiWA.\\
The time domain approach makes use of theoretical or empirical models to determine the class of an astrophysical source.  An example of this type of classification is the light curve fitting. For each new object we can fit empirical light curves to the data and the fits determine whether or not the object belong to a given class. This technique can be used by restricting to a limited number of cases. For example, assuming that a source is a certain type of supernova, a well sampled group of observed light curves can be used to to classify a new one.\\
\noindent Another type of classification is based on features, that is information derived from time series images and contextual data. The set of features represents a multidimensional space, id est an ideal working environment for machine learning algorithms. Features can be arbitrarily simple or complex. They can belong to time domain or be contextual information. The most common time domain features are based on the distribution of detected fluxes and can be derived from the light curves of the sources, such as flux ratios, amplitude, skewness or significant frequencies (in case of periodic light curves). Contextual information are instead characteristics of the source which do not change with time, like object coordinates, distance from the nearest detected galaxy and the parameters available for that galaxy, etc. Contextual information are often difficult to include in the classification process, due to the heterogeneity of the data. However, they are useful to discriminate different types of objects (e.g. an event which can be a cataclysmic variable or a supernova is more likely a supernova if it is close to a galaxy). Classification methods based on features  can be either supervised or unsupervised depending on whether or not we have a training sample of labeled data. The former ones aim to assign to a given source its class (or class probabilities), while the latter instead try to recognize clusters in the features space.\\


\noindent There are many supervised methods which can be used for classification of astrophysical variables. \\
One of the most suitable methods to deal with a sparsely populated knowledge base are Bayesian classifiers. These networks can be used to compute the  the probabilities of the object belonging to different classes of transients. Then, by making use of objective criteria, the we can determine whether or not the probability for the class of interest is high enough and if the object needs follow-up observations. In order to use the Bayesian approach we have to generate a library of prior distributions. Each distribution has to take into account several factors such as brightness changes in a certain filter over a certain time interval. These distributions need to be estimated for each type of variable astrophysical phenomenon that we want to classify. \\
\noindent Another method  for the classification of variable sources is to use Support Vector Machines (SVMs). These algorithms try to find in the features space the hyperplane which best separate the components of each pair of classes. If the two classes are not linearly separable, the SVMs make use of different kernel functions which map points of the input space into a higher dimension space in which the classes become linearly separable. SVMs have been used in recent works for variable stars classification (\citealt{willemsen2007}, \citealt{richards2011}). \\
\noindent Other popular algorithms for supervised classification are the Artificial Neural Networks (ANNs). ANNs are, in their simplest form, non-linear regression algorithms in which the classification is the result of a non linear combination of the input features. These algorithms have been used in particular to separate real transient sources from a variety of data artifacts, with a classification rate of $\mathrm{\sim 90\%}$ (\citealt{donalek2008}).\\
A new method developed by the CRTS team is based on decision trees (\citealt{graham2012c}). The use a set of 60 features extracted from the light curve of the object to build a set of decision trees which are able to discriminate between different classes of variable object.
\noindent When we do not have a set of labeled data the only classification method we can apply is an unsupervised algorithm. The most famous unsupervised algorithms in time domain astronomy are the Gaussian Mixture Modeling (GMM) algorithm and the Self Organizing Map (SOM). In case of GMM each cluster is represented by a parametric gaussian distribution and then the entire data set is modeled by a mixture of these distributions. A SOM is an unsupervised type of ANN which aim to map the feature space into a space usually two-dimensional or three dimensional, without loosing the topology of the input space.\\
\noindent From this simple review of the methods above it is clear that the classification of variable objects poses many problems: how to characterize them (using light curves or statistical indicators), in case of supervised method which type of knowledge base use ( built on the data themselves or on simulated ones), how to solve the computational challenge, how to find the unknown, throwing away all the known or clustering. \\
\noindent The strategy of our project is to use a hierarchical approach to classification. Different types of classifiers perform better for some event classes than for the others, so we propose to testing different  data mining algorithms for variable objects classification in order to find the one optimized for each type of variable object. Our approach has the typical decision tree structure and aims at a classification which becomes finer and finer as we to higher level of branching.
\begin{figure}
\centering
\includegraphics[width=12cm]{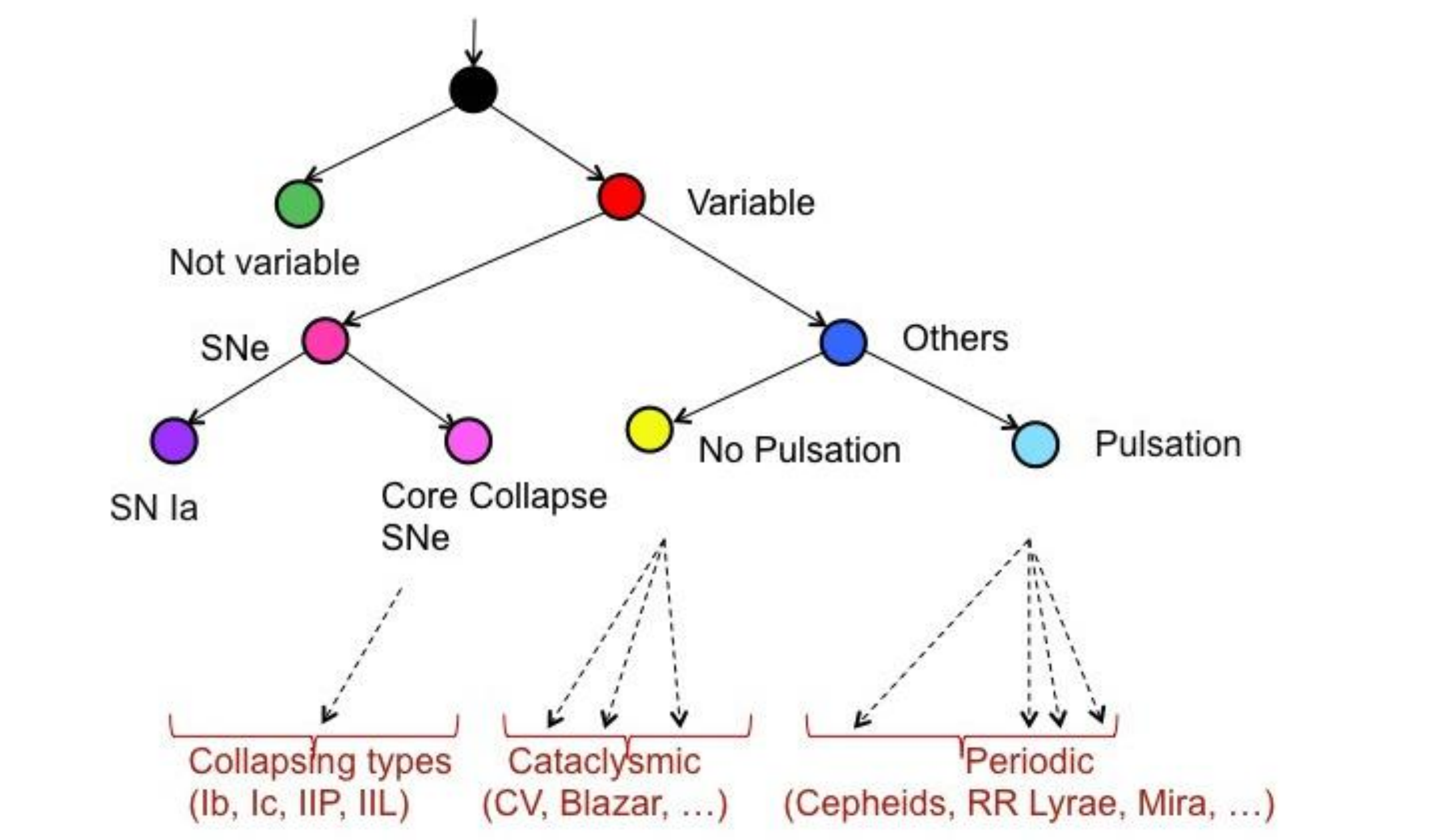}
\caption{Classification Scheme.}
\label{marianna:sbs}
\end{figure}

\noindent In our opinion, the best way to test these classifiers is through simulated data since they allow to better control the various sources of systematics.
Real data are not always suitable because they are often incomplete. A transient, in fact, detected by an increase in brightness is often missing in archival sky surveys and may have just a couple of relatively closely spaced observations in a couple of epochs to go by.
A similar approach has been undertaken by the LSST Image Simulation Group.  They have implemented a simulation pipeline which is very useful to test detection algorithms. In their first simulations transients were in fact generated randomly, loosing comprehensive theoretical treatment.
Our proposal, instead, is produce template images in which variable objects are added according to specific data models.  \\

\subsection{Simulation Pipeline}
\label{marianna:2.1}

To test the classifiers we developed a simulation pipeline which allows us to build a time series of image and from each image extracts a catalog of sources and their properties. The objects detected for each epoch, then, have to be merged in a single catalog which contains the information for all the epochs.
The flowchart of our simulation pipeline is shown in \ref{marianna:flowchart}. The pipeline makes use of three astronomical software  \textbf{Stuff}, \textbf{SkyMaker}, and SExtractor developed by E. Bert\`{\i}n and available at  \href{http://www.astromatic.net/software}{http://www.astromatic.net/software}. Stuff is responsible for the creation of a catalog of background galaxies, SkyMaker produces an image starting from a catalog of galaxies produced by STUFF, adding a random stellar field, while SExtractor is the software chosen for the catalog extraction. The reasons why we chose SExtractor instead of other similar astronomical software are explained in section~\ref{chap:comparison}.
\begin{figure}
\centering
\includegraphics[width=12cm]{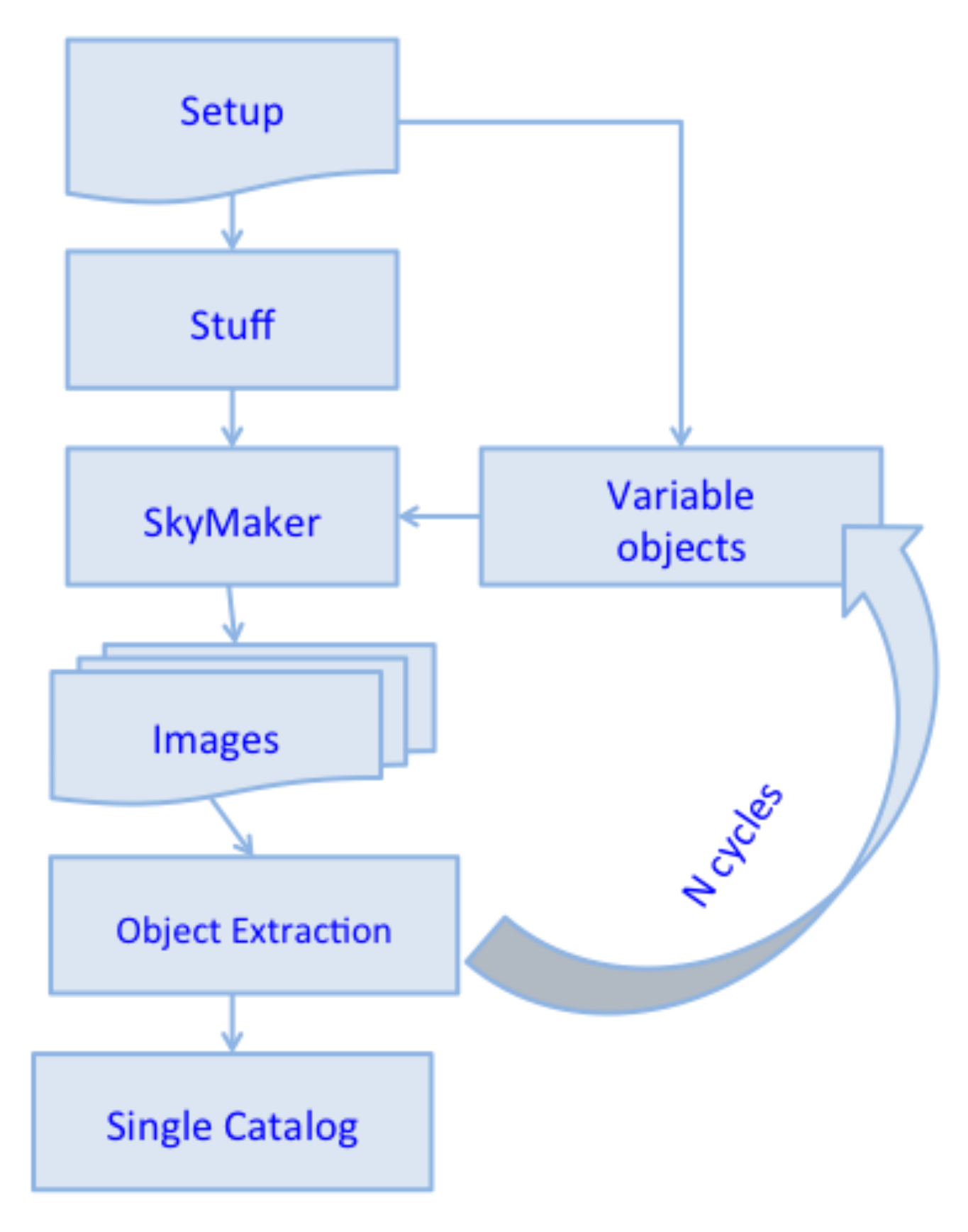}
\caption{Flowchart of the proposed simulation pipeline.}
\label{marianna:flowchart}
\end{figure}

\subsection{Setup Phase}
\label{marianna:2.2}
In the setup phase we have to set all the information needed to reproduce realistically an astronomical image. These factors are related to the objects distribution and properties, to the survey strategy, to the instrumental setup, and to the observing conditions. \\ The setup can be set through a configuration file, named STraDiWA.config. For reasons of comprehensibility in the STraDiWA configuration file we integrated and assembled all the common parameters to Stuff and SkyMaker and SExtractor. Beside these common parameters the user has also to set the configuration file needed for Stuff and SkyMaker in which there are the parameters specific for each software.\\
In the setup phase the user must choose mainly the survey strategy and the observing conditions. The survey strategy is determined by the number of the filter in which we want to produce the images (\texttt{PASSBAND\_OBS}), the limiting in magnitude of these filters (\texttt{MAG\_LIMITS}) and the sampling rate (\texttt{SAMPLING}).
\noindent When simulating (or observing) a time series of images, the sampling rate can be:
\begin{itemize}
\item Uniform. In this case the user can choose the number of days for which the same region of the sky is observed, how often in a night and must specify the length of the night in hours.
\item Uneven. In this case the user must provide a time series (in hours).
\end{itemize}
The observing conditions are ruled by the seeing full with half maximum (\texttt{SEEING$\_$ FWHM}). During several observations spaced in more days the seeing vary following two possible options:
\begin{itemize}
\item The\texttt{ SEEING$\_$FWHM } can assume each day a different value between a series specified by the user, where each value can be repeated during the simulation. A special case is that of constant seeing, which is however unrealistic.
\item The \texttt{SEEING$\_$FWHM} varies randomly each day between a given minimum and maximum.
\end{itemize}
\noindent Last but not least, the user has also to choose the type and distribution of variable objects (\texttt{VARIABLE}). The user can either choose to define for each object its parameters, or can choose to assign them randomly between their range of variation.

\subsection{Stuff: creation of the static sky}
\label{marianna:2.3}

Stuff is a software that combines spectra, luminosity functions and physical parameters to generate artificial catalogs of the deep extragalactic sky in a standard universe driven by ($\Omega_M, \Omega_{\Lambda}$). The program distributes galaxies in redshift space that is subdivided into bins. For every bin the number of galaxies for the Hubble types E, S0, Sab, Sbc, Scd and Sdm/Irr is determined from a Poisson distribution assuming a non-evolving Schechter luminosity function. The different galaxy types are simulated by linearly adding exponential (disk component) and de Vaucouleurs profiles (bulge components) in different ratios. To each galaxy are assigned a random disk inclination angle and a position angle which define the intrinsic ellipticity of the object. The output of the program is a catalog of galaxy positions, apparent magnitudes, semi-minor and major axes, position angles for disks and bulges, de Vaucouleurs type and redshift.\\
The key input parameters that have to be modified by the user, according to his own purposes, are the image dimensions (\texttt{IMAGE\_SIZE}), the pixel size (\texttt{PIXEL\_SIZE}), the allowed range of apparent magnitudes  (\texttt{MAG\_LIMITS}), the detector gain  (\texttt{GAIN}), and the required filters (\texttt{PASSBAND\_OBS}).  There are 120 filters available, covering a wavelength range from 0.29 to 87.74831 $\mu$m.The others input keys are mainly related to the cosmological model taken into account, like the value of the Hubble constant, or to the Schechter's functions. \\

\subsection[SkyMaker]{SkyMaker: Instrumental simulation and image production}
\label{marianna:2.4}

SkyMaker started out as a testing tool for the SExtractor source extraction software developed by the same author. It works by taking as input files a list of sources, which can be produced by Stuff, and a setup file. To produce an image this software first build a Point Spread Function model (PSF), that is distribution of the light from a point source, then reads the input catalog and renders sources at the specified pixel coordinates in the frame. Finally SkyMaker provide to add a uniform sky background, with surface brightness provided by the user through the input {\texttt{BACK$\_$MAG}} parameter, and applies to the image Poissonian photon white noise and Gaussian read-out noise of the detector. \\
\subsubsection{PSF Modeling}
The PSF used by SkyMaker can be internally generated or loaded through an external fits file. The PSF internal generator has to be able to represent with decent accuracy the PSF of typical astronomical instruments. This means that has to take into account the atmospheric blurring, telescope motion blurring, instrument diffraction and aberrations, optical diffusion effects and intra-pixel response. The produced PSF is a convolution between these components.\\
We want to focus only on these components that the user can control by modifying the configuration file: the instrument diffraction and aberrations and the optical diffusion. The instrumental PSF become dominated by diffusion beyond a few FWHMs from the center. This effect produces a so called `aureole" that has to be taken into account when simulating deep and wide galaxy fields, as it reproduces the background variations found on real images around bright stars.
The SkyMaker PSF simulator reproduces the effects of diffraction and aberrations in the Frauhofer regime of Fourier optics by manipulating a virtual entrance pupil function $p(\rho,\theta)$. The amplitude part of p is mainly determined by the characteristic of the primary mirror M1, by the effects caused by the presence of  spider arms, and by the obscuration of the secondary mirror on the primary.

Optical aberrations may be added by introducing changes of phase $\phi(\rho, \theta)$ of the complex pupil function. SkyMaker can simulate a wide range of aberrations:
\begin{itemize}
\item[-] defocus: $\mathrm{\phi_{defocus}\propto \rho^2}$,
\item[-] astigmatism: $\mathrm{\phi_{asti}\propto \rho^2cos^2(\theta-\theta_{asti})}$,
\item[-] coma: $\mathrm{\phi_{coma}\propto \rho^3cos(\theta-\theta_{coma})}$,
\item[-] spherical: $\mathrm{\phi_{spher}\propto \rho^4}$,
\item[-] tri-coma: $\mathrm{\phi_{tri}\propto \rho^3cos^3(\theta-\theta_{tri})}$, and
\item[-] quad-ast: $\mathrm{\phi_{quad}\propto \rho^4cos^4(\theta-\theta_{quad})}$.
\end{itemize}
\noindent Phase terms are individually normalized following the ESO $d_{80}$ convention: phase coefficients represent the diameter of a circle enclosing 80$\%$ of the total flux of an aberrated spot on the focal plane.\\
A set of input parameters, like the diameters of M1 and the central obscuration, the number, position angle and thickness of the spider arms, makes it possible to simulate with reasonable accuracy the diffraction pattern of most common telescope configurations.

\subsubsection{Source Modeling}
After build the PSF, SkyMaker deals with the source modeling. So far SkyMaker can model only galaxies and stellar objects.
The galaxies are modeled as a sum of a bulge profile and an exponential disk. The bulge follows a de Vaucouleurs profile:
\begin{equation}
\mu_B(r) = m -2.5log(B/T) + 8.3268\bigg(\frac{r}{r_{eff}}\bigg )^{1/4}.
\label{marianna:spher}
\end{equation}
where $\mu_B$ is expressed in $mag/arcsec^2$, m is the apparent magnitude, $B/T$ the apparent bulge-to-total ratio and $r_{eff}$ the effective radius of the spheroid in arcseconds. The disk component is given as exponential profile:
\begin{equation}
\mu_D(r) = m -2.5log(1-B/T) + 1.0857\bigg(\frac{r}{r_{h}}\bigg )+5logr_h + 1.9955,
\label{marianna:disk}
\end{equation}
where $r_h$ is  is the disk scalelength in arcseconds. The parameters m, $B/T$ , $r_{eff}$ , $r_h$ as well as independent aspect ratios and position angles for both components must be read from the input list. \\
\noindent There are many parameters to be set in the SkyMaker configuration file. The most important are related to \textit{i)} pupil features, e.g. the size of the mirrors and the aberration coefficients, \textit{ii) }the  detector characteristics, e.g. gain, saturation level and image size, \textit{iii)} PSF model, e.g radius and surface brightness of the aureole and \textit{iv} observing condition, full width half maximum of the seeing and exposure time. \\
We wish to stress that the simulations obtained with the combination of Stuff and SkyMaker do not include any artefact such as bad pixels, ghosts or bad columns or other effects which.\\

\subsection{Rules for variable objects}
\label{marianna:2.5}
An important aspect of the project was to identify relevant group of variable objects and derive sets of rules for their definition. The models developed for each variable object must be seen as particular instances of a general template. A variable object must be defined by a series of parameters. he number and type of parameters can vary for each class of variable object, for example for a periodic variable we can specify the amplitude, the period, etc., while for a cataclysmic variable can be important to specify the time of the explosion. Furthermore, each class has to take into account the different behavior of the objects at various wavelength.
The modules so far implemented are the Classical Cepheids and the type Ia Supernovae. We choose to start implementing the Cepheids because they are the classical example of periodic objects and are rather simple to model. Type Ia Supernovae were the following choice according to the classification scheme proposed in Figure~\ref{marianna:sbs}.\\
Furthermore we have implemented also a module for random objects. These objects in fact were verify the simulation setup fixed at the beginning of the project. Their magnitude vary randomly in the magnitude limits set in Stuff and SkyMaker in an unrelated way in each band.\\

\subsection{Classical Cepheids}
\label{marianna:2.5.1}

Classical Cepheids are pulsating stars whose magnitude vary periodically. Their light curves is generally approximates as a sinusoid with a constant phase term. The amplitude of their variation ranges from 0.2 to 2 mag, as estimated by American Association of Variable Star (AAVSO)\footnote{\href{http://www.aavso.org/}{http://www.aavso.org/}}.
We begin to simulate Classical Cepheids in our Galaxy.
In order to model this class the steps are:
\begin{itemize}
\item assign the Period;
\item use P-L relation in order to find the mean absolute magnitude;
\item assign the phase;
\item assign the amplitude;
\item assume a sinusoidal law and evaluate the temporal evolution of the absolute magnitudes;
\item estimate correction for stellar extinction, using the values of absorption coefficients given by \citet*{tammann2003} and extracting randomly the color excess.
\end{itemize}
Using different calibrations of PL relationships, we can model different types of Classical Cepheid, for example discriminating between pulsation mode or take into account their metallicity. So far, we used the coefficients for the mean  PL relation calibrated in \citealp{bono2010} and reference therein valid for Galactic Cepheids. The number of bands for which we have a calibration for PL relation limits the number of band in which we can model our objects. So far this relation has been calibrated mainly in Johnson bands BVRIJHK.\\
\noindent For a classical Cepheid we have four free parameters: the initial apparent magnitude $m_i$, the period P, the phase $\phi$, and the amplitude variation A . These can by defined by the user or extracted randomly in the appropriate ranges. These are:
\begin{itemize}
\item $[0,\, 2\pi] $ for the phase;
\item $[0,\, 2]$ mag for the amplitude;
\item $[0,\, 70]$ days for the period;
\item magnitude limits of Stuff and SkyMaker for the initial magnitude.
\end{itemize}
\noindent Coordinates of the objects are extracted randomly within the image.\\
Actually simulations of Classical Cepheid have been made in Johnson B, V and I band using coefficients for period-luminosity relation calibrated in \citet{tammann2003}.

\subsection{Type Ia Supernovae}
\label{marianna:2.5.2}

\citealt*{contardo2000} used an empirical model to fit the light curve of a sample of type Ia Supernovae in the UBVRI Johnson filter.
They found an analytical form of the light curves consisting of a Gaussian (for the peak phase) atop a linear decay (late-time decline), a second Gaussian (to model the secondary maximum in the V, R, and I band light curves), and an exponentially rising function (for the pre-maximum segment):
\begin{equation}
m(t) = \frac{f_0 + \gamma (t-t_0)+g_0e^{\frac{(t-t_0)^2}{2\sigma_0^2}}+g_1e^{\frac{(t-t_1)^2}{2\sigma_1^2}}}{(1-e^{\frac{\tau -t}{\theta}})}.
\label{marianna:cont}
\end{equation}
We decided to use this expression to model type Ia Supernovae. As we can see in Eq. \ref{marianna:cont} we have to set eight parameters ($f_0,\, \gamma,\, t_0,\, g_0,\, \sigma_0,\, g_1,\, t_1,\,$\\
$ \sigma_1,\, \tau,\, \theta$) for each band. Actually, some of these parameters are related to each other.
To obtain realistic groups of values for the simulations we searched for the range within which those parameters vary in the Contardo PhD Thesis. Analyzing the data we found the variation ranges reported in Tab. \ref{marianna:contardo_tab}. We excluded the U band because of the lack of information.  \\
\begin{table}
\small
\centering
\begin{tabular}{|>{\columncolor{lightgrey_1}{\bf}}l|c|c|c|c|}
\hline

\cellcolor{lightgrey} Parameter & \multicolumn{1}{|>{\columncolor{lightgrey_1}{\bf}} c}{B Band} & \multicolumn{1}{|>{\columncolor{lightgrey_1}{\bf}} c} {V Band}  &  \multicolumn{1}{|>{\columncolor{lightgrey_1}{\bf}} c} {R band} &  \multicolumn{1}{|>{\columncolor{lightgrey_1}{\bf}} c|} {I Band}\\
\hline
$f_0$ (mag) & [14,  20] & [13,  19]  & [12, 18]  & [9, 18]\\
\hline
$\gamma$ (mag/days) & [0.01, 0.025] & [0.02, 0.03] & [0.025, 0.04] & [0.015, 0.06]\\
\hline
$g_0$ (mag) & [-3.5, -2] & [-2.5, -1] & [-1.5, -0.5]  & [-3, -0.5 ]\\
\hline
$\sigma_0$ (days) & [10, 18] & [5, 30] & [5, 10] & [5, 15]\\
\hline
$g_1$ (mag) & & [-0.5, -0.1] & [-0.6, -0.2] & [-0.8, -0.5]\\
\hline
$\sigma_1$ (days) & & [4, 10] & [4, 10] & [5, 15]\\
\hline
$\theta$ (days) & [1, 10] & [1, 10] & [1, 10] & [1, 10]\\
\hline
\end{tabular}
\caption[Ranges of variation chosen for the parameters $f_0,\, \gamma,\, g_0,\, \sigma_0, \, g_1,\, \sigma_1,\, \theta$]{Ranges of variation chosen for the parameters $f_0,\, \gamma,\, g_0,\, \sigma_0, \, g_1,\, \sigma_1,\, \theta$ of Eq. \ref{marianna:cont}.}
\label{marianna:contardo_tab}
\end{table}

\noindent There are not variation ranges for $g_1$ and $\sigma_1$ in the B band since in this band the light curve of a Supernova Ia does not show the secondary maximum.
Temporal parameters, $\mathrm{t_0\, t_1\, and\, \tau } $ are not reported in Table~\ref{marianna:contardo_tab}. Instead, we choose to study the correlation between those parameters, focusing in particular on the relation between $\mathrm{t_{0}}$ in B band and  $\mathrm{t_0\, t_1\, and\, \tau } $ in other bands and between $\mathrm{t_{0}}$ and $\mathrm{\tau_{0}}$ in B band. Relations between temporal parameters result to be linear. Fig. \ref{marianna:tv} - \ref{marianna:ti} show $\mathrm{t_0,\, t_1\, \tau} $ in each band as function of $\mathrm{t_{0}}$ in B band. Fig. \ref{marianna:tauBt0B} shows $\mathrm{t_{0}}$ in function of $\mathrm{\tau}$ in B band. In the caption of each figure is reported the result of the linear regression.
All temporal values are measured in Julian Dates.\\

\clearpage
\begin{figure}
\centering
\includegraphics[scale=0.42]{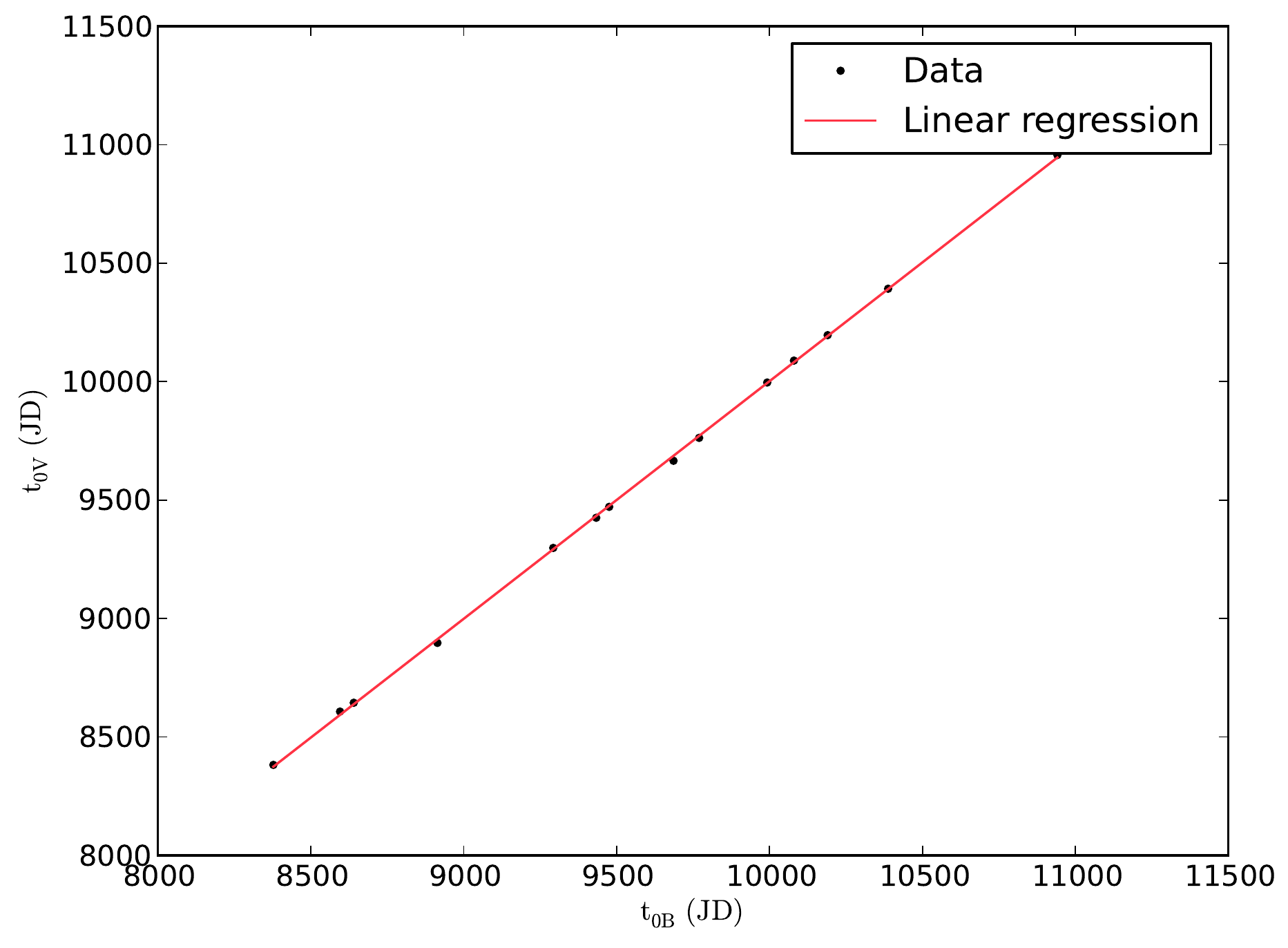} (a)
\includegraphics[scale=0.42]{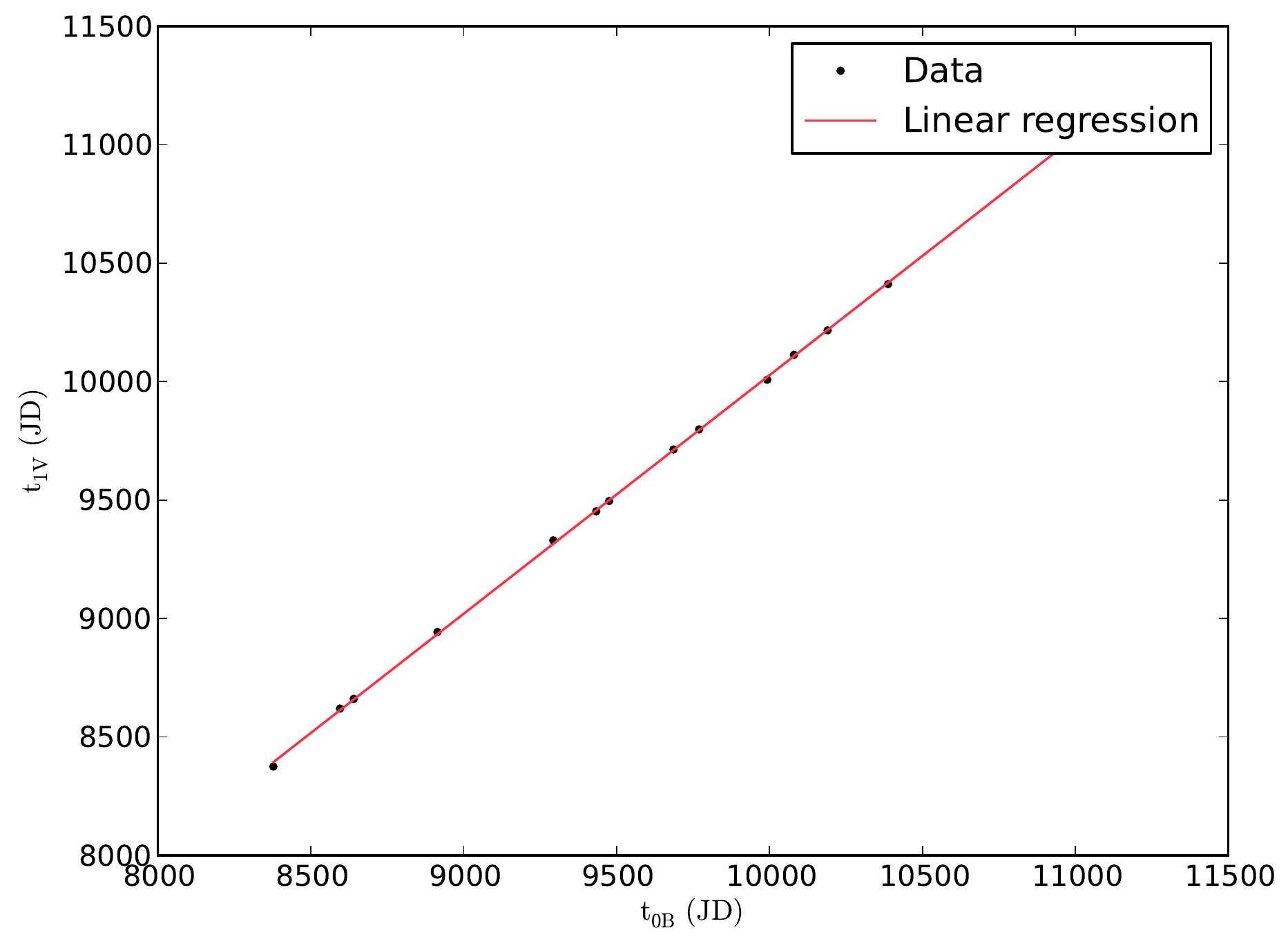} (b)
\includegraphics[scale=0.42]{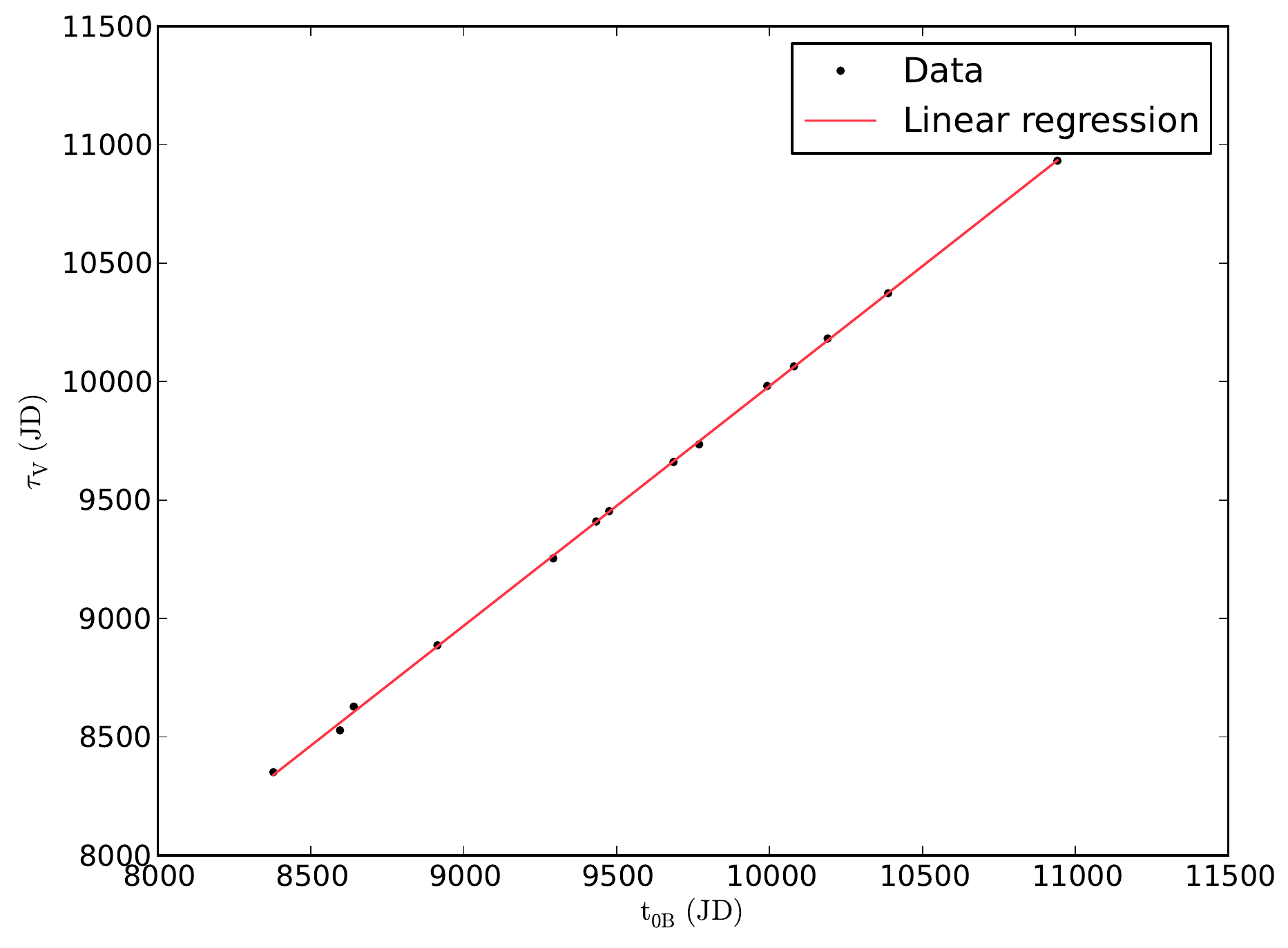} (c)
\caption[On the x-axis: $t_0$ in B band.  On the y-axis $t_0$ in V band (panel a) , $t_1$ in V band (panel b), $\tau$ in V band (panel c).]{On the x-axis: $t_0$ in B band.  On the y-axis $t_0$ in V band (panel a) , $t_1$ in V band (panel b), $\tau$ in V band (panel c). The equation of the best fits are:\\
$t_{0V}=1.003*t_{0B} -28.80$. The r-square of the fit is: 0.99981.\\
$t_{1V}=1.007*t_{0B}-42.09$.The r-square of the fit is: 0.99981.\\
$\tau_{V}=1.012*t_{0B}-139.69$. The r-square of the fit is: 0.99970.}
\label{marianna:tv}
\end{figure}

 \begin{figure}
 \centering
\includegraphics[scale=0.42]{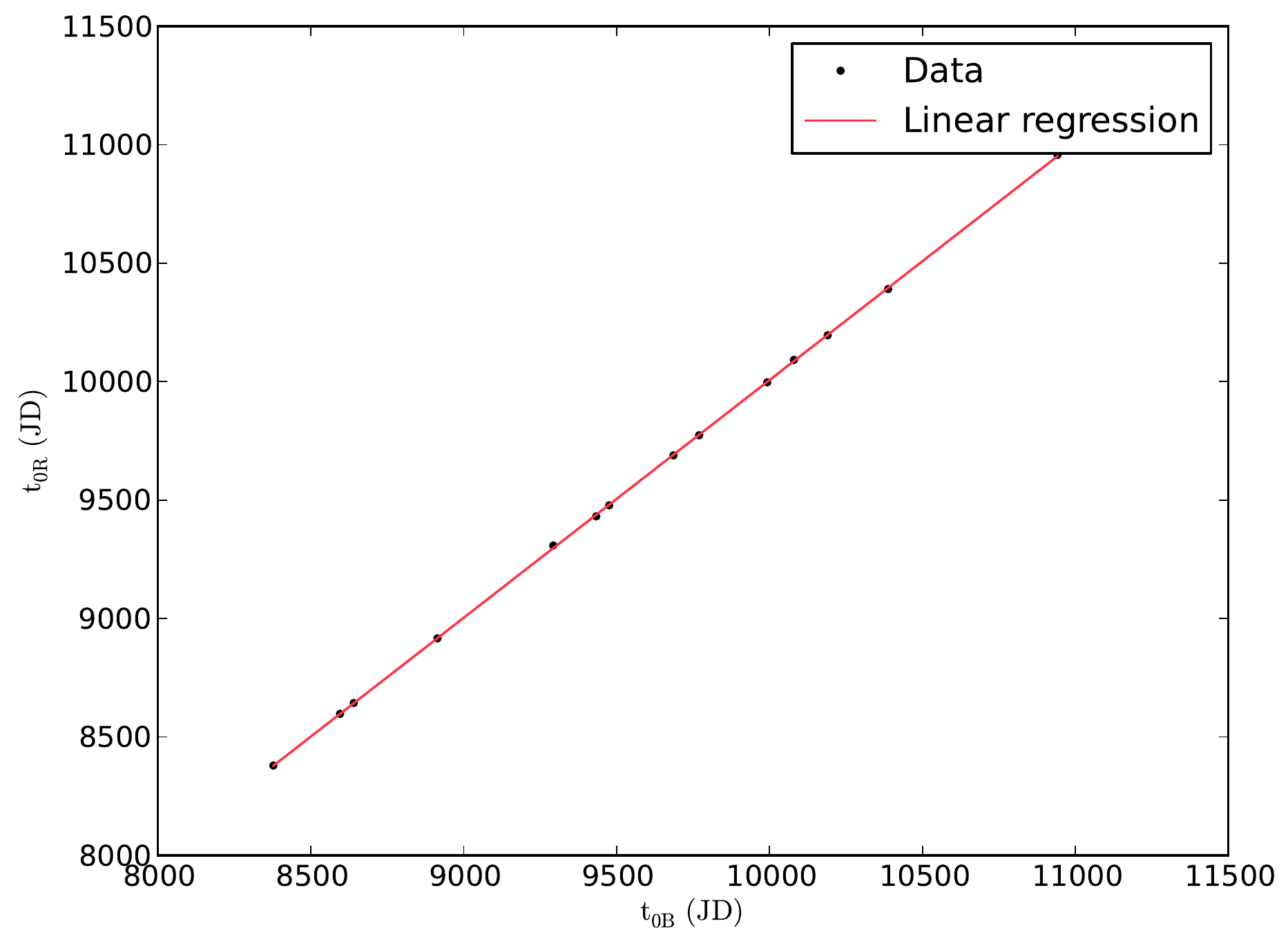} (a)
\includegraphics[scale=0.42]{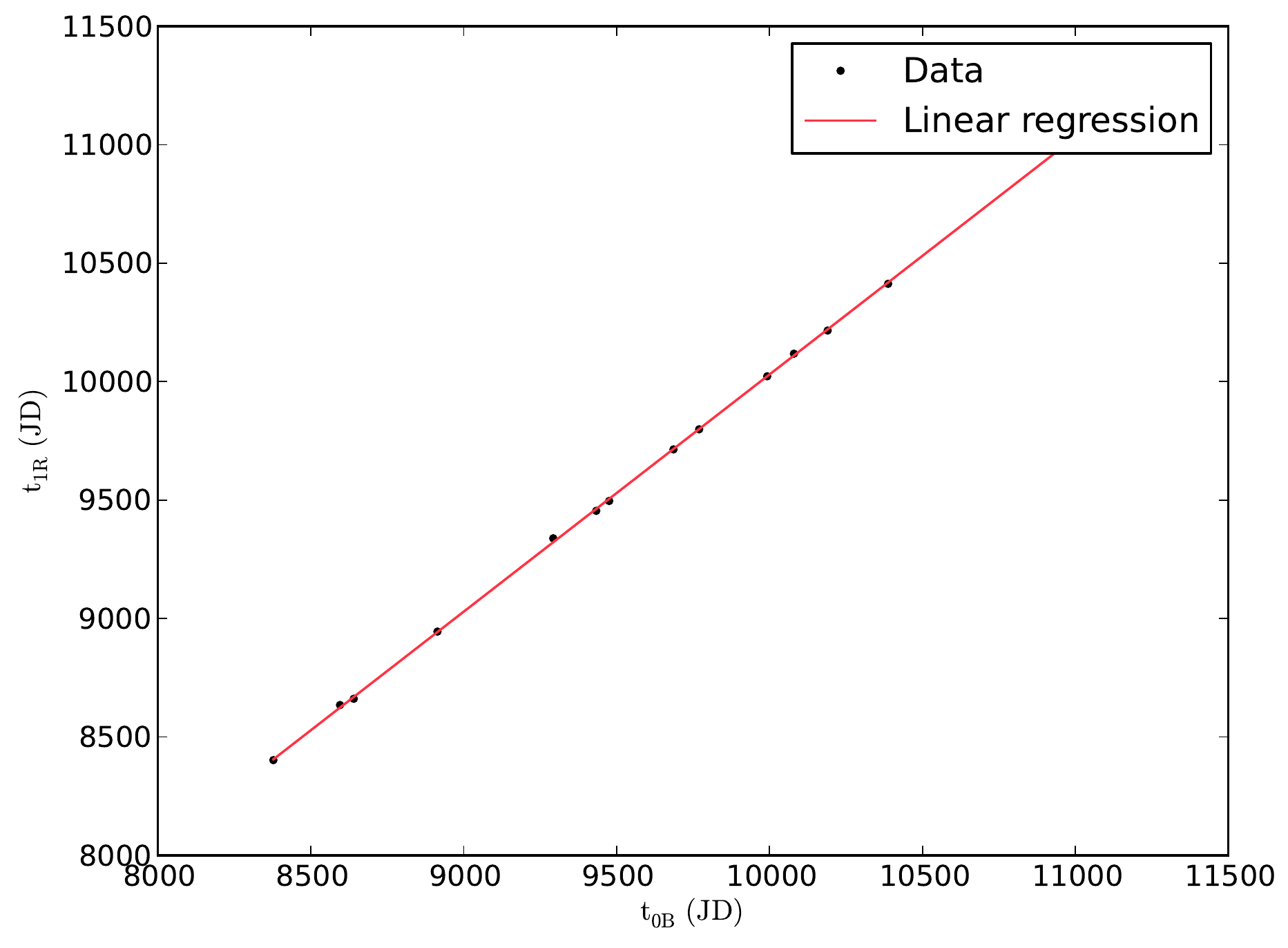} (b)
\includegraphics[scale=0.42]{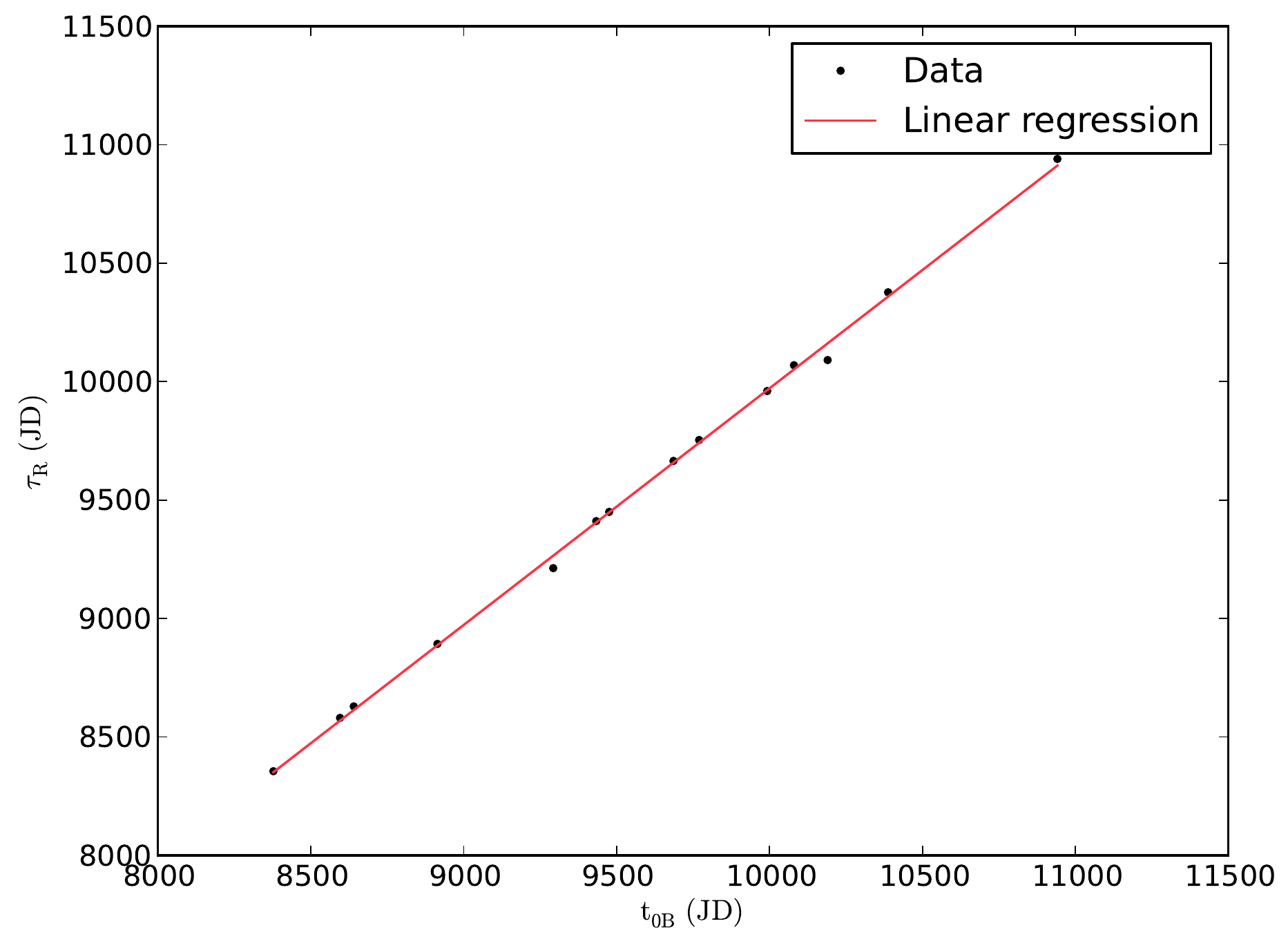} (c)

 \caption[On the x-axis: $t_0$ in B band.  On the y-axis $t_0$ in R band (panel a) , $t_1$ in R band (panel b), $\tau$ in R band (panel c).]{On the x-axis: $t_0$ in B band.  On the y-axis $t_0$ in R band (panel a) , $t_1$ in R band (panel b), $\tau$ in R band (panel c). The equation of the best fits are:\\
$t_{0R}=1.003*t_{0B}-28.23$. The r-square of the fit is: 0.99996.\\
$t_{1R}=1.001*t_{0B}+15.80$.The r-square of the fit is: 0.99989.\\
 $\tau_{R}=0.999*t_{0B}-17.25$. The r-square of the fit is: 0.99862.
}
\label{marianna:tr}
\end{figure}

 \begin{figure}
\centering
\includegraphics[scale=0.42]{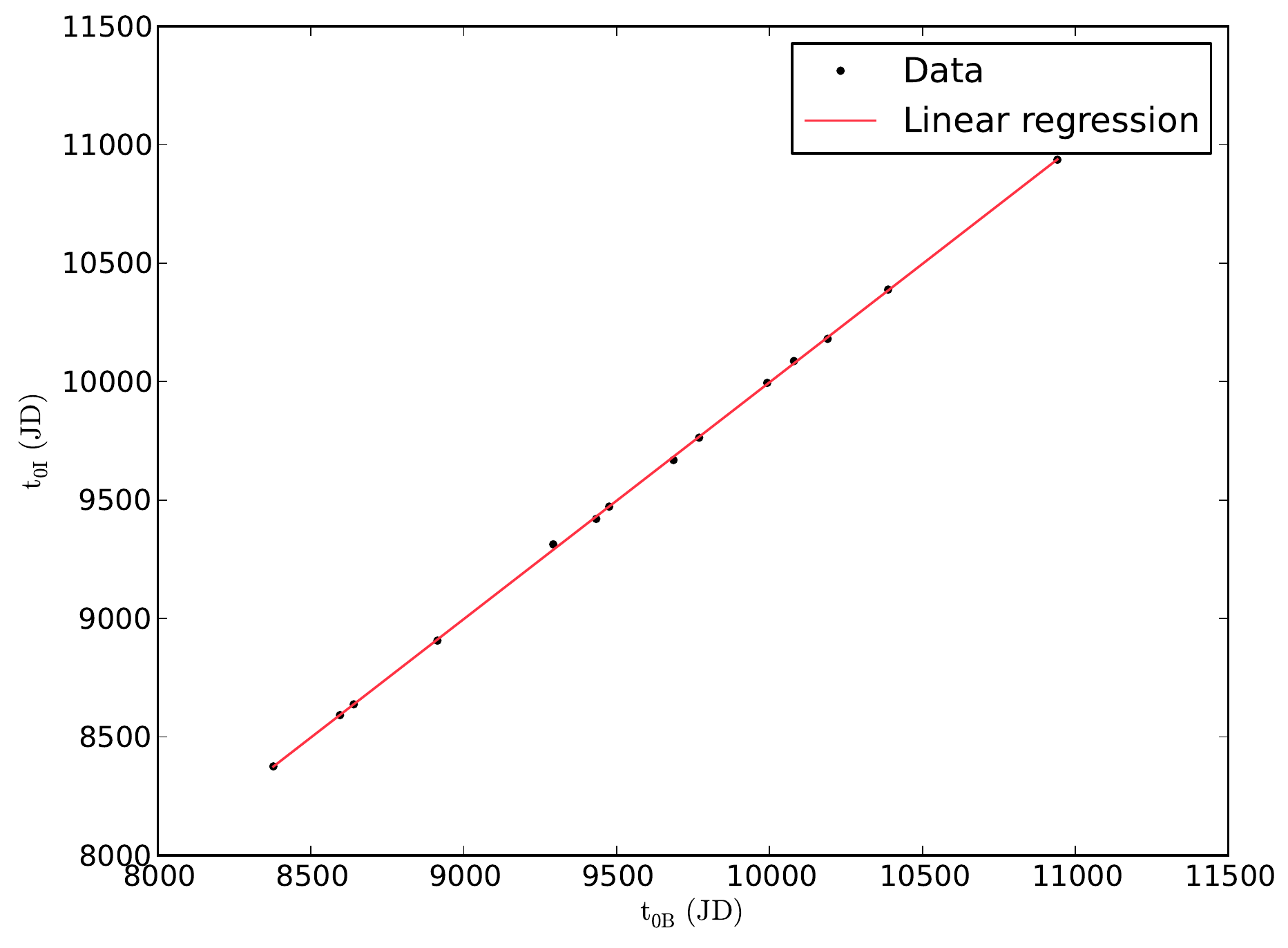} (a)
\includegraphics[scale=0.42]{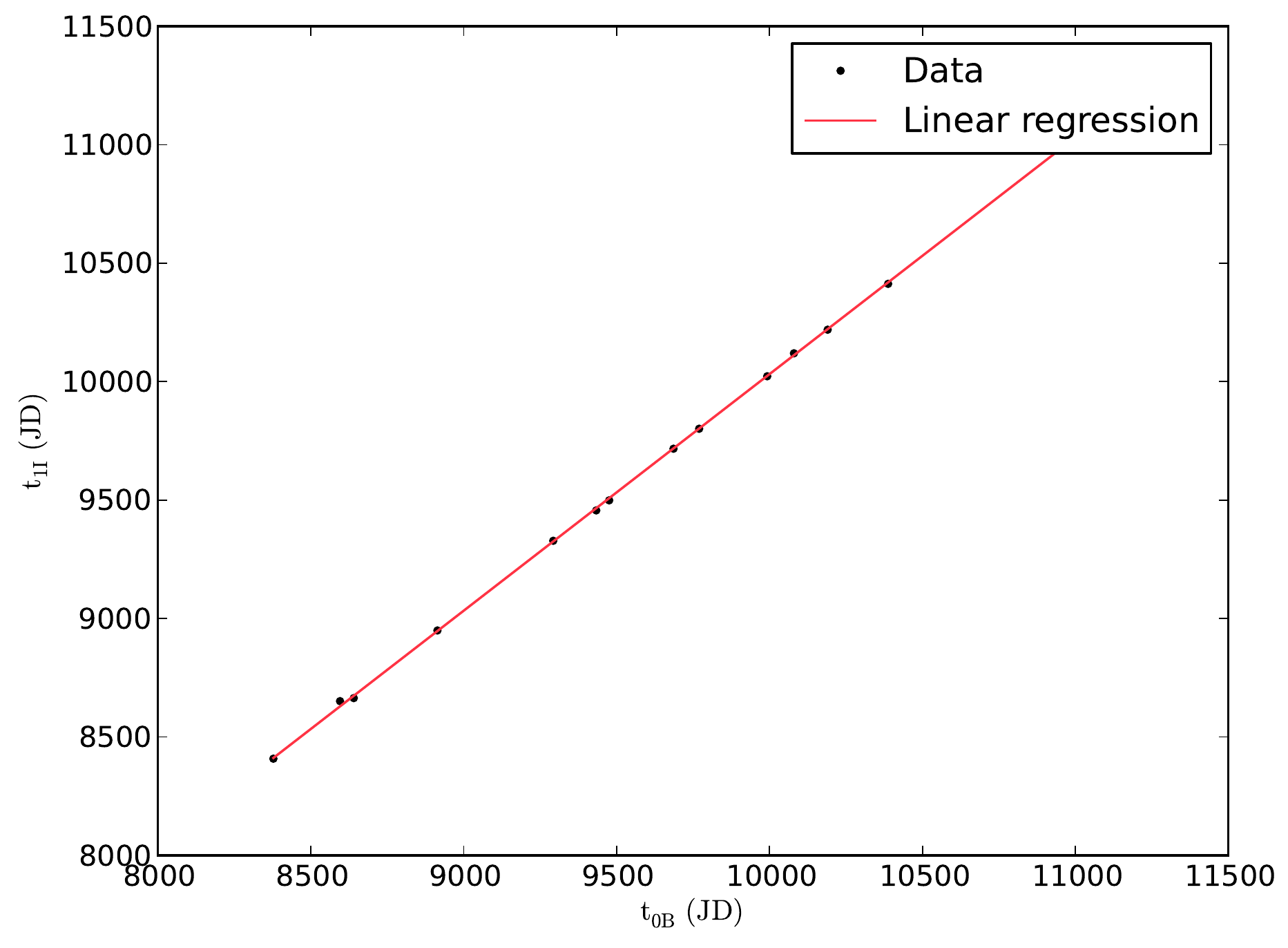} (b)
\includegraphics[scale=0.42]{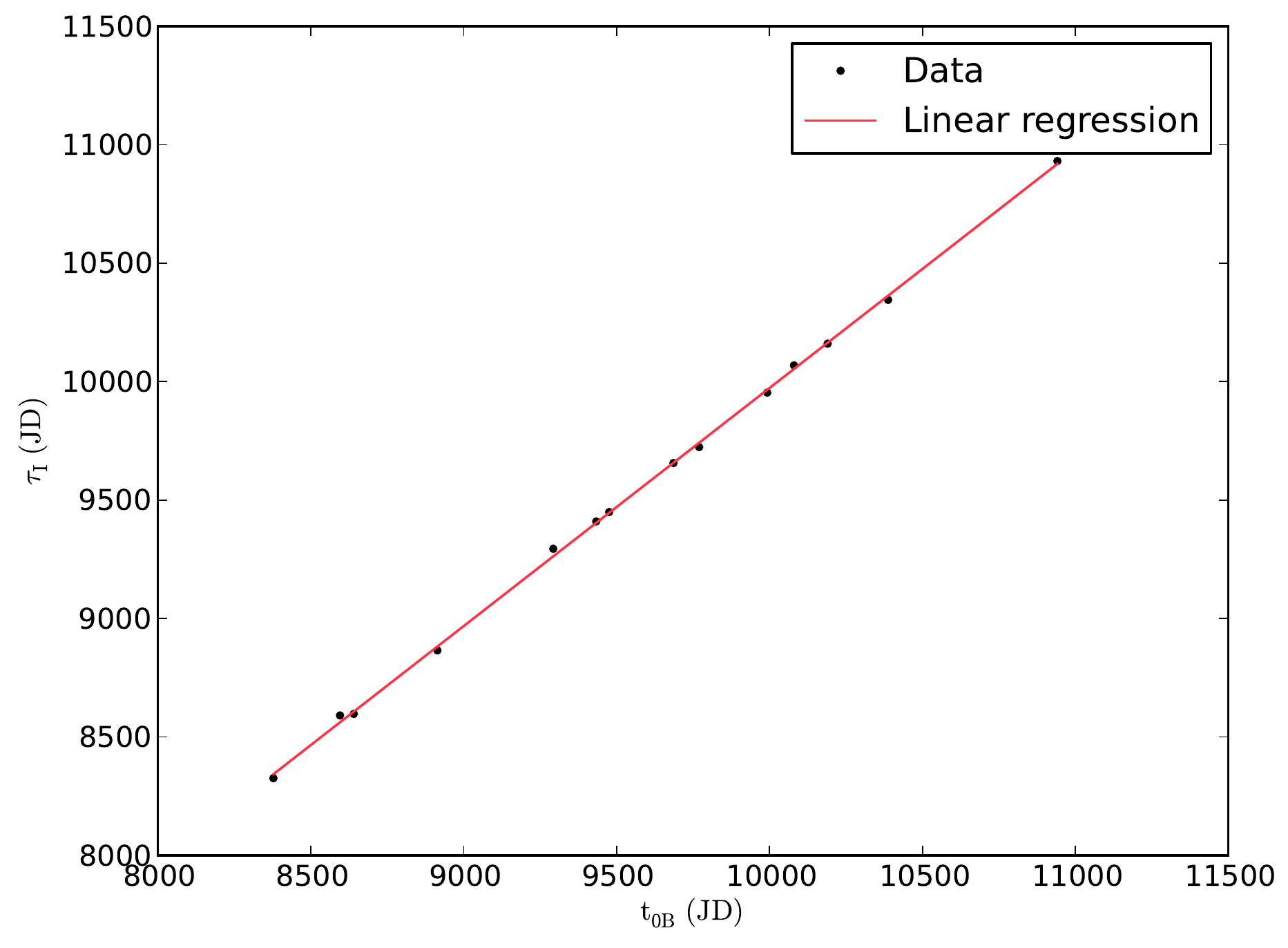} (c)
 \caption[On the x-axis: $t_0$ in B band.  On the y-axis $t_0$ in I band (panel a) , $t_1$ in I band (panel b), $\tau$ in I band (panel c).]{On the x-axis: $t_0$ in B band.  On the y-axis $t_0$ in I band (panel a) , $t_1$ in I band (panel b), $\tau$ in I band (panel c). The equation of the best fits are:\\
$t_{0I}=1.000*t_{0B}-1.024$. The r-square of the fit is: 0.99985.\\
$t_{1I}=0.999*t_{0B}+43.54$.  The r-square of the fit is: 0.9986.\\
$\tau_{I}= 1.005*t_{0B}-75.22$. The r-square of the fit is: 0.99948.
}
\label{marianna:ti}
\end{figure}
\clearpage
\begin{figure}
\centering
  \includegraphics[width=10cm]{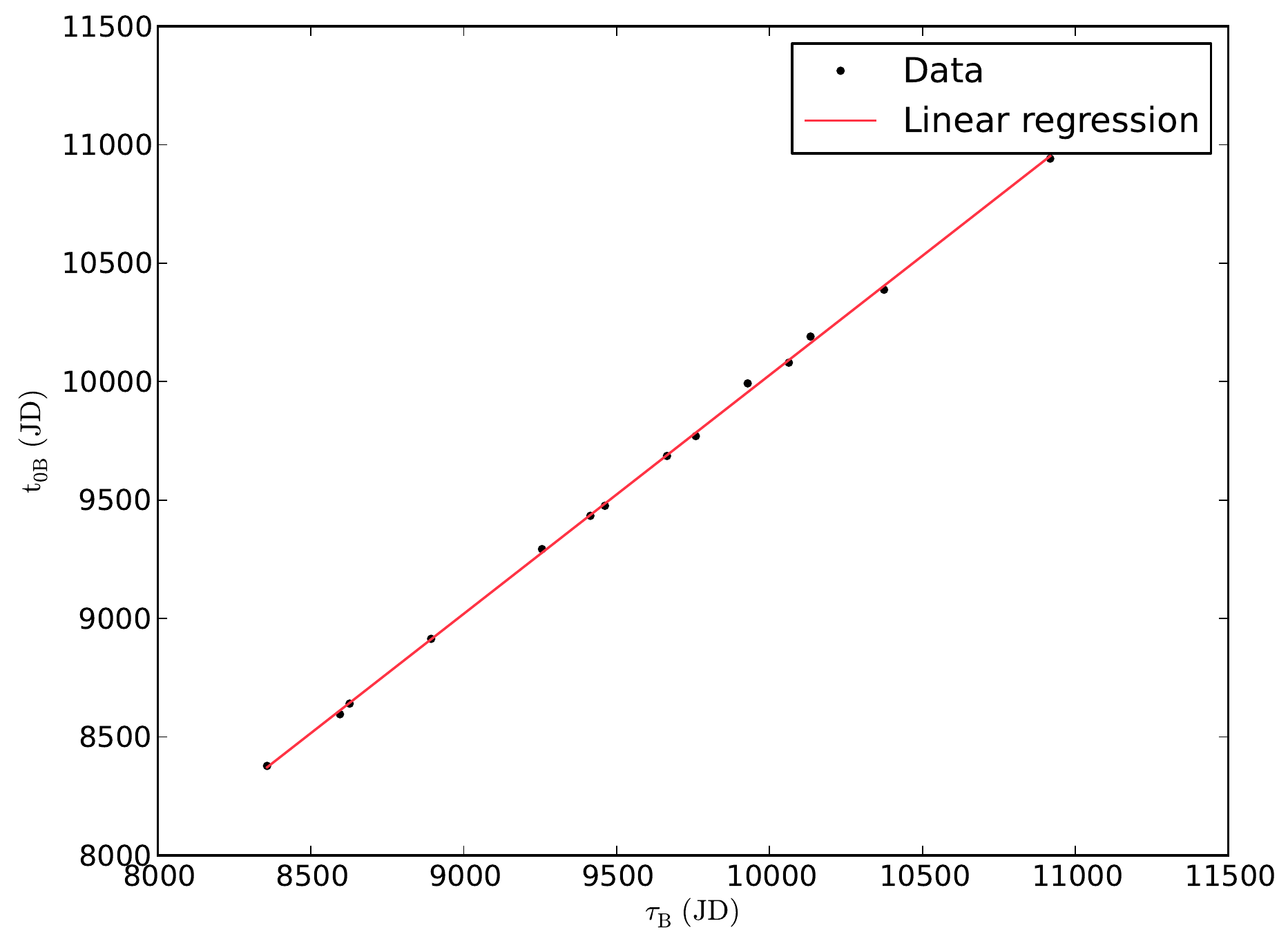}
  \caption[On the x-axis: $\tau$ in B band, on the y-axis: $t_0$ in I band.]{On the x-axis: $\tau$ in B band, on the y-axis: $t_0$ in I band. \\
The equation of the best fit line is: $\tau_{B}= 1.007*t_{0B}-49.06$. The r-square of the fit is: 0.99951.}
  \label{marianna:tauBt0B}
\end{figure}

\noindent These relations seem to represent temporal shifts, and in fact all the fits have slopes almost equal to one. \\
Handling the data, we also found that there seems to be a relationship between $f_0$ in B and in the other bands and, in first approximation, we can suppose that it is linear. Fig.~\ref{marianna:f0} shows $f_0$ in V, R and I band against $f_0$ in B band.
In the caption of each figure is reported the result of the linear regression. \\
\noindent In practice, in order to to simulate a type Ia Supernova we used the Contardo model with $\gamma,\, g_0,\, \sigma_0,\, g_1,\, \sigma_1,\, \theta$  extracted randomly from the ranges in Tab. \ref{marianna:contardo_tab}, fixating  $\tau_{B}$ in order to derive other temporal parameters using the relations shown above. At the user is given the possibility to choose how many days the Supernova is before or after the maximum. Furthermore the user can set the value of $f_{0B}$, while $f_{0V},\, f_{0R}$ and $f_{0I}$ result from the previous relations. Once modeled, the Supernova is associated to a galaxy having an integrated magnitude comparable with the maximum luminosity of the SN in the B band.\\

\begin{figure}
\centering
 \includegraphics[width=10cm]{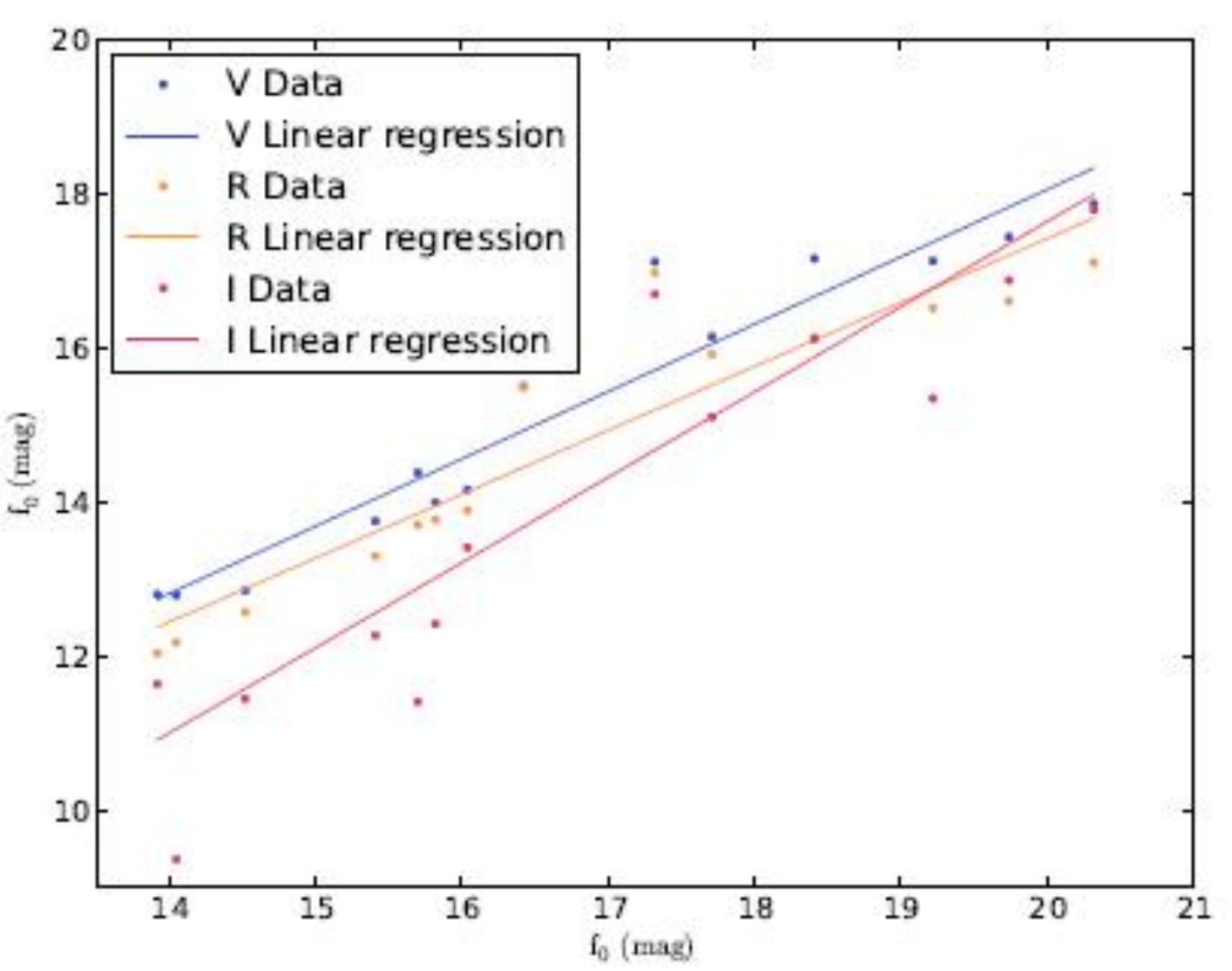}
  \captionof{figure}[On the x-axis: $f_0$ in B band. On the y-axis $f_0$ in the other bands.]{On the x-axis: $f_0$ in B band. On the y-axis $f_0$ in the other bands. Green points show $f_0$ in V band. The equation of the best fit line is: $f_{0V}= 0.872*f_{0B}+0.595$. The r-square of the fit is: 0.922.\\
  Orange points show $f_0$ in R band. The equation of the best fit line is: $f_{0R}= 0.828*f_{0B}+ 0.849$. The r-square of the fit is: 0.872.\\
  Red points show $f_0$ in I band. The equation of the best fit line is: $f_{0I}= 1.10*f_{0B}- 4.46$. The r-square of the fit is: 0.755.}
  \label{marianna:f0}
 \end{figure}

\subsection{Catalog extraction}
\label{marianna:2.6}

At the bottom of the simulation flowchart there is the extraction, from each image of a catalog of sources, containing as much information as possible on their properties. As we said before this task is achieved by SExtractor. The reasons behind this choice are explained in details in section \ref{chap:comparison}.\\

\subsection{Simulation example}
 \label{marianna:2.7}

\noindent As one of the first simulations, we produced 50 images, as observed by the VST\footnote{\url{http://www.eso.org/public/teles-instr/surveytelescopes/vst/surveys.html}} (VLT Survey Telescope) telescope and the OmegaCAM camera\footnote{\url{http://www.astro-wise.org/~omegacam/index.shtml}}. The Field of View (FoV) of OmegaCAM@VST is 1 square degree with a pixel scale of 0.213 arcsec/pixel. Therefore the size of the images was set to 16kx16k.  The aberration coefficients, the tracking errors, the positions of the spiders were set properly according to the VST technical specifications. The observations are spaced within 90 days, with an uneven sampling rate and with the FWHM of the seeing varying between 0.6 and 1.0 arcsec, according to the ESO statistics at Cerro ParAnal. The magnitude range was set to 14-26 mag and the exposure time of each image to 1500 s. \\
The total number of simulated objects is around 10000. This include non variable stars and galaxy, a sample of classical Cepheid, a sample of type Ia Supernovae with their host galaxy and a sample of randomly variable objects. \\
Fig.~\ref{marianna:SNall} and Fig.~\ref{marianna:Call} show a section of the B-band image produced at t=0d. The stamps below each image show the evolution of the variable object in the green box ( a type Ia Supernova in Fig.~\ref{marianna:SNall} and a Classical Cepheid in Fig.~\ref{marianna:Call} ).\\
\noindent The B-band light curves of the objects selected in Fig.~\ref{marianna:SNall} and Fig.\ref{marianna:Call} are shown in Fig.~\ref{marianna:SN_lc} and Fig.~\ref{marianna:C_lc}. The magnitude of the objects are the Kron magnitudes obtained by running SExtractor
on each image.

\begin{sidewaysfigure}
\centering
\subfloat[]{
\label{marianna:SN}
\includegraphics[scale=0.45]{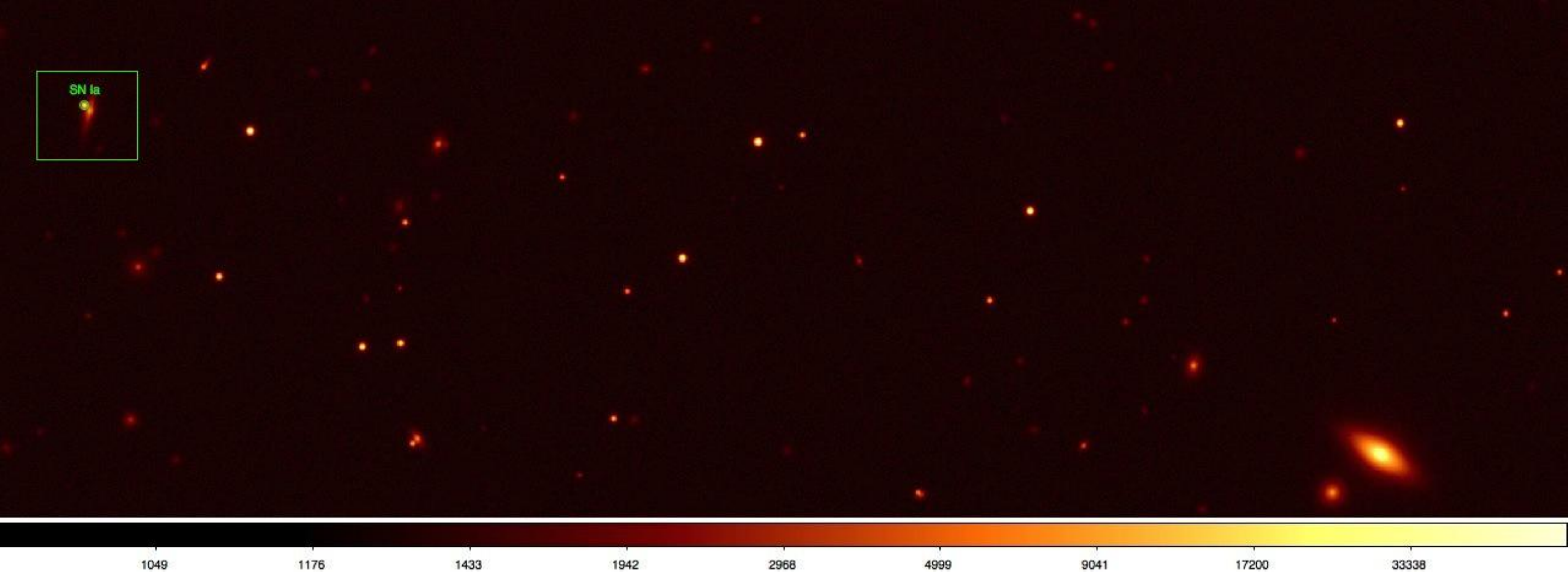}}\\
\subfloat[]{
\label{marianna:SN0}
\includegraphics[width=3 cm, height=3 cm]{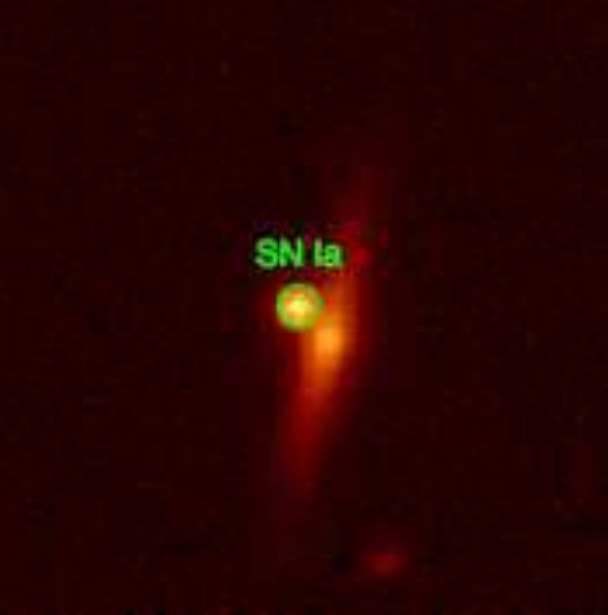}}
\subfloat[]{
\label{marianna:SN175}
\includegraphics[width=3 cm, height=3 cm]{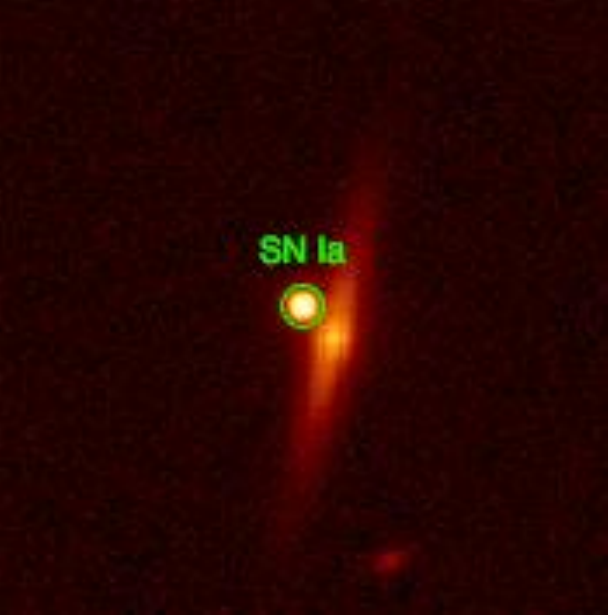}
}
\subfloat[]{
\label{marianna:SN436}
\includegraphics[width=3 cm, height=3 cm]{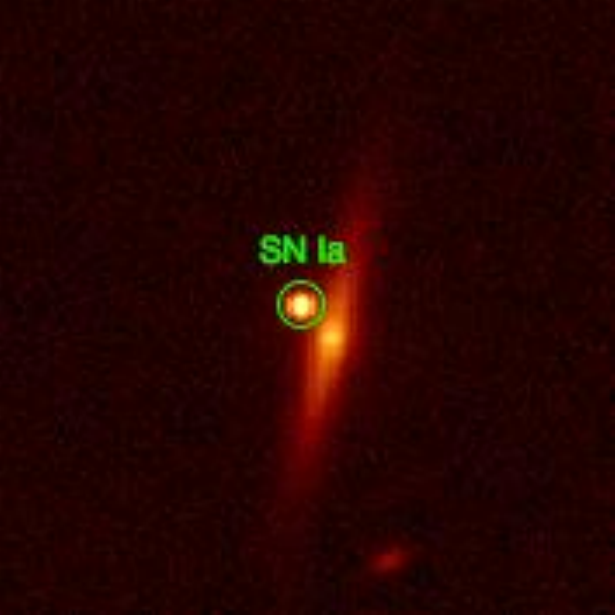} }
\subfloat[]{
\label{marianna:SN819}
\includegraphics[width=3 cm, height=3 cm]{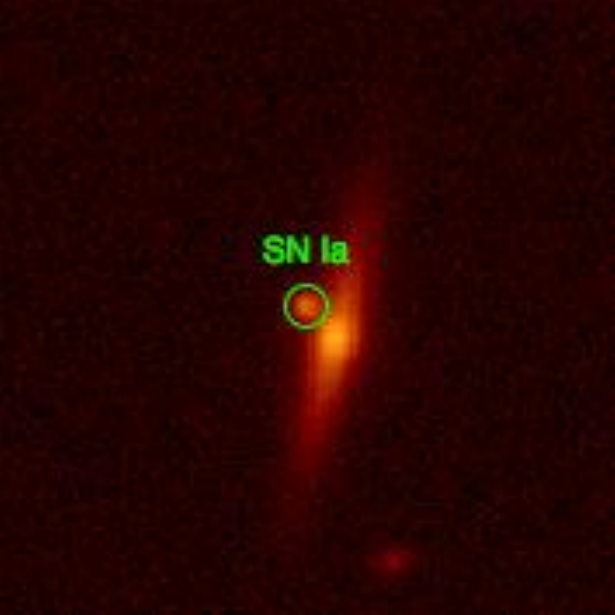} }
\subfloat[]{
\label{marianna:SN1417}
\includegraphics[width=3 cm, height=3 cm]{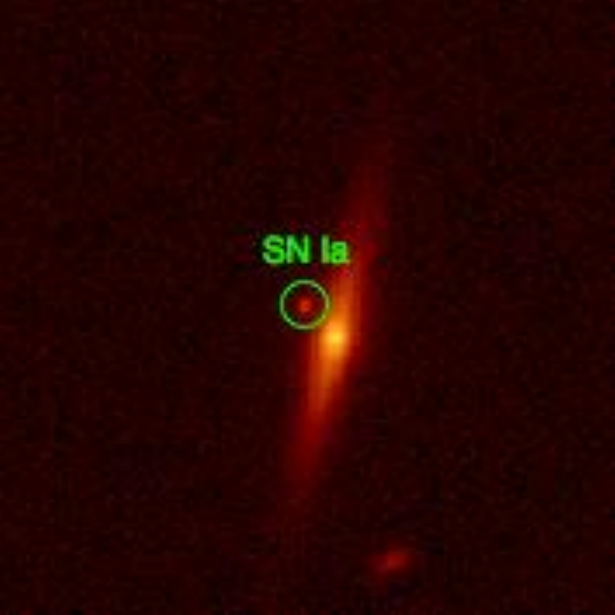}}
\subfloat[]{
\label{marianna:SN2137}
\includegraphics[width=3 cm, height=3 cm]{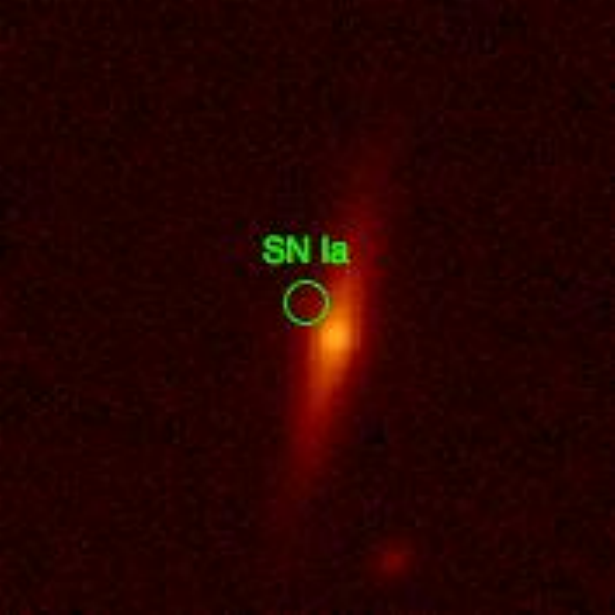}}
\caption[Stamp of the image.]{Stamp of the image: the green box in~\ref{marianna:SNall} is a type Ia Supernova at -9.34 days from its maximum light within its host galaxy. Figs. \ref{marianna:SN0} - \ref{marianna:SN2137} show a close up image of the Supernova at the beginning of the observation (t=0 days), t=7 days, t=18 days, t=34 days, t=59 days, and t=89 days respectively.
}
\label{marianna:SNall}
\end{sidewaysfigure}

\begin{sidewaysfigure}
\centering
\subfloat[]{
\label{marianna:C}
\includegraphics[scale=0.45]{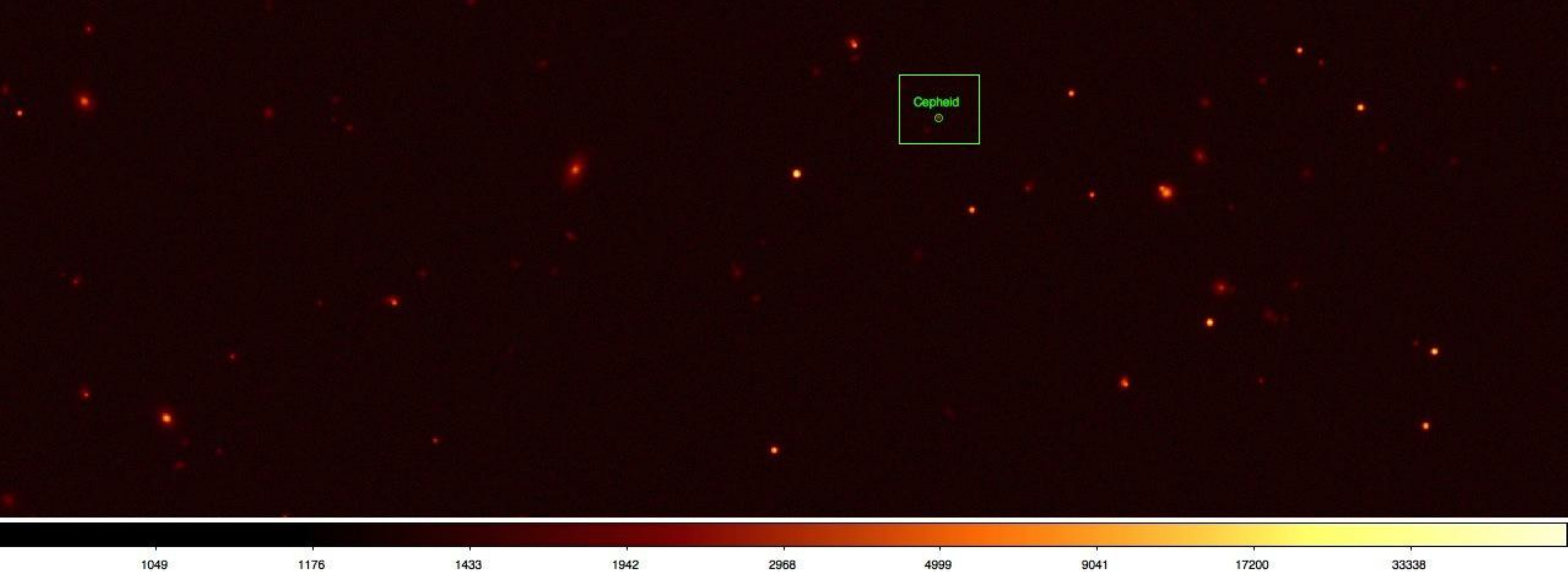}}\\
\subfloat[]{
\label{marianna:C0}
\includegraphics[width=3 cm, height=3 cm]{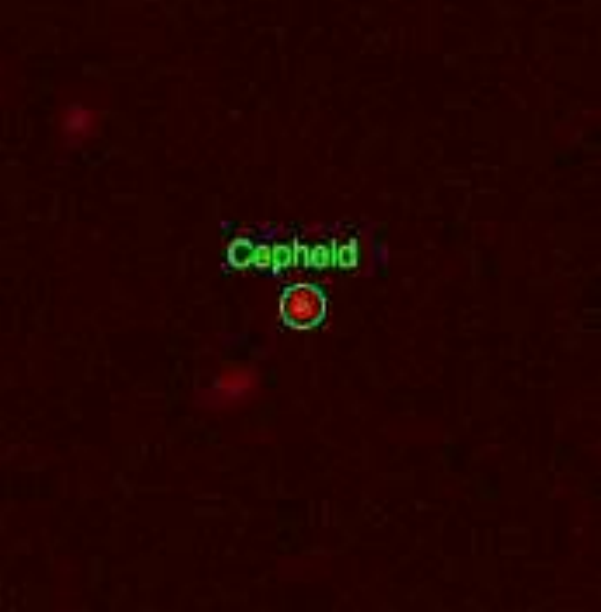}
}
\subfloat[]{
\label{marianna:C175}
\includegraphics[width=3 cm, height=3 cm]{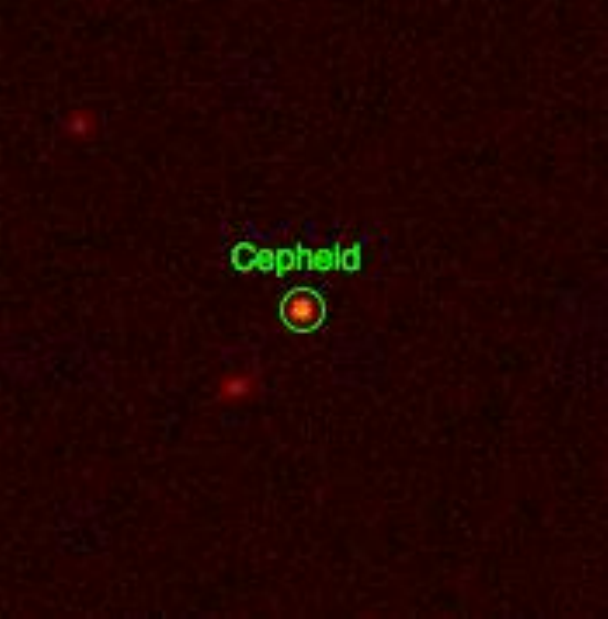}
}
\subfloat[]{
\label{marianna:C436}
\includegraphics[width=3 cm, height=3 cm]{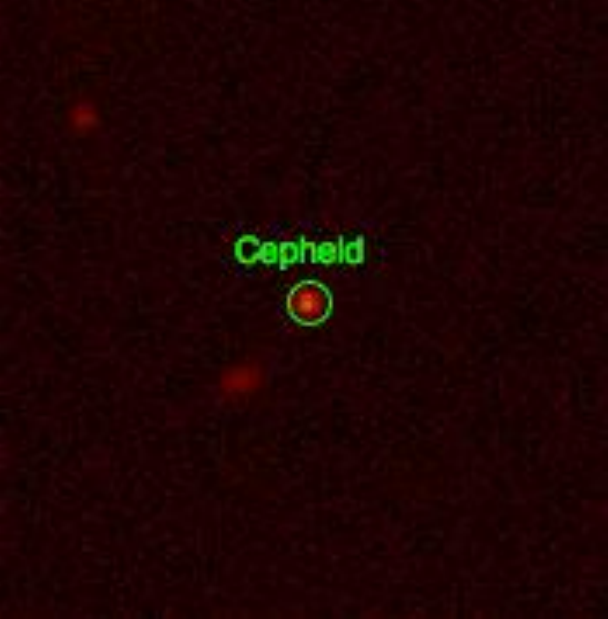} }
\subfloat[]{
\label{marianna:C819}
\includegraphics[width=3 cm, height=3 cm]{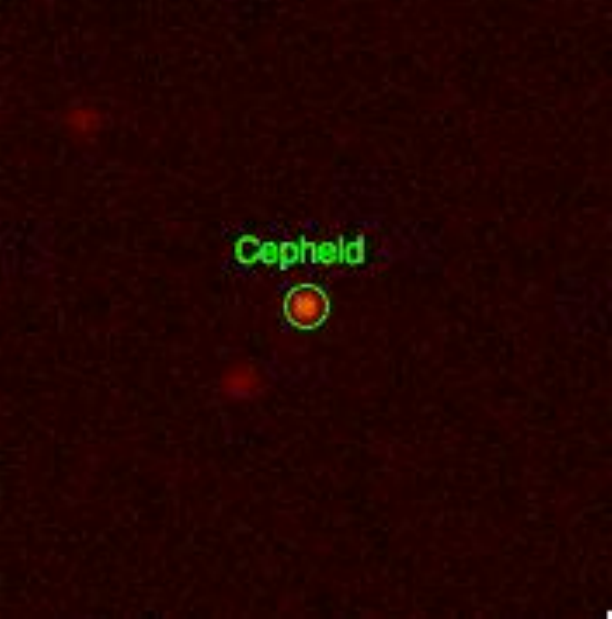} }
\subfloat[]{
\label{marianna:C1417}
\includegraphics[width=3 cm, height=3 cm]{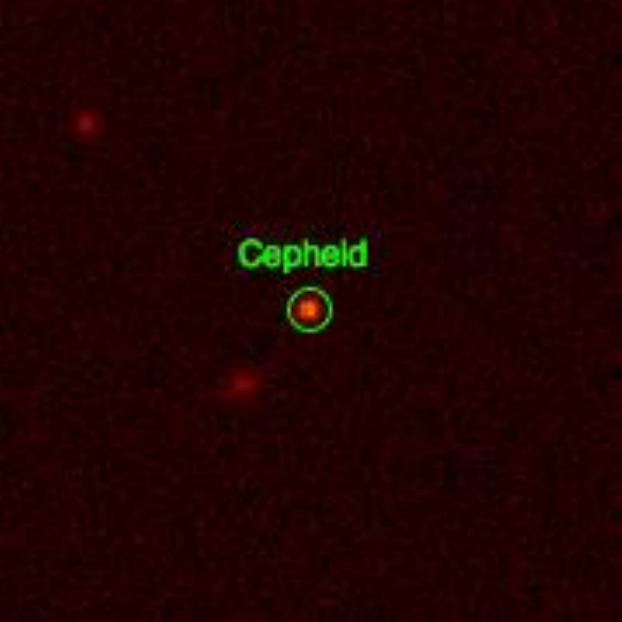}}
\subfloat[]{
\label{marianna:C2137}
\includegraphics[width=3 cm, height=3 cm]{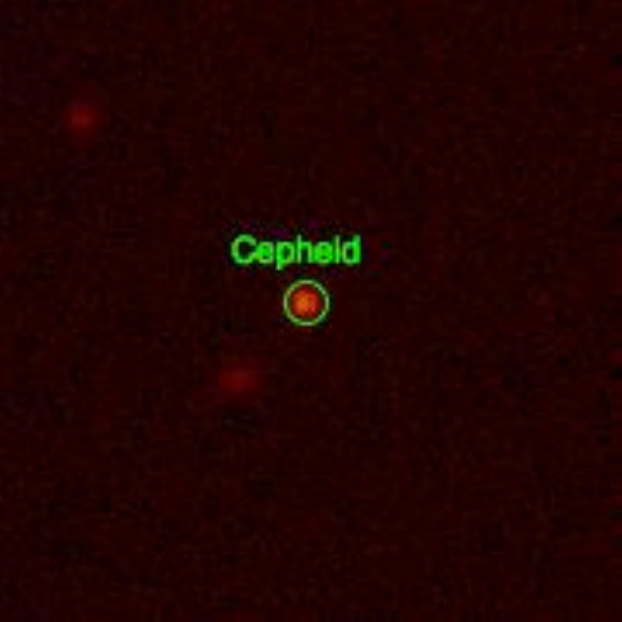}}
\caption[Stamp of the image.]{Stamp of the image: the green box in ~\ref{marianna:C} is a Classical Cepheid with a period of 25.39 days. Figs.~\ref{marianna:C0}-~\ref{marianna:C2137} show a close up image of the Cepheid at the beginning of the observation (t=0 days), t=7 days, t=22 days, t=34 days, t=45 days, and t=89 days respectively.
}
\label{marianna:Call}
\end{sidewaysfigure}

\begin{figure}
\centering
  \includegraphics[scale=0.6]{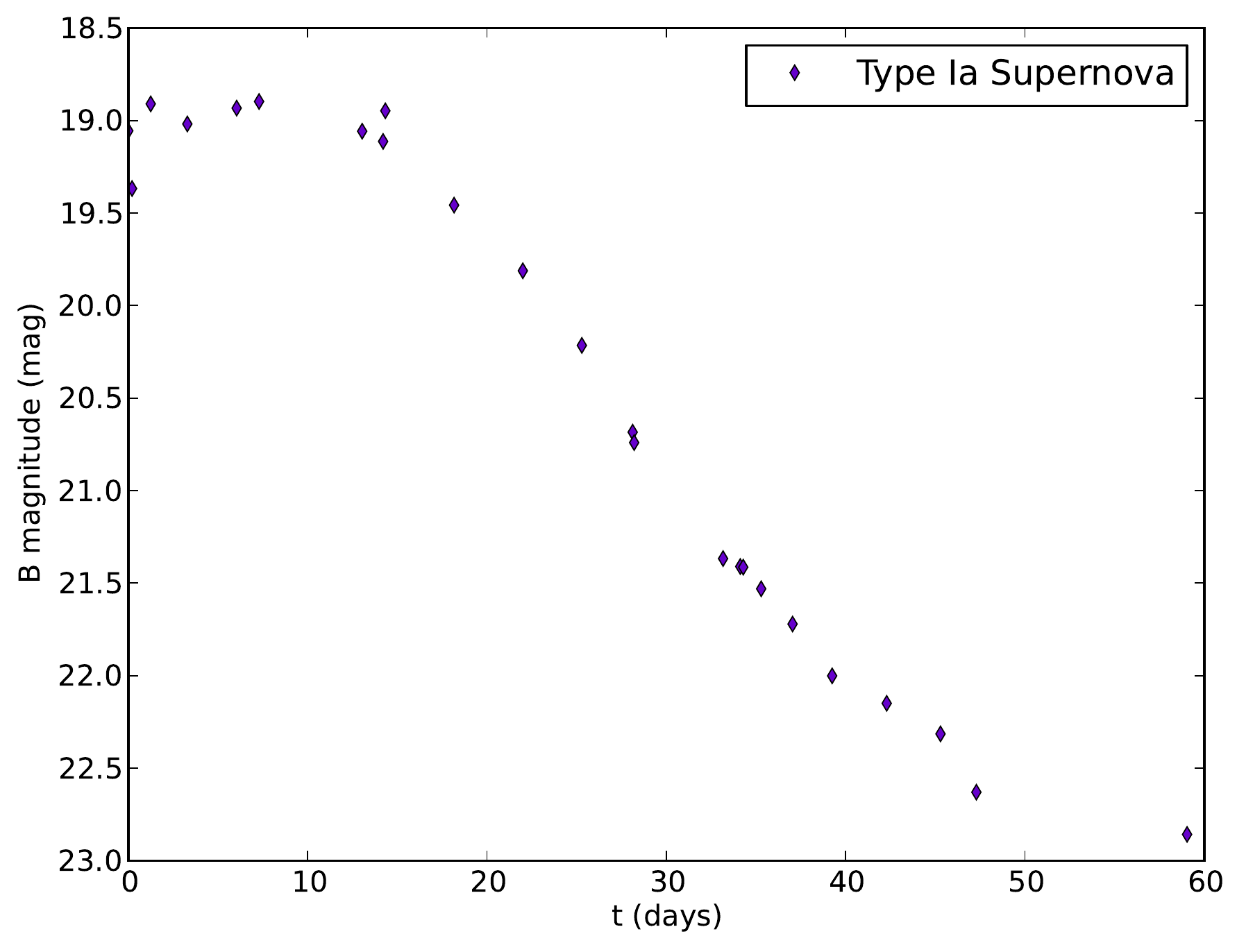}
  \caption{Light curve of the type Ia Supernova in Fig.~\ref{marianna:SNall}c}
  \label{marianna:SN_lc}
\end{figure}

\begin{figure}
\centering
  \includegraphics[scale=0.6]{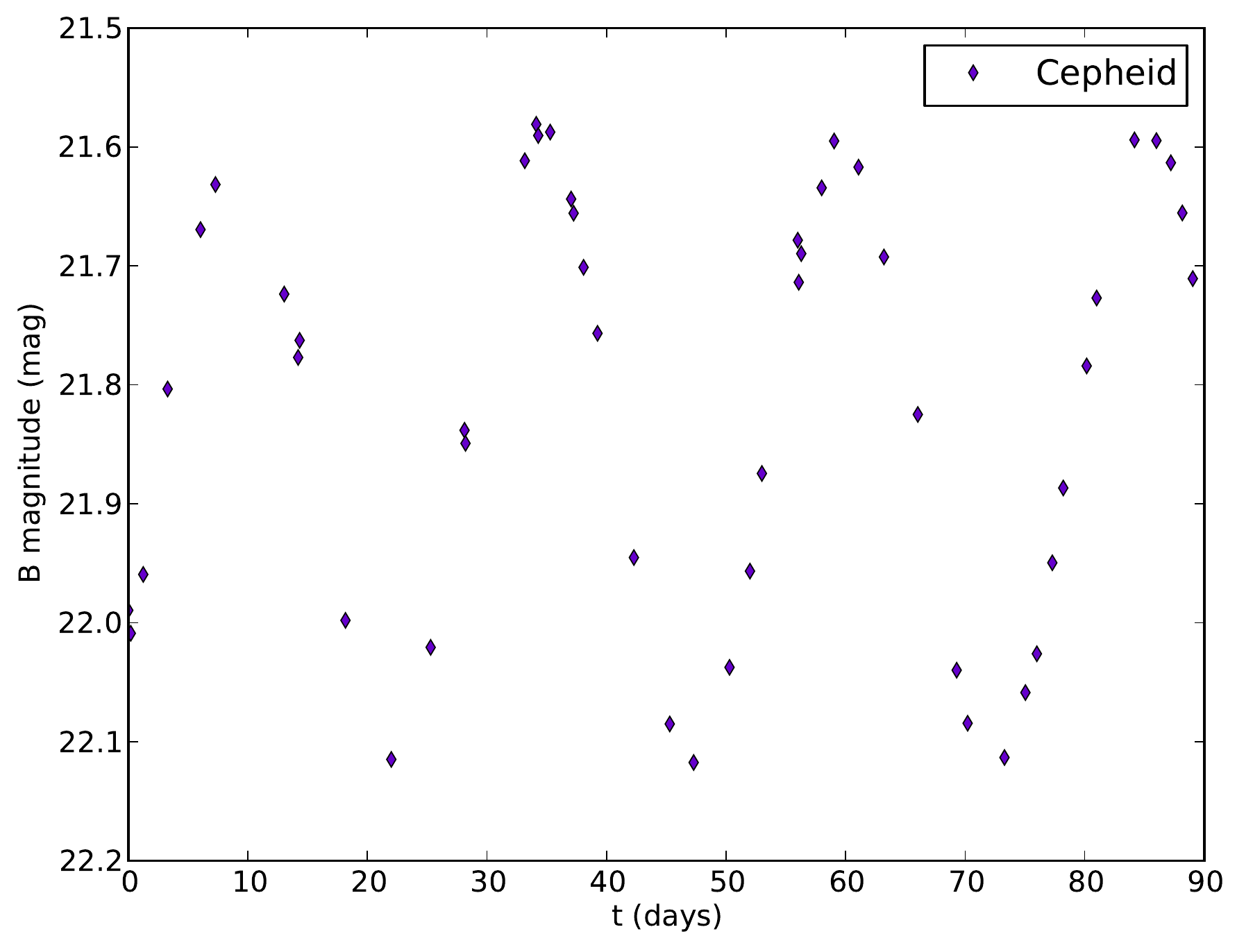}
    \caption{Light curve of the Classical Cepheid in Fig.~\ref{marianna:Call}c}
  \label{marianna:C_lc}
\end{figure}

Some application are discussed in section {marianna:4}.
    \section{Euclid Data Quality}\label{sec:dataquality}
I participated in the definition of proposal to implement a data quality environment for the Euclid Mission


Scope of this following description is to introduce the concepts of Data quality (DQ) and Data Quality Mining (DQM) and to outline how the EUCLID mission can benefit from them.

Given the present status of definition of the project it is impossible to enter into the details of the possible technical solutions but whereas possible I shall provide general examples which can help the reader to understand how DQM could fit into the proposed Euclid SGS (Science Ground Segment) operations.

In the Euclid Science Management Plan is declared that: \textit{``The Euclid Science Ground Segment will provide the resources necessary to analyse the Euclid Data and derive science data products. The Euclid Science Ground Segment is in charge of the production of the science ready calibrated images and source catalogues, and all relevant quality control and meta-data that are necessary for the scientific exploitation of the Euclid mission. This includes the data from the Euclid VIS and NISP channels, as well as all complementary External data from ground based wide field surveys..[...]...A complete Euclid data release is estimated to have a size of $2400 TB$ of Euclid data $+ 100 TB$ of Euclid catalogues $+ 8200 TB$ of non-Euclid data, which sums up to $\approx 11 PB$. If we now add a $20 \%$ margin, that reaches $13 PB$."}

Always in the Euclid Science Management Plan, the Science Ground Segment (SGS) processing is subdivided in five levels:
\begin{enumerate}
\item	Level S: pre-mission data (catalogues, satellite and mission modelling data, etc.) used before and during the mission for calibration and modelling purposes mainly. The data for this processing level are prepared before the mission (and refined/updated during the in-flight commissioning and initial calibration phase) and are used as appropriate, before and during the mission.
\item	Level E: external data (images, catalogues, all relevant calibration and meta-data, observational data in science-usable format) derived from other missions and/or external survey projects, reformatted to be handled homogeneously with Euclid data. This data is required to allow the EC to provide its final data products at the expected level of accuracy. The data at this level is delivered by the EC.
\item	Level 1: is composed of three separate processing levels, namely Level 1a, Level 1b and Level 1c.
\begin{enumerate}[a.]
\item	Level 1a refers to telemetry checking and handling, including real-time assessment (RTA) on housekeeping;
\item	Level 1b comprises quick-look analysis (QLA) on science telemetry, production of daily reports, trend analysis on instruments performance and production of weekly reports. The data for this processing level come from the satellite via MOC and are used to perform quality control.
\item	Level 1c refers to the high-quality removal of instruments signatures which provides data that will used to process Level 2 data.
\end{enumerate}
\item	Level 2: instrumental data processing, including the calibration of the data as well as the removal of instrumental features in the data. The data processing at this level is under the responsibility of the SDCs in charge of the instruments monitoring.
\item	Level 3: data processing pipelines for the production of science-ready data. The Level 3 data are also produced by SDCs
\end{enumerate}
It goes without saying that the most valuable asset of Euclid are the data and, due to the huge data volume, the quality control becomes a crucial aspect of all five items listed above and over the entire lifetime of the experiment: not only scientific data available at all various intermediate stages of the acquiring and processing workflows and pipelines, as it is foreseen during normal operations, but also telemetry, diagnostic, control, monitoring, calibration information coming in the ground segment from the instrument.

We refer hereinafter as EDW (Euclid Data Warehouse) to the total amount of Euclid SGS data to be treated in order to ensure its quality.

But what has to be intended with term ``quality"? Well, strictly connected with data quality there would be general questions like: Does the inventory data accurate represent what exists in the warehouse?
When dealing with data from any data intensive project, such as Euclid, anomalies and/or useless redundancies (duplications) may arise that often, if introduced into the warehouse framework, could
mislead the interpretation or obstruct the normal data flow affecting the steady flow bandwidth. The consequences being poor scientific performance, increased risk and even project failure.
To deal with these critical issues, the concept of DQ needs to be included in the project data warehouse organization as a crucial item.

Generally speaking, DQ is the process of rejecting or, better, transforming data that is incomplete, inaccurate, non-conforming or redundant into data that makes the project scientific solutions more successful. As the project builds its global data management strategy, DQ brings re-use and portability, making it more accessible by scientific communities, partners and data governance teams.
In other words, DQ entails more than helping instrument data governance and scientists to get correct data into their information systems; it also means identifying and dealing with corrupted, wrong or duplicate data.

In particular, clean data is a key element when integrating information across different systems, because misinformation can quickly proliferate. With interconnected information systems, as foreseen in Euclid, poor quality data spread the same way viruses are spread by travellers: erroneous information can spread quickly to other applications. The cost of compromised data is incalculable, including lost information, wrong instrument calibration, wasted scientific knowledge, loss of reputation or goodwill, and missed science target opportunities.

According to a study\footnote{\url{http://www.information-management.com/issues/20060801/1060128-1.html}} published by The Data Warehousing Institute (TDWI) entitled Taking Data Quality to the Enterprise through Data Governance, some issues are primarily technical in nature, such as the extra time required for reconciling data ($85\%$) or delays in deploying new systems ($52\%$). Other problems originally coming from the business community but easily exportable to the scientific one, are the customer dissatisfaction ($69\%$), compliance problems ($39\%$) and revenue loss ($35\%$). Poor-quality data can also cause problems with costs ($67\%$) and credibility ($77\%$).

Data quality affects all data-related projects and refers to the state of completeness, validity, consistency, timeliness and accuracy that makes data appropriate for a specific use. This means that in any kind of project related to data –being them images, tables, text, etc. - one has to ensure the best possible quality by checking the right syntax of metadata and columns, detecting missing values, optimizing relationships, and correcting any other inconsistencies.

\subsection{Data Quality General Requirements and Targets}

In order to ensure the long term reliability of an intensive data warehouse it is needed to  pay attention to input transactions in the archives and to avoid unqualified data. In order to achieve this goal, there should be evaluated the quality of every transaction that inters some data to the archive, as well as poor quality rollbacks transactions.

For the purposes of the present project we adopt the standard vision of DQ as a comprehensive data quality management solution, including data profiling, correcting, matching and interfacing concepts:
\begin{itemize}
\item	Data Profiling identifies the problems. It should provide snap-shots of a project data quality and measures its incremental evolution over time;
\item	Data Cleansing corrects incomplete or inconsistent data by crosschecking against other databases and reference data (in Euclid this could be the case of used external data [AD1]). It also enriches data by providing value-add information that actually improves the quality and usefulness of existing data;
\item	Data Matching consisting of specific matching algorithms that could be provided in the form of fully featured yet easy-to-use components, that help to find and repair duplicates and near duplicates in data volumes. Unlike other DQ solutions, matching is configured in a discrete set of components allowing for identification of duplicate records across multiple sources using confidence weights and probability, to help group or direct records into a pipeline process, for example.
\item	Data Enrichment  filling in the missing pieces in data volumes in order to reach scientific analytical goals. The variety of this information is limitless, i.e. it can include incorporating a variety of information to help plan delivery routes, or collecting data to targets or other categories.
\item	Data Interfacing includes user tools able to offer advanced reporting and analysis features.
\item	In what follows we shall also specify how the above topics fit into the Euclid scenario.
\end{itemize}

\subsection{Technical Aspects of Data Quality}
More technically speaking, in a warehouse based on intensive and diversified engineering and scientific data governance, like Euclid, some errors and inconsistencies could arise (in raw images and data, metadata, processed data) in terms of:
\begin{itemize}
\item	Blank, NaN and Null values;
\item	Hidden relations, missed or bad (in terms of its ontology) foreign keys;
\item	Wrong or out-of-range values and constants;
\item	Unknown values (symbols);
\item	Misleading instrument signature parameters and/or metadata;
\item	Duplicate fields, parameters, matrix rows and columns, values;
\end{itemize}

Such quality bugs are very general but represent a good starting point to clarify the concept of general DQ in a data warehouse. Of course, they can be extended and well defined over both different data types (images, telemetry data, metadata, etc.) project definition time (especially according to the Euclidisation process). Samples of requirements/targets of a first stage of DQ framework are the following:

\begin{itemize}
\item	\textbf{Quickly Browsing Data Structures}: the goal is to understand metadata for columns (size, comment) and table keys, including mechanisms to detect and identify missing or erroneously documented data;
\item	Getting an Overview of Database Content: the goal is to get statistics regarding data volumes. This feature familiarizes the user with a database, regardless of complexity, with no need of a guide. It also lets him detect some modeling errors, such the absence of keys.
\item	\textbf{Do Columns Contain Null or Blank Values?}: Detect and clean null and empty values from the data. It can help to check the presence of null and empty values in the data volume.
\item	\textbf{About Redundant Values in a Column}: to verify that we have one and only one record for each value in the loaded column.
\item	\textbf{Is a Max/Min Value for a Column Expected?}: Quickly detect data anomalies. It should also offer functionalities to Set data thresholds for controlling max/min values.
\item	\textbf{What is the best data sample?}: given a user customized function (mathematical, statistical, empirical, etc.), it is possible to locate the winner datum in the volume.
\item	\textbf{Using Statistics}: options to get a quick, global view of the frequency of the data and identify potential errors or misleading information inside. It can also get frequency statistics which let user to identify most present values in data or to check advanced statistics indicators for a complex analysis.
\item	\textbf{Analyzing Intervals in Numeric Data}: to get a frequency table with a specified aggregation of interval values. It can also create frequency tables for numerical columns specifying the number of desired bins.
\item	\textbf{Targeting Intervals}: to understand customer distribution to better target advertising. It enables to show the distribution of any numerical variable, by also detecting min and max, and also the most frequent data intervals.
\item	\textbf{Identifying and Correcting Bad Data}: of course there should be present any kind of prerequisite, such as any understanding of pattern usage and its standardization (Euclidisation, VO like or whatever).
\item	\textbf{Getting a Column Pattern}: the goal is to identify a data column structure and/or pattern. After having defined a pattern frequency indicator, column values can be grouped in intervals.
\item	\textbf{Detecting Keys in Tables}: to discover all candidate keys in a table. This could be used as information to optimize data distribution.
\item	\textbf{Are There Duplicate Records in Data?}: the goal is to detect and clean data from duplicate records, possibly starting from prerequisites based upon a default/custom correlation analysis. This is in principle a crucial function due to huge amount of data to be collected inside Euclid data centres.
\item	\textbf{Column Comparison Analysis}: the goal is to check that no redundancy exists in columns and optimize space reserved for data. For example to ensure that values used in a table are present in related dimension tables, or to discover new foreign keys.
\item	\textbf{Recursive Relationships}: this is a functionality connected with the exploration inside data about possible relations that recursively can be occur at different levels of data processing system.
\item	\textbf{Creating a Report}: the DQ system should be able to summarize any created and performed analysis, by specifying also the desired level of details. Of course its prerequisites are the creation of a DQ database (TBD).
\item	\textbf{Monitoring}: Monitoring helps supervise and control data quality over time. It also allows monitoring the duration of an analysis and to identify those that are taking longer than normal. The monitoring can include:
\begin{itemize}
\item	Reporting on Data Quality: monitor how the data quality indicators evolve over time. It has as prerequisites the scheduled reports foreseen to be executed regularly on the data warehouse. With a report on changes in data quality, it is easier to detect any change in data without performing the entire analyses one after the other.
\item	Tracking Data: Track development and revise old data. This feature lets users control data quality in time.
\item	Alert: DQ should generate alerts when certain values are outside of a range defined by custom rules or thresholds. This feature can be automated and closed by generating a report.
\end{itemize}
\item	\textbf{Mining Rules for Data Quality procedures}: the goal is to provide data exploration, analysis and mining rules in order to control and/or to test the accuracy of the data and to help decision making. These rules can be in principle based on deterministic, statistical or machine learning methods, able to find correlations.
\end{itemize}
The last element of the above list was intentionally left as the last item, because it represents the more complex but at the same time interesting and powerful functionality for the data quality control in a PB project such as Euclid. It directly connects flexibility of potential customized data quality analysis with powerful methods usually employed in Data Mining (DM), or equivalently in Knowledge Discovery in Databases (KDD).
Their inclusion into the data quality domain permits to overcome or at least to optimize the usual DQ analysis rules, based on:
\begin{itemize}
\item	Image (science $\&$ calibration) analysis
\begin{itemize}
\item	Detection of abnormal patterns
\item	Anomalous Instrumental signature
\end{itemize}
\item	Column analysis
\begin{itemize}
\item	Simple analysis
\item	Simple statistics evolution
\item	Frequency statistics
\item	Pattern matching
\end{itemize}
\item	Overview analysis
\begin{itemize}
\item	Connection analysis
\item	Catalogue analysis
\item	Scheme analysis
\end{itemize}
\item	Redundancy analysis
\item	Duplication analysis
\end{itemize}

Here we would like to mention technical data requirements, as arisen from the SRS document of the mission, that we intend to take strongly under consideration throughout the DQ analysis and design.

Concerning the general data processing and archiving system specifications, there are important issues to be verified and validated by DQ systems, directly coming from the requirements:
\begin{itemize}
\item	All information stored into official catalogue database should be accessible through standard SQL (or equivalent) queries;
\item	SQL queries could be able to search and retrieve all Euclid stacked images and/or individual exposures;
\item	The pipelined data should always be composed by catalogs of objects (selected by photometry, weak lensing, spectroscopy in wide or deep surveys) within a standard queryable database. For instance, for the cluster catalogues, the database should always include consistent information, like sky coordinates (RA/Dec), best estimate of redshift (either spectroscopic or photometric), richness, S/N ratio, number of objects, velocity dispersion, magnitude in different bands and morphology;
\item	The data analysis should include source detection and photometry extraction for time-series data, including DQ flags used for validation;
\item	The results of any data analysis should be provided in a queryable database providing master variable source catalogues as well as all relevant metadata.
\item	The database should include verification and validation tools to check the presence and consistency of periodogram spectral densities for periodic variables, flags indicating object variability and other particular features;
\end{itemize}
\subsection{Scientific Aspects of Data Quality}
From the scientific data point of view, the mission is subject to different layers of goals and requirements, organized into a hierarchical structure:
\begin{itemize}
\item	Payload and Mission Requirements;
\item	Survey Requirements;
\item	External Data Requirements;
\item	Data Processing Requirements;
\end{itemize}
Between this layered structure, what is directly involving DQ, it is basically the last layer (data processing), in which all specifications, in terms of data integrity and science expectations that must be checked and validated by the DQ systems, are included.
The first two layers, in fact, are mainly focused on, respectively, observing instrument design and procurements (payload and mission) and science involved group requirements (survey).
A special mention is needed for the third layer (external data), where several external packages and tools should aim at improve the general qualitative requirements for primary science (such as visible photometry and spectroscopy follow-ups and samples). At this layer, the data quality requirements should be intrinsically ensured by the providers of such systems.
So far, for what it concerns the last relevant layer for DQ constraint and architecture design, the following is a preliminary list of mandatory specifications about the data processing requirements. There are also other further details that are voluntarily omitted here, waiting for a confirmation coming from a next system trade-off and scientific analysis.
DQ specifications (both quality requirements and consistent information) for data processing assessment includes:

\begin{itemize}
\item	VIS standard image products should be validated in terms of their quality, availability and completeness (presence of artifacts, individual and stacked exposures, flat fields, bias and sky subtractions, for each field, star masks, weight and noise maps, astrometric exposure solutions). Standard products should also involve catalogues of objects to the limiting magnitude, with star/galaxy classification with a contamination less than $1\%$;
\item	The whole PSF (including telescope and optical plane instrument) should be provided matching the exposure time, position in the FOV and wavelength (especially in terms of ellipticity and size stability for all galaxy type objects) and verified in terms of intrinsic fidelity and possible undersampling in some points (eventually using also available simulations);
\item	Any additional colour information available from visible bands should be correctly present (i.e. number of filters);
\item	For galaxy type object, an estimation of the redshift (spectroscopic or photometric) should be present, together with the information about tools/technique used and their quality performance (external data and NISP/VIS information). In addition, there always should be present the related calculation error, the matching with the sky coordinates (RA/Dec) and the magnitudes in all bands;
\item	Simulations should be end-to-end and include all known relevant astrophysical, instrumental and data processing effects. In particular they should be used to calibrate any effect that cannot be directly verified with the shear catalogue or the PSF;
\item	Concerning the shear catalogue for extended (wide field) un-masked objects, its algorithm should be checked by proper simulations and verified in terms of variance between simulated and measured power spectrum. Moreover the shear should be uncorrelated with a variety of indicator quantities that are not be affected by the cosmological shear signal (such as stellar ellypticities and other optical or instrumental quantities);
\item	A variety of power spectra are required (such as tomographic cosmic shear, galaxy-shear, two-point galaxy-galaxy) as functions of different parameters (such as redshift). In all cases the power spectrum should be verified and validated in terms of minimization of overlap in the mutual redshift distribution, of consistency with mask effects, survey window, cosmological statistical effects, astrophysical and non-astrophysical sources, intrinsic alignment etc.);
\item	For the power spectra, their covariance matrix should be present and validated in terms of their source (simulations/observations), potential cosmological dependence and shape of the masks;
\item	Regarding the spectroscopic data, all patterns should be verified and validated in terms of cleaning of instrumental signatures, according to standard processing procedures;
\item	Sub-spectra extraction should be consistent with the preservation of the available information, shape and single-wavelength calibration, galaxy flux preservation, calibration and measures of energy flux and equivalent width;
\item	For any of the angular position on the sky, data must include regular information details about observations, like sky coordinates, exposure time, dithers number, wavelength covered by spectroscopy, dispersion direction etc. Moreover, in the case of survey at a chosen redshift, processing data should provide the expected density for a sample of objects matching the spatial sampling;
\item	For all observed and catalogued objects there should be verified, if available, the consistency with the same objects in external data (simulations and/or ground-based observations);
\item	There should be verified the possibility in the SGS (Science Ground Segment) to perform ``non-standard" spectral analysis to increase the efficiency of redshift measurements for special classes of objects;
\end{itemize}

\subsection{Data Quality enhancement with Data Mining}
In the traditional DQ methodology, briefly touched in the previous section, the statistical approach is usually employed for measuring the quality of data, in many common cases with good results (for example financial, enterprise, medical warehouses). But dealing with much more complex cases, especially in data warehouses designated as repositories of high precision scientific experiment results (like in the Euclid case), the traditional approach appears to be quite insufficient.

The major limit of statistical methods, when applied directly on data quality control, is the fact that traditionally DQ modifies the data themselves while for scientific data this needs to be avoided. Data Mining, on the contrary, is a methodology for measuring the quality of data, preserving their intrinsic nature. DM algorithms extract some knowledge, that can be used to measure the quality of data, with particular reference to the quality of input transactions and then, eventually flag the data of poor quality.
A typical procedure to measure DQ of data transactions should be based on three steps:
\begin{enumerate}
\item	Extract all association rules, which depend on input transactions;
\item	Select compatible association rules;
\item	Add confidence factor of compatible rules as criteria of data quality of transaction.
\end{enumerate}

There are two important challenging issues. First, the extraction of all association rules needs a lot of time and next, in most cases there is no exact mathematical formula for measuring data quality.

So far, a more effective DM approach to DQ should be alternative to find exact deterministic or statistical formulas. Therefore, for us, the answer is in employing methodologies derived from Machine Learning (ML) paradigms, such as (a) active on-line learning, which addresses the issue of optimizing the combination and trade-oﬀ of losses incurred during data acquisition; (b) associative reinforcement learning, connected with the predictive quality of the ﬁnal hypothesis.
Moreover, one of the guidelines of our proposed approach is to conjugate these machine learning paradigms with features coming from biological adaptive systems.

The key principles are to process information systems using a connectionist approach to computation, in order to emulate the powerful correlation ability at the base of the cognitive learning engine of human brain, together with the optimization process at the base of biological evolution (Darwin's law)

Our experience in such methodology has produced the DAME Program, which includes several projects, mostly connected with Astrophysics, although spread into various of its scientific branches and sub-domains. Data Mining is usually conceived as an application (deterministic/stochastic algorithm) to extract unknown information from noisy data. This is basically true but in some way it is too much reductive with respect to the wide range covered by mining concept domains. More precisely, in DAME, data mining is intended as techniques of exploration on data, based on the combination between parameter space filtering, machine learning, soft computing techniques associated to a functional domain. In the data mining scenario, the machine learning model choice should always be accompanied by the functionality domain. To be more precise, some machine learning models can be used in a same functionality domain, because it represents the functional context in which it is performed the exploration of data.

Examples of such domains are: Dimensional reduction, classification, regression, clustering, segmentation, statistical data analysis, forecasting, data Mining model filtering.

From the technological point of view, the employment of state of the art web 2.0 technology, allows the end user (i.e. the data centres) to be in the best condition to interact with the DQ process by making use of a simple web browser.

The approach outlined above has three immediate advantages:
\begin{itemize}
\item	DQ controls can be approached by remote, through homogeneous and interoperable interfaces, federated whereas possible under VO standards.
\item	Different DQ models and algorithms available by remote web applications can be tested by the end user (SDC) in a standard and intuitive way. In other words, the SDC does not need to be particularly skilled with DM methodologies to create and configure workflows on data;
\item	DM applications could be executed by remote cloud/grid frameworks, embedding all the complex management issues of the distributed computing infrastructure.
\end{itemize}

However, another indirect positive issue for our approach arises by considering that, in a massive data centric project like EDW, one of the unavoidable constraints is to minimize data flow traffic and down/up-load operations from remote sites. DQ tools should therefore be installed and maintained at the SDC.

It is worth to stress that this approach fits perfectly within the recently emerging area of interest named DQM (Data Quality Mining).  DQ uses information attributes as a tool for assessing quality of data products. The goal of DQM is to employ data mining methods in order to detect, quantify, explain and correct DQ deficiencies in very large databases. For this reason there is a reciprocal advantage between the two application fields (DQ is crucial for many applications of KDD, which on the other side can improve DQ results).

\subsection{DQ, DQM and scalability issues}
DQ and DQM are computing intensive and their computational cost grows quickly with the size and complexity of the data to be analyzed. In what follows we shortly describe how Graphic Processing Units (GPUs) could offer an effective and inexpensive way to deal with such problem even in the framework of a mission as complex as Euclid is.

In Euclid SGS warehouse the scientific quality control is particularly referred with data and metadata related to both images and spectra. Most of the KDD techniques based on machine learning that could directly be employed on such kind of data can be considered naturally parallel in terms of their analysis computation.

As an example let us consider the Multi Objective Genetic Algorithm (MOGA), based on the linkage between feature selections and association rules, that is one of the key concepts in the DQ methodology. The main motivation for using GA in the discovery of high-level prediction rules is that they perform a global search and cope better with attribute interaction, often used in DM problems. Therefore a parallel GA further promotes the performance of computing, particularly required on massive data warehouse quality control.
A traditional parallel computing environment is very difficult and expensive to set up. This can be circumvented by recurring to graphics hardware, inexpensive, more powerful, and perfectly comparable with other more complex HPC mainframes in terms of computing power (many frameworks based on GPU architecture are already included in the top 500 HPC worldwide supercomputer ranking).

The DAME Program (see section \ref{chap:dame}) has already started the investigation on the design and implementation of a hierarchical parallel genetic algorithm, implemented on new technology based on multi-core Graphics Processing Unit (GPU) provided by NVIDIA Company, by using the Compute Unified Device Architecture (CUDA) parallel programming SDK. CUDA is a platform for massively parallel high-performance computing on the company's powerful GPUs. At its cores are three key abstractions: (a) a hierarchy of thread groups, (b) shared memories, and (c) barrier synchronization, that are simply exposed to the programmer as a minimal set of language extensions. These abstractions provide fine-grained data parallelism and thread parallelism, nested within coarse grained data parallelism and task parallelism.

According to the above concepts, we intend to contribute to DQ control of SGS data warehouse by introducing and manage a specific Work Package (WP) devoted to quality control of incoming data to the SGS. This could be achieved by providing a DQM framework able to support data quality management within each system team connected with data handling (both in terms of storage, retrieval and maintaining). In order to do so, we can provide both software and specific hardware infrastructures, located into official EC SDCs, in order to minimize data flow requirements, available to end users via web interfaces.

The DQ WG should interact closely with many other WGs since it needs input from many different teams. A detailed WBS for DQ and DQM management will be provided in the next future and will be progressively adjusted to match the evolution of the project organization structure.

\chapter{Classification Problems}\label{part:science}
    \hfill\begin{tabular}{@{}p{.7\linewidth}@{}}
\textit{``Oh Be A Fine Girl Kiss Me (Right Now Sweetheart)"}\\
Henry Norris Russell.\\ \phantom{aaa}
\end{tabular}

In this chapter I present present three different encountered problems that I tackled during my PhD which are gathered together because they were approached with the same data mining functionality: Classification (see sec. \ref{sec:functionality}).
Classification problems are among the most relevant and commonly problems in astronomy.

A typical, and crucial, example is the extraction
of scientifically useful information from wide field astronomical images (both photographic plates and CCD
frames) and the recognition of the objects against a noisy
background (a problem which in image processing is also known as image segmentation) and their classification in unresolved (starlike) and resolved (galaxy).

This chapter is structured as follow:\\
In section \ref{chap:comparison} I show how the use of neural networks for object classification plus the novel {\tt{SPREAD\_MODEL}} parameter push down to the limiting magnitude the possibility to perform a reliable star/galaxy separation. In section \ref{chap:GC} I present an application o to the identification of candidate Globular Clusters in deep, wide-field, single band HST images on a small dataset while in section \ref{chap:agn} I show an application to the detection of Active Galactic Nuclei on a medium dataset.

    \section{Comparison of source extraction software}\label{chap:comparison}

\captionsetup[figure]{labelformat=simple, labelsep=period}
\captionsetup[table]{labelformat=simple, labelsep=period}

\blfootnote{this section is largely extracted from: \tiny
\begin{itemize}
\item Annunziatella, M.; Mercurio, A.; Brescia, M.; \textbf{Cavuoti, S.}; Longo, G.; 2013, Inside catalogs: a comparison of source extraction software, \textbf{PASP, 125, 68}
\end{itemize}}

In order to obtain a better understanding of the data, part of my work was to compare the catalog extraction performances obtained using the new combination of SExtractor with PSFEx, against the more traditional and diffuse application of DAOPHOT with ALLSTAR; therefore, the present section may provide a guide for the selection of the most suitable catalog extraction software. Both software packages were tested on two kinds of simulated images having, respectively, a uniform spatial distribution of sources and an overdensity in the center. In both cases, SExtractor is able to generate a deeper catalog than DAOPHOT. Moreover, the use of neural networks for object classification plus the novel {\tt{SPREAD\_MODEL}} parameter push down to the limiting magnitude the possibility of star/galaxy separation. DAOPHOT and ALLSTAR provide an optimal solution for point-source photometry in stellar fields and very accurate and reliable PSF photometry, with robust star-galaxy separation. However, they are not useful for galaxy characterization, and do not generate catalogs that are very complete for faint sources. On the other hand, SExtractor, along with the new capability to derive PSF photometry, turns to be competitive and returns accurate photometry also for galaxies. We can assess that the new version of SExtractor, used in conjunction with PSFEx, represents a very powerful software package for source extraction with performances comparable to those of DAOPHOT. Finally, by comparing the results obtained in the case of a uniform and of an overdense spatial distribution of stars, we notice, for both software packages, a decline for the latter case in the quality of the results produced in terms of magnitudes and centroids.\\


When extracting a catalog of objects from an astronomical image the main aspects to take into account are: to detect as many sources as possible; to minimize the contribution of spurious objects; to correctly separate sources in their classes (e.g. star/galaxy classification); to produce accurate measurements of photometric quantities; and, finally, to obtain accurate estimates of the positions of the centroids of the sources.\\
Among the main source extraction software packages used by the astronomical community, there are SExtractor \citep{bertin1996} and DAOPHOT II \citep{stetson1987}, and the latter is often used in combination with its companion tool ALLSTAR \citep{stetson1994}.
SExtractor is commonly used in extragalactic astronomy and has been designed to extract a list of measured properties from images, for both stars and galaxies. DAOPHOT and ALLSTAR were designed to perform mainly stellar photometry.
So far, only DAOPHOT II, used together with ALLSTAR, has been able to produce more accurate photometry for stellar objects using the Point Spread Function (PSF) fitting technique, while the PSF fitting photometry in SExtractor has,
instead, become possible only in the recent years.
The first attempt was in the late 90s, when the PSFEx (PSF Extractor) software package became available within the TERAPIX ``consortium".
This tool extracts precise models of the PSF from images processed by SExtractor. Only after 2010, through the public release of PSFEx (\citealp{bertin2011})\footnote{Available at \url{http://www.astromatic.net/software/psfex}.}, and with the recent evolution of computing power, has PSF fitting photometry become fully available in SExtractor.\\
The scope of this section is to compare the results obtained using the combination of SExtractor with PSFEx, and DAOPHOT with ALLSTAR, by focusing, in particular, on the completeness and reliability of the extracted catalogs, on the accuracy of photometry, and on the determination of centroids, both with aperture and PSF-fitting photometry.
A previous comparison among extraction software tools was performed by \cite{becker2007}. They,
in pursuit of LSST science requirements, performed a comparison among DAOPHOT, two versions of SExtractor (respectively 2.3.2 and 2.4.4), and DoPhot \citep{mateo1989}. However, differently from the present work where simulations are used, they evaluated as ``true" values the measurements obtained with the SDSS imaging pipeline \textit{photo} \citep{lupton2001}. Furthermore, we wish to stress that their results were biased by the fact that in 2007 the PSF fitting feature had not yet been implemented in SExtractor.\\
The present work performs, for the first time, a comparison between DAOPHOT and SExtractor PSF photometry, providing a guide for the selection of the most suitable catalog extraction software packages.\\
The simulations used for the comparison are described in Sect.~\ref{comparison:2}. In Sect.~\ref{comparison:3}, the main input parameters of the software packages are overviewed and the adopted values are specified in Sect.~\ref{comparison:4}. In Sect.~\ref{comparison:5}, the obtained results are shown. In order to better evaluate the performances of both software packages on crowded fields, we describe: in Sect.~\ref{comparison:6}, a test performed on an image showing an overdensity in the center. Finally, the results are summarized in Sect.~\ref{comparison:7}, together with our conclusions.\\

\subsection{Image Simulations}
\label{comparison:2}
Image simulations are suitable in testing performances of various software packages. Simulations, in fact, allow us to know exactly the percentage and the type of input sources and their photometric properties. \\
In the present work, simulations have been obtained by using two software packages: Stuff\footnote{Available at \url{http://www.astromatic.net/software/stuff}.} (described in section \ref{marianna:2.3}) and SkyMaker\footnote{Available at \url{http://www.astromatic.net/software/skymaker}.} (described in section \ref{marianna:2.4}) , developed by E. Bert\`{\i}n.
With these tools, it is possible to reproduce the real outcome of a CCD observation, once the characteristics of the telescope and the camera are known.
In practice, Stuff can be used to produce a realistic simulated galaxy catalog, while SkyMaker uses this catalog to produce an optical image under realistic observing conditions, allowing also to add a stellar field. \\
The galaxy catalog simulated by Stuff can be produced so to be consistent with the assumed cosmological model and with the statistical distributions of stars and galaxies in terms of redshift, luminosity and color. In a binned redshift space, Stuff produces galaxies of different Hubble types E, S0, Sab, Sbc, Scd and Sdm/Irr. The number of galaxies in each bin is determined from a Poisson distribution, by assuming a non-evolving Schechter luminosity function \citep{schechter1976}. Cosmological parameters, luminosity function, as well as instrumental parameters are specified by the user in the input configuration file.
In particular, the size of the image, the pixel scale, the detector gain, and the observed passbands must be provided as input information. Filters can be selected among the many available in the wavelength range [0.29, 87.74831] $\mu$m. Finally the magnitude range of simulated galaxies has to be fixed.\\
The image is then created by SkyMaker, by rendering sources of the input catalog in the frame at the specified pixel coordinates, to which is added a uniform sky background, Poissonian noise, and Gaussian read-out noise. Stellar sources are modeled using a PSF internally generated, while the various types of galaxies are modeled as differently weighted sums of a bulge profile and an exponential disk. The PSF profile takes into account both atmospheric seeing and optical aberrations.
There are many parameters to be set in the SkyMaker configuration file. The most important are related to: the pupil features, e.g., the size of the mirrors and the aberration coefficients; the detector characteristics, e.g., gain, saturation level and image size; and the observing conditions, e.g, full width at half maximum of the seeing (FWHM) and exposure time.\\
For our work, we simulated images as they would be observed by the VST\footnote{\url{http://www.eso.org/public/teles-instr/surveytelescopes/vst/surveys.html}} (VLT Survey Telescope) and the OmegaCAM camera\footnote{\url{http://www.astro-wise.org/~omegacam/index.shtml}}. The field of view (FoV) of OmegaCAM@VST is 1 square degree, with a pixel scale of 0.213 arcsec/pixel. In order to reduce the computational time, we limited our simulations to a FoV of 1/4 of VST.
The aberration coefficients, the tracking errors, and the positions of the spiders were set properly, according to the VST technical specifications.
The allowed magnitude range was set to 14-26 mag and the exposure time was fixed at 1500 s.
Finally, according to the ESO statistics at Cerro Paranal, the FWHM of the seeing was set to 0.7 arcsec.
In this section, we report results relative to B-band simulated images.
We obtain a catalog of N=4120 sources down to the input magnitude limit. This corresponds to an image with an average surface number density of $\sim$ 20 sources/arcmin$^\mathrm{2}$. \\
In order to perform a more complete comparison of both software packages, we also simulated a crowded image with an overdensity in the center (see Sect.~\ref{comparison:6}). To this purpose, the parameters of Stuff and SkyMaker are equal to those used for images with uniform source distribution, except for the image size and the exposure time. For these second group of tests, we reduced the exposure time to 300 s, with an input magnitude limit of 14-25, while the image size is set to 2048x2048 pixels ($\mathrm{\sim}$ 7x7 arcmin$^\mathrm{2}$).
We wish to stress that, in the simulations, we did not include any artifact such as bad pixels, ghosts or bad columns or other effects that, while being crucial in other applications, are of no interest here, since we are comparing the performances of two different software packages. \\

\subsection{Source extraction software}
\label{comparison:3}

In the following subsection, we briefly discuss how DAOPHOT works in combination with ALLSTAR, and how SExtractor works in combination with PSFEx, and give a brief overview of the main parameters that are needed to be set in order to optimally run the selected software packages.

\subsubsection{DAOPHOT II}
\label{comparison:3.1}

DAOPHOT II is composed of a set of routines mainly designed to perform stellar photometry and astrometry in crowded fields. \\
It requires several input parameters, listed in the file \textit{daophot.opt}, including detector gain, readout noise ({\tt{GAIN, READ NOISE}}), saturation level ({\tt{HIGH GOOD DATUM}}), approximate size of unresolved stellar sources in the frame ({\tt{FITTING RADIUS}}), PSF radius ({\tt{PSF RADIUS}}), PSF model ({\tt{ANALYTIC MODEL PSF}}), and a parameter designed to allow the user to visually inspect the output of each routine ({\tt{WATCH PROGRESS}}).\\
The first step performed by DAOPHOT II is to estimate the sky background and to find the sources above a fixed input threshold through the FIND routine.\\
This threshold represents the level (in ADU), above the sky background, required for a source to be detected. In order to ignore smooth large-scale variations in the background level of the frame, the image is convolved with a lowered truncated Gaussian function, whose FWHM is equal to the input value set by the {\tt{FWHM}} parameter. After the convolution, the program searches for the local maxima sky enhancement. \\
Once the sources are detected, DAOPHOT II performs aperture photometry via the PHOTO routine.
Aperture photometry usually requires the definition of at least two apertures. The first one is usually circular, centered on the source and with a radius of a few times its FWHM. The second one is, instead, ring-shaped; usually this is concentric to the first one and it has an inner radius equal to the radius of the first aperture. The ring-shaped aperture is used to estimate the sky contribution and it usually covers a number of pixels equal to or at least comparable with that of the aperture.
Then, the flux of the source is obtained by subtracting the sky flux from the aperture flux. The aperture size must be chosen thoroughly. In fact, if the radius of the inner aperture is too small, there will be a flux loss; while, if it is too large, too much sky is included and the measurements will become too noisy. The radii of the aperture and sky annulus for DAOPHOT can be specified in an input file: \textit{photo.opt}. The inner and outer radii of the sky annulus, which is centered on the position of each star, must also be specified.\\
Beside the source magnitude, the PHOTO routine produces the coordinates of the source centroids, corresponding to the barycenters of the intensity profile around the source. \\
Aperture photometry performs rather well under the hypothesis of bright and isolated stars. However, the stars in crowded fields are faint and tend to overlap. In these cases, the PSF fitting photometry can produce better results. This measurement requires a PSF model to be derived from the stars in the image of interest. The normalized PSF model is then fitted to each star in the image, in order to obtain the intensity and magnitude. \\
DAOPHOT II can build a PSF model from a sample of stars obtained with the PHOTO routine in an iterative procedure, intended to subtract neighboring stars that might contaminate the profile.
Among them, DAOPHOT will exclude stars within one radius from the edges of the image and the stars too close to saturated stars. The analytical formula of the PSF is chosen by the user among several available models: a Gaussian function, two implementations of a Moffat function, a Lorentz function, and two implementations of a Penny function (\citealp{penny1995}). The PSF routine produces a PSF model and a list of the PSF stars and their neighbors. The modeled PSF stars can be visually inspected by setting properly the aforementioned {\tt{WATCH PROGRESS}} parameter. \\
Although DAOPHOT is designed for stellar photometry, extended sources are likely always present in real images and, therefore, a reliable method is required to separate galaxies from stars.\\
The sharpness parameter (\texttt{SHARP}) can be used as star/galaxy classifier. \texttt{SHARP} describes how much broader is the actual profile of the object compared to the profile of the PSF.
The sharpness is, therefore, dependent on the model of the PSF, and can be easily interpreted by plotting it as a function of apparent magnitude. Objects with \texttt{SHARP} significantly greater than zero are probably galaxies.\\
Throughout the section, I indicate the stand-alone DAOPHOT~II version 1.3-2\footnote{Available at \url{http://starlink.jach.hawaii.edu/starlink}} simply as DAOPHOT.

\subsubsection{ALLSTAR}
\label{comparison:3.2}
After having derived PSF models with DAOPHOT, ALLSTAR simultaneously fits multiple overlapping point-spread functions to all the detected sources in the image. At each iteration, ALLSTAR subtracts all the stars from a working copy of the input image, according to the current best guesses about their positions and magnitudes. Then, it computes the increments to the positions and magnitudes by examining the subtraction residuals around each position. Finally, it checks each star to see whether it has converged or it has become insignificant. When a star has converged, its coordinates and magnitude are written in the output file, and the star is permanently subtracted from the working copy of the image; when a star has disappeared, it is simply discarded.\\
The input parameters for ALLSTAR, listed in \textit{allstar.opt}, are similar to those in \textit{daophot.opt} and \textit{photo.opt}.\\
Moreover, it is possible to optimize the determination of centroids, by applying a PSF correction and setting the option {\tt{REDETERMINE CENTROIDS}}.\\
An improvement of the star/galaxy classification is possible by using the aforementioned sharpness measure obtained by ALLSTAR (\texttt{SHARP}). This value may be used also in conjunction with another ALLSTAR output parameter, for instance $\mathrm{\chi}$, which is the observed pixel-to-pixel scatter from the model image profile, divided by the expected pixel-to-pixel scatter from the image profile.
Throughout the section, I indicate simply with ALLSTAR the PSF-fitting software package that comes together with DAOPHOT II version v. 1.3-2.

\subsubsection{SExtractor}
\label{comparison:3.3}

SExtractor is a software package mainly designed to produce photometric catalogs for a large number of sources, both point-like and extended. Sources are detected in four steps: \textit{i)} sky background modeling and subtracting, \textit{ii)} image filtering, \textit{iii)} thresholding and image segmentation, \textit{iv)} merging and/or splitting of detections. The final catalog is, indeed, extracted according to the input configuration file, in which parameters are set by the user.\\
The first step of the background estimation can be skipped if the user provides manually an input estimation of sky background.
For the automatic background estimation, the most critical input parameters to be set are {\tt{BACK\_SIZE}}, the size of each mesh of the grid used for the determination of the background map, and {\tt{BACK FILTERSIZE}}, the smoothing factor of the background map.\\
Once the sky background is subtracted, the image must be filtered. This implies convolving the signal with a mask, shaped according to the characteristics that the user wants to enhance in the image data.
In fact, there are different filters available in SExtractor. The more suitable are ``top-hat" functions, optimized to detect extended, low-surface brightness objects, Gaussian functions usually used for faint object detection, and ``Mexhat" filters, which work with a high value of detection threshold, suitable for bright detections in very crowded star fields.\\
The detection process is mostly controlled by the thresholding parameters ({\tt{DETECT\_ THRESHOLD}} and {\tt{ANALYSIS\_THRESHOLD}}). The choice of the threshold must be carefully considered. A too high threshold determines the loss of a high number of sources in the extracted catalog, while a too low value leads to the detection of spurious objects. Hence it is necessary to reach a compromise by setting these parameters according to the image characteristics, the background RMS, and also to the final scientific goal of the analysis.\\
Two or more very close objects can be detected as a unique connected region of pixels above threshold and, in order to correct for this effect, SExtractor adopts a deblending method based on a multi-thresholding process. Each extracted set of connected pixels is re-thresholded at N levels, linearly or exponentially spaced between the initial extraction threshold and the peak value. Also, here a compromise is needed to be found, since a too low value for the deblending parameter leads to a lack of separation between close sources, while a too high value leads to split extended faint sources in more components.
Alternatively, it is possible to extract the catalog with different deblending parameters and to merge detections for extended sources or close pairs.\\
Once sources have been detected and deblended, the software tool starts the measurement phase. SExtractor can produce measurements of position, geometry, and of several types of photometric parameters, including different types of magnitudes.
Among photometric quantities, there are: the aperture magnitude ({\tt{MAG\_ APER}}), having the same meaning as explained in Sect.~\ref{comparison:3.1}, the Kron magnitude ({\tt{MAG\_AUTO}}), which is the magnitude estimated through an adaptive aperture \citep{kron1980}, and the isophotal magnitude ({\tt{MAG\_ISO}}), computed by considering the threshold value as the lowest isophote.\\
Among the available position parameters, it is important to mention the barycenter coordinates ({\tt{X\_IMAGE}}, {\tt{Y\_IMAGE}}), computed as the first order moments of the intensity profile of the image, and windowed positional parameters ({\tt{XWIN\_IMAGE}}, {\tt{YWIN\_IMAGE}}), computed in the same way as the barycenter coordinates, except that the pixel values are integrated within a circular Gaussian window as opposed to the object's isophotal footprint.\\
To separate extended and point-like sources, it is possible to use the \textit{stellarity index} ({\tt{CLASS\_STAR}}), which results from a supervised neural network that is trained to perform a star/galaxy classification. {\tt{CLASS\_STAR}} can assume values between 0 and 1. In theory, SExtractor considers objects with {\tt{CLASS\_STAR}} equal to zero to be galaxies, and those with value 1 as a star. In practice, stars are classified by selecting a {\tt{CLASS\_STAR}} value above 0.9.
Two other parameters, often used to discriminate between star and galaxies, are the half-light radius ({\tt{FLUX\_RADIUS}}), and the peak surface brightness above background (\texttt{$\mathrm{\mu_{max}}$}). When plotted against the Kron magnitude, these two parameters identify a so-called stellar locus.\\
Throughout the section, we indicate simply with SExtractor the software version 2.14.7 (trunk.r284).

\subsubsection{PSFEx}
\label{comparison:3.4}

The last version of SExtractor can work in combination with PSFEx, a package able to build a model of the image PSF. The latter is expressed as a sum of N$\mathrm{\times}$N pixel components, each one weighted by the appropriate factor in the polynomial expansion (see \citealt{mohr2012}). Then, SExtractor takes the PSFEx models as input and uses them to carry out the PSF-corrected model fitting photometry for all sources in the image.\\
PSFEx requires, as input, a catalog produced by SExtractor to build a model of the image PSF, which can be read back in a second run by SExtractor itself. In order to allow PSFEx to work, the first catalog produced by SExtractor must contain at least a given number of parameters, as explained in the PSFEx manual\footnote{\url{https://www.astromatic.net/pubsvn/software/psfex/trunk/doc/psfex.pdf}}.	
In particular, the catalog must contain the parameter {\tt{VIGNET}}, a small stamp centered on each extracted source, used to model the PSF. The size of {\tt{VIGNET}} must be chosen accordingly to the size of the photometric apertures defined by {\tt{PHOT\_APERTURES}}. \\
PSFEx models the PSF as a linear combination of basis vectors. These may be the pixel basis, the Gauss-Laguerre or Karhunen-Lo\`eve bases derived from a set of actual point-source images, or any other user-provided basis. The size of the PSF and the number and type of the basis must be specified in the configuration file.\\
By using SExtractor combined with PSFEx, it is possible to obtain various estimates of the magnitude, in addition to those described in the previous subsection: the magnitude resulting from the PSF fitting ({\tt{MAG\_PSF}}), the point source total magnitude obtained from fitting ({\tt{MAG\_POINTSOURCE}}), the spheroidal component of the fitting ({\tt{MAG\_SPHEROID}}), the disk component of the fitting ({\tt{MAG\_DISK}}), and the sum of the spheroid and disk components ({\tt{MAG\_MO\-DEL}}).
Moreover, it is also possible to measure morphological parameters of the galaxies, such as spheroid effective radius, disk aspect ratio, and disk-scale length. \\
The model of the PSF may be employed to extract a more accurate star/galaxy classification using the new SExtractor classifier, {\tt{SPREAD\_\-MODEL}}, which is a normalized, simplified linear discriminant between the best fitting local PSF model and a more extended model made by the same PSF convolved with a circular exponential disk model with scalelength = FWHM/16, where FWHM is the full-width at half maximum of the PSF model \citep{desai2012}. \\
A more detailed description of PSFEx and the new SExtractor capabilities can be found in \citet{bertin2011} and \citet{armstrong2010}.\\
Throughout the section, we indicate simply with PSFEx the software version 3.9.1.

\subsection{Catalog extraction}
\label{comparison:4}

In this subsection we provide a general discussion on how to set the input parameters in order to extract the catalogs with the software tools presented above.\\

\subsubsection{DAOPHOT and ALLSTAR}
\label{comparison:4.1}

Besides instrumental parameters, such as gain, saturation level, and readout noise, which are set according to the values used for the simulations (Sect.~\ref{comparison:2}), in DAOPHOT the detection and photometric options must be configured by means of input setup files.
In the present analysis, the threshold value was chosen to detect as many possible sources, while avoiding as much as possible spurious detections. In fact, as the threshold decreases, the number of detected sources increases up to a certain value, for which the relation changes in steepness. Thus it is possible to choose a reasonable value for the threshold, by plotting the number of extracted sources for different threshold values and by choosing the threshold near the ``elbow" of the function. Moreover, in order to avoid spurious detections, due to Poissonian noise, the extracted catalog was visually inspected.  In fact, by using the input parameters as reported in Tab~\ref{comparison:tab1}, only the 5\% of the objects extracted in the catalog is spurious. By decreasing the threshold, although the number of extracted input sources increases, the percentage of spurious detections rapidly grows up to 30\% at 4$\sigma$ threshold, even reaching more then 88\% at 3$\sigma$. \\
The {\tt{FITTING RADIUS}} was set equal to the FWHM of the image (see Tab.~\ref{comparison:tab1}). \\
To obtain the best aperture radius, we have derived the growth curve for input stellar sources. Then, we fixed the aperture radius to 12.5 pixel (see \texttt{a1} in Tab.~\ref{comparison:tab1}), producing the better coverage of the sources input magnitude.
Thus, values of {\tt{INNER RADIUS}} and {\tt{OUTER RADIUS}} were chosen accordingly, smaller and greater than the aperture radius, respectively.\\
The PSF analytical model was chosen with the higher level of complexity, that is, the implementations of the Penny function with five free parameters (option 6). We chose to visually inspect the image of the PSF produced by DAOPHOT for all the PSF stars. ALLSTAR parameters were set accordingly to those established for DAOPHOT and, furthermore, we required the redetermination of the centroids. \\
The main parameters set for DAOPHOT and ALLSTAR are reported in Tab.~\ref{comparison:tab1}.

\begin{table}
\centering
 \begin{tabular}{ l r }
 \hline
 \hline
\rule[-1.0ex]{0pt}{2.5ex}  Parameter & Values\\
\hline
{\tt{FITTING RADIUS}} & 3.38\\
{\tt{THRESHOLD (in sigmas)}} & 5\\
{\tt{ANALYTIC MODEL PSF}} & 6\\
{\tt{PSF RADIUS}} & 7.5\\
{\tt{a1}} & 12.5\\
{\tt{INNER RADIUS}} & 20\\
{\tt{OUTER RADIUS}} & 35\\
{\tt{REDETERMINE CENTROIDS}} & 1.00\\
\hline
\end{tabular}
\caption[Main input parameters set in the DAOPHOT and ALLSTAR configuration files.]{Main input parameters set in the DAOPHOT and ALLSTAR configuration files. {\tt{FITTING RADIUS}}, {\tt{PSF RADIUS}}, {\tt{a1}}, {\tt{INNER RADIUS}} and {\tt{OUTER RADIUS}} are expressed in pixels.}
\label{comparison:tab1}
\end{table}

\subsubsection{SExtractor and PSFEx}
\label{comparison:4.2}

As for DAOPHOT, SExtractor instrumental parameters are set accordingly to those defined as input in the simulations (see Sect.~\ref{comparison:2}). \\
Concerning the sky background modeling and subtraction, we decided to automatically estimate the background within the software package adopting the global background map. Given the average size of the objects, in pixels, in our images, we chose to leave {\tt{BACK\_SIZE}} to the default value 64.
The choice of the filter was more complex. We performed several tests with various filters obtaining the best results by Gaussian and top-hat masks.
However, the choice between the various filters, although affecting the number of detected sources, does not alter their measurements.\\
For the thresholding parameters, we followed the same procedure approached with DAOPHOT, as described in Sect.~\ref{comparison:4.1}, that is, by choosing a value near to the change in gradient of the relation between the number of extracted sources and the threshold value for detections. Moreover, the catalog was visually inspected to avoid residual spurious detections and to verify the deblending parameters.\\
The size of the aperture for photometry, is fixed according to the one setted in DAOPHOT, to 25 pixels of diameter (\texttt{PHOT\_APERTURES}). For PSFEx parameters we used a set of 20 pixel basis and a size for the PSF image of 25 pixels according with the aperture size. We adopted a 25$\mathrm{\times}$25 pixel kernel following PSF variations within the image up to $\mathrm{2^{nd}}$ order. The main values set for SExtractor and PSFEx are reported in Tab.~\ref{comparison:tab2}.\\

\begin{table}
\centering
\begin{tabular}{ l r }
\hline
\hline
\rule[-1.0ex]{0pt}{2.5ex}  Parameter & Values\\
\hline
{\tt{DETECT\_MINAREA}} & 5\\
{\tt{DETECT\_THRESH}} & 1.5$\sigma$ \\
{\tt{ANALYSIS\_THRESH}} & 1.5$\sigma$ \\
{\tt{FILTER\_NAME}} & tophat\_3.0\_3x3.conv\\
{\tt{DEBLEND\_NTHRESH}} & 64\\
{\tt{DEBLEND\_MINCON}}T & 0.001\\
{\tt{BACK\_TYPE}} & GLOBAL\\
{\tt{BACK\_SIZE}} & 64\\
{\tt{BACK\_FILTERSIZE}} & 3\\
{\tt{PHOT\_APERTURES}} & 25\\
{\tt{BASIS\_TYPE}} & PIXEL\_AUTO\\
{\tt{BASIS\_NUMBER}} & 20\\
{\tt{PSF\_SIZE}} & 25,25\\
\hline
\end{tabular}
\caption[Main input parameters set in the SExtractor and PSFEx configuration files.]{Main input parameters set in the SExtractor and PSFEx configuration files. {\tt{DETECT\_MINAREA}}, {\tt{BACK\_SIZE}}, {\tt{PHOT\_APERTURES}} and {\tt{PSF\_SIZE}} are expressed in pixels.}
\label{comparison:tab2}
\end{table}

\subsection{Results}
\label{comparison:5}

In this subsection, we compare the results obtained using the two software packages. We will focus on four aspects of the extracted catalog namely: photometric depth, reliability, accuracy of the derived photometry, and determination of the positions of the centroids.\\
All the quantities and the statistics shown in this subsection are obtained by excluding saturated sources. In Fig.~\ref{comparison:fig1}, it is shown $\mathrm{\mu_{max}}$ as a function of the Kron magnitude of the objects extracted by SExtractor. As evidenced by the flattening of star sequence, sources with magnitude B$\mathrm{\le}$19 mag are saturated in the simulated images. Starting from Tab.~\ref{comparison:tab3} and Fig.~\ref{comparison:fig2}, we report the comparison among results obtained for the whole input magnitude range of unsaturated sources: 19-26 mag. However, since we consider only input stellar sources recovered by both software packages and since the completeness limit of the DAOPHOT star catalog is B=24 mag (see Sect.~\ref{comparison:5.1}), the last two reported magnitude bins are underpopulated and the results may be affected by catalog incompleteness.

\begin{figure}
\centering
\includegraphics[width=11.4cm]{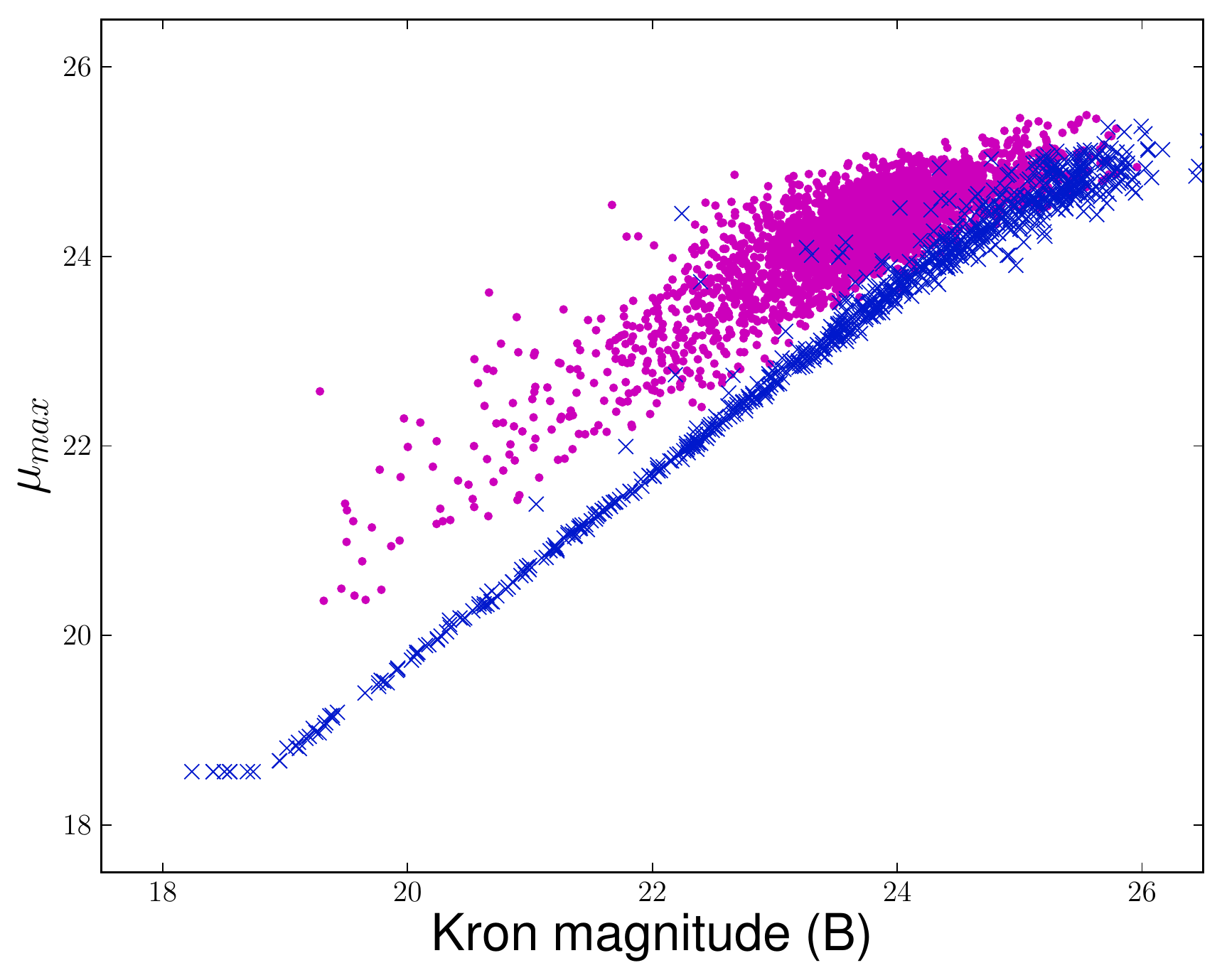}
\caption{$\mathrm{\mu_{max}}$ as a function of the Kron magnitude for stars (diagonal crosses) and galaxies (points) in the SExtractor catalog.}
\label{comparison:fig1}
\end{figure}

\subsubsection{Photometric depth}
\label{comparison:5.1}

\begin{figure}
\centering
\includegraphics[width=10.9cm]{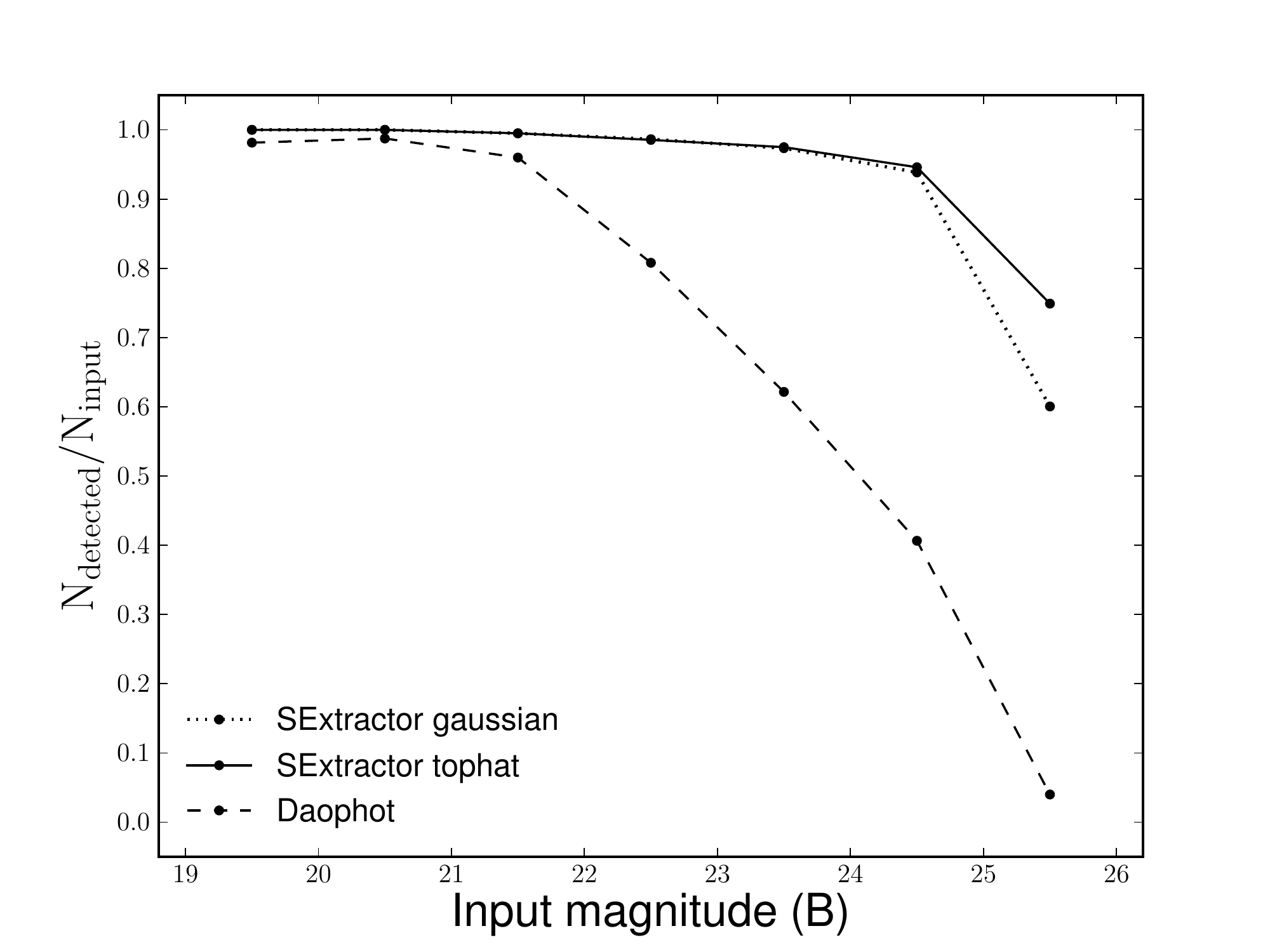}
\caption[Ratio between detected and input sources for different magnitude the bins.]{Ratio between detected and input sources for different magnitude the bins. The dotted and the solid lines refer to SExtractor used with a Gaussian and a top-hat filter respectively, while the dashed line refers to values obtained with DAOPHOT.}
\label{comparison:fig2}
\end{figure}

The photometric limiting magnitude of the extracted catalog is defined as the magnitude limit below which the completeness drops down to 90\%, where the completeness is the ratio of number of detected sources, $\mathrm{N_{detected}}$, and number of input sources, $\mathrm{N_{input}}$.\\
With DAOPHOT, the photometric depth depends mainly on the threshold applied, while for SExtractor, it depends also on the deblending of the sources, and on the filter used for the detection (see Sect.~\ref{comparison:3}). As discussed in Sect~\ref{comparison:4.1} and~\ref{comparison:4.2}, in order to fix the thresholding and deblending parameters, we performed several tests, by visually inspecting the extracted sources, and finally we fixed the values reported in Tab.~\ref{comparison:tab1} and~\ref{comparison:tab2}. Then, we compared the results of source extraction obtained using two different filters: a Gaussian (dotted line in Fig.~\ref{comparison:fig2}) and a ``top-hat" function (continuous line in Fig.~\ref{comparison:fig2}). As shown in Fig.~\ref{comparison:fig2}, using SExtractor with a top-hat filter, we can improve the detection of faint sources. In this case, the depth of the catalog is $\sim$ 25.0 mag. Hence, we refer to this filter in all the tests performed with SExtractor and reported below. Figure~\ref{comparison:fig2} shows also the percentage of extracted sources per magnitude bin obtained using DAOPHOT (dashed line). With this software package, the completeness drops rapidly to very low values for magnitudes fainter than B = 22.0 mag. However, this comparison is misleading. In fact, DAOPHOT is not designed to work with extended sources.
For this reason, in Fig.~\ref{comparison:fig3a} we report the ratio between the detected sources, which are \textit{a priori} known to be stars ($\mathrm{S_{detected}}$), and the input stars ($\mathrm{S_{input}}$). In Fig.~\ref{comparison:fig3b} we show the same quantities but for galaxies ($\mathrm{G_{detected}}$, $\mathrm{G_{input}}$).
\begin{figure*}
\centering
\subfloat[]{
\label{comparison:fig3a}
\includegraphics[width=11.cm]{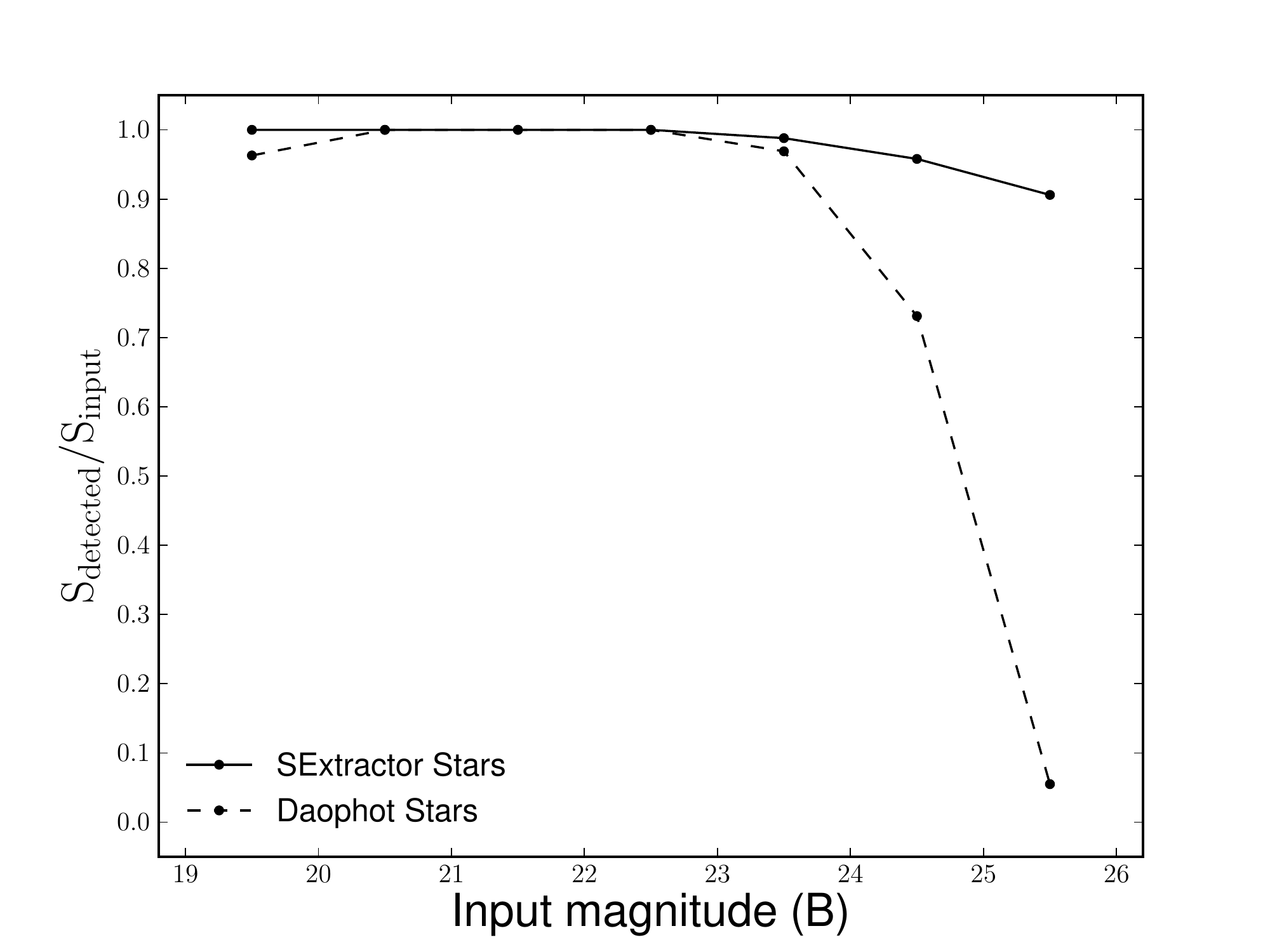}}
\\ \subfloat[]{
\label{comparison:fig3b}
\includegraphics[width=11.cm]{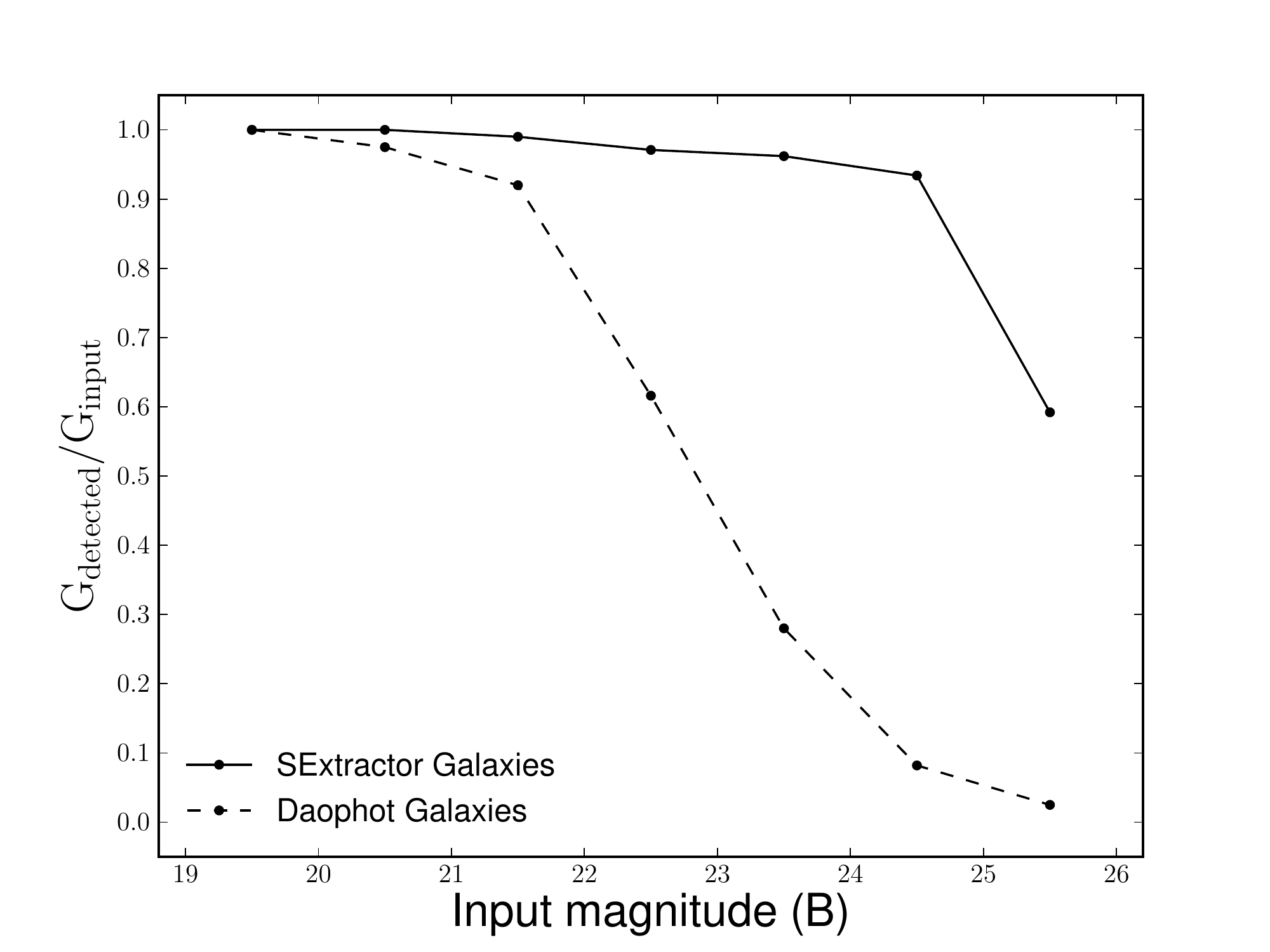}}
\caption[the ratio between the numbers of detected and input objects (star or galaxies) as a function of magnitude bins, as obtained by SExtractor and by DAOPHOT.]{The left panel shows the ratio between the numbers of detected and input stars as a function of magnitude bins, as obtained by SExtractor (solid line) and by DAOPHOT (dashed line); in the right panel are plotted the same quantities, but for galaxies.}
\end{figure*}
We can see that the fraction of detected source is higher for stars for both SExtractor (B = 26.0 mag) and DAOPHOT (B = 24.0 mag). Hence, in conclusion, considering only stars, the final depth returned by DAOPHOT is $\sim$ 2 mag brighter than those produced by SExtractor. \\

\subsubsection{Reliability of the catalog}
\label{comparison:5.2}

\begin{figure*}
\centering
\subfloat[]{
\centering
\label{comparison:fig4a}
\includegraphics[width=10.cm]{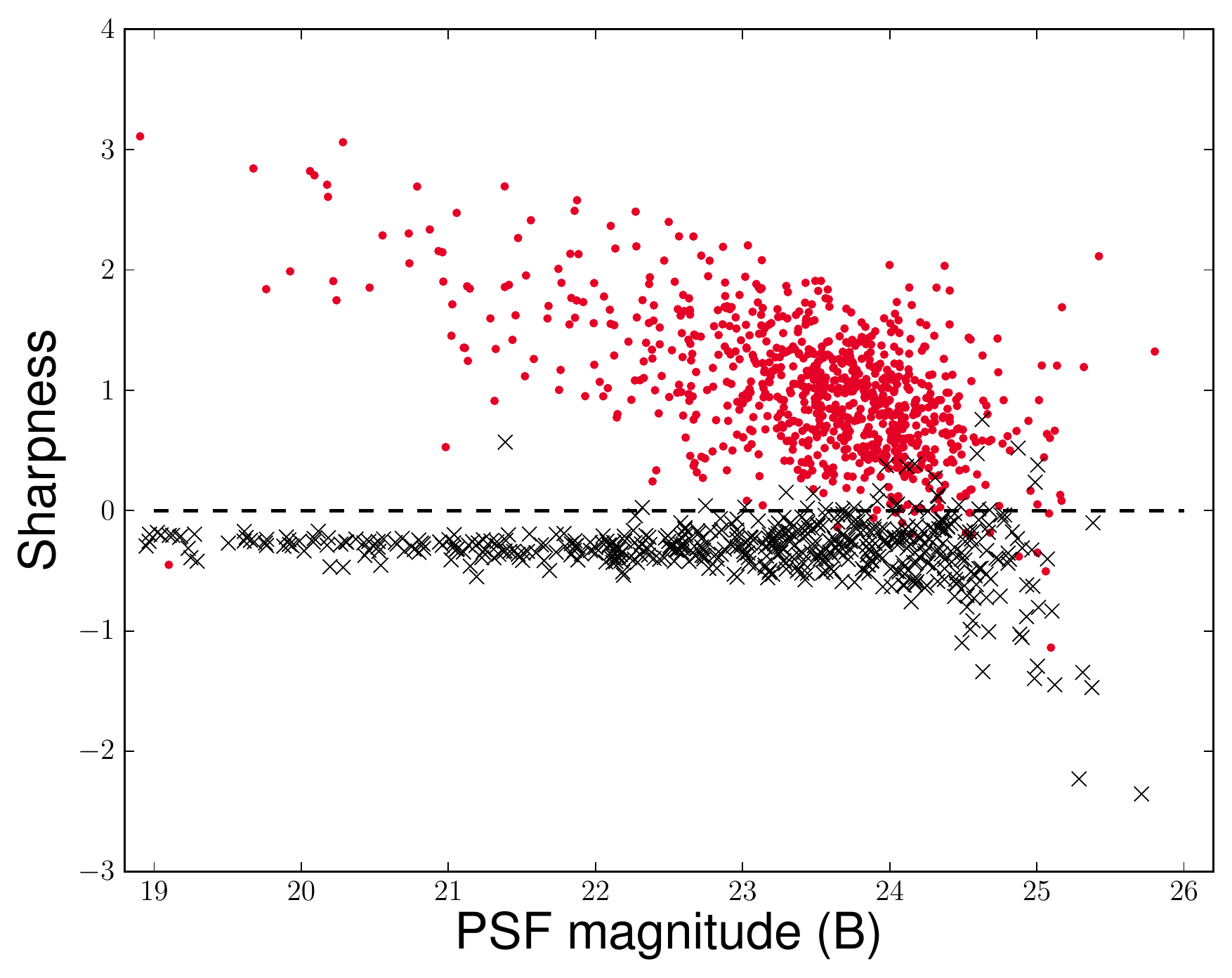}}
\\ \subfloat[]{
\centering
\label{comparison:fig4b}
\includegraphics[width=10.cm]{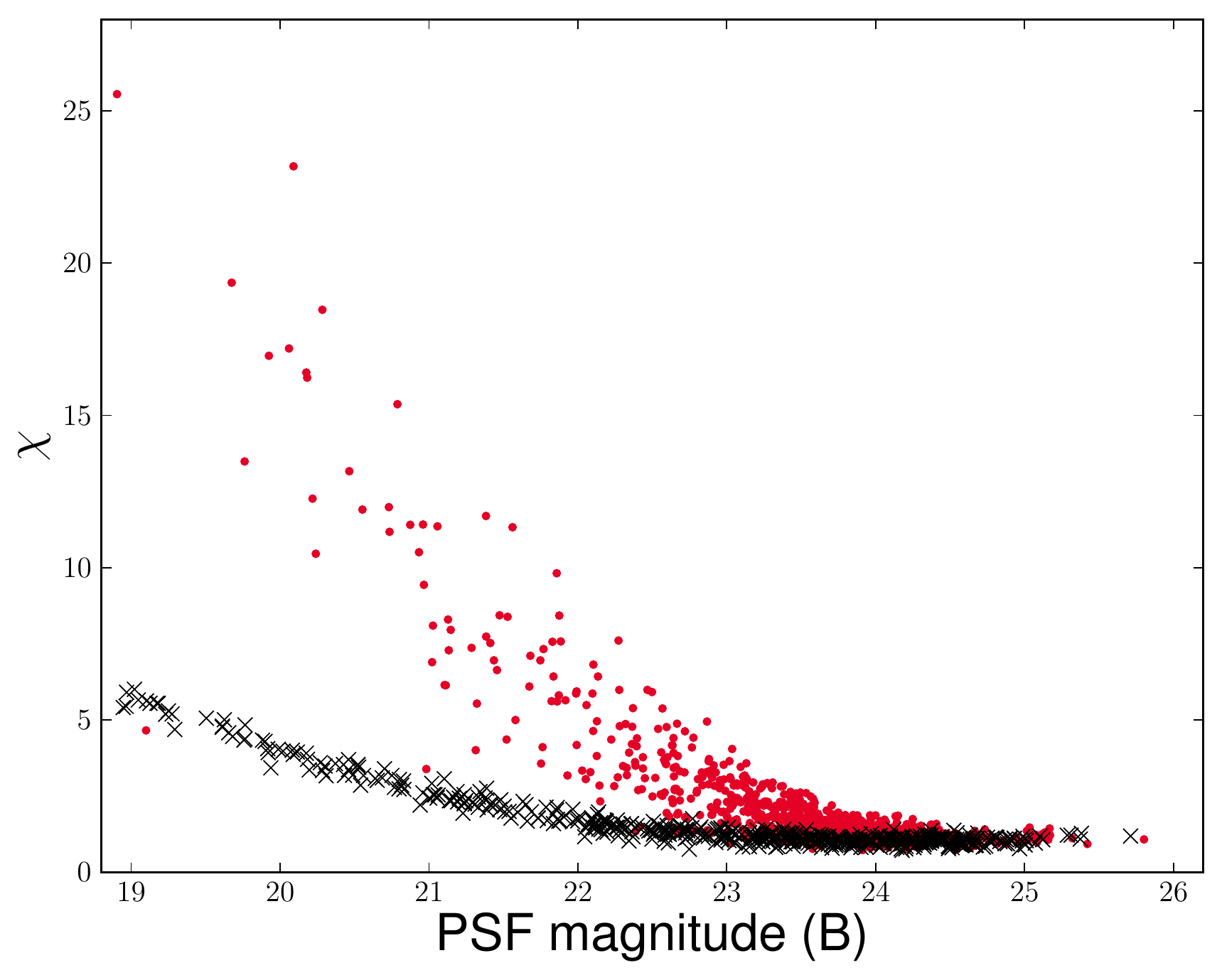}}
\label{comparison:fig4}
\caption[Distribution of DAOPHOT sharpness and $\chi$  as a function of the PSF magnitude for simulated stars and galaxies.]{Distribution of DAOPHOT sharpness ({\it left panel}) and $\chi$ ({\it right panel}) as a function of the PSF magnitude for simulated stars (diagonal crosses) and galaxies (points). In the left panel the dashed line at zero sharpness is the adopted separation limit for the star/galaxy classification (see Sect.~\ref{comparison:5.2}).}
\end{figure*}

The reliability of the catalog is defined as the ratio between the number of well-classified sources and the number of the sources detected by the software packages, see Eq. (6) of \cite{laher2008}.
For these tests we use only the set of stars detected ($\mathrm{S_{detected}}$) and well-classified ($\mathrm{S_{classified}}$) by both SExtractor and DAOPHOT. We compared results obtained with several methods to classify the sources. In fact, each method leads to a different estimate of the reliability. \\
As far as DAOPHOT is concerned, we used the output parameters \texttt{SHARP} (see Fig.~\ref{comparison:fig4a}) and $\chi$ (see Fig.~\ref{comparison:fig4b}), made the determination with ALLSTAR (see Sect.~\ref{comparison:3.1}). Fig.~\ref{comparison:fig4a} shows the distribution of ALLSTAR sharpness {\tt{SHARP}} for our data. The separation between the two classes seems to be well defined. On the other hand, Fig.~\ref{comparison:fig4b} shows that the use of the $\chi$ parameter does not improve the star/galaxy classification. For this reason, we classified as stars all the sources with {\tt{SHARP}} lower than 0.\\
In order to investigate the reliability of the catalog made with SExtractor, we used both traditional methods as well as the new parameter {\tt{SPREAD\_MODEL}}. In Fig.~\ref{comparison:fig5a}, we plot {\tt{CLASS\_STAR}} as a function of the Kron magnitude for our data. As shown, the lower the established limit to separate stars and galaxies, the higher will be the contamination of the star subsample from galaxies. A reasonable limit for the separation is 0.98.\\

\begin{figure*}
\centering
\subfloat[]{
\includegraphics[width=6.2cm]{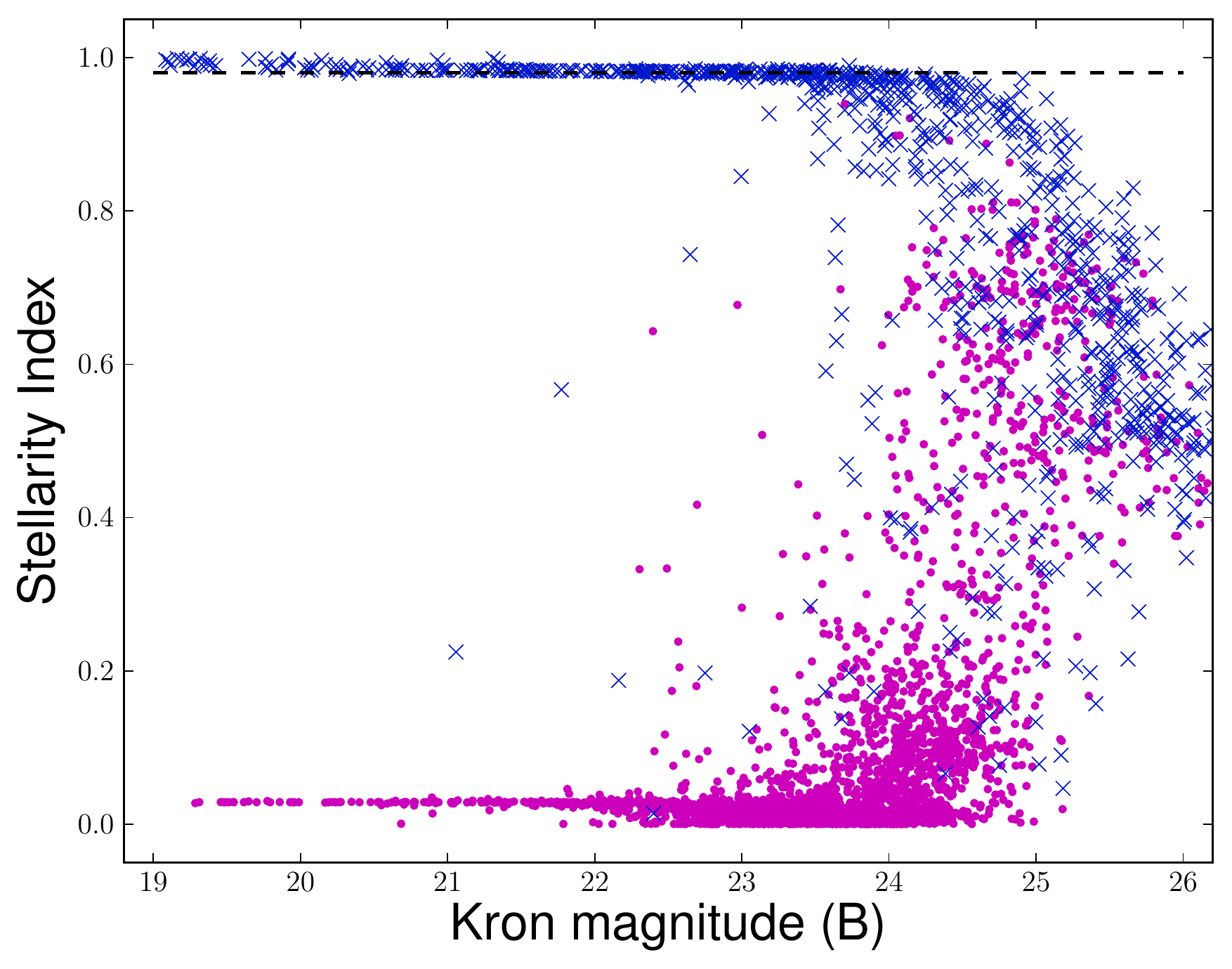}
\label{comparison:fig5a}}
\subfloat[]{
\includegraphics[width=6.2cm]{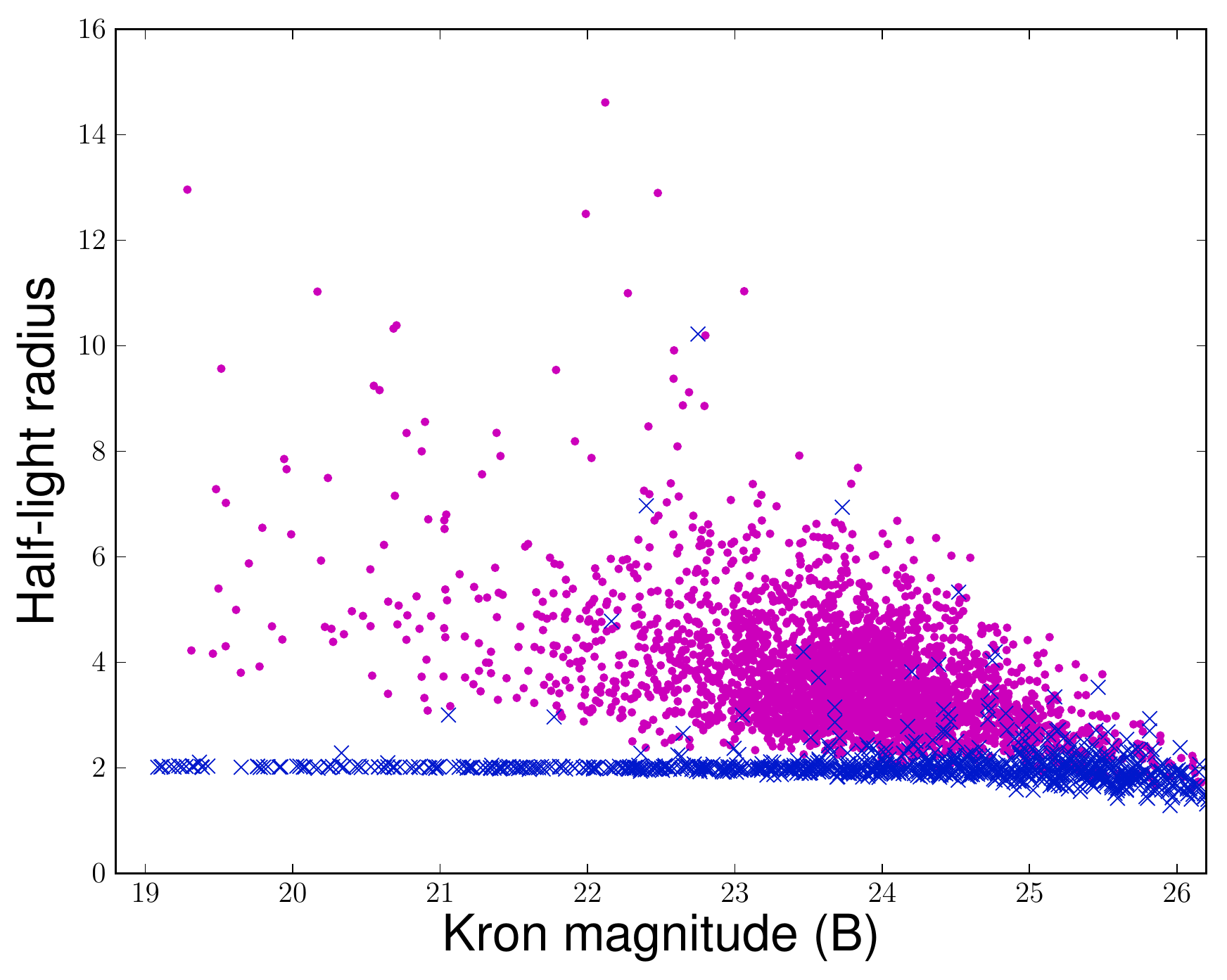}
\label{comparison:fig5b}}
\\
\subfloat[]{
\includegraphics[width=6.2cm]{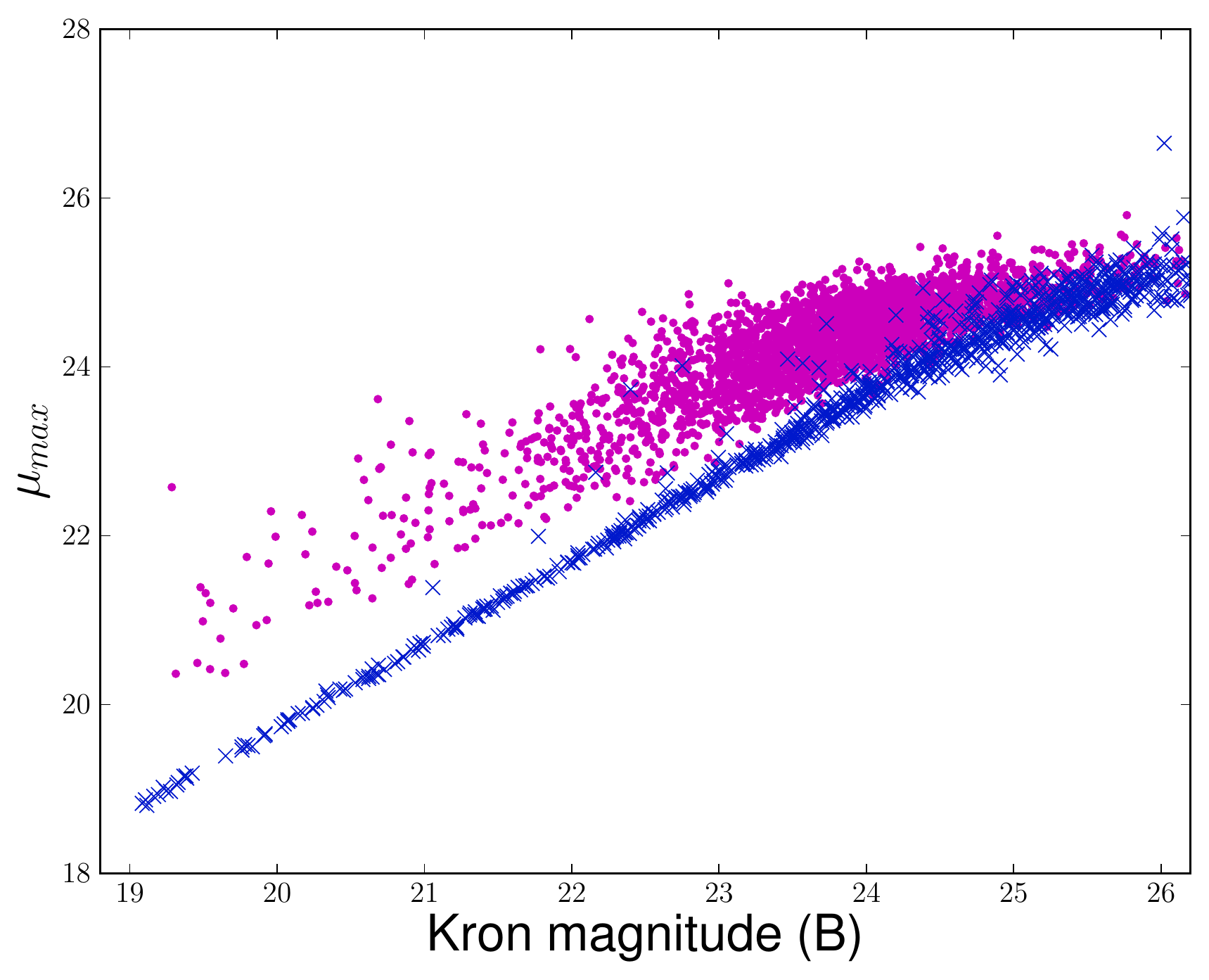}
\label{comparison:fig5c}}
\subfloat[]{
\includegraphics[width=6.2cm]{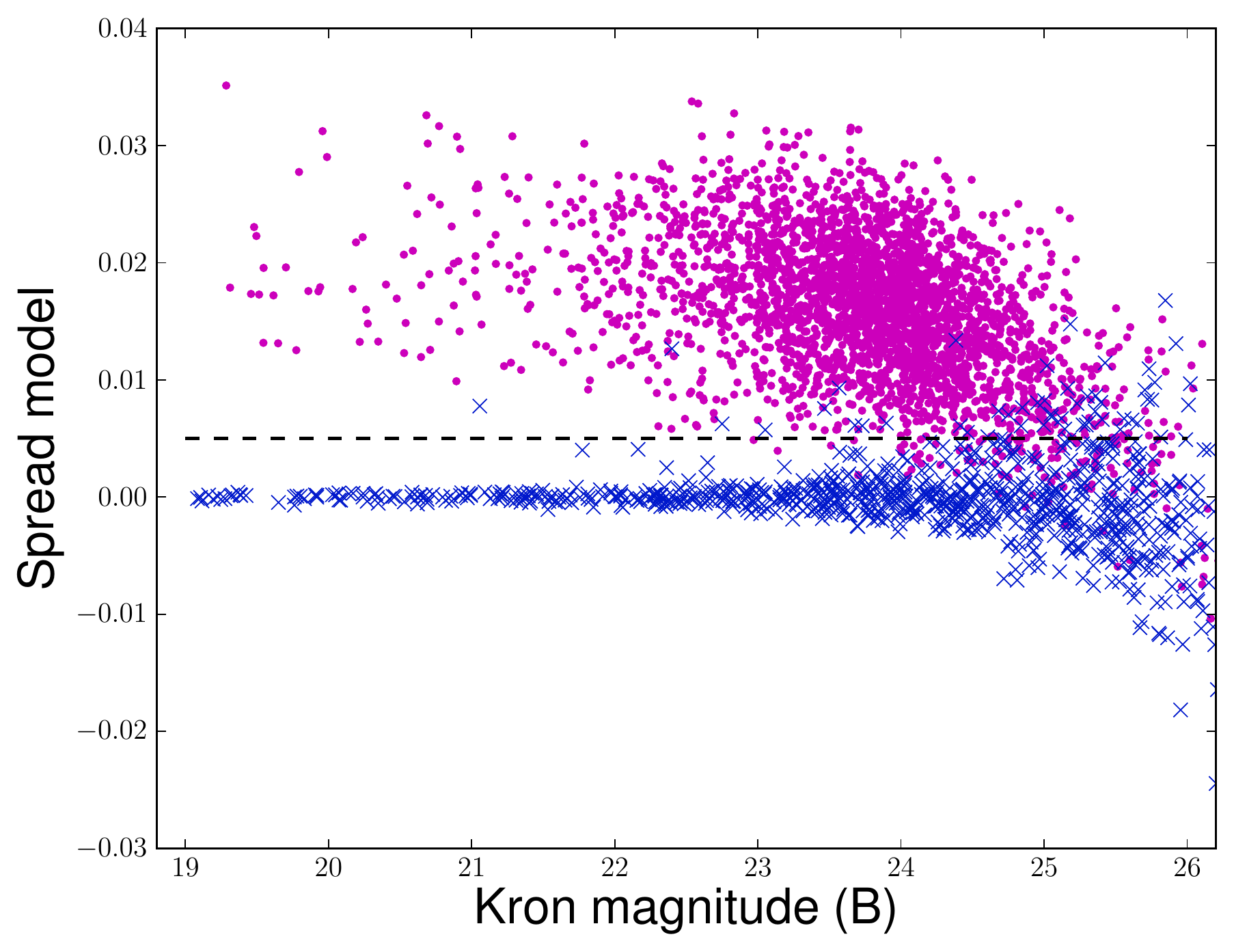}
\label{comparison:fig5d}}
\caption[Distribution of SExtractor stellarity index, half-light radius, $\mathrm{\mu_{max}}$ and spread model for simulated stars and galaxies.]{Distribution of SExtractor stellarity index ({\it{panel a}}), half-light radius ({\it{panel b}}), $\mathrm{\mu_{max}}$ ({\it{panel c}}) and spread model ({\it{panel d}}), as a function of the Kron magnitudes for simulated stars (diagonal crosses) and galaxies (points). The dashed line in {\it{panels a}} and \textit{d} is the adopted separation limit for the star/galaxy classification (see Sect.~\ref{comparison:5.2}). }
\end{figure*}

Figures~\ref{comparison:fig5b} and~\ref{comparison:fig5c} show the locus of stars, selected according to the relation between half-light radius and $\mu_{max}$, respectively, as a function of the Kron magnitude. There is an improvement of the source classification compared to the use of {\tt{CLASS\_STAR}} parameter, allowing a reliable star/galaxy separation down to B $\mathrm{=}$ 23.5 mag.\\
Finally, Fig.~\ref{comparison:fig5d} shows {\tt{SPREAD\_MODEL}} values as a function of Kron magnitude.
Stars and galaxies tend to arrange themselves in two distinct places on the plot. Also in this case, the higher we choose the separation limit, the higher will be the contamination of the stellar sequence from galaxies. A good compromise between a reliable classification and a low contamination is the value 0.005. \\
In Fig.~\ref{comparison:fig6}, it is shown the ratio between the sources correctly classified as stars using the stellarity index (dotted line), spread model (continuous line) and sharpness parameter (dashed line), as function of input magnitude. \\
In conclusion, if we define a classification with a reliability of at least 90\% with these methods we can acceptably classify the stars in DAOPHOT down to about 24 mag, which is the photometric depth of the extracted catalog, while in the case of SExtractor, the classifier {\tt{SPREAD\_MODEL}} allows us to obtain a reliable star/galaxy separation down to B $\mathrm{=}$ 26 mag.\\

\subsubsection{Photometry}
\label{comparison:5.3}

In this subsection, we compare the results obtained with aperture and PSF photometry on the sample of stars detected by both SExtractor and DAOPHOT. We also investigate the results obtained with Kron, isophotal and model-fitting photometry for galaxies detected by SExtractor. \\
In Tab.~\ref{comparison:tab3}, we report the mean difference and the standard deviation between aperture and PSF magnitudes, as estimated by DAOPHOT (parts a and b, respectively) and SExtractor (parts c and d, respectively), against input magnitude.\\
Figure~\ref{comparison:fig7} shows the residuals between aperture and input magnitudes (top panels), and the residuals between PSF and input magnitudes (bottom panels), as estimated by DAOPHOT (left panels) and SExtractor (right panels).\\
Table~\ref{comparison:tab3} and Fig.~\ref{comparison:fig7} show that there is a characteristic broadening of the residuals at fainter magnitudes, as we expect when measurements become sky-noise dominated, but the spread in the case of PSF photometry remains smaller than for aperture measurements.
This behavior is well known (e.g. \citealp{becker2007}) for DAOPHOT, but it is worth underlining that SExtractor has reached this level of accuracy in PSF photometry only after the release of PSFEx.\\
In the top part of the Tab.~\ref{comparison:tab3}, we also report the mean difference and the standard deviation between Kron (part a), isophotal (part b) and model magnitudes (part c), respectively, and input magnitudes for stars.\\
For completeness, since SExtractor is designed also to obtain accurate galaxy photometry, we report in the bottom part of the Tab.~\ref{comparison:tab4} the mean difference and the standard deviation between Kron (part d), isophotal (part e) and model magnitudes (part f), and input magnitudes for the ``true" galaxies detected by the software package (See Sect.~5.2).\\
By considering only stellar photometry, both software packages are able to deliver acceptable performances for both aperture and PSF photometry, up to a threshold two magnitudes brighter than the limiting magnitudes of the input simulated images, which is the completeness limit of the DAOPHOT catalog. Furthermore, the Kron magnitude yields $\sim$ 94\% of the total source flux within the adaptive aperture (\citealp{bertin1996}) so, accordingly, we see a shift of $\sim$ 0.07 mag even in the brightest magnitude bin. On the other hand, the isophotal magnitude depends on the detection threshold and the model magnitudes (obtained through a sum of bulge plus disk), and produce an unbiased estimate of the total magnitude also for stars.\\
In conclusion, the new PSF modeling of SExtractor produces photometric measurements as accurate as those obtained with DAOPHOT.\\

\begin{figure}
\centering
\includegraphics[width=12.5cm]{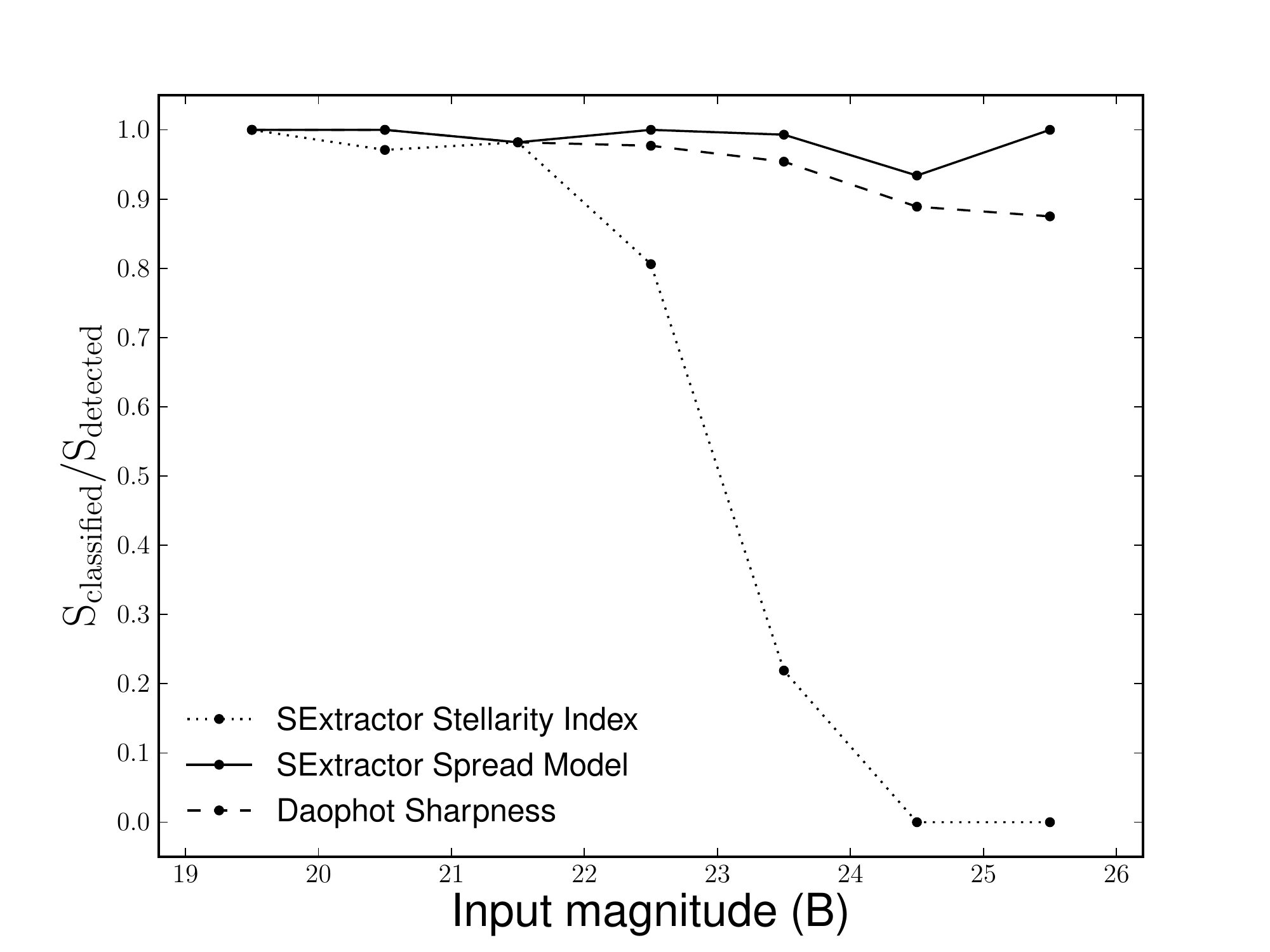}
\caption[Ratio between stars classified by Stellarity Index and Spread Model from SExtractor and by DAOPHOT sharpness.]{Ratio between stars classified by Stellarity Index (dotted line) and Spread Model (solid line) from SExtractor with threshold values respectively to 0.98 and 0.005, and by DAOPHOT sharpness (dashed line) with a threshold value equal to zero and detected stars, as function of input magnitude.}
\label{comparison:fig6}
\end{figure}

 \subsubsection{Centroids}
 \label{comparison:5.4}

 \begin{figure*}
 \centering
 \subfloat[]{
 \includegraphics[width=6.2cm]{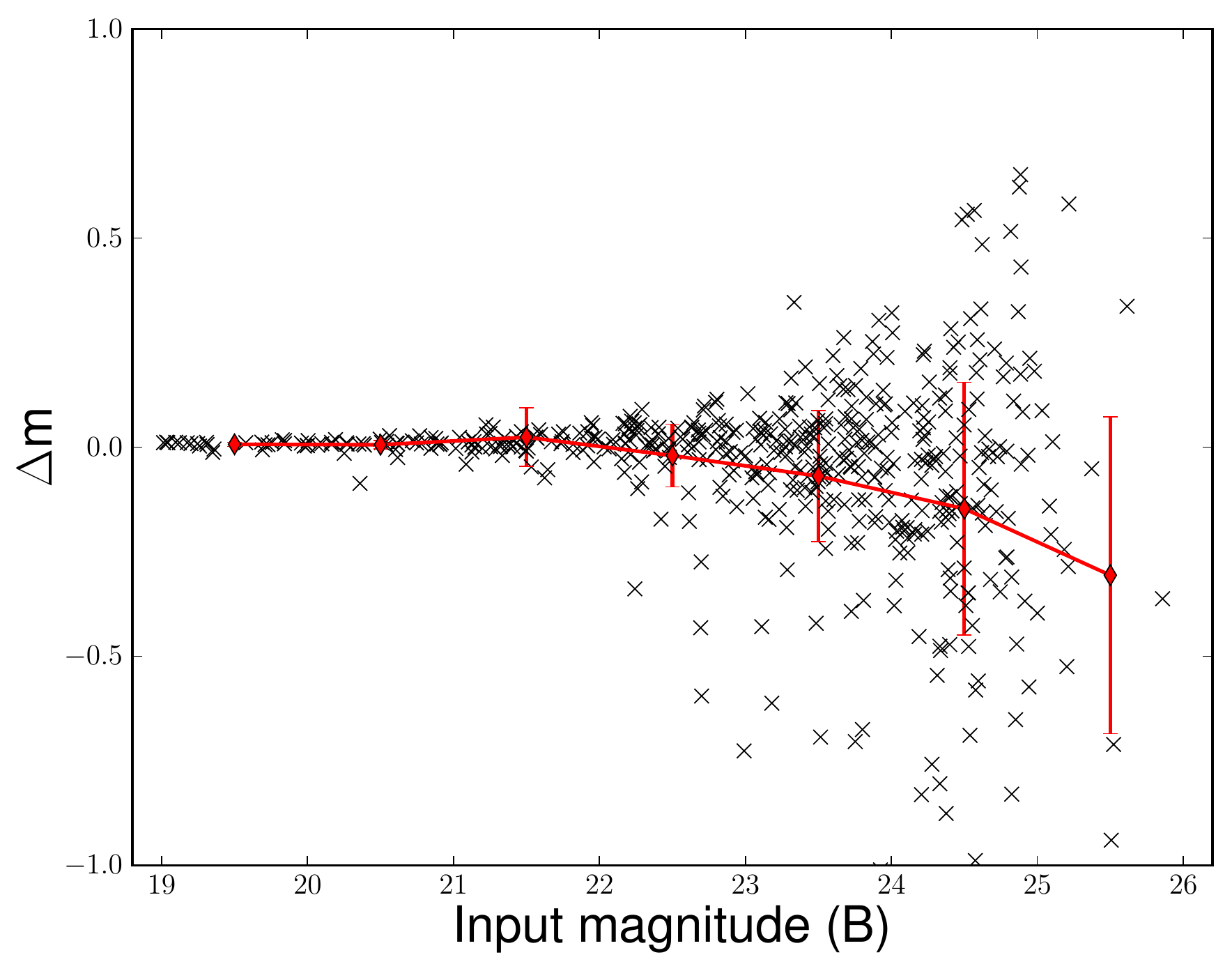}
 \label{comparison:fig7a}
 }
 \subfloat[]{
 \includegraphics[width=6.2cm]{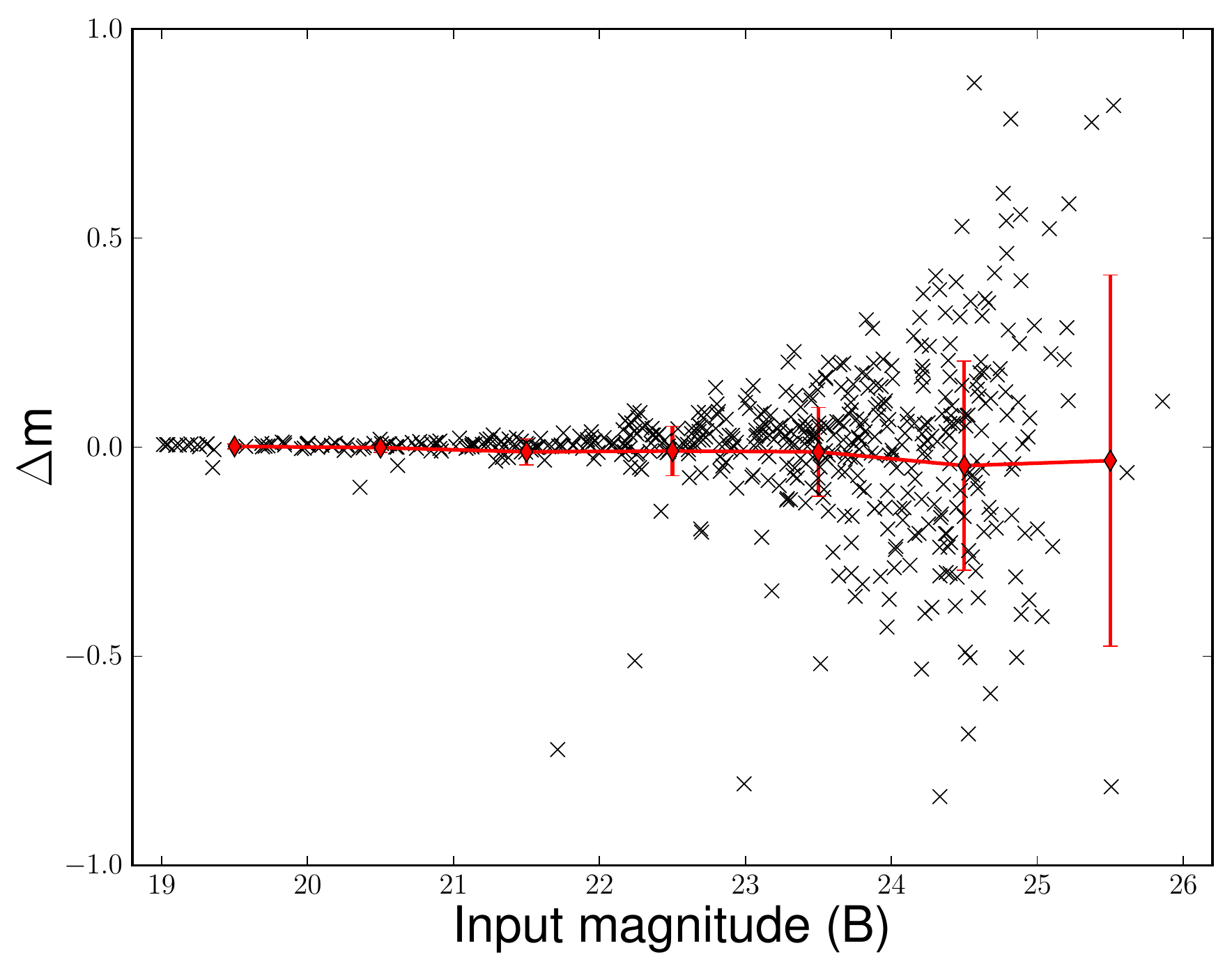}
 \label{comparison:fig7b}
 }
 \\
 \subfloat[]{
 \includegraphics[width=6.2cm]{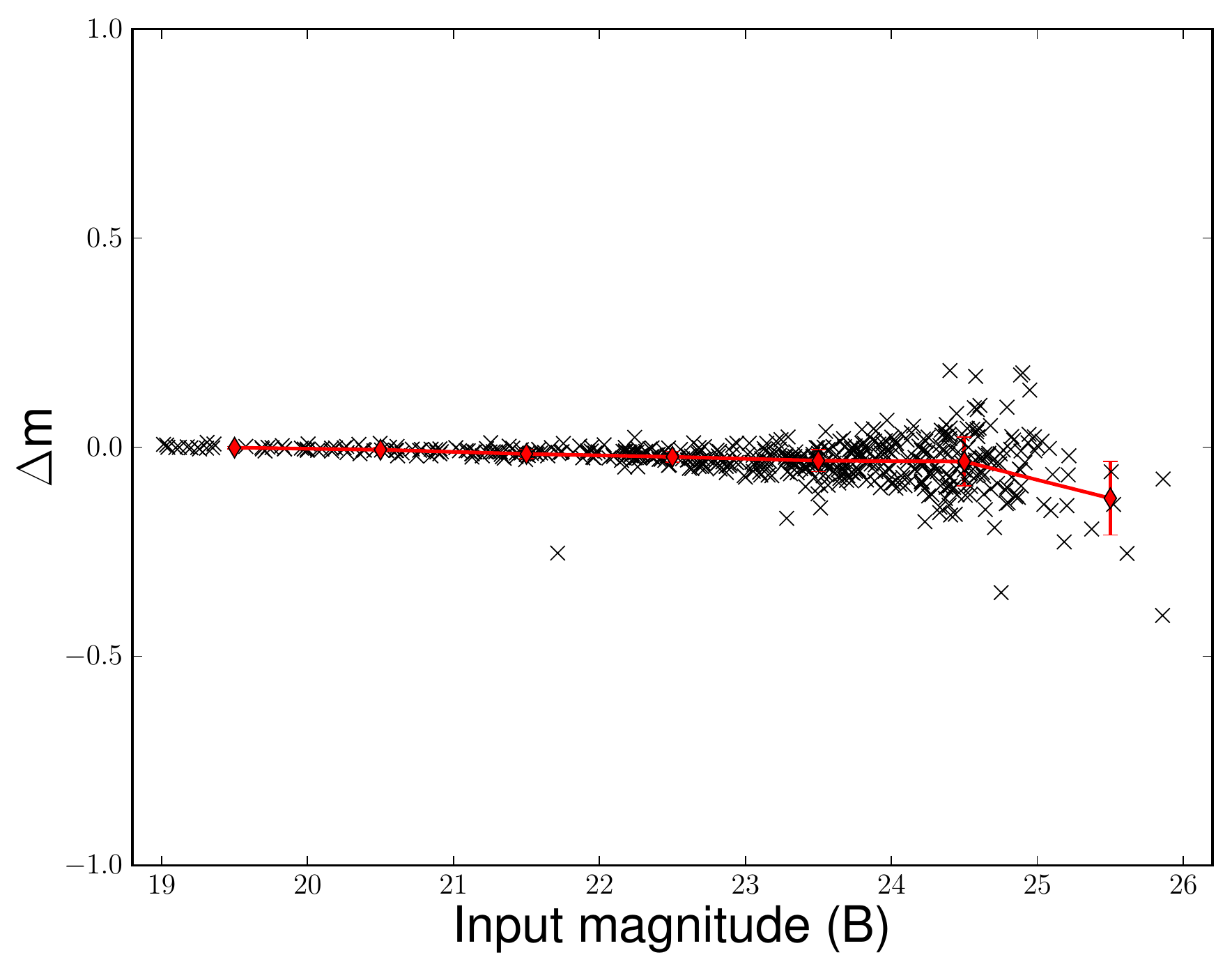}
 \label{comparison:fig7c}
 }
 \subfloat[]{
 \includegraphics[width=6.2cm]{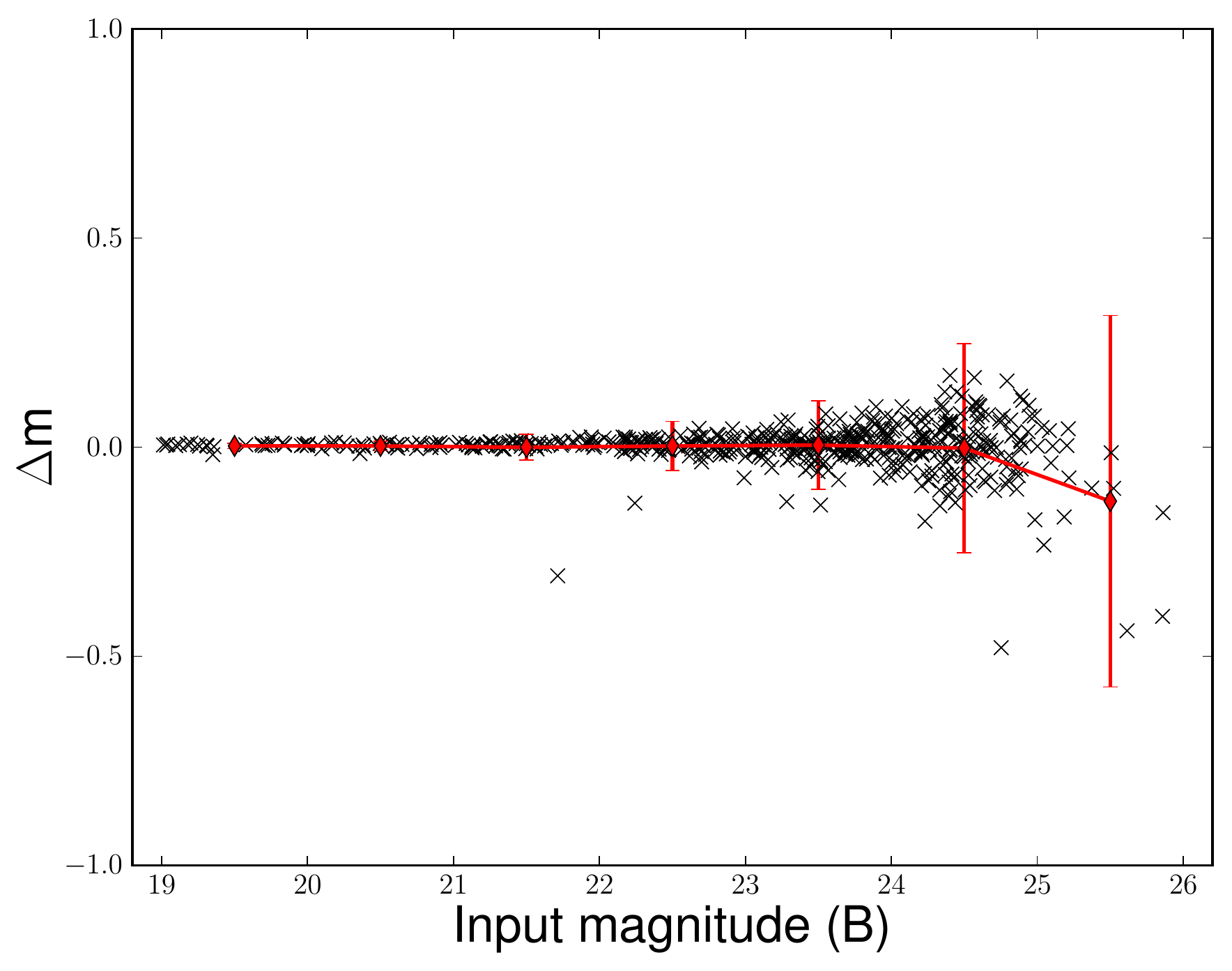}
 \label{comparison:fig7d}
 }
 \caption[Residuals between aperture magnitudes and PSF estimated by DAOPHOT and by SExtractor]{\textit{Top panels}: Residuals between aperture magnitudes estimated by DAOPHOT ({\it{left panel}}) and by SExtractor ({\it{right panel}}), and input magnitudes for detected stars. \textit{Bottom panels}: Residuals between PSF magnitude estimated by DAOPHOT ({\it{left panel}}) and by SExtractor ({\it{right panel}}), and input magnitude for detected stars. Superimposed red points and solid red lines draw the mean and standard deviation values reported in Tab.~\ref{comparison:tab3}.}
 \label{comparison:fig7}
 \end{figure*}

The last comparison is among extracted and input positions. There are different ways to obtain centroid measurements. As stated above, DAOPHOT can provide two different measurements for centroids. The simplest are the coordinates of the source barycenter, derived during the thresholding process. These coordinates can be redetermined by ALLSTAR, once DAOPHOT has built a PSF model, by applying a PSF correction.\\
Concerning SExtractor, we chose to compare the results obtained using the barycenter and the PSF corrected coordinates, as for DAOPHOT, and the results obtained by using the windowed positions along both axes. These coordinates are obtained by integrating pixel values within a circular Gaussian window.
In Tab.~\ref{comparison:tab5}, it is reported the mean difference between barycenter coordinates and PSF-corrected coordinates, estimated respectively with DAOPHOT (parts a and b) and SExtractor (parts c and d) and input coordinates.\\
Finally, Tab.~\ref{comparison:tab6} shows the difference among input and windowed coordinates estimated by SExtractor. \\
Fig.~\ref{comparison:fig8} and ~\ref{comparison:fig9} show the difference between the input and barycenter coordinates and between the input and PSF corrected coordinates.\\
Both software packages show a bias between output centroid coordinates $\le$ 0.01 arcsec (equal to $\sim$ 0.47 pixel) and input X and Y, with an average deviation of $\le$ 0.02 arcsec (equal to $\sim$ 0.94 pixel), down to the DAOPHOT completeness magnitude limit. These values are in particular improved in terms of average deviation ($\sigma_{\Delta X(Y)} \le$ 0.01 arcsec), when PSF correction is applied. Hence, we can conclude that the results for centroids are satisfactory in both cases.\\

 \begin{figure*}
 \centering
 \subfloat[]{
 \includegraphics[width=10cm]{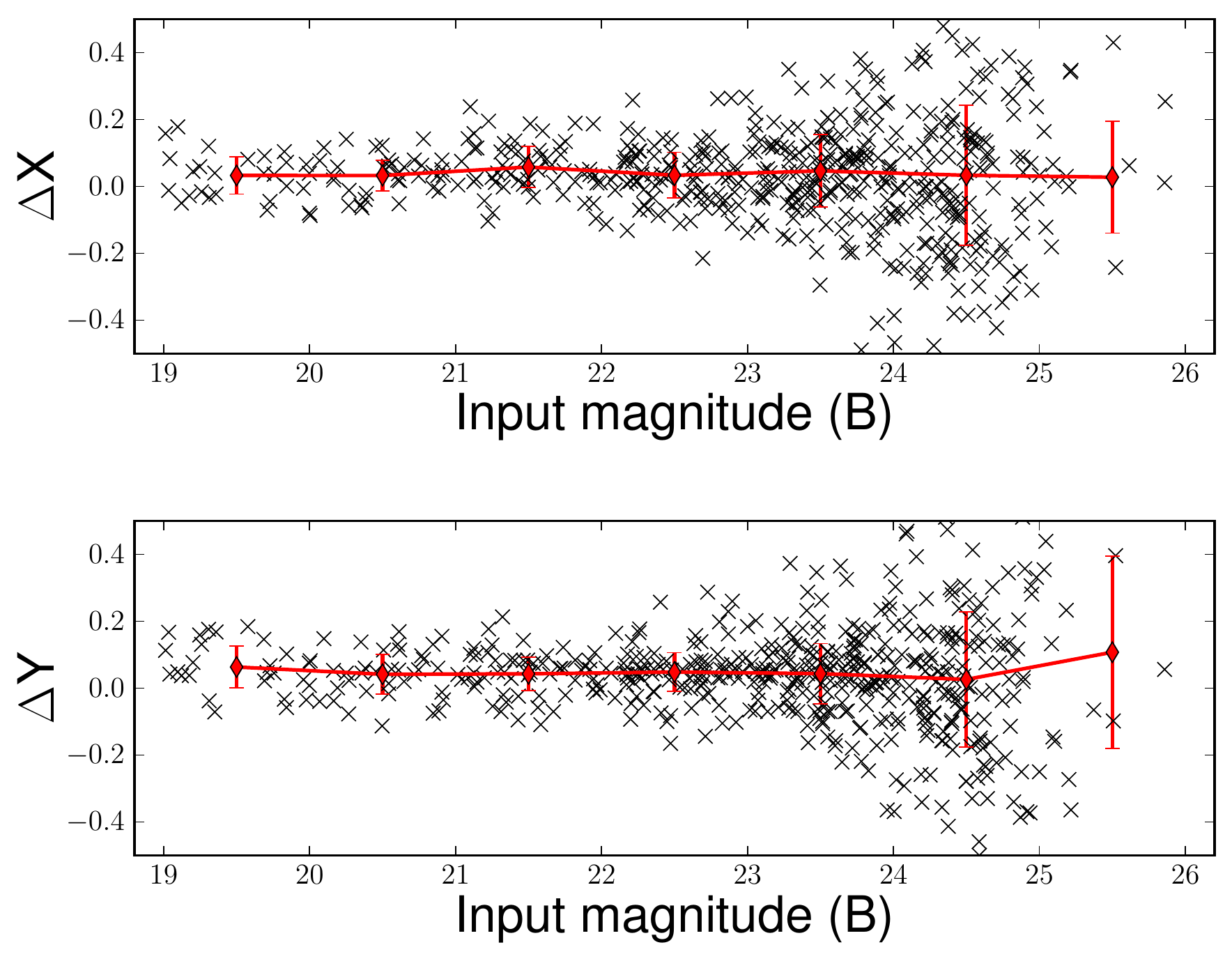}
 \label{comparison:fig8a}}\\
 \subfloat[]{
 \includegraphics[width=10cm]{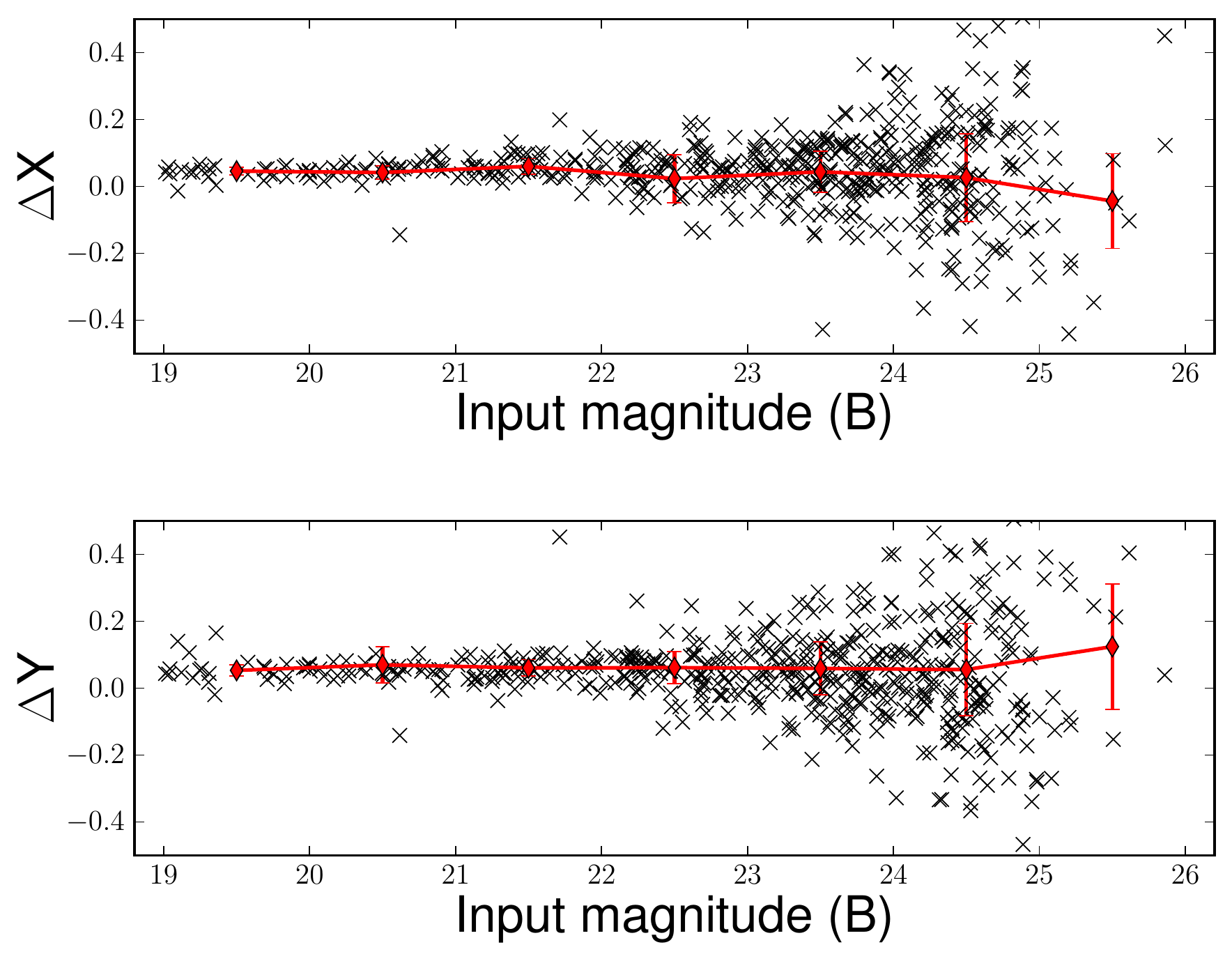}
 \label{comparison:fig8b}}
 \caption[Difference between barycenter coordinates estimated by DAOPHOT and by SExtractor.]{Difference between barycenter coordinates estimated by DAOPHOT ({\it{upper panels}}) and by SExtractor ({\it{lower panels}}), and input coordinates and as a function of input magnitude for detected stars. Superimposed red points and solid red lines draw the mean and standard deviation values reported in top part of Tab.~\ref{comparison:tab5}.}
 \label{comparison:fig8}
 \end{figure*}

 \begin{figure*}
 \centering
 \subfloat[]{
 \includegraphics[width=10.cm]{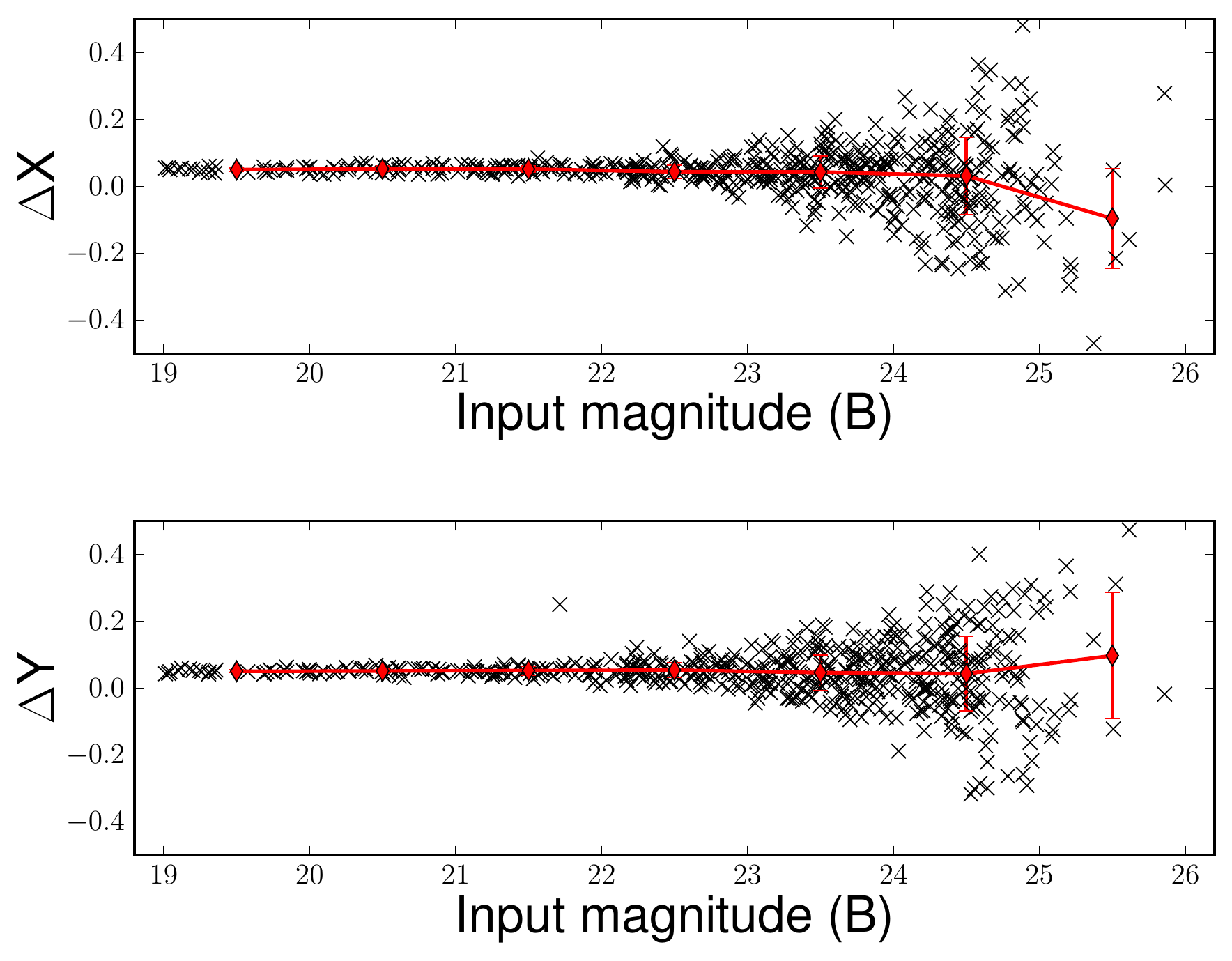}
 \label{comparison:fig9a}
 }\\
 \subfloat[]{
 \includegraphics[width=10.cm]{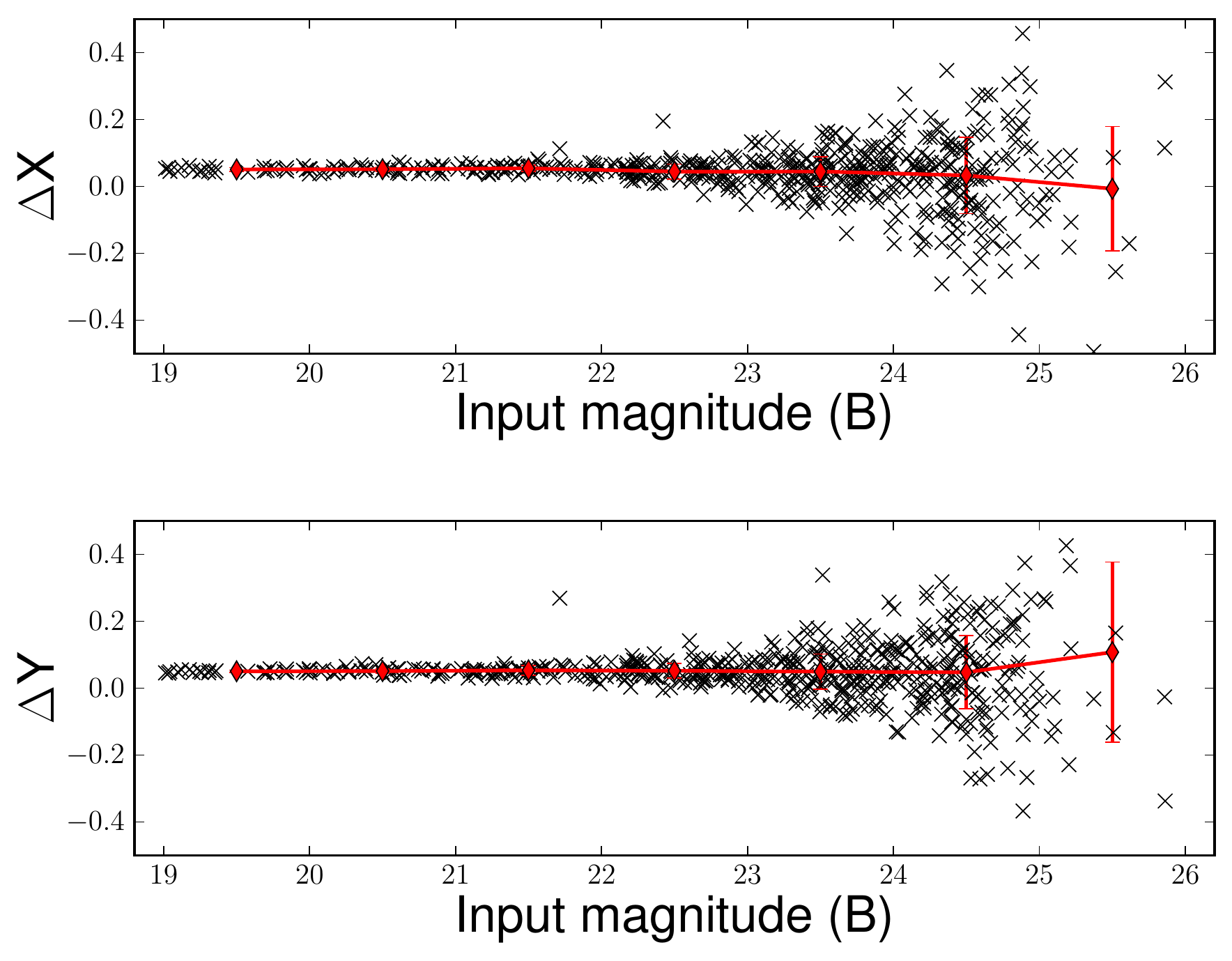}
 \label{comparison:fig9b}
 }
 \caption[Difference between PSF-corrected and input coordinates estimated by DAOPHOT and by SExtractor.]{Difference between PSF-corrected and input coordinates estimated by DAOPHOT ({\it{left panels}}) and by SExtractor ({\it{right panels}}), as a function of input magnitude for detected stars. Superimposed red points and solid red lines draw the mean and standard deviation values reported in bottom part of Tab.~\ref{comparison:tab5}.}
 \label{comparison:fig9}
 \end{figure*}

 \begin{table*}
\centering
\small
\setlength{\tabcolsep}{3pt}
 \begin{tabular}{|l c c || c c || c c || c c|}
 \hline
 \hline
\rule[-1.0ex]{0pt}{1.0ex}  \tt{Bin} & \tt{$\mathrm{\Delta m_{mean}}$} &  \tt{$\mathrm{\sigma_{\Delta m}}$} & \tt{$\mathrm{\Delta m_{mean}}$} &  \tt{$\mathrm{\sigma_{\Delta m}}$} & \tt{$\mathrm{\Delta m_{mean}}$} &  \tt{$\mathrm{\sigma_{\Delta m}}$}& \tt{$\mathrm{\Delta m_{mean}}$} &  \tt{$\mathrm{\sigma_{\Delta m}}$} \\
\rule[-1.0ex]{0pt}{1.0ex}  (mag) & \tt{$\mathrm{(mag)}$} &  \tt{$\mathrm{(mag)}$} & \tt{$\mathrm{(mag)}$} &  \tt{$\mathrm{(mag)}$} & \tt{$\mathrm{(mag)}$} &  \tt{$\mathrm{(mag)}$} & \tt{$\mathrm{(mag)}$} & \tt{$\mathrm{(mag)}$}\\
\hline
\rule[-1.0ex]{0pt}{1.0ex}  & \multicolumn{2}{c ||}{\tt{$\mathrm{(a)}$}} & \multicolumn{2}{c ||}{\tt{$\mathrm{(b)}$}} & \multicolumn{2}{c ||}{\tt{$\mathrm{(c)}$}}  & \multicolumn{2}{c|}{\tt{$\mathrm{(d)}$}}\\
\hline
\hline
 \rule[-1.0ex]{0pt}{1.0ex}  19 - 20 & 0.007 & 0.005 & 0.002 & 0.006 & -0.001 & 0.003 & 0.003 & 0.003 \\
 \rule[-1.0ex]{0pt}{1.0ex}  20 - 21 & 0.006  & 0.010 & -0.001 & 0.011 & -0.006  & 0.005 &  0.003  & 0.004 \\
 \rule[-1.0ex]{0pt}{1.0ex}  21 - 22 & 0.024 & 0.070 & -0.011 & 0.031 & -0.016 & 0.011 & 0.000 & 0.012 \\
 \rule[-1.0ex]{0pt}{1.0ex}  22 - 23 & -0.020 & 0.075 & -0.009 & 0.059 & -0.023 & 0.013 & 0.003 & 0.014 \\
 \rule[-1.0ex]{0pt}{1.0ex}  23 - 24 & -0.069 & 0.157 & -0.011 & 0.106 & -0.032 & 0.026 & 0.005 & 0.025 \\
 \rule[-1.0ex]{0pt}{1.0ex}  24 - 25 & -0.147 & 0.302 & -0.044 & 0.250 & -0.034 & 0.058 & -0.002 & 0.055 \\
 \rule[-1.0ex]{0pt}{1.0ex}  25 - 26 & -0.306 & 0.379 & -0.032 & 0.444 & -0.122 & 0.088 &  -0.129 & 0.129 \\
 \hline
 \end{tabular}
\caption[The mean difference, and the standard deviation between aperture and input magnitudes as estimated by DAOPHOT (part a) and by SExtractor (part b). Parts c and d report the mean difference and the standard deviation between PSF and input magnitudes as obtained by using DAOPHOT and SExtractor, respectively.]{The table reports, as a function of the magnitude bin (col.~1), the mean difference {\tt{$\mathrm{\Delta m_{mean}}$}} (columns~2, 4, 6 and 8), and the standard deviation {\tt{$\mathrm{\sigma_{\Delta m}}$}} (columns~3, 5, 7 and 9) between aperture and input magnitudes as estimated by DAOPHOT (part a) and by SExtractor (part b). Parts c and d report the mean difference and the standard deviation between PSF and input magnitudes as obtained by using DAOPHOT and SExtractor, respectively.}
\label{comparison:tab3}
\end{table*}

\begin{table*}[!ht]
\centering
\small
 \begin{tabular}{|l c c || c c || c c|}
 \hline
 \hline
\rule[-1.0ex]{0pt}{1.0ex}  \tt{Bin} & \tt{$\mathrm{\Delta m_{mean}}$} &  \tt{$\mathrm{\sigma_{\Delta m}}$} & \tt{$\mathrm{\Delta m_{mean}}$} &  \tt{$\mathrm{\sigma_{\Delta m}}$}  & \tt{$\mathrm{\Delta m_{mean}}$} &  \tt{$\mathrm{\sigma_{\Delta m}}$} \\
\rule[-1.0ex]{0pt}{1.0ex}  (mag) & \tt{$\mathrm{(mag)}$} &  \tt{$\mathrm{(mag)}$} & \tt{$\mathrm{(mag)}$} &  \tt{$\mathrm{(mag)}$} &  (mag) & \tt{$\mathrm{(mag)}$} \\
\hline
\rule[-1.0ex]{0pt}{1.0ex}  & \multicolumn{2}{c ||}{\tt{$\mathrm{(a)}$}} & \multicolumn{2}{c ||}{\tt{$\mathrm{(b)}$}} & \multicolumn{2}{c|}{\tt{$\mathrm{(c)}$}}\\
\hline
\hline \rule[-1.0ex]{0pt}{2.5ex}  19 - 20 & 0.074 & 0.005 & 0.058 & 0.007 & 0.057 & 0.023\\
 \rule[-1.0ex]{0pt}{1.0ex}  20 - 21 & 0.077 & 0.009 & 0.073 & 0.012 & 0.048 & 0.025\\
 \rule[-1.0ex]{0pt}{1.0ex}  21 - 22 & 0.076 & 0.027 & 0.093 & 0.024 & 0.028 & 0.046\\
 \rule[-1.0ex]{0pt}{1.0ex}  22 - 23 & 0.076 & 0.046 & 0.128 & 0.039 & -0.002 & 0.074\\
 \rule[-1.0ex]{0pt}{1.0ex}  23 - 24 & 0.087 & 0.065 & 0.216 & 0.052 & -0.034 & 0.091\\
 \rule[-1.0ex]{0pt}{1.0ex}  24 - 25 & 0.051 & 0.143 & 0.389 & 0.121 & -0.098 & 0.146\\
 \rule[-1.0ex]{0pt}{1.0ex}  25 - 26 & 0.147 & 0.195 & 0.749 & 0.143 & -0.145 & 0.151\\
\hline
\hline
\end{tabular}
\caption[The mean difference, and the standard deviation between Kron (part a), isophotal (part b), model (part c) and and input magnitudes as obtained by using SExtractor.]{The table reports, as a function of the magnitude bin (col.~1), the mean difference {\tt{$\mathrm{\Delta m_{mean}}$}} (columns~2, 4 and 6), and the standard deviation {\tt{$\mathrm{\sigma_{\Delta m}}$}} (columns~3, 5 and 7) between Kron (part a), isophotal (part b), model (part c) and and input magnitudes as obtained by using SExtractor.}
\label{comparison:tab4}
\end{table*}

\begin{table*}
\centering
\small
\setlength{\tabcolsep}{3pt}
 \begin{tabular}{|l c c | c c || c c | c c|}
 \hline
 \hline
\rule[-1.0ex]{0pt}{1.0ex}  \tt{Bin}
& \tt{$\mathrm{\Delta X_{mean}}$} &  \tt{$\mathrm{\sigma_{\Delta X}}$}
& \tt{$\mathrm{\Delta Y_{mean}}$} &  \tt{$\mathrm{\sigma_{\Delta Y}}$}
& \tt{$\mathrm{\Delta X_{mean}}$} &  \tt{$\mathrm{\sigma_{\Delta X}}$}
& \tt{$\mathrm{\Delta Y_{mean}}$} &  \tt{$\mathrm{\sigma_{\Delta Y}}$} \\
\rule[-1.0ex]{0pt}{1.0ex}  (mag)
& \tt{$\mathrm{(pixel)}$} &  \tt{$\mathrm{(pixel)}$}
& \tt{$\mathrm{(pixel)}$} &  \tt{$\mathrm{(pixel)}$}
& \tt{$\mathrm{(pixel)}$} &  \tt{$\mathrm{(pixel)}$}
& \tt{$\mathrm{(pixel)}$} &  \tt{$\mathrm{(pixel)}$} \\
\hline
\rule[-1.0ex]{0pt}{1.0ex}  & \multicolumn{4}{c ||}{\tt{$\mathrm{(a)}$}} & \multicolumn{4}{c| }{\tt{$\mathrm{(b)}$}}\\
\hline
\hline
 \rule[-1.0ex]{0pt}{2.5ex}  19 - 20 & 0.033  & 0.055 & 0.063 & 0.062 &  0.046 & 0.011 & 0.053 & 0.016\\
 \rule[-1.0ex]{0pt}{1.0ex}  20 - 21 & 0.032  & 0.045 & 0.041 & 0.059 &  0.041 & 0.019 & 0.070 & 0.054\\
 \rule[-1.0ex]{0pt}{1.0ex}  21 - 22 & 0.058  & 0.061 & 0.043 & 0.050 &  0.060 & 0.024 & 0.061 & 0.024\\
 \rule[-1.0ex]{0pt}{1.0ex}  22 - 23 & 0.033  & 0.068 & 0.048 & 0.058 &  0.023 & 0.072 & 0.061 & 0.047\\
 \rule[-1.0ex]{0pt}{1.0ex}  23 - 24 & 0.046  & 0.108 & 0.043 & 0.090 &  0.043 & 0.062 & 0.059 & 0.079\\
 \rule[-1.0ex]{0pt}{1.0ex}  24 - 25 & 0.033  & 0.209 & 0.026 & 0.202 &  0.026 & 0.132 & 0.055 & 0.138\\
 \rule[-1.0ex]{0pt}{1.0ex}  25 - 26 & 0.027  & 0.167 & 0.107 & 0.288 & -0.044 & 0.142 & 0.124 & 0.187\\
\hline
\rule[-1.0ex]{0pt}{1.0ex}  & \multicolumn{4}{c ||}{\tt{$\mathrm{(c)}$}} & \multicolumn{4}{c| }{\tt{$\mathrm{(d)}$}}\\
\hline
\hline
 \rule[-1.0ex]{0pt}{2.5ex}  19 - 20 &  0.050 & 0.005 & 0.050 & 0.004 & 0.051 & 0.005 & 0.050 & 0.003\\
 \rule[-1.0ex]{0pt}{1.0ex}  20 - 21 &  0.052 & 0.009 & 0.051 & 0.006 &  0.051 & 0.007 & 0.051 & 0.006\\
 \rule[-1.0ex]{0pt}{1.0ex}  21 - 22 &  0.052 & 0.009 & 0.053 & 0.014 &  0.054 & 0.009 & 0.053 & 0.014\\
 \rule[-1.0ex]{0pt}{1.0ex}  22 - 23 &  0.044 & 0.020 & 0.054 & 0.022 &  0.044 & 0.022 & 0.052 & 0.022\\
 \rule[-1.0ex]{0pt}{1.0ex}  23 - 24 &  0.043 & 0.048 & 0.046 & 0.053 &  0.044 & 0.044 & 0.049 & 0.052\\
 \rule[-1.0ex]{0pt}{1.0ex}  24 - 25 &  0.031 & 0.116 & 0.043 & 0.111 &  0.033 & 0.114 & 0.047 & 0.109\\
 \rule[-1.0ex]{0pt}{1.0ex}  25 - 26 & -0.096 & 0.149 & 0.097 & 0.189 & -0.007 & 0.186 & 0.107 & 0.269\\
\hline
\hline
\end{tabular}
\caption[The mean difference between DAOPHOT X,Y barycenter measure and input X, Y in the part a, while in the part b there are the mean difference between SExtractor X, Y barycenter measure and input X, Y. In parts c and d are reported the mean difference between X, Y PSF corrected and input measurements obtained by using DAOPHOT and SExtractor, respectively.]{The table reports, as a function of the magnitude bin (col.~1), the mean difference between DAOPHOT X (col.~2),Y (col.~4) barycenter measure and input X, Y with the relative standard deviation (columns~3 and 5) in the part a, while in the part b there are the mean difference between SExtractor X (col.~6), Y (col.~8) barycenter measure and input X, Y with the relative standard deviation (columns~7 and 9). In parts c and d are reported the mean difference between X (columns~2 and 6), Y (columns~4 and 8) PSF corrected and input measurements obtained by using DAOPHOT and SExtractor, respectively, with the relative standard deviation (columns~3, 5, 7 and 9).}
\label{comparison:tab5}
\end{table*}

\begin{table*}
\centering
\small
 \begin{tabular}{|l c c | c c|}
 \hline
 \hline
\rule[-1.0ex]{0pt}{1.0ex}  \tt{Bin} & \tt{$\mathrm{\Delta X_{mean}}$} &  \tt{$\mathrm{\sigma_{\Delta X}}$} & \tt{$\mathrm{\Delta Y_{mean}}$} &  \tt{$\mathrm{\sigma_{\Delta Y}}$} \\
\rule[-1.0ex]{0pt}{1.0ex}  (mag) & \tt{$\mathrm{(pixel)}$} &  \tt{$\mathrm{(pixel)}$} & \tt{$\mathrm{(pixel)}$} &  \tt{$\mathrm{(pixel)}$} \\
\hline
\hline
 \rule[-1.0ex]{0pt}{2.5ex}  19 - 20 & 0.050 & 0.004 & 0.050 & 0.004\\
 \rule[-1.0ex]{0pt}{1.0ex}  20 - 21 & 0.051 & 0.006 & 0.051 & 0.007 \\
 \rule[-1.0ex]{0pt}{1.0ex}  21 - 22 & 0.054 & 0.010 & 0.056 & 0.016\\
 \rule[-1.0ex]{0pt}{1.0ex}  22 - 23 & 0.032 & 0.036 & 0.054 & 0.025\\
 \rule[-1.0ex]{0pt}{1.0ex}  23 - 24 & 0.044 & 0.046 & 0.052 & 0.057\\
 \rule[-1.0ex]{0pt}{1.0ex}  24 - 25 & 0.028 & 0.127 & 0.040 & 0.120\\
 \rule[-1.0ex]{0pt}{1.0ex}  25 - 26 & -0.043 & 0.124 & 0.134 & 0.173\\
\hline
\hline
\end{tabular}
\caption[The mean difference between X and Y windowed and input measurements as estimated by SExtractor with the relative standard deviation.]{The table reports, as a function of the magnitude bin (col.~1), the mean difference between X (col.~2) and Y (col.~4) windowed and input measurements as estimated by SExtractor with the relative standard deviation (columns~3 and 5).}
\label{comparison:tab6}
\end{table*}

\subsection{Non uniform star distribution}
\label{comparison:6}
In order to evaluate the performances of both software in crowded fields, we tested them on a simulated image with a non uniform stellar distribution, i.e., showing an overdensity of stars in the center. In the left panel of Fig.~\ref{comparison:fig10}, we plot the spatial distribution of the simulated stars, while the stellar density of the field, as a function of the distance from the center, is shown in the right panel.
\begin{figure*}
\centering
\subfloat{
\includegraphics[width=12 cm ]{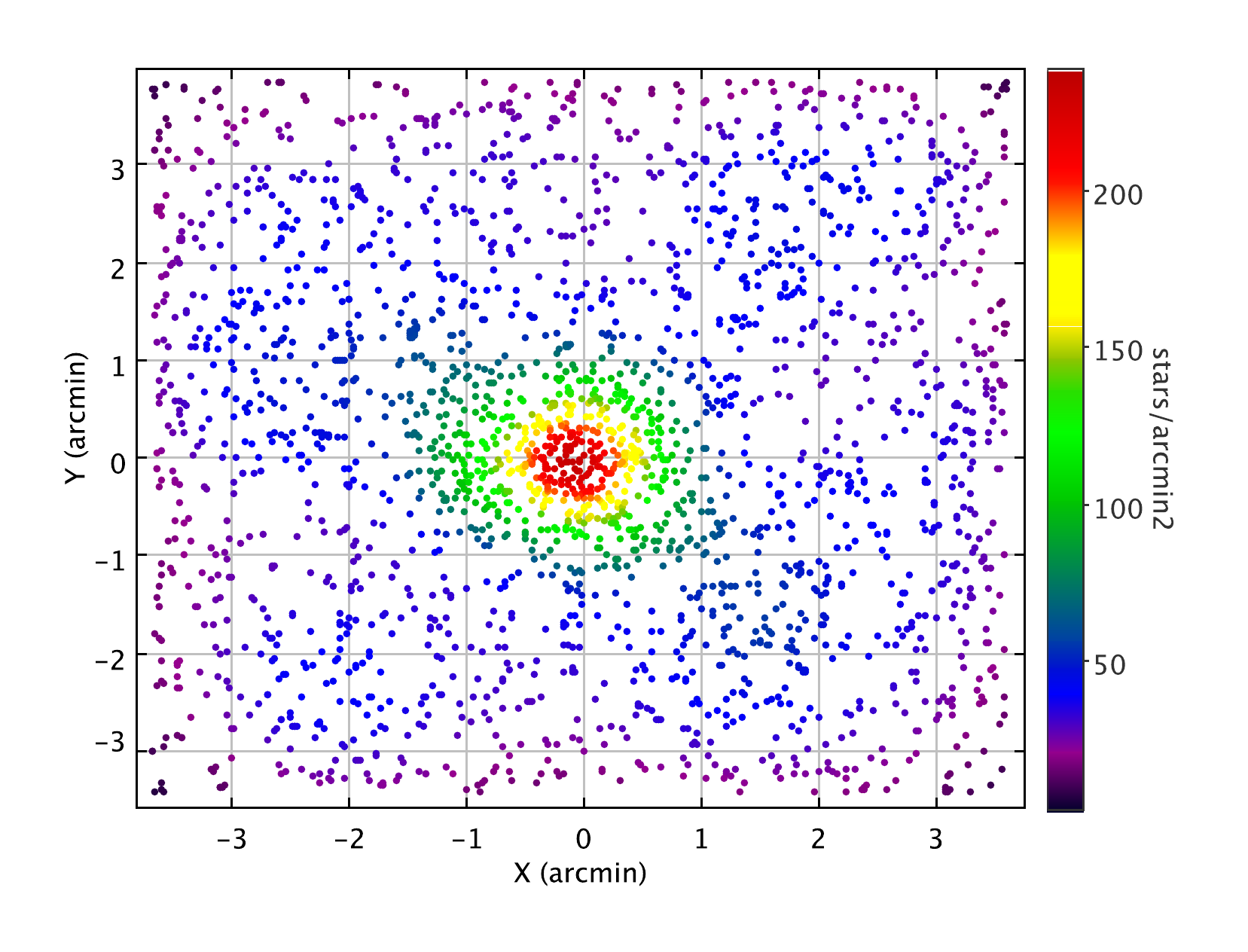}
\label{comparison:fig10a}}
\\ \subfloat{
\includegraphics[width=10 cm]{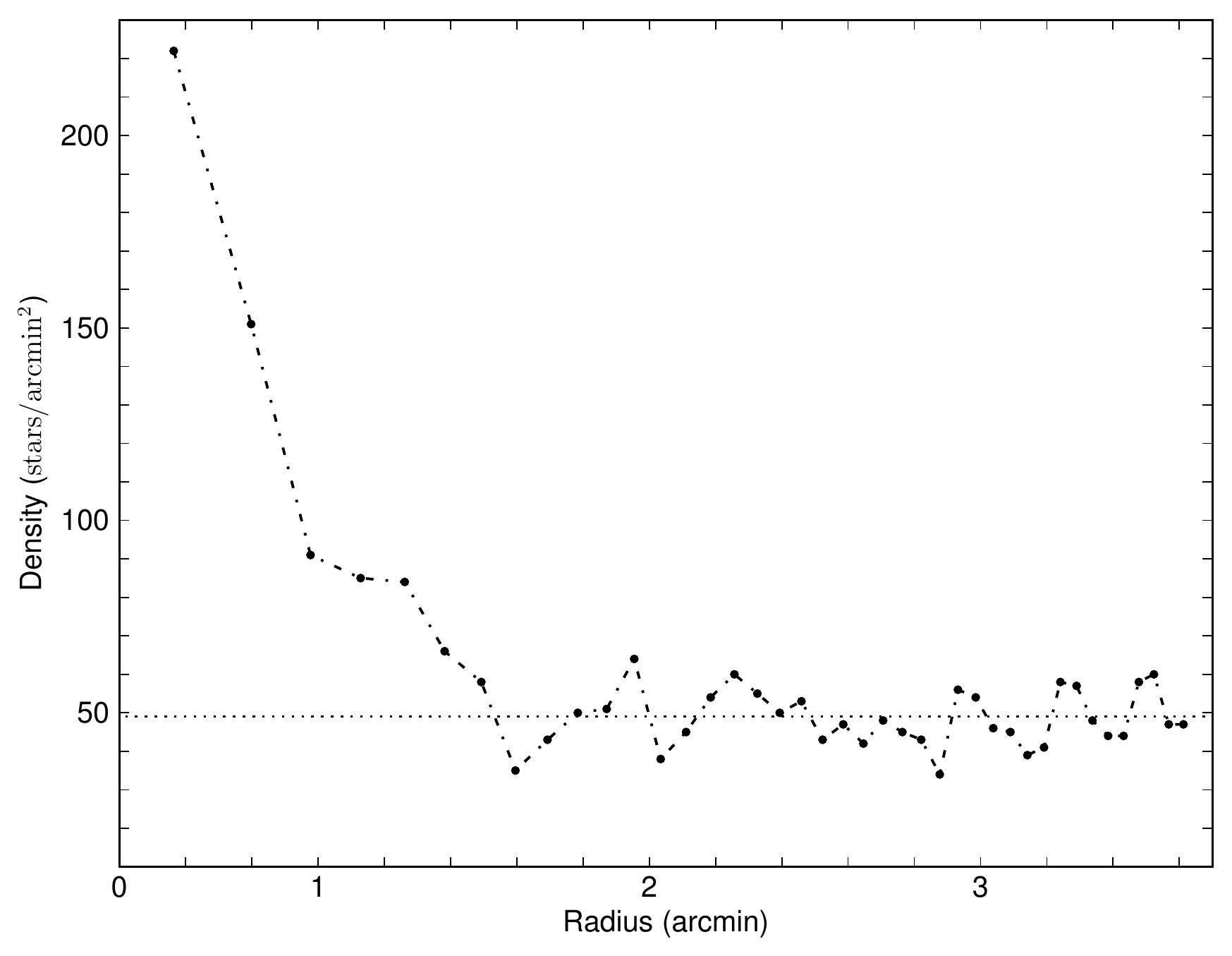}
\label{comparison:fig10b}}
\caption[Spatial distribution of input stars for the over-dense stellar field and numerical stellar density.]{\textit{Upper panel}: Spatial distribution of input stars for the over-dense stellar field. Colors of points refer to the surface number density of stars. Density values are reported in the vertical color-bar. \textit{Lower panel}: Numerical stellar density as a function of the distance from the center. Dotted line represents the average stellar density ($\sim$ 40 stars/arcmin$\mathrm{^2}$) obtained excluding the circular region within a radius of 1.5 arcmin from the center. }
\label{comparison:fig10}
\end{figure*}

In order to carry out the test, we fixed the values for DAOPHOT, ALLSTAR, SExtractor and PSFEx as in Sect.s~\ref{comparison:4.1} and \ref{comparison:4.2}.\\
The obtained results were evaluated by following the same criteria as in Sect.~\ref{comparison:5}. For all tests, we excluded saturated stars, which are stars with magnitude brighter than B = 17 mag.\\
Also, in case of a non uniform stellar field, we obtain that, for point-like sources, the final depth returned by DAOPHOT is $\sim$ 2 mag brighter than those produced by SExtractor. In fact, SExtractor can recover the input star catalog, resulting in a photometric depth of B = 25 mag, while for DAOPHOT the resulted limit in magnitude is around 23 mag.\\
Moreover, by considering the reliability of the extracted catalogs, it is possible to separate stars and galaxies down to the magnitude limit for each software package: B = 25 mag for SExtractor and B = 23 mag for DAOPHOT.\\
In order to not bias the comparison of photometric measurements, we consider only the input stellar sources recovered by both software packages.
As a first step, we analyze results for aperture photometry. Comparing the values of the mean and the standard deviation of the difference between aperture and input magnitudes with those reported in Tab.~\ref{comparison:tab3}, we obtained a decline in the quality of the results, in particular, those produced by SExtractor. In fact, in the magnitude bin 22-23, we go from $\mathrm{\Delta m} \pm \sigma_{\Delta m}$ = -0.009 $\pm$ 0.059 mag in the case of non-crowded field, to -0.165$\pm$ 0.287 mag for SExtractor and from $\mathrm{\Delta m} \pm \sigma_{\Delta m}$ -0.020 = 0.075 $\pm$ mag to -0.487 $\pm$ 0.474 for DAOPHOT.\\
By using PSF-fitting photometry, there is an improvement in the determination of magnitude with respect to aperture photometry, but the results show a quality decline as compared to the case of less crowded stars.\\
By considering the last bin, for DAOPHOT PSF-fitting photometry, the results change from $\mathrm{\Delta m} \pm \sigma_{\Delta m}$ = -0.023 $\pm$ 0.013 mag to -0.079 $\pm$ 0.080 mag. While, by using SExtractor PSF-fitting photometry, we pass from $\mathrm{\Delta m} \pm \sigma_{\Delta m}$ = -0.003 $\pm$ 0.014 mag to -0.063 $\pm$ 0.110 mag. \\
Finally, we compare the results among those obtained with barycenter and PSF corrected centroids and those reported in Tab.~\ref{comparison:tab5}.
In the last bin of magnitude, DAOPHOT measurements of PSF centroids go from $\mathrm{\Delta X} \pm \sigma_{\Delta X}$; $\mathrm{\Delta Y} \pm \sigma_{\Delta Y}$ = -0.044 $\pm$ 0.020;  -0.054 $\pm$ 0.022 pixel to $\mathrm{\Delta X} \pm \sigma_{\Delta X}$; $\mathrm{\Delta Y} \pm \sigma_{\Delta Y}$ = -0.045 $\pm$ 0.146;  -0.040 $\pm$ 0.148 pixel.\\
On the other hand, in the same bin, SExtractor measurements of PSF centroids go from $\mathrm{\Delta X} \pm \sigma_{\Delta X}$; $\mathrm{\Delta Y} \pm \sigma_{\Delta Y}$ = -0.044 $\pm$ 0.022;  -0.052 $\pm$ 0.022 pixel to $\mathrm{\Delta X} \pm \sigma_{\Delta X}$; $\mathrm{\Delta Y} \pm \sigma_{\Delta Y}$ = -0.063 $\pm$ 0.133;  -0.055 $\pm$ 0.150 pixel.\\
On the basis of this comparison, we can conclude that, as shown in Sect~\ref{comparison:5}, the results obtained by both packages, in the case of an overdense region, are completely equivalent both for PSF fitting photometry and PSF fitting determination of centroids, and with regard of the reliability of the extracted catalog. With SExtractor, however, we can optimize the detection of the sources to recover the whole input catalog.

\subsection{Summary}
\label{comparison:7}

The advent of new photometric surveys and the need to deal with fainter sources have led to an increase in the demand for quality and accuracy of photometric measurements. Moreover, the analysis of massive data sets needs a single software tool in order to minimize the number of required reprocessing steps.
Therefore, a crucial aspect in this context is the choice of source detection software package. An important innovation was introduced since 2010 with the public release of PSFEx, which can work in combination with SExtractor. This new software package has deeply improved the performances of SExtractor, filling the gap against other available procedures, such as DAOPHOT and PHOTO, concerning PSF photometry. \\
In the present work, for the first time, the performances of DAOPHOT and SExtractor are compared to check the quality and accuracy of extraction procedures with respect to the requirements of modern wide surveys. The software packages are tested on two kinds of B-band simulated images having, respectively, a uniform spatial distribution of sources and an overdensity in the center.\\
In the first case, by considering only the number of extracted sources, it appears that the limiting magnitude for the extracted catalog is extremely low, in particular for DAOPHOT is B = 22 mag. If we limit to consider only stellar sources, the photometric depth is improved down to 24 mag for DAOPHOT and 26 mag for SExtractor. This could be related to the fact that only SExtractor gives the possibility to choose the filter for catalog extraction, according to the image characteristics (see Sect.~\ref{comparison:4.2}). In fact, as shown in Sect~\ref{comparison:5.1}, the filter choice affects the photometric depth of the extracted catalog.\\
A relevant aspect of the catalog extraction is the capability to discriminate between extended and point-like sources. As we have seen, within the different software packages, there are various methods to perform the star/galaxy classification. In particular the sharpness parameter available in DAOPHOT, improved by using ALLSTAR, returns a reliable star/galaxy classification down to the photometric depth of the catalog (B = 24 mag). All the traditional methods available in SExtractor, instead, limit the star/galaxy classification, at least one magnitude above the completeness magnitude of the catalog. The new parameter $\mathrm{\tt{SPREAD\_MODEL}}$, which is a discriminant between the best fitting local PSF and a more extended model, has largely improved the classification, allowing to separate extended and point-like sources down to the completeness limit of the catalog (B = 26 mag for stellar sources).\\
Since DAOPHOT is mainly designed to perform stellar photometry, in order to not bias the comparison of photometric measurements, we considered only input stellar sources recovered by both software packages.
Both software tools are able to deliver acceptable performances in both aperture (with a $\mathrm{\sigma_{\Delta m}}\, <$ 0.2 mag) and PSF photometry (with a $\mathrm{\sigma_{\Delta m}}\, <$ 0.03 mag), down to B = 24 mag, the completeness limit of the DAOPHOT catalog.\\
Moreover, since SExtractor allows us to derive different estimates of the total magnitudes of sources, which we can also compare among themselves: Kron, isophotal and model magnitudes. The isophotal magnitude is highly dependent on the detection threshold. In fact, in the [23-24] magnitude bin, there is a higher shift of $\mathrm{\Delta m}$ (0.216 mag) than in other magnitudes. The Kron magnitude yields $\sim$ 94\% of the total source flux within the adaptive aperture (\citealp{bertin1996}). Accordingly, we find a shift of $\sim$ 0.07 mag even in the brightest magnitude bin. The model magnitude results a good estimate of the input magnitude also for stars, with an error of 0.091 mag in the [23 - 24] magnitude bin.  \\
An accurate determination of the object's centroids is crucial for the relative astrometry and thus, also for matching sources in different bands. Both software packages show a bias between output centroids and input X and Y coordinates $\le$ 0.01 arcsec, with an average deviation of $\le$ 0.02 arcsec down to B = 24 mag. These values are improved in terms of average deviation ($\sigma_{\Delta X(Y)} \le$ 0.01 arcsec), when PSF correction is applied. So, we can conclude that these results are satisfactory in both cases.\\
We also tested both software packages on simulated images with a non uniform stellar distribution, i.e., showing an overdensity of stars in the center. By analyzing the results obtained in this last case, we can confirm the conclusions described above. In particular, DAOPHOT provides a catalog $\sim$ 2 mag shallower than the one extracted by SExtractor.
On the other hand, by analyzing the extracted catalogs in terms of the mean difference and the standard deviation among output and input magnitudes and centroids, we notice a decline in the quality of the results for both software packages, with respect to the case of a uniform spatial distribution of stars.
Finally, DAOPHOT and ALLSTAR provide very accurate and reliable PSF photometry, with a robust star-galaxy separation. However, it is not useful for galaxy characterization.
On the other hand SExtractor, associated with PSFEx, turns competitive in terms of PSF photometry. It returns acceptable aperture photometry and accurate PSF modeling also for faint sources. The windowed centroids are as good as PSF centroids. Moreover, SExtractor allows to go very deep in source detection through a properly choice of image filtering masks. The deblending model is very extensible, and the use of neural networks for object classification, plus the novel {\tt{SPREAD\_MODEL}} parameter, push down to the limiting magnitude the capability of star/galaxy separation. Considering that SExtractor returns accurate photometry also for galaxies, we can conclude that the new version of SExtractor, used in combination with PSFEx, represents a very powerful software tool for source extraction, with performances comparable to DAOPHOT also for overdense stellar fields. In the next years, it will be important to extensively test SExtractor plus PSFEx on real crowded stellar fields in order to definitively assess the performances of this software tool. However, an important aspect for the use of PSFEx and SExtractor, that we cannot avoid to mention, is the processing time. Without considering problems, such as degradation in performances during periods of heavy disk access, on average, SExtractor requires 0.5 $s$ per detection to perform PSF photometry and source modeling, by using one single CPU with 6GB of RAM. This suggests that currently, the only disadvantage of using SExtractor and PSFEx on wide field images is the processing time. However, on the other hand, although DAOPHOT is more efficient in terms of processing time just for the calculation, it requires more time if the user visually inspects the modeled PSF stars.

    \section{Globular Cluster Classification}\label{chap:GC}
        \blfootnote{this section is largely extracted from: \tiny
\begin{itemize}
			\item Brescia, M.; \textbf{Cavuoti, S.}; Paolillo, M.; Longo, G.; Puzia, T.; 2012, The detection of Globular Clusters in galaxies as a data mining problem, \textbf{MNRAS, Volume 421, Issue 2, pp. 1155-1165}, available at \href{http://arxiv.org/abs/1110.2144v1}{arXiv:1110.2144v1}.
			\item \textbf{Cavuoti, S.}; Garofalo, M.; Brescia , M.; Paolillo, M.; Pescape', A.; Longo, G.; Ventre, G.; GPUs for astrophysical data mining. A test on the search for candidate globular clusters in external galaxies, \textbf{Submitted to New Astronomy, accepted}
			\item  \textbf{Cavuoti, S.}; Garofalo, M.; Brescia, M.; Pescape', A.; Longo, G.; Ventre, G., 2012, Genetic Algorithm Modeling with GPU Parallel Computing Technology, 22nd WIRN, Italian Workshop on Neural Networks, Vietri sul Mare, Salerno, Italy, May 17-19 \textbf{refereed proceeding}
\end{itemize}}
In this paragraph I present an application of self-adaptive supervised learning classifiers derived from the Machine Learning
paradigm, to the identification of candidate Globular Clusters in deep, wide-field, single band HST images.
Several methods provided by the DAME web application, were tested and compared
on the NGC1399 HST data described in \citet{paolillo2011}.
The best results were obtained using a Multi Layer Perceptron with Quasi Newton learning rule which achieved
a classification accuracy of $98.3 \%$, with a completeness of $97.8\%$ and $1.6\%$ contamination. An extensive set of experiments revealed that
the use of accurate structural parameters (effective radius, central surface brightness) does improve the final result, but only by $\sim 5 \%$.
It is also shown that the method is capable to retrieve also extreme sources (for instance, very extended
objects) which are missed by more traditional approaches.

%
In this work I used a variety of methods provided by DAMEWARE for the identification of Globular Clusters (GCs) in the galaxy NGC1399 using single band photometric
data obtained with the Hubble Space Telescope (HST).

The identification and physical characterization of Globular Cluster (GC) populations in external galaxies is
of interest to many astrophysical fields: from cosmology, to the evolution of star clusters and galaxies, to the formation
and evolution of binary systems.
The identification of Globular Clusters in external galaxies usually requires the use of wide-field, multi-band
photometry since in galaxies located more than a few Mpc away they appear as unresolved sources
in ground-based astronomical images and are thus hardly distinguishable from background galaxies which introduce
significant contamination problems.
For such reason, GCs are traditionally selected using methods based on their colors and luminosities.
However, in order to minimize contamination and to measure GC properties such as sizes and structural parameters
(core radius, concentration, etc.), high-resolution data are required as well which,  for star clusters outside the Local Group, are
available only through the use of space facilities (i.e. Hubble Space Telescope, HST).
Obtaining suitable HST data is however challenging in terms of observing time since the optimal datasets should
be: i) deep, in order to sample the majority of the GC population and ensure the high S/N required to measure
structural parameters \citep[see e.g.][]{carlson2001}; ii) with wide-field coverage, in order to minimize projection
effects as well as to study the overall properties of the GC populations, which often differ from those inferred
from observations of the central region of a galaxy only; and iii) multi-band, to effectively select GC based on
colors.

It is apparent that, in order to reduce observing costs, it would be much more effective to use single-band HST
data.
Such approach however requires to carefully select the candidate GCs based on the available photometric and
morphological parameters in order to avoid introducing biases in the final sample (see below).
Here I intend to show that the use of properly tuned DM algorithms can yield very complete datasets with low
contamination even with single band photometry, thus minimizing the observing time requirements and allowing to
extend such studies to larger areas and to the outskirts of nearby galaxies.

\subsection{The data}\label{gc:thedata}
The dataset used in this experiment consists of wide field HST observations of the giant elliptical NGC1399 located
at the heart of the Fornax cluster.
This galaxy represents an ideal test case since, due to its distance \citep[$D=20.13\pm0.4$ Mpc, see][]{dunn2006}, it is possible to cover a large fraction
of its GC system (out to $>5 R_e$) with a limited number of observations.
Furthermore, it is particularly challenging because, at this distance, GCs are only marginally resolved even by HST; in fact at NGC~1399 distance 1 ACS pixel corresponds to 2.93 pc (1" = 97.7 pc).
This dataset was used by \citet{paolillo2011} to study the GC-LMXB connection and by Puzia et al. (2012, in preparation) to study the structural properties of the GC population.
We summarize below the main properties of the dataset, and refer to these works for a more detailed description of the observations and of data analysis.

The optical data were taken with the HST Advanced Camera for Surveys (ACS, program GO-10129, PI T.Puzia), in the
F606W (broad $V$ band) filter, with integration time of 2108 seconds for each field. The observations were arranged in a 3x3 ACS mosaic,
and combined into a single image using the MultiDrizzle routine \citep{koekemoer2003}.
The final scale of the images is 0.03"/pix, providing Nyquist sampling of the ACS PSF.
The field of view of the ACS mosaic covers ~100 square arcmin (Figure \ref{gc:FOV}) extending out to a projected galactocentric
distance of $\sim 55$ kpc, i.e. $4.9 r_e$ of the GC system ($\sim 5.7 r_e^{gal}$).
The source catalog was generated with \textsc{SExtractor} by imposing a minimum area of 20 pixels: it contains 12915 sources and
reaches $7\sigma$ detection at $m_V=27.5$, i.e. 4 mag below the GC luminosity function turnover, thus allowing to sample the
entire GC population (see Figure \ref{gc:maghisto}).
The catalog astrometric solution was registered to the USNO-B1 reference frame, obtaining a final accuracy of 0.2" r.m.s.

For 4239 sources we were able to measure structural parameters (which require very high S/N, see \citealt{carlson2001} and Puzia et al. 2012), fitting King surface brightness profile models with the \textsc{Galfit} software \citep{peng2002}, and deriving tidal, core, effective radii
and central surface brightness values for each cluster. The accuracy of these measurements was estimated simulating several thousand artificial GCs with the \textsc{Multiking} code\footnote{available at: \url{http://people.na.infn.it/paolillo/Software.html}} specifically written to account for ACS field distortion, PSF variation, dithering pattern \citep[Puzia et al. 2012]{paolillo2011}.

\begin{figure}
\centering
 \includegraphics[width=12cm]{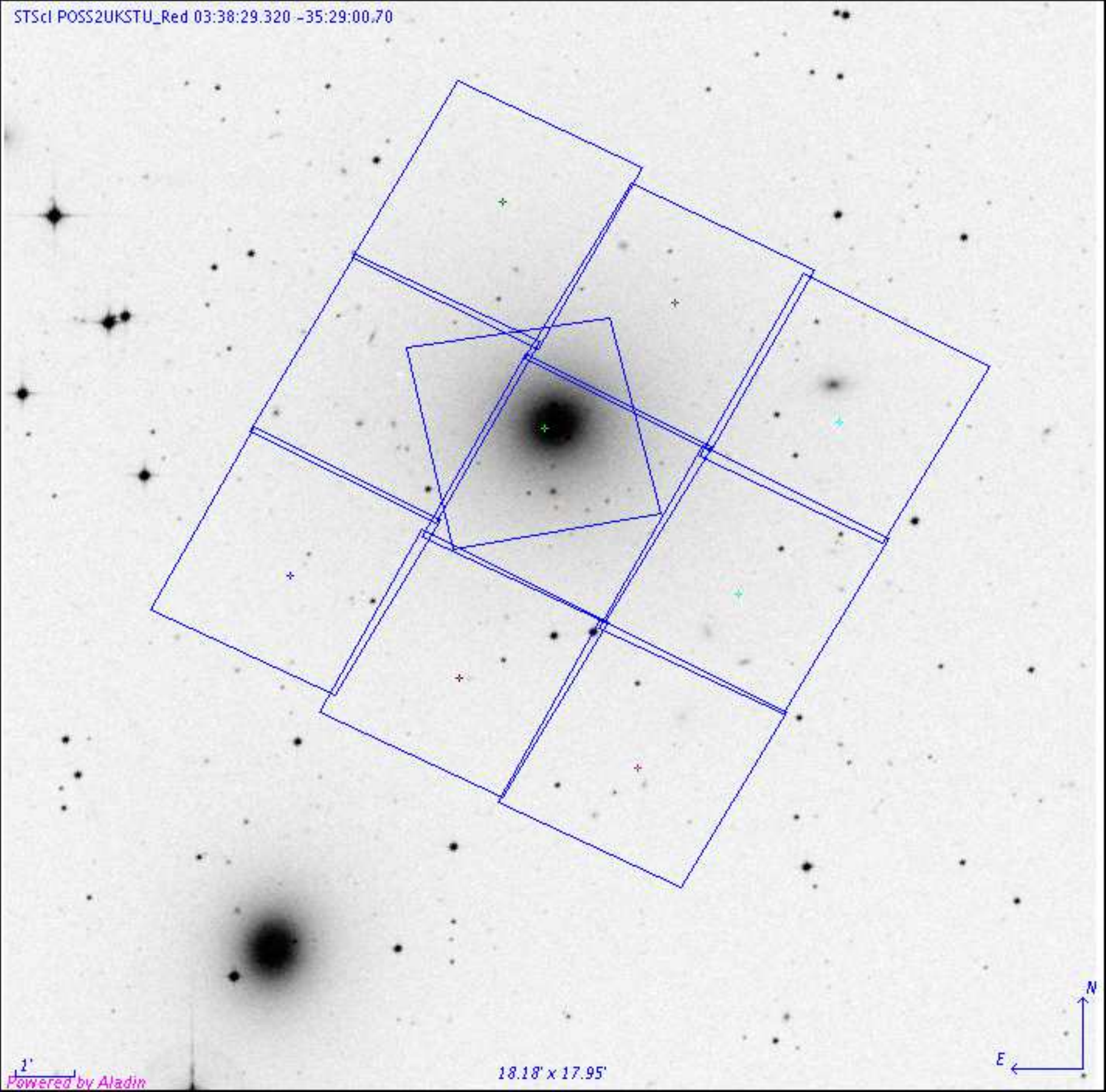}
 \caption[The field of view covered by the 3x3 HST/ACS mosaic. The central field shows the region covered by previous archival ASC observations.]{The field of view covered by the 3x3 HST/ACS mosaic in the F606W band. The central field, with a different orientation, shows the region covered by previous archival ASC observations in $g$ and $z$ bands.}
 \label{gc:FOV}
\end{figure}

The NGC1399 region covered by our mosaic lacks color information for all HST F606W sources.
In this work I make use of two ancillary multiwavelength datasets: archival HST $g-z$ observations \citep{kundu2005}, which cover the very central region of the galaxy ($10\%$ of the sample, see Figure \ref{gc:FOV}), and
$C-T1$ ground based photometry from \cite{bassino2006}, covering the whole mosaic.
The latter is only available for $\sim  14\%$ of our sources, and due to background light contamination it is very
incomplete in the proximity of the galaxy center. In total 2740 sources of the catalog have multi-band (either $g-z$ or $C-T1$) photometry.

Finally, the subsample of sources used to build our Knowledge Base (KB) to train the DM algorithms, is composed by the 2100 sources with all photometric and morphological informations: isophotal magnitude, kron radius, aperture magnitudes within a 2, 6 and 20 pixels (corresponding to $0.06^{\prime\prime}$, $0.18^{\prime\prime}$, and $0.6^{\prime\prime}$) diameter, ellipticity, position angle, FWHM, \textsc{SExtractor} stellarity index, King's tidal and core radii, effective radii, central surface brightness, and either $g-z$ or $C-T1$ color. The magnitude distribution of such subsample is shown in Figure \ref{gc:maghisto} as a dashed line.

The typical choice to select GCs based on multi-band photometry would be to adopt the magnitude and color cuts
reported in Table \ref{gc:col_sel}, and highlighted in Figure \ref{gc:col_mag} with a dashed line; the magnitude limit $z<22.5$ does not exploit the full depth of the HST data but is adopted to be consistent with the $T1<23$ limit used for the ground-based colors, thus ensuring a uniform limit across the whole field-of-view. In the following we thus assume that bona-fide GCs are represented by such sources, in order to explore how well different selection methods based on single band photometry are able to retrieve the correct population of objects. The F606W magnitude distribution of color-selected GCs is shown in Figure \ref{gc:maghisto} as a black solid line.

\begin{figure}
\centering
\includegraphics[width=12cm]{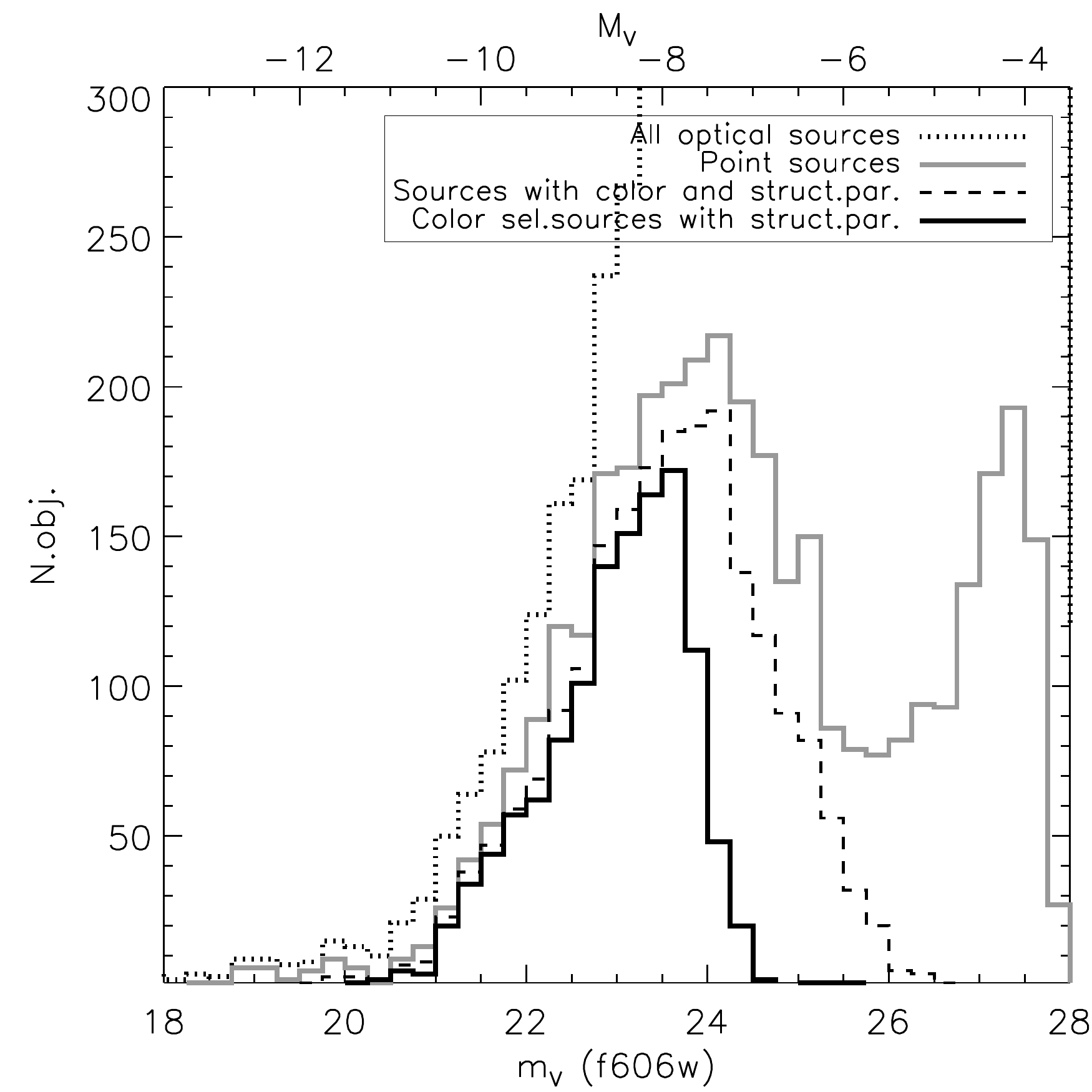}
\caption[Luminosity distributions of all detected sources within the HST FOV]{Luminosity distributions of all detected (dotted line) and point-like (e.g. with stellarity index $>0.9$, solid gray line) sources§ within the HST FOV. Also shown are the two additional subsamples discussed in section \ref{gc:thedata} and section \ref{gc:exp}: the KB composed of sources with both color and structural parameters (dashed line), and the subset of bona-fide color-selected GCs based on Table \ref{gc:col_sel} (solid black line).}
\label{gc:maghisto}
\end{figure}

\begin{table}
\centering
\caption{Photometric selection criteria for GC candidates}
\label{gc:col_sel}
\begin{tabular}{lcc}
\hline
 & color cut & magnitude cut\\
\hline
Ground-based & $1.0\leq C\!-\!T1<2.2$ & $T1<23$\\
data &  & \\\\
HST data & $1.3\leq g\!-\!z<2.5$ & $z<22.5$ \\
\hline
\end{tabular}
\end{table}

\begin{figure*}
\centering
\includegraphics[width=8.3cm]{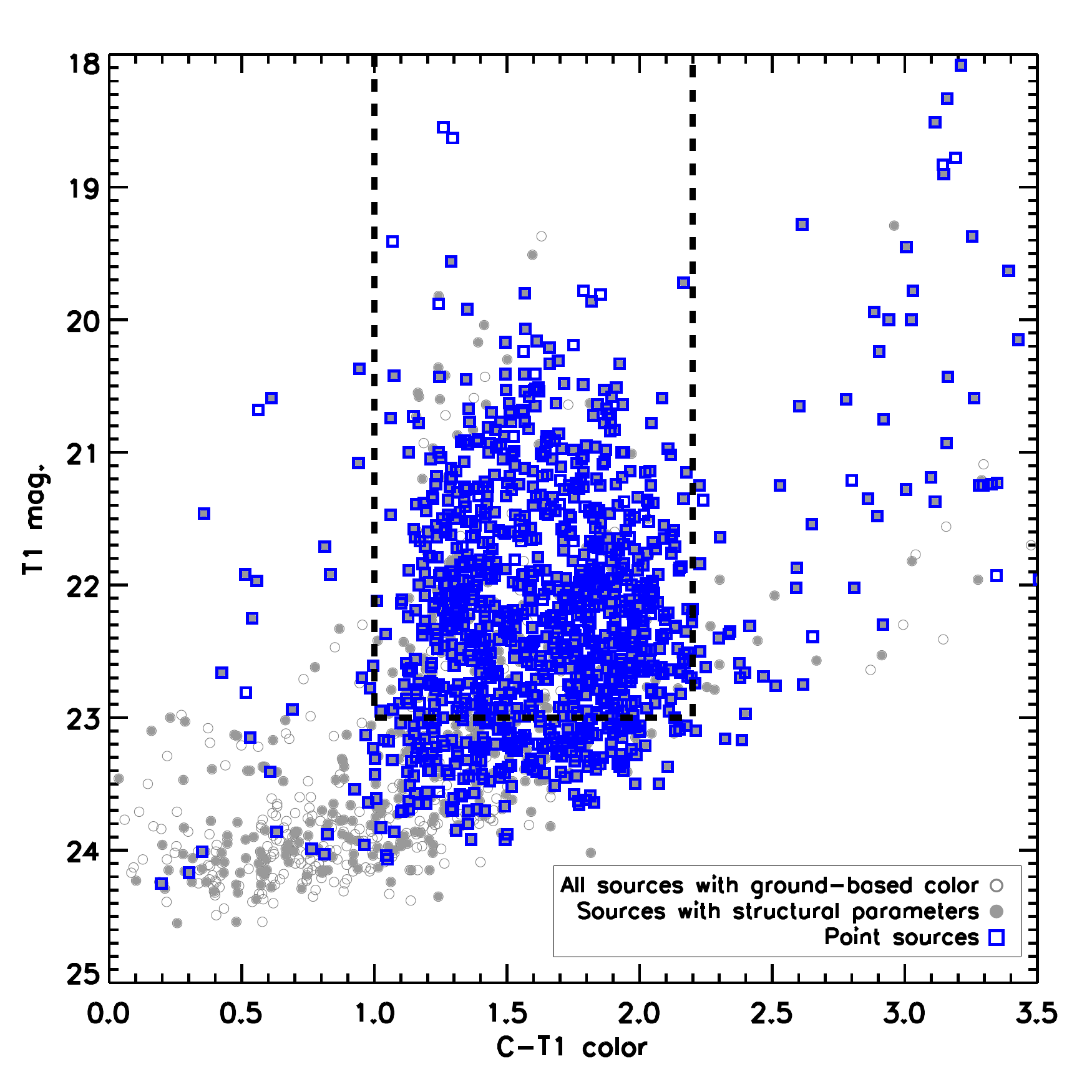}\\
\includegraphics[width=8.3cm]{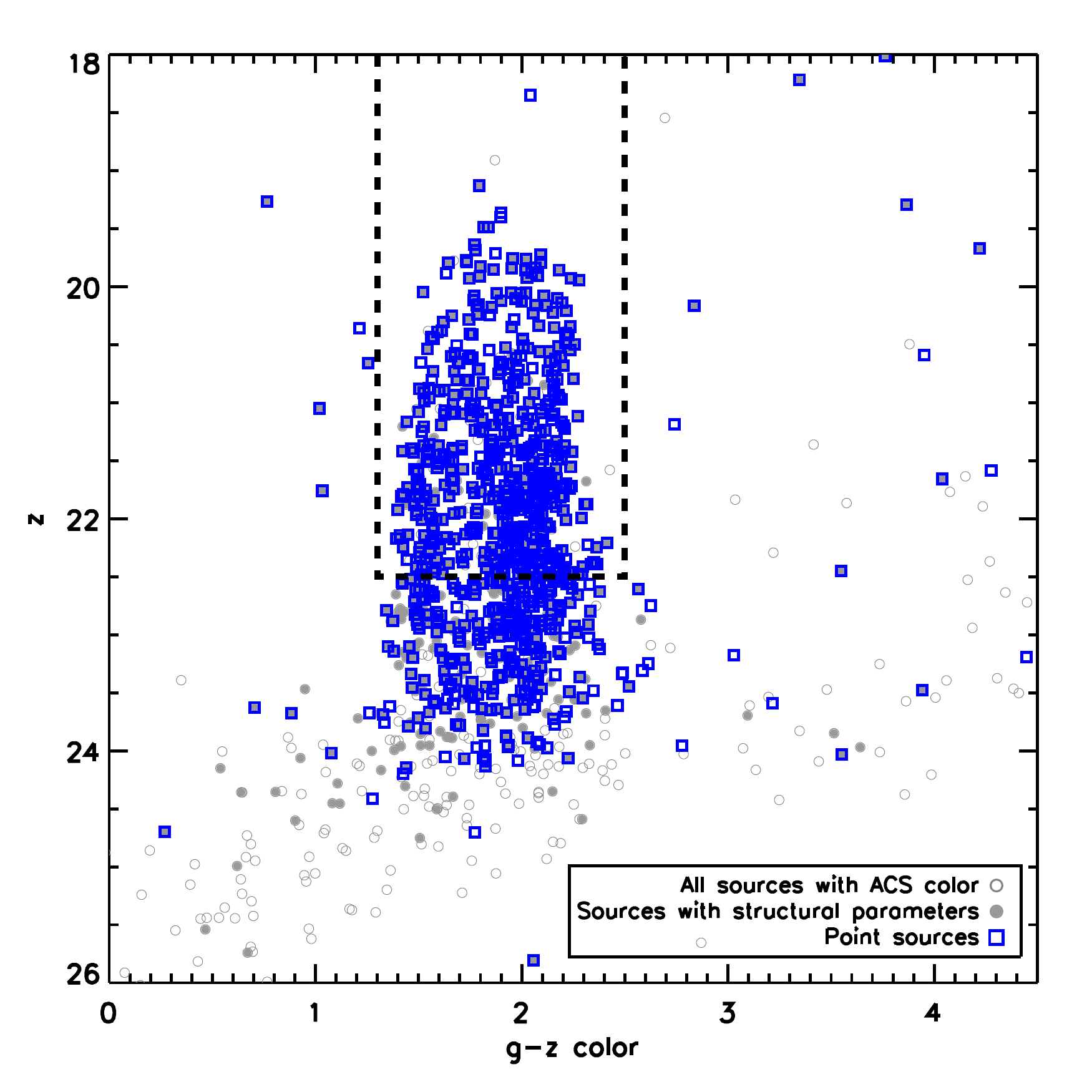}
\caption[Color-magnitude diagrams using $C\!-\!T1$ ground-based (left panel) and $g\!-\!z$ HST photometry (right panel).]{Color-magnitude diagrams using $C\!-\!T1$ ground-based (left panel) and $g\!-\!z$ HST photometry (right panel). Ground-based photometry covers the whole FOV of our ACS mosaic, while HST colors are limited to the central ACS field ($\sim\!200"\!\times\!200"$, Figure \ref{gc:FOV}). Open grey dots represent all sources in color catalogs while solid ones refer to the subsample with both color and structural parameters that represents our Knowledge Base. Blue squares mark pointlike sources, i.e. sources with stellarity index $>0.9$, while the dashed line highlights the parameter space (Table \ref{gc:col_sel}) used to select bona-fide GC.}
\label{gc:col_mag}
\end{figure*}

\subsection{Results}\label{gc:exp}

\begin{sidewaystable*}
\centering
\begin{tabular}{l|lrccccc}
\hline
type of experiment &	 missing features & figure of merit &	MLPQNA & 	GAME & 	SVM  &	MLPBP & 	MLPGA \\
\hline
complete patterns  & --	 \\
&                   & \textit{class.accuracy}		    &    \textbf{98.3} &	82.1 &	90.5 &	59.9 &	66.2\\
 &                  & \textit{completeness}		     &    \textbf{97.8} &  73.3 &  89.1 &  54.1 &  61.4\\
  &                 & \textit{contamination}		      &     \textbf{1.8} &  18.7 &   7.7 &  42.2 &  35.1\\
\hline
no par. 11         & 11 \\
&                 & \textit{class.accuracy}		    &    \textbf{98.0} &	81.9 &	90.5 &	59.0 &	62.4\\
  &                 & \textit{completeness}		      &    \textbf{97.6} &  79.3 &  88.9 &  56.1 &  62.2\\
   &                & \textit{contamination}		       &     \textbf{1.6} &  19.6 &   7.9 &  43.1 &  38.8\\
\hline
only optical &   8, 9, 10, 11      \\
& 	    & \textit{class.accuracy}		       &    93.9 &  86.4 &  90.9 &  70.3 &  76.2\\
  &                 & \textit{completeness}		      &    91.4 &  78.9 &  88.7 &  54.0 &  65.1\\
   &                & \textit{contamination}		       &     5.9 &  13.9 &	 8.0 &  33.2 &  24.6\\
\hline
mixed      &  5, 8, 9, 10, 11        \\
& 	    & \textit{class.accuracy}		     &    94.7 &  86.7 &  89.1 &	68.6 &  71.5\\
  &                 & \textit{completeness}		      &    92.3 &  81.5 &  88.6 &  52.8 &  63.8\\
   &                & \textit{contamination}		       &     5.0 &  16.6 &   8.1 &  37.6 &  30.1\\
\hline
\end{tabular}
\caption[Summary of the performance of the five classifiers.]]{Summary of the performances (in percentage) of the five classifiers.
For each entry the first line refers to the classification accuracy,
while the second and third refer to completeness and contamination, respectively.}
\label{gc:comp}
\end{sidewaystable*}

As discussed in the Introduction, the purpose of this work was to implement an alternative, DM based, method to
select globular clusters in single band HST images, thus saving the observing time needed to obtain complete sets
of multiband data.
In this subsection we shortly summarize the results of the series of (numerical) "experiments" which were performed to determine the best model and the best combination of features, while in next subsection
we discuss the overall properties of the sample obtained with the DM algorithms, in comparison with traditional selection methods.

Terms like completeness, contamination, accuracy etc. are differently defined by astronomers and ``data miners".
In what follows we use the following definitions. Classification accuracy: fraction of patterns (objects) which are correctly classified (either GCs or non-GCs) with respect
to the total number of objects in the sample;
completeness: fraction of GCs which are correctly classified as such;
contamination: fraction of non-GC objects which are erroneously classified as GCs.

All experiments were performed on the KB sample presented in section \ref{gc:thedata}, assuming that bona-fide GCs are represented by
sources selected according to the color cuts in Table \ref{gc:col_sel}. We used as features the following quantities:

\begin{itemize}
\item The isophotal magnitude (Feature 1). 	
\item Three aperture magnitudes (features 2--4) obtained through  circular apertures of radii 2, 6, and 20 arcsec respectively;
\item	The Kron radius, the ellipticity and the FWHM of the image (features 5-7);
\item The structural parameters	(features 8-11) which are, respectively, the central surface brightness, the	core radius, the effective radius and the tidal radius.
\end{itemize}

By making an exhaustive pruning test on all 11 dataset parameters, with the 5 machine learning models previously introduced,
we collected a total of 425 experiments (85 per model). The details of the experiment setup can be found in Appendix \ref{gc:AppII}.

Table \ref{gc:comp} summarizes the most relevant results: in terms of classification accuracy and completeness, the best results (98.3\% and 97.8\% respectively) are obtained by MLPQNA using all parameters; using all available features but the number 11 (the tidal radius) we obtain marginally worse results, as can be expected given the high noise present in this last parameter, which is affected by the large background due to the host galaxy light. In terms of contamination comparable results ($\lesssim 2\%$) are obtained with the same model both with or without feature 11. We point out that since the experiment without feature 11 provides results comparable to the one using all features, but requires less information and is less computationally demanding, we consider the latter to be the case providing the highest overall performance, as usually done in DM experiments. In other words the experiment without feature 11 represents the best compromise between required overall performance and complexity of the KB.

The best result obtained without using the structural parameters is 93.9\% (classification accuracy) thus indicating that
the availability of detailed structural parameters does indeed help to improve the results, but only by $ \sim 5\%$.
Moreover, the pruning in the mixed cases (by excluding some structural and optical features) revealed a similar
behavior in all models, in terms of quantity of correlated information introduced by individual features in the patterns.
Five optical features (namely the isophotal and aperture magnitudes and the FWHM of the image) were recognized as the most relevant by
all models.
Among the structural parameters, the central surface brightness and the core radius were recognized as relevant by all models but the SVM and MLPGA models.
In all other cases, other residual optical and structural parameters were evaluated low carriers of correlated information.

\begin{figure*}
\centering
\includegraphics[width=8.3cm]{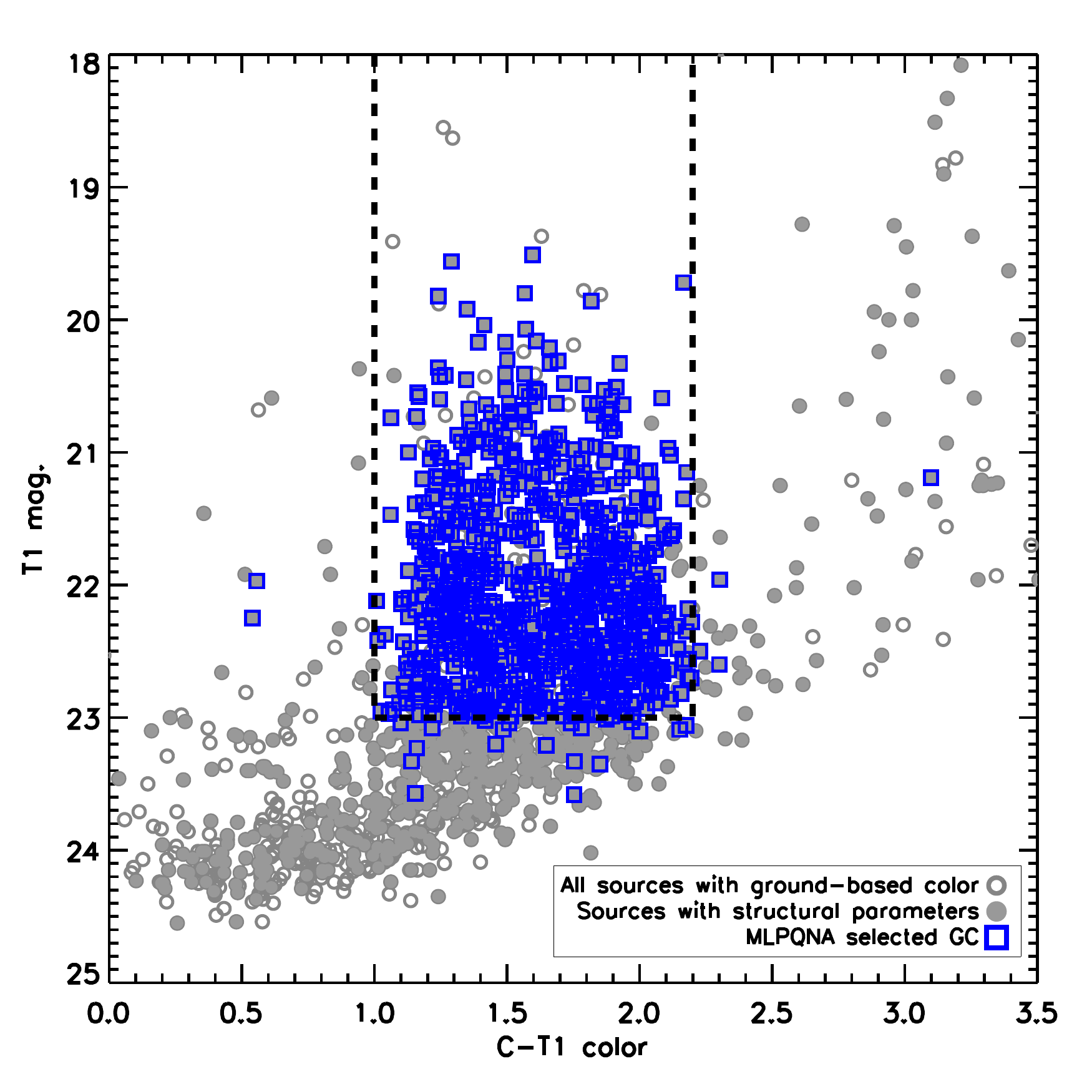}\\
\includegraphics[width=8.3cm]{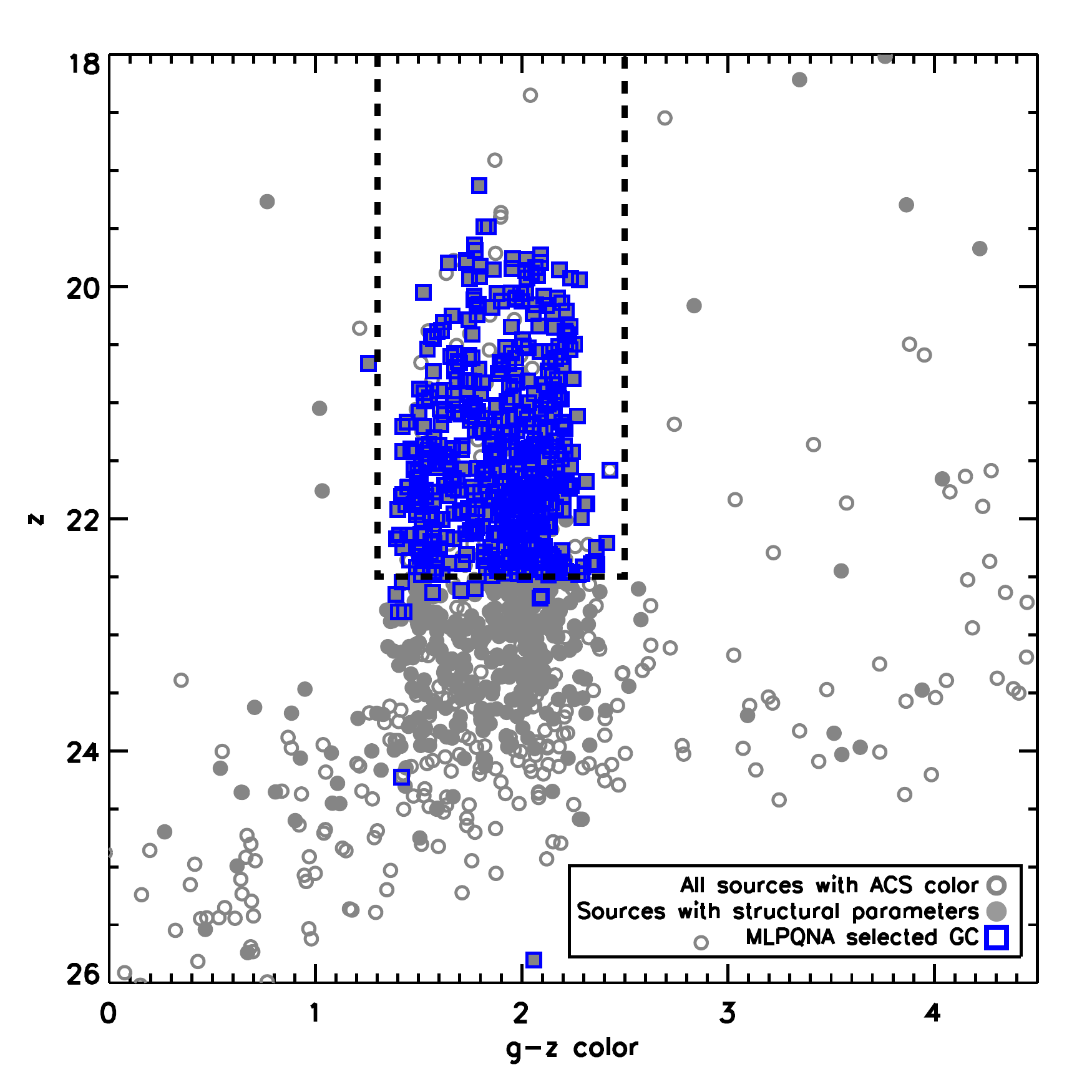}
\caption[Same as Figure \ref{gc:col_mag} showing the color distribution of the MLPQNA selected sample.]{Same as Figure \ref{gc:col_mag} showing the color distribution of the MLPQNA selected sample. The MLPQNA sample (blue squares) reproduces the properties of the color-selected GC population (i.e. the KB) with much less contaminants than, e.g., the pointlike population shown in Figure
\ref{gc:col_mag}.} \label{gc:col_mag_NN}
\end{figure*}

It is worth pointing out that the performances, in terms of completeness and contamination, quoted above are all derived with respect to the test sample (and thus ultimately from the KB), and do not include possible biases affecting the KB itself. Such biases will be propagated to the final sample by any DM algorithm, as these rely on the assumption that the KB is a fair and complete representation the ``real" population that we want to identify. Thus if the KB is severely incomplete or contaminated, this is a separate issue that has to be addressed in the training sample selection phase.

\subsection{Discussion}\label{gc:disc}
In order to test the effectiveness of our method, we need to compare its performances with those
offered by more traditional approaches. For homogeneity (same data set) we shall use as template
the method discussed in \citet{paolillo2011} which used a selection criteria based on magnitude
and morphology.
Figure \ref{gc:maghisto} shows that sources with \textsc{SExtractor} stellarity index $>0.9$ (grey solid line)
are distributed as the GC luminosity function down to $m_V=26$, while at fainter magnitudes background unresolved
sources dominate the overall sample.
 Based on these considerations  Paolillo et al. choose as GC candidates sources having stellarity index $>0.9$ and $m_V<26$ mag.
Clearly a more sophisticated selection process, based on complex combinations of photometric and structural
parameters \citep[see for instance][]{puzia2012}, could be adopted but any such approach requires anyway extensive
testing to verify what biases are introduced in the final sample and it is not clear how such biases can be evaluated and
corrected for without the availability of additional data (e.g. more uniform color coverage or random background fields to compare with).

From Figure  \ref{gc:col_mag} it can be seen that although the use of the stellarity and magnitude criteria effectively selects the bulk of the color-selected GC population, there are sources consistent with GC colors, which are missed by this approach; on the other hand this subsample includes many objects outside the allowed color range.
We can calculate the level of completeness and contamination resulting from the
simple approach of \citet{paolillo2011}, as done in section \ref{gc:exp} for the DM methods. We derive two different estimates: i) for the central region covered by the more accurate $g$ and $z$ HST photometry and,
ii) for the entire field covered by the ground based $C$ and $T1$ data.
Within the central region $92\%$ of our GC candidates (within $m_V<26$ by definition) are consistent with the $1.3\leq g-z<2.5$
color cut and $z<22.5$. Using the $C-T1$ photometry instead, which extends over the whole HST mosaic we find that $82\%$ of the GC candidates are consistent with the $1.0\leq C\!-\!T1<2.2$ color and $T1<23$ magnitude cuts.
On the other hand, $\sim 4\%$ and $\sim 9\%$ of the GC candidates have respectively $g-z$ and $C-T1$ colors outside the allowed range as given in Table \ref{gc:col_sel}.

When these numbers are compared with those presented in Table 1, we see that the MLPQNA outperforms the simpler approach
used by \citet{paolillo2011} both in the central region and across the whole field, in the sense that it results in higher completeness, retrieving a larger fraction of the color-selected sources using only single band photometry.
GAME and SVM may still perform better in the galaxy outskirts, although in the galaxy center they are slightly less accurate.
In terms of contamination the MLPQNA again performs better than the \citet{paolillo2011} approach, yielding $<2\%$ spurious sources in the two best experiments (\textit{complete patterns} and \textit{no par.11}). The other MLPQNA experiments and all SVM cases are still competitive in the galaxy outskirts.

The performance of the MLPQNA method is better understood looking at the color-magnitude plot shown in Fig. \ref{gc:col_mag_NN}.
The MLPQNA sample reproduces the properties of the color-selected GC population with much less contaminants than, e.g.,
the pointlike population shown in Fig. \ref{gc:col_mag}, and less outliers.
In Fig. \ref{gc:maghisto_NN} we show the luminosity distribution of the MLPQNA sample:
the MLPQNA approach (dashed red line) is able to retrieve almost the entirety of the color-selected GC population (solid black line). We point out that the luminosity limit at $m_V\sim 24$ is due to the magnitude threshold imposed on the color-selected sample (Table \ref{gc:col_sel}) in order to get a uniform limit across the whole color range (Figure \ref{gc:col_mag_NN}) and FOV, and is thus not an intrinsic feature of the GC luminosity function which extends down to $m_V \gtrsim 26$ mag.

A detailed investigation of the properties of the spurious sources is difficult since the strength of DM algorithms is to detect \textit{hidden} correlations among the parameters, and use them to classify unknown sources; this however means that such correlations are hard to identify through a simple (and low-dimensional) view of the source distribution in the parameter space. In our specific case we found that most contaminants are indeed GCs which fail the color-magnitude classification technique by only one criteria (e.g. they lie just outside the chosen color or magnitude range, see Figure \ref{gc:col_mag_NN}). It is thus unsurprising that the MLPQNA identifies such sources as GCs as all other parameters obey the correlations identified in the training phase. A few more extreme objects are found to be affected by photometric or structural problems in the data, such as the overlap with a nearby source which may introduce severe contamination in low-resolution ground based data, or a position close to the chip gap in HST data. A few can also be expected to be foreground stars which are misclassified due to the small angular size of some GCs in the training sample, lying at the resolution limit of HST data (gray region in Figure \ref{gc:size_distro}).

An additional advantage in the use of DM techniques can be see comparing the structural parameters of pointlike sources
with $m_V<26$ mag with those of the
color-selected subsample (Figure \ref{gc:size_distro}, left panel): we find that the \citet{paolillo2011} selection criteria misses
extended sources with $R_{eff}\gtrsim 5$ pc, as it can be expected given the compactness requirement (stellarity index $>0.9$).
The right panel of the same Figure shows that the MLPQNA methods is instead able to retrieve also the most extended GCs.
While some of these extended sources may be background galaxies, we point out that the most extended GCs, such as the Galactic GC $\omega$-Cen, do fall in this range. In fact we used the subset of GC confirmed through radial velocity measurements \citep{Dirsch2004} to verify that a significant fraction ($\sim 10\%$) of the NGC1399 GC population has $R_{eff}\gtrsim 5$ pc, and that the size distribution of this subsample is statistically indistinguishable from both the color- and  MLPQNA-selected populations. Obviously we cannot confirm that \textit{all} extended sources are genuine GCs but, as already discussed in section \ref{gc:exp}, we emphasize that the performance of the method has to be evaluated only by its ability to retrieve the same sources included in the training sample, i.e. the color-selected GCs in our experiment.

\begin{figure}
\centering
\includegraphics[width=12cm]{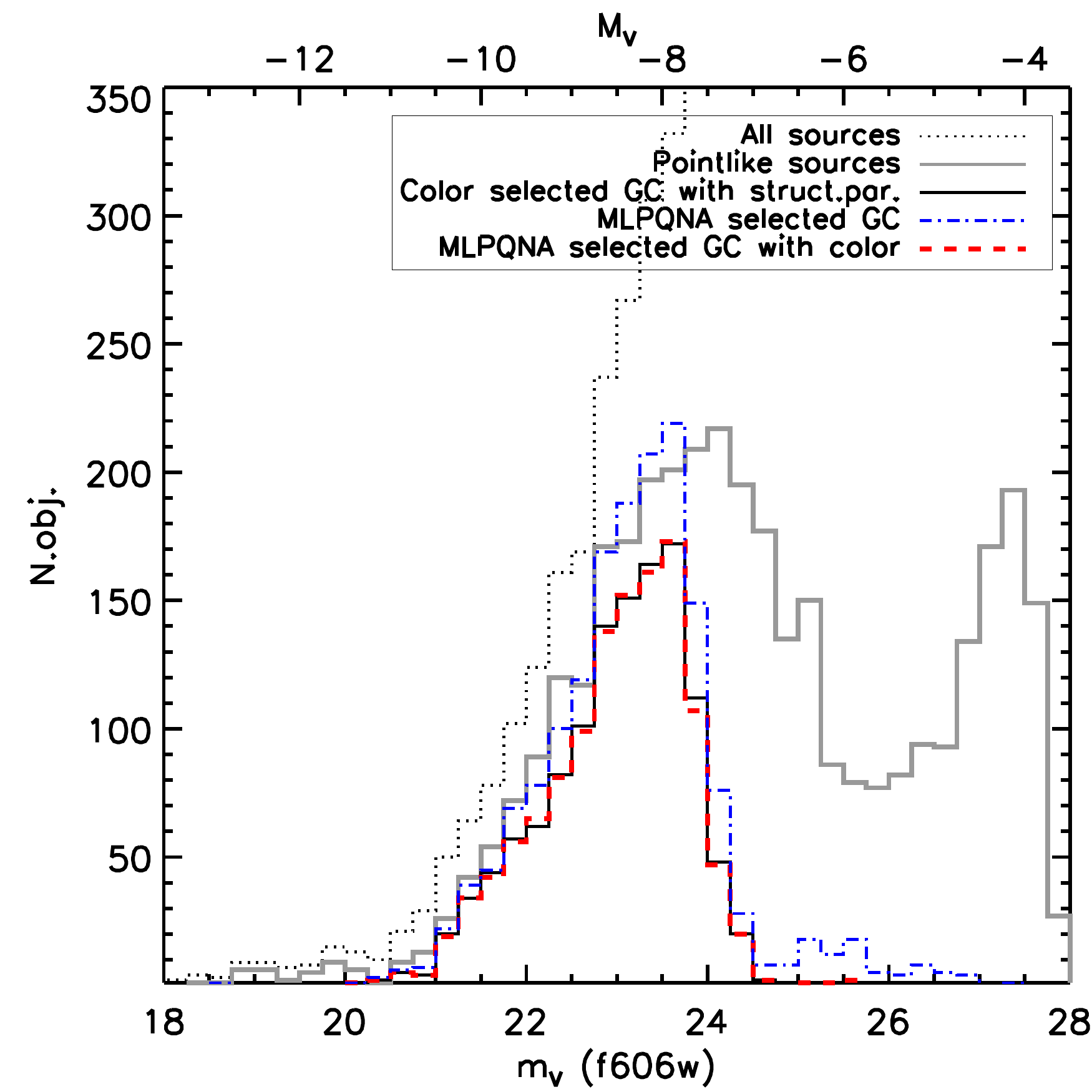}
\caption[Same as Figure \ref{gc:maghisto} but for the MLPQNA selected samples.]{Same as Figure \ref{gc:maghisto} but for the MLPQNA selected samples.
The MLPQNA approach (dashed red line) is able to retrieve almost the entirety of the color-selected GC population (solid black line);
applying the same algorithm to all sources with structural parameters (but no color, blue dot-dashed line) we can thus retrieve many more objects than available in the color-selected subsample, sharing the same luminosity distribution of the latter population.}
\label{gc:maghisto_NN}
\end{figure}

Applying the same algorithm to the larger ensemble of sources with structural parameters (but no color information) we are now able to retrieve more objects than available in the color-selected subsample, sharing very similar properties to the latter population. The population of MLPQNA selected GCs identified within the whole population is shown in Figure \ref{gc:maghisto_NN} and Figure \ref{gc:size_distro} (right panel) as a dot-dashed line. In our specific test case (e.g. NGC1399) this method allows to identify $\sim 30\%$ more GCs than relying the subsample of sources with color; this larger sample closely follow the GC LF down to the magnitude limit imposed by the color selection, as well as the structural properties of the bona-fide GC population. Thus the gain with respect to other selection techniques is in the ability to retrieve a larger population with well defined properties, at lower observational cost. In other programs the gain can be much larger: for instance in cases of large surveys where DM algorithms can be trained on a KB consisting on a limited number of multi-band observations covering only a small fraction of the FOV; the trained algorithm will then allow to extract statistically equivalent samples from the entire survey.



Finally we note that each experiment was not really time consuming. It was executed on a common desktop multi-core PC in a multi-threading environment, resulting in about 3600 sec (1 hour) of duration for the training phase in the worst case (i.e. on the whole dataset patterns with all 11 features). The test phase is instantaneous, since the trained network acts like a one-shot function.
Of course the complexity and indeed the execution time depends in a quadratic form on the dataset dimension. But in case of small datasets, like the present one, this is not an issue.
Besides computing time, the relevant result is that the proposed MLPQNA model revealed a strong performance also in case of small datasets where, as known, machine learning method performances are usually degrading, due to the limited size of the training samples. This is demonstrated by the poorer results obtained by other methods, shown in Table \ref{gc:comp}, which usually perform significantly better on larger datasets.

\subsection{Conclusion}
We performed an experiment showing that the use of Data Mining (DM) techniques on small datasets, allows to solve complex astronomical problems such as the selection of Globular Cluster candidates in external galaxies, from single band images, provided that a subsample of sources can be used to train the DM algorithm. Since such methods do not assume any a-priori model of the population we are looking for, they allow to retrieve samples which share the same properties of the training sample and are affected by less biases than results using simpler selection techniques.

In principle we could use more refined approaches than those tested here, such as the use of radial velocity (RV) measurements to improve the the reliability of the Knowledge Base (KB), but any such approach would require the availability of additional data i.e., in this particular case, spectroscopic observations. Such type of data are difficult to obtain and expensive in terms of observing time, thus justifying the DM methods proposed in this work. Obviously in some instances these data could already be available in the archives, as for the NGC1399 case were they have been used to verify some of our results (section \ref{gc:disc}).

As a closing remark, we can safely state that, in the emerging scenario of the data-driven science, a Data Mining based approach to data
analysis and interpretation seems to provide a large competitive edge over classical methods in particular for what concerns the ability to recognize patterns and derive correlations in high dimensionality dataset that are not easily handled by human perception.


\begin{figure*}
\centering
\includegraphics[width=7.4cm]{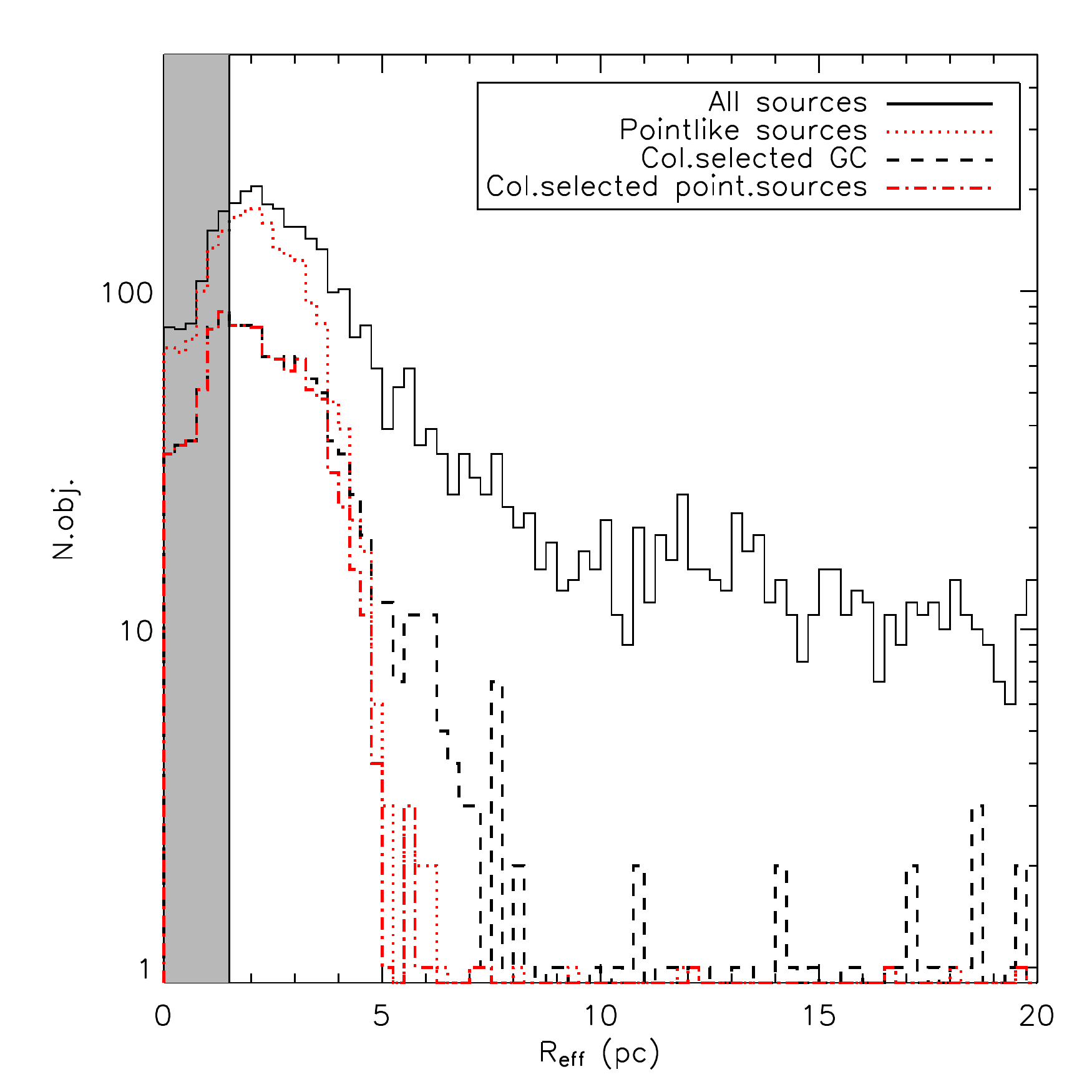}
\includegraphics[width=7.4cm]{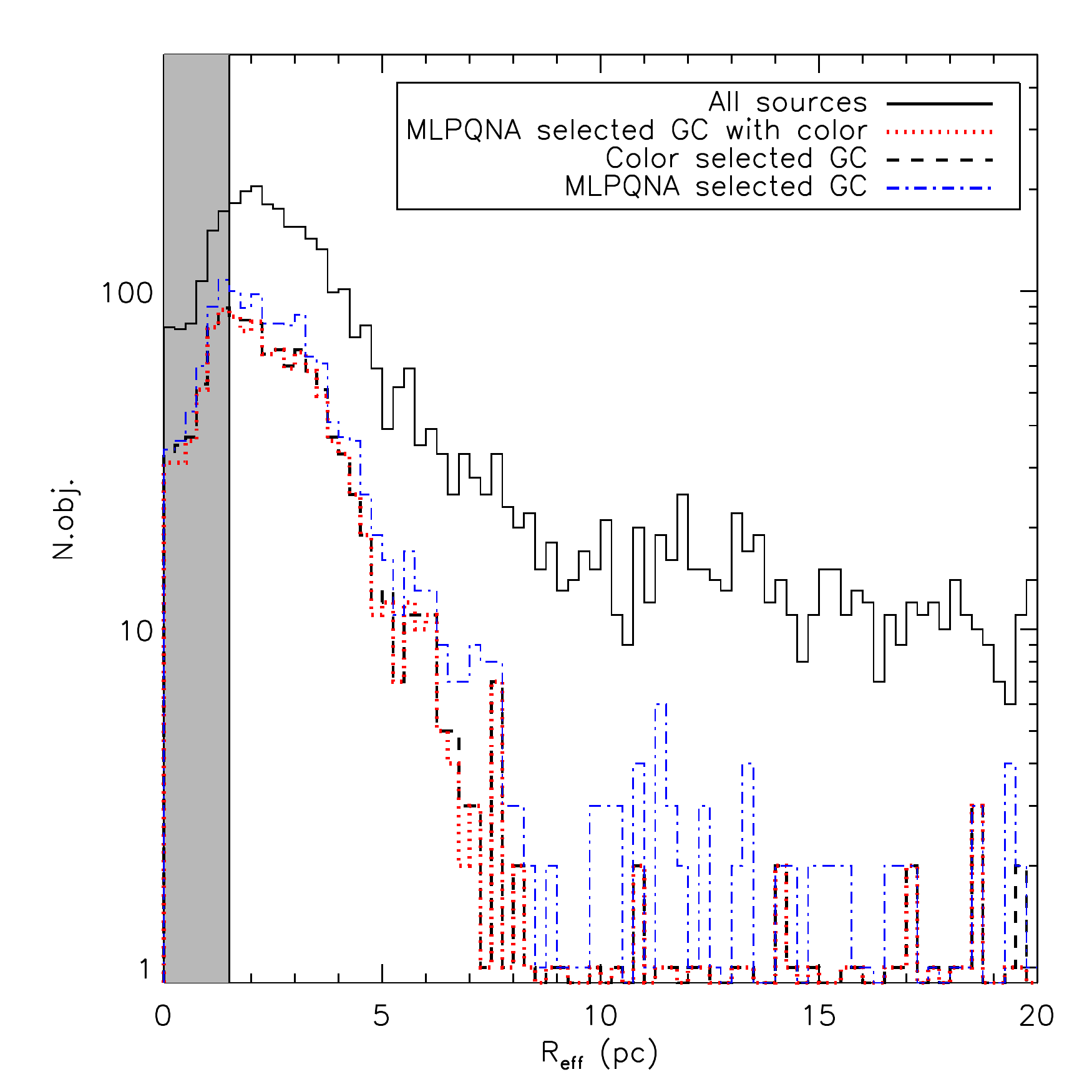}
\caption[Half-light radius distribution for the ACS catalog, compared to \citet{paolillo2011} GC candidates and for the MLPQNA selected samples.]{\textit{Upper panel:} Half-light radius distribution for the entire ACS optical catalog (solid line), compared to \citet{paolillo2011} GC candidates, i.e. pointlike sources with $m_V<26$ (dotted line). Restricting the sample to color-confirmed GCs (dashed and dot-dashed lines) shows that the \citet{paolillo2011} selection criteria misses very extended GCs with $R_{\rm eff}>5$ pc. The shaded region highlights the region where our size measurement are poorly constrained (see \citealt{paolillo2011, puzia2012}). \textit{Lower panel:} Same as left panel but for the MLPQNA selected samples.
The MLPQNA selected sample (dotted red line) reproduces the size distribution of the the color-selected GC population (dashed black line), thus avoiding the size biases resulting from the simpler \citet{paolillo2011} selection criteria; the same is true when applying the MLPQNA algorithm to the larger subsample with structural parameters (blue dot-dashed line).}
\label{gc:size_distro}
\end{figure*}

    \section[Photometric AGN Classification]{Photometric AGN CLassification in the SDSS with Machine Learning Methods}\label{chap:agn}

\blfootnote{this section is largely extracted from: \tiny
\begin{itemize}
\item \textbf{Cavuoti, S.}; Brescia, M.; D'Abrusco, R.; Longo, G.; Photometric AGN Classification in the SDSS with Machine Learning Methods \textbf{to be Submitted to MNRAS}	
\end{itemize}}
%
Active galactic nuclei (AGNs) and quasars are known as the
most luminous long-lived discrete objects in the Universe. One of
the major scientific goals in extragalactic astronomy is to
investigate the physical conditions in the proximity of the central
power source (the so called ``central engine"), almost certainly a
supermassive black hole with a surrounding accretion disk.

This section is aimed at showing the feasibility and the performances of a method based on
self-adaptive learning techniques to classify AGNs using photometric data instead of spectroscopic information.
AGN selection is usually performed by observing the overall spectral distribution and based on some
spectroscopic indicators, like equivalent line width FWHM of specific lines, or lines flux ratios. A reliable
and accurate AGN classification tool based only on photometric information would allow to save precious
instrument observing time and engage several studies based on statistically significant samples of objects.

Modern surveys, like SDSS, produce a great quantity of data, several
TB, in which the usual statistical approach is not suitable because it may be
applied just to few dimensions, so we want to engage a different paradigm
of analysis. Throughout the present work we show how Machine Learning (ML) methodology may be useful
to distinguish strong candidates AGN from SDSS data using just
photometric parameters.

We apply some DAMEWARE algorithms to a dataset obtained by the
joining of three catalogues of objects within $0.02<z<0.3$. The resulting dataset contains photometric parameters (hereinafter named as input features), supervised by a flag, based on the spectroscopic
information (hereinafter named as target vector) used only in the training steps of the procedure.

Over the principal target we apply the algorithm first to distinguish between Type 1 and Type 2 objects, while the last application has been based on Seyfert vs. LINERs classification.

\subsection{The data}

The experiments in this section have been performed using objects belonging to at least one
of the catalogues of galaxies and candidate AGNs provided by \cite{sorrentino2006},
and by \cite{kauffmann2003}.

According to the
standard unified model of active galactic nuclei \citep{antonucci1993}
AGNs can be classified into two categories,
depending on whether the central black hole and its associated
continuum and broad emission-line region are viewed directly (a
``type 1" AGN) or are obscured (a ``type 2" AGN) by the dust torus
surrounding the black hole. The AGNs have been selected from the normal
galaxies using the method devised by Baldwin, Phillips and Terlevich
(hereafter BPT, \citealt{baldwin1981}), i.e. considering the
intensity ratios of two pairs of relatively strong emission lines,
and classifying objects according to their position in the so-called
BPT diagram.

Three different experiments have been performed using
three different samples of photometric objects for which a classification
based on spectroscopy, is available. All three samples have been drawn from
the SDSS DR4 catalogue: they belong to the PhotoSpecAll table, containing
all the objects for which both photometric and spectroscopic observations
have been carried out.

Our sample is made by the join of the three catalogues, described in the following.

\subsubsection{Catalogue of Sorrentino et al.}

The first selected catalogue has been obtained by \cite{sorrentino2006} which operates a separation between ($0.05<z<0.095$)
objects into Seyfert 1, Seyfert 2 and Not AGN.
That catalogue contains 24293 objects.

One object is an AGN if it is over one of the Kewley's
line \citep{kewley2001}:

\begin{equation}\label{AGN:kewley1}
\log \frac{{[OIII]\lambda 5007}}{{H_\beta  }} = \frac{{0.61}}{{\log
\frac{{[NII]\lambda 6583}}{{H_\alpha  }} - 0.47}} + 1.19
\end{equation}
\begin{equation}\label{AGN:kewley2}
\log \frac{{[OIII]\lambda 5007}}{{H_\beta  }} = \frac{{0.72}}{{\log
\frac{{[SII]\lambda \lambda 6717,6731}}{{H_\alpha  }} - 0.32}} +
1.30
\end{equation}
\begin{equation}\label{AGN:kewley3}
\log \frac{{[OIII]\lambda 5007}}{{H_\beta  }} = \frac{{0.73}}{{\log
\frac{{[OI]\lambda 6300}}{{H_\alpha  }} - 0.59}} + 1.33
\end{equation}

\begin{figure}
   \centering
   \resizebox{\hsize}{!}{\includegraphics{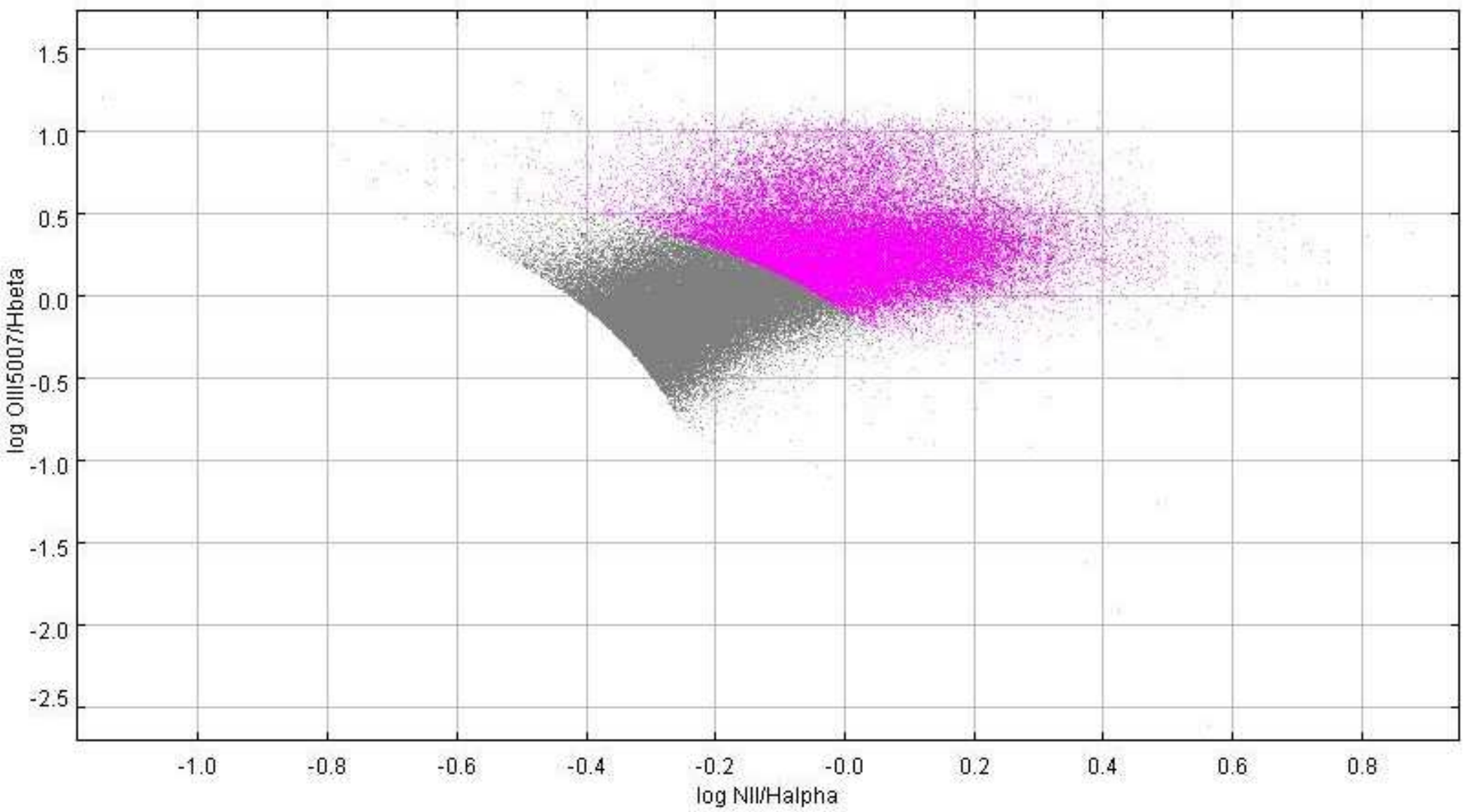}}
     \caption{The division obtained according to the Kewley's line.}
\label{AGN:Fig:bptkewley}
\end{figure}
The division obtained according the Kewley's line is shown in figure \ref{AGN:Fig:bptkewley}.
Once one object is declared as AGN, then it is declared to be a Seyfert 1 if

\begin{equation}\label{AGN:classificazione1} FWHM(H_\alpha)
> 1.5 FWHM([OIII]\lambda 5007)
\end{equation}

or

\begin{equation}\label{AGN:classificazione2} FWHM(H_\alpha)
>1200 Km s^{-1}\end{equation}

and

\begin{equation}\label{AGN:classificazione3}  FWHM([OIII]\lambda 5007) < 800 Km s-1\end{equation}

The rest of the objects is classified as Seyfert 2.
 They found:
\begin{itemize}
\item 1829 AGN:

    \begin{itemize}
        \item 725 Sy 1
        \item 1105 Sy 2
    \end{itemize}
    \item 22464 Not AGN
\end{itemize}

\subsubsection{Catalogue of Kauffman et al.}

The second selected catalogue\footnote{\url{http://www.mpa-garching.mpg.de/SDSS/DR4/}} contains
spectra lines and ratio for 88178 galaxies ($0.02<z<0.3$).
According to the work of \cite{kauffmann2003}, we define
a zone where there are just AGNs located over the Kewley's line,
(eq. \ref{AGN:kewley1}) \citealt{kewley2001}, and a zone where objects are not AGNs, i.e. under the Kauffman's line, \citealt{kauffmann2003,
kewley2006}):
\begin{equation}\label{AGN:kauffman}
\log \frac{{[OIII]\lambda 5007}}{{H_\beta  }} = \frac{{0.61}}{{\log
\frac{{[NII]\lambda 6583}}{{H_\alpha  }} - 0.05}} + 1.3
\end{equation}
The division obtained according the Kauffman's line is shown in figure \ref{AGN:Fig:bptkauff},
with an overlap mixed zone in which both AGNs and not AGNs objects are present.
\begin{figure}
   \centering
   \resizebox{\hsize}{!}{\includegraphics{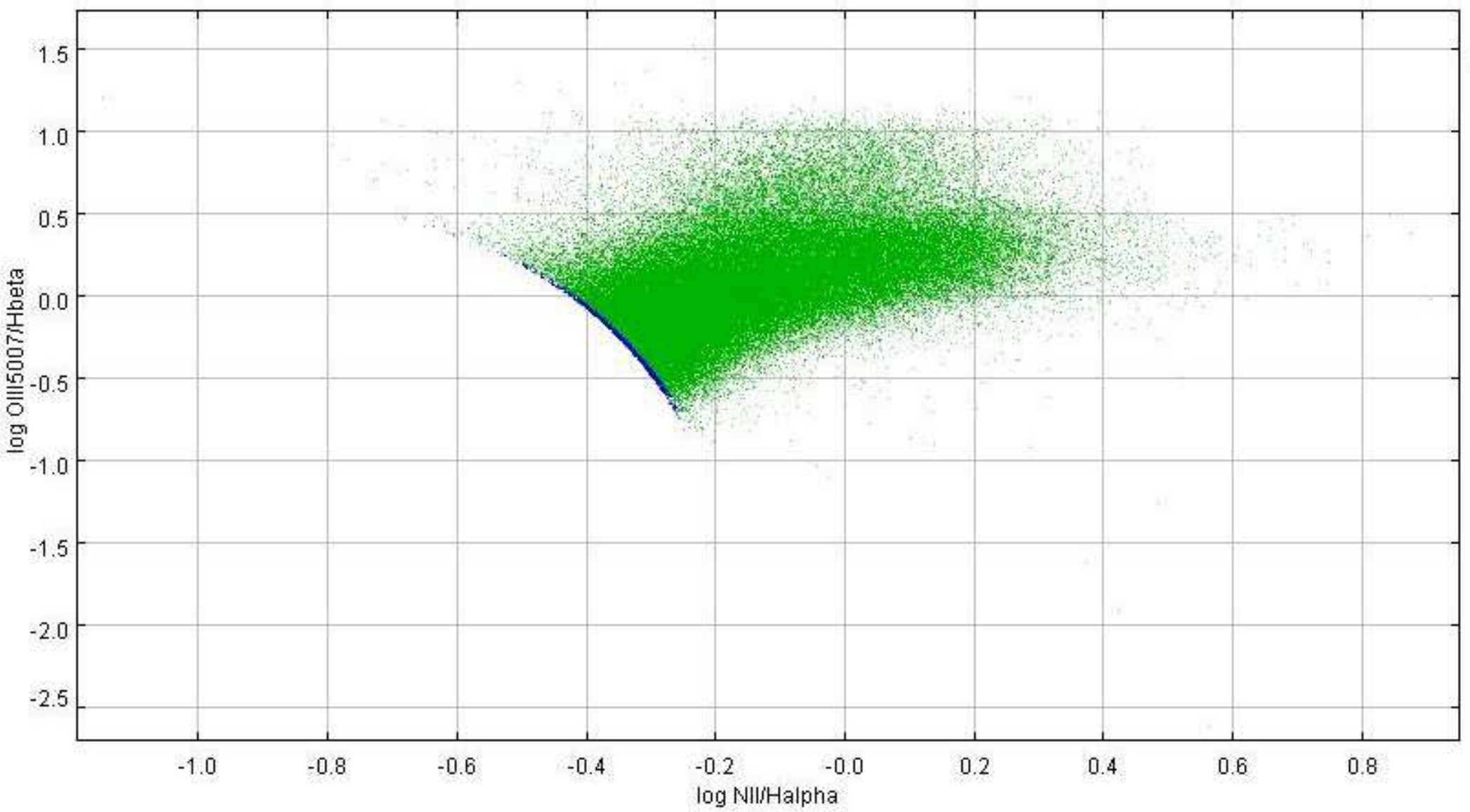}}
     \caption{The division obtained according to the Kauffman's line.}
\label{AGN:Fig:bptkauff}
\end{figure}

The mixed and pure AGN zones have been further divided into Seyfert and LINER zones by following the Heckman's line \citep{heckman1980, kewley2006}:

\begin{equation}\label{AGN:LINERS}
    \frac{{[OIII]\lambda 5007}}{{H_\beta  }}=2.1445 \frac{{[NII]\lambda 6583}}{{H_\alpha  }} + 0.465
\end{equation}
The division obtained according to that line is shown in figure \ref{AGN:Fig:bptheck}.

\begin{figure}
   \centering
   \resizebox{\hsize}{!}{\includegraphics{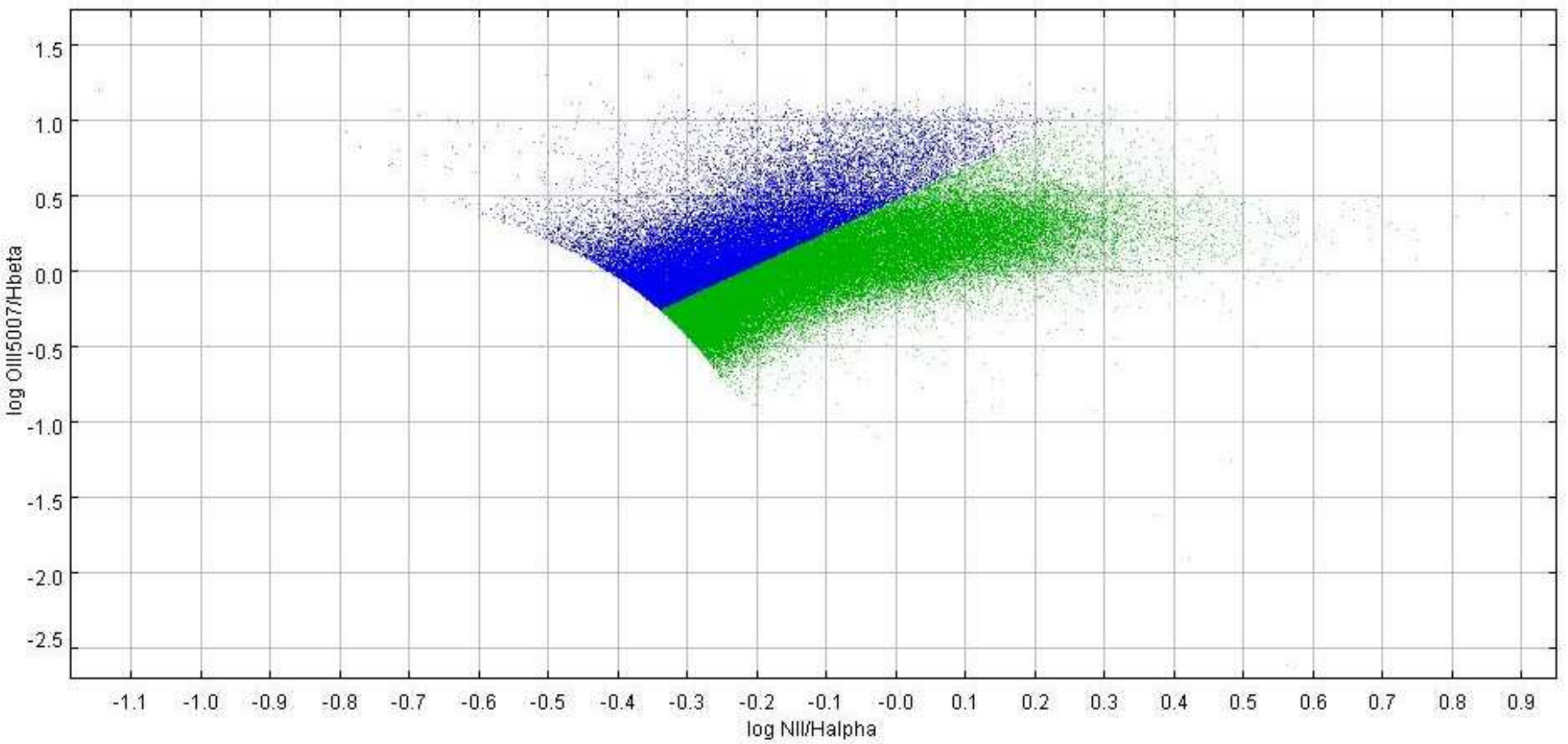}}
     \caption{The division obtained according to the Heckman's line.}
\label{AGN:Fig:bptheck}
\end{figure}

The resulting five areas of the AGN map are shown in figure \ref{AGN:Fig:bpt}.
\begin{figure}
   \centering
   \resizebox{\hsize}{!}{\includegraphics{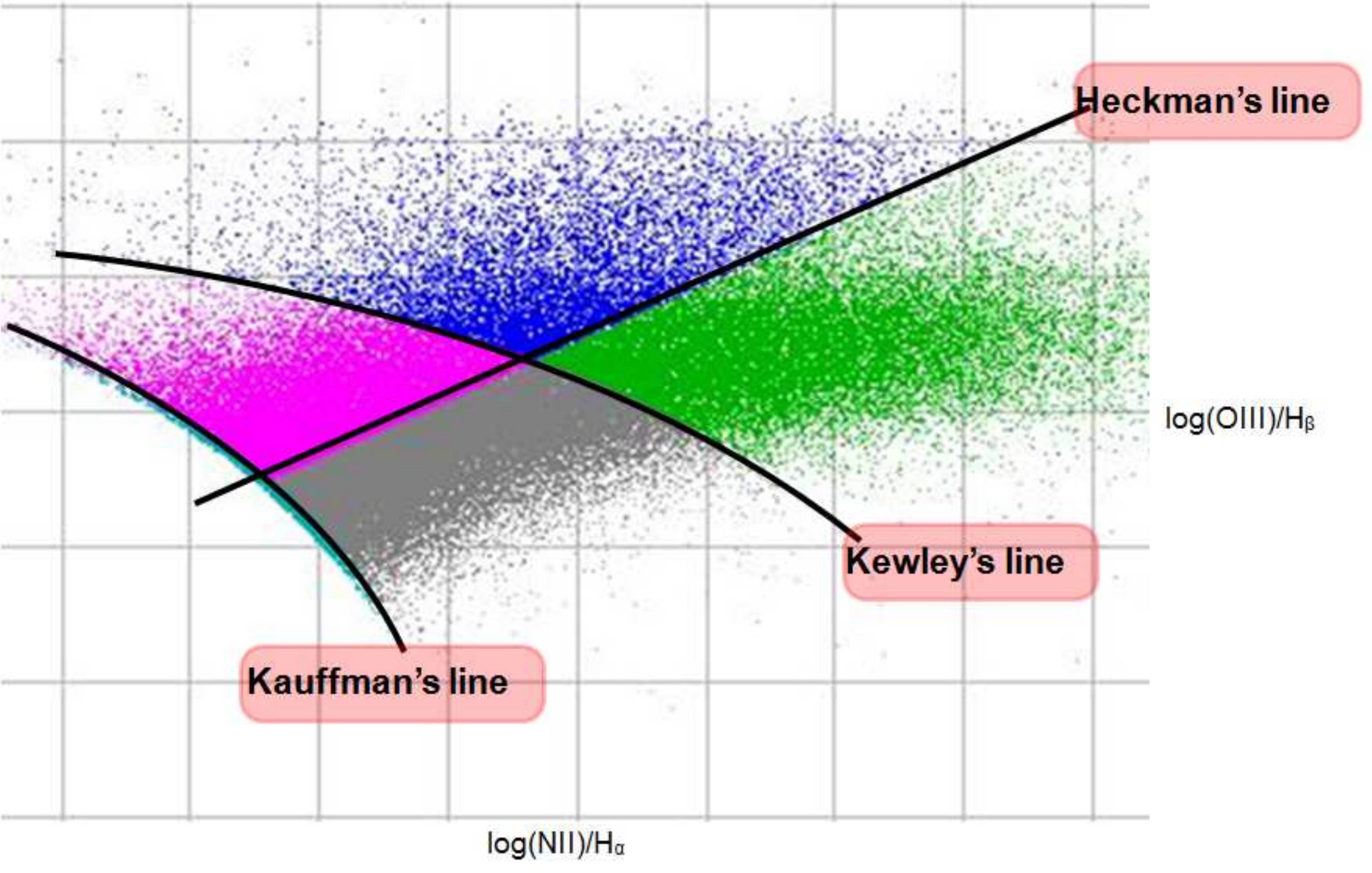}}
     \caption{A representation of the catalogue in the BPT diagram.}
\label{AGN:Fig:bpt}
\end{figure}

\subsubsection{Catalogue of D'Abrusco et al.}

The catalogue made by \cite{dabrusco2007}
contains photometric redshift with an accuracy estimated by
$\sigma_{rob} = 0.02$, evaluated by ($z_{phot} - z_{spec}$).\\

\subsubsection{The final catalogues}
The process performed in order to obtain the final catalogues for our experiments was divided into four phases: the first was to join catalogs of Kauffman and Sorrentino. In case of overlap, the information of both type 1 and 2 objects has been taken from Sorrentino, while all other types from Kauffman. By merging these two catalogs, we obtained a total of 108162 objects. The second phase was to merge the outcoming catalogue with the one of D'Abrusco by matching and including the photometric redshift information, obtaining a final catalogue of 100069 objects. The third stage was to extract the corresponding photometric information for the selected objects from the SDSS archive.

The information retrieved from the SDSS DR4 database for each object was:
\begin{itemize}
\item $fiberMag\_u$
\item $fiberMag\_g$
\item $fiberMag\_r$
\item $fiberMag\_i$
\item $fiberMag\_z$
\item $petroMag\_u$
\item $petroMag\_g$
\item $petroMag\_r$
\item $petroMag\_i$
\item $petroMag\_z$
\item $petroR50\_u$
\item $petroR50\_g$
\item $petroR50\_r$
\item $petroR50\_i$
\item $petroR50\_z$
\item $petroR90\_u$
\item $petroR90\_g$
\item $petroR90\_r$
\item $petroR90\_i$
\item $petroR90\_z$
\item $dered\_u$
\item $dered\_g$
\item $dered\_r$
\item $dered\_i$
\item $dered\_z$
\end{itemize}

After a statistical pruning just a subsample of those parameters, described more in detail in the section \ref{AGN:dataset}, has been used for the final experiments.

From the obtained dataset, all the objects with undefined values for some of their parameters (named also as NaN) have been removed. This was because machine learning empirical models may suffer from the potentially misleading information coming from such unknown terms.

 The last phase was the creation of three different catalogues for the three different experiments:

\begin{enumerate}
  \item AGN vs Not AGN
  \item Type 1 vs Type 2
  \item Seyferts vs LINERs
\end{enumerate}

The dataset for the first experiment is the whole dataset itself, with 84885 objects; the dataset for the second experiment contains just the objects belonging to the dataset of Sorrentino and that are pure AGN, resulting into 1570 objects; the last dataset contains the objects, belonging to the catalogue of Kauffman, that are pure AGN divided into LINERs and Seyferts, obtaining 30380 objects; the datasets obtained are summarized in table \ref{AGN:tabledataset}.

\begin{table}

\centering
\tabcolsep 5.8pt
\small
$\begin{array}{@{}|c|c|c|c|c|@{}} \hline
CLASS & CATALOGUE & Exp. 1 & Exp. 2 & Exp. 3 \\ \hline \hline
Not AGN & All & Class 0 & - & - \\ \hline
Type 1 & Sorrentino & Class 1 & Class 1 & - \\ \hline
Type 2 & Sorrentino & Class 1 & Class 0 & - \\ \hline
Mix-LINER & Kauffman & Class 0 & - & - \\ \hline
Mix-Seyfert & Kauffman & Class 0 & - & - \\ \hline
Pure-LINER & Kauffman & Class 1 & - & Class 0 \\ \hline
Pure-Seyfert & Kauffman & Class 1 & - & Class 1 \\ \hline
Mix-LINER-Type1 & overlap & Class 0 & - & - \\ \hline
Mix-Seyfert-Type1 & overlap & Class 0 & - & - \\ \hline
Pure-LINER-Type1 & overlap & Class 1 & Class 1 & Class 0 \\ \hline
Pure-Seyfert-Type1 & overlap & Class 1 & Class 1 & Class 1 \\ \hline
Mix-LINER-Type2 & overlap & Class 0 & - & - \\ \hline
Mix-Seyfert-Type2 & overlap & Class 0 & - & - \\ \hline
Pure-LINER-Type2 & overlap & Class 1 & Class 0 & Class 0 \\ \hline
Pure-Seyfert-Type2 & overlap & Class 1 & Class 0 & Class 1 \\ \hline
SIZE: 24293 & Sorrentino  & 84885 & 1570 & 30380 \\
SIZE: 88178 & Kauffman  &  &  &  \\ \hline
 \end{array}$
\caption[The dataset composition after the merging from original catalogues. The empty fields indicate the unused typology.]{The dataset composition after the merging from original catalogues. The empty fields indicate the unused typology. The division between class $0$ and class $1$ are referred to the target vector (used during training). The final sizes fo the three experiment datasets are obtained after the D'Abrusco photo-z catalogue matching and the whole NaN removal process.}\label{AGN:tabledataset}
\end{table}

\subsection{The experiments}

In order to evaluate the performances after the training we need to explain the evaluation terminology:
\begin{itemize}
\item \textbf{total efficiency}: the ratio between the amount of correctly classified objects and the total number of objects in the dataset.
\item \textbf{purity of a class}: the ratio between the number of correctly classified objects of a class and the number of objects classified in that class, also known as efficiency of a class.
\item \textbf{completeness of a class}: the ratio between the amount of correctly classified objects of that class and the total number of objects of that class in the dataset.
\item \textbf{contamination of a class}: is the dual of the pureness, the ratio of misclassified object of a class and the number of objects classified in that class.
\end{itemize}

Each experiment was performed by dividing the dataset in two parts, a train set containing $80\%$ of the dataset and a test set containing the remaining $20\%$. All the results presented in the following sections are evaluated on the test set that is never used to train the ML models.

\subsubsection{The dataset}\label{AGN:dataset}

In order to perform our experiments we use the following groups of
parameters (also named as input features) to feed our nets:

\begin{enumerate}
\item $petroR50\_u$
\item $petroR50\_g$
\item $petroR50\_r$
\item $petroR50\_i$
\item $petroR50\_z$
\item $concentration\_index\_r$
\item $z\_phot\_corr$
\item $fibermag\_r$
\item $(u-g) dered)$
\item $(g-r) dered)$
\item $(r-i) dered)$
\item $(i-z) dered)$
\item $dered_r$
\end{enumerate}

Parameters i-v are the radii which contain 50\% of the
petrosian magnitude. \\

The SDSS has adopted a modified form of the Petrosian (1976) system, by
measuring galaxy fluxes within a circular aperture whose radius is
defined by the shape of the azimuthally averaged light profile.

In order to define the Petrosian radii, it is necessary to report what it is intended as petrosian ratio. the ``Petrosian ratio" $\Re _P$ at a radius r from the
center of an object to be the ratio of the local surface brightness
in an annulus at r to the mean surface brightness within r \citep{blanton2001, yasuda2001}:
\begin{equation}
\Re _P (r) = \frac{{\int_{0.8r}^{1.25r} {dr'2\pi r"I(r')/[\pi
(1.25^2  - 0.8^2 )r^2 ]} }}{{\int_0^r {dr'2\pi r'I(r')/(\pi r^2 )}
}}
\end{equation}

where $I(r)$ is the azimuthally averaged
surface brightness profile.

The Petrosian radius $r_P$ is defined as the radius at which $\Re
_P(r_P)$ equals some specified value $\Re _{P,lim}$, set to 0.2 in
our case. The Petrosian flux in any band is then defined as the flux
within a certain number $N_P$ (equal to 2.0 in our case) of $r$
Petrosian radii:

The concentration index, parameter vi, is the ratio between radii
that contain 50\% and 90\% of the total flux from the object in the
approximation of symmetric source, is a good index of how much the
flux is bounded in the center of the source.

The photometric redshift, parameter vii, is the redshift derived by
the catalogue of \cite{dabrusco2007}.\\

The fiber magnitude in the r-band, parameter viii, is the flux
contained in 3".\\

Parameters from ix to xii are deredded colors, while the parameter xiii is the
deredded r-band magnitude.

The presence of parameters not measured, usually known as ``Not a Number" or NaN, can be misleading for our models so we remove each object that has at least one parameter not measured.

\subsubsection{Experiment Nr. 1 - AGN Classification}

Concerning the first of the three mentioned experiments, i.e. classification between AGN and not AGN, the described ML models are feeded using a target vector whose values are labeled as 1 for each object that is over the Kewley's line(pure AGN) and 0 for the object under this line (mixed zone and certainly not AGN), resulting in 84885 objects after the removal of the patterns affected by NaN presence.
According to the mentioned strategy the train set ($80\%$ of the whole dataset) contains $67908$ patterns while the test set ($20\%$ of the dataset) contains $16977$ patterns.

The best result has been obtained by the MLP with the Quasi Newton learning rule.

The following values summarize the validation performances after the training for the best experiment. \\

\begin{itemize}
\item total efficiency : $e = 75.99\%$
\item AGN pureness:  $p_{agn} = 71.38\%$	
\item AGN completeness: $c_{agn} = 55.64\%$
\item not AGN completeness: $c_{mix} = 87.44\%$
\item AGN contamination: $1-p_{agn}=28.62\%$
\end{itemize}

The percentage of false positives that we know spectroscopically to
be surely not AGN is $0.89\%$; The percentage of object that
spectroscopically are certainly not AGN that became false positive is
$0.82\%$.

As we have seen the contamination due to galaxies is very small,
and contamination caused by objects in the Mixed zone could be
partially attributable to unrecognized AGN. Anyway we tried
to maximize pureness of the AGN class. This was done
by varying the confidence threshold value. The MLP classifier is not crisp, so the output represents a probability of belonging to a class; in this sense a confidence threshold value major than $0.5$ is considered as class label $1$ (in that case, a class label 1 stands for AGN objects).
In order to avoid statistical fluctuations the results on the test set have been splitted by considering alternately the first half, the second one and then the whole test set, as shown in
figures respectively \ref{AGN:Fig:thres1}, \ref{AGN:Fig:thres2} and \ref{AGN:Fig:thres3}.

\begin{figure}
   \centering
   \resizebox{\hsize}{!}{\includegraphics{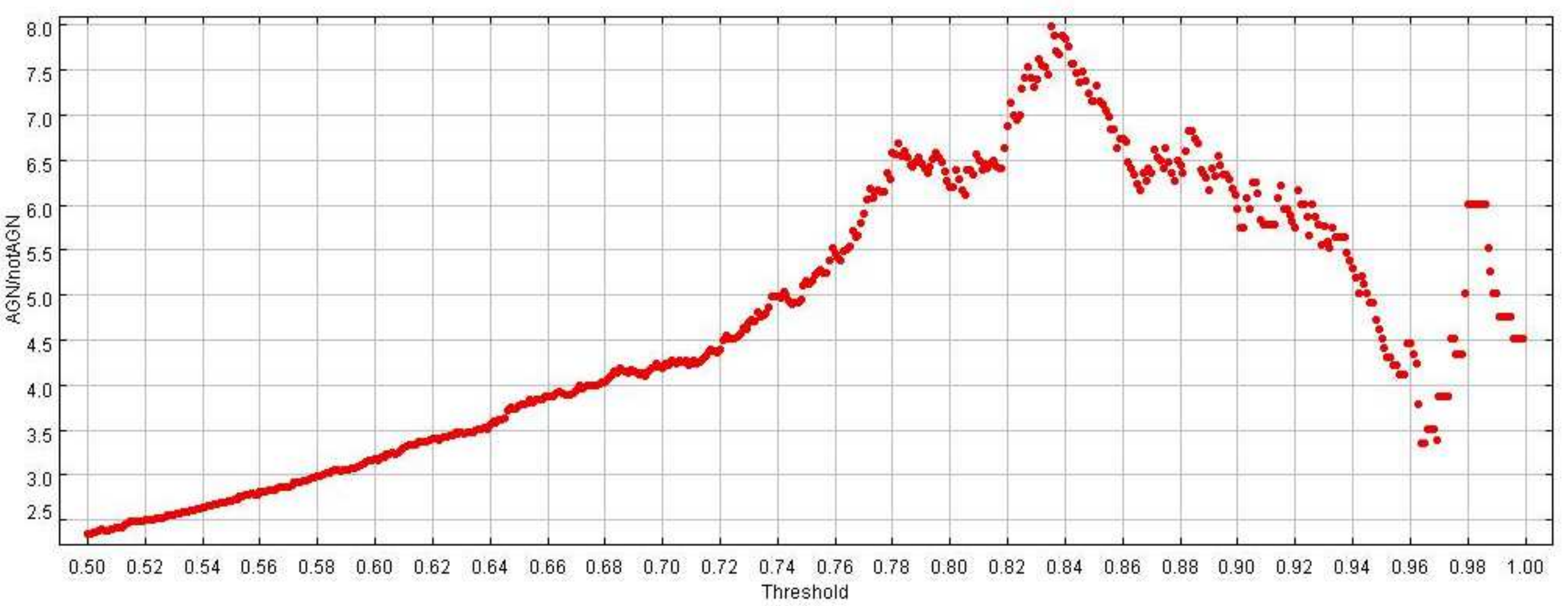}}
     \caption{Ratio of correct evaluated AGN and not-AGN misclassified, first half of the Test Set.}
\label{AGN:Fig:thres1}
\end{figure}

\begin{figure}
   \centering
   \resizebox{\hsize}{!}{\includegraphics{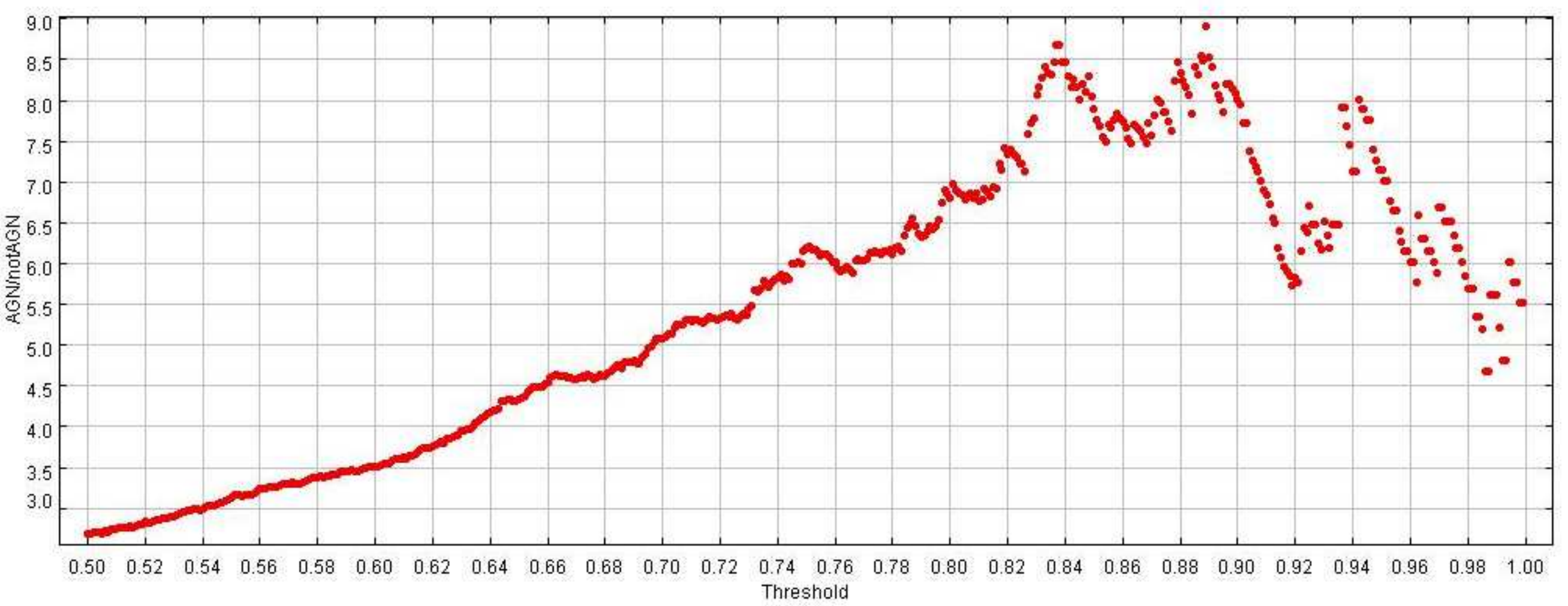}}
     \caption{Ratio of correct evaluated AGN and not-AGN misclassified, first half of the Test Set.}
\label{AGN:Fig:thres2}
\end{figure}

\begin{figure}
   \centering
   \resizebox{\hsize}{!}{\includegraphics{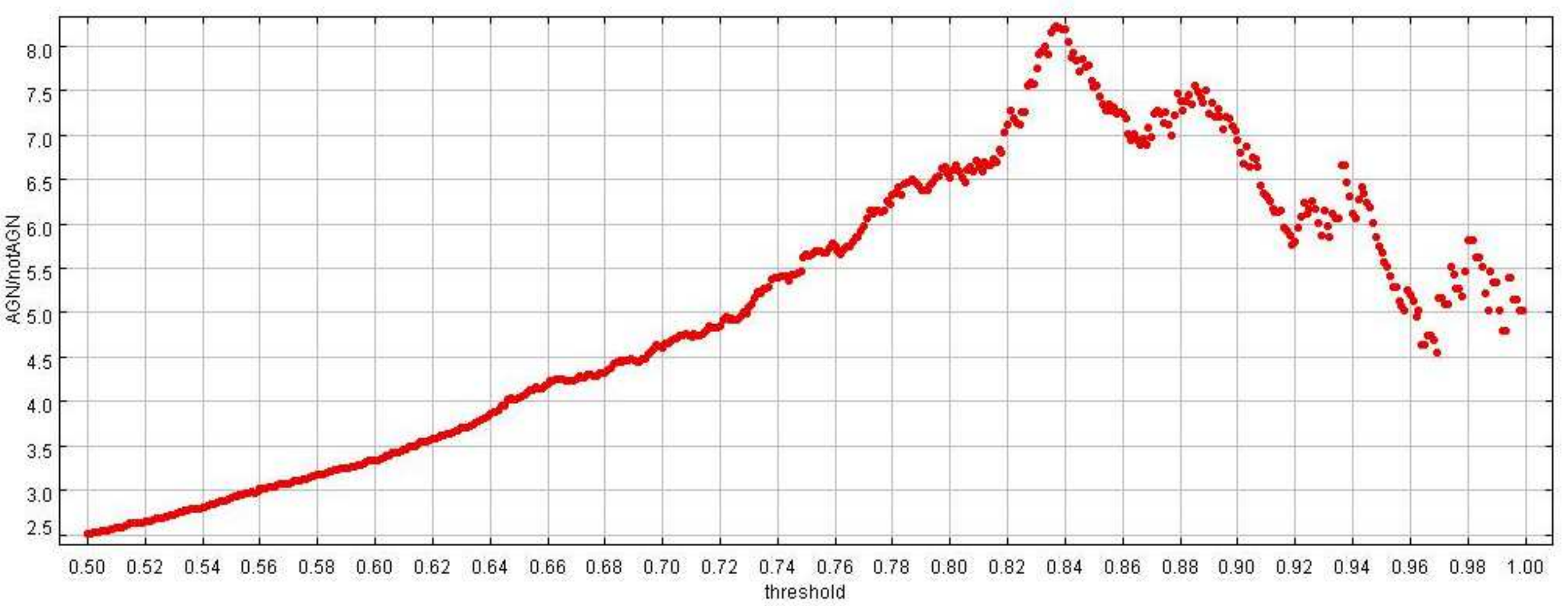}}
     \caption{Ratio of correct evaluated AGN and not-AGN misclassified, whole Test Set.}
\label{AGN:Fig:thres3}
\end{figure}

By following such strategy we found a common maximum at a threshold of around $0.837$; with such threshold we obtained a very low completeness, about $9.13\%$ but a great pureness, about $89.14\%$.

\begin{table}
  \centering
  $\begin{array}{|c|c|c|c|c|c|} \hline
$rule$	& $total$ \%	& $agn$ \% & $not agn$ \%	& $pureness$	& $contamin.$ \\ \hline \hline
$CG$	& $75.54\%$	& $55.63\%$	& $86.74\%$	& $68.49\%$	& $31.51\%$ \\ \hline
$SCG$	& $75.74\%$	& $55.12\%$	& $87.23\%$	& $68.36\%$	& $31.64\%$ \\ \hline
$QNA$	& $75.99\%$	& $55.64\%$	& $87.44\%$	& $71.38\%$	& $28.62\%$ \\ \hline
$SVM$   & $75.76\%$ & $55.41\%$ & $87.20\%$ & $70.91\%$ & $29.09\%$ \\ \hline
\end{array}$
 \caption[Experiment nr. 1, results.]{Experiment nr. 1, results; the first column is the model used, while the others are the percentages of respectively, total efficiency, AGN completeness, Not-AGN completeness, AGN pureness and AGN contamination.}\label{AGN:riepilogoAGN}
\end{table}

\begin{figure}
   \centering
   \resizebox{\hsize}{!}{\includegraphics{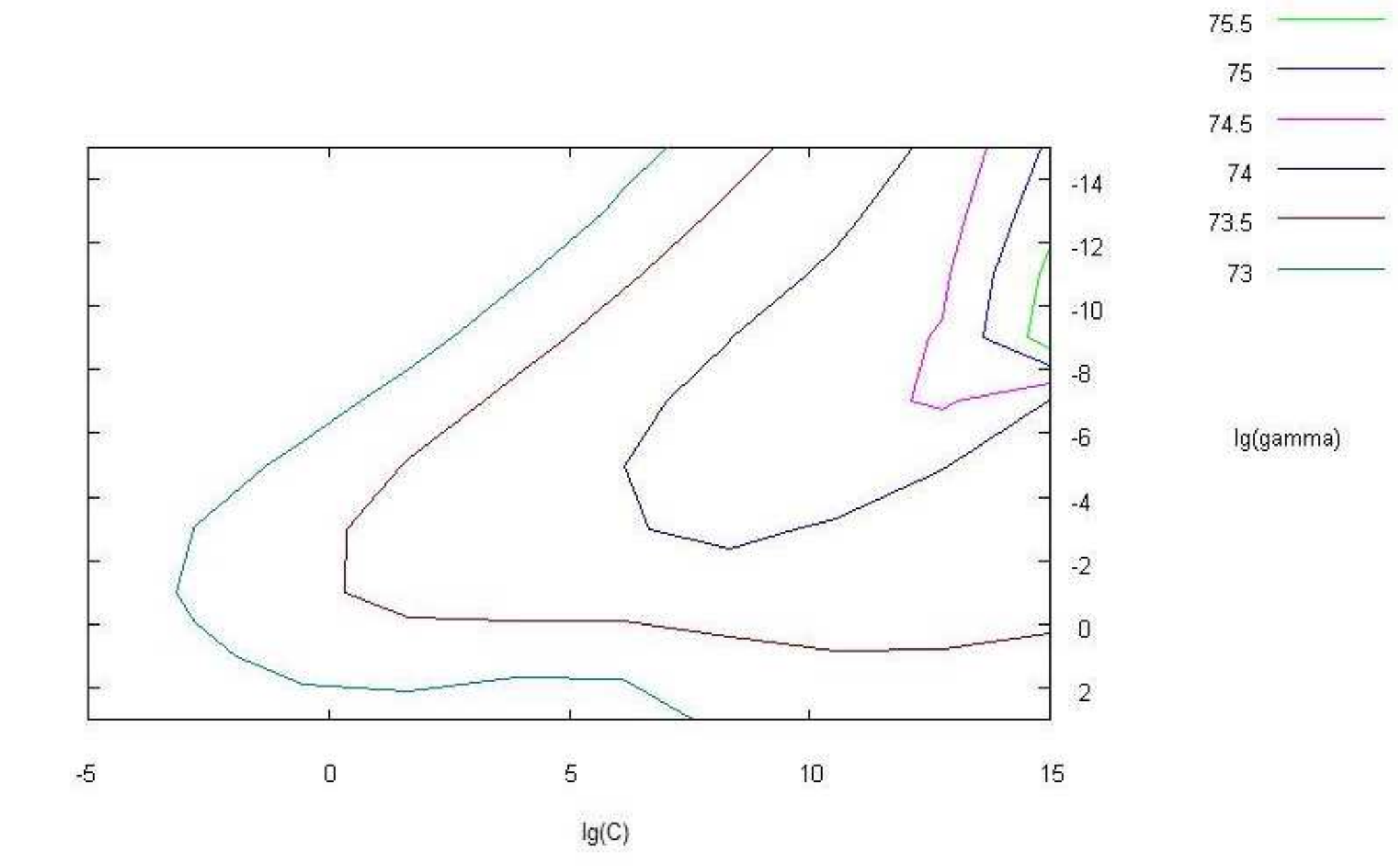}}
     \caption{AGN classification, contour graph of the grid search for the best configuration of $C$, and $\gamma$; $\gamma$ vary as: $C=2^{-5}, 2^{-3}, ... 2^{15}, \gamma  = 2{-15}, 2^{-13}... 2^{3}$}
\label{AGN:Fig:svm1}
\end{figure}

With the SVM we reached a lower efficiency that in the best experiment is equal to $75.76\%$ obtained with  $C = 32768$ and $\gamma
 = 0.001953125$; see figure \ref{AGN:Fig:svm1}.

A single SVM training experiment, in that case, had a huge computational cost (about one week), thus, in order to be able to perform the foreseen $110$ experiments, as described in the section \ref{svmstrategy}, we launched them in parallel by exploiting the SCoPE\footnote{\url{http://www.scope.unina.it}} GRID infrastructure resources \citep{brescia2009} available.

The table \ref{AGN:riepilogoAGN} reports the complete results by using the three MLP learning rules and the SVM.

\subsubsection{Experiment Nr. 2 - Type 1 - Type 2 AGNs separation}

Concerning the second of mentioned three experiments, i.e. classification between type 1 and type 2 objects, the described ML models are feeded using a target vector whose values are labeled as 1 for each object that are classified as seyfert 1 in the catalogue of \cite{sorrentino2006} and as 0 the objects classified as seyfert
2, resulting in 1570 objects after the removal of the patterns affected by NaN values.

According to the mentioned strategy the train set ($80\%$ of the whole dataset) contains $1256$  patterns while the test set ($20\%$ of the dataset) contains $314$ patterns.

The best result has been obtained by the MLP with the Quasi Newton learning rule.\\
The following values summarize the validation performances after the training for the best experiment. \\
\begin{itemize}
\item total efficiency: $e = 99.4\%$
\item type 1 pureness: $p_{type 1} = 98.4\%$
\item type 2 pureness: $p_{type 2} =  100\%$
\item type 1 completeness: $c_{type 1} = 100\%$
\item type 2 completeness: $c_{type 2} = 98.9\%$
\end{itemize}

\begin{table}
  \centering
  $\begin{array}{|c|c|c|c|} \hline
$learning rule$	& $total efficiency$	& $type 1 \% $& $type 2 \%$	\\ \hline \hline
$CG$	& $97.1\%$	& $94.7\%$	& $98.9\%$	\\ \hline
$SCG$	& $95.9\%$	& $93.1\%$	& $97.8\%$	\\ \hline
$QNA$	& $99.4\%$	& $98.4\%$	& $100\%$	\\ \hline
$SVM$   & $81.5\%$  & $75.8\%$  & $75.7\%$       \\ \hline
\end{array}$
 \caption[Experiment nr. 2, results.]{Experiment nr. 2, results; the first column is the model used, while the others are respectively, the total efficiency, type 1 pureness and type 2 pureness.}\label{AGN:riepilogotipo12}
\end{table}

\begin{figure}
   \centering
   \resizebox{\hsize}{!}{\includegraphics{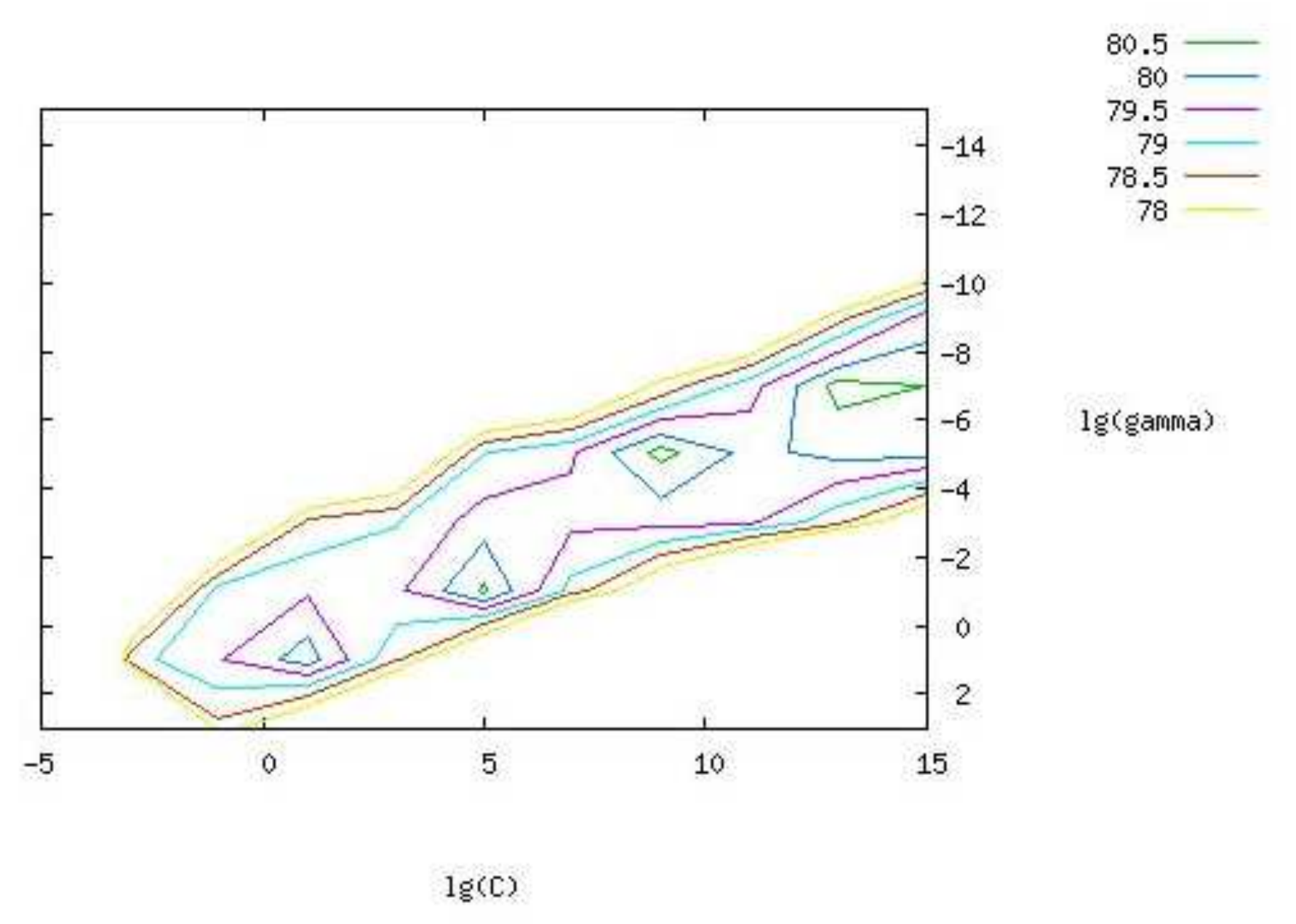}}
     \caption{Type 1 - Type 2 separation, contour graph of the grid search for the best configuration of $C, and \gamma$; $\gamma$ vary as: $C=2^{-5}, 2^{-3}, ... 2^{15}, \gamma  = 2{-15},2^{-13}...2^{3}$}
\label{AGN:Fig:svm2}
\end{figure}

Using the SVM, the best result produced a total efficiency equal to
$81.5\%$ obtained with  $C = 512$ and $\gamma
 = 0.03125$; see figure \ref{AGN:Fig:svm2}.

The table \ref{AGN:riepilogotipo12} reports the complete results by using the three MLP learning rules and the SVM.

\subsubsection{Experiment Nr. 3 - Seyferts - LINERs separation}

Concerning the last experiment, i.e. classification between Seyferts and LINERS objects, the described ML models are feeded using a target vector whose values are labeled as 1 for objects under the Heckman's line and 0 for objects over this line, resulting in 30380 objects after the removal of the patterns affected by NaN presence.

According to the mentioned strategy the train set ($80\%$ of the whole dataset) contains $24304$  patterns while the test set ($20\%$ of the dataset) contains $6076$ patterns.

The best result has been obtained by the MLP with the Quasi Newton learning rule.\\
The following values summarize the validation performances after the training for the best experiment. \\
\begin{itemize}
\item total efficiency: $e = 79.69\%$
\item Seyfert pureness: $p_{Seyfert} = 74.76\%$
\item LINER pureness: $p_{LINER} =  81.09\%$
\item Seyfert completeness: $c_{Seyfert} = 52.77\%$
\item LINER completeness: $c_{LINER} = 91.69\%$
\end{itemize}

\begin{table}
  \centering
  $\begin{array}{|c|c|c|c|} \hline
$learning rule$	& $total efficiency$	& $Seyfert \% $& $LINER \%$	\\ \hline \hline
$CG$	& $78.09\%$	& $72.34\%$	& $79.53\%$	\\ \hline
$SCG$	& $79.36\%$	& $74.37\%$	& $80.74\%$	\\ \hline
$QNA$	& $79.69\%$	& $74.76\%$	& $81.09\%$	\\ \hline
$SVM$   & $78.18\%$ & $71.01\%$ & $80.24\%$ \\ \hline
\end{array}$
 \caption[Experiment nr. 3, results]{Experiment nr. 3, results; the first column is the model used, while the others are respectively, the total efficiency, Seyfert pureness and the LINER pureness.}\label{AGN:riepilogoseyfertliner}
\end{table}

\begin{figure}
   \centering
   \resizebox{\hsize}{!}{\includegraphics{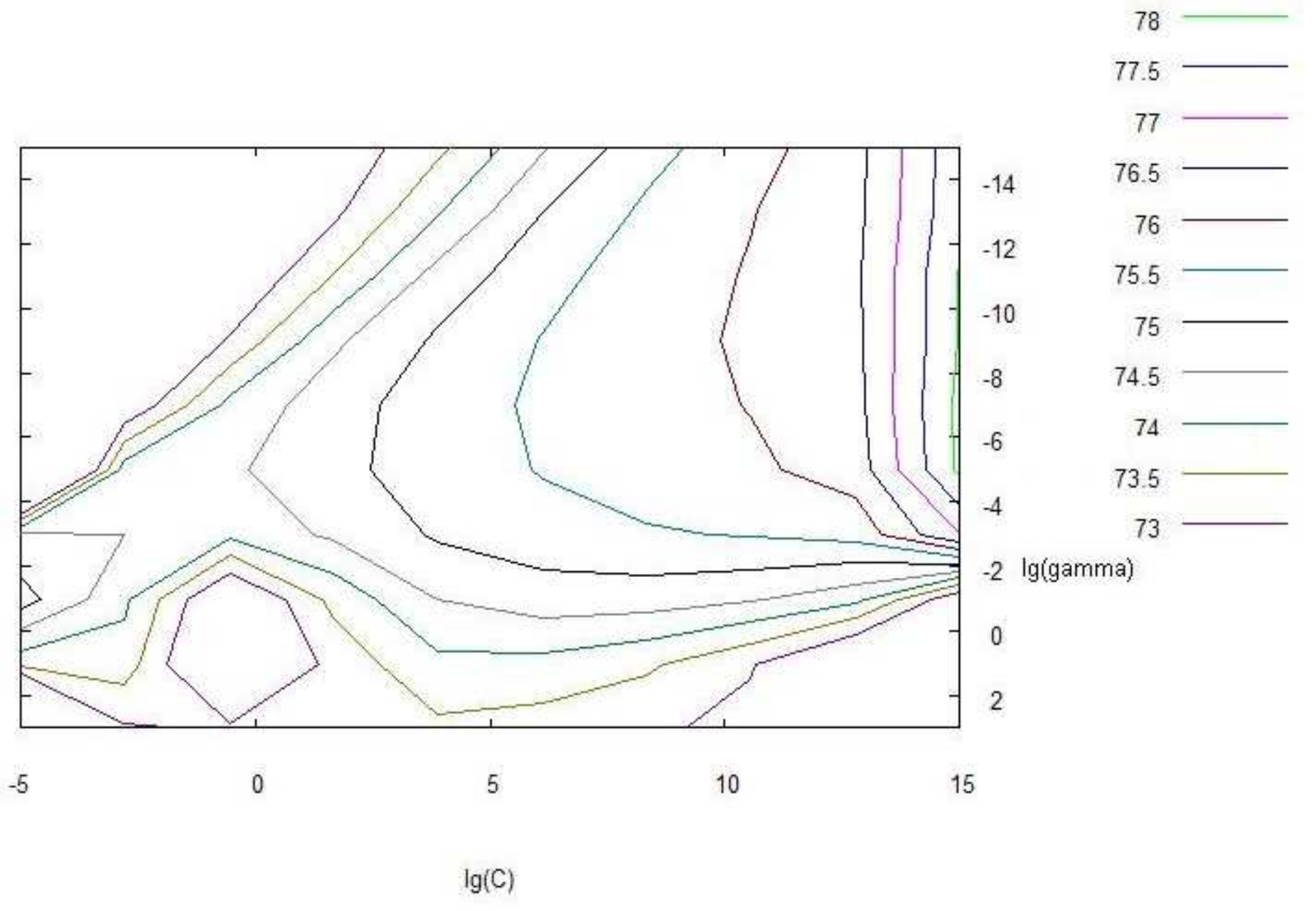}}
     \caption{Seyferts - LINERs separation, contour graph of the grid search for the best configuration of $C, and \gamma$; $\gamma$ vary as: $C=2^{-5}, 2^{-3}, ... 2^{15}, \gamma  = 2{-15},2^{-13}...2^{3}$}
\label{AGN:Fig:svm3}
\end{figure}

Using SVM we reached a total efficiency equal to $78.18\%$ obtained with  $C = 8192$ and $\gamma
 = 0.03125$; see figure \ref{AGN:Fig:svm3}.

The table \ref{AGN:riepilogoseyfertliner} reports the complete results by using the three MLP learning rules and the SVM.

\subsection{Discussion}

In general terms, the three different classification experiments, despite their intrinsic diversity, basically in terms of dataset dimension, heterogenous source and complexity of classification task, have revealed a better behavior of the MLP with QNA model, both in terms of performance and robustness. For instance, it is well known that empirical self-adaptive methods, such as neural networks, particularly suffer a low dimension of data samples, causing a poor learning capability and consequently resulting in a poor efficiency and robustness. On the contrary, the MLP trained by QNA shows a wider capability to optimize the poor information in such cases, due to its fine approximation in the calculation of the inverse Hessian of the training error.

Regarding the experiment nr. 1 we obtain a classifier able to select strong candidates AGN on purely photometric data.
The network is not intended to find all the candidates (great completeness), but to have
very high confidence on the selected candidates (great pureness). The presence of the mixed zone in the KB has a relevant effect on the overall performances of the methods.
The experiments performed show minimum variations in terms of efficiency (less than $1\%$), and the three MLP learning rules show that the different degree of approximation improves the results.
A similar comment may be formulated for the contamination and the purity. \\

Concerning the experiment nr. 2 the SVM seems to perform much worse than the three MLPs, where the separation shows a very high efficiency, that may be caused by the relatively small dimension of the dataset. By concerning the comparison between the three MLP variants, the best result was achieved by the QNA learning rule, although their difference in terms of overall efficiency is quite small, for instance about $4\%$ from the best to the worse MLP learning rule.\\

For what it concerns the last experiment, the separation between Seyfert and LINER types is not so evidently sharp. This probably depends on a low difference between the two categories, as it can be seen in figures \ref{AGN:Fig:bptheck} and \ref{AGN:Fig:bpt}, where it is evident that the separation also in the spectroscopic parameters is far over the Kewley's line. Hence a different kind of KB is probably needed. The results of different models show minimum changes, about $1.5\%$ from the best to the worse result. The consideration about the degree of approximation in the MLP rules is the same as in the experiment nr.1.
Moreover, in terms of purity, there is a bigger difference between the CG learning rule and the other MLP variants; in fact, the CG method performs worse of about $5\%$, which may be due to the lack of capability of separating the two classes in presence of many contaminants.

\section{Summary}
The three different classification experiments, despite their intrinsic diversity, basically in terms of dataset dimension, heterogenous source and complexity of classification task, have revealed a better behavior of the MLP with QNA model, both in terms of performance and robustness.
Moreover we obtained a classifier able to select strong candidates AGN on purely photometric data.
The network is designed not to look for all the candidates (great completeness), but to have
very high confidence on the selected candidates (great pureness)
Concerning to the separation between objects of type 1 and type 2 are
have been also obtained good results although few data in our
possession.
The separation between Seyfert and LINER is not clear and results
 probably depend on a difference between the two categories that is not particularly strong, which in effect
one may observe in figures \ref{AGN:Fig:bptheck} and \ref{AGN:Fig:bpt}, where it is evident that the separation
also in the spectroscopic parameters is far above the line
Kewley regarding the phenomenological aspects.

    \chapter{Regression Problems: Photometric redshifts}\label{chap:photoz}
            \hfill\begin{tabular}{@{}p{.6\linewidth}@{}}
\textit{``Space is big. Really big. You won't believe how hugely mind boggling big it really is."}\\ Douglas Adams.\\ \phantom{aaa}
\end{tabular}

\blfootnote{this chapter is largely extracted from: \tiny
\begin{itemize}
\item Brescia, M.; \textbf{Cavuoti, S.}; D'Abrusco, R.; Longo, G.; Mercurio, A.; 2013, Photo-z prediction on WISE - GALEX - UKIDSS - SDSS Quasar Catalogue, based on the MLPQNA model, \textbf{Submitted to ApJ}
\item \textbf{Cavuoti, S.}; Brescia, M.; Longo, G.; Mercurio, A.; 2012, Photometric Redshifts with Quasi Newton Algorithm (MLPQNA). Results in the PHAT1 Contest, \textbf{A\&A, Vol. 546, A13, pp. 1-8}
\end{itemize}}
Redshift happens when light seen coming from an object that is moving away is proportionally
increased in wavelength, either by doppler effect or by cosmological expansion.


In the early part of the twentieth century, Slipher, Hubble and others made the first measurements of the ``red" and ``blue" shifts of galaxies beyond the Milky Way. They initially
interpreted these redshifts and blueshifts as due solely to the Doppler effect, but later first Lundmark and then Hubble
discovered a rough correlation between the increasing redshifts and the increasing distance of
galaxies. Theorists almost immediately realized that these observations could be explained by a
different mechanism for producing redshifts. Hubble's law of the correlation between redshifts
and distances was in fact required by cosmological models derived from general relativity that have a
metric expansion of space. As a result, photons propagating through the expanding space
are stretched, creating a cosmological redshift from the observational point o view.

When a spectrum can be obtained, determining
the redshift is rather straight-forward: if you can
localize the spectral fingerprint of a common
element, such as hydrogen, then the redshift
can be computed using simple arithmetic by comparing the observed wavelength with the expected in laboratory.
But similarly to the case of Star/Quasar classification,
the task becomes much more difficult when only
photometric observations are available.
In fact in this case, no spectral features can be observed and an estimate of the redshift must be done only with photometrical features.
Because of the spectrum shift, an identical source at different redshifts will have a different color through each pair of filters as illustrated in the figure \ref{photoz:fig1}.

\begin{figure}
  \centering
  \includegraphics[width=12cm]{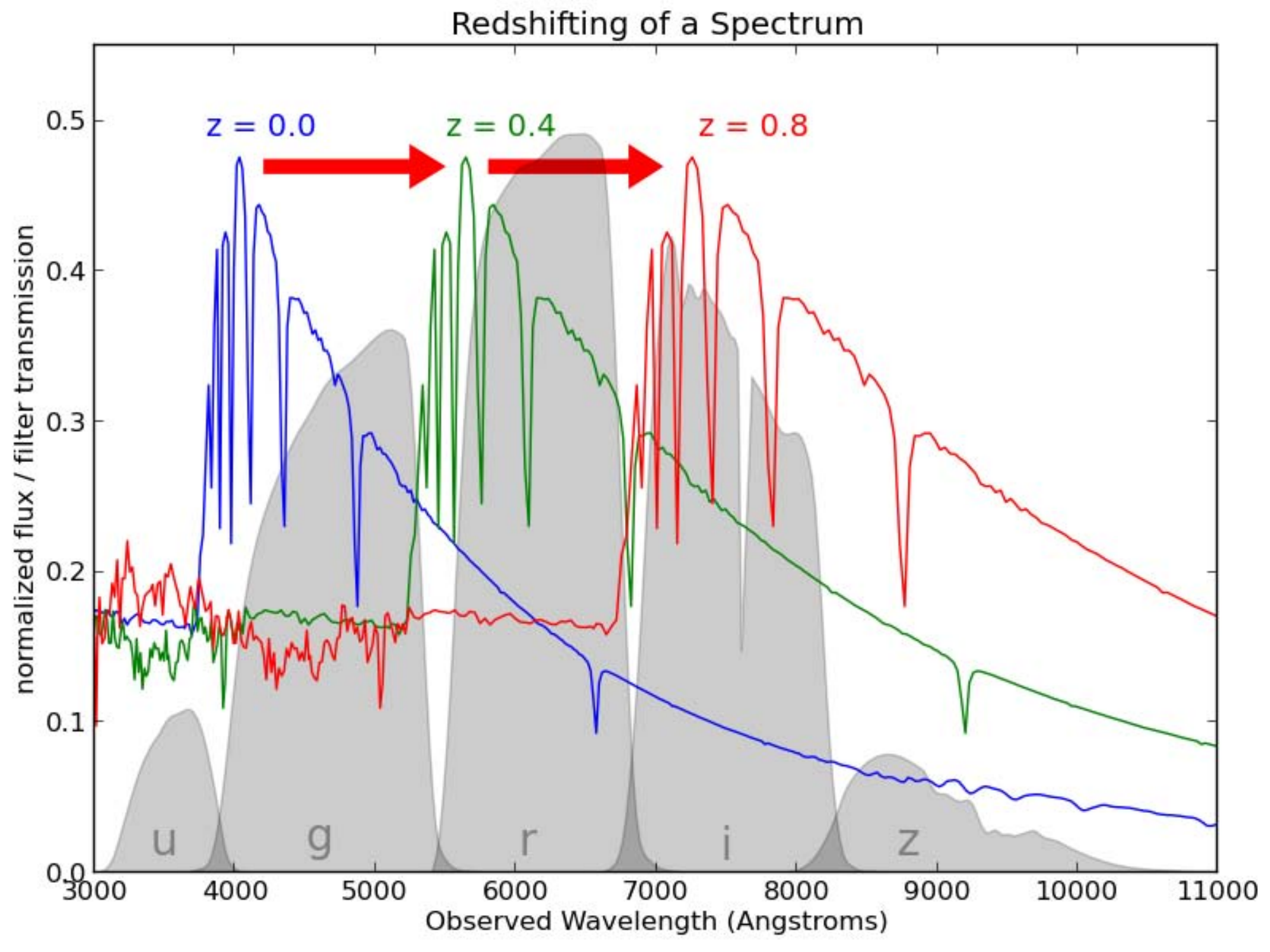}\\
  \caption[The spectrum of the star Vega ($\alpha$-Lyr) at three different redshifts.]{The spectrum of the star Vega ($\alpha$-Lyr) at three different redshifts. The SDSS ugriz filters are shown in gray for reference.\\ At redshift $z=0$, the spectrum is bright in the u and g filters, but dim in the i and z filters. At redshift $z=0.8$, the opposite is the case. This suggests the possibility of determining redshift from photometry alone. The situation is complicated by the fact that each individual source has unique spectral characteristics, but nevertheless, these photometric redshifts are often used in astronomical applications.\\ Credit of: \cite{pedregosa2011}} \label{photoz:fig1}
\end{figure}

Photometric redshifts are a typical example of the data mining functionality called regression (see sec. \ref{sec:functionality}).

Estimating redshifts of celestial objects is one of the most
  pressing technological issues in observational astronomy and, since
  the advent of modern multiband digital sky surveys, photometric
  redshifts (photo-z) have become fundamental when it is necessary
  to know the distances of million of objects over large cosmological
  volumes. Photo-z's provide redshift estimates for objects fainter
than the spectroscopic limit and turn out to be much more efficient in terms
of the number of objects per telescope time with respect to spectroscopic ones (spec-z).
For these reasons, after the advent of modern panchromatic digital surveys, photo-z's have
become crucial. For instance, they are essential in constraining dark
matter and dark energy studies by means of weak gravitational lensing for the identification of galaxy clusters and groups
(e.g. \citealt{capozzi2009}), for type Ia supernovae, and for the study of the
mass function of galaxy clusters
\citep{albrecht2006,peacock2006,keiichi2012}.

The need for fast and reliable methods of photo-z evaluation will become even greater in
the near future for exploiting ongoing and planned surveys.
In fact, future large-field public imaging projects, such as KiDS
  (Kilo-Degree
  Survey\footnote{http://www.astro-wise.org/projects/KIDS/}), DES
  (Dark Energy Survey\footnote{http://www.darkenergysurvey.org/}),
  LSST (Large Synoptic Survey
  Telescope\footnote{http://www.lsst.org/lsst/}), and Euclid
  (\citealt{euclidredbook}), require extremely accurate photo-z's to obtain
  accurate measurements that do not compromise the survey's scientific goals. This explains the very rapid growth in the
number of methods that can be more or less effectively used to derive
photo-z estimates and in the efforts made to better understand and
characterize their biases and systematics.  The possibility of achieving
a very low level of residual systematics \citep{huterer2006,dabrusco2007,laurino2011} is in fact strongly
influenced by many factors: the observing strategy, the accuracy of
the photometric calibration, the different point spread function in
different bands, the adopted de-reddening procedures, etc.

The evaluation of photo-z's is made possible by the existence of a rather
complex correlation existing between the fluxes, as measured in broad
band photometry, the morphological types of the galaxies, and their
distance. The search for such a correlation (a nonlinear mapping
between the photometric parameter space and the redshift values) is
particularly suited to data mining methods.  Existing methods can be
broadly divided into two large groups: theoretical and empirical.
Theoretical methods use templates, such as libraries of either observed
  galaxy spectra or model spectral energy distributions (SEDs). These templates can be
  shifted to any redshift and then convolved with the transmission
  curves of the filters used in the photometric survey to create the
  template set for the redshift estimators (e.g. \citealt{koo1999}, \citealt{massarotti2001a}, \citealt{massarotti2001b}, \citealt{csabai2003}). However, for datasets in which accurate and multiband photometry for a large number of objects are complemented by  spectroscopic redshifts, and for a statistically significant subsample of the same objects, the empirical methods offer greater accuracy, as well as being far more efficient. These methods use the
  subsample of the photometric survey with spectroscopically-measured
  redshifts as a \textit{training set} to constrain the fit of a polynomial
function mapping the photometric data as redshift estimators.

Several template-based methods have been developed
to derive photometric redshifts with increasingly high precision
such as \textit{BPZ\footnote{http://acs.pha.jhu.edu/~txitxo/bpzdoc.html}, HyperZ\footnote{http://webast.ast.obs-mip.fr/hyperz/},
Kcorrect\footnote{http://cosmo.nyu.edu/blanton/kcorrect/}, Le PHARE\footnote{http://www.cfht.hawaii.edu/~arnouts/LEPHARE/lephare.html}, ZEBRA\footnote{http://www.exp-astro.phys.ethz.ch/ZEBRA},
LRT Libraries\footnote{http://www.astronomy.ohio-state.edu/~rjassef/lrt/},  EAzY\footnote{http://www.astro.yale.edu/eazy/}, and Z-PEG\footnote{http://imacdlb.iap.fr:8080/cgi-bin/zpeg/zpeg.pl}}. Moreover there are also training-based methods, such as \textit{AnnZ\footnote{http://www.homepages.ucl.ac.uk/~ucapola/annz.html} and RFPhotoZ\footnote{http://www.sdss.jhu.edu/~carliles/photoZ/RFPhotoZ/}.}
The variety of methods and approaches and their application to
different types of datasets, as well as the adoption of different and
often not comparable statistical indicators, make it difficult to
evaluate and compare performances in an unambiguous and homogeneous
way. Blind tests of photo-z's that one useful but limited in scope have been
performed in \cite{hogg1998} on spectroscopic data from the Keck
telescope on the Hubble Deep Field (HDF), in \cite{hildebrandt2008}
on spectroscopic data from the VIMOS VLT Deep Survey (VVDS;
\citealt{lefevre2004}) and the FORS Deep Field (FDF; \citealt{noll2004},
and in \citealt{abdalla2008}) on the sample of luminous red galaxies
from the SDSS-DR6.

A significant advance in comparing different methods has been introduced by
Hildebrandt and collaborators (\citep{hildebrandt2010}), with the so-called PHAT (PHoto-z Accuracy Testing) contest, which adopts a
black-box approach that is typical of benchmarking. Instead of
insisting on the subtleties of the data structure, they performed a
homogeneous comparison of the performances, concentrating the analysis
on the last link in the chain: the photo-z's methods themselves.

As pointed out by the authors, in fact, ``{\it it is clear that the two regimes - data and method - cannot be separated cleanly
because there are connections between the two. For example, it is highly likely that one method of photo-z estimation will
perform better than a second method on one particular dataset while the situation may well be reversed on a different data
set.}" (cf. \citealt{hildebrandt2010}).\\

Considering that empirical methods are trained on real data and do not
require assumptions on the physics of the formation and evolution of stellar populations, neural networks (hereafter NNs) are
excellent tools for interpolating data and extracting patterns and trends (cf. the standard textbook by \citealt{bishop2006}). In
this section I show the application both to the galaxy (in particular the PHAT1 contest) and to the quasars of the MLPQNA that has been employed for the first time to interpolate the photometric redshifts.



        \section{The PHAT contest}\label{sec:phat}

First results from the PHAT contest were presented in
\cite{hildebrandt2010}, but the contest still continues on the
project's web site. PHAT provides a standardized test environment
that consists of simulated and observed photometric catalogs complemented by additional materials like filter curves
convolved with transmission curves, SED templates, and training
sets. The PHAT project has been conceived as a blind contest, still open to host new participants who want to test their own regression method performances, as in our case, since we developed our model in the past two years.
However, the subsets used to evaluate the performances are
still kept secret in order to provide a more reliable comparison of
the various methods.  Two different datasets are available
(see \citealt{hildebrandt2010} for more details).

The first one, indicated as PHAT0, is based on a very limited
template set and a long-wavelength baseline (from UV to
mid-IR). It is composed of a noise-free catalog with accurate
synthetic colors and a catalog with a low level of additional
noise.
PHAT0 represents an easy case for testing the most basic elements
of photo-z estimation and identifying possible low-level discrepancies
between the methods.

The second one, which is the one used in the present work, is
the PHAT1 dataset, which is based on real data originating in
the Great Observatories Origins Deep Survey Northern field (GOODS-North;
\citealt{giavalisco2004}). According to \cite{hildebrandt2010}, it
represents a much more complex environment to test methods to
estimate photo-z's, pushing codes to their limits and revealing
more systematic difficulties.  Both PHAT test datasets are made
publicly available through the PHAT
website\footnote{\url{http://www.astro.caltech.edu/twiki_phat/bin/view/Main/GoodsNorth}},
while in \cite{hildebrandt2010} there is a detailed description of
the statistical indicators used for comparing the
results provided by the 21 participants who have so far participated by
submitting results obtained with 17 different photo-z codes.

The PHAT1 dataset consists of photometric observations, both from
ground and space instruments, presented in \cite{giavalisco2004},
complemented by additional data in other bands derived from
\cite{capak2004}. The final dataset covers the full UV-IR range and
includes 18 bands: U (from KPNO), B, V, R, I, Z (from SUBARU),
F435W, F606W, F775W, F850LP (from HST-ACS), J, H (from ULBCAM), HK
(from QUIRC), K (from WIRC), and 3.6, 4.5, 5.8, and 8.0 $\mu$ (from IRAC
Spitzer).

The photometric dataset was then cross correlated with spectroscopic
data from \cite{cowie2004}, \cite{wirth2004}, \cite{treu2005}, and \cite{reddy2006}.
Therefore, the final PHAT1 dataset consists of 1984 objects with
18-band photometry and accurate spectroscopic redshifts.  In the
publicly available dataset a little more than one quarter of the
objects comes with spectroscopic redshifts and can be used as the
knowledge base (KB) for training empirical methods. In this contest, in fact, only 515 objects were made available with the corresponding spectroscopic redshift, while for the remaining 1469 objects the related spectroscopic redshift has been hidden from all participants. The immediate consequence is that any empirical method exploited in the contest was
constrained to using the 515 objects as training set (knowledge base) and the 1469 objects as the test set, to be delivered to PHAT contest board in order to receive the statistical evaluation results back.
While it is clear that the limited amount of objects in the knowledge base is not
enough to ensure the best performances of most empirical methods,
the fact that all methods must cope with similar difficulties makes
the comparison consistent.
\subsection{The experiment workflow}
\label{phat:experiments}
In this subsection we describe the details of the sequence of concatenated computational steps performed in order to determine photometric redshifts. This is what we intended as a workflow, which can be seen also as the description of the procedure building blocks. The MLPQNA method was applied by following the standard machine learning (ML) workflow (\citealt{bishop2006}),
which is summarized here: {\it i)}
extraction of the KB by using the 515 available spectroscopic
redshifts; {\it ii)} determination of the ``optimal" model parameter
setup, including pruning of data features and training/test with the
available KB; {\it iii)} application of the tuned model to measure
photometric redshifts on the whole PHAT1 dataset of N=1984 objects, by
including also the re-training on the extended KB.  We also follow the
rules of the PHAT1 contest by applying the new method in two different
ways, first to the whole set of 18 bands and then only to the 14 non-IRAC
bands.
In order to better clarify what is discussed more in the next subsections, it is important to stress that the 515 objects, the publicly available spectroscopic redshifts, have been used to tune our model. In practice, 400 objects have been used as a training set and the remaining 115 as a test/validation set (steps {\it i)} and {\it ii)} of the workflow, see Sec.s~\ref{phat:sub:i},~\ref{phat:sub:ii}).
After having tuned our model, we performed a full training on all 515 objects, in order to exploit all the available knowledge base (see Sec.~\ref{Phat:sec:4.3}).

\subsubsection{Extraction of the knowledge base}
\label{phat:sub:i}

For supervised methods it is common praxis to split the KB into at least
three disjoint subsets: one (training set) to be used for training
purposes, i.e. to teach the method how to perform the regression; the
second one (validation set) to check against loss of generalization
capabilities (also known as overfitting); and the third one (test set)
to evaluate the performances of the model.  As a rule of
thumb, these sets should be populated with 60\%, 20\% and 20\% of the
objects in the KB. In order
to ensure a proper coverage of the parameter space (PS), objects in
the KB are divided up among the three datasets by random extraction, and
usually this process is iterated several times to minimize the biases introduced by fluctuations in the coverage of the PS.

In the case of MLPQNA described here, we used cross-validation
(cf. \citealt{geisser1975}) to minimize the size of the
validation set ($\sim 10\%$). Training and validation were therefore
performed together using $\sim 80\%$ of the objects as a training set
and the remaining $\sim 20\%$ as test set (in practice 400 records in
the training set and 115 in the test set). To ensure proper coverage of the PS, we checked that the randomly extracted
populations had a spec-z distribution that is compatible with that of the
whole KB.  The automatized process of the cross-validation was done by
performing ten different training runs with the following procedure: (i) we split the training set into ten
random subsets, each one composed of 10\% of the dataset; (ii) at each training run we
apply the 90\% of the dataset for training and the excluded 10\% for validation.
This procedure is able to avoid overfitting on the training set (\citealt{bishop2006}).
There are several variants of cross validation methods \citep{sylvain2010}. We  have chosen the k-fold cross validation in particular,
because it is particularly suitable in the presence of a scarcity of known data samples \citep{geisser1975}.

\subsubsection{Model optimization}
\label{phat:sub:ii}

As is known, supervised machine learning models are powerful methods for learning the hidden correlation between input and
output features from training data.  Of course, their generalization and prediction
capabilities strongly depend on the intrinsic quality of data
(signal-to-noise ratio), level of correlation inside of the PS, and the amount of missing data present in the dataset.  Among the factors
that affect performances, the most relevant is that most ML
methods are very sensitive to the presence of Not a Number (NaN)
in the dataset to be analyzed (\citealt{vashist2012}).  This is especially relevant in
astronomical dataset where NaNs may either be nondetections
(i.e. objects observed in a given band but not detected
since they are below the detection threshold) or related to patches of
the sky that have not been observed.  The presence of features with a
large fraction of NaNs can seriously affect the performances of a
given model and lower the accuracy or the generalization capabilities
of a specific model.  It is therefore good praxis to analyze the
performance of a specific model in presence of features with large
fractions of NaNs. This procedure is strictly related to the so
called feature selection or ``pruning of the features" phase which
consists in evaluating the significance of individual features to the
solution of a specific problem.  In what follows we briefly
discuss the outcome of the ``pruning" performed on the PHAT1 dataset.

\subsubsection{Pruning of features}
\label{phat:subsub:pru}

It is also necessary to underline that especially in the presence of small datasets, there is a need for  compromise. On the one hand, it is necessary to minimize the effects of NaNs; on the other, it is not possible to simply remove each record containing NaNs, because otherwise too much information would be lost.

In Table \ref{phat:Tab:pruning} we list the percentage of NaNs in each photometric band, both in the training and the full datasets. Poor features, namely the fluxes in the K and m5.8 bands, were not used for the subsequent analysis. As shown this difference remains always under 3\%, demonstrating that the two datasets are congruent in terms of NaN quantity.

\begin{table}
\centering
\tiny
\begin{tabular}{|c|c|c|c|c|c|c|}
  \hline
BAND & Dataset column ID & \% NaN in whole set & \% NaN in Training & NaN \% absolute difference
\\ \hline m5.8 & 17 & 19.35 & 17.28 & 2.07
\\K & 14 & 17.14 & 18.64 & 1.5
\\HK & 13 & 5.65 & 6.21 & 0.57
\\m8 & 18 & 3.48 & 3.5 & 0.02
\\F435W & 7 & 2.67 & 1.75 & 0.92
\\H & 12 & 2.37 & 2.52 & 0.16
\\J & 11 & 1.16 & 1.55 & 0.39
\\U & 1 & 1.01 & 1.17 & 0.16
\\R & 4 & 0.15 & 0.19 & 0.04
\\B & 2 & 0.1 & 0.19 & 0.09
\\V & 3 & 0.05 & 0.19 & 0.14
\\F606W & 8 & 0.05 & 0 & 0.05
\\m 3.6 & 15 & 0.05 & 0 & 0.05
\\I & 5 & 0 & 0 & 0
\\Z & 6 & 0 & 0 & 0
\\F775W & 9 & 0 & 0 & 0
\\F850LP & 10 & 0 & 0 & 0
\\m4.5 & 16 & 0 & 0 & 0
\\
  \hline
\end{tabular}
  \caption[The percentages of Not a Number in the datasets.]{The percentages of Not a Number (NaN) in the whole dataset (col 3), with 1984 objects and in the trainset (col 4) with 515 objects, for each band. \newline The last column reports the absolute differences between the two NaN percentages.} \label{phat:Tab:pruning}

\end{table}

The pruning was performed separately on the two PHAT1 datasets (18-bands and 14-bands). A total of 37 experiments was run on the two datasets, with the various experiments differing in the groups of features removed. We started by considering all features (bands), removing the two worst bands, for instance K and m5.8, whose outlier quantity was over the 15\% of patterns. Then a series of experiments was performed by removing one band at a time, by considering the NaNs percentage shown in Table \ref{phat:Tab:pruning}.

\subsubsection{Performance metrics}
\label{phat:subsub:met}

The performances of the various experiments were evaluated (as done in the PHAT contest) in terms of
\begin{itemize}
\item {\it scatter}: the RMS of $\Delta z$
\item {\it bias}: the mean of $\Delta z$
\item {\it fraction of outliers}: where outliers are defined by the condition: $\left|\Delta z \right| > 0.15$,
\end{itemize}
where
\begin{equation}
\Delta z \equiv \frac{z_{spec} - z_{phot}}{1+z_{spec} }.
\label{PHAT:eq8}
\end{equation}

\begin{table}
\centering
\tiny
\begin{tabular}{|l|rrrrrrrrrrr|}
\hline
exp. n & missing features & feat. & hid.& step	&res. &	dec.   &	MxIt&	CV&	scatter&	outliers$\%$&	bias  \\
\hline
37	   & m5.8,K, HK, m8	 & 14	   & 29	  & 0.0001&	30	& 0.1	 & 3000	  & 10& 0.057	 & 22.61\%  & -0.0077\\
26	   & m5.8, K, m3.6, m4.5, HK, m8& 12 & 25	  & 0.0001&	30	& 0.1	 & 3000	  & 10&	0.062	 & 17.39\% 	& 0.0078 \\
\hline
\end{tabular}
\caption[Description of the best experiments for the  18 bands and the 14 bands datasets.]{Description of the best experiments for the  18 bands (Exp. n. 37) and the 14 bands datasets (Exp. n. 26).
\newline Column 1: sequential experiment identification code; column 2: features not used in the experiment; columns 3-4: number
of input (features) and hidden neurons; column 5--9: parameters of the MLPQNA used during the experiment; column 10: scatter error evaluated as described in the text; column 11: fraction of outliers; column 12: bias.}\label{phat:Tab:exp1}
\end{table}
At the end of this process, we obtained the best results, reported in Table \ref{phat:Tab:exp1}.

\subsection{Application to the PHAT1 dataset}\label{Phat:sec:4.3}

We performed a series of experiments in order to fine tune the model
parameters, whose best values are\\
MLP network topology parameters (see Table ~\ref{phat:Tab:exp1}):
\begin{itemize}
\item feat: 14 (12) input neurons (corresponding to the pruned number of input band magnitudes listed in Table ~\ref{phat:Tab:pruning}),
\item hid: 29 (25) hidden neurons,
\item 1 output neuron.
\end{itemize}
QNA training rule parameters (see Table ~\ref{phat:Tab:exp1}):
\begin{itemize}
\item step: 0.0001 (one of the two stopping criteria. The algorithm stops if the approximation error step size is less than this value. A step value equal to zero means to use the parameter MxIt as the unique stopping criterion.);
\item res : 30 (number of restarts of Hessian approximation from random positions, performed at each iteration);
\item dec : 0.1 (regularization factor for weight decay. \\The term $dec*||network weights||^2$ is added to the error function, where $network weights$ is the total number of weights in the network. When properly chosen, the generalization error of the network is highly improved);
\item MxIt: 3000 (max number of iterations of Hessian approximation. If zero the step parameter is used as stopping criterion);
\item CV: 10 (k-fold cross validation, with k=10. This parameter is described in Sec. 4.1).
\end{itemize}
With these parameters, we obtained the statistical results (in terms of scatter, bias, and outlier percentage) as reported in the last three columns of Table ~\ref{phat:Tab:exp1}.

\noindent Once the model optimization described above had been
determined, the MLPQNA was re-trained on the
whole KB (515 objects) and applied to the whole PHAT1 dataset
(1984 objects), which was then submitted to the PHAT contest
for final evaluation (see below).

Details of the experiments can be found on the DAME web site\footnote{\url{http://dame.dsf.unina.it/dame_photoz.html}}, while the parameter settings and the results for the best models are summarized in Table \ref{phat:Tab:results}.

\subsection{The PHAT1 results and comparison with other models}\label{phat:PR}

With the model trained as described in the above subsection, we calculated photometric redshifts for the entire PHAT1 dataset, i.e. also for the remaining 1469 objects, for which the corresponding spectroscopic redshift was hidden to the contest participants, obtaining a final photometric catalog of 1984 objects. This output catalog has finally been delivered to the PHAT contest board, receiving the statistical results (scatter, bias and outlier's percentage) as feedback coming from the comparison between spectroscopic and photometric information, in both cases (18 and 14 bands).\\
So far, the statistical results and plots have referred to the whole data sample, which is kept secret from all participants as required by the PHAT contest, were provided by H. Hildebrandt and also reported in the PHAT Contest wiki site \footnote{\url{http://www.astro.caltech.edu/twiki_phat/bin/view/Main/GoodsNorthResults\#Cavuoti_Stefano_et_al_neural_net}}.
So far, the results obtained by analyzing the photometric redshifts calculated by MLPQNA, are shown in Table \ref{phat:Tab:results}.

The most significant results can be summarized as follows:

\begin{description}
\item[i)] 18-band experiment: 324 outliers with $\left|\Delta_z\right| > 0.15$, corresponding to a relative fraction of $16.33\%$.
For the remaining 1660 objects bias and rms are $0.000604251 \pm 0.0562278$;
\item[ii)] 14-band experiment: 384 outliers with $\left|\Delta_z\right| > 0.15$, corresponding to a relative fraction of $19.35\%$;
1600 objects with bias and variance $ 0.00277721 \pm  0.0626341$.
\end{description}

A more detailed characterization of the results can be found in the
first line of parts A, B, and C in Table \ref{phat:Tab:results}, while
Fig. \ref{PHAT:scatter}, provided by H. Hildebrandt, gives the scatter plots (spec-z's vs
photo-z's) for the 18 and 14 bands.

 \begin{figure*}
   \centering
   \includegraphics[width=7cm]{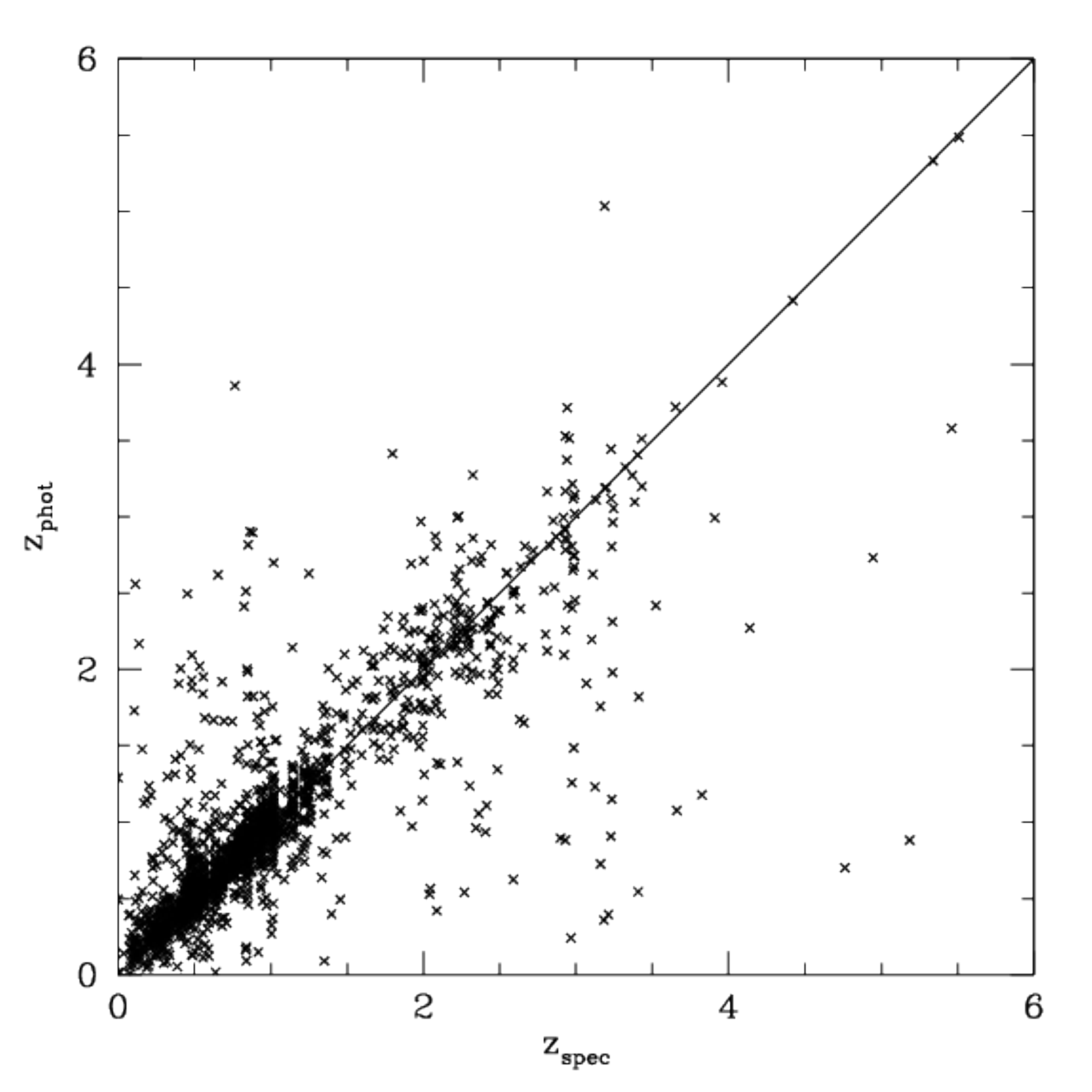} (a)
   \includegraphics[width=7cm]{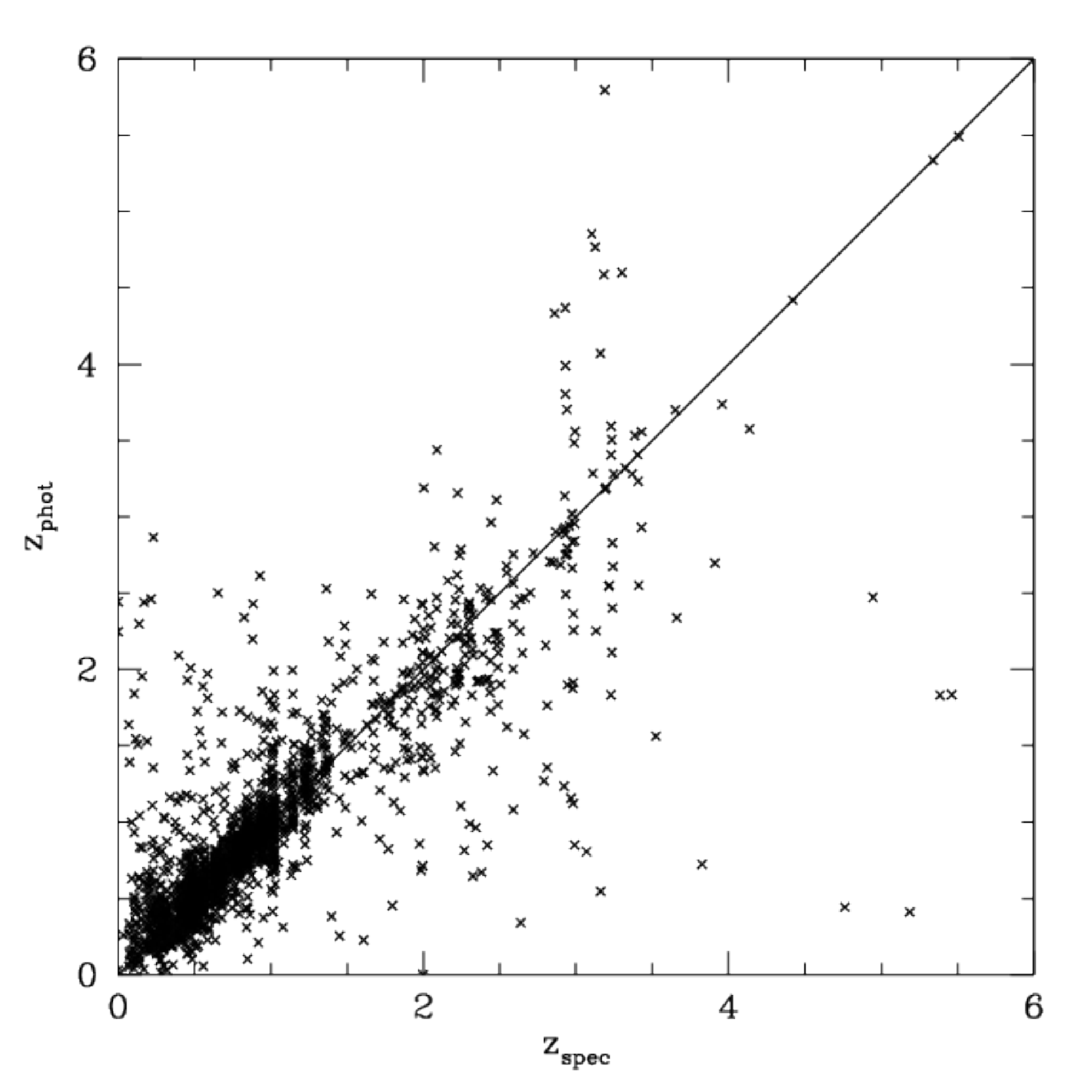} (b)
   \caption[Results obtained by our model and provided by the PHAT contest board.]{Results obtained by our model and provided by the PHAT contest board in terms of direct comparison between our photometric and blind spectroscopic information.  The (a) panel plots the photometric vs. spectroscopic redshifts for the whole dataset
     using 10 photometric bands (Experiment 37). In panel (b) the same
     but using only 14 photometric bands (Experiment
     26). (Courtesy of H. Hildebrandt).}\label{PHAT:scatter}
   \end{figure*}

\begin{sidewaystable}
\tiny
\centering
\begin{center}

\begin{tabular}{|l|rrr|rrr|rrr|rrr|}
\hline
\hline
{\bf A}	    &\multicolumn{3}{|c|}{18-band; $\left|\Delta z\right| \leq 0.15$}
            &\multicolumn{3}{|c|}{14-band; $\left|\Delta z\right| \leq 0.15$}				
            &\multicolumn{3}{|c|}{18-band; R$< 24$; $\left|\Delta z\right| \leq0.15$}
            & \multicolumn{3}{|c|}{14-band; R$< 24$; $\left|\Delta z\right| \leq 0.15$}\\
\hline		
Code	& bias	  &scatter& outliers \% &	bias	&scatter &outliers \% &bias	  &scatter &outliers \% &bias	    &scatter &outliers \% \\
\hline
QNA	    & 0.0006 &0.056	&16.3	& 0.0028 &0.063	&19.3	& 0.0002 &0.053	&11.7	& 0.0016	&0.060	&13.7\\
AN-e	&-0.010	 &0.074	&31.0	&-0.006 &0.078	&38.5	&-0.013	&0.071	&24.4	&-0.007	&0.076	&32.8\\
EC-e	&-0.001	 &0.067	&18.4	& 0.002 &0.066	&16.7	&-0.006	&0.064	&14.5	&-0.003	&0.064	&13.5\\
PO-e	&-0.009	 &0.052	&18.0	&-0.007 &0.051	&13.7	&-0.009	&0.047	&10.7	&-0.008	&0.046	& 7.1\\
RT-e	&-0.009	 &0.066	&21.4	&-0.008 &0.067	&24.2	&-0.012	&0.063	&16.4	&-0.012	&0.064	&18.4\\
\hline
{\bf B}     &\multicolumn{3}{|c|}{18-band; $\left|\Delta z\right| \leq 0.5$}
            &\multicolumn{3}{|c|}{14-band; $\left|\Delta z\right| \leq 0.5$}
            &\multicolumn{3}{|c|}{18-band; R$< 24$; $\left|\Delta z\right| \leq 0.5$}
            &\multicolumn{3}{|c|}{14-band; R$< 24$; $\left|\Delta z\right| \leq 0.5$}\\
\hline		
Code	& bias	  &scatter & outliers \% &	bias	&scatter & outliers \% &bias	  &scatter &outliers \% &bias	    &scatter & outliers \% \\
\hline
QNA	  &-0.0028	&0.114	&3.8	&-0.0046 &0.125	&3.8	 &-0.0039  &0.101	&1.7	&-0.0039 &0.101	&1.7\\
AN-e	&-0.036	&0.151	&3.1	&-0.035	&0.173	&4.2	 &-0.047	&0.130	&1.4	&-0.047	&0.130	&1.4\\
EC-e	&-0.007	&0.120	&3.6	&-0.003	&0.114	&3.6	 &-0.015	&0.106	&1.9	&-0.015	&0.106	&1.9\\
PO-e	&-0.013	&0.124	&3.1	&0.001	&0.107	&2.3	 &-0.020	&0.098	&1.2	&-0.020	&0.098	&1.2\\
RT-e	&-0.031	&0.126	&3.2	&-0.028	&0.137	&3.6	 &-0.034	&0.111	&1.4	&-0.034	&0.111	&1.4\\
\hline
{\bf C}      &\multicolumn{3}{|c|}{18-band; $z_{sp}\leq 1.5$, $\left|\Delta z\right| \leq 0.15$}
             &\multicolumn{3}{|c|}{14-band; $z_{sp}\leq 1.5$,  $\left|\Delta z\right| \leq 0.15$}				
             &\multicolumn{3}{|c|}{18-band; $z_{sp}>1.5$,      $\left|\Delta z\right| \leq 0.15$}
             &\multicolumn{3}{|c|}{14-band; $z_{sp}>1.5$,      $\left|\Delta z\right| \leq 0.15$}\\
\hline
Code	& bias	  &scatter& outliers \% &	bias	 & scatter & outliers \% &	bias	&scatter &outliers \% &bias	  &scatter & outliers \% \\
\hline
QNA	  &-0.0004	&0.053	&14.6	&0.0001   &0.061	&16.6	&0.0074	&0.072	&26.3	&0.0222	&0.070	&35.0\\
AN-e	&-0.017  &0.070  &27.6	&-0.010	  &0.076	&33.6	&0.051	&0.078	&50.7	&0.045	&0.077	&66.4\\
EC-e	&-0.003	&0.065  &16.1	&-0.000	  &0.064	&14.5	&0.015	&0.077	&32.3	&0.015	&0.077	&29.5\\
PO-e	&-0.012	&0.049  &12.6	&-0.011	  &0.047	&9.4	&0.019	&0.075	&48.3	&0.026	&0.074	&37.7\\
RT-e	&-0.016	&0.062	&19.6	&-0.014	  &0.064	&21.1	&0.040	&0.072	&31.8	&0.039	&0.071	&41.9\\
\hline
\end{tabular}
\end{center}
\caption[Comparison of the performances of our MLPQNA method against all other empirical methods analyzed by PHAT board. ]{Comparison of the performances of our MLPQNA (here labeled as QNA) method against all other empirical methods analyzed by PHAT board. \newline For a description of other methods (namely AN-e, EC-e, PO-e and RT-e) see the text. The table is divided into three parts
(namely A, B and C). Data for the other empirical method have been extracted from \cite{hildebrandt2010}.
In each part of the table we list the results (on both the 18 and the 14 bands datasets) for a specific subsample of the PHAT objects.
Part A: statistical indicators (bias and scatter) for the 18 and 14 bands computed on objects with $\left|\Delta z\right| \leq 0.15$ and for objects with $\left|\Delta z\right|\leq 0.15$ and $R<24$. The column ``outliers" gives the fraction of outliers defined as objects with $\left|\Delta z\right| >0.15$.
Part B: the same but for $\left|\Delta z\right| \leq 0.5$.
Part C: the same but for objects with spectroscopic redshift $z_{sp} \leq 1.5$ and $\left|\Delta z\right| \leq 1.5$, and for $z_{sp} > 1.5$ and $\left|\Delta z\right| \leq 1.5$. The definitions of bias, scatter, and outliers fraction are given in the text.
Values were computed by the PHAT collaboration on the whole PHAT1 dataset.}\label{phat:Tab:results}
\end{sidewaystable}

To compare our results with other models, we also report in
Table \ref{phat:Tab:results} the statistical indicators for the other
empirical methods that competed in the PHAT1 contest.  The methods
are

\begin{itemize}
\item {AN-e}: ANNz, artificial neural network, an empirical photo-z code
  based on artificial neural networks  \citep{collister2004};
\item {EC-e}: Empirical $\chi^2$, a subclass of kernel regression
  methods; which mimics a template-based technique with the main
  difference that an empirical dataset is used in place of the
  template grid  \citep{wolf2009};
\item {PO-e}: Polynomial fit, a ``nearest neighbor" empirical
  photo-z method based on a polynomial fit so that the galaxy redshift
  is expressed as the sum of its magnitudes and colors \citep{li2008};
\item {RT-e}: Regression Trees, based on random forests which
  are an empirical, non-parametric regression technique \citep{carliles2010}.
\end{itemize}
More details can be found in the quoted references and in
\cite{hildebrandt2010}.

For each of the datasets (18 and 14 bands), statistics in Table
\ref{phat:Tab:results} refer to several regimes: the first one (A) defines all objects having $\left| \Delta z \right| > 0.15$ as outliers and it
is divided into two subsections: the left hand side includes all objects,
while the right hand side includes objects brighter than R = 24; the second
one (B) defines objects having $\left| \Delta z \right| >
0.50$ as outliers; the third one (C) defines as
outliers objects having $\left| \Delta z \right| > 0.50$ and divided
into a left side, for objects with $z \leq 1.5$ and a right side having
$z > 1.5$.

By analyzing the MLPQNA performance in the different regimes, we obtained:

{\it All objects}: in the 18 bands experiment, QNA scores the
  best results in term of bias, and gives comparable results with PO-e
  in terms of scatter and number of outliers. In fact, while the scatter is slightly larger in Part A than those of PO-e method (0.052
  against 0.056), the number of outliers is lower (18.0\% against
  16.3\%), and in Part. B is the viceversa (0.124 against 0.114 and
  3.1\% against 3.8\%). In the 14-band experiment QNA obtains values
  slightly higher than PO-e in terms of scatter (0.051 against 0.063)
  and than EC-e in terms of bias (0.002 against 0.0028). For the fraction of outliers, QNA scores turn out to be larger than PO-e
  and EC-e (13.7\% and 16.7\% against 19.3\%).

{\it Bright objects}: for bright objects (R$<$24), the QNA resulting bias is again the best within the different empirical methods,
  while for scatter and number of outliers, QNA obtains slightly
  higher values than PO-e in both the 18 (0.047 against 0.053 and 10.7\%
  against 11.7\%) and the 14 band datasets (0.046 against 0.060 and
  7.1\% against 13.7\%).

{\it Distant vs near objects}: in the distant sample
  ($z_{sp}>1.5$) QNA scores as first in terms of bias, scatter, and
  number of outliers for 18 bands. In the 14-band dataset case, it is the best method in
  terms of scatter, but with a bias (0.015 against 0.0222)
  and number of outliers (29.5\% against 35.0\%) higher than EC-e.
  In the near sample ($z_{sp}<1.5$) QNA is the best in terms of bias. The scatter is slightly higher than PO-e's for both 18 (0.049 against 0.053) and 14 bands (0.047 against 0.061). For outliers, PO-e performs better at
  18 bands (12.6\% against 14.6\%), while PO-e and EC-e perform
  better at 14 bands (9.4\% and 14.5\% against 16.6\%).

\subsection{Summary}
\label{phat:discus}
The MultiLayer Perceptron with Quasi Newton learning rule was applied on the whole PHAT1 dataset of N=1984 objects \cite{hildebrandt2010} ) to determine photometric redshifts after an optimization of the model performed by using the 515 available spectroscopic redshifts as a training set.

The statistics obtained by the PHAT board, by analyzing the photometric redshifts derived with MLPQNA, and the comparison with other
empirical models are reported in Table \ref{phat:Tab:results}. From a quick inspection of Table \ref{phat:Tab:results}, it exists no empirical method that can be regarded as the best in terms of all the indicators (e.g.
bias, scatter, and number of outliers) and that EC-e (Empirical $\chi^2$ method), PO-e (Polynomial Fit method), and MLPQNA produce comparable results.
However, the MLPQNA method, on average, gives the best result in terms of bias in any regime.

By considering the dataset with 18 bands reported in Parts A and B of Table \ref{phat:Tab:results}, MLPQNA obtains
results for the scatter comparable to the PO-e method.
In fact, in Part A, PO-e's scatter is better than MLPQNA, but with more outliers, while the trend is reversed in Part B.
In the other cases both the scatter and number of outliers are slightly worse than with PO-e and EC-e
methods.

In general, MLPQNA seems to have better generalization capabilities
than most other empirical methods especially in the presence of
underpopulated regions of the knowledge base.
In fact, $\sim 500$ objects with spectroscopic redshifts spread over such a large redshift interval are by far not sufficient to train most other empirical codes on the data.
This has also been pointed out by \cite{hildebrandt2010}, who noticed that the high fraction of
outliers produced by empirical methods is on average higher than what is
currently found in the literature ($\sim 7.5 \%$) and explained it
as an effect of the small size of the training sample, which poorly maps the very wide range in redshifts and does not include enough objects with peculiar SEDs.

In this respect we wish to stress that, as already shown in
another application (cf. \citealt{brescia2012a}) and as will be more extensively discussed in section \ref{sec:qso},
MLPQNA enjoys the very rare prerogative of being able to obtain good performances, also when
the KB is small and thus undersampled (Brescia et al. in preparation).

        \section{Redshifts for quasars}\label{sec:qso}

In the last few years it has in fact been demonstrated that, provided an accurate enough photometry and significant wavelength coverage, it is possible to obtain samples of photometrically selected quasars matching the low contamination and high completeness (cf. \citep{dabrusco2009, bovy2012}) required by many fields of modern cosmology.
The relevance of these photometrically selected samples will increase more and more in the near future,
when the new generation of deeper and more accurate surveys will allow to access larger and more complete samples of QSOs. These \textit{photometric} samples are in fact already being used for a variety of applications such as the measurement of the integrated Sachs--Wolfe effect \citep{giannantonio2008}, the cosmic magnification bias \citep{scranton2005},  the clustering of quasars on large \citep{myers2006,myers2007a} and small \citep{hennawi2006} scales, to quote just a few. Since both candidate selection and photometry redshift estimates are performed on the same data (colors in many bands), it is also apparent that for the same samples, photometric data alone should carry enough information to characterize in an almost univocal way the SED and therefore also to derive accurate estimates of photometric redshifts \citep{dabrusco2009,laurino2011,bovy2012}.\\ It goes without saying that the utility of this photometrically selected samples goes hand in hand with the development of photo-z methods capable to provide accurate enough estimates of the redshifts.\\

In this section we present the application of the MLPQNA to the evaluation of photometric redshift of quasars.


\subsection{The Dataset}
\label{qso:sec:thedata}

\begin{sidewaystable*}
\centering
\begin{tabular}{|l|l|l|l|}
  \hline
           Survey & Bands & Name of feature & Synthetic description\\
\hline\hline
GALEX     & NUV, FUV & mag, mag\_iso                                                          & Near and Far UV total and isophotal mags \\
                   & &mag\_Aper\_1 mag\_Aper\_2 mag\_Aper\_3& phot. through 3, 4.5 and 7.5 arcsec apertures\\
                   & & mag\_auto and  kron\_radius & magnitudes and Kron radius in units of A or B\\
\hline
SDSS       & u, g, r, i, z  & psfMag& PSF fitting magnitude in the u g, r, i, z bands.\\
\hline
UKIDSS   & Y, J, H, K  & PsfMag                                & PSF fitting magnitude in $Y, J, H, K$ bands\\
              & & AperMag3, AperMag4, AperMag6                 & aperture photometry through 2,   2.8 \& 5.7$^{\prime\prime}$\\
                    &&&  circular aperture in each band\\
             & & HallMag, PetroMag                                               & Calibrated magnitude within circular \\
                                                                                                                                  &&&aperture r\_hall and Petrosian magnitude\\
             & & & in $Y, J, H, K$ bands\\
\hline
WISE           & W1, W2, W3, W4 & W1mpro, W2mpro, W3mpro, W4mpro
                                                                    & W1: 3.4 $\mu m$ and 6.1$^{\prime\prime}$ angular resolution;\\
                                                                &&& W2: 4.6 $\mu m$ and 6.4$^{\prime\prime}$ angular resolution; \\
                                                                &&&W3: 12 $\mu m$ and 6.5$^{\prime\prime}$ angular resolution;\\
                                                                &&&W4: 22 $\mu m$ and 12$^{\prime\prime}$ angular resolution.\\
                                                                &&& Magnitudes measured with profile-fitting photometry \\	
                                                                &&& at the 95\% level. Brightness upper limit if the flux \\
                                                                &&&measurement has SNR$<2$\\
 \hline\hline  SDSS & - & $z_{spec}$ & Spectroscopic redshift\\
 \hline
\end{tabular}
\caption[Summary of the data extracted from the four surveys databases and merged to form our final catalogue.]{Summary of the data extracted from the four surveys databases and merged to form our final catalogue. Even though most names of the parameters are self explanatory, we wish to remind that the various \textit{PSFMag} are magnitudes derived by integrating fluxes over the best fitting point spread function. The aperture sizes refer to the radii.}\label{qso:tab:features}
   \end{sidewaystable*}

\begin{figure*}
\centering
  \includegraphics[width=10.cm]{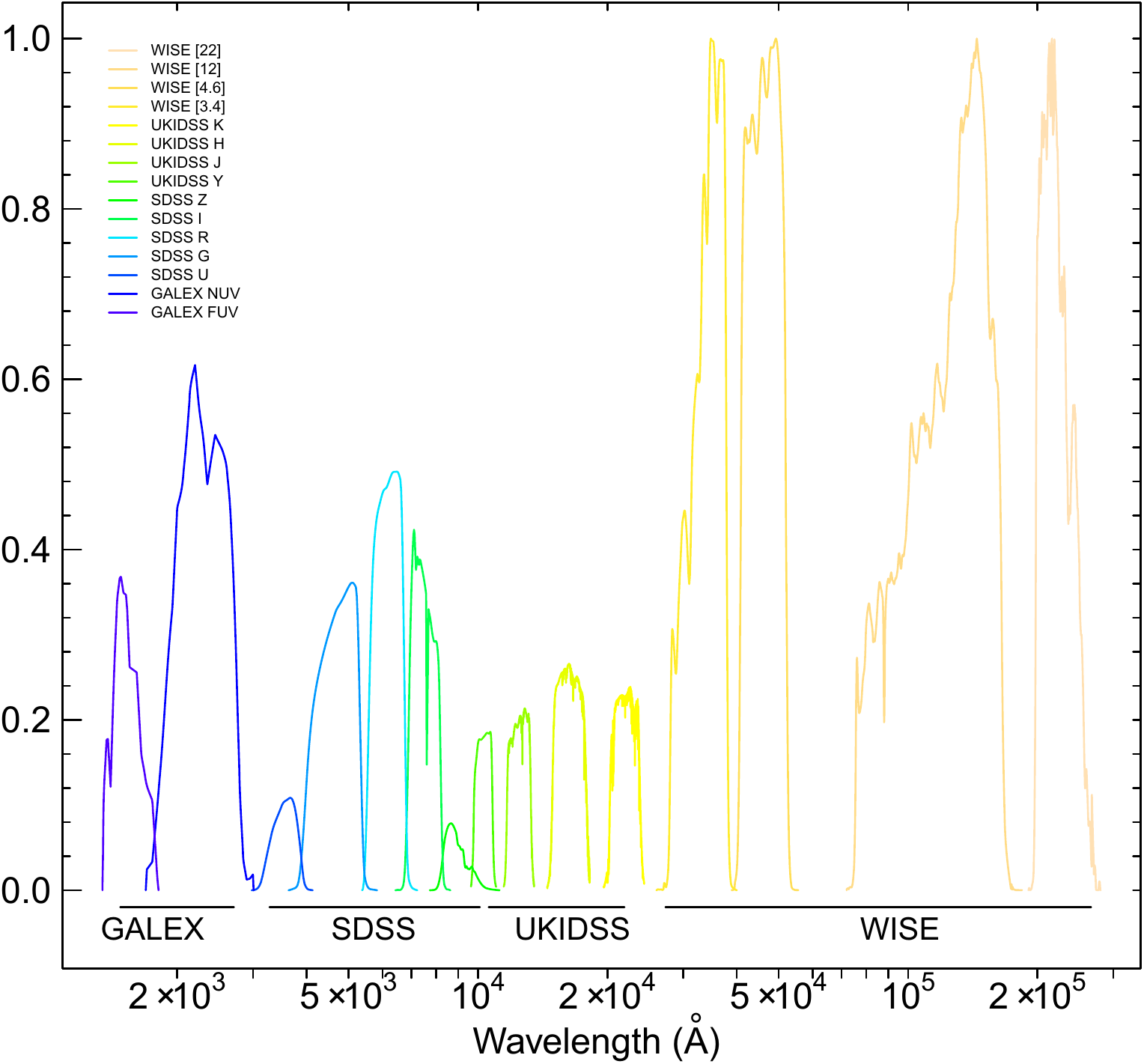}\\
  \caption{Transmission curves for all filters in the four surveys considered.}\label{qso:fig:transmission_curves}
\end{figure*}

The sample of quasars, used in the experiments described in this section, is based on the spectroscopically selected quasars from the SDSS-DR7 database (table "Star" of the SDSS database). According to the spectroscopic classification index ({\it index SP} or {\it specClass}) provided in the SDSS-DR7 release, objects are divided in six different classes: $SP=0$ unknown sources, $SP=1$ stars, $SP=2$ galaxies, $SP=3$ nearby AGN, $SP=4$ quasars; $SP=5$ sky and $SP=6$ late type stars. In order to minimize contamination from other sources, we selected only those with {\it specClass} = \{3,4\}, and for which a reliable measure of the spectroscopic redshifts (with {\it zConf} $>$ 0.90) is available.\\
We then cross-matched the SDSS quasars sample identified as point sources with clean measured photometry in all filters (\textit{ugriz}), with the latest versions of the datasets from: GALEX~\citep{martin2005}, UKIDSS~\citep{lawrence2007} and WISE~\citep{wright2010}. These three surveys observed large fractions of the sky in the ultraviolet, near infrared and middle infrared spectral intervals, respectively.\\
{\bf SDSS - (DR7)}~\citep{aihara2011} has observed $\sim\!1.4\times10^4$  deg$^2$ of the sky in 5 bands (\textit{ugriz}) covering the [3551, 8931] \AA \phantom{ }wavelengths range. Photometric SDSS observations reach the limiting magnitude of 22.2 in the \textit{r} band ($95\%$ completeness for point sources; \cite{abazajian2009}).\\
{\bf GALEX - (DR6/7)}~\citep{martin2005} is a 2-band survey (\textit{nuv, fuv} for near and far ultraviolet respectively) covering the [1300,3000] \AA \phantom{ }wavelength interval. GALEX photometric survey has observed the whole sky to the near ultraviolet limiting magnitude $\textit{nuv}\!=\!20.5$.\\
{\bf UKIDSS - (DR9)}~\citep{lawrence2007} has been designed to be the SDSS infrared counterpart and covers $\sim$7500 deg$^2$ of the sky in the \textit{YJHK} near-infrared bands covering the $\sim$ 0.9 to 2.4 $\mu$m spectral range down to the limiting magnitude \textit{K}=18.3. The Large Area Survey (LAS) has imaged $\sim4000$ deg$^2$ (overlapping with the SDSS), with the additional \textit{Y} band down to the limiting magnitude of 20.5.\\
The {\bf WISE} mission~\citep{wright2010} has observed the entire sky in the mid-infrared spectral interval at 3.4, 4.6, 12, and 22 $\mu$m with an angular resolution of 6.1$^{\prime\prime}$, 6.4$^{\prime\prime}$, 6.5$^{\prime\prime}$ \& 12.0$^{\prime\prime}$ in the four bands, achieving 5$\sigma$ point source sensitivities of 0.08, 0.11, 1 and 6 mJy in unconfused regions on the ecliptic, respectively. The astrometric accuracy of WISE is $\sim 0.50^{\prime\prime}, 0.26^{\prime\prime}, 0.26^{\prime\prime}$, and 1.4$^{\prime\prime}$ for the four WISE bands respectively.\\

All these surveys present a large common overlap region and  overall good astrometry with comparable astrometric accuracy. In order to cross-match the catalogue we used a maximum radius $r=1.5^{\prime\prime}$ to associate the optical quasars to counterparts in each of the three catalogs.
Afterwords we rejected all sources with missing data (or NaN) in any of the photometric parameters.
This last step is crucial in empirical methods since the presence of missing data might affect the generalization capabilities \cite{marlin2008}.

The resulting number of objects in the datasets used for the experiments are:
\begin{itemize}
\item SDSS: $\sim1.0\times10^5$;
\item SDSS $\cap$ GALEX: $\sim4.5\times10^4$;
\item SDSS $\cap$ UKIDSS:  $\sim3.1\times10^4$;
\item SDSS $\cap$ GALEX $\cap$ UKIDSS:  $\sim1.5\times10^4$;
\item SDSS $\cap$ GALEX $\cap$ UKIDSS $\cap$ WISE:  $\sim1.4\times10^4$;
\end{itemize}
An additional dataset was produced by decimating the final \textit{four-surveys} cross-matched catalogue. This dataset was used to perform the preliminary feature-selection or \textit{pruning} phase (see sect.~\ref{qso:featureselection}) and consisted of $\sim 3.8 \times 10^3$ objects, each observed in 15 bands (4 UKIDSS, 2 GALEX, 5 SDSS and 4 WISE) and with accurate spectroscopic redshift estimates.
The decimation was needed to reduce the computational time needed to perform the large number of experiments needed.
For some bands there are multiple measurements (i.e. magnitude measured accordingly to different
definitions) and therefore we are left with a total of 43 different features.

Finally, we wish to emphasize that in producing test and training sets we made sure that they had compatible spectroscopic redshifts distributions.

\begin{figure*}
\centering
\includegraphics[width=12cm]{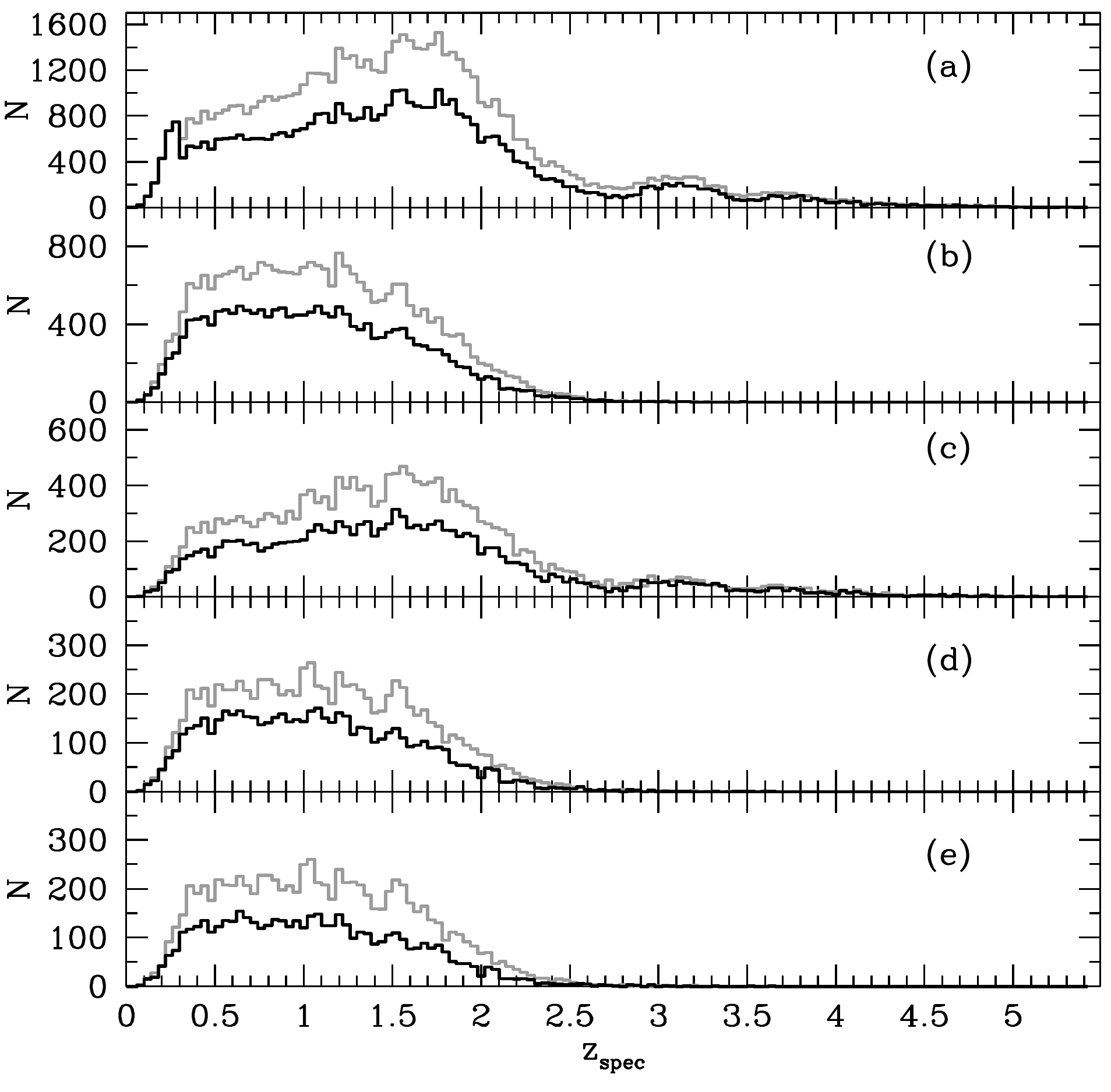}
\caption[Histograms of spectroscopic redshift distribution in the five survey cross-matched samples as derived from the SDSS spectroscopic data.]{Histograms of spectroscopic redshift distribution in the five survey cross-matched samples as derived from the SDSS spectroscopic data. (a) SDSS; (b) SDSS+GALEX; (c) SDSS+UKIDSS; (d) SDSS+GALEX+UKIDSS; (e) SDSS+GALEX+UKIDSS+WISE. Gray dotted line is the training sample. Black line is the test sample.}
\label{qso:fig:zhisto}
\end{figure*}

\subsection{Statistical Indicators}
In order to evaluate and reciprocally compare the experiments described in the next section we adopted the following definitions: \\

\begin{equation}
\mbox{bias} = \frac{{\sum\limits_{i = 1}^N {x_i } }}{N}
\label{qso:eq21}
\end{equation}

\begin{equation}
\sigma  = \sqrt {\frac{{\sum\limits_{i = 1}^N {\left[ {x_i  - \left( {\frac{{\sum\limits_{i = 1}^N {x_i } }}{N}} \right)} \right]^2 } }}{N}}
\label{qso:eq22}
\end{equation}

\begin{equation}
\mbox{MAD } = Median\left( \left| x \right| \right )
\label{qso:eq21bis}
\end{equation}

\begin{equation}
\mbox{NMAD } = 1.48 \times Median\left( \left| x \right| \right )
\label{qso:eq21tris}
\end{equation}

\begin{equation}
\mbox{RMS} = \sqrt {\frac{{\sum\limits_{i = 1}^N {x_i^2 } }}{N}}
\label{qso:eqrms}
\end{equation}

where $\sigma$ is the standard deviation, MAD is the Median Absolute Deviation and RMS is the Root Mean Square. The term $x$ may be either $\Delta z$ defined as :

\begin{equation}
\Delta z = (z_{spec}-z_{phot})
\label{qso:eq20}
\end{equation}

or the normalized residuals $\Delta z'$ defined as:

\begin{equation}
\Delta z' = (z_{spec}-z_{phot}) / (1+z_{spec})
\label{qso:eqdznorm}
\end{equation}

In the latter case the statistical indicators are labeled as $normalized$ quantities (e.g. $\sigma_{norm}$, $bias_{norm}$, $MAD_{norm}$, $RMS_{norm}$).

\subsection{The experiments}

\label{qso:sec:experiments}
In what follows we shall discuss the general experiment workflow and the outcome of the experiment phases: feature selection (see Sec. \ref{qso:featureselection}), color magnitude selection (see Sec. \ref{qso:magnitudecolor}) and then the complexity selection (see Sec. \ref{qso:complexityselection}).

As it is common practice in machine learning experiments, the MLPQNA model has been applied by following a series of concatenated steps (what we call a \textit{workflow}):  i)
extraction of the proper KB by using the available spectroscopic redshifts related to the 15 bands from the 4 used surveys (see Sec. \ref{qso:sec:thedata}); ii) tuning of the \textit{optimal} model parameter setup, including the selection of data features and training/test subsets within the available KBs (see Sec. \ref{qso:featureselection}) and the validation of training; iii) analysis of the  tests.

\subsubsection{The knowledge base and model setup}\label{qso:sec:kbsetup}

For machine learning supervised methods it is common praxis to use the available KB to obtain at least three disjoint subsets for every experiment: one (training set) for training purposes, i.e. to train the method in order to acquire the hidden correlation among the input features which is needed to perform the regression;
the second one (validation set) to check the training, in particular against loss of generalization capabilities (a phenomenon also known as overfitting); and the third one (test set) to evaluate the overall performances of the model.  As a rule of thumb, these sets should be populated with respectively 60\%, 20\% and 20\% of the objects in the KB \citep{kearns1996}. In order to ensure a proper coverage of the KB Parameter Space (PS), the data objects are indeed divided up among the three datasets by random extraction, and usually this process is iterated several times to minimize the possible biases induced by fluctuations in the coverage of the PS.\\
The first two criteria used to decide the stopping condition of the algorithm, as mentioned at the end of Sec. \ref{qso:sec:mlpqna}, are mainly sensitive to the choice of specific parameters and may lead to poor results if the parameters are improperly set. The cross validation does not suffer of such drawback. It can avoid overfitting the data and is able to improve the generalization performance of the model. However, if compared to the traditional training procedures, it is much more computationally expensive. By exploiting the cross validation technique (see Sec. \ref{qso:sec:mlpqna}), training and validation were indeed performed together using $\sim 80\%$ of the objects as a training + validation set, and the remaining $\sim 20\%$ as test set (in practice 3029 records in the training/validation set and 758 in the test set). \\
The automatized process of the cross-validation was done by performing ten different training runs with the following procedure: (i) splitting of the training/validation set into ten random subsets, each one composed of 10\% of the dataset; (ii) at each training run we applied the 90\% of the dataset and the excluded 10\% for validation.\\
As remarked in \ref{qso:sec:mlpqna}, the k-fold cross validation is able to avoid overfitting on the training set \citep{bishop2006}, with an increase of the execution time estimable around $k-1$ times the total number of runs \citep{cavuoti2012b}.\\
In terms of the internal parameter setup of the MLPQNA, we need to consider the following topological parameters:
\begin{itemize}
\item input layer: a variable number of neurons (corresponding to the pruned number of survey parameters used in all experiments , up to a maximum number of 43 nodes (all available features);
\item neuron on the first hidden layer: a variable number of hidden neurons, depending on the number $N$ of input neurons (features in the dataset), equal to $2N+1$ as rule of thumb;
\item neuron on the second hidden layer: a variable number of hidden neurons, ranging from 0 (to ignore the second layer) to  $N-1$ which usually is the default value;
\item output layer: one neuron, returning the calculated photo-z value.
\end{itemize}
For the QNA learning rule, after several trials, we fixed the following values as best parameters for the final experiments:
\begin{itemize}
\item step: $0.0001$ (one of the two stopping criteria. The algorithm stops if the approximation error step size is less than this value. A step value equal to zero means to use the parameter MaxIt as the unique stopping criterion);
\item res : $40$ (number of restarts of Hessian approximation from random positions, performed at each iteration);
\item dec : $0.1$ (regularization factor for weight decay. The term $dec*||network weights||^2$ is added to the error function, where $network weights$ is the total number of weights in the network, directly depending on the total number of neurons inside. When properly chosen, the generalization performances of the network are highly improved);
\item MaxIt: $8000$ (max number of iterations of Hessian approximation. If zero the step parameter is used as stopping criterion);
\item CV (k): $10$ (k-fold cross validation, with $k=10$);
\item Error evaluation: Mean Square Error (between target and network output).
\end{itemize}
With these parameters, we obtained the statistical results reported in Sec. \ref{qso:complexityselection}.

\subsubsection{Selection of features}\label{qso:featureselection}

As it is known, supervised machine learning models are powerful methods for learning the hidden correlation between input and output features from training data.  Of course, their generalization and prediction capabilities strongly depend on: the intrinsic quality of data (signal-to-noise ratio), the level of correlation among different features; the amount of missing data present in the dataset \citep{ripley1996}. It is obvious that some, possibly many, of the $43$ parameters listed in Tab. \ref{qso:tab:features} may not be independent and that their number needs to be reduced in order to speed up the computation (which scales badly with the number of features). This is a common problem in data mining and there is a wide literature on how to optimize the selection of features which are most relevant for a given purpose \citep{lindeberg1998, guyon2003, guyon2006, brescia2012d}. This process is usually called \textit{Feature selection} or pruning, and consists in finding a compromise between the number of features (and therefore the computational time) and the accuracy of the final results. In order to do so, we extracted from the main catalogue several subsets containing different groups of variables (features). Each one of these subsets was then analyzed separately in specific runs of the method (runs which in the data mining community are usually called experiments), in order to allow the comparison and evaluation.
We wish to stress that our main concern was not only to disentangle which bands carry the most information but also, for a given band, which type of measurements (e.g. Point Spread Function, Kron or isophotal magnitude) are more effective.\\

We performed a series of regression experiments to evaluate the performances obtained by the pruning of photometric quantities on the small dataset described in \ref{qso:sec:thedata}.

The pruning experiments consisted into several combinations of surveys and their features:
\begin{itemize}
\item a \textit{full} features experiment to be used as a benchmark for all the other experiments;
\item some \textit{service} experiments  used to select the best combination of input features in order to eliminate redundancies in the flux measurmemnts (i.e., petrosian magnitudes
against isophotal magnitudes).
\item \textit{three-survey} experiments for all possible combinations of three (out of four) surveys;
\item \textit{two-survey} experiments with all possible combinations of two (out of four) surveys;
\item \textit{single-survey} experiments.
\end{itemize}

The output of the experiments consisted of lists of photometric redshift estimates for all objects in the KB. All experiments were performed using $\sim 3000$ objects in the training set and
$\sim 800 $ in the test set.
In Table \ref{qso:tab:experiments}, we list the outcome of the experiments for the feature selection. Both bias and $\sigma \left(\Delta z \right)$
were computed using the objects in the test set alone. As it can be seen, among the various types of magnitudes available for GALEX and UKIDSS
the best combination is obtained using the isophotal magnitudes for GALEX and the calibrated magnitudes ($all\_mag$) for UKIDSS.

Therefore at the end of the pruning phase the best compbination of features turned out to be: the five SDSS $PSF_MAG$, the two isophotal magnitudes of GALEX,
the four $all\_mag$ for UKIDSS, the four magnitudes for WISE.

\begin{table*}
\centering
\begin{tabular}{|c|c|c|c|r|r|}
\hline
 {\bf GALEX}	&	 {\bf SDSS}	&	 {\bf UKIDSS}&	{\bf WISE}	&	{\bf $BIAS$}&{\bf $\sigma$} \\ \hline \hline
\multicolumn{6}{|c|}{Service Experiments} \\ \hline
 X				&	 X			&	 X			&	 X			&	    0.0033 	&	 0.174 		 \\ \hline
 X$^{1,2}$ 		&	 X			&	 X$^{1}$ 	&	 X			&	  -0.0001 	&	 0.152 		 \\ \hline
 X$^{3}$ 		&	 X			&	 X$^{1}$ 	&	 X		   	&	 -0.0016 	&	 0.165 		 \\ \hline
 X$^{1}$ 		&	 X			&	 X$^{1}$ 	&	 X			&	  0.0054 	&	 0.151 		 \\ \hline
 X$^{2}$ 		&	 X			&	 X$^{1}$ 	&	 X         	&	 -0.0026 	&	 0.151 		 \\ \hline
 X$^{4,5}$ 		&	 X			&	 X$^{1}$ 	&	 X			&	 -0.0008 	&	 0.152 		 \\ \hline
 X$^{1,2,3}$ 	&	 X			&	 X$^{1}$ 	&	 X			&	  0.0041 	&	 0.163 		 \\ \hline
 X$^{2,3}$		&	 X			&	 X$^{1}$ 	&	 X			&	  -0.0033 	&	 0.155 		 \\ \hline
 				&	 			&	 X$^{1,2}$ 	&	        	&	 -0.0091 	&	 0.299 		 \\ \hline
 				&	 			&	 X$^{2}$  	&	           	&	 0.0111 	&	 0.465 		 \\ \hline	
 				&	 			&	 X$^{1}$  	&	           	&	 -0.0081 	&	 0.294 		\\ \hline \hline
\multicolumn{6}{|c|}{Four Survey Experiment} \\ \hline
 X$^{2}$ 		&	 X			&	 X$^{1}$ 	&	 X         	&	 -0.0026 	&	 0.151 		 \\ \hline \hline
\multicolumn{6}{|c|}{Three Survey Experiment} \\ \hline
 X$^{2}$ 		&	 X			&	 X$^{1}$ 	&	           	&	 -0.0046 	&	 0.152 		 \\ \hline
 X$^{2}$ 		&	 X			&	 			&	 X         	&	 0.0025 	&	 0.162 		 \\ \hline
  				&	 X			&	 X$^{1}$ 	&	 X         	&	 -0.0032 	&	 0.179 		 \\ \hline
 X$^{2}$ 		&	 			&	 X$^{1}$ 	&	 X         	&	 0.0110 	&	 0.203 		 \\ \hline \hline
\multicolumn{6}{|c|}{Two Survey Experiment} \\ \hline
		 		&	 			&	 X$^{1}$ 	&	 X         	&	 0.0045 	&	 0.236 		 \\ \hline
 X$^{2}$ 		&	 			&	  			&	 X         	&	 0.0175 	&	 0.288 		 \\ \hline
  				&	 X			&	 X$^{1}$ 	&	         	&	 -0.0027 	&	 0.210 		 \\ \hline
  				&	 X			&	 			&	 X        	&	 -0.0039 	&	 0.197 		  \\ \hline
 X$^{2}$ 		&	 X			&	  			&            	&	 -0.0055 	&	 0.240 		 \\ \hline
 X$^{2}$ 		&	  			&	 X$^{1}$ 	&	         	&	 0.0133 	&	 0.238 		\\ \hline \hline
\multicolumn{6}{|c|}{One Survey Experiment} \\ \hline
 				&	 			&	 			&	 X         	&	 0.0165 	&	 0.297 		\\ \hline
 				&	 X  		&		 		&	           	&	 -0.0162 	&	 0.338 		 \\ \hline
 X$^{1,2}$ 		&	 			&	  			&			  	&	 0.0550 	&	 0.419 		 \\ \hline	
 				&	 			&	 X$^{1}$  	&	           	&	 -0.0081 	&	 0.294 		\\ \hline
\hline
\end{tabular}
\caption[List of the experiments performed during the feature selection phase in order to evaluate the best possible combination of parameters.]{List of the experiments performed during the feature selection phase in order to evaluate the best possible combination of parameters.  Column 1 through 4: surveys used for the experiment where superscript index indicates the magnitude used in our experiments. For istance the index for GALEX means: 1 \textit{mag}; 2 \textit{mag$\_$iso}; 3 \textit{mag\_Aper 1, 2 and 3}; 4 \textit{mag$\_$auto} and 5 \textit{kron\_radius}. While for UKIDSS: 1 stands for \textit{HallMag} and 2 for \textit{PetroMag}. A cross in a column means that all the bands of the survey corresponding to that column were used for the experiment. }
\label{qso:tab:experiments}
\end{table*}

\subsubsection{Magnitudes vs Colors}\label{qso:magnitudecolor}

Once the most significant features had been identified we had to check which type of flux combinations were more effective.
More in detail, we wanted to check whether  it is better
to propose to the network, magnitudes, colors or colors plus one
reference magnitude (e.g. $R$ for SDSS, $nuv$ for GALEX, $K$ for UKIDSS and $W4$ for WISE).
Experiments were performed on all 5 datasets listed in section \ref{qso:sec:thedata}.

As it could be expected, the optimal combination turned out to be always the mixed one, i.e the one
including colors and one reference magnitude for each of the included surveys  (r for SDSS, nuv  for GALEX,  K for UKIDSS).
From the data mining point of view this is rather surprising since the amount of information should not change by applying linear combinations
between features. But from the physical point of view this can be easily understood by noticing that
even though colors are derived as a simple subtraction of magnitudes, the content of
information is  different since  an ordering relationship is
implicitly assumed thus increases the amount of information in the final output (gradients instead of fluxes).
The additional magnitude removes instead the degeneracy in the luminosity class for a specific galaxy type.
It was however surprising that experiments with just colors or just magnitudes did not provide the same \textit{winner}
in all experiments.

\subsubsection{Network Topology}\label{qso:complexityselection}
The final check was about the complexity of the network, whether \textit{shallow} or \textit{deep} \cite{bengio2007}.
The above quoted datasets were therefore processed through both a three layers (input-hidden-output)
layer and a four (input - 2 hidden - output) layers network.
In all cases the four layers network performed significantly better.

\subsection{Discussion and conclusions}\label{qso:sec:discussion}

Beginning in 2002 \cite{tagliaferri2002} some people in this group begun to explore the usage of MLPs for the evaluation of photo-z's both
for \textit{normal} galaxies and quasars. In 2007 \citealt{dabrusco2007}  a combination of two MLPs was used to correct for the degeneracy introduced by the disomogeneity in the
knowledge base.
In 2011 \citealt{laurino2011} it was demonstrated that the subtleties in the mapping function could be more easily captured using the so called Weak Gated Experts method: a combination of MLPs each specialised in a specific partition of the parameter space, with the individual outputs combined by an additional MLP.
In 2007 \citealt{bengio2007} published a seminal paper which somehow disproved the Haykin-pseudo theorem pointing out that problems with a high variation of density in the parameter space, are better treated by
what they called \textit{deep} networks, i.e. networks with more than one computational (hidden) layer.  In this section we exploited \citealt{bengio2007} findings
by using the new machine learning based method MLPQNA to evaluate photometric redshifts of quasars using multiband data obtained from the intersection of the
GALEX, SDSS, UKIDSS and WISE surveys.

In the Tables \ref{qso:tab:comparison} and \ref{qso:tab:compoutliers} we compare our best results to those presented by other authors in the
recent literature. In order to be as fair as possible, we applied our method to datasets which, at the best
of our knowledge, matched the dataset used by the other teams and used an homogeneous set
of statistical indicators (also in normalized form).
Unfortunately, the whole set of indicators was not available for all bibliographical sources and, actually, in several cases we could only use a few
not normalized quantities.
Results are listed according to the combinations of surveys used in the experiment.

The larger the number of surveys (bands) used, the more accurate are the results. This result - which might seem obvious - is not obvious at all
since the higher amount of information carried by the additional bands implies also a strong decrease in the number of objects which are contained
in the training set and should therefore imply a decrease in the performances of the network.
This result, together with the fact that MLPQNA performs well also with small KBs (cf. \cite{cavuoti2012b}),
seems particularly interesting, since it has far reaching implications on ongoing and future surveys:
a better photometric coverage is much more relevant than an increase of spectroscopic templates in the KB.

We wish to stress that whatever are the statistical indicators used, it is good praxis
to evaluate them on data (i.e. the test set) which have never been presented to the network.
The usage of \textit{test plus training} data as it is often encountered in the literature, introduces an obvious
positive systematic bias which masks reality.
As already mentioned above, all statistical indicators used in this section were evaluated using the test data only.

\noindent As it can be seen, in  all cases MLPQNA outperforms the other methods.
Only in the SDSS+GALEX case, the non-normalized quantities show a substantial agreement between our results and those by \citealt{laurino2011}.
The better performances of MLPQNA in the normalized indicators is a consequence of the better performances of the MLPQNA method in terms of
fraction of outlayers.
We wish to stress that both MLPQNA and the WGE method discussed in \citealt{laurino2011}
take advantage of a substantial improvement in complexity with respect to the traditional
single layer MLP networks used in the literature, and therefore deal better with the
complexity of the multi-color parameter space.

Average statistical indicators such as bias and standard deviation, however, provide only part of the information which allows to correctly evaluate the performances of a method and, for instance, they provide only very little evidence of the systematic trends which are observed as a sudden increase in the residuals spread over specific regions of the redshift space \citep{laurino2011}.

\noindent In the worst cases, these regions correspond to degeneracies in the parameter space and, as it could be expected, the relevance of such
degeneracies decreases for increasing number of bands.
In  we list the percentage of catastrophic outliers, defined as objects with

Let us know briefly discuss the problem posed by the catastrophic outliers as summarised in Table \ref{qso:tab:compoutliers}.
According to \cite{mobasher2007}, the parameter $D_{95} \equiv \Delta_{95}/\left(1+z_{phot}\right)$
enables the identification of outliers in photometric redshifts derived through SED fitting methods (usually
evaluated through numerical simulations based on mock catalogues). In fact, in the hypothesis that the redshift error $\Delta z = \left( z_{spec}-z_{phot}\right)/\left(1 + z_{spec}\right)$ is Gaussian, the catastrophic redshift error limit would be limited by the width of the redshift probability distribution corresponding to the
$95\%$ confidence interval, i.e. with $\Delta_{95} = 2\sigma \left( \Delta z_{norm} \right)$.

In our case, however, phot-z's are empirical, i.e. not based on any specific fitting model and it is preferable to use the standard deviation
value $\sigma \left( \Delta z_{norm} \right) $  derived from the photometric cross matched samples, altough it could overestimate the theoretical Gaussian $\sigma$,
due to the residual spectroscopic uncertainty as well as to the method training error. Therefore, we consider as catastrophic outliers the objects
with $\left| \Delta z_{norm} \right| > 2 \sigma \left( \Delta z_{norm} \right)$.

It is also important to notice that for empirical methods it is useful to analyze the correlation between the
$NMAD (\Delta z_{norm}) = 1.48 \times median \left( \left| \Delta z_{norm} \right| \right)$ and the standard deviation $\sigma_{clean}(\Delta z_{norm})$ calculated on the data samples cleaned for which $\left| \Delta z_{norm} \right| \le 2 \sigma (\Delta z_{norm})$.
In fact, the quantity $NMAD$ would be comparable to the value of $\sigma_{clean}$. 

As it is shown in Table \ref{qso:tab:catastrop}, in our data, the $\sigma_{clean}(\Delta z_{norm})$ is always slightly larger than the corresponding $NMAD(\Delta z_{norm})$, which is exactly what is expected due to the overestimate induced by the above considerations.

\begin{table*}
\small
\centering
\setlength{\tabcolsep}{3pt}
\begin{tabular}{lccccc}
\hline
Exp & n. obj. & $\sigma \left( \Delta z_{norm} \right) $ & \% outliers & $\sigma_{clean}\left( \Delta z_{norm} \right)$ & $NMAD\left( \Delta z_{norm} \right)$\\
                        &           &   &  $\left| \Delta z_{norm} \right| > 2 \sigma \left(\Delta z_{norm} \right)$&\\
\hline
SDSS                    & 41431     & 0.15 & 6.53 & 0.062 & 0.058 \\
SDSS + GALEX            & 17876     & 0.11 & 4.57 & 0.045 &0.043\\
SDSS+UKIDSS             & 12438     & 0.11 & 3.82 & 0.041 & 0.040 \\
SDSS+GALEX+UKIDSS       & 5836      & 0.087 & 3.05 & 0.040 & 0.032\\
SDSS+GALEX+UKIDSS+WISE  & 5716      & 0.069 & 2.88 & 0.035 & 0.029 \\
\hline
\end{tabular}
\caption{Catastrophic outliers evaluation and comparison between the residual $\sigma_{clean}\left( \Delta z_{norm} \right)$ and $NMAD\left( \Delta z_{norm} \right)$.}\label{qso:tab:catastrop}
\end{table*}

\begin{figure*}
\centering
\includegraphics[width=10.cm]{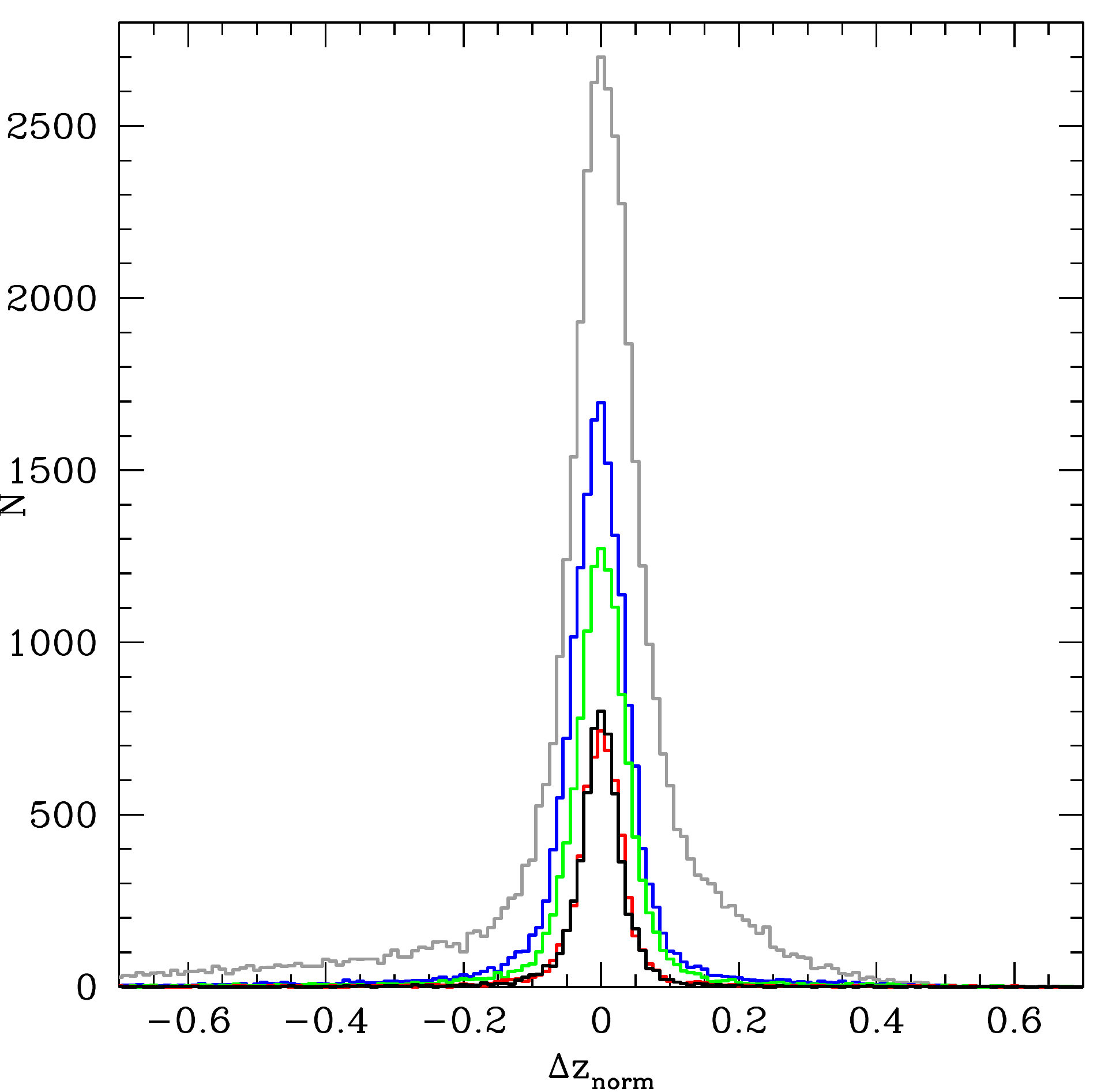}
\caption[$\Delta z_{norm}$ distributions for all five cross-matched test data sets.]{$\Delta z_{norm}$ distributions for all five cross-matched test data sets. Lines are referred to, respectively, SDSS (gray), SDSS+GALEX (blue), SDSS+UKIDSS (green), SDSS+GALEX+UKIDSS (red) and SDSS+GALEX+UKIDSS+WISE (black).}\label{qso:fig:histogram}
\end{figure*}

Finally, we would like to stress that the difficulties  encountered by us and by other teams in comparing different methods, especially in light of the crucial role that photo-z play in the scientific exploitation of present and future large surveys (cf. \citealt{des2005}, \citealt{chambers2011}, \citealt{refregier2010}), confirm that it would be desirable to repropose an upgraded version of the extremely useful PHAT contest \citep{hildebrandt2010}, which allowed a direct, effective and non ambiguous comparison3 of different methods applied on the same datasets and through the same set of statistical indicators.
This new contest should be applied to a much larger dataset, with a more practical selection of photometric bands, and should take into account also other parameters such as scalability, robustness of the algorithms as well as a better characterization of the degeneracies.

\begin{table*}
\centering
\small
\setlength{\tabcolsep}{2pt}
\begin{tabular}{ccccccccccc}
\hline
Exp. & $BIAS$ & $\sigma$ & $MAD$ & $RMS$ & $BIAS_{norm}$ & $\sigma_{norm}$ &
$MAD_{norm}$ & $RMS_{norm}$ & $NMAD_{norm}$\\
\hline\hline
\multicolumn{10}{|c|}{SDSS } \\
\hline
MLPQNA          & 0.007  & 0.25 & 0.102 & 0.26 & 0.032  & 0.15 & 0.039 & 0.17 &0.058\\
Bovy et al.     & -      & 0.46 & -     & -    & -      & -    & -     & -    &-\\
Laurino et al.  & 0.210  & 0.28 & 0.110 & 0.35 & 0.095  & 0.16 & 0.041 & 0.19 &-\\
Ball et al.     & -      & 0.35 & -     & -    & 0.095  & 0.18 & -     & -    &-\\
Richards et al. & -      & 0.52 & -     & -    & 0.115  & 0.28 & -     & -    &-\\
\hline
\multicolumn{10}{|c|}{SDSS + GALEX } \\
\hline
MLPQNA          & 0.003  & 0.21 & 0.060 & 0.22 & 0.012  & 0.11 & 0.029 & 0.11 &0.043\\
Bovy et al.     & -      & 0.26 & -     & -    & -      & -    & -     & -    &-\\
Laurino et al.  & 0.13   & 0.21 & 0.061 & 0.25 & 0.058  & 0.29 & 0.029 & 0.11 &-\\
Ball et al.     & -      & 0.23 & -     & -    & 0.06   & 0.12 & -     & -    &-\\
Richards et al. & -      & 0.37 & -     & -    & 0.071  & 0.18 & -     & -    &-\\
\hline
\multicolumn{10}{|c|}{SDSS + UKIDSS } \\
\hline
MLPQNA          & 0.001  & 0.25 & 0.066 & 0.26 & 0.008  & 0.11 & 0.027 & 0.11 &0.040\\
Bovy et al.     & -      & 0.28 & -     & -    & -      & -    & -     & -    &-\\
Laurino et al.  & -      & -    & -     & -    & -      & -    & -     & -    &-\\
Ball et al.     & -      & -    & -     & -    & -      & -    & -     & -    &-\\
Richards et al. & -      & -    & -     & -    & -      & -    & -     & -    &-\\
\hline
\multicolumn{10}{|c|}{SDSS + GALEX + UKIDSS } \\
\hline
MLPQNA          & 0.0009 & 0.18 & 0.043 & 0.19 & 0.005  & 0.087 & 0.022 & 0.088&
0.032\\
Bovy et al.     & -      & 0.21 & -     & -    & -      & -     & -     & -     &-\\
Laurino et al.  & -      & -    & -     & -    & -      & -     & -     & -     &-\\
Ball et al.     & -      & -    & -     & -    & -      & -     & -     & -     &-\\
Richards et al. & -      & -    & -     & -    & -      & -     & -     & -     &-\\
\hline
\multicolumn{10}{|c|}{SDSS + GALEX + UKIDSS + WISE } \\
\hline
MLPQNA          & 0.002  & 0.15 & 0.040 & 0.15 & 0.004  & 0.069 & 0.020 & 0.069 &
0.029\\
Bovy et al.     & -      & -    & -     & -    & -      & -     & -     & -     &-\\
Laurino et al.  & -      & -    & -     & -    & -      & -     & -     & -     &-\\
Ball et al.     & -      & -    & -     & -    & -      & -     & -     & -     &-\\
Richards et al. & -      & -    & -     & -    & -      & -     & -     & -     &-\\
\hline
\end{tabular}
\caption[Comparison among the performances of the different references.]{Comparison among the performances of the different references. Column 1:
reference; Column 2-9, respectively: bias, standard deviation, MAD, RMS, normalized
bias, normalized standard deviation, normalized MAD, normalized RMS. For the
definition of the parameters and for discussion see
text.}\label{qso:tab:comparison}
\end{table*}

\begin{table*}
\centering
\begin{tabular}{ccccc}
\hline
Exp  			&			 \multicolumn{2}{c}{Outliers ($|\Delta z|$)} 			&			 \multicolumn{2}{c}{Outliers ($|\Delta z_{norm}|$)} \\
				&	 	$>2\sigma$ 			&	 $> 4\sigma$ 	&	 $>2\sigma$ 	&			 $> 4\sigma$ \\ \hline\hline
\multicolumn{5}{|c|}{SDSS } \\ \hline
MLPQNA 			&				 7.68 		&	 0.38 			&	 6.53 			& 			 1.24\\
Bovy et al. 	&							&	 0.51	\\ \hline
\multicolumn{5}{|c|}{SDSS+GALEX} \\ \hline
MLPQNA 			&				 4.88 		&	 1.61 			&	 4.57			&			1.37\\
Bovy et  al. 	&							&	 1.86	\\ \hline
\multicolumn{5}{|c|}{SDSS+UKIDSS} \\ \hline
MLPQNA 			&				4.00 		&	 1.73			&	 3.82			&			 1.38\\
Bovy et al. 	&							&	 1.92	\\ \hline
\multicolumn{5}{|c|}{SDSS+GALEX+UKIDSS} \\ \hline
MLPQNA 			&				2.86 		& 	 1.47 			&	 3.05			&			0.23\\
Bovy et al. 	&							&	 1.13 	\\ \hline
\multicolumn{5}{|c|}{SDSS+GALEX+UKIDSS+WISE} \\ \hline
MLPQNA 			&				 2.57		&	 0.87			&	 2.88			&			 0.91\\
\hline
\end{tabular}

\caption[Comparison in terms of outliers percentages among the different references.]{Comparison in terms of outliers percentages among the different references. Column 1: reference; Column 2-5, fractions of outliers at different $\sigma$ based on deltaz; Column 6-9, fractions of outliers at different $\sigma$ based on normalized deltaz. For the definition of the parameters and for discussion see text.}\label{qso:tab:compoutliers}
\end{table*}


\begin{figure*}
\centering
\includegraphics[width=12.cm]{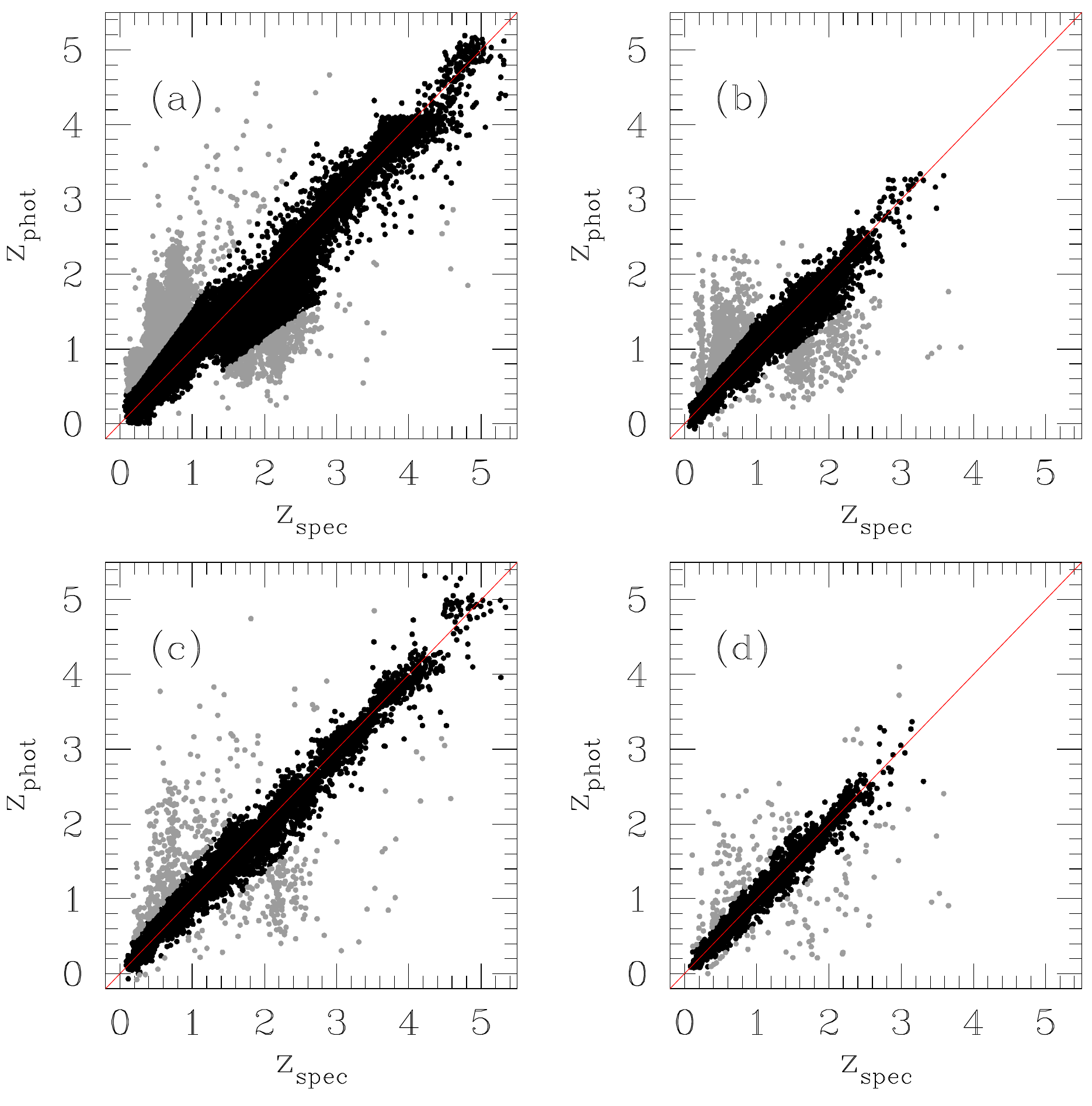}
  \includegraphics[width=6.cm]{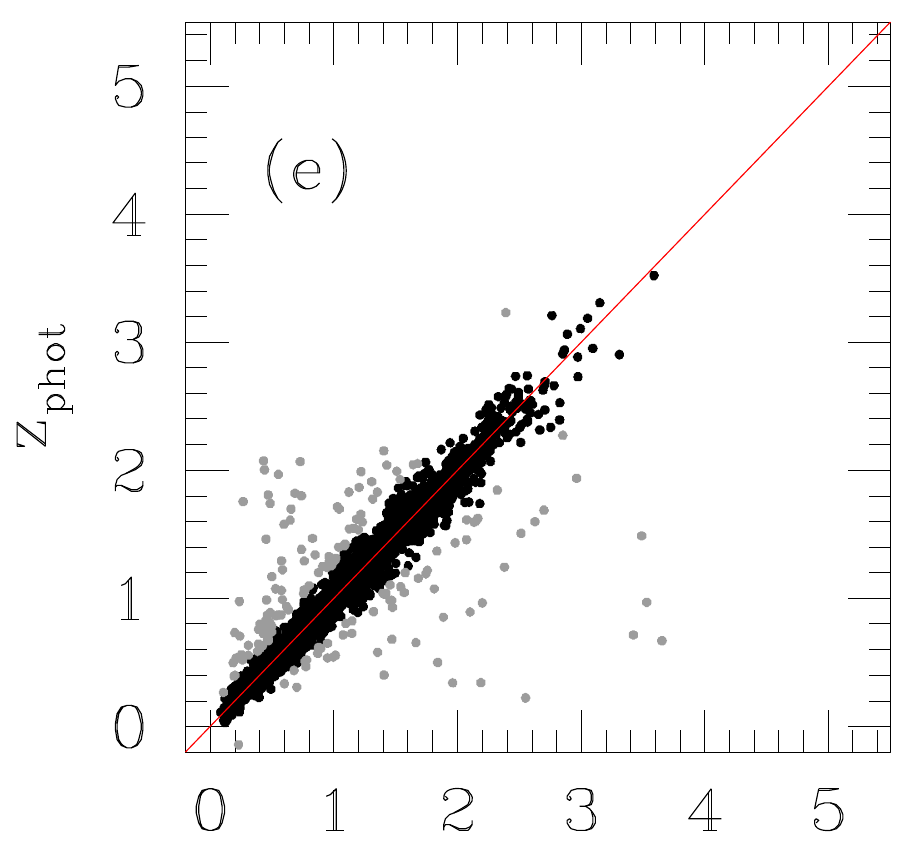}
\caption[Scatter plots ($z_{spec}$ vs $z_{phot}$); all diagrams refer to results on test sets.]{Scatter plots ($z_{spec}$ vs $z_{phot}$); (a) SDSS, (b) SDSS+GALEX, (c) SDSS+UKIDSS, (d) SDSS+GALEX+UKIDSS and (e) SDSS+GALEX+UKIDSS+WISE. All diagrams refer to results on test sets. Gray points are catastrophic outliers (defined in Tab.~\ref{qso:tab:compoutliers}).}\label{qso:fig:SCATTERtestset}
\end{figure*}

    \chapter{The variable sky}\label{chap:transients}
        \hfill\begin{tabular}{@{}p{.3\linewidth}@{}}
\textit{``I am not now}\\
\textit{That which I have been."}\\ Lord Byron.\\ \phantom{aaa}
\end{tabular}

The exploration of the Time Domain allow us to study numerous types of astrophysical phenomena. Targets of Time Domain Astronomy are in fact all those sources which show some kind of variability. As mentioned before, imaging surveys allow us to find and observe astrometric and photometric variables. Astrometric variables, also defined transits, are objects whose position in the sky changes with time. Photometric variability instead takes place when a source show an intrinsic change in brightness at different epochs. In this work we shall focus only on photometric variables. \\
Photometric variability may take place at any wavelength range. In this thesis, however, we shall focus only on those phenomena which have at least an optical manifestation.
A schematic representation of the different types of photometric optical transients is given in Figure~\ref{marianna:SemTree}.
\begin{sidewaysfigure}
\centering
\includegraphics[width=19cm]{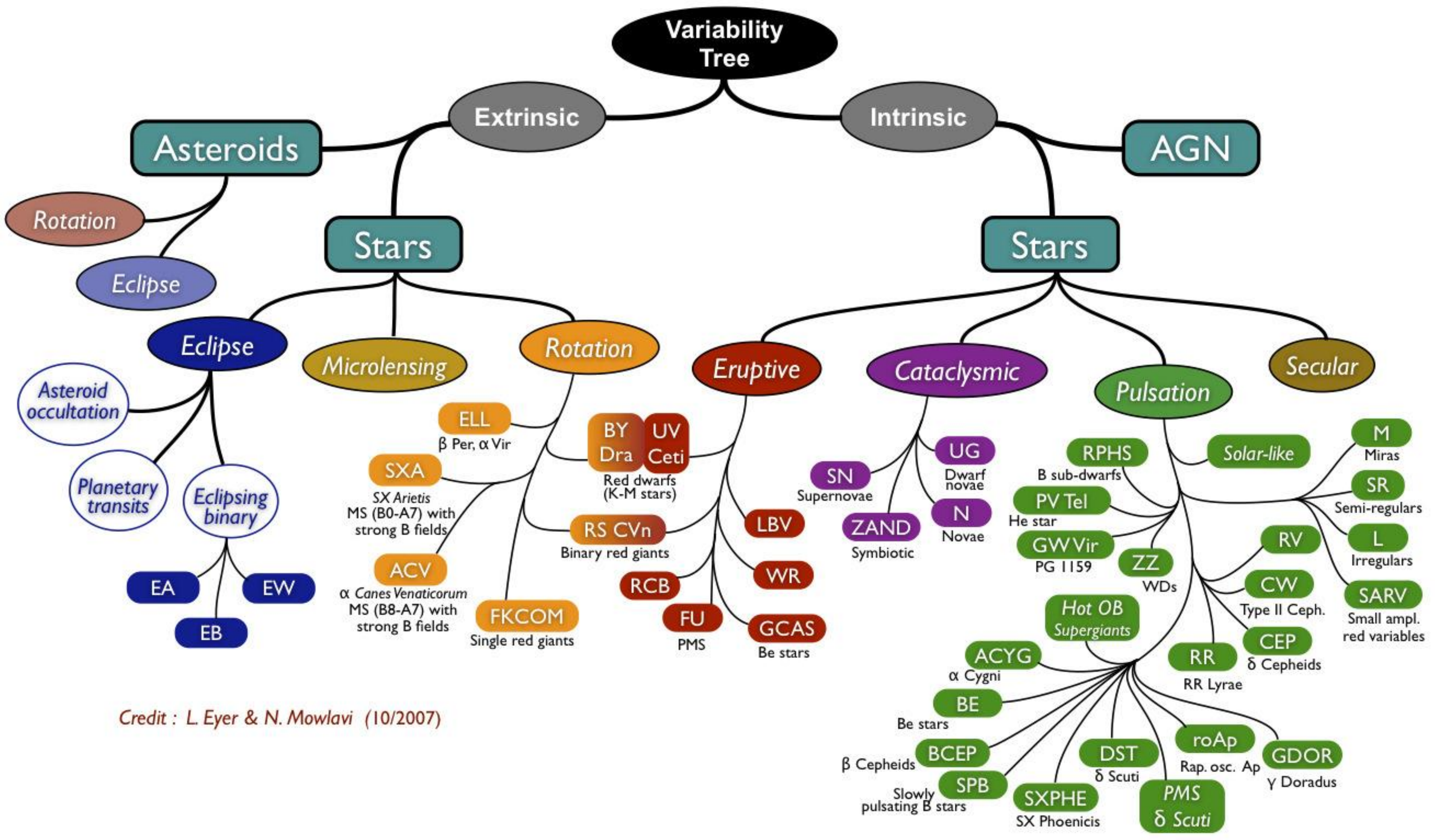}
\caption[Semantic Tree of Astronomical Transient Objects]{Semantic Tree of Astronomical Transient Objects, credit to \citealt*{eyer2008}.}
\label{marianna:SemTree}
\end{sidewaysfigure}

\noindent Photometric variability can be both extrinsic or intrinsic. \\
\textit{Extrinsic variables} show change in brightness due to the eclipse of one object by another or to the effect of rotation. They can be asteroids or stellar objects.  Among the second group there are the microlensing events, the eclipsing binary systems or the Rotating stars. In an eclipsing system a star can change its brightness due to an asteroid occultation, to a planetary transit or to the interaction with another star. in the latter case we talk about Eclipsing Binaries. These systems are formed by physically bound stars having an orbital plane lying near the line-of-sight of the observer. The components periodically eclipse each another, causing a decrease in the apparent brightness of the system as seen by the observer. The period of the eclipse, can range from minutes to years. Rotating stars, instead, show small changes in light that may be due to dark or bright spots on their stellar surfaces.\\
In this thesis we focus on intrinsic variables. \textit{Intrinsic variables} show brightness variations caused by changes  in the physical parameters of the object. This group can be divided in stellar objects, and galaxies. \\
Intrinsic variable stars are usually divided in three more classes, pulsating, cataclysmic, and eruptive variables depending on which phenomenon is at the origin of their variability. Another class is then formed by stars displaying secular evolution, which are usually stars in the post-AGB (Asymptotical Giant Branch) of the H-R (Herztsprung-Russell) diagram. \\
Eruptive variable stars vary in brightness because of violent processes and flares occurring in their chromospheres and coronae. The light changes are usually accompanied by shell events or mass outflow in the form of stellar winds of variable intensity and/or by interaction with the surrounding interstellar medium. The most famous example of eruptive variables are the Wolf-Rayet and the R Coronae Borealis stars. A R Coronae Borealis variable is a luminous, hydrogen-poor, carbon-rich, supergiant star which spend most of its time at maximum light, occasionally fading even nine magnitudes at irregular intervals. Wolf-Rayet stars are very luminous hot Population I stars of effective temperatures between 30000 and 50000 K. They have a characteristic high mass-loss rate ($\mathrm{\sim 10^{-5}M_{\odot}yr^{-1}}$). They show  light variations with amplitudes of several hundredths of a magnitude and time scales from milliseconds to years. \\
Of the other two types of intrinsic stellar variables we shall discuss in detail in the next subsections focusing in particular on the most representative object of each class,  Cepheids for Pulsating variables and Supernovae for Cataclysmic variables.\\
Galaxies hosting Active Galactic Nuclei (AGNs) are also usually variable.
AGNs, however, are very particular variables. In fact they emit strongly over a wide range of wavelengths, from X-ray to radio. Many AGNs vary in brightness by substantial amounts over timescales as short as, months, days, or even hours. AGNs are conveniently divided in two main classes called radio-loud and radio-quiet, depending on whether or not they emit in the radio portion of the electromagnetic spectrum.\\

\begin{figure}
\centering
\includegraphics[width=11cm]{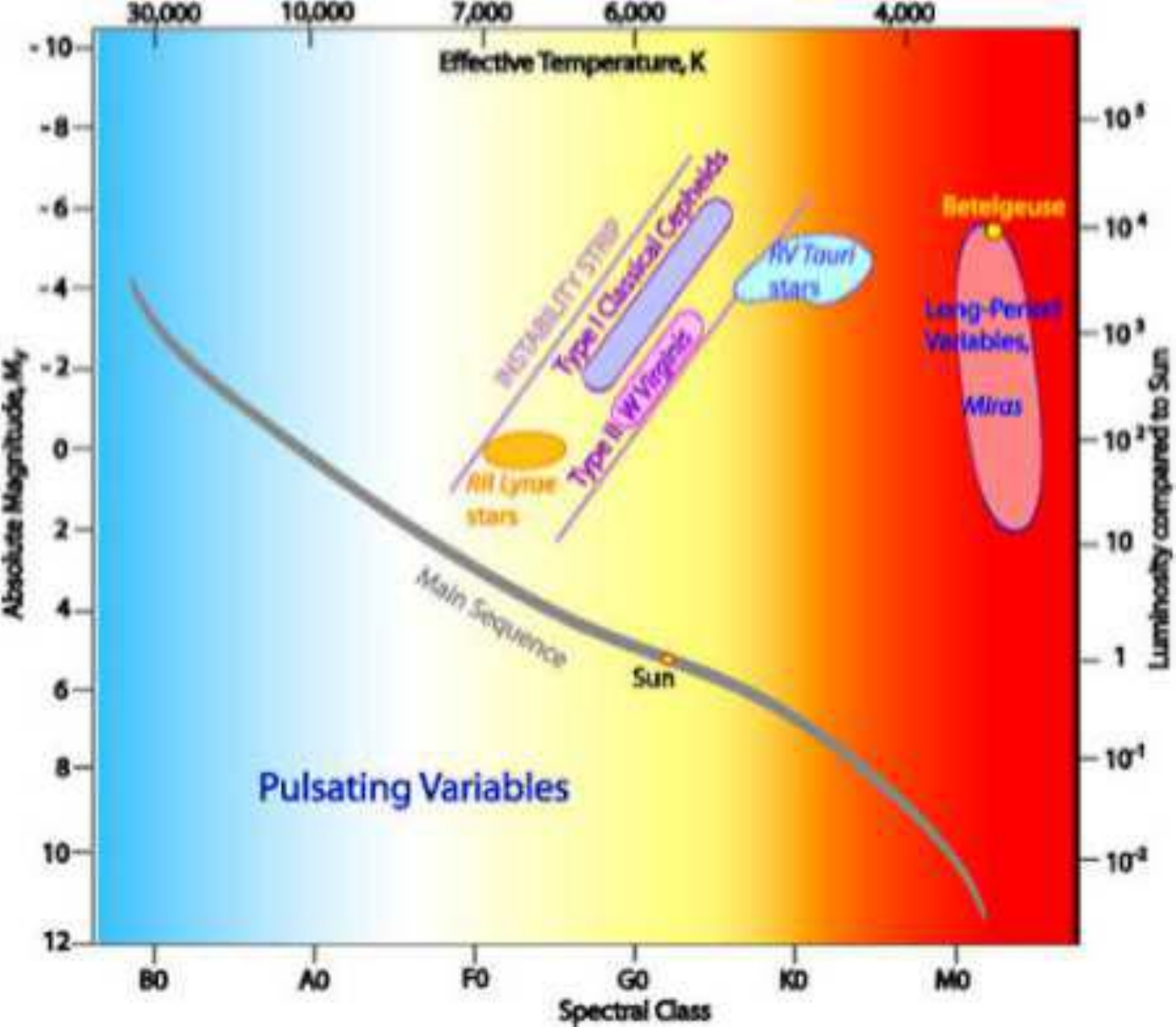}
\caption{Position of some Pulsating Variables in the H$-$R diagram. }
\label{marianna:InStrip}
\end{figure}

\section{Types of Pulsating Variables}
Types of pulsating variables may be identified on the basis of their pulsation period, mass and evolutionary status of the star, and the characteristics of their pulsations.
\begin{itemize}
\item RR Lyrae stars. These are short-period (0.05 to 1.2 days), pulsating, blue giant stars, usually of spectral class A. The amplitude of variation of RR Lyrae stars is generally from 0.3 to 2 magnitudes.
\item $\delta$ Scuti. This variable stars exhibit variations in their luminosity due to both radial and non-radial pulsations of their surface. Typical brightness fluctuations are from 0.003 to 0.9 magnitudes in V over a period of a few hours, although the amplitude and period of the fluctuations can vary greatly. These stars are usually A0 to F5 type giant or main sequence stars.
\item RV Tauri. These stars are yellow supergiants having a characteristic light variation with alternating deep and shallow minima. Their periods, defined as the interval between two deep minima, range from 30 to 150 days. The light variation can be up to 3 magnitudes. Some of these stars show long-term cyclic variations from hundreds to thousands of days. Generally, the spectral class ranges from G to K.
\item Pulsating white dwarf. The luminosity of these white dwarf varies due	to	non-radial gravity wave pulsations. These variables all exhibit small (1$\%-$ 30$\%$) variations in light output, arising from a superposition of vibration modes with periods of hundreds to thousands of seconds.
\item Long Period Variables. These stars  are pulsating red giants or supergiants in which variations in brightness occur over long timescales of months or years. The two major subclasses are Mira and Semiregular variables.
\item Irregular Variable Star. These are usually red supergiants with little or no periodicity. They are often poorly studied semi-regular variables that, upon closer scrutiny, should be reclassified.
\end{itemize}
The main example of pulsating stars are Cepheid variables. They are massive stars of spectral type changing during the pulsation and varying from F at maximum luminosity  to a G or K at minimum. These stars are mostly radial pulsators.
There are four classes of Cepheid variables:
\begin{itemize}
\item Classical Cepheids, or type I Cepheids, fundamental mode pulsators with periods running form 1 to 70 days.
\item Beat Cepheids, which display the presence of two or more simultaneously operating pulsation modes, usually the fundamental one and the first overtone. The have periods between 2 and 7 days.
\item \textit{S} Cepheids, which are probably first-overtone pulsators, with periods in the same range of Beat Cepheids.
\item W Virginis, population II Cepheids with periods between 1 and 30 days. These stars are fundamental mode pulsators.
\end{itemize}
Although Cepheids exhibit strong correlations between their periods, luminosities and colors, the amplitudes of Cepheids do not appear to correlate with other observables. Cepheids, as well as most of the other pulsating variables, exhibit periodic light curves with a sinusoidal form. An example of light curve for each type of Cepheids are reported in Fig. \ref{marianna:DCEPlc}, Fig. \ref{marianna:BCEPlc}, Fig. \ref{marianna:SCEPlc} and Fig. \ref{marianna:WVirlc}.\\
\begin{figure}[ht]
\centering
\includegraphics[width=11cm]{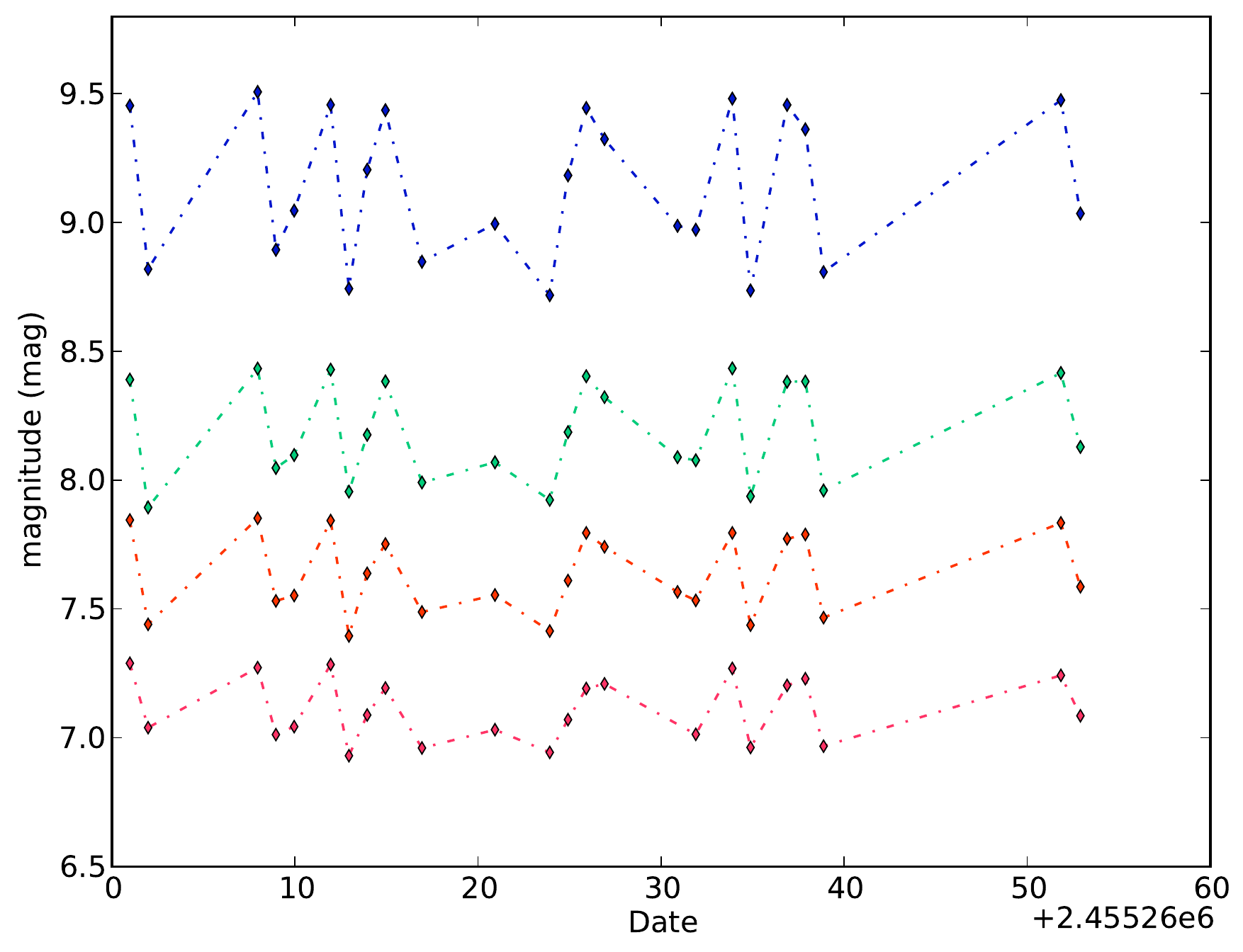}
\caption[Pre-calibrated BVRI light curve for the Classical Cepheid SS Sct.]{Pre-calibrated BVRI light curve for the Classical Cepheid SS Sct. On the y-axis there is the apparent magnitude of the star and on the y-axis the Julian date of the observation. Blu points are the values of the magnitude in B band. Green points are the values of the magnitude in B band. Blu points are the values of the magnitude in B band. Orange points are the values of the magnitude in R band. Red points are the values of the magnitude in I band.}
\label{marianna:DCEPlc}
\end{figure}

\begin{figure}[ht]
\centering
\includegraphics[width=11cm]{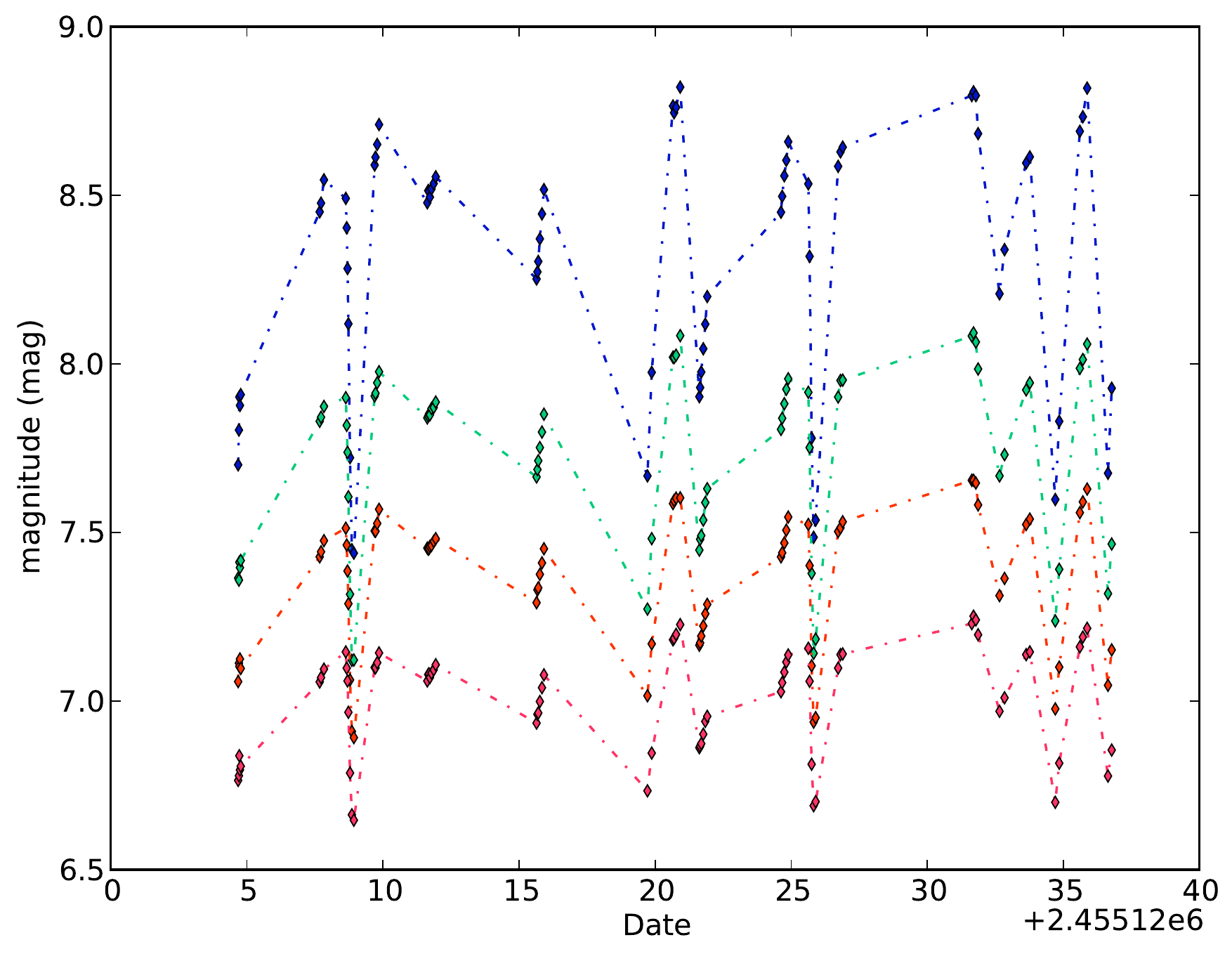}
\caption[Pre-calibrated BVRI light curve for the Beat Cepheid TU Cas.]{Pre-calibrated BVRI light curve for the Beat Cepheid TU Cas. On the y-axis there is the apparent magnitude of the star and on the y-axis the Julian date of the observation. Blu points are the values of the magnitude in B band. Green points are the values of the magnitude in B band. Blu points are the values of the magnitude in B band. Orange points are the values of the magnitude in R band. Red points are the values of the magnitude in I band.}
\label{marianna:BCEPlc}
\end{figure}

\begin{figure}
\centering
\includegraphics[width=\textwidth]{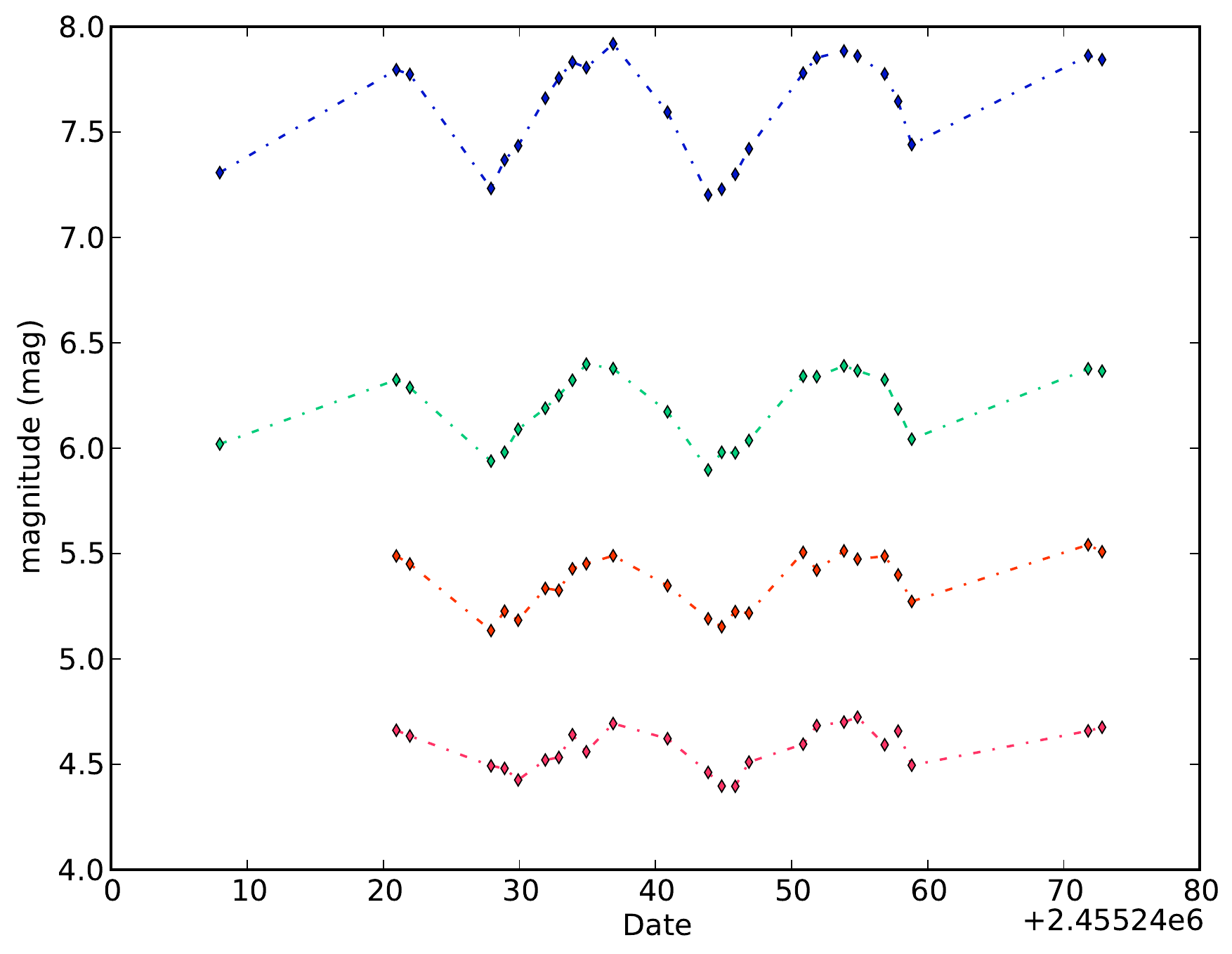}
\caption[Pre-calibrated BVRI light curve for the \textit{S} Cepheid Y Oph.]{Pre-calibrated BVRI light curve for the \textit{S} Cepheid Y Oph. On the y-axis there is the apparent magnitude of the star and on the y-axis the Julian date of the observation. Blu points are the values of the magnitude in B band. Green points are the values of the magnitude in B band. Blu points are the values of the magnitude in B band. Orange points are the values of the magnitude in R band. Red points are the values of the magnitude in I band.}
\label{marianna:SCEPlc}
\end{figure}

\begin{figure}
\centering
\includegraphics[width=\textwidth]{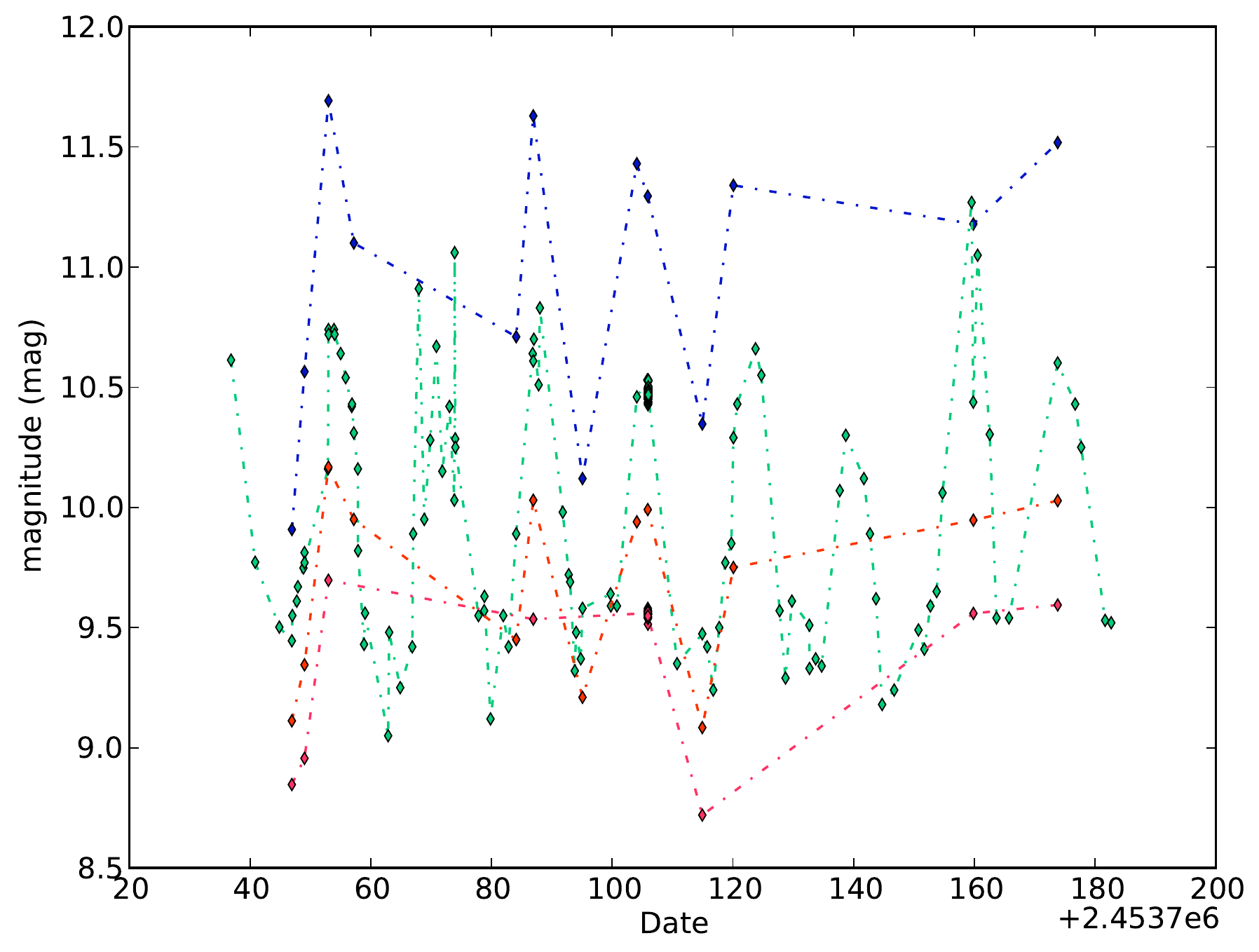}
\caption[Pre-calibrated BVRI light curve for the prototype of W Virginis variables, W Vir.]{Pre-calibrated BVRI light curve for the prototype of W Virginis variables, W Vir. On the y-axis there is the apparent magnitude of the star and on the y-axis the Julian date of the observation. Blu points are the values of the magnitude in B band. Green points are the values of the magnitude in B band. Blu points are the values of the magnitude in B band. Orange points are the values of the magnitude in R band. Red points are the values of the magnitude in I band.}
\label{marianna:WVirlc}
\end{figure}

\noindent All the light curves have been produced with the data available on the site of the AAVSO (American Association of Variable Star Observers)\footnote{\href{http://www.aavso.org/}{http://www.aavso.org/}}.

\subsection{Period-Luminosity relation}
In 1912 the American astronomer Henrietta Swan Leavitt found that for a sample of Classical Cepheids in the Large Magellanic Cloud there was a linear correlation between the apparent magnitude of the star and the logarithm of its period. Since all the Cepheids in the LMC, can be considered at the same distance from us, this relation is valid also for the absolute magnitude, up tp a zeropoint magnitude. Leavitt's discovery is known as the "Period-Luminosity relation" and can be expressed as:
\begin{equation}
M = a + b*log_{10}P.
\end{equation}
The original P-L relationship obtained by Leavitt is shown is Fig.~\ref{marianna:PL}.\\
Once it has been properly calibrated, the Period-Luminosity relation allow us to derive from the measured period of a Cepheid, its absolute magnitude and therefore, via the comparison with the observed one, its distance module. \\
\begin{figure}
\centering
\includegraphics[scale=0.6]{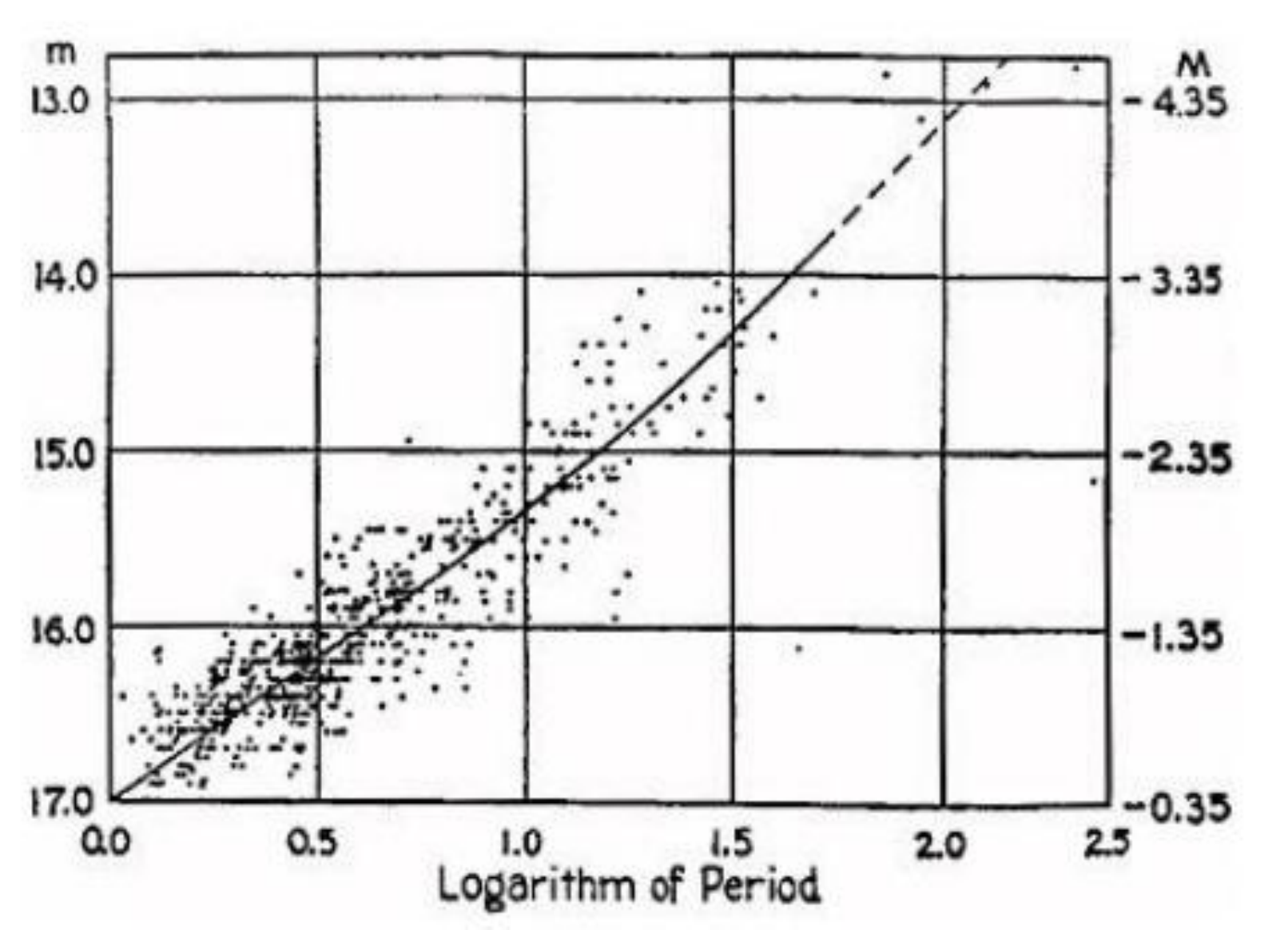}
\caption[The first period-luminosity diagram for the Cepheids.]{The first period-luminosity diagram for the Cepheids. This diagram shows Henrietta Leavitt's graph of data for the Small Magellanic Cloud. On the x-axis there is the Logarithm of the Period of the stars. On the y-axis on the left there is the average apparent magnitude of the variable as observed, on the right the absolute magnitude of the variable stars.}
\label{marianna:PL}
\end{figure}

\subsection{Cataclysmic variables: Supernovae}\label{marianna:CVs and SNe}
Cataclysmic variables are usually close binary stars in which the most massive component is usually a white dwarf and the companion is  commonly a main sequence star. The majority of these systems steadily transfer mass from the companion to the white dwarf through a surrounding accretion disk. This accreted material powers symbiotic activity, including occasional eruptions and jets. Components of this class of objects are:
\begin{itemize}
\item Novae. These systems are composed by a white dwarf and a main-sequence low mass star. A classical nova shows an increase of brightness from 7 to 15 magnitude in a range of 1 to several hundred days.
\item Dwarf Novae. These consist of a white dwarf and a red dwarf star slightly cooler of our sun. They  show semi-regular outbursts with a typical timescale ranging from weeks to years and a typical amplitude of 4-5 magnitudes.
\item Symbiotic Stars. These systems are interacting binary stars composed of an evolved red giant and a hot companion star. The hot component can be a main sequence star, a white dwarf, or a neutron star. Most symbiotics have orbital periods of a few years; some systems orbit over several decades.
\end{itemize}
The most famous type of cataclysmic variables still remain the Supernovae (SNe).\\
With the term Supernova we refer to the catastrophic explosion occurring in the later stages of the life of a massive star. During these explosions a mass of $\sim 10-100 M_{\odot}$ is ejected with velocities of about 0.01-0.1c. The explosion commonly ejects heavy elements. 
The burst of radiation in a Supernova often briefly outshines the luminosity of the  host galaxy, before fading from view over several weeks or months. \\
\begin{figure}
\centering
\begin{tabular}{cc}
\includegraphics[width=5.85 cm, height=4.55 cm]{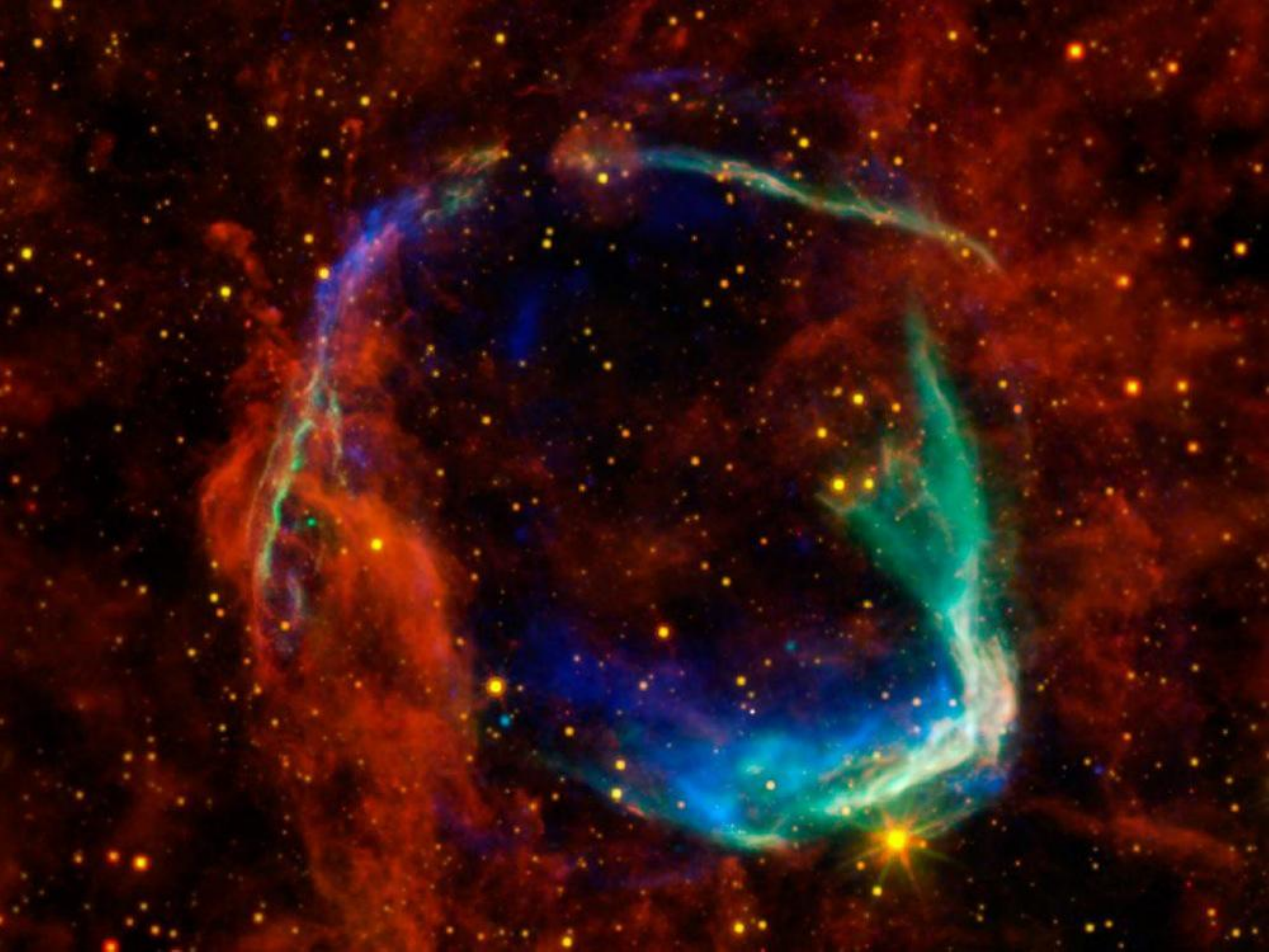}&
\includegraphics[width=5.85 cm, height=4.55 cm]{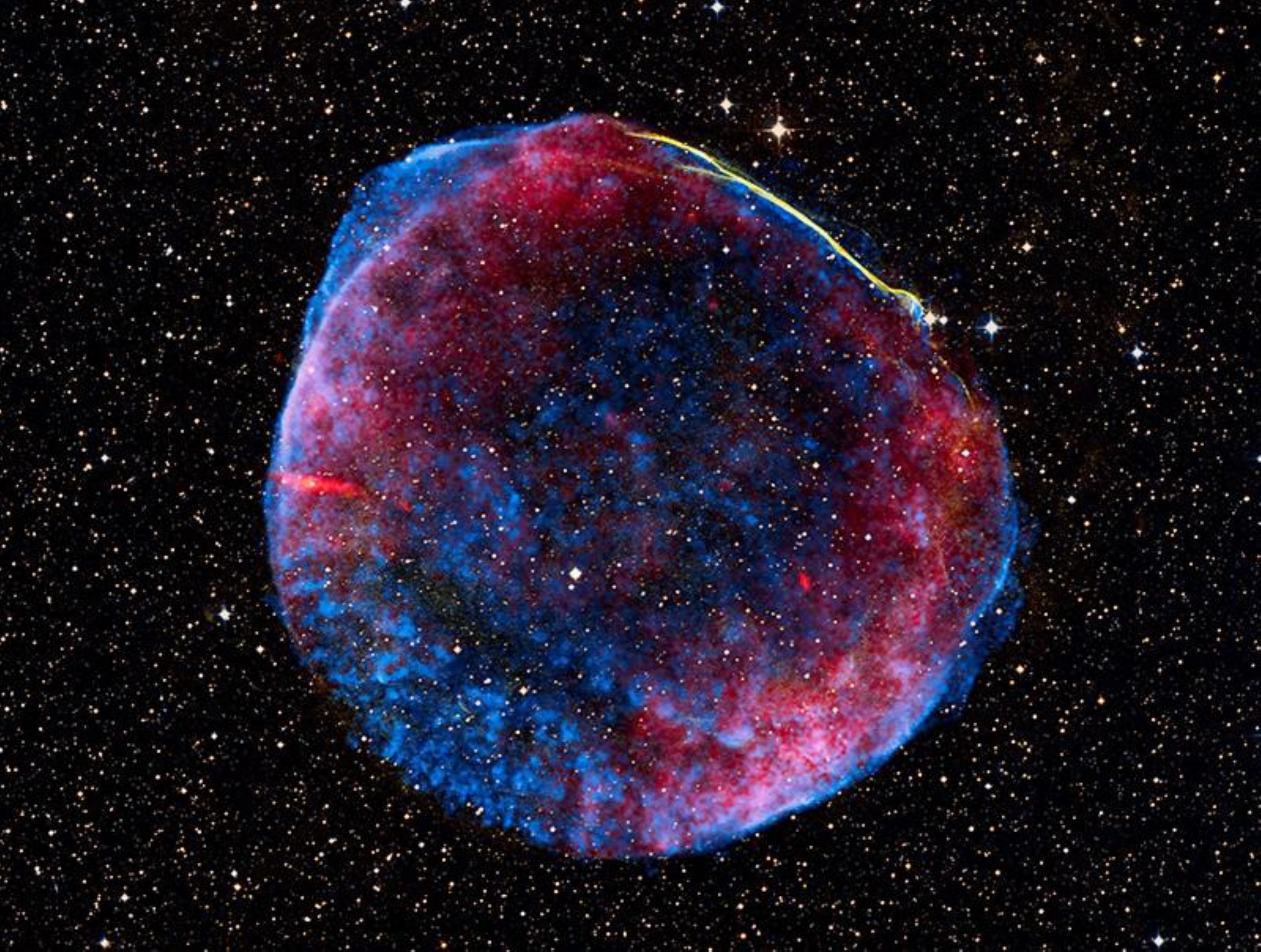} \\
\includegraphics[width=5.85 cm, height=4.55 cm]{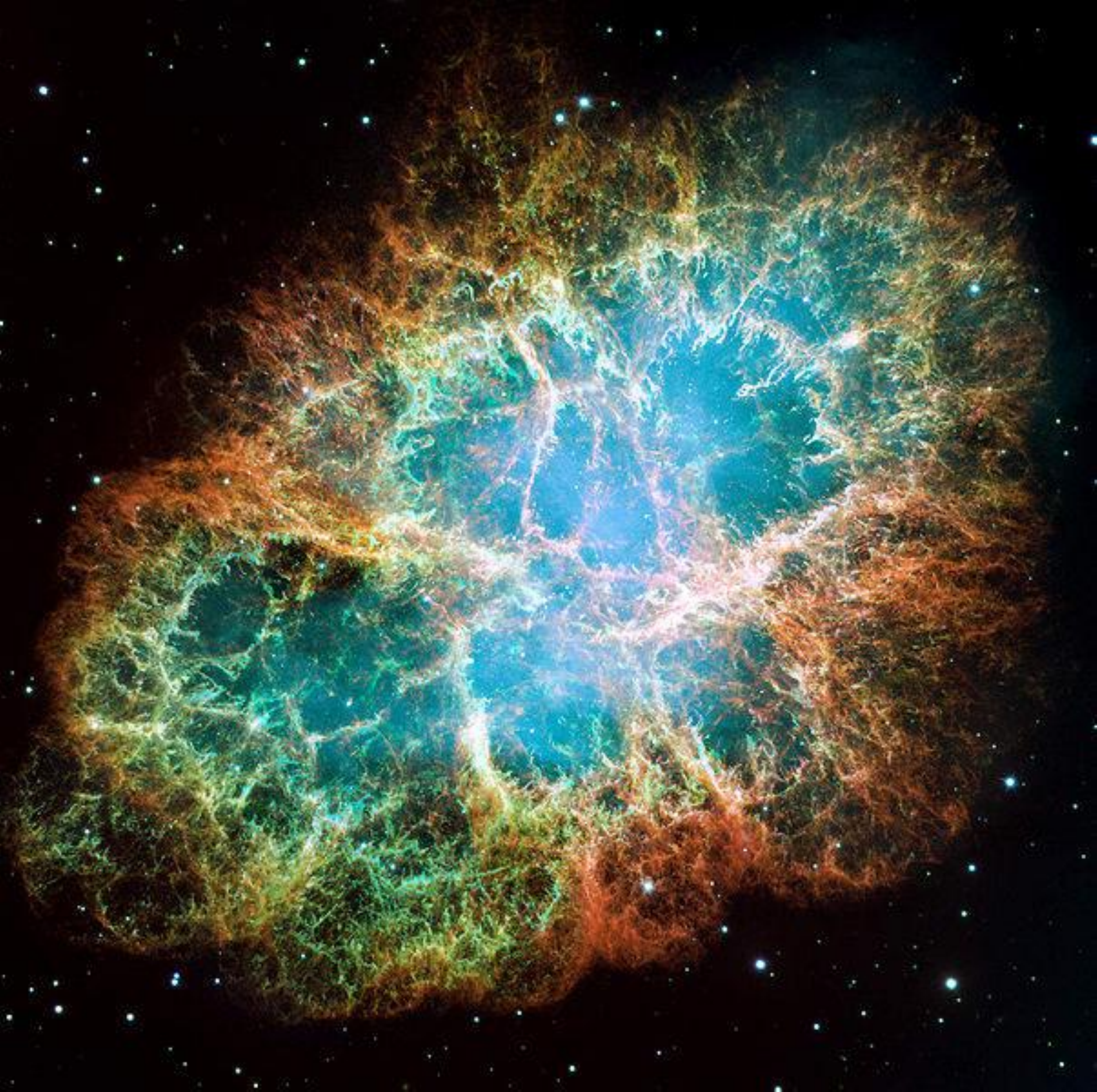} &
\includegraphics[width=5.85 cm, height=4.55 cm]{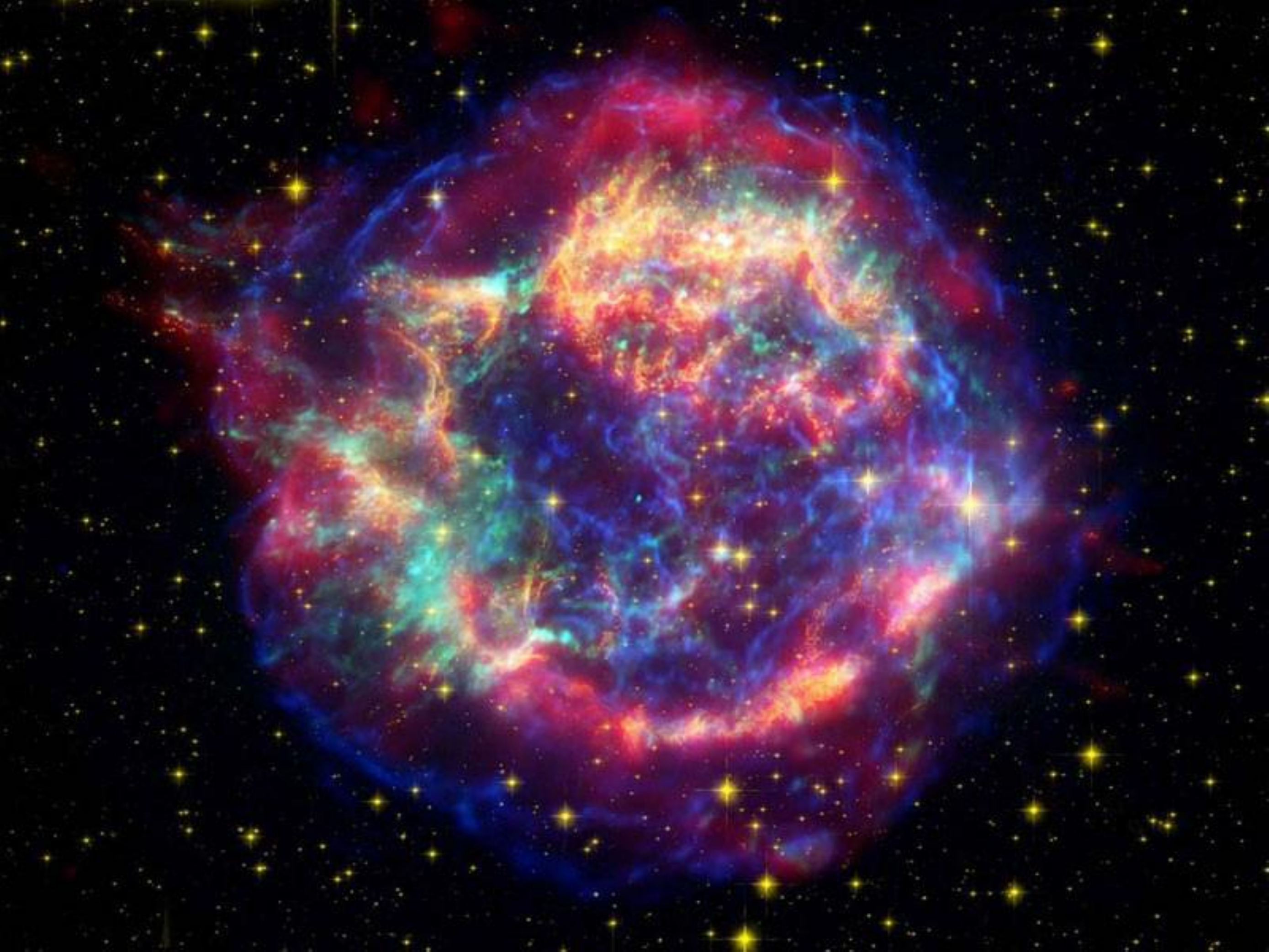}
\end{tabular}
\caption{Remnants of the Supernovae SN185, SN1006, SN1054, SN1604.}
\label{marianna:SNrem}
\end{figure}

\noindent Since no Supernova has been observed in our Galaxy in the telescopic era, it is clear that almost all we know about this phenomenon, has been derived from Supernovae in other galaxies.

\subsection{Classification of Supernovae}
Categories of Supernovae are traditionally defined by the features of their optical spectra near the maximum light and, at later stages, by the characteristics of their light curves. Supernovae were first categorised in 1941, by R. Minkowski, in two main types, type I and type II. The main difference between them being the lack of hydrogen emission line, $H_{\alpha}$ in type I Supernovae. Type Ia Supernovae are further divided in three sub classes: Type Ia, Ib and Ic, according to their spectral characteristics. Type Ia Supernovae show the absorption line of the  Si II$\lambda$6355, type Ib show, instead, the absorption line of He I$\lambda$5876 together with emission lines of Oxygen and Calcium, while type Ic Supernovae do not show any of the previous adsorption lines. type II Supernovae are divided in two further categories based on the resulting light curve following the explosion. type II-L  show a steady (Linear) decline after the maximum, whereas type II-P display a period of slower decline (a plateau) followed by a normal decay.\\
\noindent Type Ia Supernovae seem to be present in all kind of galaxies, that is ellipticals, spirals and irregulars. They show characteristic elements in their spectrum, such as magnesium, silicon, sulphur, and calcium near maximum light and iron later on. Their presence in elliptical galaxies, where there is no evidence of stellar formation, means that their progenitors must be long-lived stars.\\
\noindent Type Ib and Ic Supernovae only seem to explode in the arms of spiral gala\-xies, that is instellar-formation zones. This indicates that their progenitors must be short-lived stars. The composition of these objects is similar to that expected in the core of a massive star that has been stripped of its hydrogen. In the case of Type Ic, most of the helium is gone as well. \\
Type II Supernovae occur mostly in stellar formation zones, like HII regions of Spiral's disks, or in Irregular galaxies. Their progenitors are also short-lived stars, hence massive stars.
\subsection{Type Ia Supernovae light curves}
The light curves of all types of Supernovae, can all be explained by the energy released by the decay of the $^{56}$Ni in $^{56}$Co and subsequently in $^{56}$Fe.\\
The type Ia light curves in optical and near-infrared bands have all similar shapes. We can identify four phases.
\begin{itemize}
\item Rise time. The SN rises to maximum very fast. Only in very lucky occasions have early observations been recorded.
\item Maximum phase.
\item Second Maximum. A pronounced second maximum has been observed in redder light curves about from 20 to 40 days after the first maximum.
\item Late decline. After about 50 days the light curves settle onto a steady decline, which is exponential in luminosity.
\end{itemize}
The peak luminosity is directly linked to the amount of radioactive $^{56}$Ni produced during the explosion. The rise time of the light curve is determined primarily by the explosion energy and by the manner in which the ejecta become optically thin to thermalized radiation, while the late decline of the light curve is governed by the combination of the energy input by the radioactive material and the rate at which this input energy is converted to optical photons in the ejecta.

An example of the light curve of the Supernova SN 1998bu in the M96 galaxy in the filters of the Johnson photometric system is reported below.
\begin{figure}
\centering
\includegraphics[scale=0.45]{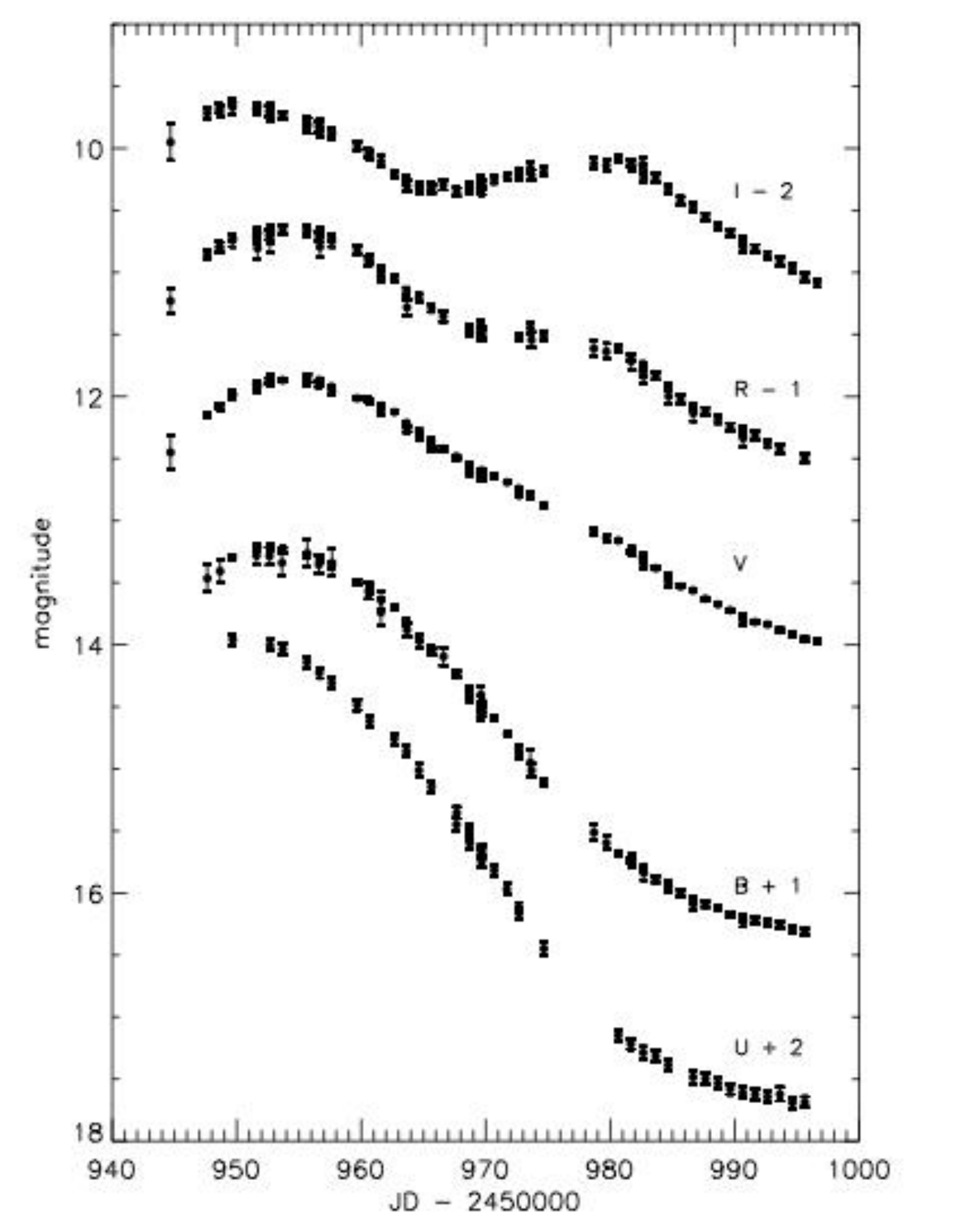}
\caption[UBVRI light curves of SN 1998bu.]{UBVRI light curves of SN 1998bu, credit to \citealt{jha1999}. The light curves in U,B,R,I have been shifted to avoid overlapping.}
\label{marianna:UBVRISN}
\end{figure}

\section{The detection experiments}
\label{marianna:4}

As we said in section~\ref{marianna:2} among the scopes of this work there was also to test several data mining algorithms in order to find the one optimized for each type of variable object. \\
The algorithms we used in this work belong to the rather wide category of Neural Networks which have been used for classification tasks in a variety of scientific and non scientific domains.\\
The term Neural Network refers to an artificial system of information processing methods that attempt to simulate the functional mechanisms at the base of the human brain (\citealt{bishop2006}).\\
Neural Networks are mathematical models which define a function f: \textbf{X} $\rightarrow$ \textbf{Y}, between a set of input variables (also called features) and a set of output variables (the targets).
This function f(x) can be defined as a composition of other functions  $\mathrm{g_i(x)}$. A widely used type of composition is the nonlinear weighted sum, $\mathrm{f(\textbf{x}) = K \left(\sum_i w_i g_i(\textbf{x})\right)}$ , where  K (commonly referred to as the activation function) is some predefined function.\\
What is most interesting of the Neural Networks is their possibility to learn.  Given a specific task to solve, and a class of functions F, learning means using a set of observations to find $\mathrm{f^* \in F}$ which solves the task in some optimal sense. There are two different learning paradigms for a neural network: supervised and unsupervised. We can use a supervised method when we have a training set including typical examples of the inputs and the corresponding outputs: in this way the network can learn how to infer the relation between the input and the output variables. Then, the network is trained by using a variety of suitable learning rules (such as the well known Back Propagation; \citealt{bishop2006}), which use the input-output data samples (also called patterns) in order to modify the internal weights and other parameters of the network itself in order to minimize an error function, representing the training error.
If the training is successful, the network learns to recognize the unknown relationship between the input and the output variables, and is therefore able to make predictions on new input samples even if their output is not known a priori (generalization capability).\\
An unsupervised learning method, instead, is based on training algorithms that modify the weights of the network making reference only to a set of data that includes the only input variables. These algorithms attempt to group the input data by making use of topological or probabilistic methods.\\

\subsection{The experiments}
\label{marianna:4.2}
As we explained in section~\ref{marianna:2} the strategy of our project is to use a hierarchical approach to variable object classification. This approach has the typical decision tree structure and aims at a classification which becomes finer and finer as we to higher level of branching. The first level of this approach is to perform the crispy classification
based on the variable/not variable object dichotomy. We choose to use the MLPQNA algorithm, since it is the one which provides best results to several astrophysical problems (\citealp{brescia2012a}, \citealp{brescia2012e}) and the one which deals better with poorly populated datasets.\\
In the following subsections there are presented the data used for these tests and the strategy behind the choice of the parameters for the classifier.

\subsection{The data}
\label{marianna:4.2.1}

To test the MLPQNA algorithm for the variable object classification we built a set of four simulations, each of them consisting in 50 images, corresponding to 50 different epochs, spaced within 90 days with an uneven sampling rate. The instrumental characteristics are the same for each simulation and fine tuned the characteristics of VST optics and instrumentation (see Sect. \ref{marianna:2.7}). The characteristics of the detector, such as the gain and saturation level are chosen equal to those defined in Sect. \ref{marianna:2.7}, while the image size varied between the simulations.
The magnitude range is set to 14-25 mag, in order to remain within the magnitude limit of SExtractor. The seeing FWHM is chosen to vary randomly between 0.6 and 1.0 arcsec (respectively medium and worst conditions at VST site, Cerro Paranal, Chile), while the exposure time is set to 1500s.\\
The types objects simulated in our images are: non variable stars, and galaxies, Cepheids, type Ia Supernovae with their host galaxies and random variable objects approximating the behavior of eruptive variables and Active galactic nuclei.
As described in Sect.~\ref{marianna:2.5.2} every SN is associated to a nearby galaxy. This implies that in some cases the SN is so close to the nucleus of the parent galaxy that the extraction software fails in deblending the Supernova. In these cases the whole galaxy appears to br variable due to the contribution of the SN (Fig.~\ref{marianna:hg}).\\
\begin{figure}[t]
\centering
\includegraphics[width=12cm]{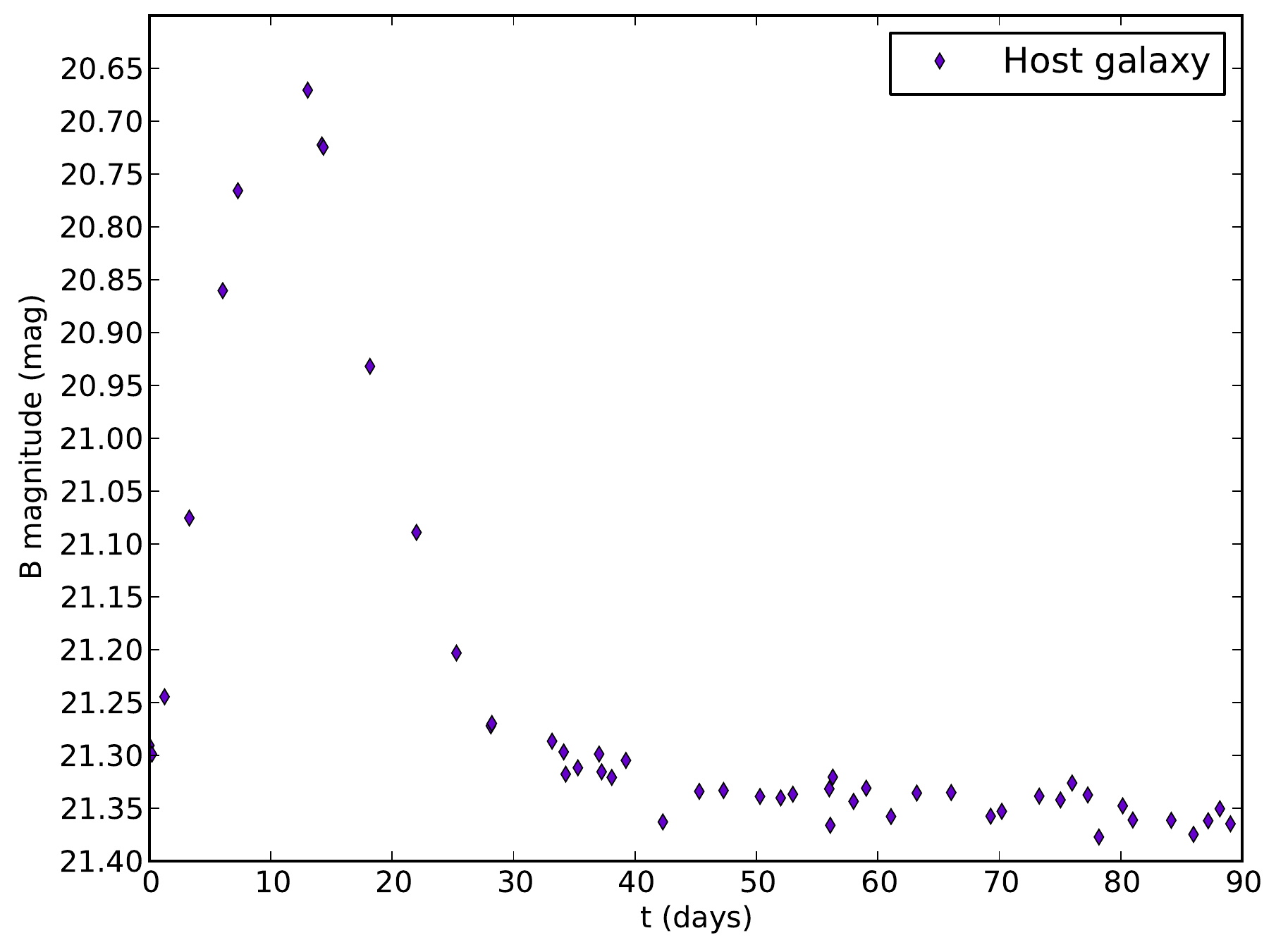}
\caption{Light curve of a Host galaxy, whose Supernova is not detected by the extraction software. }
\label{marianna:hg}
\end{figure}

\noindent Tables~\ref{marianna:sim1}-~\ref{marianna:sim4} summarize the number of the objects in each simulation, divided in their categories, i.e. the total catalog, the train and test sets respectively. The number of the objects is obviously different in each simulation, as result of the different image size.  Fig~\ref{marianna:sim1image}-\ref{marianna:sim4image} show a stamp of the B image at t=0d for each simulation.\\
\noindent For each simulation we obtain a train and a test catalog. The train set contains $\sim$ 80 \% of the total number of the objects, being careful to assign a Supernova and its host galaxy to the same set. \\

\begin{table}
\small
\centering
\begin{tabular}{|%
>{\columncolor{lightblue}{\bf}}  l%
|>{\bf}l %
|c|c|c|}
\hline
\cellcolor{blue_2}{OBJECTS} &\cellcolor{blue_2} {TYPE}  &  \cellcolor{blue_2}  {\textbf{FULL}} &  \cellcolor{blue_2}  {\textbf{TRAIN}}&  \cellcolor{blue_2}  {\textbf{TEST}}\\
\hline
& SN Ia                                             & 80 & 64 & 16\\
& Cepheids                                          & 80 & 64 & 16\\
& Random                                            & 80 & 64 & 16\\
& Host Galaxy with SN                               & 80 & 64 & 16\\
\multirow{-5}{*}{Variable} & Host Galaxy without SN & 11 & 8 & 3\\
\hline
& Stars                                             & 216 & 172 & 44\\
\multirow{-2}{*}{Not variable}& Galaxies            & 1259 & 1007 & 252\\
\hline
TOTAL &                                             & 1806 & 1443 & 363\\
\hline
\end{tabular}
\caption[Number of objects in the first simulation.]{Number of objects in the first simulation. For each class of objects, the col. 3 shows the quantities in the entire simulation, while col. 4 and 5 show the number of the objects in train and test set respectively.}
\label{marianna:sim1}
\end{table}

\begin{figure}
\centering
\includegraphics[width=12cm]{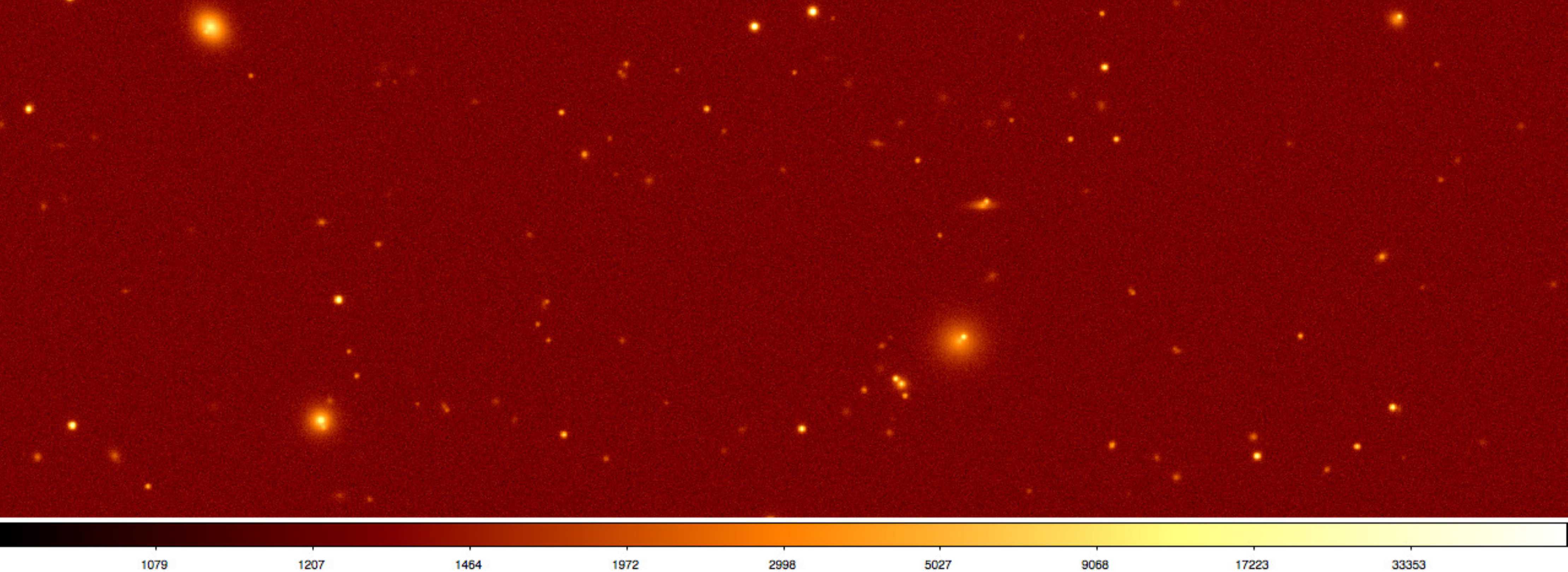}
\caption{Stamp of the B band image at t=0d for the first simulation.}
\label{marianna:sim1image}
\end{figure}

\begin{table}
\small
\centering
\begin{tabular}{|%
>{\columncolor{green3}{\bf}}  l%
|>{\bf}l %
|c|c|c|}
\hline
\cellcolor{green4}{OBJECTS} &\cellcolor{green4} {TYPE}  &  \cellcolor{green4}  {\textbf{FULL}} &  \cellcolor{green4}  {\textbf{TRAIN}}&  \cellcolor{green4}  {\textbf{TEST}}\\
\hline
& SN Ia & 206 & 160 & 46\\
& Cepheids & 200 & 160 & 40\\
& Random & 200 & 160 & 40\\
& Host Galaxy with SN & 206 & 160 & 46\\
\multirow{-5}{*}{Variable} & Host Galaxy without SN & 23 & 18 & 5\\
\hline
& Stars & 617 & 493 & 124\\
\multirow{-2}{*}{Not variable}& Galaxies & 3510 & 2808 & 702\\
\hline
TOTAL & & 4956 & 3959 & 1003\\
\hline
\end{tabular}
\caption[Number of objects in the second simulation.]{Number of objects in the second simulation. For each class of objects, the col. 3 shows the quantities in the entire simulation, while col. 4 and 5 show the number of the objects in train and test set respectively.}
\label{marianna:sim2}
\end{table}

\begin{figure}
\centering
\includegraphics[width=12cm]{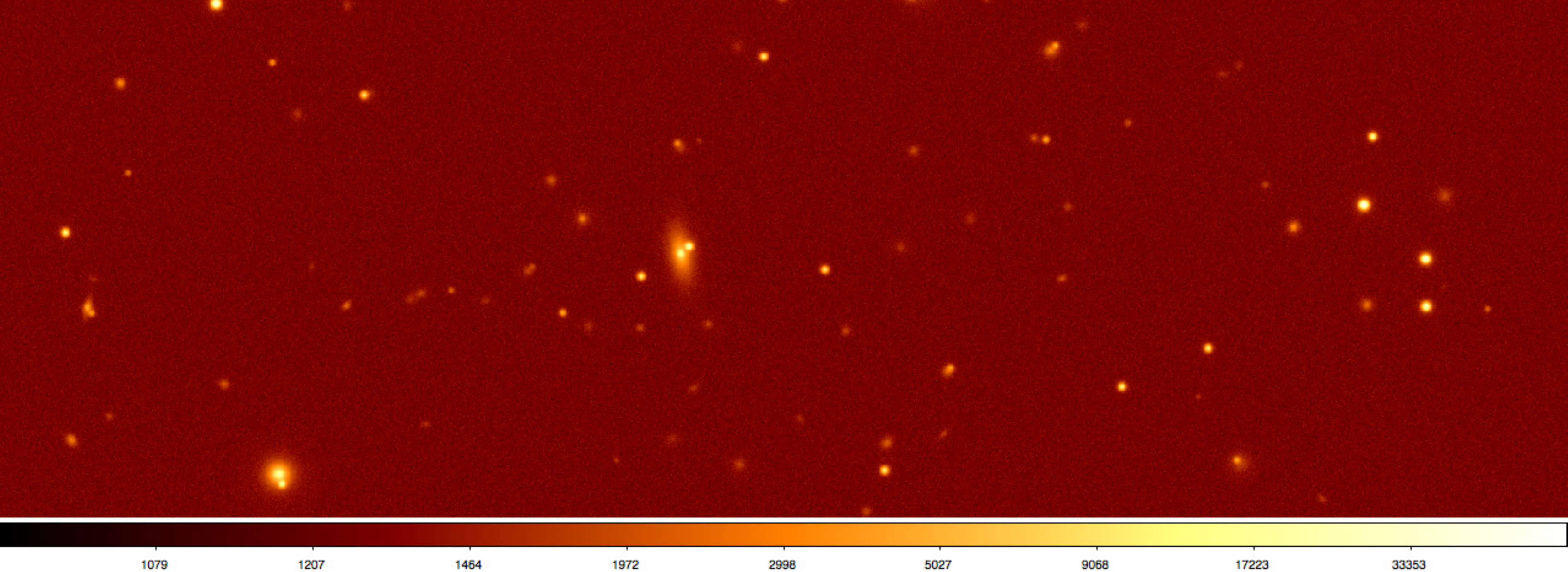}
\caption{Stamp of the B band image at t=0d for the second simulation.}
\label{marianna:sim2image}
\end{figure}

\begin{table}
\small
\centering
\begin{tabular}{|%
>{\columncolor{yellow3}{\bf}}  l%
|>{\bf}l %
|c|c|c|}
\hline
\cellcolor{orange2}{OBJECTS} &\cellcolor{orange2} {TYPE}  &  \cellcolor{orange2}  {\textbf{FULL}} &  \cellcolor{orange2}  {\textbf{TRAIN}}&  \cellcolor{orange2}  {\textbf{TEST}}\\
\hline
& SN Ia & 1081 & 678 & 144\\
& Cepheids & 1100 & 677 & 173\\
& Random & 1100 & 702 & 148\\
& Host Galaxy with SN & 1081 & 678 & 172\\
\multirow{-5}{*}{Variable} & Host Galaxy without SN & 18 & 11 & 3\\
\hline
& Stars & 1374 & 1093 & 281\\
\multirow{-2}{*}{Not variable}& Galaxies & 6580 & 5245 & 1333\\
\hline
TOTAL & & 12334 & 9084 & 2254\\
\hline
\end{tabular}
\caption[Number of objects in the third simulation.]{Number of objects in the third simulation. For each class of objects, the col. 3 shows the quantities in the entire simulation, while col. 4 and 5 show the number of the objects in train and test set respectively.}
\label{marianna:sim3}
\end{table}

\begin{figure}
\centering
\includegraphics[width=12cm]{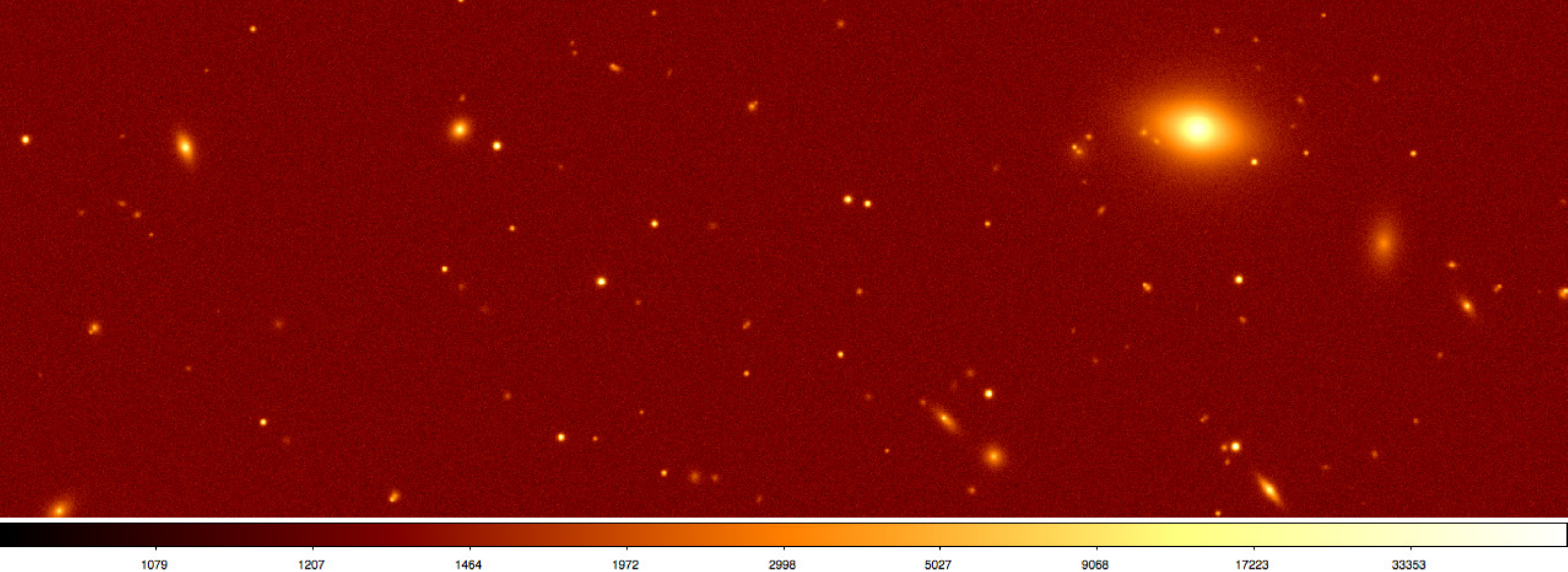}
\caption{Stamp of the B band image at t=0d for the third simulation.}
\label{marianna:sim3image}
\end{figure}
\vspace{1cm}
\begin{table}
\small
\centering
\begin{tabular}{|%
>{\columncolor{lilla}{\bf}}  l%
|>{\bf}l %
|c|c|c|}
\hline
\cellcolor{viola2}{OBJECTS} &\cellcolor{viola2} {TYPE}  &  \cellcolor{viola2}  {\textbf{FULL}} &  \cellcolor{viola2}  {\textbf{TRAIN}}&  \cellcolor{viola2}  {\textbf{TEST}}\\
\hline
& SN Ia & 1079 & 681 & 169\\
& Cepheids & 1099 & 670 & 180\\
& Random & 1100 & 705 & 145\\
& Host Galaxy with SN & 1079 & 681 & 169\\
\multirow{-5}{*}{Variable} & Host Galaxy without SN & 19 & 12 & 7\\
\hline
& Stars & 1387 & 1099 & 288\\
\multirow{-2}{*}{Not variable}& Galaxies & 6576 & 5263 & 1313\\
\hline
TOTAL & & 12339 & 9111 & 2271\\
\hline
\end{tabular}
\caption[Number of objects in the fourth simulation.]{Number of objects in the forth simulation. For each class of objects, the col. 3 shows the quantities in the entire simulation, while col. 4 and 5 show the number of the objects in train and test set respectively.}
\label{marianna:sim4}
\end{table}

\begin{figure}
\centering
\includegraphics[width=12cm]{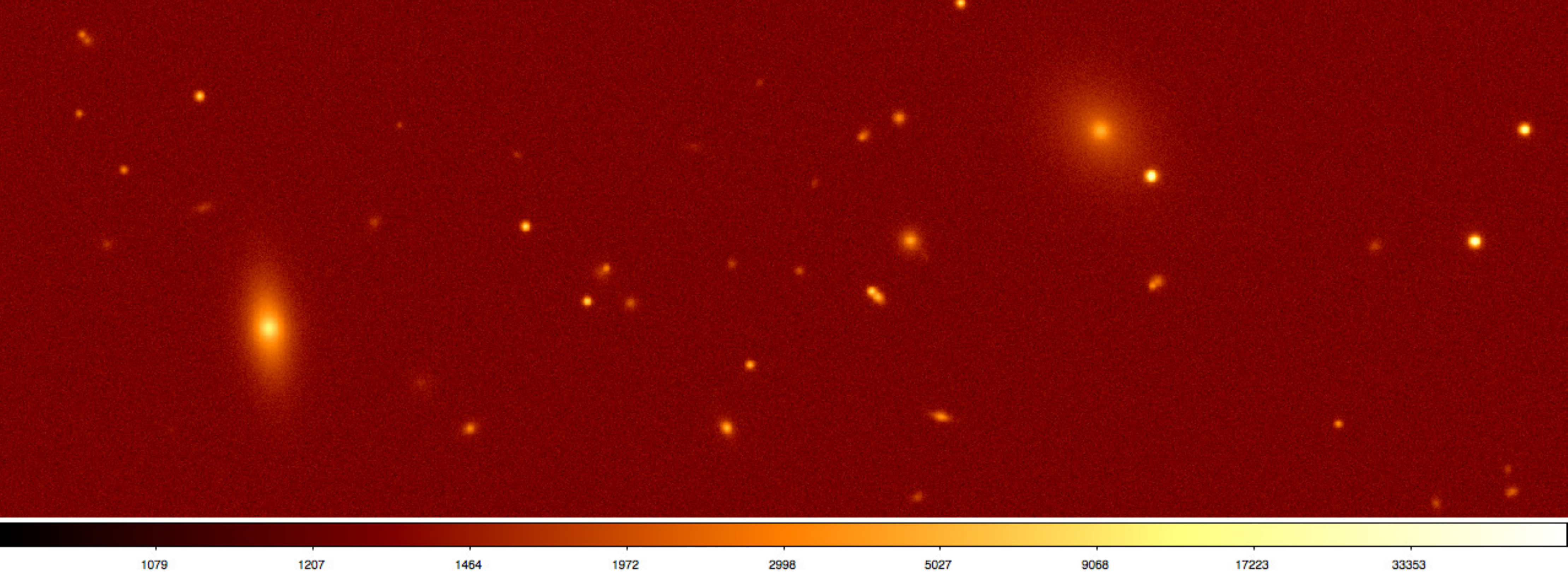}
\caption{Stamp of the B band image at t=0d for the fourth simulation.}
\label{marianna:sim4image}
\end{figure}

\subsection{Choice of parameters for MLPQNA}
\label{marianna:4.2.2}

In this first phase of the project we choose to use as parameters for MLPQNA a set of magnitudes, (Kron magnitudes as estimated by the source extraction software), and times at which they are measured ($\mathrm{m_i,\,  t_i}$) . In order to be more congruent with a real case, and in order to reduce the computational time, we can not use ($\mathrm{m_i,\,  t_i}$) for all the available epochs, but we have to select a subset of epochs. How many epochs, and how they must be chosen, must be carefully evaluated. There are at least three possibilities to take into account:
\begin{enumerate}
\item N epochs randomly extracted equal for each object;
\item N epochs randomly extracted different for each objects;
\item N epochs equally spaced equal dor each objects.
\end{enumerate}
\noindent Once selected the epochs according to one of these possibilities, we have to be sure that among them each object has at least one measure of magnitude. It is possible, in fact, that the object is not always detected from source extraction software. This happens when the magnitude of the object is near the limit in magnitude of the source extraction software. It is possible that due to their magnitude variation a variable object is detected only in some epochs.
If an object in the train or in test set in the chosen epochs does not have any measure for the magnitude, it is rejected.
\subsection{Results}
\label{marianna:4.2.3}

The results of each test performed by the MLPQNA on each simulation are discussed below in terms of three evaluation criteria: \textit{accuracy, purity} and \textit{contamination}. The accuracy (CA) is the fraction of objects correctly classified (either variable or not-variable), with respect to the total number of objects in the sample. The purity (CO) is the fraction of variable objects correctly classified as variable. The contamination is the fraction of not variable objects erroneously classified as variable.\\

\subsubsection{Test 1}
In the first test we used a simulation dataset of 1806 objects, consisting of 1443 objects as training set and 363 objects as test set, as shown in Tab.~\ref{marianna:sim1}.\\

\noindent In terms of the internal parameter setup of the MLPQNA, we used the following topological parameters:
\begin{itemize}
\item \textbf{MLP network topology}: a three layer MLP;
\item  \textbf{input layer}: 20;
\item  \textbf{hidden layer}: 41;
\item  \textbf{output layer}: 2.
\end{itemize}

For the QNA learning rule we fixed the following values as best parameters:
\begin{itemize}
\item  \textbf{step}: $0.0001$;
\item  \textbf{res}: $30$;
\item  \textbf{dec}: $0.01$;
\item  \textbf{MaxIt}: $3000$;
\item  \textbf{CV(k)}: $10$;
\item  \textbf{Error evaluation}: Cross Entropy.
\end{itemize}

\noindent The numerical results are shown in the confusion matrices referred to respectively, training phase in Tab.~\ref{marianna:sim1trainconfmat} and test phase in Tab.~\ref{marianna:sim1testconfmat}. \\

\begin{table}
\small
\centering
\begin{tabular}{c|c|c} %
& Predicted class 1 & Predicted class 2\\
\hline
Target class 1 & \cellcolor{lightblue} 1173 & 6\\
\hline
Target class 2 & 9 & \cellcolor{lightblue} 255\\
\end{tabular}\\
\caption[Confusion matrix of the training performed with the data of the first simulation.]{Confusion matrix of the training performed with the data of the first simulation (Tab.~\ref{marianna:sim1}). Each column of the matrix represents the instances in a predicted class, while each row represents the instances in an actual class. All correct guesses are located in the diagonal of the table.}
\label{marianna:sim1trainconfmat}
\end{table}

\noindent As we can see in the training case, we obtain:
\begin{itemize}
\item total classification percentage: 98.96\%
\item class 1 (NOT VARIABLES) classification percentage: 99.49\%
\item class 2 (VARIABLES) classification percentage: 96.59\%
\end{itemize}

\begin{table}
\small
\centering
\begin{tabular}{c|c|c} %
& Predicted class 1 & Predicted class 2\\
\hline
Target class 1 & \cellcolor{lightblue} 280 & 16\\
\hline
Target class 2 & 29 & \cellcolor{lightblue} 30\\
\end{tabular}\\
\caption[Confusion matrix of the test performed with the data of the first simulation.]{Confusion matrix of the test performed with the data of the first simulation (Tab.~\ref{marianna:sim1}). Each column of the matrix represents the instances in a predicted class, while each row represents the instances in an actual class. All correct guesses are located in the diagonal of the table.}
\label{marianna:sim1testconfmat}
\end{table}

\noindent As we can see in the test case, in terms of statistical indicators, the \textit{accuracy} of the network, which is the ratio between the number of the objects on the diagonal of the matrix and the total number of the objects of the test set, is $\sim$88 \% , while the \textit{purity} is $\sim$ 57 \%. The \textit{contamination} of the experiment is about $\sim$ 5\%.

\subsubsection{Test 2}
In the second test we used a simulation dataset of 4956 objects, consisting of 3959 objects as training set and 1003 objects as test set, as shown in Tab.~\ref{marianna:sim2}.\\
\noindent The internal parameter setup of the MLPQNA has be set as in Test 1.\\
\noindent The numerical results are shown in the confusion matrices referred to respectively, training phase in Tab.~\ref{marianna:sim2trainconfmat} and test phase in Tab.~\ref{marianna:sim2testconfmat}. \\

\begin{table}
\small
\centering
\begin{tabular}{c|c|c} %
& Predicted class 1 & Predicted class 2\\
\hline
Target class 1 & \cellcolor{green3} 3274 & 27\\
\hline
Target class 2 & 16 & \cellcolor{green3} 655\\
\end{tabular}\\
\caption[Confusion matrix of the training performed with the data of the second simulation.]{Confusion matrix of the training performed with the data of the second simulation (Tab.~\ref{marianna:sim2}). Each column of the matrix represents the instances in a predicted class, while each row represents the instances in an actual class. All correct guesses are located in the diagonal of the table.}
\label{marianna:sim2trainconfmat}
\end{table}

\noindent As we can see in the training case, we obtain:
\begin{itemize}
\item total classification percentage: 98.92\%
\item class 1 (NOT VARIABLES) classification percentage: 99.18\%
\item class 2 (VARIABLES) classification percentage: 97.61\%
\end{itemize}

\begin{table}
\small
\centering
\begin{tabular}{c|c|c} %
& Predicted class 1 & Predicted class 2\\
\hline
Target class 1 & \cellcolor{green3} 766 & 60\\
\hline
Target class 2 & 58 & \cellcolor{green3} 123\\
\end{tabular}
\caption[Confusion matrix of the test performed with the data of the second simulation.]{Confusion matrix of the test performed with the data of the second simulation (Tab.~\ref{marianna:sim2}). Each column of the matrix represents the instances in a predicted class, while each row represents the instances in an actual class. All correct guesses are located in the diagonal of the table.}
\label{marianna:sim2testconfmat}
\end{table}

\noindent The \textit{accuracy} of the network in this case remains $\sim$ 88 \% , while the \textit{purity} increases of about $\sim$ 10\% by doubling the dimension of the dataset, reaching a value of $\sim$ 67 \%. The \textit{contamination} of the experiment is about $\sim$ 7\%.

\subsubsection{Test 3}
In the third test we used a simulation dataset of 12334 objects, consisting of 9084 objects as training set and 2254 objects as test set, as shown in Tab.~\ref{marianna:sim3}.\\
\noindent The internal parameter setup of the MLPQNA has be set as in Test 1.\\
\noindent The numerical results are shown in the confusion matrices referred to respectively, training phase in Tab.~\ref{marianna:sim3trainconfmat} and test phase in Tab.~\ref{marianna:sim3testconfmat}. \\

\begin{table}
\small
\centering
\begin{tabular}{c|c|c} %
& Predicted class 1 & Predicted class 2\\
\hline
Target class 1 & \cellcolor{yellow3} 2511 & 235\\
\hline
Target class 2 & 162 & \cellcolor{yellow3} 6176\\
\end{tabular}\\
\caption[Confusion matrix of the training performed with the data of the third simulation.]{Confusion matrix of the training performed with the data of the third simulation (Tab.~\ref{marianna:sim3}). Each column of the matrix represents the instances in a predicted class, while each row represents the instances in an actual class. All correct guesses are located in the diagonal of the table.}
\label{marianna:sim3trainconfmat}
\end{table}

\noindent As we can see in the training case, we obtain:
\begin{itemize}
\item total classification percentage: 95.63\%
\item class 1 (NOT VARIABLES) classification percentage: 91.44\%
\item class 2 (VARIABLES) classification percentage: 97.44\%
\end{itemize}

\begin{table}
\small
\centering
\begin{tabular}{c|c|c} %
& Predicted class 1& Predicted class 2\\
\hline
Target class 1 & \cellcolor{yellow3} 384 & 393\\
\hline
Target class 2 & 244 & \cellcolor{yellow3} 1233\\
\end{tabular}
\caption[Confusion matrix of the test performed with the data of the third simulation.]{Confusion matrix of the test performed with the data of the third simulation (Tab.~\ref{marianna:sim3}). Each column of the matrix represents the instances in a predicted class, while each row represents the instances in an actual class. All correct guesses are located in the diagonal of the table.}
\label{marianna:sim3testconfmat}
\end{table}

\noindent The \textit{accuracy} of the network decreases to $\sim$ 72 \% . The \textit{purity} is $\sim$ 83 \%. The \textit{contamination} increases to $\sim$ 51\%.

\subsubsection{Test 4}
In the forth test we used a simulation dataset of 12339 objects, consisting of 9111 objects as training set and 2271 objects as test set, as shown in Tab.~\ref{marianna:sim4}.\\

\noindent In terms of the internal parameter setup of the MLPQNA, we used the following topological parameters:
\begin{itemize}
\item \textbf{MLP network topology}: a four layer MLP;
\item \textbf{input layer}: 20;
\item \textbf{first hidden layer}: 41;
\item \textbf{second hidden layer}: 20;
\item \textbf{output layer}: 2.
\end{itemize}
For the QNA learning rule, after several trials, we fixed the following values as best parameters:
\begin{itemize}
\item \textbf{step}: $0.001$;
\item \textbf{res}: $30$;
\item \textbf{dec}: $0.01$;
\item \textbf{MaxIt}: $4000$;
\item \textbf{Error evaluation}: Cross Entropy.
\end{itemize}

\noindent The numerical results are shown in the confusion matrices referred to respectively, training phase in Tab.~\ref{marianna:sim4trainconfmat} and test phase in Tab.~\ref{marianna:sim4testconfmat}. \\

\begin{table}
\small
\centering
\begin{tabular}{c|c|c} %
& Predicted class 1 & Predicted class 2\\
\hline
Target class 1 & \cellcolor{lilla} 2645 & 104\\
\hline
Target class 2 & 83 & \cellcolor{lilla} 6278\\
\end{tabular}\\
\caption[Confusion matrix of the training performed with the data of the fourth simulation.]{Confusion matrix of the training performed with the data of the fourth simulation (Tab.~\ref{marianna:sim4}). Each column of the matrix represents the instances in a predicted class, while each row represents the instances in an actual class. All correct guesses are located in the diagonal of the table.}
\label{marianna:sim4trainconfmat}
\end{table}

\noindent As we can see in the training case, we obtain:
\begin{itemize}
\item total classification percentage: 97.9473\%
\item class 1 (NOT VARIABLES) classification percentage: 96.22\%
\item class 2 (VARIABLES) classification percentage: 98.69\%
\end{itemize}

\begin{table}
\small
\centering
\begin{tabular}{c|c|c} %
& Predicted class 1 & Predicted class 2\\
\hline
Target class 1 & \cellcolor{lilla} 439 & 231\\
\hline
Target class 2 & 266 & \cellcolor{lilla} 1335\\
\end{tabular}
\caption[Confusion matrix of the test performed with the data of the fourth simulation.]{Confusion matrix of the test performed with the data of the fourth simulation (Tab.~\ref{marianna:sim4}). Each column of the matrix represents the instances in a predicted class, while each row represents the instances in an actual class. All correct guesses are located in the diagonal of the table.}
\label{marianna:sim4testconfmat}
\end{table}

\noindent The \textit{accuracy} of the network is $\sim$ 78 \%, while the \textit{purity} is $\sim$ 65 \%. The \textit{contamination} is $\sim$ 40\%.\\
\noindent In Tab.~\ref{marianna:train_test_results} there are summarized the results obtained in all four simulation. For completeness we report also the results obtained with the train set, although less relevant for the quality evaluation. In fact, it is clear that the only relevant results are those obtained with the test sets, that have not been used for the network training process.\\
\newcolumntype{g}{>{\columncolor{lightblue}}c}
\begin{table}
\centering
\small
\begin{tabular}{|l|g|g|g|c|c|c|} \hline
        & \multicolumn{3}{g|}{\textbf{TRAIN}}   & \multicolumn{3}{c|}{\textbf{TEST}}\\ \cline{2-7}
        & CA \% & CO \% & CN \%                 & CA \% & CO \% & CN \% \\ \hline
SIM 1   & 98,96 & 96,59 & 0,51                  & 87,60 & 56,72 & 5,41  \\ \hline
SIM 2   & 98,92 & 97,61 & 0,82                  & 88,28 & 67,96 & 7,26  \\ \hline
SIM 3   & 95,63 & 97,44 & 8,56                  & 71,74 & 83,48 & 50,58 \\ \hline
SIM 4   & 97,95 & 96,22 & 1,30                  & 78,12 & 65,52 & 39,70 \\ \hline
\end{tabular}
\caption[Train and test recognition rates for the four simulations]{Train and test recognition rates for the four simulations described in Sect.~\ref{marianna:4.2}.}
\label{marianna:train_test_results}
\end{table}

\noindent As we can see from the previous tests and in Tab.~\ref{marianna:train_test_results}, when increasing the sample of objects there is a large increase in the contamination, while not obtaining a significant improvement in terms of accuracy and purity of the network. This also by exploring slight differences in the model setup. \\
\noindent These results, although not exalting, are to be considered very preliminary. In particular, what affects the classification is the choice of the features, that in this case is carrying poor information in terms of feature correlation. We also decided to use MLPQNA, one of the more robust existing classification empirical models, based on the machine learning supervised paradigm, in order to exclude, with a high confidence, the possibility that poor results could be associated to the selected model. So far, future developments of the work will consists basically into the investigation whether the classification may be improved by using more fine statistical features, indirectly derived from the light curves, and not simply the light curves themselves. \\

    \section{Euclid Supernovae Working Group}\label{sec:supernovae}
        
I was involved in the definition of the requirements for the proposal for a Supernova Pipeline
What follows is largely extracted from our proposal to the Euclid Board for the Supernova Pipeline.
I wish to stress that in the near future I expect to use the above described simulation pipeline to produce simulated Euclid data streams.
Generally speaking, transient science requires rapid reduction of data in order to trigger follow-up observations. In the case of Euclid, transient searching is likely to be limited to very specific datasets, for example repeated observations of the deep fields and/or a dedicated SN survey taking place over a limited period during the mission. Thus our goals do not need to be met for the entire Euclid dataset, since it concerns one (or several) special deep survey field(s) observed many times.

Below we use the term ``Supernova Trigger Pipeline" (STP) to mean the rapid inspection of the data that is needed to trigger follow up observations. The STP is addressing the treatment of images taken each time the SN field is surveyed.
The ``Supernova Final Pipeline" (SFP) addresses the general (re)-processing of all the images associated with a given supernova, with the final reference images built from all images, and the analysis of light-curves.
The STP data does not need to be calibrated to the same precision as the final reductions. Whether or not the STP and SFP reductions are performed using the same pipeline software is left open at this stage. In particular, in the early stages of the transient search it is important that the pipeline software be quickly adaptable, as lessons are learned from the first subtractions of real data. Not only can the reference image be updated, but also the code might need optimization. The software should be engineered to allow easy maintenance/debugging upgrade (especially during commissioning).

In case of long, possibly single exposures for our dedicated survey, it would be necessarily to consider not standard dither patterns. However other constraints may mean to split the exposures, allowing sub-pixel dithering. This is also likely to make clean image subtraction more difficult. For both STP and SFP analysis, we could need to create an oversampled reference image for subtraction (made from several dithers). Note that this is specific to Euclid, so existing SN search software will not work straightforwardly.

\subsection{Summary of Euclid Science Ground Segment specifications}
In this section a quick overview of Science Ground Segment (SGS) specifications is recalled, by focusing the attention to some aspects strictly related to the proposed SN pipelines and detection strategies.

Concerning the official data level hierarchy, shown in \ref{euclid:SN1}, there are several constraints to the availability of data at different processing levels, that must be taken into account at this stage of the preliminary design of the SN pipelines.

\begin{figure}
\centering
\includegraphics[width=7.5cm]{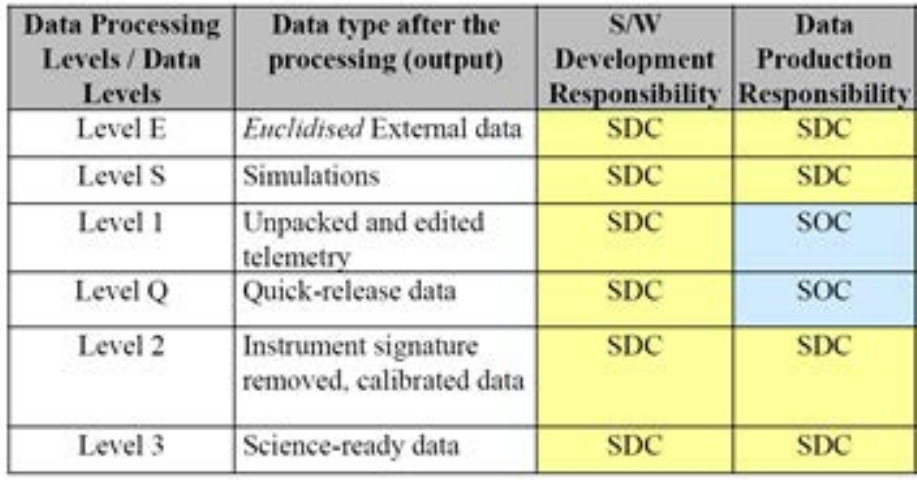}
\caption{The SGS data processing levels, indicating data products and responsibilities}
\label{euclid:SN1}
\end{figure}
In terms of internal (to the EC) data release, it is planned that Level 1 would be daily available, Level Q TBC hours after L1, while fully calibrated data (L2) would be ready within 3 days after their observation and transfer to the archive. The L2 processing will have not to wait the completion of L1 production. It begins immediately after L1 metadata start to be registered in the EMA.

In the following, we report the definitions of data processing levels for completeness (cf. EC SGS Science Implementation Plan, issue 2, November 2011, EUCL-OTS-SGS-PL-00003), in which we have highlighted (bold typing) the interesting parts for the SN pipelines:

\begin{itemize}
\item	Level E (LE): external data (images, catalogues, all relevant calibration and meta-data, observational data in science-usable format) derived from other missions and/or external survey projects, reformatted to be ``euclidised"\footnote{\textbf{Euclidisation}: Function of the Euclid pipeline processing that resample all Euclid data (internal and external) in a common reference system: pixel ($0.1\prime\prime/pixel$), sky (astrometry catalogue, NIR  catalogue), and physical units. Euclidisation is a two-step process: step-1 delivers External images and External calibration data and meta-data with the same meaning as Euclid data; step-2 provides the homogenization and merging of all Euclid data into a single Euclid reference system, based on the Euclid VIS imaging instrument} , i.e. handled homogeneously with Euclid data. These data are required to allow the EC to provide its final data products at the expected level of accuracy. The official strategy within SGS is not yet defined.

\item	Level S (LS): pre-mission simulated data (catalogues, satellite and mission modeling data, etc.) used before and during the mission, mainly for calibration and modeling purposes. This level will be implemented in parallel with respect to levels 1, 2 and 3 and provide inputs to their development. The data for this processing level are prepared before the mission (and refined/updated during the in-flight commissioning and initial calibration phase) and are used as appropriate, before and during the mission. The data are delivered by the EC.
\item	Level 1  (L1): is composed of three separate sub processing functions: telemetry checking and handling, including real-time assessment (RTA) on housekeeping; telemetry unpacking and decompression (edited telemetry); quick-look analysis (QLA) on science telemetry and production of daily reports. The input data for this processing level come from the satellite via MOC and are used to perform quality control. The data are delivered by the SOC. The EMA L1 contains:
\begin{itemize}
\item	Raw VIS and NISP images;
\item	Processed housekeeping telemetry and associated ancillary information such as pointing history files;
\end{itemize}
\item	Level Q  (LQ): Quick-Release data. Basic removal of instruments signatures and production of calibrated images for rapid distribution to the scientific community. The Euclid Science Team will define the exact content of LQ. Level Q is a subset of Level 2, integrated and run at SOC. The EMA LQ contains:
\begin{itemize}
\item	Products defined so that they are suitable for most purposes in Astronomy, except for the main cosmological goals of the mission.
\end{itemize}
\item	Level 2 (L2): instrumental data processing, including the calibration of the data as well as the refined removal of instrumental features in the data; trend analysis on instruments performance and production of weekly reports. The data processing at this level is under the responsibility of the SDCs in charge of the instruments monitoring. The EMA L2 contains:
\begin{itemize}
\item	Calibrated and co-added images from VIS and NISP – validated for cosmology analysis
\item	PSF model and optical distortion maps
\item	Co-added spectra
\item	Transients: Transient events data products: include derived transient category (e.g. supernova candidates , solar system object, etc) and brightness, target position and possible finding chart.
\end{itemize}
\item	Level 3 (L3): data processing pipelines for the production of science-ready data. The Level 3 data are also produced by SDCs. The EMA L3 contains:
\begin{itemize}
\item	Catalogues (including redshift, ellipticity, shear, etc)
\item	Dark matter mass distribution
\item	Shear and galaxy correlation functions and covariance errors
\item	Additional science catalogues
\item	Ground based information which was used in the derivation of the data products
\end{itemize}
\end{itemize}
\subsection{SN Pipeline description}
By taking into account the above specifications, the diagrams shown in figure \ref{euclid:SN2} and \ref{euclid:SN3} the double SN pipeline strategy workflows.

\begin{figure}
\centering
\includegraphics[width=12cm]{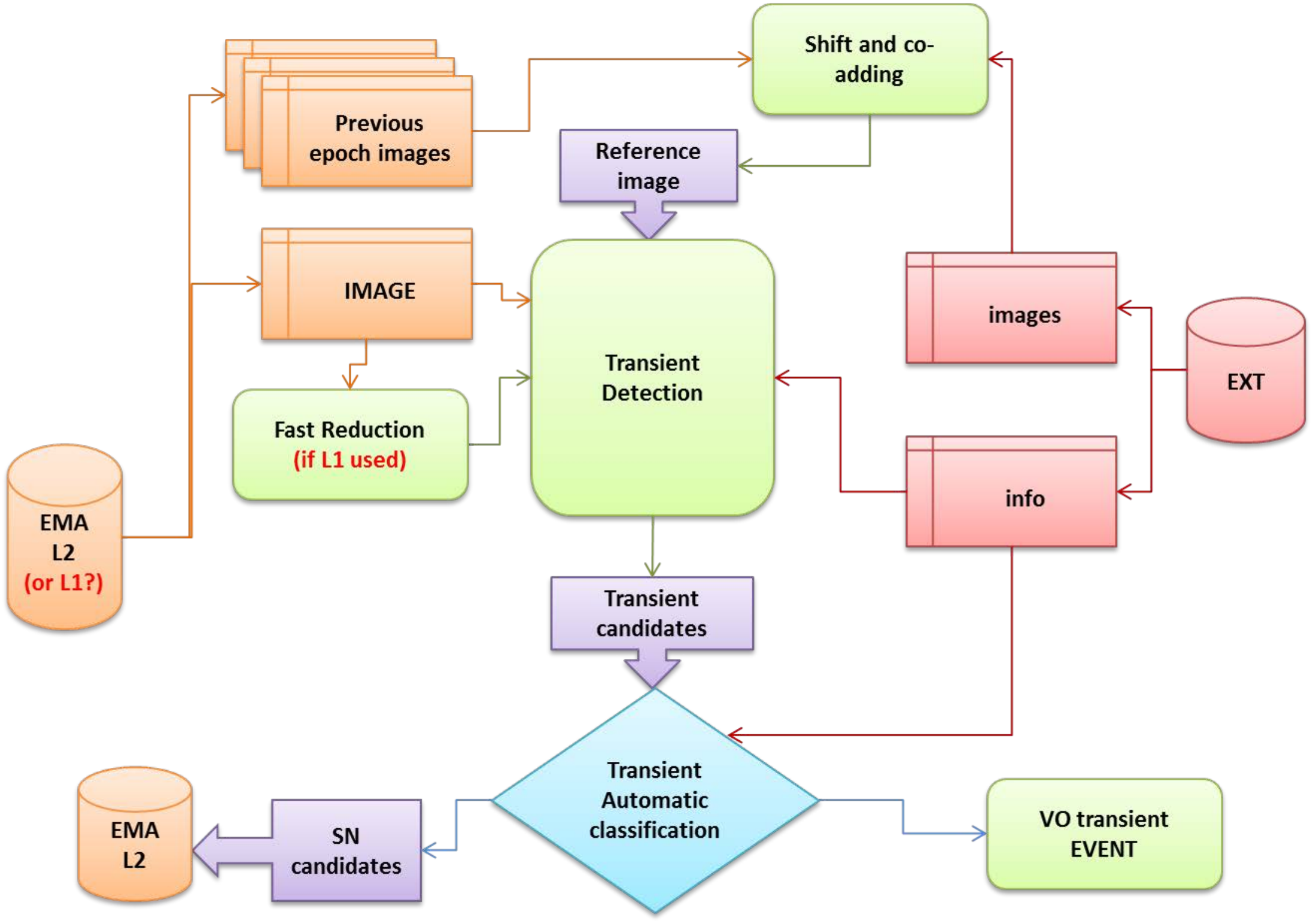}
\caption{The SN Trigger Pipeline (STP) workflow.}
\label{euclid:SN2}
\end{figure}

\begin{figure}
\centering
\includegraphics[width=12cm]{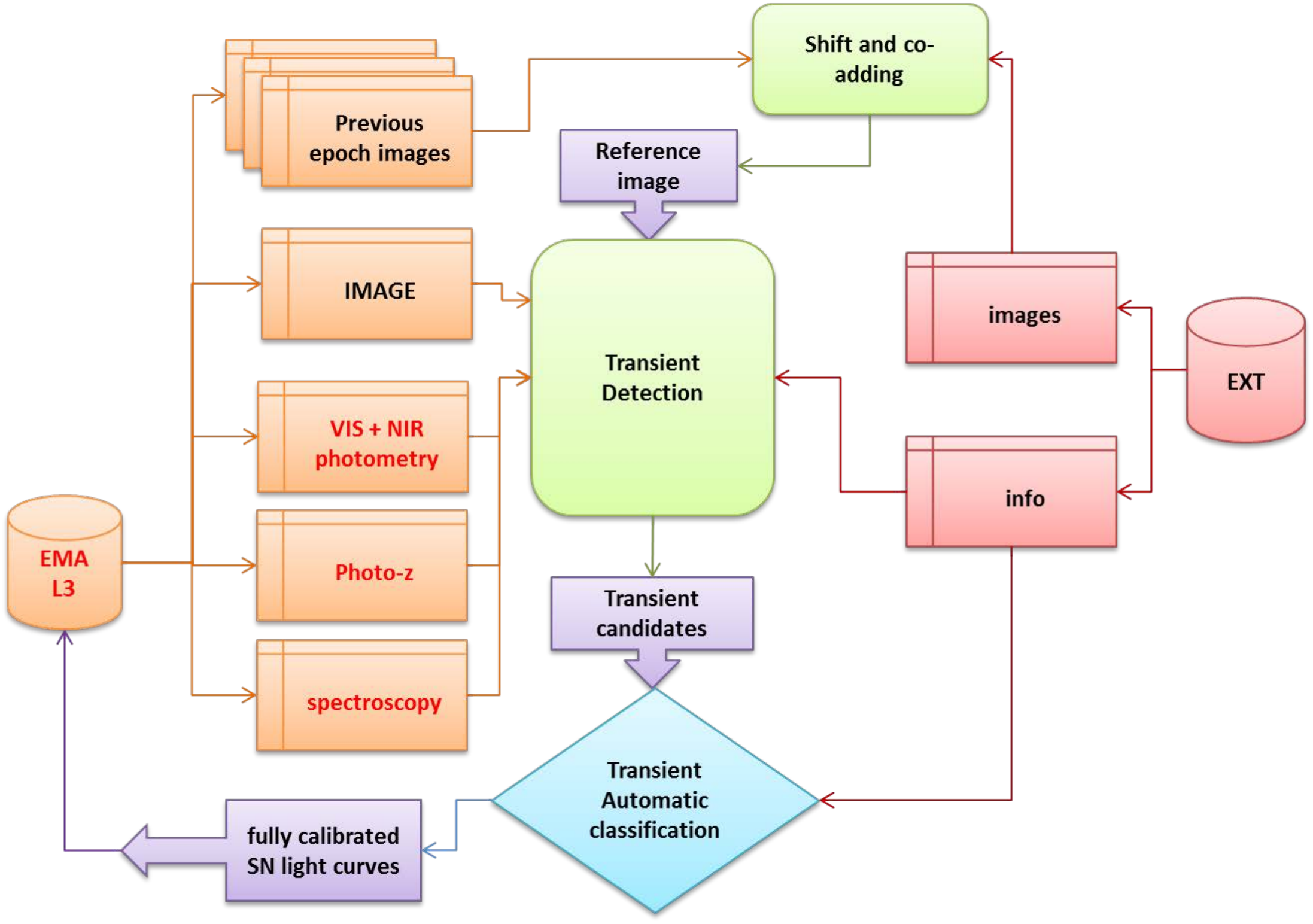}
\caption[The SN Final Pipeline (SFP) workflow.]{The SN Final Pipeline (SFP) workflow. The blocks with red text inside are those different from the Trigger Pipeline and require a final decision. In fact, this diagram implicitly intends to use L3 products, as outcoming from official SGS pipelines, as input for transient detection. These blocks may be replaced by a dedicated reduction pipeline. This issue must be discussed and decided at the requirement definition stage.}
\label{euclid:SN3}
\end{figure}

As shown in the figure \ref{euclid:SN2} and \ref{euclid:SN3}, the SFP processing flow would start from the final Euclid data release (EMA L3), having the possibility to use already reduced images, as well photometry and spectroscopy with high accuracy, coming from the EC main pipeline, for each epoch and also for euclidised external data. However, in case of frequent re-processing of data reduction related to candidates, the SFP would also require a dedicated reduction process to be implemented outside the official SGS pipelines, of course by re-using as much as possible their software modules. Hopefully the dedicated part could only be from the image subtraction step onwards.

The green block called ``Transient Detection" is referred to the different strategy adopted to extract transient candidates, which is mainly affected also by the Euclid observing strategy and by the data type available. It is possible to guess different types of detection, such as:
\begin{itemize}
\item	A homogeneous detection: a simple comparison between single-band reference and L1/LQ image. In this case the detection block should include traditional statistical methods (such as chi-square, threshold etc.);
\item	A heterogeneous detection: specific methods based on multi-band comparison between external or L1/LQ and reference images.
\end{itemize}
Moreover in this transient detection block all available expertise is expected to be re-used and optimized for Euclid needs. This block could be designed as a modular workflow, able to perform multiple types of transient analysis in parallel. It could also be organized to host different SN extraction methods, all converging in a final pattern matching system.

By considering the above schemes, main role of the SFP is basically consisting in a fine tuning of the STP and in a re-processing of the reduction steps, using more accurate data (L3 instead of L1/L2) or requiring a more accurate calibration of raw data.


    \chapter{Conclusions?}\label{chap:conclusionstech}
        \hfill\begin{tabular}{@{}p{.5\linewidth}@{}}
\textit{``It is the end that crowns us, not the fight."}\\
Robert Herrick.\\ \phantom{aaa}
\end{tabular}

May be there was no need for a concluding chapter since the  various parts of my work were already summarized at the end of the respective sections.
But I feel that a thesis of almost 250 pages cannot close
without a final chapter and therefore I will try to explain why there are no conclusions, but just a series of
starting points.

As already mentioned before, my thesis spans a quite variegate spectrum of topics:
from algorithms to information and communication technologies (ICT),
to observational astronomy and cosmology;
the main drivers being the interest in cosmology and the fascination to be able to cope with the methodological revolution that is currently
taking place in astronomy.

While a few years ago in astronomy there were not enough data, nowadays new technologies have solved the problem by creating the opposite one: a data tsunami of unprecedented proportions which requires an extreme use of ICT not only to produce and store the data but also to understand them.
The real scope of my thesis, therefore was not so much {\it ``to produce the best photometric redshift available in literature,
nor to obtain a great classification for globular cluster"} but rather to contribute via specific science cases and publications to the growth and to the social understanding and acceptance of a new paradigm which I like to call \textbf{Astronomy 2.0};
a paradigm which finds its implementation in the new discipline of Astroinformatics.

While the romantic figure of a lonely astronomer that takes and reduces his own data is being buried,
 contemporary astronomy requires multidisciplinary groups where astroinformaticians
are needed to communicate with and to integrate  all different souls of an heterogeneous group
of expertises: data base experts, data analysts, data miners, theoreticians etc.

From the technological point of view, what I learned is that on the one end it is crucial to continuously  monitor what up-to-date technology is offering to us, and that, on the other side, the only way to be
truly effective and innovative is  to have the courage and the stamina to abandon the old path and start a
new adventure every time a new technology is appealing enough. This, for instance, is the reason why at the end of my thesis, we decided to discontinue DAMEWARE and
to start a new project: in the course of my Ph.D. the technologies adopted at the beginning became obsolete and in order to exploit what is now available, to stick patches on an old software would not be enough. \\

From the practical point of view we still have a deep problem in terms of Knowledge Bases, since
the catalogues that are available at the moment are still suffering from the fact that
they were produced either during the ``dark ages" of Astronomy 1.0 or during the transition period when the
requirements of Astronomy 2.0 had not yet been well defined.
In other words most catalogues  available nowadays are still  inhomogeneous and built in a haphazard fashion  without a physical ontology behind.
I am sure, however, that this problem will be solved in a few years due to a kind of \textit{snowball effect}: each new catalogue that will be produced will allow the production of some better catalogues each of which will allow... in a virtuous circle.

At that point it will become possible to approach a redefinition of the concept of astronomical classification and to obtain a \textit{physical driven} classification rather than \textit{eyeball driven} classifications which are both not effective and too much subjective, as it was shown, for instance, in a famous paper by \cite{naim1995}. \\

For what photometric redshifts are concerned,  besides obtaining what we believe are excellent results for
both quasars and galaxies, I was intrigued by the fact that the new methodology outlined here allows
accurate feature selection which can be used both to optimize computations and to optimize future
surveys or even large queries on massive data sets. These ``by-products" were for me a clear indication of how astroinformatics will shape the new future of Astronomy 2.0. These findings will perhaps prove crucial for
Euclid which will be among the first surveys where the data quality aspects
will have a relevance with no precedents.\\

In closing my work I should be able to give some answers, but I found more questions than answers and
my supervisors told me that this is an healthy attitude:

\begin{itemize}
\item Are we truly at the point-break from astronomy 1.0 to 2.0? and if so what is still missing?
\item Will the community have the courage to leave behind decades of \textit{institutionalised working
practice} and try something else?
\item Are astronomers ready to share each step of their own workflows with the community,
thus allowing not only data interoperability but also the much more challenging ``knowledge
interoperability"?
\item Are we going towards real programming and communication standards?
\item and so on...
\end{itemize}

The only real answer I found is that it is not possible to deal with the new amount of data that new survey are producing as it has been done till now and that new surveys would be just a huge waste of time and money if astronomers do not change their \textit{mental habits}:
a new age of discovery is at the door but someone has to open that door.

I said that this is the only real and long-term answer because all the other answers I found
are not definitive (and it would have been surprising otherwise...). I  mean, for instance,
that GPUs are a great idea at the moment but I am quite sure that tomorrow (even before)
there will be a new and better idea to pursue!      \\

For what future work is concerned I can just say that not only I do not want to abandon
the scientific problems which I have tackled (such as the redshifts, trying to produce accurate photometric redshift for new surveys) so far, but I also want new problems, new algorithms, new technologies and...  paraphrasing a famous Rudyard Kipling poem:\\

\noindent{\it Making each day one heap of all mine winnings}\\
\noindent{\it And risk it on one turn of pitch-and-toss,}\\
\noindent{\it And if I lose, start again at the beginnings}\\
\noindent{\it And never breathe a word about this loss.}\\

That is my plan for the future.

    \chapter*{Acknowledgements}\addcontentsline{toc}{chapter}{Acknowledgements}  \markboth{\MakeUppercase{Acknowledgements}}{}
        \hfill\begin{tabular}{@{}p{.45\linewidth}@{}}
\textit{``Good company in a journey makes the way to seem the shorter."}\\ Izaak Walton.\\ \phantom{aaa}
\end{tabular}

It would not have been possible to obtain this doctoral thesis without the help
and support of the kind people around me, to only some of whom it is possible to
give particular mention here.\\

First and foremost I want to thank my advisors Giuseppe (Peppe, the Boss)
Longo and Massimo (Max) Brescia. It has been an honour to be their Ph.D. student.\\

Peppe has taught me, both consciously and unconsciously, how good experimental
physics is done. I appreciated all his contributions in terms of time, ideas, and fundings
to make my Ph.D. experience productive and stimulating. The joy and enthusiasm
he has for his research was contagious and motivational for me, even during tough
times in the Ph.D. pursuit.\\

The good advice, support and friendship of Max has been invaluable on both an
academic and a personal level, for which I am extremely grateful. I am also thankful
for the excellent example he has provided in how to approach the work. There are not enough words
to explain how much his help has been fundamental during this thesis, and his involvement went well beyond what I could ever expect.\\

The experience of Amata and Maurizio also were precious during this years: and I wish to thank
both of you.\\

The members of the DAME group have contributed immensely along my personal
and professional timeline. The group has been a source of friendships as well as good
advice and collaboration. I am especially grateful for all the pleasant moments to the DAME group members: Alessandro, Alfonso, Bojan, Civita,
Ettore, Francesco, Giuseppe, Luca, Marianna, Marisa, Michelangelo, Pamela, Sabrina,
Sandro. In particular I want to thank for his patience Mauro who has endured
to talk about work also in the evenings when we, officially speaking, went out to have fun.\\
I must also thank all the people in the lab who have endured my bad mood while
writing and rewriting this thesis without groping to kill me, or at least without me
noticing of their trial.\\


Last but not least George Djorgovsky whose seminal work in the field of Astroinformatics has triggered many of my interests;
then Ciro and Raffaele for their support, advices, patience and least but not last their friendship during this years.\\

The Boss made me know and work with all these people, thanks again.\\

\begin{flushright}
Stefano Cavuoti\\
March 2013
\end{flushright}

    \listoffigures \addcontentsline{toc}{chapter}{List of figures}
    \listoftables \addcontentsline{toc}{chapter}{List of tables}
    \appendix
        \chapter{Setup of Globular Clusters Experiments}
\label{gc:AppII}

\noindent In the following sections the feature are referred to the cardinal number (feature 1: MAG\_ISO, etc).
For each model we choose the configuration parameters in order to perform the best results.\\

\section[MLP-BP]{Multi Layer Perceptron trained by Back Propagation}
\begin{itemize}
\item Input Nodes (equivalent to the number of features considered in the dataset patterns)		
max number: 11	(complete patterns);	min number: 4 (pruning on optical features);nominal number: 7 (complete optical dataset);	

\item Hidden Nodes (depending on the number of features considered in the dataset patterns).
Max number: 23	(with input nodes in [8, 11]);	min number: 15 (with input nodes in [4, 7]);	

\item Output Nodes (based on crispy classification): 2 (1 0 GC, 0 1 not GC);
	
\item Activation Functions (neuron function type, used to provide its output, by processing inputs).
input layer: (no input processing, just propagate it);		
hidden layer: nonlinear hyperbolic tangent of input;		
output layer: linear with softmax normalization  (outputs  sums  up  to  1.0  and  converge to posterior probabilities);		

\item Learning Rule Parameters. Output Error Type: Cross Entropy;	Training Mode: Batch (weights update after each whole dataset patterns calculation);	Training Rule: Back Propagation with Conjugated Descent Gradient;	Error Loop Threshold: 0.001 (one of the stopping criteria);	Number of Iterations: 10000 (one of the stopping criteria);
\end{itemize}
			
\section[SVM]{Support Vector Machines}
\begin{itemize}
\item Model: C-Support Vector Classification (C-SVC); Kernel: Radial Basis Function;

\item Gamma (for each experiment we have a multiplicative step). Min number: $2^{-15}$; max number: $2^{23}$;
step: $4$(multiplicative). $C$ (for each experiment we have a multiplicative step). Min number: $2^{-5}$;
max number: $2^{15}$; step: $4$ (multiplicative);

\item Error Tolerance: 0,001;

\item Cache: 100MB;

\item Shrinking: On;

\item Probability Estimates: Off;

\item Cross Validation: k-fold (k = 5);

\item Weights: 1;
\end{itemize}

\section[GAME]{Genetic Algorithm Model Experiment}
\begin{itemize}
\item Model: Genetic algorithm with fitness based on trigonometric polynomial expansion;

\item Topology: population of chromosomes, each of them composed by genes;	

\item Input features (depending on the number of features considered in the dataset patterns).
Max number: 11 (complete dataset); min number: 4 (pruning on optical features); nominal number: 7 (complete optical dataset);

\item Genetic Population Size (depending on the number of features and polynomial order).
Max number: 67 (with 11 features); min number: 25 (with 4 features);

\item population size: $(polynomial_{order} * num_{features}) + 1$;

\item Genetic Chromosome Size (depending on the polynomial order). Number: 13 (with polynomial order = 6);
chromosome size: $(2 \times polynomial_{order}) + 1$;

\item Output (based on crispy classification). Number in KB: 1 (0 if no GC; 1 else);

\item Output Error Type: TMSE (Thresholded Mean Square Error) with threshold 0,4;

\item Error Loop Threshold: 0,001 (one of the stopping criteria);

\item Polynomial Order: 6;

\item Tournament Selection (based on the Wheel Roulette, max probability on the entire population fitness).
Number of Tournament Chromosomes: 2;

\item Genetic Operators. Crossover Probability: 0,9; Mutation Probability: 0,2; Elitism Factor: 2;

\item Initial Population Distribution: gaussian standard, with all values generated into range [-1, +1];

\item Number of Iterations: 10000 (one of the stopping criteria);
\end{itemize}

\section[MLPQNA]{Multi Layer Perceptron trained by Quasi Newton}
\begin{itemize}
\item Input Nodes	(depending on the number of features considered in the dataset patterns). Max number: 11 (complete dataset);
min number: 4 (pruning on optical features);
nominal number: 7 (complete optical dataset);

\item Hidden Nodes (depending on the number of features considered in the dataset patterns). Max number: 23 (with input nodes in
[8, 11]);	min number: 15 (with input nodes in [4, 7]	);
\item Output Nodes (based on crispy classification): number in KB: 1 (0 if no GC; 1 else);
\item Activation Functions (neuron function type used to provide its output, by processing inputs). Input layer: no input processing, just propagate it; hidden layer: not linear hyperbolic tangent of input; output layer: linear with softmax normalization  (outputs  sums  up  to  1.0  and  converge to posterior probabilities).
\end{itemize}

\noindent Learning Rule Parameters
\begin{itemize}
\item Output Error Type: Cross Entropy;
\item Training Mode: Batch (weights update after each whole dataset patterns calculation);
\item	Training Rule: Quasi Newton (inverse hessian approximation by error function gradients);
\item QNA Implementation Rule: based on L-BCFG method (L is for Limited memory);
\item QNA Parameters. Decay: 0,001 (weight decay during gradient approximation);
Restarts: 20 (random restarts for each Approximation Step);
Wstep: 0,01 (stopping threshold, min error for each Step);
MaxIts: 1500 (max number of  Iterations for each Approx. Step);
\end{itemize}

\section[MLPGA]{Multi Layer Perceptron trained by Genetic Algorithms}
\begin{itemize}
\item Input Nodes (depending on the number of features considered in the dataset patterns).
Max number: 11 (complete dataset);	
min number: 4 (pruning on optical features	);
nominal number: 7 (complete optical dataset);
\item Hidden Nodes (depending on the number of features considered in the dataset patterns)
max number: 23 (with input nodes in [8, 11]);	
min number: 15 (with input nodes in [4, 7]);	
\item Output Nodes (based on crispy classification).
number in KB: 1 (0 if no GC; 1 else);	
\item Activation Functions (neuron function type used to provide its output, by processing inputs).
Input layer: no input processing, just propagate it;		
hidden layer: nonlinear hyperbolic tangent of input;		
output layer: nonlinear hyperbolic tangent of input;		
\item Learning Rule Parameters.
Output Error Type: MSE;			
Training Mode: Batch (weights update after each whole dataset patterns calculation);	
Training Rule: Genetic Algorithm with Roulette Wheel selection function and fitness based on the MSE between target and output of dataset patterns;
\item MLPGA Parameters.		
Genetic Population Size: 25;
Genetic Chromosome Size: 13;
Error Loop Threshold: 0,001;
Tournament Selection: based on the Wheel Roulette method (max probability on the entire population fitness);	
Number of Tournament Chromosomes: 2;
Crossover Probability: 0,9;
Mutation Probability: 0,2;
Elitism Factor: 2;
Initial Population Distribution: gaussian standard, with all values generated into range [-1, +1];
Number of Iterations: 10000 (one of the stopping criteria).
\end{itemize}

        \newpage\phantom{ciao}
\end{document}